\documentclass [11pt, proquest] {uwthesis}[2020/02/24]
 
\usepackage{graphicx}
\usepackage{dcolumn}
\usepackage{bm}

\usepackage{braket}
\usepackage{physics}
\usepackage{float}
\usepackage{graphicx}
\usepackage{placeins}
\usepackage{amsmath,amssymb}
\usepackage{amsfonts}
\usepackage{xcolor}
\usepackage{appendix}

\usepackage{wrapfig}
\usepackage{dsfont}
\usepackage{tabu}
\usepackage{tabularx}
\usepackage{bbold}
\usepackage{bm}
\usepackage{tikz}
\usepackage{qcircuit}
\usepackage{textgreek}
\usepackage{multirow}
\usepackage{makecell}
\usepackage{comment}
\usepackage{enumitem}
\usepackage{soul}

\usepackage{alltt}  %

\makeatletter
\newcommand{\fmarki}{*}
\newcommand{\fmarkii}{\ensuremath{\dagger}}
\newcommand{\fmarkiii}{\ensuremath{\ddagger}}
\newcommand{\fmarkiv}{\ensuremath{\mathsection}}
\newcommand{\fmarkv}{\ensuremath{\mathparagraph}}
\newcommand{\fmarkvi}{\ensuremath{\|}}
\newcommand{\nub}{\overline{\nu}}
\newcommand{\eb}{\overline{e}}
\newcommand{\ub}{\overline{u}}
\newcommand{\db}{\overline{d}}

\def\@fnsymbol#1{{\ifcase#1\or \fmarki\or \fmarkii\or \fmarkiii\or \fmarkiv\or \fmarkv\or \fmarkvi \else\@ctrerr\fi}}
\makeatother

\renewcommand{\fmarkvi}{\$}

\usepackage[hypertexnames=false]{hyperref}
\hypersetup{
    colorlinks=true,       
    linkcolor=blue,          
    citecolor=blue,        
    filecolor=blue,      
    urlcolor=blue           
}

\def\pslash{{ p\hskip-0.5em /}}

\newcolumntype{Y}{>{\centering\arraybackslash}X}

\makeatletter
\pretocmd\frontmatter@thefootnote{\color{black}}{}{}
\makeatother

\begin{document}
 
%
%

\prelimpages
 
%
%
\Title{Developing techniques for Simulation of SU(3) Quantum Field Theories on State-of-the-Art Quantum Devices }
\Author{Ivan Alexandrovich Chernyshev}
\Year{2025}
\Program{Physics}

\Chair{Martin J. Savage}{}{Physics}
\Signature{Stephen R. Sharpe}
\Signature{Silas R. Beane}

\copyrightpage

\titlepage

%
%

%
%

\setcounter{page}{-1}
\abstract{%
Quantum computing has long been an experimental technology with the potential to simulate, at scale, phenomena which on classical devices would be too expensive to simulate at any but the smallest scales. Over the last several years, however, it has entered the NISQ era, where the number of qubits are sufficient for quantum advantage but substantial noise on hardware stands in the way of this achievement. This thesis details NISQ device-centered improvements to techniques of quantum simulation of the out-of-equilbrium real-time dynamics of lattice quantum chromodynamics (LQCD) and of dense 3-flavor neutrino systems on digital quantum devices. 

The first project concerning LQCD is a comparison of methods for implementing the variational quantum eigensolver (VQE) that initializes the ground state of an SU(3) plaquette-chain. The thesis then pivots to a 1+1D lattice of quarks interacting with an SU(3) gauge-field. A VQE-based state-preparation for the vacua and a Trotterized time-evolution circuit is designed and applied to the problems of simulating beta and neutrinoless double beta decay.  Finally, these circuits are adapted to a version useable on quantum devices with nearest-neighbor connectivity with minimal overhead, with an eye towards utilizing the higher qubit count of such devices for hadron dynamics and scattering. 

This thesis covers two projects that concern dense 3-flavor neutrino systems. The first details design and testing of Trotterized time-evolution circuits on state-of-the-art quantum devices. The second, motivated by the Gottesman-Knill theorem's result that deviation from stabilizer states ("magic") is necessary for a problem to exhibit quantum advantage, details results with implications for the Standard Model in general that the 3 flavor ultradense neutrino systems with the highest, most-persistent magic are those that start with neutrinos in all 3 flavors.
}
 
%
%
\tableofcontents
\listoffigures
\listoftables  

%
%
\acknowledgments{

  Science is not a solitary pursuit. To quote many a syllabus of the courses I took as an undergraduate, “no one knows it all, but together you know lots.” To me, this extends beyond advice merely to work together on problem-sets to a broader principle that none of us could have gotten to where we are now without the collaboration, exchange of ideas, and mentorship we got from our colleagues. I would like to acknowledge the people who have given me these things in my own professional and personal life.

First, I would like to thank my advisor, Professor Martin Savage. He has done everything from taking me on as an RA in his group, the InQubator for Quantum Simulation (IQuS), during my first year and training me on what is expected of me in papers, project-work, and all of the other things that I need to do as a researcher in general and a Ph.D. student in particular to organizing all of the IQuS workshops and seminars which exposed me to the perspectives and work of a wide variety of people in the field and helping me pick projects that I found to be of interest. Through all this, he has helped me grow from someone who has a baseline background knowledge of physics but whose day-job is entry-level coding to an independent scientist ready to equipped to decide and follow through on the direction of projects.

I would also like to thank my groupmates in the InQubator for Quantum Simulation (IQuS) for the discussions and collaborations we have had. In particular, I would like to thank Anthony Ciavarella, Roland Farrell, and Francesco Turro for letting me join their projects and teaching me much about how research in our field works through said projects. I would also like to thank Marc Illa and Dorota Grabowska in particular for being there if I had any questions. 

Additionally, I would like to thank Katie Hennessey for doing the unseen work which made my trips to conferences and all of IQuS’s seminars and workshops.

I am grateful to everyone on my doctoral supervisory committee and my reading committee for taking time out of their busy schedules to attend to my career milestones and give the occasional advice. I would like to extend special thanks to Professor Emily Levesque of the Astronomy department for stepping up as my GSR after my original GSR’s schedule turned out to be too busy.

Lastly, I would like to thank Catherine Provost for her help with all of the logistical matters related to my graduate program on the end of class-registration, degree milestones, and matters related to TA and RA positions.

I would like to acknowledge the QuantumX initiative and the AQET program at UW and their leadership, particularly I-Tung Chen and Professor Kai-Mei Fu. Their program has exposed me to many opportunities to connect with other quantum information science and engineering specialists throughout academia, national laboratories, and industry and learn about their perspectives and the problems that they attempt to solve.

Additionally, Los Alamos National Laboratory’s Quantum Computing Summer School (QCSS), which I attended during the summer of 2023, has made a profound impact on my career, resulting in my first sole-author paper and setting the stage for me to continue at Los Alamos as a postdoc this coming winter. It has also broadened my horizons by introducing me to new topics within my field, particularly adiabatic quantum computing and quantum magic. I would like to thank Lukasz Cincio, Marco Cerezo, and Yigit Subasi for handling the logistics of the summer school. I would like to thank Joseph Carlson and others in his group, particularly Ionel Stetcu and Ronen Weiss, for their mentorship at QCSS and their support in my ongoing process of joining their group as a postdoc. Additionally, I would like to thank Carleton Coffrin and Zachary Morell for helping me with the quantum device I used in my summer school project. Lastly, I am grateful for the people whom I met from all over the world at this summer school. I enjoyed hearing a variety of diverse perspectives from them, both on matters of quantum computing and on life outside of work, as well as exploring New Mexico and Colorado together. I’m especially thankful to my roommates Manuel Rudolph and Samuel Tan, as well as to Joel Rajakumar for letting us stay at his flat when we had to leave our inn for a day.

In addition, I have my old colleagues and mentors from UC Berkeley to thank for getting me to the point where I can start research a quarter after beginning graduate school.
I am particularly grateful to Dr. Andr{\'e} Walker-Loud from LBNL for introducing me to the field of lattice gauge theory, which my work is still adjacent to, as well as mentoring me through a senior thesis project within that field despite delays due to me having to switch groups and the COVID-19 pandemic. I would also like to thank Professor Barbara Jacak, my first research advisor, and everyone in her group, especially Dr. Miguel Arratia, for introducing me to my first scientific projects and how to apply skills I learned in my classes to real-world scientific research, as well as teaching me how to read scientific literature. Outside of research, I would like to thank UC Berkeley SPS and all of my classmates from undergraduate, especially Nicholas Rui for working together on tough problem sets and for being good friends who have helped me to maintain a semblance of life outside of work.

Speaking of friends, my cohort-mates at UW have proven to be great friends who have made starting graduate school in the middle of a pandemic a little bit more bearable and who have made my life outside of work post-pandemic a lot more interesting. I would like to thank Chong Xia for teaching me how to cook beyond the basics, Jimmy Sinnis and Kent Wilson for our interesting philosophy discussions, Obinna Ukogu and Eli Lileskov for bonding over chess, movies, card games, and all the other things we did together, and Nikita Zemlevskiy and Murali Saravanan for teaching me how to ski last winter.

I would also like to thank my friends outside of grad school. In particular, I would like to thank Ben Simila, Parker Martin, Alex dePolo, and everyone else in their friend group for our games of poker and Magic: the Gathering and for bringing me to their live music events, and everyone from the Pacific Northwest Jugger clubs for giving me people to play a very niche sport that I’ve been interested in since undergrad while I was in Seattle.

I would also like to acknowledge my union, UAW 4121, for both big things, such as pushing for a collective bargaining agreement with reasonable working conditions and wages and coverage of out-of-network therapy at 80\% (which I personally ended up needing), to small things such as advice on navigating the healthcare system.

I would also like to thank Karl Reichert, my therapist, who has taught me several techniques on relieving stress and improving my executive function, and has given me many insights which have helped me understand myself. 

I would also like to thank my immediate for always being there for me to go to. In particular, I would like to thank my father for inspiring me to enter a career in research in the first place and for giving me many pieces of much-needed advice which have helped me navigate the world of academia.

Finally, I would like to acknowledge the financial support for the research enclosed herein from the U.S. Department of Energy, Office of Science, Office of Nuclear Physics, InQubator for Quantum Simulation (IQuS) under Award Number DOE (NP) Award DE-SC0020970 via the program on Quantum Horizons: QIS Research and Innovation for Nuclear Science, and from the Department of Physics and the College of Arts and Sciences at the University of Washington.

}

%
%
\dedication{To my mother, father, and brother}

%

%
%

\textpages
\chapter{Introduction}
\label{chap:intro}

\section{Classical simulation of QCD}
Quantum Chromodynamics (QCD), the quantum field theory behind interactions between quarks and gluons, underlies many unsolved problems in nuclear and particle physics, such as the nucleon spin problem \cite{1988364, ASHMAN19891, LIU201499, PhysRevLett.118.102001, PhysRevD.93.074005} and the nucleon mass problem \cite{PhysRevLett.74.1071}.
Thanks to asymptotic freedom \cite{gross:1973id}, QCD can be treated perturbatively at high energies. Thus, perturbative QCD (pQCD) calculations have been instrumental in high-energy physics.  However, QCD is strongly coupled in the low-energy limit. This precludes many interesting phenomena, such as quark confinement \cite{physrevd.10.2445, jaffe1976unconventional}, dynamical chiral symmetry breaking \cite{pagels1979dynamical, weinberg1979phenomenological}, and hadron formation \cite{fodor2012light, metz2016parton}, from a perturbative treatment. Since the mid-1970s, the field has addressed this problem through numerical simulations of-nonperturbative QCD. A recent literature review of QCD, both perturbative and nonperturbative, can be found in Ref. \cite{gross202350}. Reviews of non-perturbative quantum field theory in general can be found in Ref. \cite{Frishman:2010tc,Frishman2014book}.

 First, in order to conduct numerical simulations, QCD is discretized onto a lattice. Besides serving as a method of translation into computer language, the lattice formalism also serves as a regularization scheme - the volume of the lattice would control infrared divergences while the spacing between points on the lattice (``lattice spacing") would control ultraviolet divergences. The downside is that the physical continuum, infinite volume theory is out of reach and would need to be extrapolated to. 

\subsection{Monte Carlo Lattice QCD}
 
The leading method of lattice QCD works by first quantizing the theory on a discrete Euclidean spacetime \cite{physrevd.10.2445}. In a Euclidean spacetime, time (labeled $\tau$) is treated as a spatial dimension, so its metric is as follows:
\begin{equation}
    ds^2 = d\tau^2 + \sum_{i \geq 1} dx_i dx^i 
    \label{eq:intro:euclideanspacetimemetric}
\end{equation}

\noindent with $dx_i$ being the canonical spatial dimensions.  Thus, in order to map the physical Minkowski spacetime (with a time labeled $t$), whose metric is as follows:
\begin{equation}
    ds^2 = -dt^2 + \sum_{i \geq 1} dx_i dx^i ,
    \label{eq:intro:minkowskispacetimemetric}
\end{equation}
\noindent to Euclidean spacetime, a Wick rotation, which is the mapping of Minkowski time $t$ to Euclidean time $\tau$ using the relation $t \rightarrow i \tau$ \cite{wick1954properties}, is used. The choice of Euclidean spacetime means that the Feynman path integral \cite{feynman1948space},
\begin{equation}
    \psi (y, t) = \frac{1}{Z} \int_{ {\bf x} (0) = x }^{ {\bf x} (t) = y } \mathcal{D} {\bf x} e^{i S({\bf x})} \psi (x, 0)
    \label{eq:minkowskifeynmanpathintegral}
\end{equation}
\noindent for a set of paths ${\bf x}$ from space-time coordinates (x, 0) to space-time coordinates (y, t), with $S({\bf x})$ being the action of a path and $\psi$ an amplitude at a given space-time coordinate, would be mapped to \cite{physrevd.10.2445}
\begin{equation}
    \psi (y, t) = \frac{1}{Z} \int_{ {\bf x} (0) = x }^{ {\bf x} (t) = y } \mathcal{D} {\bf x} e^{- S_{Euclidean}({\bf x})} \psi (x, 0)
    \label{eq:euclideanfeynmanpathintegral}
\end{equation}

\noindent where $S_{Euclidean}$ is the action of the theory quantized in the Euclidean spacetime. $Z$ is a normalization factor in both equations. The path integral in Eqn. \ref{eq:euclideanfeynmanpathintegral} can then be calculated using Monte Carlo integration\cite{physrevd.21.2308}. Today, this lattice QCD (LQCD) procedure is run on high-performance computing facilities and is used extensively for a wide variety of applications within nuclear and particle physics, such as as finding the nucleon isovector charges and form factors which are necessary in particle physics experiments to estimate weak interaction matrix elements for purposes including, but not limited to, nucleon-pion interactions, neutrino-nucleus interactions, values of parameters that are expected by various Beyond-the-Standard-Model theory candidates  \cite{park2022precision, PhysRevD.109.094505, universe10030135, PhysRevD.109.034503, PhysRevD.109.014503, petti2024nucleon, smail2023constraining, barca2023toward, aoki2023b, physrevd.96.054505,PhysRevLett.115.132001,PhysRevLett.112.112001,PhysRevD.99.114509,gamiz2014,PhysRevD.98.074512,PhysRevD.90.074509,2018axial,PhysRevD.94.054508}, and the radii of protons and neutrons\cite{djukanovic2024precision}. Other physics problems that Monte Carlo LQCD has been applied to includes bounds on the QCD equation of state \cite{fujimoto2024bounds},  hadronic spectra \cite{doi:10.1126/science.1257050,PhysRevLett.105.252002,2003resonance}, and multi-hadron interactions \cite{PhysRevLett.119.062002,PhysRevD.77.094507,BEANE200762,PhysRevLett.124.032001}. For recent reviews, see Ref. \cite{Joo:2019byq,Aoki:2021kgd,osti_1369223,Habib:2016sce,Joo:2019byq,DAVOUDI20211,Aoki_2020}.

The above-cited LQCD experiments use lattices with dimensions ranging from $16^3 \times 48$ to $160^4$ lattice-spacings, and this will likely grow as high-performance computing improves in capacity.

The choice of the path integral in Eqn. \ref{eq:euclideanfeynmanpathintegral} is important, since in this form the Monte Carlo integration will converge to the correct result in a minimum number of iterations. This is because the applying Monte Carlo integration to Eqn. \ref{eq:euclideanfeynmanpathintegral} is equivalent to sampling from a statistical ensemble in thermal equilibrium \cite{physrevd.21.2308}. By contrast, applying Monte Carlo integration to Eqn. \ref{eq:minkowskifeynmanpathintegral} is a textbook case of the numerical sign problem. This means that it is not feasible to use Monte Carlo methods to conduct simulations in real time. The sign problem is also present for systems with a finite fermion density \cite{barbour1998results, PhysRevD.66.074507, DEFORCRAND2002290, SEILER2013213, HASENFRATZ1992539,Density2020}.  Additionally, the theta vacuum term of the QCD Lagrangian \cite{PhysRevLett.37.172, CALLAN1976334, PhysRevD.86.105012} is imaginary in Euclidean space, so it also exhibits a sign problem when included in Monte Carlo calculations. There have been attempts to implement the Minkowski path integral without the sign problem \cite{NAGATA2022103991, usqcd2019hot, alexandru2022complex, berger2021complex, cristoforetti2012new, boyda2022applications}, and in the event of an emergent SU(4) spin-flavor symmetry (which is present in low-energy nuclear forces, SU($N_c$) systems in the limit of $N_c \rightarrow \infty$, and low-entangling-power interactions) the sign problem is suppressed \cite{physrev.51.106, physrev.51.947, kaplan:1995yg, lee:2016sxr, beane:2018oxh, wagman:2017tmp}. However, a solution to the numerical sign problem in the general case is an NP-hard proposition \cite{troyer_2005, 2020arxiv200602660s}. Thus, out-of-equilibrium real-time evolution such as quark-gluon hadronization, problems that require a chemical potential, such as full {\it ab initio} calculations of QCD inside neutron stars, and calculations with the $\theta$ vacuum are generally out of reach for Monte Carlo-based LQCD.

\subsection{Tensor Network Methods}
\label{sec:intro_tnm}

Realization of the ability to model out-of-equilibrium real-time dynamics has been a long-time goal of Standard Model physics research~\cite{Glashow:1961tr,Higgs:1964pj,Weinberg:1967tq,Salam:1968rm,Politzer:1973fx,gross:1973id}. Thus, since Monte Carlo LQCD is most likely precluded from delivering such a result in the general case by the numerical sign problem, alternative methods are needed. Tensor network methods can be used in lieu of Monte Carlo LQCD in the event of a sign error that cannot be circumvented. In tensor network methods, the quantum state $\ket{\psi}$ of a system decomposable into $N$ subcomponents is represented in the form of tensors $T^{i_1, \ldots, i_N}$, where each element of the set ${i_1, \ldots, i_N}$ is an integer index representing a basis state of on, such that 

\begin{equation}
    \ket{\psi} = \sum_{i_1, \ldots, i_N} T^{i_1, \ldots, i_N} \ket{i_1} \otimes \ldots \otimes \ket{i_N}
    \label{eq:tensornet_generalstate}
\end{equation}

\noindent where an index $i_j$ enumerates a basis state within the Hilbert space of the $j^{th}$ subcomponent. $T^{i_1, \ldots, i_N}$ would then be broken down into products of smaller tensors (``tensor networks") in order to efficiently use compute resources. The QCD Lagrangian would be discretized and re-written into a Hamiltonian $H_{TN}$, turning QCD into a many-body quantum system that takes the form of a lattice with the fermions on the lattice sites and the gauge field on the links between the sites and is characterized by $H_{TN}$. Operations on the states, which include gauge-transformations, projection onto subspaces, time-evolution (expressed via the operator $e^{i H_{TN} t}$), and computation of expectation values of observables and of $H_{TN}$,  are expressed in terms of gauge-invariant operators that are themselves also expressed in terms of tensor networks. In order to efficiently use the computational resources provided, the Hilbert space of $H_{TN}$ is, as a rule, truncated. \cite{tagliacozzo2014tensor, buyens2014matrix, PhysRevLett.112.201601, PhysRevLett.113.091601, silvi2014lattice} 

Recently, tensor network methods have been used to study phenomena such as real-time evolution, low-lying spectra, string-breaking, phase diagrams, chiral symmetry breaking, entanglement entropy, and the theta vacuum of non-Abelian gauge theories on 1+1D \cite{silvi2019tensor, rigobello2023hadrons, sala:2018dui, Silvi2017finitedensityphase, hayata2024dense, banuls:2017ena, kuhn2015non, PhysRevLett.112.201601} and 2+1D lattices \cite{cataldi2024simulating}. One limitation of this approach is poor scaling, up to and including exponential with the size of the system, in highly-correlated systems. This is reflected in the small scale of state-of-the-art tensor network simulations: the references in the previous sentence studied 1+1D lattices that were up to approximately 200 matter sites long and 2+1D lattices that had up to 32 matter sites. These lattices are much smaller than the $16^3 \times 48$ lattices commonly found in Monte Carlo LQCD. This relatively small scale makes problems that would be simulated using tensor network methods but whose size and extent of correlations are out of reach for tensor network methods for the foreseeable future a potential opportunity for finding a problem that could be simulated on a quantum computer but not on a classical computer.

\section{Simulation using Quantum Computers}

The exploration of utilizing quantum mechanical systems to construct computational hardware and algorithms dates back to work by Paul Benioff, Yuri Manin, and Richard Feynman in the early 1980s \cite{benioff1980, manin1980, feynman:1981tf, feynman1986}. These works found that the simulation of quantum mechanical systems using classical systems has, in the general case, a complexity that scales exponentially with the size of the simulated system, and proposed construction of quantum computers to correct this inefficiency. In 1996, Seth Lloyd proved that quantum computers are capable of simulating any local quantum system with computational resource requirements that scale polynomially at worst with the size of the system ~\cite{Lloyd1073}. In the nearly 30 years that followed, there has been much development of both quantum hardware and algorithms. Today, digital quantum computers are in the NISQ (Noisy Intermediate-Scale Quantum) era~\cite{preskill2018quantumcomputingin}. Devices built by companies such as Google \cite{GoogleQuantumAI}, IBM \cite{ibmq}, Quantinuum \cite{quantinuum}, IonQ \cite{IonQ}, and QuEra \cite{QuEraComputing} that researchers can access now have a number of qubits that is within the range of system-sizes probed by tensor network methods and higher than the maximum, even on the most powerful classical supercomputers, that a brute-force mapping of every element in their Hilbert space to a bit-register can simulate. However, quantum devices currently exhibit significant noise that limits the number of gates that can be run on them and stands in the way of quantum computers offering an advantage in performance over classical supercomputers. The qubit-counts and post-noise gate fidelities (the odds of a quantum state being in the correct state after an operation is applied) for a selection of state-of-the-art quantum devices can be found in Tab. \ref{tab:intro_devices_performances}. Meanwhile, the search for quantum utility continues. Last year, IBM's quantum team published a result where it obtained expectation values from the kicked Ising model in regimes where the two tensor network methods it used failed to produce results \cite{kim2023evidence}, only for several solutions to this result using classical methods to appear within months \cite{beguvsic2024fast, begušić2023fastclassicalsimulationevidence, rudolph2023classicalsurrogatesimulationquantum, liao2023simulationibmskickedising, PRXQuantum.5.010308, PhysRevResearch.6.013326, torre2023dissipativemeanfieldtheoryibm, anand2023classicalbenchmarkingzeronoise}. IBM \cite{ibmq}, Quantinuum \cite{quantinuum} and IonQ \cite{IonQ} all have roadmaps aimed at fault-tolerant quantum computing, where the errors on quantum devices are suppressed to the point where one can design circuits on said quantum devices with little to no regard for noise, within a decade.

Analog quantum simulators, which are quantum-mechanical systems whose Hamiltonian is specifically tailored to the target problem, achieved quantum advantage back in 2012 when a relaxation to equilibrium of the one-dimensional Bose-Hubbard model was studied for longer times on an optical lattice of ultracold $^{87}Rb$ atoms than what was possible on classical computers at the time \cite{trotzky2012probing}. Since then, analog quantum simulators have been deployed to the task of solving problems difficult to address with classical computers \cite{ebadi2021, scholl2021quantum}. There have also been attempts to use analog quantum simulators to simulate both Abelian and non-Abelian lattice gauge theories in the past decade\cite{physrevlett.112.120406, gonzalez-cuadra:2017lvz, Luo:2019vmi, Davoudi:2019bhy, kasper:2015cca, PhysRevA.107.042404, Gonzalez-Cuadra:2022hxt, bazavov:2015kka, su2024cold, Yang_2020, aidelsburger:2021mia, Knaute:2021xna, Illa:2022jqb}. However, the Hamiltonians of analog quantum simulators are limited in their ability to solve problems different than the ones they are tailored to. Thus, the scope of the work in this thesis is focused primarily on digital quantum computers, which are designed to be able to run any possible unitary operation on a register of qubits. For recent reviews of simulation using quantum computers, see Refs. \cite{banuls:2019bmf, Preskill:2021apy, klco:2021lap, Funcke:2021aps, Bauer:2022hpo, bauer2023quantum, funcke2023review, beck2023quantum}. 

\begin{table}[]
    \centering
    \begin{tabular}{c|c|c|c}
         Device & \# of qubits & 1-qubit gate fidelity & 2-qubit gate fidelity \\
         \hline
        Quantinuum H2-1 & 56 & 99.997 \% & 99.87 \%\\ 
        \hline
        IonQ Forte & 36 & 99.98 \% & 99.6 \%\\
        \hline
        IBM {\tt ibm\_torino} & 133 & 99.96 \% (median) & 99.15 \% \\
        
    \end{tabular}
    \caption{The qubit-counts and gate-fidelities of a few superconducting (IBM) and trapped-ion (Quantinuum and IonQ) devices available as of Nov. 26, 2024. Data is from Refs. \cite{quantinuum, IonQ, ibmq}}
    \label{tab:intro_devices_performances}
\end{table}

\subsection{Magic, Entanglement, and Quantum Advantage}
\label{sec:intro_magic_entanglement_quantumadvantage}
From the work of Daniel Gottesman, Emanuel Knill, and Scott Aaronson\cite{gottesman:1998hu, aaronson_2004}, high entanglement is a necessary but not a sufficient condition for a problem to be sufficiently difficult to simulate classically to warrant the use of a quantum computer. This is reflected in the classical simulation of the quantum advantage that the IBM quantum team had at first assumed they had found in the kicked Ising model. Many of the algorithms that successfully simulated IBM's kicked Ising model finding were designed to be limited not by low entanglement or low number of correlation, but by a low level of deviation of the system from being a stabilizer state, where a stabilizer state is a state that can be obtained from a post-measurement state entirely through Clifford gates. The work of Gottesman, Knill, and Aaronson support the proposition that this is no mere coincidence, and that a stabilizer state can be simulated efficiently on a classical device \cite{gottesman:1998hu, gottesman:1997zz, Rattacaso:2023kzm,frau2024nonstabilizerness}. This finding points to the idea that problems where quantum advantage may be found must exhibit both high entanglement and high non-stabilizerness (or ``magic"). Studies characterizing the magic of systems have been done recently for lattice gauge theories \cite{Tarabunga:2023ggd}, nuclear physics \cite{Robin:2024bdz, Brokemeier:2024lhq}, and various other systems \cite{Oliviero_2022, Haug:2023hcs, Rattacaso:2023kzm, frau2024nonstabilizerness, Catalano:2024bdh}. 

Chapter \ref{chap:neutrinomagic} studies the time evolution of magic, parametrized in the form of a Renyi entropy \cite{Leone:2021rzd}, in the context of collective oscillations of neutrinos with the physical three flavors. Collective neutrino oscillations happen when neutrinos are sufficiently dense that neutrino-neutrino interactions affect their flavor dynamics. Environments where this is the case include core-collapse supernovae \cite{Pantaleone:1992xh,Pantaleone:1992eq,Qian:1994wh,Pastor:2002we,Mirizzi:2015eza}, mergers involving compact stellar objects \cite{Malkus:2012ts,Malkus:2015mda,Zhu:2016mwa,Frensel:2016fge,Chatelain:2016xva,Wu:2017qpc,Tian:2017xbr,Purcell:2024bim}, and the nucleosynthesis that happens in the two aformentioned cases \cite{Fuller:1995ih,Balantekin:2004ug,Duan:2010af,Xiong:2020ntn,Balantekin:2023ayx}. See Refs.~\cite{Duan:2009cd,Duan:2010bg,Chakraborty:2016yeg,Tamborra:2020cul,Capozzi:2022slf,Richers:2022zug,Patwardhan:2022mxg,Volpe:2023met,Balantekin:2023qvm} for reviews on this topic.

Collective neutrino oscillations have been studied for decades for their potential astrophysical impact, but recent results suggest that they could also have applications for the Standard Model as a whole \cite{Suliga:2024oby}. The astrophysical scenarios where collective neutrino oscillations happen are all highly out-of-equilibrium scenarios and collective neutrino oscillations exhibit extensive entanglement \cite{Illa:2022zgu, Martin:2023ljq}. Thus in principle three-flavor collective neutrino oscillations are promising ground to look for quantum advantage in. There have been several simulations of collective neutrino oscillations on digital quantum devices in the literature in recent years, all of which have modeled neutrinos as having 2 flavors~\cite{Hall:2021rbv,Yeter-Aydeniz:2021olz,Amitrano:2022yyn,Illa:2022zgu,Siwach:2023wzy}. Following the lead of these simulations, methods for studying 3-flavor collective neutrino oscillations on digital quantum devices are laid out in Chapter~\ref{chap:qutritqubitneutrino}. Additionally, the behavior of the magic in the context of collective neutrino oscillations is checked in Chapter \ref{chap:neutrinomagic}. As a note, neutrino flavors are in the same SU(3) fundamental irreducible representation that quark colors are in, and I find it of interest to assess the similarities and differences of SU(3) system behavior in these two dissimilar environments. 

\subsection{Digital quantum device formalism}
\begin{figure}
    \centering
    \includegraphics[width=0.8\linewidth]{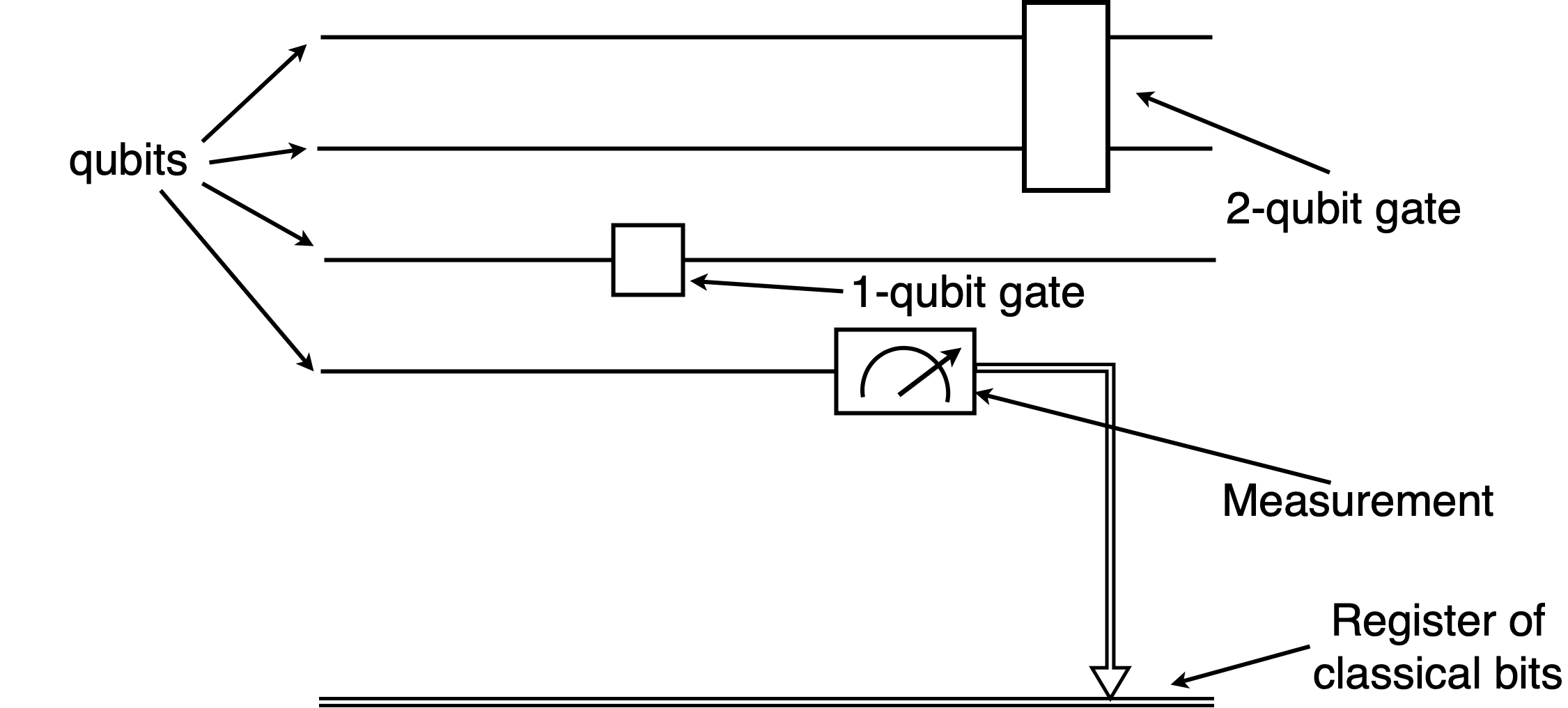}
    \caption{An example of a quantum circuit diagram, with each component labeled. The 1-qubit and 2-qubit gates here do not have anything written on them, but customarily they do come with a label indicating which operation they are applying to the qubits they act on.}
    \label{fig:intro_circuitdiagramlegend}
\end{figure}

Digital quantum devices accomplish the ability to run any possible unitary operation on their qubits by implementing a set of quantum operations called ``gates". Customarily, this universal gate-set is composed of a set of gates on 1 qubit that can implement an arbitrary SU(2) transformation on the qubit and at least one 2-qubit gate that is transformable into a CNOT gate using only single-qubit operations. According to the theorems of Sec. 4.5 of Ref. \cite{nielsen_chuang_2010}, such a gate-set is universal, meaning that it can implement any unitary gate.  Designs of circuits are typically represented in the form of diagrams. The layout of the diagrams used in most places in the field, including this thesis, is shown in Fig. \ref{fig:intro_circuitdiagramlegend}.

\noindent As seen in Tab. \ref{tab:intro_devices_performances}, fidelities of 2-qubit gates are, as a rule, significantly less than the fidelities of 1-qubit gates. For this reason, the number of 2-qubit gates on a circuit and the 2-qubit gate depth (defined as the largest number of 2-qubit gates a line drawn through a quantum circuit can go through, given that this line can only be drawn along qubits and through 2-qubit gates) are used in this thesis (and throughout the field) as a measure of circuit complexity.

\subsection{Technical details of the quantum hardware}

There is a variety of choices of quantum hardware available today, each with their advantages and disadvantages. A workshop report that lists several hardware technologies can be found in Ref. \cite{brown20245yearupdatesteps}. The hardware used in the work discussed in this thesis consists of superconducting devices developed by IBM and trapped-ion devices developed by Quantinuum. Detailed reviews of how superconducting quantum devices work can be found in Ref. \cite{huang2020superconducting, gao2021practical} and a detailed review of how trapped-ion quantum devices work can be found in Ref. \cite{2019trap}. In summary, Quantinuum's trapped ion devices work by having a set of $^{171}Yb^+$ ions in a trap that holds the ions in place through a series of electric fields that oscillates at RF frequencies. Quantum operations are applied using a series of lasers and measurement is done through state-dependent resonance flourescence. The device has all-to-all connectivity that is accomplished by physically moving the ions, a process that tends to occupy approximately 60 \% of the runtime. Two-qubit operations are executed in one of the four gate-zones on the device. This has the side effect of allowing up to four two-qubit operations at a time.\cite{moses2023race} Quantinuum's devices have the best gate fidelity out of devices currently available on the cloud. IBM's superconducting devices work by having a Josephson junction in parallel with a capacitor as a qubit and having a circuit of these qubits on each quantum device. Quantum operations and measurements are done by firing microwave-frequency photons at the qubits. \cite{ibmq} Generally, superconducting devices have more qubits than trapped ion devices and unlike trapped ion devices can execute an arbitrary number of gates in parallel (provided all of the gates executed in parallel target different qubits). However, they have lower gate fidelities and their connectivity is effectively nearest-neighbor for most practical intents and purposes.

\subsection{Quantum error mitigation}

Quantum error correction (QEC), defined as mapping of several of a device's qubits to one ``logical" qubit and obtaining a level of noise for the logical qubit that is less than the noise on the device's (``physical") qubits, is in its early stages. Simple demonstrations of QEC have recently been conducted on IBM's \cite{bravyi2024high}, Google's \cite{google2023suppressing}, Quantinuum's \cite{ryan2021realization, Hong_2024}, IonQ's \cite{delfosse2024lowcostnoisereductionclifford, egan2021fault}, and QuEra's \cite{xu2023constantoverheadfaulttolerantquantumcomputation} devices. However, quantum error correction is not yet at the stage where it can be utilized for physical quantum simulations. 

Instead, quantum error mitigation (QEM), which is a set of techniques that minimize the impact of errors on the hardware on the final results, is used. One of the most commonly used QEM methods in the work in this thesis is measurement error mitigation \cite{Nation:2021kye, qiskit,ibmmeaserror}, used in Chapters \ref{chap:prepofsu3latyangmilsvacvarqumethods}, \ref{chap:1p1dQCD}, and \ref{chap:qutritqubitneutrino}. In this method, the final measurement results are multiplied by a matrix meant to map from the noisy probabilities of each possible measurement to the ideal probabilities of each possible measurement. This matrix is obtained from a previous run when the expected ideal results were trivially known. Physical space post-selection is just as commonly-used, appearing in Chapters \ref{chap:1p1dQCD}, \ref{chap:1p1dSM}, and \ref{chap:qutritqubitneutrino}. It works by removing any measurement result that corresponds to an unphysical state from the pool of measurement result. After this, the most commonly used QEM method used in this thesis is operator decoherence renormalization (ODR), originally introduced in Refs. \cite{urbanek:2021oej, arahman:2021ktn} and implemented in Chapters \ref{chap:1p1dQCD} and \ref{chap:qutritqubitneutrino} of this thesis as well as Refs. \cite{Ciavarella:2023mfc, Farrell:2023fgd, Ciavarella:2024fzw}. Decoherence renormalization assumes that all errors are depolarizing (i.e. they map a state to a mixture of itself and the maximally mixed state). It works by running the one circuit repeatedly and taking measurements, then running a second circuit with the exact same complexity as the original circuit but whose result is manifestly a return of the input, and then taking the same measurements. The measurements on each of the two circuits are used to estimate the probability of the circuit returning a given result $A$. The first circuit's measurement-estimated probability of producing $A$ is denoted $P^{noisy}_{phys}$ and the second circuit's measurement-estimated probability of producing $A$ is denoted $P^{noisy}_{id}$. Because the second circuit simply returns its input, its error-free-case probability of producing $A$, $P^{ex}_{id}$, is trivial to calculate. $P^{ex}_{phys}$, the estimated error-free probability of the first circuit producing $A$, can then be extrapolated through a ratio relation of the differences between each of the four probabilities and the probability of $A$ given a maximally mixed state, which is equal to $\frac{1}{2^{N_q}}$. $N_q$ is the number of qubits involved a measurement that can produce $A$. This ratio relation is as follows:
\begin{equation}
    P^{ex}_{phys} - \frac{1}{2^{N_q}} = \frac{P^{ex}_{id} - \frac{1}{2^{N_q}}}{P^{noisy}_{id} - \frac{1}{2^{N_q}}} ( P^{noisy}_{phys} - \frac{1}{2^{N_q}})
    \label{eq:intro_ODRformula}
\end{equation}

\noindent Also used in both Chapters \ref{chap:1p1dQCD} and \ref{chap:qutritqubitneutrino} is Pauli twirling \cite{physreva.94.052325}. Pauli twirling is designed to effectively transform coherent errors into a random noise, and it does so by introducing within each run of a circuit a set of Pauli-matrix operators to the 2-qubit gates that would do nothing in the event that there is no noise. This set would be selected at random each time the circuit is run. Dynamical decoupling (DD) \cite{physreva.58.2733, 2012RSPTA.370.4748S, Ezzell:2022uat, duan1999139, zanardi199977, physrevlett.82.2417} is another method used in both Chapters \ref{chap:1p1dQCD} and \ref{chap:qutritqubitneutrino}, and it attempts to cause the noise to average out to zero by applying a periodic set of inversion pulses to the qubits. Finally, there is zero-noise extrapolation (ZNE) \cite{physrevx.7.021050, physrevlett.119.180509, 2020arXiv200510921G}, used in Chapter \ref{chap:prepofsu3latyangmilsvacvarqumethods}. In ZNE noisy circuit components would be substituted for circuit components with the same result in a noiseless scenario but with progressively higher complexities. The data samples would then be plotted with respect to complexity, and a model (often linear) would be fitted to the data. This model would be used to predict what the result would be in a hypothetical zero-complexity scenario, and that hypothetical result would be returned as the error-mitigated result. ZNE is not used in other chapters largely due to its overhead in circuit complexity and number of quantum device runs in cases where these resources are at a premium even in the base-case circuit.

\section{Putting a Hamiltonian on quantum devices}

\subsection{Fermion doubling problem}

Based on the Nielsen–Ninomiya theorem \cite{NIELSEN1981219, NIELSEN198120,NIELSEN1981173}, a local, translationally invariant lattice gauge theory described by a Hermitian Hamiltonian and with a charge that is locally defined, quantized, and exactly conserved, there must be an equal number of left-handed and right-handed fermions. This implies that unless one violates one of the above assumptions or violates chiral invariance, placing a fermion on a lattice will result in that fermion being doubled at least once. This fermion-doubling can cause complications in interacting theories, where there is the possibility of the doubling-artifact fermions interacting with the fermions intended to be in the system. Thus, several schemes have been devised in order to either eliminate or reduce the doubling problem to the point where the interactions involving the doubling-artifact fermions can be made to conform the theory relatively easily. These schemes involve reducing chiral invariance, translational invariance, and/or locality from symmetries expected of the lattice to something that is approached in the continuum limit. These include Wilson fermions\cite{physrevd.10.2445}, Ginsparg-Wilson fermions\cite{PhysRevD.25.2649, NEUBERGER1998141, NEUBERGER1998353}, and domain-wall fermions\cite{kaplan_1992, SHAMIR199390}. The scheme used in the construction of the lattice gauge theories discussed in this work is John Kogut's and Leonard Susskind's Hamiltonian formalism \cite{physrevd.11.395}. This approach uses a staggered lattice, which is defined as a lattice which places fermions and antifermions on alternating lattice sites. It is more computationally efficient than the previously mentioned schemes. For 3 or more dimensions, there are doublers, but in principle it's sufficiently few doublers to work with. In principle, it is possible to use the fermions that come into being as a result of the doubling to represent different quark flavors with different masses. This idea has been investigated in the Lattice QCD literature for decades \cite{banks1977strong,susskind1977lattice, GOLTERMAN198461, HOELBLING2011422, de2012numerical, de2011numerical, misumi2012strong, durr2022topological, adams2010theoretical, adams2011pairs, hoelbling2011single}, and holds the promise of combining the computational efficiency of the staggered lattice with the advantages of the other schemes, as well as providing a convenient method of representing topological charge. However, it is seldom used, because as a rule implementing it requires the breaking of most of the symmetries that lattice gauge theories follow, resulting in many complications not present in other approaches. Nonetheless, my colleague Anthony Ciavarella recently implemented this idea on a demonstration of a 3+1D QCD lattice on a quantum computer \cite{Ciavarella:2023mfc}.


\subsection{Kogut-Susskind Hamiltonian}
\label{sec:intro_kogutsusskind}
There have been many attempts to simulate lattice gauge theories on quantum devices  \cite{physreva.98.032331,kokail:2018eiw,kharzeev:2020kgc,lu:2018pjk,chakraborty2020digital,shaw:2020udc,Bender_2018, physrevd.101.074512,Zohar_2013,zohar:2015hwa,zohar_2015_2,Lamm_2019,alexandru:2019nsa,Ba_uls_2020,Tagliacozzo_2013,tagliacozzo:2012vg,PhysRevLett.105.190404,Tavernelli20_094501,physreva.73.022328,physrevd.103.094501,kan:2021xfc,stryker:2021asy,Paulson_2021,davoudi:2021ney,Zohar:2012ay,Zohar:2012xf,physrevlett.110.125303,Banerjee:2012pg,Martinez2016,Muschik:2016tws,Zohar:2016iic,banuls:2017ena,Kaplan:2018vnj,Zache:2018jbt,Stryker:2018efp,Raychowdhury:2018tfj,Haase:2020kaj,Paulson:2020zjd,Davoudi:2020yln,Buser:2020cvn,Raychowdhury:2018osk,Raychowdhury:2019iki,Tagliacozzo:2012df,Ji:2020kjk,Chandrasekharan:1996ih,Brower:2003vy,Wiese:2006kp,atas:2021ext,dejong2021quantum,meurice:2021pvj,zohar:2021nyc,armon2021photonmediated,andrade:2021pil,Hauke:2013jga,Banuls:2013jaa,Martinez:2016yna,Avkhadiev:2019niu,Magnifico:2019kyj,Mishra:2019xbh,Halimeh:2020ecg,Halimeh:2020djb,VanDamme:2020rur,arahman:2021ktn,PhysRevLett.122.050403, VanDamme:2021njp,Halimeh:2021lnv,Thompson:2021eze,Yeter-Aydeniz:2021mol,rahman:2022rlg,Bauer:2021gek,Mildenberger:2022jqr, atas:2022dqm,grabowska:2022uos,carena:2022hpz,gustafson:2022xdt,Davoudi:2022uzo,avkhadiev:2022ttx}. Many of these works use the Kogut-Susskind staggered fermion Hamiltonian to express the theories that they study, as do many of the tensor network studies mentioned in Sec. \ref{sec:intro_tnm}.
The lattice gauge theories discussed in this work follow the lead of these works. The Kogut-Susskind Hamiltonian characterizes a staggered lattice obeying a gauge theory, with the fermion field on the sites and the gauge field on the links. Its full form is as follows\cite{physrevd.11.395, RevModPhys.51.659, banks:1975gq, robson:1980nt, ligterink2000983c, physrevd.31.2020, physreva.73.022328}:

\begin{multline}
    H = m \sum_r \left[(-1)^r \psi^{\dagger} (r) \psi (r) \right] + \frac{1}{2 a} \left[ \sum_{r, \mu} \psi^{\dagger} (r) ((\vec{\sigma} \cdot \mu) \otimes U(r, \mu)) \psi (r + \mu)  \right] +\\
    \frac{g^2}{2 a^{d-2}} \left[ \sum_{r, \mu} \sum_{\alpha} |E^{\alpha}(r, \mu)|^2 \right] + \frac{2}{g^2 a^{4-d}} \sum_{p} \left[ 2 N_c - (Z(p) + Z^{\dagger}(p)) \right]
    \label{eq:KogutSuskindHam_intro}
\end{multline}

\noindent where $g$ is a gauge coupling strength, $a$ a lattice spacing, $d$ the number of spatial dimensions in the system, $\psi (r)$ is a fermion field spinor on site $r$, $U(r, \mu) = e^{i \frac{g^2}{2} \tau^{\alpha} A^{\alpha} (r, \mu)}$, where $A^{\alpha} (r, \mu)$ is the gauge field on the link with one end on site $r$ and whose other end is $\mu$ (which is a unit vector) away from site $r$, $E^{\alpha}(r, \mu)$ is $A^{\alpha} (r, \mu)$'s conjugate variable and represents the electric field, $\vec{\sigma}$ is the vector $(\sigma_x, \sigma_y, \sigma_z)$ of Pauli matrices, and $Z(p)$ is a Wilson loop on the plaquette $p$, defined like so:

\begin{equation}
    Z(p) = Tr[U(r, \mu)U(r + \mu, \nu)U^{\dagger}(r + \nu,\mu)U^{\dagger}(r, \nu)]
    \label{eq:plaquettedef_intro}
\end{equation}

\noindent with $\nu$ and $\mu$ being the perpendicular unit vectors which define the plaquette. The simplest implementation of the Kogut-Susskind Hamiltonian is in U(1) lattice gauge theory. On a 1+1D lattice, this is a discretization of the Schwinger model, which is essentially Quantum Electrodynamics (QED) mapped to 1 spatial dimension \cite{PhysRev.82.914, PhysRev.91.713}. There have been several simulations of the discretized Schwinger model on quantum devices~\cite{Martinez:2016yna,physreva.98.032331,lu:2018pjk,kokail:2018eiw,Nguyen:2021hyk,Thompson:2021eze} and there have been several attempts to extend these simulations to multiple spatial dimensions~\cite{Zohar:2011cw,Zohar:2012ay,tagliacozzo:2012vg,Zohar:2012ts,Wiese:2013uua,Marcos:2014lda,Kuno:2014npa,bazavov:2015kka,kasper:2015cca,Brennen:2015pgn,Kuno:2016xbf,Zohar:2016iic,Kasper:2016mzj,gonzalez-cuadra:2017lvz,Ott:2020ycj,Paulson:2020zjd,Kan:2021nyu,aidelsburger:2021mia,Bauer:2021gek}. Motivations for this work with the lattice Schwinger model include the fact that the Schwinger model has many of the same features that QCD has and use of the lattice Schwinger model to benchmark similar efforts with non-Abelian lattice gauge theory, which are discussed in Chapters Chapters \ref{chap:prepofsu3latyangmilsvacvarqumethods}, \ref{chap:1p1dQCD}, \ref{chap:1p1dSM}, and \ref{chap:0vBBandlargeQCD}. Prior to the work conducted in the aforementioned chapters, the state of non-Abelian lattice gauge theory simulation on quantum devices consisted of single- and few-plaquette systems with an SU(2) or SU(3) gauge field without fermions \cite{physrevd.101.074512,physrevd.103.094501,rahman:2022rlg,arahman:2021ktn} and a simulation of 1+1D SU(2) lattice gauge theory with fermions \cite{atas:2021ext}. Shortly after the completion of the work in Chapter \ref{chap:1p1dQCD}, which details the first quantum simulation of SU(3) 1+1D lattice gauge theory with quarks to our knowledge, a study using similar methods of the dynamics of tetraquarks and pentaquarks on a 1+1D SU(3) lattice was released \cite{atas:2022dqm}.

As a convention, the work in Chapters \ref{chap:prepofsu3latyangmilsvacvarqumethods}, \ref{chap:1p1dQCD}, \ref{chap:1p1dSM}, and \ref{chap:0vBBandlargeQCD} hold the lattice spacing $a$ to be 1. In Chapter \ref{chap:prepofsu3latyangmilsvacvarqumethods}, which only factors in the gauge field and its respective terms, the basis states of the gauge field on each link of the lattice are mapped onto qubit-states in the eigenbasis of the third (``electric") term of the right side of Eq. \ref{eq:KogutSuskindHam_intro}. Each link's gauge field has a theoretically infinite number of states, so, for purposes of mapping to qubits, a truncation scheme that keeps only the basis states with the lowest electric-term eigenvalues is applied. This method was developed in Refs. \cite{physreva.73.022328, zohar_2015_2, banuls:2017ena, physrevd.103.094501, physrevd.101.074512}. In Chapter \ref{chap:1p1dQCD}, a gauge transformation is described which is closely related to the one used in Ref. \cite{atas:2021ext} and which eliminates the need to explicitly represent the gauge field on qubits. The gauge-transformed Kogut-Susskind Hamiltonian is then used in Chapters \ref{chap:1p1dQCD}, \ref{chap:1p1dSM}, and \ref{chap:0vBBandlargeQCD}. In Chapters  \ref{chap:1p1dSM} and \ref{chap:0vBBandlargeQCD}, where the intention of the setup is to simulate beta decay and neutrinoless double beta decay, there must also be a lepton term in the Hamiltonian for the electrons and neutrinos involved in the decay. The leptons are represented on a separate register within the lattice and are described by the same Kogut-Susskind Hamiltonian, except without the gauge field. The only gauge field used in this work is the QCD gauge field.






\subsection{Jordan-Wigner mapping}

In order to simulate a system on a quantum device, its degrees of freedom must be mapped onto those of a quantum device. Several algorithms for mapping scalar field theories to the degrees of freedom of quantum devices can be found in Ref. \cite{preskill2018simulating, jordan2019quantum,10.5555/3179430.3179434,PhysRevA.98.042312,Klco:2018zqz, Klco:2019yrb,Yeter_Aydeniz_2019,physreva.103.042410}, a few algorithms for mapping non-linear $\sigma$ models can be found in Ref. \cite{Alexandru_2019_2,Singh:2019uwd,Bhattacharya:2020gpm,Hostetler_2021}, and one algorithm for mapping superstring theory can be found in Ref. \cite{Gharibyan_2021}. One challenge facing the mapping of fermions to quantum devices in particular is that fermions are anticommuting particles. Thus, in order to represent operators present in the Kogut-Susskind Hamiltonian that act on quarks in terms of qubit operators (which are spin-systems), this anticommutation must also be mapped. In this work, the occupation state of each possible fermion is encoded onto a qubit, with the qubit's $\ket{0}$ state standing in for an occupied fermion or an unoccupied antifermion state and the qubit's $\ket{1}$ state standing in for an unoccupied fermion or an occupied antifermion state. The Jordan-Wigner (JW) transformation \cite{jordan:1928wi} is then used in order to map fermionic creation ($\psi^{\dagger}$) and annihilation ($\psi$) operators (which the Kogut-Susskind Hamiltonian's fermion-dependent parts can written in terms of) onto Pauli operations on qubits like so:

\begin{subequations}
\label{eq:JWmappingdef_intro}
    \begin{align}
        \psi^{\dagger}_i = \prod_{j < i} \left(-\sigma^{(z)}_j\right) \sigma^{+}_i \label{eq:JWmappingdef_creation_intro} \\
        \psi_i = \prod_{j < i} \left(-\sigma^{(z)}_j\right) \sigma^{-}_i \label{eq:JWmappingdef_annihilation_intro}
    \end{align}
\end{subequations}

The subscripts denote integer indices assigned to each fermion. $\sigma^{(z)}_i$ is a Pauli-Z operator applied to the qubit corresponding to the index-$i$ fermion, $\sigma^{(+)}_i$ is the operator $\sigma^{(+)} = \frac{\sigma^{(x)} + i \sigma^{(y)}}{2}$ applied to the qubit corresponding to the index-$i$ fermion, and $\sigma^{(-)}_i$ is the operator $\sigma^{(-)} = \frac{\sigma^{(x)} - i \sigma^{(y)}}{2}$ applied to the qubit corresponding to the index-$i$ fermion. One alternative to the Jordan-Wigner transformation is the Bravyi-Kitaev transformation \cite{bravyi2002210, doi:10.1063/1.4768229}. It is not used in the projects discussed in this manuscript, as we have found that for our purposes, the Jordan-Wigner method is simpler to work with. A few problems unrelated to the work in this thesis that have used the Jordan-Wigner and/or Bravyi-Kitaev transformations can be found in Ref. \cite{jordan2014quantum,Lamm_2020,Mazza_2012}.



\section{Quantum algorithms implemented}

So far, we have motivated the simulation of SU(3) quantum field theories on quantum devices, gave an overview of the capacity of state-of-the-art quantum devices, how one would obtain a lattice Hamiltonian for a gauge theory and translate it into a form that can be processed on a quantum device, and introduced several error mitigation techniques. In this section, we discuss the algorithms implemented and improved on by the work in this thesis for simulation of real-time evolution and the state preparation methods needed to produce the initial states for aforementioned time-evolution as well as explore the low-lying spectra and phase transitions of the simulated systems.

\subsection{Time-evolution}

It is not trivial to simply express the Kogut-Susskind Hamiltonian's time evolution in terms of gates on a digital quantum computer. Thus, following the direction of Ref. \cite{physreva.73.022328, stetina:2020abi}, time evolution on quantum devices is executed separately for the individual terms in the Hamiltonian (whose time evolution is trivially expressible on digital quantum devices) using the Suzuki-Trotter formula \cite{10.2307/2033649,suzuki1976generalized}:
\begin{equation}
    e^{i \sum\limits_{i=1}^{N_{t}} (H_i t)} \approx \left(\prod_{i=1}^{N_{t}} e^{i H_i \frac{t}{n}} \right)^n
    \label{eq:intro_trotter_formula}
\end{equation}

\noindent with $H_i$ being the terms of the Hamiltonian, $N_{t}$ being the number of terms in the Hamiltonian, $t$ being the time evolved for, and $n$ being the number of times that the sequence of executing the time evolution of each individual $H_i$ is repeated. 

\begin{figure}
    \centering
    \includegraphics[width=1.0\linewidth]{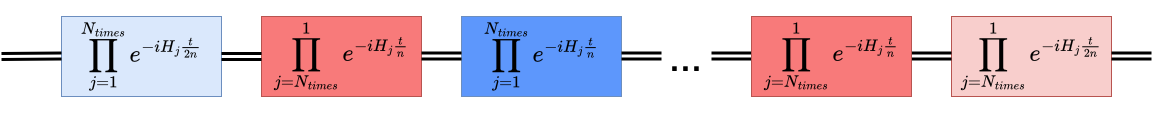}\\
    \includegraphics[width=0.7\linewidth]{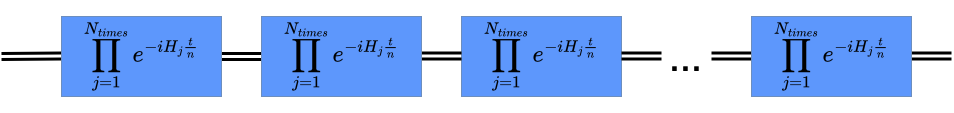}
    \caption{Top: a graphic of a quantum circuit whose form is specified by Eq. \ref{eq:intro_trotter_formula_secondorder_reverse}, with neighboring identical terms merged. Bottom: a graphic of a quantum circuit whose form is specified by Eq. \ref{eq:intro_trotter_formula}. All rectangle-elements have the same complexity}
    \label{fig:intro_firstandsecondordertrottergraphics}
\end{figure}

If noise on devices wasn't a limitation, $n$ would be taken to the limit of infinity. This is due to Trotter errors. That is, in its naive form (``first order Trotterization"), Eqn. \ref{eq:intro_trotter_formula} picks up an extra factor of $e^{- \sum_{i,j} [H_i, H_j] \frac{t^2}{2 n} + O(t^3) }$ from the Baker-Campbell-Hausdorff formula \cite{baker1901further, campbell1896law, hausdorff1906symbolische, Poincare1899, schur1889neue} in addition to the desired $e^{i \sum_i H_i t}$. The premise of Trotterization is that by making $\frac{t}{n}$ as small as possible, the resulting Trotter error would vanish because $- \sum_{i,j} [H_i, H_j] \frac{t^2}{2 n} + O(t^3) \ll i \sum_i H_i t$. One simple way of making the Trotter error vanish faster than with the first order form is with second-order Trotterization:

\begin{equation}
    e^{i \sum\limits_{i=1}^{N_{t}} (H_i t)} \approx \left(\left(\prod_{i=1}^{N_{t}} e^{i H_i \frac{t}{2n}}\right)\left(\prod_{i=N_{t}}^{1} e^{i H_i \frac{t}{2n}}\right)\right)^n ,
    \label{eq:intro_trotter_formula_secondorder}
\end{equation}

\noindent in which case the $O(t^2)$ terms would be cancelled out and the additional factor from the Trotter error would be  $e^{i E_{trott}\frac{t^3}{n^2} + O(t^5)}$, which would converge to zero with increasing $n$ faster than its counterpart in the first order Trotter error. For asymptotically large $n$, second-order Trotterization can be made to cost effectively the same as first-order Trotterization by reversing the order of the terms after each iteration. Then, the second-order Trotter formula would become like so:
\begin{equation}
    e^{i \sum\limits_{i=1}^{N_{t}} (H_i t)} \approx \left(\left(\prod_{i=1}^{N_{t}} e^{i H_i \frac{t}{2n}}\right) \left(\prod_{i=N_{t}}^{1} e^{i H_i \frac{t}{2n}}\right)\left(\prod_{i=N_{t}}^{1} e^{i H_i \frac{t}{2n}}\right)\left(\prod_{i=1}^{N_{t}} e^{i H_i \frac{t}{2n}}\right)\right)^\frac{n}{2} ,
    \label{eq:intro_trotter_formula_secondorder_reverse}
\end{equation}

\noindent and all of the $\left(\prod\limits_{i=1}^{N_{t}} e^{i H_i \frac{t}{2n}}\right)$ and $\left(\prod\limits_{i=N_{t}}^{1} e^{i H_i \frac{t}{2n}}\right)$ terms except for the first and the last one would be identical to one of its neighbors and so could be merged with that neighbor. The resulting general-case circuit is shown in Fig. \ref{fig:intro_firstandsecondordertrottergraphics}, with the corresponding general-case circuit for first-order Trotterization shown for comparison.

Due to noise on the quantum hardware, the $n$ in the projects discussed in this thesis ranges from 1 to 4, with the goal of scaling up $n$ once circuits with the capacity to do so are built. Nevertheless, the technique of the second-order Trotter with the reversed order of terms was still helpful in our low-$n$ regime, and it is used in Chapter \ref{chap:qutritqubitneutrino} in particular.

In Chapter \ref{chap:1p1dQCD}, the first implementation of SU(3) lattice gauge theory with fermions on a quantum device is presented. As part of this, time-evolution circuits for all of the individual terms in the 1+1D SU(3) Kogut-Susskind Hamiltonian are designed, put together into a Trotterization, and tested using IBM's devices {\tt ibm\_jakarta} and {\tt ibm\_perth} on a simple model where time evolution under the Kogut-Susskind Hamiltonian is applied to the trivial vacuum. 

In Chapter \ref{chap:1p1dSM}, the circuits from Chapter \ref{chap:1p1dQCD} applied to a simulation of the time-evolution of $\beta$-decay. It is motivated as a prelude to use of the formalism from Chapter \ref{chap:1p1dQCD} or a similar one to conduct real-time simulations of less-well-understood nuclear reactions. The $\beta$-decay is modeled as a new term in the Hamiltonian that follows the form of an effective field theory point-interaction. Additionally, a lepton register is added and with it the appropriate Hamiltonian terms. The full Trotterized time evolution is then done using Quantinuum's {\tt H1-1} trapped ion device. Finally, a Majorana mass term for future simulations of neutrinoless double beta decay is derived.

In Chapter \ref{chap:0vBBandlargeQCD}, a family of Trotterization circuits is derived for time evolution on devices with nearest-neighbor connectivity with minimal overhead from SWAP gates is developed, in order to take advantage of the larger number of qubits on IBM's superconducting deviced than on the all-to-all connectivity trapped ion devices. Additionally, the Trotterization circuits envisioned in Chapter \ref{chap:1p1dSM} for simulation of neutrinoless double beta decay are further developed from the state they were in in Chapter \ref{chap:1p1dSM} and tested on classical devices in order to inform future design choices.

In Chapter \ref{chap:qutritqubitneutrino}, the techniques for Trotter circuit design laid out in previous chapters are applied to the problem of the real-time dynamics of collective oscillations of neutrinos with all 3 physical flavors. First, a two-neutrino Trotterization circuit with a circuit-depth of only 4 two-qutrit gates is devised. A qubit counterpart to this circuit is also designed, and that counterpart is used to implement a Trotterization of the 3-flavor collective neutrino oscillation Hamiltonian which is implemented on Quantinuum's {\tt H1-1} and IBM's {\tt ibm\_torino} devices. Collective neutrino oscillations are an all-to-all interaction, but this is nonetheless implemented on {\tt ibm\_torino}'s nearest-neighbor connectivity by taking advantage of the structure of the neutrino-neutrino interactions to incorporate SWAP operations into said interaction at no extra cost in circuit complexity.

\subsection{State-preparation}
\label{sec:intro_stateprep}
The projects in this thesis rely in all but the most trivial cases on some variant of the Variational Quantum Eigensolver (VQE), first introduced in Ref. \cite{peruzzo_2014}, for state-preparation. VQE has been used extensively in quantum chemistry \cite{peruzzo_2014,mcclean_2016,physrevx.6.031007,Kandala_2017,PhysRevX.8.031022,physrevx.8.011021,PhysRevA.100.022517,PhysRevA.100.010302,grimsley_2019,Kandala_2019,2020,gyawali2021insights,yamamoto2019natural,2020QNG} and has seen applications to problems in quantum field theory such as the Abelian Higgs model with a topological $\theta$ term \cite{zhang:2021bjq}, as well as ground-state preparation \cite{physreva.98.032331,PhysRevLett.126.220501,kokail:2018eiw} and calculation of forces between mesons \cite{lu:2018pjk} in the Schwinger model. Most relevant to the work in this thesis, VQE was recently used to prepare hadron ground states in SU(2) 1+1D lattice gauge theory \cite{atas:2021ext}. Alternatives to VQE include adiabatic time-evolution \cite{2000quant.ph..1106f, revmodphys.90.015002}, projection-based state-preparation \cite{motta_2019, kosugi2022imaginary,turro:2021vbk, PhysRevA.106.062435, ge2019faster, PRXQuantum.3.040305,keen2021quantumalgorithmsgroundstatepreparation,choi:2020pdg,stetcu2023projection}
VQE does have the disadvantage of, by design, only approximating the correct ground state as opposed to performing an exact mapping to it, which many of its alternatives are designed to do. VQE is used nonetheless because it has proven to be much more resilient to error and generally requires a much lower circuit-complexity than do the aforementioned alternatives \cite{physrevx.6.031007, mcclean_2016,physrevx.8.011021}. It also has the advantage of not requiring ancilla qubits or throwing out of shots on the quantum circuit based on mid-circuit measurements, both of which strain limited resources on NISQ devices and are present on projection-based algorithms.
There also exist state-preparation algorithms that depend on the ground state being known beforehand, many of which do initialize the exact ground state with relatively low depth \cite{PhysRevA.71.052330,PhysRevA.83.032302,PhysRevResearch.3.043200,Zhang_2024,rosenthal2023querydepthupperbounds,PhysRevLett.129.230504,sun2023asymptotically}. These are not used, because the ground-state of SU(3) lattice gauge theory is not trivially known. Even in the case of the lepton ground state present in Chapter \ref{chap:0vBBandlargeQCD}, whose ground state is in principle possible to solve for, VQE initializes the ground state sufficiently well that it is not worth the effort to do so. The periodic boundary conditions lepton initialization circuits in Sec. \ref{sec:pbclephamiltonianinitializatiocircuits} of Chapter \ref{chap:0vBBandlargeQCD} do initialize the target lepton ground state exactly, but even they are based on VQE ansatz circuits (specifically the SC-ADAPT-VQE building-block from Ref. \cite{Farrell:2023fgd}) in all but the most trivial cases. Knowledge of the symmetries of the ground states is nonetheless used to simplify VQE ans{\"a}tze throughout this thesis.

VQE is a hybrid quantum-classical algorithm that uses a quantum circuit which implements a variational ansatz controlled by several parameters whose values are set by a classical optimizer whose job it is to optimize an observable, often the expectation value of the Hamiltonian, that can be obtained by measuring the quantum circuit. The procedure for executing VQE is as follows\cite{peruzzo_2014}:
\begin{enumerate}
    \item Use a collection of priors to set the values of the parameters.
    \item Use the parameters to initialize the quantum circuit.
    \item Run the circuit for as many times as is necessary to measure all of the observables that the classical optimizer needs for the next step.
    \item Submit the measurement values for the observables to the classical optimizer and receive a new set of values for the parameters from the classical optimizer.
    \item Set the parameters equal to the new values from the classical optimizer.
    \item Repeat Steps 2-5 until the value of the optimized observable converges.
\end{enumerate}

In the application where the optimized observable is the expectation value of the Hamiltonian, once the aforementioned procedure is complete one has created a reusable circuit which initializes the ground state of the Hamiltonian. One challenge facing VQE is the choice of variational ansatz. To address this, Ref. \cite{grimsley_2019} devises ADAPT-VQE, which unlike VQE does not fix the variational anzatz from the beginning. Instead, it iteratively grows the ansatz by adding an operator to it from a pre-selected operator pool $\hat{A}_m$, optimizes the grown ansatz, and repeats the grow-and-optimize process until convergence happens. The steps to implementation are as follows\cite{grimsley_2019}:
\begin{enumerate}
    \item Define the operator pool.
    \item Create a variational ansatz with only an initialization to a simple reference state as its component for now.
    \item Run a set of quantum circuits with the variational ansatz followed by an element of the operator pool $\hat{A}_m$, with the aim of finding the gradient of the parameters controlling $\hat{A}_m$ at the location where said parameters would set $\hat{A}_m$ to the identity. 
    \item Repeat Step 3 for each element of the operator pool. If the magnitude of all of the gradients is below a certain threshold, exit the algorithm. Otherwise, choose the operator pool element with the highest gradient and append it to the end of the variational ansatz.
    \item Optimize the new variational ansatz using VQE
    \item Go back to Step 3.
\end{enumerate}

One variation of ADAPT-VQE is TETRIS-ADAPT-VQE, which attempts to prioritize operators that can be completed in parallel with the rest of the ansatz.\cite{PhysRevResearch.6.013254}. Another, one which is directly relevant to the work in Chapter \ref{chap:0vBBandlargeQCD}, is SC-ADAPT-VQE, developed in Ref. \cite{Farrell:2023fgd}. Its procedure is as follows \cite{Farrell:2023fgd}:
\begin{enumerate}
    \item Given a system that we want to conduct state-preparation on, create several subsystems of it with different sizes, all smaller than the original.
    \item Use ADAPT-VQE to create a state-preparation circuit for each of the smaller subsystems, with the operator-pool fed to ADAPT-VQE being one where all of the operators automatically scale with system size.
    \item Scale the operators in the subsystems' initialization-circuits up to the original system's size, and use an extrapolation-scheme to obtain the values of the parameters at the original system size.
\end{enumerate}

Using SC-ADAPT-VQE, one can in principle create an initialization-circuit too large to simulate on a classical computer by running ADAPT-VQE entirely on a classical computer for several analogous systems of smaller size, then extrapolating up to the full-size system \cite{Farrell:2023fgd}. A recently-developed variant, $(SC)^2$-ADAPT-VQE\cite{gustafson2024surrogate}, uses a classical surrogate to reduce the size of the operator pool that SC-ADAPT-VQE uses.

One question facing VQE and its relatives discussed here is the choice of classical optimizer used to select the value for the parameters for each iteration after the first. This is explored in Chapter \ref{chap:prepofsu3latyangmilsvacvarqumethods}, where a simple gradient descent and a Bayesian optimizer are compared as choices for the classical optimizer.

In Chapter \ref{chap:1p1dQCD}, the low-lying spectrum of the SU(3) Kogut-Susskind Hamiltonian, as well as phase transitions as the chromoelectric coupling constant, $g$ from Eqn. \ref{eq:KogutSuskindHam_intro}, is varied, are studied. The state-preparation in these scenarios are done using classical methods, as state-preparation using VQE on a quantum device failed to converge. Nonetheless, a VQE circuit is still used to initialize the vacuum state of the SU(3) Kogut-Susskind Hamiltonian with one physical lattice site and one quark flavor. The VQE ansatz used is based on the $\beta$-angle circuit designed in Ref. \cite{klco:2019xro}, but the knowledge about the color-singlet state in this particular situation is used to reduce both the complexity of the circuit and the number of degrees of freedom among the VQE parameters. This VQE circuit is used in Chapter \ref{chap:1p1dSM} to initialize the up-quark qubits in preparation for the Trotterization circuit that executes the real-time dynamics of the $\beta$ decay.

In Chapter \ref{chap:0vBBandlargeQCD}, a general method for constructing the color-singlet space of the SU(3) color singlet space Hamiltonian is devised and proved. This method is then used to simplify the SU(3) Kogut-Susskind Hamiltonian for purposes of VQE and all other applications restricted to the color singlet space. It is also used to compactify the Hamiltonian in all calculations in Chapter \ref{chap:0vBBandlargeQCD} that involve classical methods.  
Lepton-initialization circuits are created. The ones for a periodic boundary conditions lattice are built out of the $e^{\theta(XY \pm YX)}$ building-block circuit from \cite{Farrell:2023fgd}, while for open boundary conditions new parametrized circuits are constructed. The technique of using FSWAP networks to implement the Jordan-Wigner transformation is used in order to enable the electron-qubits and the neutrino-qubits to be initialized separately.
The mass-hierarchies of the choices of parameters in the classical simulations in Sec. \ref{sec:0vBB_classicalsim} are verified by a classical exploration of the low-lying spectrum, similar to the low-lying spectrum exploration in Chapter \ref{chap:1p1dQCD}.




\chapter{Preparation of the SU(3) Lattice Yang-Mills Vacuum with Variational Quantum Methods}
\label{chap:prepofsu3latyangmilsvacvarqumethods}

{\it This chapter is associated with Ref. \cite{ciavarella:2021lel}:}
{\it ``Preparation of the SU(3) Lattice Yang-Mills Vacuum with Variational Quantum Methods" by Anthony Ciavarella and Ivan Chernyshev}

	
	
	
	
	
	
{
}


\section{Introduction}
As mentioned in Sec. \ref{sec:intro_stateprep}, VQE has several properties, including low resource demand and high resilience to noise compared to many of its alternatives and reusability of VQE ansatz circuits once the optimization is complete, that make it an attractive option for state-preparation in quantum simulations of lattice gauge theory. However, to scale VQE calculations to situations with a useful quantum advantage, it will be necessary to understand how to connect these small lattice calculations to a calculation on a larger lattice and how the optimization procedure performs as system size is increased.

In this chapter, the application of VQE to pure SU(3) lattice Yang-Mills gauge theory is studied. This provides a starting point for understanding the resources required to simulate lattice QCD on a quantum computer. We perform a VQE calculation of the vacuum state for one and two plaquette systems using superconducting quantum processors. We use the former as a testing-ground for evaluating choices of the classical optimizer used in the VQE optimization. We also examine how to apply ideas from domain decomposition in lattice QCD calculations on classical computers to the construction of ansatz states for VQE of large lattices from the vacuum state of smaller lattices. VQE is executed both on error-free classical simulators and on IBM's {\tt Manila} device~\cite{ibmq}.

\section{Electric Multiplet Basis}
\FloatBarrier
Quantum simulation of SU(3) Yang-Mills theory on a lattice can be performed with link variables connecting neighboring sites of the lattice. The Hamiltonian, first discussed by Kogut and Susskind \cite{physrevd.11.395}, is
\begin{equation}
	\hat{H} = \frac{g^2}{2a^{d-2}} \sum_{b,\text{links}} |\hat{E}^b|^2 + \frac{1}{2 a^{4-d} g^2} \sum_{\text{plaquettes}}\left(6 - \hat{\Box}(\mathbf{x})  - \hat{\Box}^\dagger(\mathbf{x}) \right) \ \ \ ,
\end{equation}
where $g$ is the coupling constant, $a$ is the lattice spacing and $d$ is the number of spacial dimensions. The plaquette operator $\hat{\Box}(\mathbf{x})$ is defined by
\begin{equation}
	\hat{\Box}(\mathbf{x}) = \text{Tr} \left(\hat{U}(\mathbf{x}, \mathbf{x} + a \mathbf{i}) \hat{U}(\mathbf{x} + a \mathbf{i}, \mathbf{x} + a \mathbf{i}  + a \mathbf{j}) \hat{U}(\mathbf{x} + a \mathbf{i} + a \mathbf{j}, \mathbf{x} + a \mathbf{j}) \hat{U}(\mathbf{x} + a \mathbf{j}, \mathbf{x}) \right) \ \ \ ,
\end{equation}
where $\hat{U}(\mathbf{x},\mathbf{y})$ is an SU(3) matrix on the link between sites $\mathbf{x}$ and $\mathbf{y}$ and $\mathbf{i}$ and $\mathbf{j}$ are unit vectors that define the orientation of the plaquette. This theory can be described in the electric field basis, where each link's Hilbert space is spanned by the state vectors $\ket{\mathbf{R}, m_L, m_R}$, where $\mathbf{R}$ is an irreducible representation of $SU(3)$, $m_L$ labels the component of the representation on the left side of the link, and $m_R$ labels the component of the representation on the right side of the link. Physical states in this Hilbert space are subject to a constraint from Gauss's law which requires the wavefunction of the links meeting at each vertex to form a singlet state.
\begin{figure}
\centering
	\includegraphics[scale=0.25]{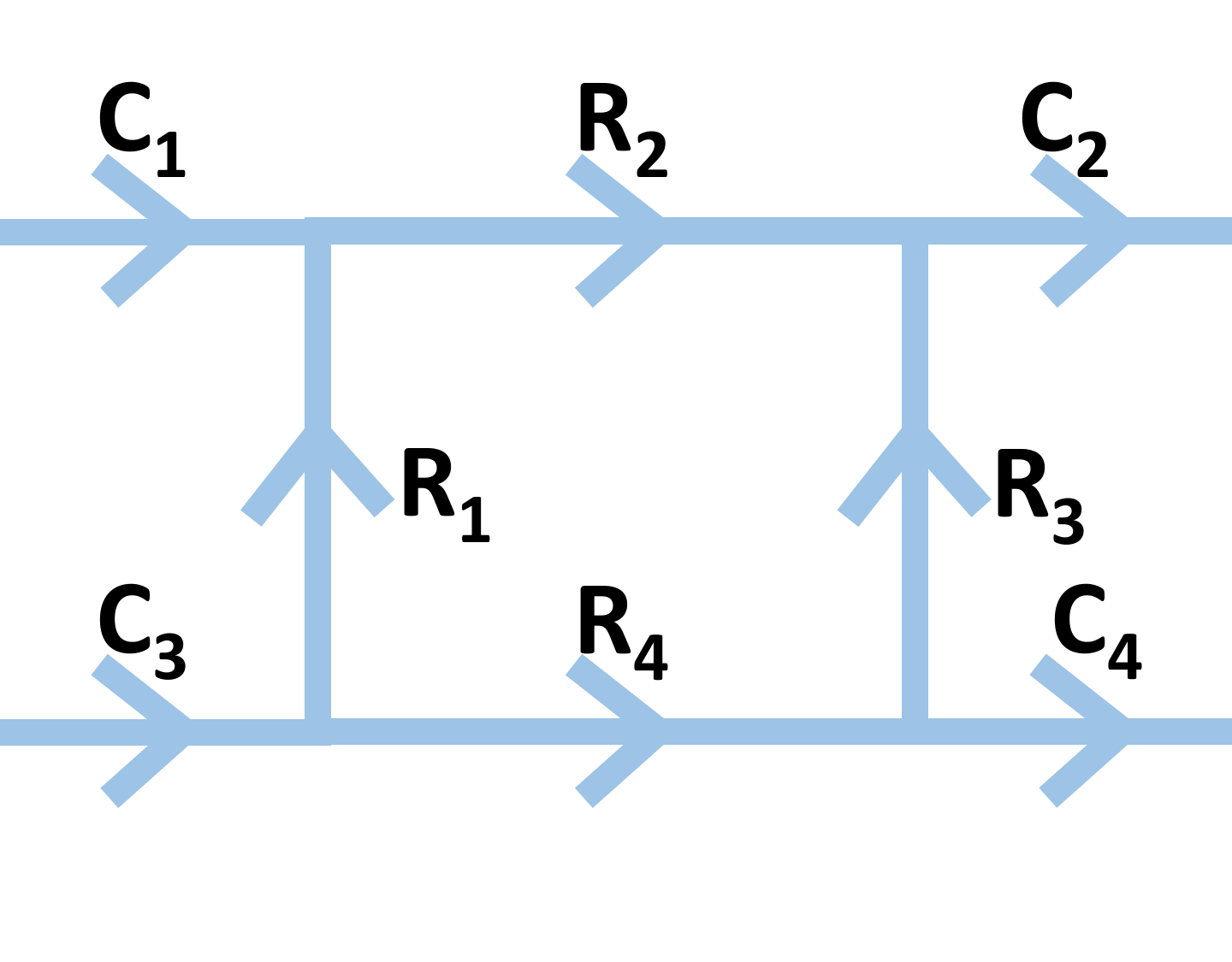}
	\caption{An SU(3) plaquette in a 1D chain of plaquettes. The electric multiplet basis states of the links in this figure are represented by $
		\Bigg | \chi\begin{pmatrix}\mathbf{C_1}, \mathbf{R_2},\mathbf{C_2} \\  \mathbf{R_1}, \mathbf{R_3} \\
			\mathbf{C_3}, \mathbf{R_4}, \mathbf{C_4}\end{pmatrix} \Bigg\rangle $ where each $\mathbf{C_i}$ and $\mathbf{R_i}$ labels the irrep on the corresponding link. }
	\label{fig:plaq_controls}
\end{figure}
In previous work, it was noted that for a lattice consisting of a chain of plaquettes, the Gauss's law constraint can be used to integrate out the irrep state labels $m_L$ and $m_R$ on each link \cite{banuls:2017ena, physrevd.101.074512,physrevd.103.094501}. Integrating out $m_L$ and $m_R$ allows basis states to be described by only specifying $\mathbf{R}$ on each link. Fig. \ref{fig:plaq_controls} shows an example of a plaquette in a chain with the basis labels necessary to specify its state.  For an SU(3) gauge theory, the representation on each link can be labeled by a pair of non-negative integers $p$ and $q$ that count the upper and lower tensor indices. These labels can be mapped onto a quantum computer in a local basis by using two registers of qubits on each link to represent $p$ and $q$ in binary. Alternatively, Gauss's law can be solved at each vertex on the lattice and the resulting physical states can be mapped onto the basis of a quantum computer. This global basis construction is not scalable to large lattices, but can be used to map small lattices onto near term devices. The number of states in the global basis that need to be considered can be reduced by making use of symmetries to study different sectors of the theory. For example, SU(3) lattice Yang-Mills theory has a color parity (CP) symmetry related to the invariance of the theory under reversal of the direction of the links. The global and CP invariant bases were studied in detail for one and two plaquettes in Ref. \cite{physrevd.103.094501}.

\FloatBarrier
	
\section{Single Plaquette}
\label{section:OnePlaq}
\FloatBarrier
A single plaquette is one of the simplest systems that can be considered in lattice gauge theory. In this work, the single plaquette system will be studied in the electric multiplet basis described in the previous section. In this formulation, Gauss's law guarantees that each link in the plaquette will have the same representation. Therefore, the basis states of the plaquette can be specified by $\ket{p,q}$, where $p$ and $q$ are specified earlier.  In units where the lattice spacing equals one, the Hamiltonian for a single plaquette is
\begin{figure}
\centering
	\includegraphics[scale=0.25]{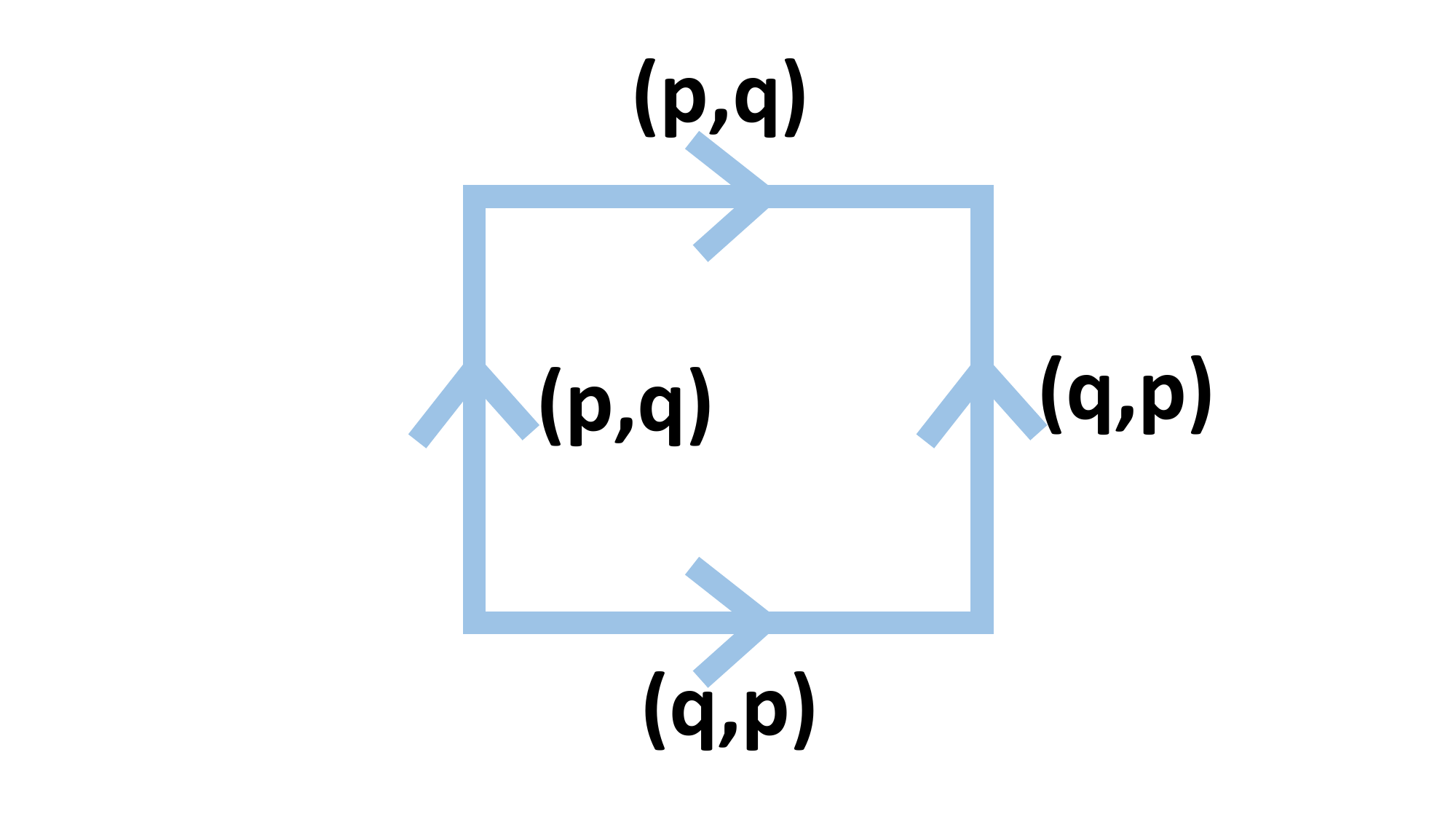}
	\caption{A single SU(3) plaquette. p and q label the chromo-electric flux on each link.}
\end{figure}
\begin{equation}
	\label{eq:HamOnePlaq}
	\hat{H} = 2g^2 \sum_b |\hat{E}^b|^2 + \frac{1}{2 g^2} \left(6 - \hat{\Box} - \hat{\Box}^\dagger \right) \ \ \ ,
\end{equation}
where $\sum_b |\hat{E}^b|^2$ is the Casimir for the chromo-electric field representation, given by
\begin{equation}
	\sum_b |\hat{E}^b|^2 \ket{p,q} = \frac{p^2 +q^2 + pq + 3p + 3q}{3} \ket{p,q} \ \ \ ,
    \label{eq:su3_singleplaquette_casimirdef}
\end{equation}
and the plaquette operator $\hat{\Box}$ acts on the basis states by
\begin{equation}
	\hat{\Box} \ket{p,q} = \ket{p+1,q} + \ket{p-1,q+1} + \ket{p,q-1} \ \ \ .
\end{equation}

While the exponential decay of correlations in gapped quantum systems is known to allow for state preparation using circuits localized in position space \cite{Klco_2020F}, the depth of the circuits needed to prepare the local color-space degrees of freedom has not been studied in as much detail. Due to Gauss's law guaranteeing every link in a single plaquette has the same chromo-electric flux, the single plaquette system can be used to study state preparation of the local color-space while avoiding the complications of spacial correlations.

\subsection{Vacuum Preparation}
\subsubsection{Initialization}
\label{section:VQEInitial}
\begin{figure}
    \centering
	\includegraphics{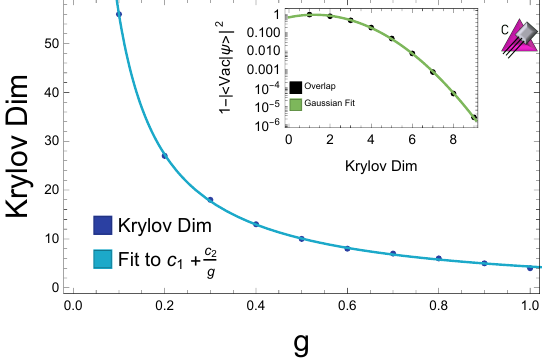}
	\caption{This figure shows the dimension of the Krylov subspace required for the overlap of the state prepared by the Lanczos algorithm, $\ket{\psi}$, with the true vacuum, $\ket{\text{Vac}}$, to satisfy $\abs{\bra{\psi} \ket{\text{Vac}}}^2 \geq 0.999999$. The inset panel shows the overlap with the true vacuum as a function of Krylov dimension for $g=0.5$.}
	\label{fig:one_plaq_krylov}
\end{figure}

VQE is a hybrid quantum algorithm that can improve the overlap of an initial state with the vacuum state. The performance of VQE has a strong dependence on the initial state used \cite{peruzzo_2014,mcclean_2016,Kandala_2017}. In applications of VQE to electronic structure problems, Hartree-Fock states and unitary coupled cluster states computed on classical computers have been used as initial starting points for VQE. However, lattice gauge theory does not have comparable classical calculations in the Hamiltonian formulation available. As an alternative, the Lanczos algorithm can used to initialize VQE for a single plaquette.\footnote{This application of Krylov subspaces to quantum simulation was developed in collaboration with other members of IQuS during the spring of 2020.} The Lanczos algorithm works by constructing the Krylov subspace spanned by $\{\ket{\psi}, \hat{H} \ket{\psi}, \dots, \hat{H}^n \ket{\psi}\}$ for some integer $n$ and initial state $\ket{\psi}$ and diagonalizing the Hamiltonian in this subspace \cite{Lanczos:1950zz}. Quantum variations of the Lanczos algorithm have also been proposed for use in the study of state preparation \cite{motta_2019}. The result of applying the Lanczos algorithm to a single plaquette using the electric vacuum as the initial state is shown in Fig. \ref{fig:one_plaq_krylov}.\footnote{The icons in the corners of the plots in this text were introduced in Ref. \cite{klco:2019xro} and are available at {\tt iqus.uw.edu/resources/icons/}. The pink icons indicate the calculations in the figure were performed on a classical computer and the blue icons indicate the calculations in the figure were performed on a quantum computer.}
For a fixed coupling, the overlap with the true vacuum is shown to scale asymptotically as a Gaussian with the Krylov dimension used in the Lanczos algorithm. The dimension of the Krylov subspace needed to reach a fixed accuracy scales as $\frac{1}{g}$. This behavior can be seen to follow from the structure of the single plaquette vacuum wavefunction. The vacuum wavefunction is asymptotically Gaussian in the chromo-electric field with a width inversely proportional to $g$. Each time $\hat{H}$ is applied to increase the dimension of the Krylov subspace, the maximum $p$ and $q$ included in the Krylov subspace is increased by 1. Therefore, the size of the vacuum wavefunction components added by increasing the Krylov dimension fall off asymptotically as a Gaussian, and the Krylov dimension needed to reach a desired accuracy $\epsilon$ scales as $\frac{\log(\frac{1}{\epsilon})}{g}$. It should be noted that an exponential convergence with field truncation has also been observed in the simulation of scalar field theories \cite{Klco:2018zqz} and $U(1)$ gauge theories \cite{zache2021achieving}, and has been proven to be a rather generic property of theories involving bosonic modes \cite{tong2021provably}.

The Lanczos algorithm provides approximate wavefunction components of the vacuum state that must be mapped into a quantum circuit to be useful for state preparation. The state prepared by using a $d$-dimensional Krylov subspace potentially spans all basis states with $p,q < d$. Therefore, a state with nontrivial support on $d^2$ basis states must be prepared, which can be done using a circuit of length $O(d^2)$ using standard state preparation procedures \cite{Nielsen:2011:QCQ:1972505}. Using the previous result on the Krylov dimension required to reach an accuracy $\epsilon$, a quantum circuit of size
\begin{equation}
	S = O \left(\left(\frac{\log(\frac{1}{\epsilon})}{g}\right)^2 \right) \ \ \ ,
\end{equation}
can be used to prepare the vacuum of a single plaquette with coupling $g$ on a quantum computer within an accuracy of $\epsilon$.

\subsubsection{Optimization}
\label{section:VQEOpt}
The VQE algorithm makes use of a classical optimizer to improve the overlap of the ansatz state with the actual vacuum. In previous work, Bayesian optimizers have been used in the VQE algorithm to prepare the ground state of the Schwinger model \cite{physreva.98.032331} and to prepare hadron states in an SU(2) gauge theory \cite{atas:2021ext} on small lattices. Bayesian optimization minimizes an objective function by iteratively constructing an interpolator, usually a Gaussian process, from existing data and optimizing the interpolator.  It is ideal for optimizations where the number of available evaluations of the objective function is limited (typically to a few hundred evaluations), the objective function is continuous, and the dimensionality of the domain is no more than 20 \cite{frazier2018tutorial}. On existing hardware that only has a handful of qubits available, circuits that can prepare a generic ansatz state can be implemented with fewer than 20 parameters. However, as quantum computers grow in qubit count and coherence time, this will no longer be true. To reach a quantum advantage, it will be important to understand when Bayesian optimization breaks down. To test the performance of a Bayesian optimizer for lattice gauge theory, VQE was simulated without noise on a classical computer for a single SU(3) plaquette with a truncation of $p,q \leq 3$. This system can be represented using 4 qubits on a quantum processor. The vacuum state of this system lies in a 10-dimensional CP-invariant subspace which can be parametrized in spherical coordinates with 9 degrees of freedom. The details of how the Bayesian optimization was performed are available in Appendix \ref{app:bayes}.

\begin{figure}
	\includegraphics{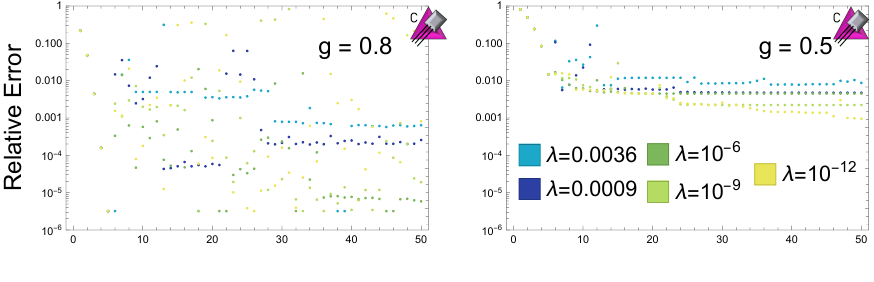} \\
	
	\includegraphics{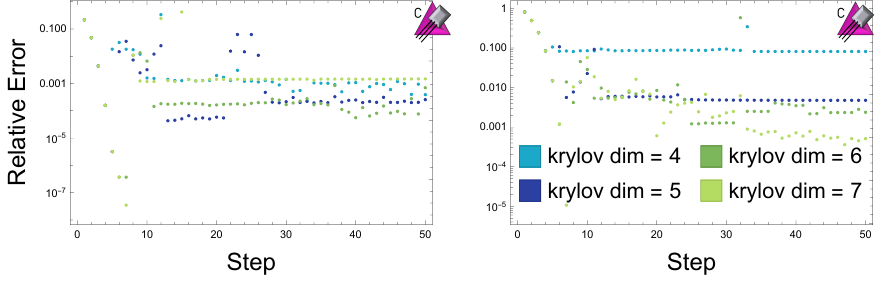}
	\caption{The relative error in the estimation of the vacuum energy obtained by performing a classical simulation of VQE using Bayesian optimization for a single plaquette with $p,q\leq 3$. The left panel is for $g=0.8$ and the right panel is for $g=0.5$. The top panel shows the results of Bayesian optimization as a function of the number of iterations of the optimization for different values of the regulator $\lambda$. Each of the calculations in the top panel was initialized with the vacuum states obtained from the Lanczos algorithm with subspace of Krylov dimension equal to 5. The bottom panel shows the result of Bayesian optimization using $\lambda=0.0009$ with different maximum Krylov dimensions.}
	\label{fig:BOpreccouplingcomparison}
\end{figure}

The results of the simulation of VQE with a Bayesian optimizer are shown in Fig. \ref{fig:BOpreccouplingcomparison}. In these calculations, the Gaussian process used to model the energy function being minimized suffered from multicollinearity. This was mitigated with Tikohonov regularization, which in this context is equivalent to adding a small constant term $\lambda$ to the covariance matrix of the energies \cite{mittikhonovnotes}. As this figure shows, the convergence of the Bayesian optimizer has a dependence on the regulator $\lambda$. The energy that the Bayesian optimizer converges to cannot be made arbitrarily close to the vacuum energy because at sufficiently small values of $\lambda$, multicollinearity returns and the covariance matrix cannot be inverted, causing the Bayesian optimizer to fail. The lower panels in Fig. \ref{fig:BOpreccouplingcomparison} show the dependence of the Bayesian optimizer's convergence on the dimension of the Krylov subspace used to initialize the calculation. For certain initializations, the Bayesian optimizer is not able to improve upon the initial state's overlap with the actual vacuum. Even for this modest system size, Bayesian optimization has limitations in how close it can get to the vacuum state.

Gradient descent is an alternative classical optimizer that can be used in VQE. Gradient descent evaluates the gradient of the energy, $\grad{f(\bf{x})}$, at the current step's ansatz parametrization ${\bf{x}}_i$, then selects the next step's ansatz parametrization ${\bf{x}}_{i+1}$ according to
\begin{equation}
	{\bf{x}}_{i+1} = {\bf{x}}_{i} - \eta \grad{f({\bf{x}}_i)} \ \ \ ,
\end{equation} 
where $\eta$ is a learning rate that controls the convergence of the gradient descent. Convergence to a local minimum can be guaranteed by the use of backtracking, where $\eta$ is steadily decreased during the course of the calculation \cite{truong2019convergence}. Alternatively, the step size can be selected by using Bayesian optimization to perform a line search \cite{tamiya2021stochastic}. In applications to VQE, the gradient can be computed on a quantum processor by making use of parameter shift formulas which give the gradient without discretization errors due to large shift size \cite{PhysRevA.101.032308}. The use of gradient descent as the classical optimizer in VQE will require the energy of the state to be calculated on the quantum processor a number of times equal to two times the number of parameters in the circuit ansatz per step in the optimization. For comparison, Bayesian optimization only requires the energy to be computed once per step. The increase in quantum resources per step in the optimization may be offset by a faster rate of convergence and ability to converge to the actual vacuum state. As an optimizer, gradient descent also requires fewer classical resources per step than Bayesian optimization. This is because with gradient descent, the classical computer only needs to perform subtraction during gradient descent. Bayesian optimization, on the other hand, requires the computations of determinants and inverses of a matrix whose dimension is equal to the number of times the energy was previously evaluated.

\begin{figure}
	\includegraphics{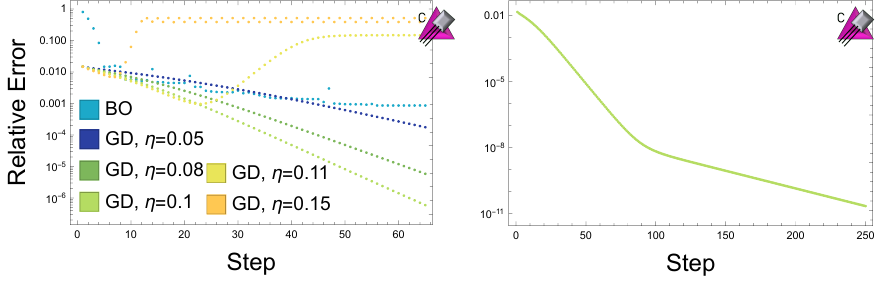}
	\caption{The relative error in the estimation of the vacuum energy obtained by performing a classical simulation of VQE for a single plaquette with $p,q\leq3$. The coupling is $g = 0.5$ and the initial state was obtained from the Lanczos algorithm using a Krylov dimension of 5.  The left panel shows a comparison of the results obtained by performing VQE using a Bayesian optimizer to those obtained by performing VQE using a numerical gradient descent for different learning rates $\eta$.  The right panel shows the results of 250 iterations of gradient descent with $\eta=0.1$.  }
	\label{fig:compareBOvsGD}
\end{figure}

\begin{figure}
	\includegraphics{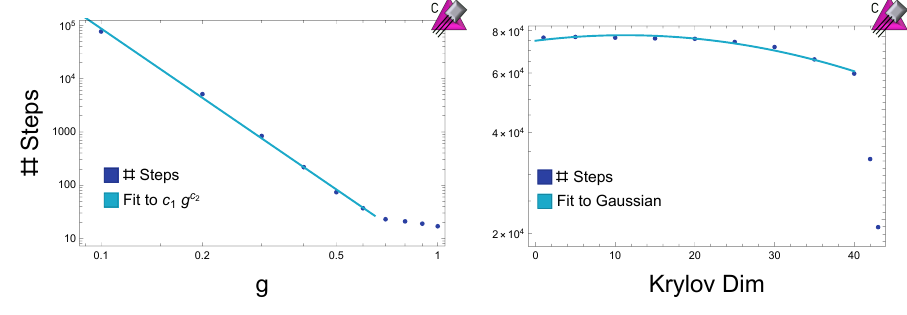}
	\caption{The left panel shows the number of steps needed for VQE using a backtracking gradient descent to converge to the true vacuum with an accuracy of 0.999 as a function of coupling for a single plaquette with $p,q \leq 31$. The right panel shows the number of steps needed for a backtracking gradient descent to converge to the true vacuum with an accuracy of 0.999  for $g=0.1$ as a function of the dimension of the Krylov subspace used to obtain the initial state.}
	\label{fig:grad_scaling}
\end{figure}

Fig. \ref{fig:compareBOvsGD} compares,  for a single plaquette truncated at $p,q\leq3$ and with $g=0.5$, the results of using Bayesian optimization for the classical optimizer to those of using numerically-computed gradient descent. The Bayesian optimizer shown in this plot was run with $\lambda=10^{-12}$. Both optimizers were initialized with the vacuum obtained using the Lanczos algorithm with a Krylov dimension of $5$. As this plot shows, the Bayesian optimizer converges above the vacuum energy, while VQE using gradient descent with a sufficiently small $\eta$ is limited only by the number of steps performed in the optimization. To understand if VQE can offer a quantum advantage, it is helpful to know how many steps in the optimizer must be performed to reach a certain level of accuracy. Fig. \ref{fig:grad_scaling} shows the number of steps needed for a backtracking gradient descent to converge for a single plaquette with a truncation of $p,q\leq 31$. This truncation was chosen so that the relative error in the mass gap and the vacuum expectation of the plaquette operator due to field truncation was $\leq 1 \%$ for each coupling studied. The left panel shows that as $g$ is decreased, the number of steps needed by the gradient descent algorithm to start from the electric vacuum and reach a state $\ket{\psi}$ with $\abs{\bra{\text{Vacuum}} \ket{\psi}}^2 \geq 0.999$ scales as $O(g^{-4})$. The number of steps needed to reach this level of accuracy can be decreased by beginning the optimization at a state closer to the vacuum, such as a state obtained from the Lanczos algorithm. The right panel in Fig. \ref{fig:grad_scaling} shows the number of steps needed by a backtracking gradient descent to converge to $\abs{\bra{\text{Vacuum}} \ket{\psi}}^2 \geq 0.999$ for a coupling $g=0.1$ as a function of the dimension of the Krylov subspace used in the Lanczos algorithm to initialize the starting state. From the fit in the right panel, it appears that the number of steps required for the gradient descent to converge scales asymptotically as a Gaussian as a function of the Krylov dimension used. This is expected, as the discussion in the previous section showed that the error in the state obtained from the Lanczos algorithm falls off asymptotically as a Gaussian as a function of the Krylov dimension. By beginning in a state obtained from the Lanczos algorithm and performing the optimization step using gradient descent, classical simulations of the VQE algorithm are able to reach the vacuum state of a single plaquette at weak couplings that are beyond the reach of Bayesian optimization. Based on these results, Bayesian optimization will not be a practical optimizer for VQE calculations at scale, while gradient based methods have a chance of reaching the vacuum state at scale.

\FloatBarrier
\subsection{Hardware Implementation}
\label{section:hardwareimp}
The discussion in the previous section suggests that VQE should be capable of preparing the vacuum state for a single plaquette. However, existing quantum hardware suffers from the effects of noise and imperfect gate implementations. This will have an impact on how VQE performs in practice. To understand how near-term hardware will perform in the simulation of SU(3) lattice Yang-Mills theory, IBM's {\tt Manila} superconducting quantum processor was used to perform a VQE calculation for a single plaquette \cite{ibmManila}.
\begin{figure}
\centering
	\includegraphics{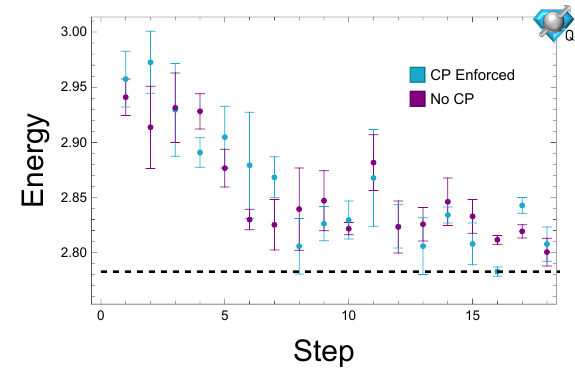}
	\caption{Variational state preparation of the vacuum state for a single plaquette truncated at $\mathbf{8}$ with $g=1$ run on the {\tt Manila} quantum processor. The blue points show the results of gradient descent with CP symmetry enforced in the rotation angles in the ansatz circuit and the purple points show the result of not explicitly enforcing CP symmetry in the state. The data in this figure is available in Table \ref{tab:OnePlaq8Data}.}
	\label{fig:VQE_one_plaq_8}
\end{figure}

The SU(3) lattice Yang-Mills Hamiltonian possesses a CP symmetry that guarantees that the amplitude of a given representation in the vacuum wavefunction will be the same as the amplitude of the conjugate representation. In principle, this symmetry can be used to restrict the state preparation circuit used in VQE which will reduce the number of free parameters. However, in the presence of noise and imperfect gate implementations, attempting to explicitly enforce the symmetry may prevent the actual state prepared on the quantum processor from respecting the symmetry. This would be the case if, hypothetically, the rotations in the circuit suffered from a constant offset error that was not corrected for. To understand if this is an issue on existing hardware, a single plaquette was simulated in the global basis truncated at a representation of $\mathbf{8}$. The Hamiltonian is given by Eq. (14) of Ref. \cite{physrevd.103.094501}. A VQE state preparation procedure described in Appendix \ref{appendix:hardware} was used to prepare the vacuum state starting from the electric vacuum and to optimize the angles using gradient descent. VQE was performed both by enforcing CP symmetry in the rotation angles in the circuit ansatz and by allowing all three of the angles to vary freely. The results of both calculations are displayed in Fig. \ref{fig:VQE_one_plaq_8}. As this figure shows, explicitly enforcing the CP symmetry in the VQE calculation does not break the symmetry in the vacuum state prepared using VQE on this hardware. The ability to explicitly enforce CP symmetry in the ansatz circuit will be helpful when performing VQE calculations on larger systems where the number of free parameters is much greater.

\begin{figure}
    \centering
    \includegraphics{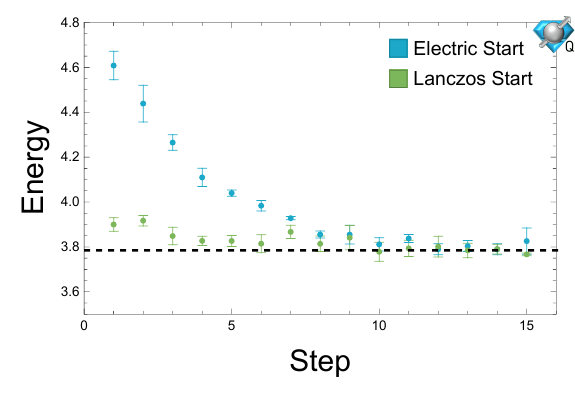}
	\caption{Variational state preparation of the vacuum state for a single plaquette truncated at $\mathbf{6^+}$ in the color parity basis with $g=0.8$ run on the {\tt Manila} quantum processor. The blue points show the result of gradient descent beginning at the electric vacuum and the green points begin at the state obtained using the Lanczos algorithm with a Krylov dimension of two. The data in this figure is available in Table \ref{tab:OnePlaq6Data}.}
	\label{fig:VQE_one_plaq_6}
\end{figure}

As discussed in Section \ref{section:VQEInitial}, the Lanczos algorithm can be used to obtain an initial ansatz for the VQE algorithm. At a coupling of $g=1$, the vacuum state obtained using a two dimensional Krylov subspace has an overlap with the true vacuum within experimental errors on the {\tt Manila} chip \cite{ibmManila}. To accurately reproduce physics at a lower coupling, more electric field representations must be included. This can be done without increasing the qubit count by performing a calculation in the color parity basis. Using two qubits, the color parity basis allows the $\mathbf{6}$ and $\mathbf{\bar{6}}$ representations to be included, which is sufficient to accurately describe a plaquette with a coupling of $g=0.8$. Fig. \ref{fig:VQE_one_plaq_6} shows the results of performing VQE for a single plaquette with $g=0.8$ in the color parity basis, beginning both at the electric vacuum and the vacuum obtained using a Krylov subspace of dimension two. As this figure shows, pre-conditioning with the vacuum obtained using the Lanczos algorithm allows one to begin closer to the actual vacuum and to converge to the true vacuum faster. Note that in both Fig. \ref{fig:VQE_one_plaq_8} and \ref{fig:VQE_one_plaq_6}, the energy computed fluctuates at late steps in the gradient descent instead of converging. This is because the gradient is computed on the {\tt Manila} chip with both statistical and systematic errors. As the optimizer approaches the vacuum state, the magnitude of the gradient vector decreases. Once the size of the gradient vector is comparable to the device errors, it can no longer be reliably computed and the updates to the circuit parameters are random noise which leads to the displayed fluctuations. This is a generic feature of having uncertainties in the computation of the gradient and will have to be considered when devising stopping criterion for VQE calculations of larger systems.

\FloatBarrier

\section{Multiple Plaquettes}
\FloatBarrier
The single plaquette calculations in Section \ref{section:OnePlaq} provide insight into the requirements of state preparation in a simple system. To perform calculations at scale, these insights need to be combined with features that only occur on larger lattices, such as Gauss's law constraints that can't be solved exactly without sacrificing locality.
\begin{figure}
        \centering
	\includegraphics[scale=0.25]{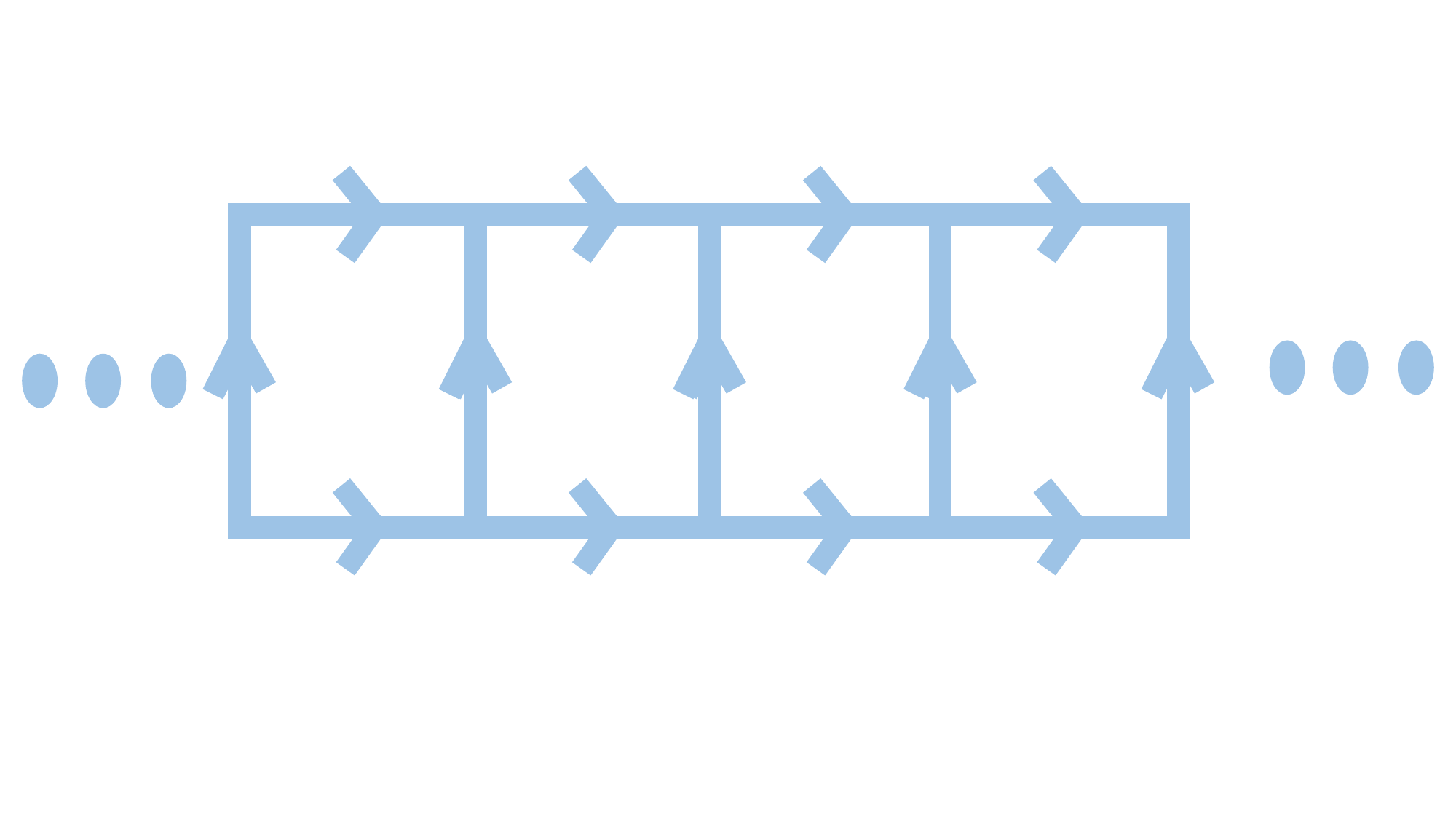}
	\caption{A lattice composed of a chain of plaquettes.}
	\label{fig:Plaq_Chain}
\end{figure}
The Lanczos algorithm provides a good starting ansatz for VQE on a single plaquette, but it is inefficient on larger lattices. This can be seen by using the electric vacuum as the initial state for a chain of $L$ plaquettes with periodic boundary conditions (PBC) as shown in Fig. \ref{fig:Plaq_Chain}. When a Krylov subspace with dimension $d$ is used, every basis state with $d$ plaquettes excited to have a loop of electric fields in the $\mathbf{3}$ representation will occur with equal amplitude. There are ${L \choose d}$ of these states and their superposition requires non-local circuits to capture the non-local correlations in the state. This leads to the circuit required to prepare the state given by the Lanczos algorithm growing exponentially in size with the Krylov dimension, and therefore no quantum advantage. An alternative approach is to use a form of domain decomposition.

In lattice QCD calculations on classical computers, a large amount of time is spent solving discretized versions of the Dirac equation. These calculations have been accelerated by making use of a domain decomposition \cite{LUSCHER2004209,frommer2014adaptive,Heybrock_2014}. Domain decomposition accelerates the calculation by solving the Dirac equation in separate sub-regions and then stitching the solutions together. Similar to solving the Dirac equation, directly preparing the vacuum state for a theory on a large lattice is difficult because the Hilbert space associated with the entire lattice is too large to efficiently work with. The ideas behind domain decomposition can be applied in a VQE calculation by splitting the lattice into separate disconnected sub-regions and preparing each sub-region in its vacuum state (note that there will be links between these regions that will remain unexcited). The vacuum state of each sub-region can be computed classically or in another VQE calculation. The VQE algorithm can then be used to excite links in-between the sub-regions and stitch together the sub-regions to form the vacuum state for the entire lattice. SU(3) Yang-Mills is a theory with spatial correlations that decay exponentially fast with distance, so it is anticipated that the domain decomposition ansatz should converge exponentially fast to the true vacuum as the domain size is increased. 

Conceptually, this approach to vacuum state preparation is similar to the density matrix renormalization group (DMRG) algorithm on classical computers \cite{PhysRevLett.69.2863}. In DMRG, the vacuum state of a lattice is prepared, and the density matrix of a sub-region is diagonalized. The eigenstates of the density matrix with largest weight are then used as the local basis for a calculation on a larger lattice. In this manner, DMRG constructs the vacuum state for a large lattice from the vacuum state for smaller regions. This is analogous to beginning the VQE optimization in a domain-decomposed vacuum, except the calculation on the quantum computer has no need to extract eigenstates of the density matrix for subregions. Once the desired lattice length is achieved, DMRG optimizes the approximation to the vacuum state by decomposing the system into left and right blocks and using the eigenstates of the density matrix of the subregions to generate a new basis for the regions. By growing and shrinking the size of the left and right blocks, DMRG is able to converge to the true vacuum state. The process of growing and shrinking the blocks used is analogous to the stitching procedure described in this work to improve the overlap with the true vacuum, except, once again, the quantum calculation does not require the diagonalization of density matrices.

While this stitching procedure will be explicitly demonstrated for a quasi one dimensional system, it can be performed in higher dimensions as well. For a system with three spatial dimensions, the sub-regions initialized in their vacuum state will be cubes of some size. Unlike in one dimension, the number of links left unexcited between the initial subregions will scale as the surface area of the subregions. A sequence of unitary transformations acting on the individual unexcited links, controlled by their neighboring links on the two cubes they connect, can be optimized using VQE to get closer to the vacuum state of the entire lattice. By limiting the number of links each unitary acts on in this manner, the number of free parameters in the VQE ansatz circuit can be restricted to grow linearly with the surface area instead of exponentially as it could if all links were allowed to be acted on simultaneously.

\subsection{Domain Decomposition on Plaquette Chains}
A lattice composed of a chain of plaquettes as shown in Fig. \ref{fig:Plaq_Chain} with PBC displays many of the complications inherent to larger lattices while still being tractable to simulate on classical computers. A domain decomposition of a plaquette chain can be performed by breaking up the lattice into separate sub-chains, preparing each subchain in its vacuum state and using VQE to optimize circuits that act on the boundaries and space between the domains to stitch them together.

\begin{table}
\centering
	\begin{tabular}{| c | c | c |} 
		\hline
		& State 1 & State 2  \\ [0.5ex] 
		\hline
		$R_1$ & $
		\Bigg | \chi\begin{pmatrix}\mathbf{1}, \mathbf{1},\mathbf{1} \\  \mathbf{1}, \mathbf{1} \\
			\mathbf{1}, \mathbf{1}, \mathbf{1}\end{pmatrix} \Bigg\rangle \nonumber$ & $
		\Bigg | \chi\begin{pmatrix}\mathbf{1}, \mathbf{3},\mathbf{1} \\  \mathbf{3}, \mathbf{\bar{3}} \\
			\mathbf{1}, \mathbf{\bar{3}}, \mathbf{1}\end{pmatrix} \Bigg\rangle \nonumber$  \\ 
		\hline
		$R_2$ & $
		\Bigg | \chi\begin{pmatrix}\mathbf{3}, \mathbf{1},\mathbf{1} \\  \mathbf{\bar{3}}, \mathbf{1} \\
			\mathbf{\bar{3}}, \mathbf{1}, \mathbf{1}\end{pmatrix} \Bigg\rangle \nonumber$ & $
		\Bigg | \chi\begin{pmatrix}\mathbf{3}, \mathbf{3},\mathbf{1} \\  \mathbf{1}, \mathbf{\bar{3}} \\
			\mathbf{\bar{3}}, \mathbf{\bar{3}}, \mathbf{1}\end{pmatrix} \Bigg\rangle \nonumber$  \\
		\hline
		$R_3$ & $
		\Bigg | \chi\begin{pmatrix}\mathbf{3}, \mathbf{1},\mathbf{1} \\  \mathbf{\bar{3}}, \mathbf{1} \\
			\mathbf{\bar{3}}, \mathbf{1}, \mathbf{1}\end{pmatrix} \Bigg\rangle \nonumber$ & $
		\Bigg | \chi\begin{pmatrix}\mathbf{3}, \mathbf{\bar{3}},\mathbf{1} \\  \mathbf{3}, \mathbf{\bar{3}} \\
			\mathbf{\bar{3}}, \mathbf{3}, \mathbf{1}\end{pmatrix} \Bigg\rangle \nonumber$  \\
		\hline
		$R_4$ & $
		\Bigg | \chi\begin{pmatrix}\mathbf{1}, \mathbf{1},\mathbf{3} \\  \mathbf{1}, \mathbf{\bar{3}} \\
			\mathbf{1}, \mathbf{1}, \mathbf{\bar{3}}\end{pmatrix} \Bigg\rangle \nonumber$ & $
		\Bigg | \chi\begin{pmatrix}\mathbf{1}, \mathbf{3},\mathbf{3} \\  \mathbf{\bar{3}}, \mathbf{1} \\
			\mathbf{1}, \mathbf{\bar{3}}, \mathbf{\bar{3}}\end{pmatrix} \Bigg\rangle \nonumber$  \\
		\hline
		$R_5$ & $
		\Bigg | \chi\begin{pmatrix}\mathbf{1}, \mathbf{1},\mathbf{3} \\  \mathbf{1}, \mathbf{\bar{3}} \\
			\mathbf{1}, \mathbf{1}, \mathbf{\bar{3}}\end{pmatrix} \Bigg\rangle \nonumber$ & $
		\Bigg | \chi\begin{pmatrix}\mathbf{1}, \mathbf{\bar{3}},\mathbf{3} \\  \mathbf{\bar{3}}, \mathbf{3} \\
			\mathbf{1}, \mathbf{3}, \mathbf{\bar{3}}\end{pmatrix} \Bigg\rangle \nonumber$  \\
		\hline
		$R_6$ & $
		\Bigg | \chi\begin{pmatrix}\mathbf{3}, \mathbf{3},\mathbf{3} \\  \mathbf{1}, \mathbf{1} \\
			\mathbf{\bar{3}}, \mathbf{\bar{3}}, \mathbf{\bar{3}}\end{pmatrix} \Bigg\rangle \nonumber$ & $
		\Bigg | \chi\begin{pmatrix}\mathbf{3}, \mathbf{1},\mathbf{3} \\  \mathbf{\bar{3}}, \mathbf{3} \\
			\mathbf{\bar{3}}, \mathbf{1}, \mathbf{\bar{3}}\end{pmatrix} \Bigg\rangle \nonumber$  \\ 
		\hline
		$R_7$ & $
		\Bigg | \chi\begin{pmatrix}\mathbf{3}, \mathbf{\bar{3}},\mathbf{\bar{3}} \\  \mathbf{3}, \mathbf{1} \\
			\mathbf{\bar{3}}, \mathbf{3}, \mathbf{3}\end{pmatrix} \Bigg\rangle \nonumber$ & $
		\Bigg | \chi\begin{pmatrix}\mathbf{3}, \mathbf{1},\mathbf{\bar{3}} \\  \mathbf{\bar{3}}, \mathbf{\bar{3}} \\
			\mathbf{\bar{3}}, \mathbf{1}, \mathbf{3}\end{pmatrix} \Bigg\rangle \nonumber$  \\ 
		\hline
	\end{tabular}
	\caption{This table enumerates the local Givens rotations required to initialize a domain vacuum on the plaquette chain truncated at an electric field representation of $\mathbf{3}$, (up to CP conjugates of the rotations listed here). The basis states are defined in the same way as the states in Fig. \ref{fig:plaq_controls}. The first column labels the rotation and the other two columns specify the basis states being rotated. $R_1$ excites a single plaquette loop of electric flux. $R_2$ through $R_5$ stretch the length of a loop of electric flux by one plaquette. $R_6$ and $R_7$ break a loop of electric flux into two loops. The basis labels used here were introduced in Ref. \cite{physrevd.103.094501}.}
	\label{tab:GivensRot3}
\end{table}
To be useful as an initial state for VQE, a quantum circuit for the preparation of these domain-decomposed vacuums must be designed. The circuit to prepare the vacuum state for a domain of length $l$ can be constructed recursively from the circuit to prepare the vacuum state for a domain of length $l-1$ as follows. A single plaquette state can be constructed by performing an $R_1$ rotation from Table \ref{tab:GivensRot3} and its CP conjugate on the qubits that make up the links in the plaquette. The two plaquette state can be prepared by applying $R_1$ rotations on two neighboring plaquettes and then applying $R_3$ and $R_4$ rotations on one of the plaquettes. The circuit that prepares the three plaquette vacuum state can then be constructed by exciting a third plaquette (i.e. apply an $R_1$ rotation), stretching over the previous two plaquettes (i.e. apply $R_3$ and $R_4$ rotations to the plaquettes that have been excited), and performing a rotation on the center plaquette to de-excite it (i.e. apply $R_6$ and $R_7$ rotations to the center plaquette). In general, the circuit to prepare a domain of size $l$ can be constructed from the circuit for a domain of size $l-1$ by exciting a neighboring plaquette, stretching it over the previous domain, and then de-exciting plaquettes in the center. In general, this approach to constructing circuits for a domain state scales exponentially with the size of the domain. 

The initial domain decomposition ansatz can be improved upon by stitching together the different domains. More specifically, in the circuit that prepares the vacuum ansatz, gates $R_1$ through $R_7$, along with their CP conjugates, can be applied to the plaquettes in-between the domains and VQE can be used to optimize the rotation angles. This stitching procedure can also be used to construct the vacuum for a larger domain instead of using the generic state preparation circuit. After performing the stitching, the overlap with the true vacuum can be increased further by layering another block of gates on the original domains and optimizing the angles with VQE again. Explicitly, if the state obtained from the VQE algorithm is $S(\vec{\theta}_2) D(\vec{\theta}_1)\ket{0}$, where $D(\vec{\theta}_1)$ prepares the states on the domain and $S(\vec{\theta}_2)$ stitches the domains together, then the ansatz state
\begin{equation}
	C(\vec{\theta}_1,\vec{\theta}_2,\vec{\theta}_3)\ket{0} = D(\vec{\theta}_3) S(\vec{\theta}_2) D(\vec{\theta}_1)\ket{0} 
\end{equation}
can be prepared on the quantum processor and the energy can be minimized as a function of $\vec{\theta}_1$, $\vec{\theta}_2$ and $\vec{\theta}_3$ using the VQE algorithm. Due to the exponentially decaying correlations in SU(3) Yang-Mills, the overlap with the true vacuum should increase exponentially with the number of additional gate layers stacked on the domains and their boundaries. 

A plaquette chain simulated in the multiplet basis with chromo-electric fields truncated at the $\mathbf{3}$ representation will be used to test the performance of the domain decomposition ansatz. Finite and infinite plaquette chains were studied using an MPS representation of states in the TEBD algorithm as described in Appendix \ref{appendix:PlaqTN}. Fig. \ref{fig:domain_decomp_finite} shows the results of optimizing different domain decomposition ansatzes for a chain of five plaquettes with $g=0.9$ and open boundary conditions. Fig. \ref{fig:Plaq5_HeatMap} shows the expectation of the electric energy for the initial single plaquette ansatz and the state obtained after stitching the boundaries together with VQE. As the size of the initial domains is increased, the overlap with the actual vacuum increases.  However, the improvement eventually saturates due to boundary effects. Due to the short correlation length at this coupling, even a single layer of stitching is able to achieve a high overlap with the actual vacuum.
\begin{figure}
	\includegraphics[width=1.\textwidth]{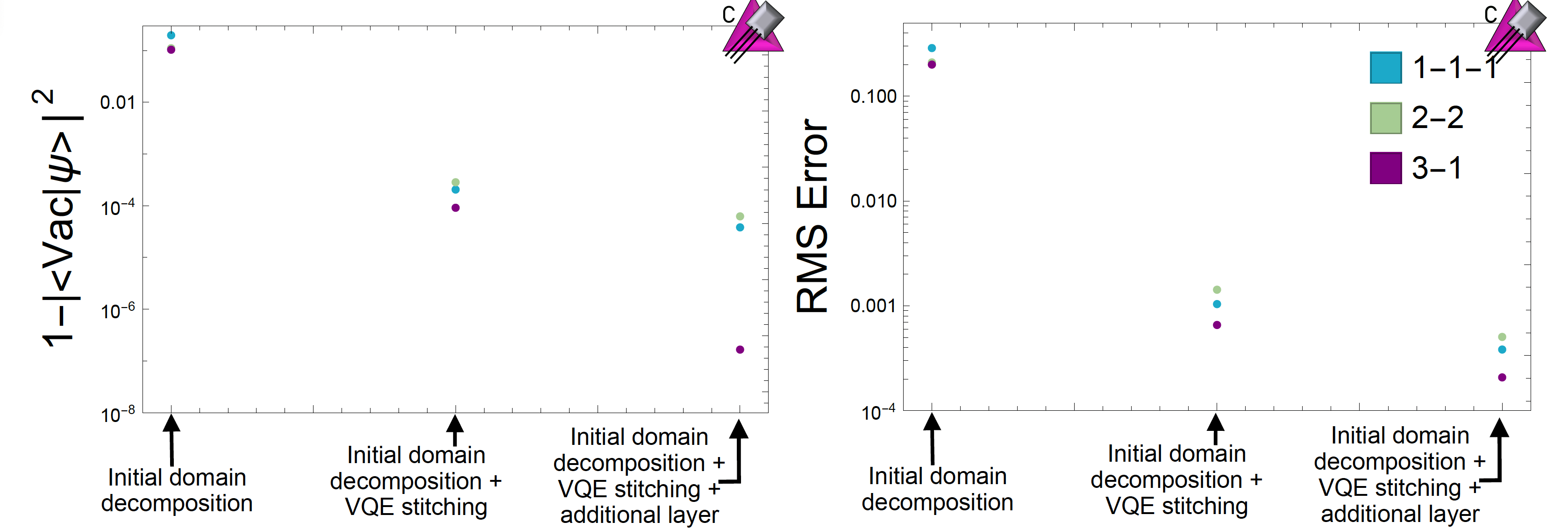}
	\caption{The left panel shows the overlap of different domain decompositions with the true vacuum. The right panel shows the RMS error in the expectation of the different single plaquette operators on the five plaquette lattice with open boundary conditions. The left-most points show the results for the initial domain decomposition, the middle points show the result after using VQE to stitch the boundaries of the domains together, and the right points show the results after using VQE to optimize another layer of circuits on the domains after stitching.}
	\label{fig:domain_decomp_finite}
\end{figure}

\begin{figure}
\centering
	\includegraphics{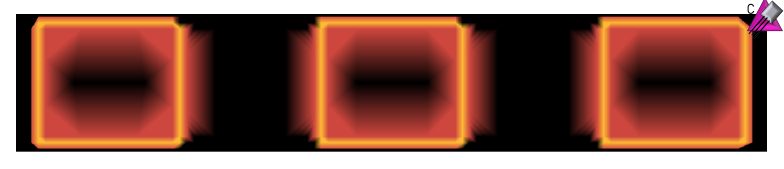}
	\includegraphics{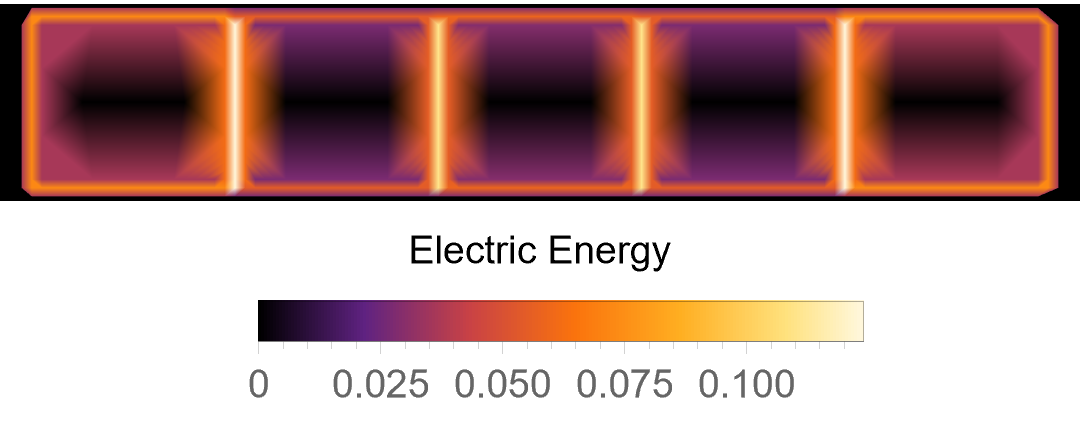}
	\caption{The top panel shows the expectation of the electric energy for a five plaquette chain with open boundary conditions where every other plaquette has been initialized to the single plaquette vacuum. The bottom panel shows the expectation of the electric energy after the boundaries of the initial domains have been stitched together with VQE.}
	\label{fig:Plaq5_HeatMap}
\end{figure}

\begin{figure}
	\includegraphics{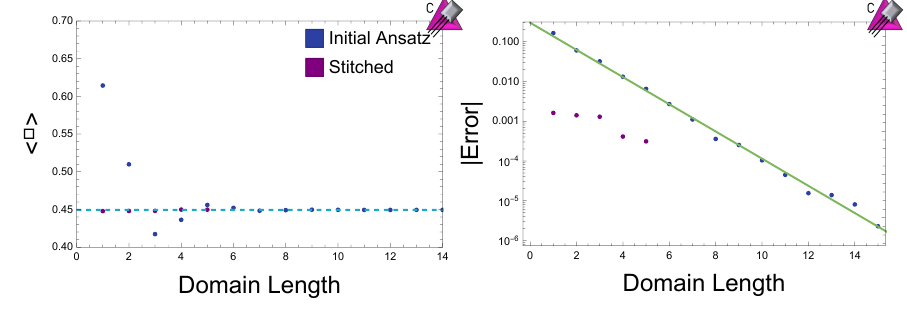}
	\caption{The left panel shows the expectation of a plaquette operator at the center of a domain as a function of domain length for both the initial ansatz and the state after using VQE to stitch domains together. The dashed blue line shows the vacuum expectation of a single plaquette operator on an infinite chain of plaquettes with $g=0.9$. The right panel shows the error in the vacuum plaquette expectation as a function of the domain size.}
	\label{fig:domain_decomp_infinite}
\end{figure}

To understand how the domain decomposition VQE ansatz performs for a large lattice, the time evolving block decimation algorithm was used to prepare the vacuum state and simulate VQE on an infinite plaquette chain as described in Appendix \ref{appendix:PlaqTN}. VQE was performed using gradient descent as the classical optimizer. The vacuum expectation of a single plaquette operator was chosen as a test observable to study the convergence to the true vacuum. As Fig. \ref{fig:domain_decomp_infinite} shows, the vacuum expectation of the plaquette operator converges exponentially fast with the domain size. A classically simulated version of VQE was used to simulate the stitching of small domains together. For domains of lengths 1-4 plaquettes, the initial domain vacuum was prepared using a generic state preparation circuit. For the initial domain of length five, the circuit to prepare the vacuum was constructed by stitching together a vacuum state preparation circuit for a domain of length three plaquettes and of length one plaquette. The circuit optimized in VQE consisted of the initial domain vacuum circuit, along with all rotations in Table \ref{tab:GivensRot3} with all rotation angles allowed to vary freely. For each domain size, the optimization of the stitching improved the estimation of the vacuum plaquette expectation by at least an order of magnitude.

\FloatBarrier
\subsection{Hardware Implementation}
As with the single plaquette case, it is instructive to study multiple plaquettes on existing quantum hardware. Unfortunately, simulating multiple plaquettes in a local basis as described in the previous section is beyond the reach of existing hardware. However, these techniques can be applied to state preparation in a global basis. IBM's {\tt Manila} quantum processor was used to simulate a two plaquette system truncated at an electric field representation of $\mathbf{3}$ in the global CP invariant basis \cite{ibmManila}. For this simple system, preparing the single plaquette vacuum is equivalent to using the vacuum state obtained using the Lanczos algorithm with a Krylov dimension of two. The results of performing VQE with the error mitigation procedures described in Appendix \ref{appendix:hardware} are shown in Fig. \ref{fig:VQE_two_plaq}. As this figure shows, the VQE algorithm is able to converge to the true vacuum energy whether it begins in the electric or single plaquette vacuum. However, by initializing the state in the single plaquette vacuum, the VQE algorithm is able to converge to the true vacuum state faster.
\begin{figure}
    \centering
	\includegraphics{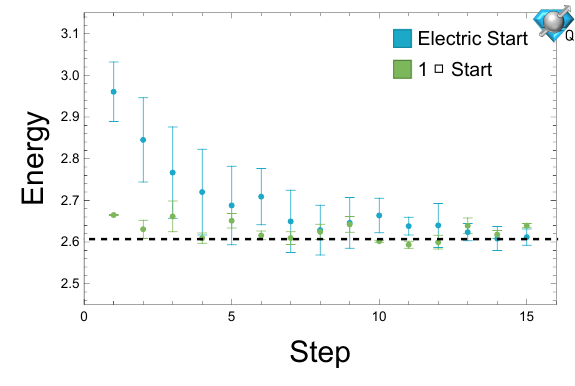}
	\caption{Variational state preparation of the vacuum state for a two plaquette system with $g=1$ and PBC run on the IBM {\tt Manila} quantum processor. The blue points show the results of performing gradient descent beginning at the electric vacuum and the green points show the results for beginning with the single plaquette vacuum.  The data in this figure is available in Table \ref{tab:TwoPlaqData}}
	\label{fig:VQE_two_plaq}
\end{figure}
While the two initial states converge to the same vacuum state, the uncertainties in the vacuum energy they converge to are quite different. This is due to the circuit ansatz used to initialize the state having redundancies in the angle parametrization of the state, leading to the two initial ansatzes converging to different sets of angles describing the same state. In the absence of noise on the quantum processor, these parametrizations would be equivalent. However, existing quantum processors are noisy and there are systematic errors with angle dependence leading to the different error bars shown in Fig. \ref{fig:VQE_two_plaq}.

\FloatBarrier
	
\section{Discussion}
Achieving a quantum advantage in the simulation of lattice gauge theories requires the preparation of physically interesting states, such as the vacuum. In the NISQ era, hybrid algorithms such as VQE will be essential. To make use of VQE, an appropriate classical optimizer and ansatz circuit must be chosen. In this work, state preparation on simple SU(3) lattice gauge theories has been performed with an eye towards scalability. In the variational state preparation of single plaquette systems, we showed that Bayesian optimization suffers from convergence issues as the coupling $g$ is decreased, while gradient descent methods suffer from no such issue. This suggests that VQE calculations at scale may need to make use of gradient descent methods in order to converge, despite the increase in computational overhead required to compute the gradient. Note that gradient based methods may converge to a local minimum instead of the true vacuum. This has not occurred for the simple systems studied in this work, but may need to be considered when performing calculations at scale. 

Calculations at scale will also require appropriate ansatz circuits to perform VQE. Due to the exponential growth of the Hilbert space with lattice size, circuits capable of preparing a generic state on the lattice will not be able to go to scale. In this work, it was demonstrated that in a quasi-1D SU(3) lattice gauge theory, VQE can be used to stitch together domains in their vacuum state to prepare the vaccum state of a larger lattice. The exponential convergence with domain size on an infinite lattice suggests that even shallow circuits may be able to achieve a large overlap with the true vacuum state at scale. The calculations on IBM's {\tt Manila} quantum processor showed that circuit ansatzes that respect a global symmetry will still respect the global symmetry on existing hardware despite the presence of noise and imperfect gates. This allows global symmetries to be used to construct circuit ansatzes that have fewer free degrees of freedom which makes them easier to optimize. 

While the computations in this work are encouraging, preparing a vacuum state for QCD with VQE will require significant developments in the application of quantum algorithms to lattice gauge theories. The calculations performed in this work were for a one dimensional string of plaquettes, but QCD is a three-dimensional theory. In a 3D theory, the domains being initialized in their vacuum state will be 3D blocks and the number of circuits required to stitch them together will scale with the surface area of the domain blocks. Additionally, a QCD calculation that can be taken to the continuum limit may require more electric field representations to be included, which will increase the number of possible local rotations in the VQE stitching circuit. It is conceivable that it is possible to reach the continuum limit without increasing the field truncation, but this remains to be investigated. Regardless, as the continuum limit is approached, the correlation length of the system will diverge and more layers of circuits will be required in the VQE stitching to accurately prepare the vacuum state. Matter will also need to be included at the sites, which will complicate the integrating out of the internal gauge space. In addition to these conceptual complications, achieving a quantum advantage in the simulation of lattice QCD will require quantum hardware with more qubits and a lower error rate, in order to enable the simulation of a large lattice in a local basis. While scaling up quantum hardware is challenging, the rapid improvement in quantum hardware and recent proposals for co-design \cite{physrevd.103.094501,zhang:2021bjq,andrade:2021pil} of quantum processors suggest that it can be done in a manner that will allow the simulation of lattice QCD on quantum computers in the near future.


\FloatBarrier

\begin{subappendices}
\section{Hardware Calculations}
\label{appendix:hardware}
\begin{figure}
    \centering
	\includegraphics{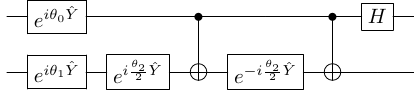}
	
	\includegraphics{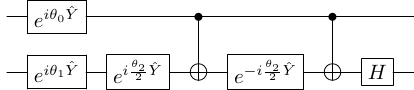}
	
	\includegraphics{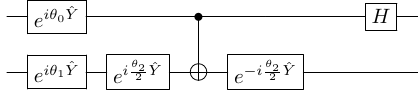}
	
	\caption{The top circuit is used to compute the expectation of $H_1$, the second circuit is used to compute the expectation of $H_2$, and the bottom circuit is used to compute the expectation of $H_3$.}
	\label{fig:VQECircuits}
\end{figure}

To perform VQE on a quantum computer, a circuit must be designed to prepare the ansatz state. For the calculations demonstrated here, only two qubits were used, so the circuit used to construct the state was capable of preparing an arbitrary 2 qubit state whose wavefunction has only real coefficients. Once the ansatz state has been prepared on the quantum computer, the energy of the state must also be computed. This can efficiently be done by breaking the Hamiltonian up into a sum over tractable terms, applying gates that diagonalize each term of the Hamiltonian, and performing measurements in the computational basis. This approach to computing the energy will require one circuit per term in the Hamiltonian. Each of the Hamiltonians studied in this work can be written in the form
\begin{align}
	\hat{H} &= \hat{H}_1 + \hat{H}_2 + \hat{H}_3 \nonumber \\
	\hat{H}_1 &= h_{11} \hat{1} \otimes \hat{Z} + h_{12} \hat{X} \otimes \hat{1}  + h_{13} \hat{X} \otimes \hat{Z}  \nonumber \\
	\hat{H}_2 &= h_{21} \hat{Z} \otimes \hat{1} +h_{22} \hat{1} \otimes \hat{X} \nonumber \\
	\hat{H}_3 &= h_{31} \hat{X} \otimes \hat{X} + h_{32} \hat{Y} \otimes \hat{Y} + h_{33} \hat{Z} \otimes \hat{Z}\ \ \ . 
\end{align}
These Hamiltonians can be diagonalized using the circuits shown in Fig. \ref{fig:VQECircuits}. To use gradient descent based methods in the classical optimization step of VQE, the gradient for the energy of the state as a function of the rotation angles in the ansatz circuit must be computed on the quantum computer. Due to the periodicity of $\sin$ and $\cos$, the gradient can be computed exactly using a symmetric finite difference formula with a shift of $\frac{\pi}{4}$. Explicitly, components of the gradient are computed using
\begin{equation}
	\partial_i E\left(\vec{\theta}\right) = E\left(\vec{\theta} + \frac{\pi}{4} \hat{i}\right) - E\left(\vec{\theta} - \frac{\pi}{4} \hat{i}\right) \ \ \ ,
\end{equation}
where $E\left(\vec{\theta}\right)$ is the energy as a function of the angles in the ansatz circuit and $\hat{i}$ is a unit vector pointing in the $i$-th direction. Therefore the gradient can be computed on the quantum computer using a number of circuits equal to two times the number of parameters in the ansatz circuit.
The calculation  of the energy on a real quantum computer suffers from systematic errors due to errors in the implementation of the gates on the computer and errors in the measurement process. The measurement errors can be mitigated by using Qiskit's {\tt measurement filter} subroutine, which removes the leading order measurement errors by optimizing an approximate inverse of the calculated all-to-all measurement matrix \cite{ibmmeaserror}. The dominant gate errors come from the implementation of CNOT gates. The errors associated with CNOT gates are mitigated using an extrapolation procedure \cite{physrevx.7.021050,physrevlett.119.180509}. Each CNOT in the circuit is replaced with an odd number $r$ of CNOT gates ($r=3,5,7$) and a linear extrapolation is performed to $r=0$.
\FloatBarrier

\section{Bayesian Optimization}
\label{app:bayes}

Bayesian optimization is a classical optimizer that can be used in the VQE algorithm.  Bayesian optimization uses the data already collected to create a Gaussian process-based surrogate function that approximates the function, $f$, being optimized.  This surrogate function is then used to create an acquisition function, which is then optimized to find a new trial point for the location of $f$'s minimum.  $f$ is then evaluated at that new point and the result is incorporated into the data for the next iteration \cite{bayesopt_presentation}. The Gaussian process used requires both a mean and covariance matrix for the function $f$. The covariance matrix used in this work is constructed from the Gaussian kernel \cite{gprocess_presentation}, which defines the covariance between $f({\bf x}_1)$ and $f({\bf x}_2)$ to be
\begin{equation}
	\label{eq:gaussker}
	K({\bf x}_1,  {\bf x}_2) = e^{-\sum\limits_{i=1}^d \frac{({x_1}_i-{x_2}_i)^2}{l_i^2}} \ \ \ , 
\end{equation}
where d is the number of dimensions of the inputted point and $l_i$ are hyperparameters specifying the width of the Gaussian for each component of $\bf{x}$. The mean of $f$ is generically unknown, but given the covariance matrix the mean can be approximated by the best linear unbiased predictor,
\begin{equation}
	\label{eq:blupmean}
	\mu = ({\bf{1}^T} {\bf C}^{-1} {\bf 1})^{-1} {\bf{1}^T}  {\bf C}^{-1} {\bf Z} \ \ \,
\end{equation}
where $\bf{1}$ is a vector with all entries equal to 1,  $\bf{C}$ is the covariance matrix with matrix elements given by ${\bf C}_{ij} = K({\bf x}_i,  {\bf x}_j)$,  and $\bf{Z}$ is a vector with entries given by the value of the function at the evaluated points, ${\bf{Z}}_i = f({\bf x}_i)$ \cite{Cressie1990}.

Given the mean and variance of the Gaussian process, the value of $f$ at a point ${\bf x_{posterior}}$ that has not already been evaluated follows a Gaussian distribution with a mean and variance given by
\begin{align}
	\label{eq:ok}
	\mu_{posterior} &= {\bf{c}^T} {\bf C}^{-1} {\bf Z} - (1 - {\bf{c}^T} {\bf C}^{-1} {\bf 1}) ({\bf{1}^T} {\bf C}^{-1} {\bf 1})^{-1} {\bf{1}^T}  {\bf C}^{-1} {\bf Z} \nonumber \\	
	\sigma^{2}_{posterior} &= K({\bf{x}_{posterior}},  {\bf{x}_{posterior}}) - {\bf{c}^T} {\bf C}^{-1} {\bf c} + {(1 - {\bf{c}^T} {\bf C}^{-1} {\bf 1})^{2}} ({\bf {1}^T} {\bf C}^{-1} {\bf 1})^{-1} \ \ \ ,
\end{align}
where $\bf{c}$ is a vector with entries ${\bf{c}}_i = K({\bf x_{posterior}},  {\bf x}_i)$ \cite{Cressie1990}. Eq. (\ref{eq:ok}) expresses the posterior mean and variance under the assumption that $f$ can be evaluated without error. In order to incorporate errors,  the variance of the data must be added to the diagonal elements of the covariance matrix $\bf{C}$ and to $\sigma^{2}_{posterior}$ \cite{gprocess_presentation}.

To use a Gaussian process in practice, the hyperparameters of the kernel must be selected. In this work, this was done by maximizing the likelihood of the data under a multivariate Gaussian model with a mean equal to the best linear unbiased predictor's mean and with a covariance equal to $\bf{C}$ (with the variance of the data added to its diagonal elements) from Eq. (\ref{eq:blupmean}). Another issue with practical implementation that arises is that $\bf{C}$ often ends up singular as the Gaussian process is iterated. This issue is known as multicollinearity and it occurs when one of the points used to construct $\bf{C}$ can be exactly predicted from the other points leading to zero being an eigenvalue of $\bf{C}$.  This can be remedied by using Tikohonov regularization where a fake ``data variance'' distinct from the real data variance is added to $\bf{C}$ but not to $\sigma^{2}_{posterior}$ \cite{mittikhonovnotes}.

The probability distribution of $f$ at unevaluated points is used to construct an acquisition function, whose job it is to balance exploration and exploitation. The acquisition function is optimized to find the minimum of $f$. In this work, probability of improvement \cite{bayesopt_presentation} was used as the acquisition function, i.e. the probability that the minimum of $f$ is smaller than the previously found minimum is maximized.  This is equivalent to minimizing
\begin{equation}
	\label{eq:acqpofi_used}
	acq({\bf x})_{PI} = \frac{\mu_{posterior}({\bf x}) - f_{min}}{\sigma_{posterior}({\bf x})}
\end{equation}
where $f_{min}$ is the previously found minimum of $f$.

\section{Plaquette Chain Tensor Network}
\label{appendix:PlaqTN}
\begin{figure}
    \centering
	\includegraphics[scale=0.25]{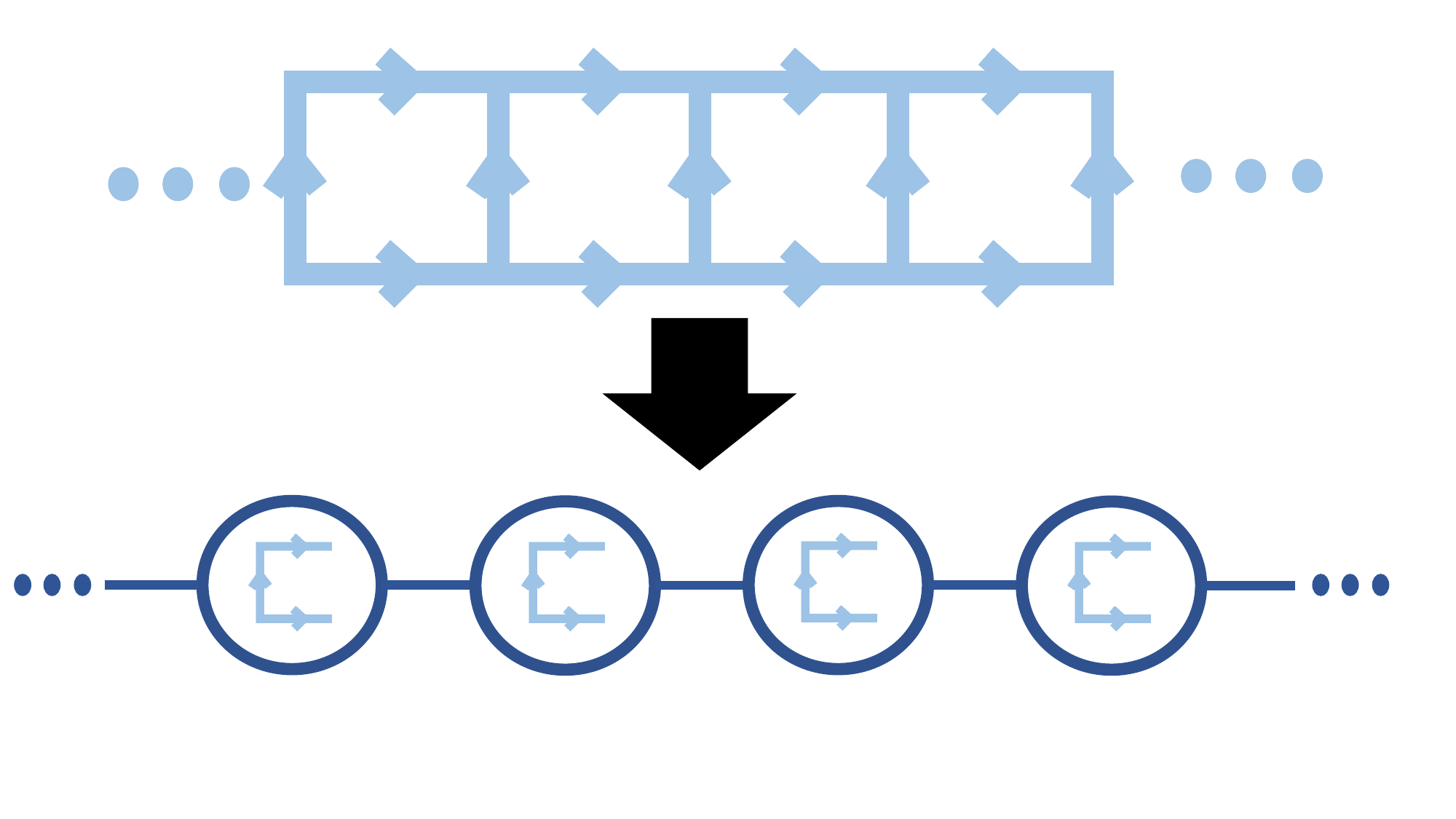}
	\caption{An infinite chain of SU(3) plaquettes can be mapped onto a 1D quantum system whose state can be represented with MPS by blocking sets of 3 links together as shown.}
	\label{fig:plaq_chain_tensor}
\end{figure}

The time evolving block decimation (TEBD) algorithm can be used to simulate the time evolution of an infinite translationally invariant quantum system by Trotterizing the time evolution operator \cite{2003,2004Eff,2004MPS}. The vacuum state of a system can be prepared by performing imaginary time evolution. This algorithm was developed for the simulation of systems whose Hamiltonian only consists of 2-site nearest-neighbor couplings, so its application to the simulation of a plaquette chain requires nonstandard modifications. Fig. \ref{fig:plaq_chain_tensor} shows how the links in the plaquette chain can be blocked together to form a 1D quantum system whose state can be described with MPS.
\FloatBarrier

In this blocking, the electric field operator on a single link becomes a single site operator, the plaquette operator becomes a three site operator, and the Gauss's law constraint become a constraint on neighboring sites. The Gauss's law constraint can be enforced by adding an energy penalty for violating Gauss's law.

\begin{figure}
    \centering
	\includegraphics[scale=0.25]{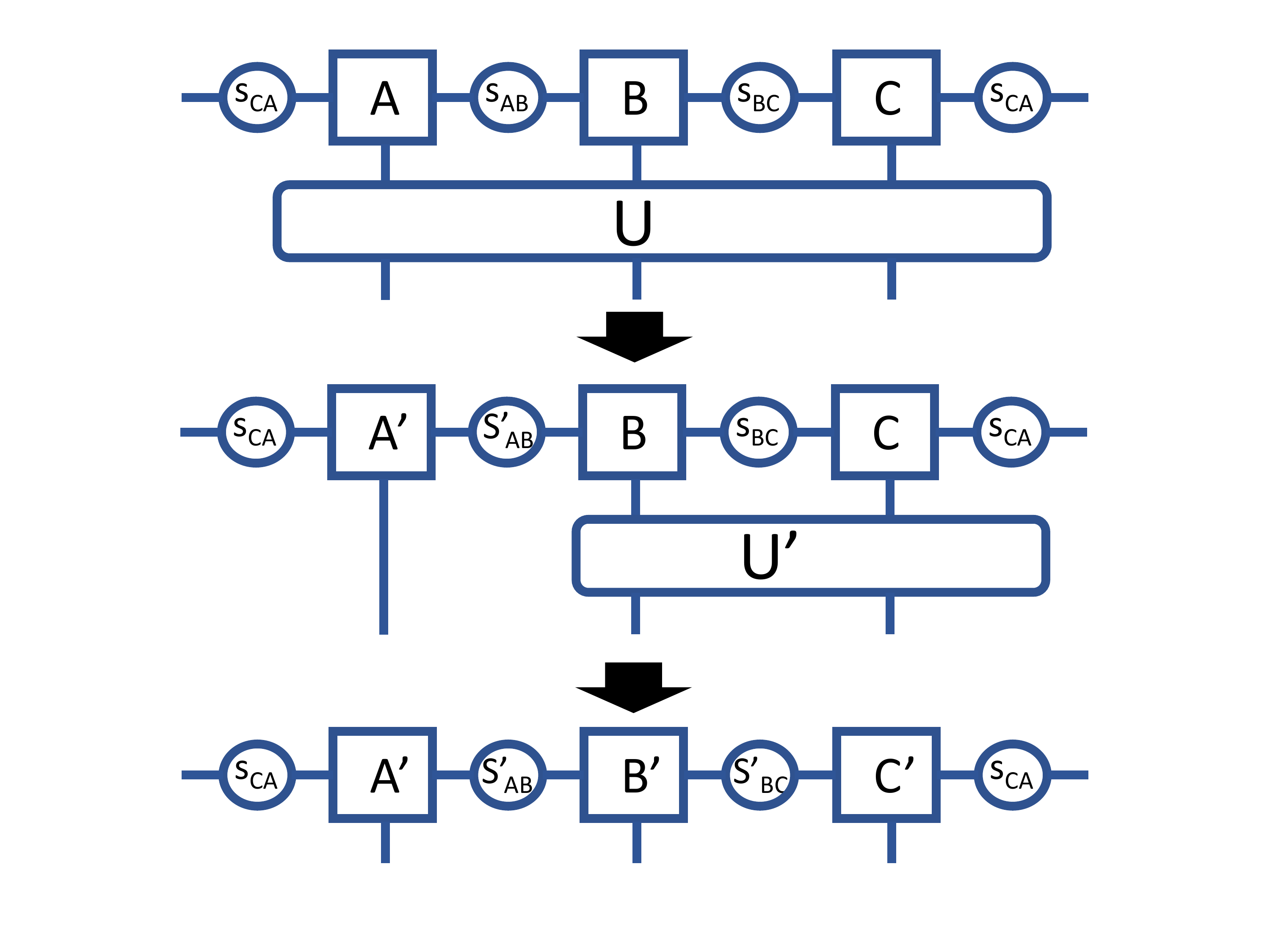}
	\caption{This figure shows the required sequence of SVDs that must be performed to return an MPS tensor network to MPS form after applying a 3 site gate.}
	\label{fig:plaq_chain_SVD}
\end{figure}

The TEBD algorithm finds the vacuum by applying a Trotterized version of the imaginary time evolution operator to a translationally invariant state. For a 2-site Hamiltonian, this is accomplished by storing a unit cell of 2 sites and performing an SVD after applying each gate to keep the most relevant states. For a 3-site Hamiltonian, such as the Hamiltonian obtained for the plaquette chain, a unit cell of 3 sites must be stored and two SVD's must be performed to obtain the most relevant local states as shown in Fig. \ref{fig:plaq_chain_SVD}. The approach used to perform time evolution in TEBD can also be used to apply arbitrary gates. To represent the ansatz states obtained using domains of $l$ plaquettes, a unit cell of length $l+1$ had to be stored and the state was prepared by applying gates and performing a SVD to return to MPS form as in the case of time evolution.
\FloatBarrier

\section{Data from IBM's Manila Processor}
\label{appendix:Data}
The following tables in this appendix contain the energies that were computed on IBM's {\tt Manila} quantum processor. All error bars were computed from the uncertainty in the linear CNOT extrapolation as described in Appendix \ref{appendix:hardware}.
\begin{table}
    \centering
	\begin{tabular}{| c | c | c |} 
		\hline
		Step Number & CP Symmetry Enforced & CP Symmetry Unenforced  \\ [0.5ex] 
		\hline
		1 & $2.957 \pm 0.025$ & $2.941 \pm 0.017$ \\ 
		\hline
		2 & $2.973 \pm 0.028$ & $2.91 \pm 0.04$ \\ 
		\hline
		3 & $2.93 \pm 0.04$ & $2.931 \pm 0.032$ \\ 
		\hline
		4 & $2.891 \pm 0.013$ & $2.928 \pm 0.016$ \\ 
		\hline
		5 & $2.905 \pm 0.028$ & $2.877 \pm 0.017$ \\ 
		\hline
		6 & $2.88 \pm 0.05$ & $2.830 \pm 0.009$ \\ 
		\hline
		7 & $2.868 \pm 0.019$ & $2.825 \pm 0.023$ \\ 
		\hline
		8 & $2.806 \pm 0.025$ &  $2.84 \pm 0.04$\\ 
		\hline
		9 & $2.826 \pm 0.016$ & $2.847 \pm 0.027$ \\ 
		\hline
		10 & $2.83 \pm 0.017$ & $2.822 \pm 0.006$ \\ 
		\hline
		11 & $2.87 \pm 0.04$ & $2.882 \pm 0.025$ \\ 
		\hline
		12 & $2.824 \pm 0.02$ & $2.823 \pm 0.024$ \\ 
		\hline
		13 & $2.806 \pm 0.026$ & $2.826 \pm 0.015$ \\ 
		\hline
		14 & $2.834 \pm 0.007$ & $2.846 \pm 0.021$ \\ 
		\hline
		15 & $2.808 \pm 0.019$ & $2.833 \pm 0.015$ \\ 
		\hline
		16 & $2.783 \pm 0.004$ & $2.812 \pm 0.004$ \\ 
		\hline
		17 & $2.843 \pm 0.007$ & $2.819 \pm 0.006$ \\ 
		\hline
		18 & $2.808 \pm 0.016$ & $2.801 \pm 0.013$ \\ 
		\hline
	\end{tabular}
	\caption{This table lists the data shown in Fig. \ref{fig:VQE_one_plaq_8}. The left column states the number of times gradient descent was applied, the center column contains the energies computed for the circuit that had the CP symmetry explicitly enforced, and the right column contains the energies computed for the circuit without the CP symmetry enforced.}
	\label{tab:OnePlaq8Data}
\end{table}

\begin{table}
    \centering
	\begin{tabular}{| c | c | c |} 
		\hline
		Step Number & Electric Start & Krylov Start  \\ [0.5ex] 
		\hline
		1 & $4.61 \pm 0.06$ & $3.900 \pm 0.030$ \\ 
		\hline
		2 & $4.44 \pm 0.08$ & $3.917 \pm 0.024$ \\ 
		\hline
		3 & $4.265 \pm 0.034$ & $3.85 \pm 0.04$ \\ 
		\hline
		4 & $4.11 \pm 0.04$ & $3.827 \pm 0.020$ \\ 
		\hline
		5 & $4.040 \pm 0.014$ & $3.826 \pm 0.025$ \\ 
		\hline
		6 & $3.984 \pm 0.024$ & $3.81 \pm 0.04$ \\ 
		\hline
		7 & $3.928 \pm 0.007$ & $3.867 \pm 0.030$ \\ 
		\hline
		8 & $3.855 \pm 0.016$ &  $3.814 \pm 0.034$\\ 
		\hline
		9 & $3.85 \pm 0.04$ & $3.84 \pm 0.05$ \\ 
		\hline
		10 & $3.811 \pm 0.030$ & $3.78 \pm 0.04$ \\ 
		\hline
		11 & $3.837 \pm 0.018$ & $3.79 \pm 0.04$ \\ 
		\hline
		12 & $3.790 \pm 0.024$ & $3.80 \pm 0.05$ \\ 
		\hline
		13 & $3.804 \pm 0.025$ & $3.785 \pm 0.032$ \\ 
		\hline
		14 & $3.789 \pm 0.024$ & $3.790 \pm 0.022$ \\ 
		\hline
		15 & $3.83 \pm 0.06$ & $3.767 \pm 0.007$ \\ 
		\hline
	\end{tabular}
	\caption{This table lists the data shown in Fig. \ref{fig:VQE_one_plaq_6}. The left column states the number of times gradient descent was applied, the center column contains the energies computed for the gradient descent that began at the electric vacuum, and the right column contains the energies computed for the gradient descent that began at the state obtained from the Lanczos algorithm with a Krylov dimension of 2.}
	\label{tab:OnePlaq6Data}
\end{table}

\begin{table}
    \centering
	\begin{tabular}{| c | c | c |} 
		\hline
		Step Number & Electric Start & One Plaquette Start  \\ [0.5ex] 
		\hline
		1 & $2.96 \pm 0.07$ & $2.6648 \pm 0.0013$ \\ 
		\hline
		2 & $2.85 \pm 0.10$ & $2.631 \pm 0.022$ \\ 
		\hline
		3 & $2.77 \pm 0.11$ & $2.66 \pm 0.04$ \\ 
		\hline
		4 & $2.72 \pm 0.10$ & $2.609 \pm 0.012$ \\ 
		\hline
		5 & $2.69 \pm 0.09$ & $2.651 \pm 0.017$ \\ 
		\hline
		6 & $2.71 \pm 0.07$ & $2.616 \pm 0.010$ \\ 
		\hline
		7 & $2.65 \pm 0.07$ & $2.610 \pm 0.015$ \\ 
		\hline
		8 & $2.63 \pm 0.06$ &  $2.625 \pm 0.018$\\ 
		\hline
		9 & $2.65 \pm 0.06$ & $2.642 \pm 0.019$ \\ 
		\hline
		10 & $2.66 \pm 0.04$ & $2.6021 \pm 0.0023$ \\ 
		\hline
		11 & $2.638 \pm 0.021$ & $2.594 \pm 0.008$ \\ 
		\hline
		12 & $2.64 \pm 0.05$ & $2.600 \pm 0.017$ \\ 
		\hline
		13 & $2.624 \pm 0.021$ & $2.639 \pm 0.019$ \\ 
		\hline
		14 & $2.608 \pm 0.029$ & $2.618 \pm 0.010$ \\ 
		\hline
		15 & $2.612 \pm 0.019$ & $2.639 \pm 0.005$ \\ 
		\hline
	\end{tabular}
	\caption{This table lists the data shown in Fig. \ref{fig:VQE_two_plaq}. The left column states the number of times gradient descent was applied. The center column contains the energies computed for the gradient descent that began at the electric vacuum.  The right column contains the energies computed for the gradient descent that began at the single plaquette vacuum.}
	\label{tab:TwoPlaqData}
\end{table}
\end{subappendices}






\chapter{Preparations for Quantum Simulations of Quantum Chromodynamics in \texorpdfstring{\boldmath$1+1$}{1+1} Dimensions:
(I) Axial Gauge}
\label{chap:1p1dQCD}

{\it This chapter is associated with Ref. \cite{physrevd.107.054512}:}

{\it ``Preparations for quantum simulations of quantum chromodynamics in dimensions. I. Axial gauge" by Roland C. Farrell, Ivan A. Chernyshev, Sarah J. M. Powell, Nikita A. Zemlevskiy, Marc Illa, and Martin J. Savage}

\section{Introduction}

This chapter follows the lead of preceding work in implementation of 1+1D QCD with classical numerical methods (discussed in Sec. \ref{sec:intro_tnm}) and, more imporantly, a study using IBM's quantum devices~\cite{ibmq} of 1+1D SU(2) lattice Yang-Mills theory (mentioned in Sec.~\ref{sec:intro_kogutsusskind})~\cite{atas:2021ext}, where VQE was applied to the Jordan-Wigner transformed SU(2) Kogut-Susskind Hamiltonian to obtain vacuum energy and baryon and meson masses. It extends these results to the quantum simulation of $1+1$D $SU(N_c)$ lattice gauge theory with quarks for arbitrary $N_c$ (number of colors in the theory) and $N_f$ (number of quark flavors).
In accordance with the gauge-transformation mentioned in Sec. \ref{sec:intro_kogutsusskind}, calculations are primarily done in $A^{(a)}_x=0$ axial (Arnowitt-Fickler) gauge,\footnote{For a discussion of Yang-Mills in axial gauge, see, for example, Ref.~\cite{Reinhardt:1996dy}.}
which leads to non-local interactions in order to define the chromo-electric field contributions to the energy density via Gauss's law.
This is in contrast to Weyl gauge, $A_t^{(a)}=0$, where contributions remain local.
The resource estimates for asymptotic quantum simulations of the Schwinger model in Weyl gauge have been recently performed~\cite{shaw:2020udc}, and also for Yang-Mills gauge theory based upon the Byrnes-Yamamoto mapping~\cite{kan:2021xfc}.
Here, the focus is on near-term, and hence non-asymptotic, quantum simulations to better assess the resource requirements for quantum simulations of non-Abelian gauge theories with multiple flavors of quarks. 
For concreteness, $N_f=2$ QCD is studied in detail, including the mass decomposition of the low-lying hadrons (the $\sigma$- and $\pi$-meson, the single baryon and the two-baryon bound state), color edge-states, entanglement structures within the hadrons and quantum circuits for time evolution.
Further, results are presented for the quantum simulation of a $N_f=1$, single-site system, using IBM's quantum computers~\cite{ibmq}.
Such quantum simulations will play a critical role in evolving the functionality, protocols and workflows to be used in $3+1$D simulations of QCD, including the preparation of scattering states, time evolution and subsequent particle detection.
As a step in this direction, the results of this work have been applied to the quantum simulation of $\beta$-decay of a single baryon in $1+1$D QCD, as outlined in Chapter \ref{chap:1p1dSM}.
Motivated by the recent successes in co-designing efficient multi-qubit operations in trapped-ion systems~\cite{andrade:2021pil,Katz:2022czu}, 
additional multi-qubit or qudit operations are identified, 
specific to lattice gauge theories,
that would benefit from being native operations on quantum devices.

\section{QCD with Three Colors and Two flavors in \texorpdfstring{\boldmath$1+1$}{1+1}D}
\label{sec:Nc3Nf2}
\noindent
In $3+1$D, 
the low-lying spectrum of $N_f=2$ QCD is remarkably rich. 
The lightest hadrons are the $\pi$s, which are identified as the pseudo-Goldstone bosons associated with the spontaneous breaking of the approximate global $SU(2)_L\otimes SU(2)_R$ chiral symmetry, which becomes exact in the chiral limit where the $\pi$s are massless. At slightly higher mass are the broad $I=0$ spinless 
resonance, $\sigma$, and the narrow $I=0$, $\omega$, and $I=1$, $\rho$, vector resonances as well as the multi-meson continuum. 
The proton and neutron, which are degenerate in the isospin limit and the absence of electromagnetism,
are the lightest baryons, forming 
an $I=J=1/2$ iso-doublet.  
The next lightest baryons, which become degenerate with the nucleons in the large-$N_c$ limit (as part of a 
large-$N_c$ tower), are the four $I=J=3/2$ $\Delta$ resonances.
The nucleons bind together to form the periodic table of nuclei, the lightest being the deuteron, an $I=0$, $J=1$ neutron-proton bound state with a binding energy of $\sim 2.2~{\rm MeV}$, which is to be compared to the mass of the nucleon $M_N\sim 940~{\rm MeV}$.  
In nature, the low-energy two-nucleon systems have S-wave scattering lengths that are much larger than the range of their interactions, rendering them unnatural. Surprisingly, this unnaturalness persists for a sizable range of light-quark 
masses, e.g., Refs.~\cite{Beane:2002xf,Epelbaum:2012iu,Berengut:2013nh,wagman:2017tmp,NPLQCD:2020lxg}. 
In addition, this unnaturalness, and the nearby renormalization-group fixed point~\cite{Kaplan:1998tg,Kaplan:1998we}, provides the starting point for a systematic effective field theory expansion about unitarity~\cite{Kaplan:1998tg,Kaplan:1998we,vanKolck:1998bw,Chen:1999tn}.
Much of this complexity is absent in a theory with only one flavor of quark.

As a first step toward $3+1$D QCD simulations of real-time dynamics of nucleons and nuclei, we will focus on preparing to carry out quantum simulations of $1+1$D QCD with $N_f=2$ flavors of quarks. While the isospin structure of the theory is the same as in $3+1$D, the lack of spin and orbital angular momentum significantly reduces the richness of the hadronic spectrum and S-matrix.
However, many of the relevant features and processes of $3+1$D QCD that are to be addressed by quantum simulation in the future are present in $1+1$D QCD.
Therefore, quantum simulations in $1+1$D are expected to provide inputs to the development of quantum simulations of QCD.

\subsection{Mapping \texorpdfstring{\boldmath$1+1$}{1+1}D QCD onto Qubits}
\label{sec:1p1dQCD_mapping1p1DQCDtoqubits}
\noindent
The Hamiltonian describing non-Abelian lattice gauge field theories in arbitrary numbers of spatial dimensions was first given by Kogut and Susskind (KS) in the 1970s~\cite{physrevd.11.395,banks:1975gq}. For $1+1$D QCD with $N_f = 2$ discretized onto $L$ spatial lattice sites, which are mapped to 2L $q$, $\overline{q}$ sites to separately accommodate quarks and antiquarks,  the KS lattice Hamiltonian is
\begin{align}
    H_{\rm{KS}} 
     = & 
    \sum_{f=u,d}\left[
        \frac{1}{2 a} \sum_{n=0}^{2L-2} \left ( \phi_n^{(f)\dagger} U_n \phi_{n+1}^{(f)}
        \ +\ {\rm h.c.} \right ) 
    \: + \: 
    m_f \sum_{n=0}^{2L-1} (-1)^{n} \phi_n^{(f)\dagger} \phi_n^{(f)} 
    \right]
    \: + \: 
    \frac{a g^2}{2} 
    \sum_{n=0}^{2L-2} 
    \sum_{a=1}^{8}
    | {\bf E}^{(a)}_n|^2
    \nonumber\\
     & - \: \frac{\mu_B}{3} \sum_{f=u,d} \sum_{n=0}^{2L-1} \phi_n^{(f)\dagger} \phi^{(f)}_n 
    \ - \: 
    \frac{\mu_{I}}{2} \sum_{n=0}^{2L-1}\left(\phi_n^{(u)\dagger} \phi^{(u)}_n \ - \
    \phi_n^{(d)\dagger} \phi^{(d)}_n  \right)
    \ .
    \label{eq:KSHam}
\end{align}
The masses of the $u$- and $d$-quarks are $m_{u,d}$,
$g$ is the strong coupling constant at the spatial lattice spacing $a$,
$U_n$ is the spatial link operator in Weyl gauge
$A_t^{(a)}=0$,
$\phi^{(u,d)}_n$ are the $u$- and $d$-quark field operators which transform in the fundamental representation of $SU(3)$
and 
${\bf E}^{(a)}_n$ is the chromo-electric field associated with the $SU(3)$ generator,
$T^a$.
For convention, we write, for example, $\phi^{(u)}_n=(u_{n,r}, u_{n,g}, u_{n,b})^T$ to denote the $u$-quark field(s) at the $n^{\rm th}$ site in terms of 3 colors $r,g,b$.
With an eye toward simulations of dense matter systems, chemical potentials for baryon number, $\mu_B$, and the third component of isospin, $\mu_I$, are included.
For most of the results presented in
this work, the chemical potentials will be set to zero, $\mu_B=\mu_I = 0$,
and there will be exact isospin symmetry, $m_u=m_d \equiv m$.
In Weyl gauge and using the chromo-electric basis of the link operator $|{\bf R},\alpha,\beta\rangle_n$\footnote{
%
%
The indices $\alpha$ and $\beta$ specify the color state in the left (L) and right (R) link Hilbert spaces respectively. 
States of a color irrep {\bf R} are labelled by their total color isospin $T$, third component of color isospin $T^z$ and color hypercharge $Y$, i.e., $\alpha = (T_L, T^z_L, Y_L)$ and $\beta = (T_R, T^z_R, Y_R)$.
\label{foot:irrep}},
the contribution from the energy in the chromo-electric field from each basis state is proportional to the Casimir of the irrep ${\bf R}$ (see Eq. \ref{eq:su3_singleplaquette_casimirdef} for the definition of the Casimir).
The fields have been latticized
such that the quarks reside on even-numbered sites, $n=0,2,4,6,\ldots$, and antiquarks reside on odd-numbered sites, $n=1,3,5,\ldots$.
Open boundary conditions (OBCs) are employed in the spatial direction, 
with a vanishing background chromo-electric field.

The KS Hamiltonian in Eq.~(\ref{eq:KSHam}) 
is constructed in Weyl gauge.
A unitary transformation can be performed on Eq.~(\ref{eq:KSHam}) to eliminate the gauge links~\cite{sala:2018dui}, with Gauss's Law 
uniquely providing the energy in the chromo-electric field in terms of a non-local sum of products of charges, i.e., the Coulomb energy. 
This is equivalent to formulating the system in axial gauge~\cite{PhysRev.127.1821,weinberg1995quantum}, $A^{(a)}_x = 0$, from the outset.
The Hamiltonian in Eq.~(\ref{eq:KSHam}), when formulated with $A^{(a)}_x = 0$, becomes
\begin{align}
    H 
    = & 
    \sum_{f=u,d}\left[ 
        \frac{1}{2} \sum_{n=0}^{2L-2} \left ( \phi_n^{(f)\dagger} \phi_{n+1}^{(f)}
        \ +\ {\rm h.c.} \right ) 
    \: + \: 
    m_f \sum_{n=0}^{2L-1} (-1)^{n} \phi_n^{(f)\dagger} \phi_n^{(f)} 
    \right]
    \: + \: 
    \frac{g^2}{2}
    \sum_{n=0}^{2L-2} 
    \sum_{a=1}^{8}
    \left ( \sum_{m \leq n} Q^{(a)}_m \right ) ^2    \nonumber\\
     & - \: \frac{\mu_B}{3} \sum_{f=u,d} \sum_{n=0}^{2L-1} \phi_n^{(f)\dagger} \phi^{(f)}_n  
    \ - \:
    \frac{\mu_{I}}{2} \sum_{n=0}^{2L-1}\left(\phi_n^{(u)\dagger} \phi^{(u)}_n \ - \
    \phi_n^{(d)\dagger} \phi^{(d)}_n  \right)
    \ ,
    \label{eq:GFHam}
\end{align}
where the color charge operators on a given lattice site are the sum of contributions from the $u$- and $d$-quarks,
\begin{equation}
    Q^{(a)}_m \ =\ 
    \phi^{(u) \dagger}_m T^a \phi_m^{(u)}\ +\ 
    \phi^{(d) \dagger}_m T^a \phi_m^{(d)}
    \ .
    \label{eq:SU3charges}
\end{equation}
To define the fields, 
boundary conditions with $A_0^{(a)}(x)=0$ at spatial infinity and zero background chromo-electric fields are used, with Gauss's law sufficient to determine them at all other points on the lattice,
\begin{equation}
   {\bf E}^{(a)}_n  = \sum_{m\leq n} Q^{(a)}_m \ .
\end{equation}
In this construction, a state is completely specified by the fermionic occupation at each site. This is to be contrasted with the Weyl
gauge construction where both fermionic occupation and the $SU(3)$ multiplet defining the chromo-electric field are required.

\begin{figure}[!ht]
    \centering
    \includegraphics[width=15cm]{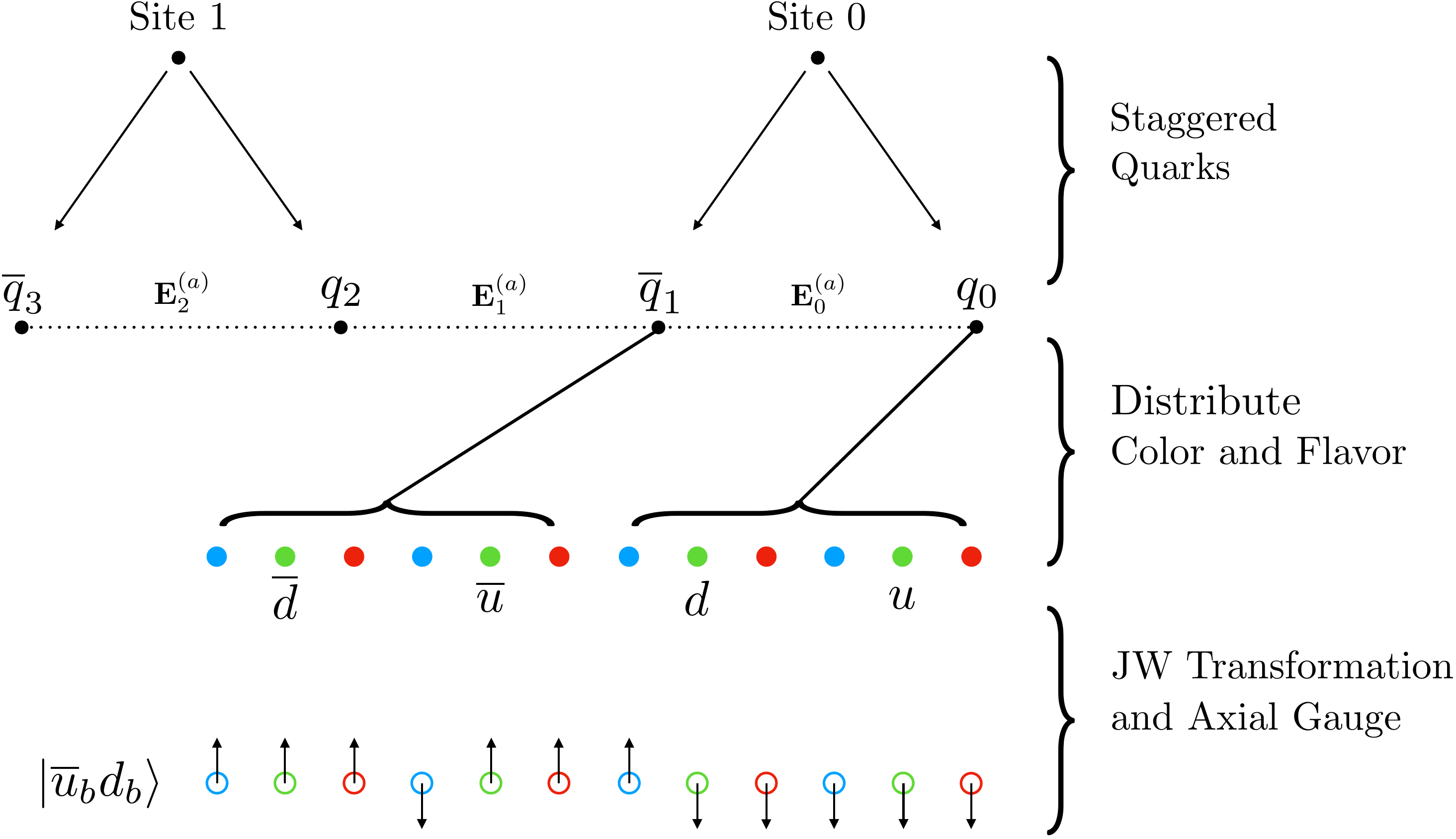}
    \caption{
    The encoding of $N_f=2$ QCD onto a lattice of spins describing $L=2$ spatial sites. 
    Staggering is used to discretize the quark fields, which doubles the number of lattice sites, with (anti)quarks on (odd) even sites.
    The chromo-electric field resides on the links between quarks and antiquarks. 
    Color and flavor degrees of freedom of each quark and antiquark site are distributed over six qubits with a JW mapping, and axial gauge along with Gauss's law are used to remove the chromo-electric fields. 
    A quark (antiquark) site is occupied if it is spin up (down), and the example spin configuration corresponds to the state $\ket{\overline{u}_b \, d_b}$.}
    \label{fig:2flavLayout}
\end{figure}

There are a number of ways that this system,
with the Hamiltonian given in Eq.~(\ref{eq:GFHam}), could be mapped
onto the register of a quantum computer.
In this work, both a staggered discretization and a JW transformation~\cite{1928ZPhy...47..631J} are chosen to map the $N_c=3$ and $N_f=2$
quarks to 6 qubits, with ordering $d_b, d_g, d_r, u_b, u_g, u_r$,
and the antiquarks associated with the same spatial site adjacent with ordering 
$\overline{d}_b, \overline{d}_g, \overline{d}_r, \overline{u}_b, \overline{u}_g, \overline{u}_r$.
This is illustrated in Fig.~\ref{fig:2flavLayout} and
requires a total of 12 qubits per spatial lattice site (see App.~\ref{app:hamConst} for more details). 
The resulting JW-mapped Hamiltonian is the sum of the following five terms:
\begin{subequations}
    \label{eq:H2flav}
    \begin{align}
    H = & \ H_{kin}\ +\ H_m\ +\ H_{el} \ +\ 
    H_{\mu_B}\ +\ H_{\mu_I} \ ,
    \end{align}
    \begin{align}
    H_{kin} = & \ -\frac{1}{2} \sum_{n=0}^{2L-2} \sum_{f=0}^{1} \sum_{c=0}^{2} \left[ \sigma^+_{6n+3f+c} \left ( \bigotimes_{i=1}^{5}\sigma^z_{6n+3f+c+i} \right )\sigma^-_{6(n+1)+3f+c} +\rm{h.c.} \right]\ ,
        \label{eq:Hkin2flav}
    \end{align}
    \begin{align}
    H_m = & \ \frac{1}{2} \sum_{n=0}^{2L-1} \sum_{f=0}^{1} \sum_{c=0}^{2} m_f\left[ (-1)^{n} \sigma_{6n + 3f + c}^z + 1\right]\ ,
        \label{eq:Hm2flav}
    \end{align}
    \begin{align}
    H_{el} = & \ \frac{g^2}{2} \sum_{n=0}^{2L-2}(2L-1-n)\left( \sum_{f=0}^{1} Q_{n,f}^{(a)} \, Q_{n,f}^{(a)} \ + \
        2 Q_{n,0}^{(a)} \, Q_{n,1}^{(a)}
         \right)   \nonumber \\[4pt]
        & + g^2 \sum_{n=0}^{2L-3} \sum_{m=n+1}^{2L-2}(2L-1-m) \sum_{f=0}^1 \sum_{f'=0}^1 Q_{n,f}^{(a)} \, Q_{m,f'}^{(a)} \ ,
         \label{eq:Hel2flav}
    \end{align}
    \begin{align}
    H_{\mu_B}  = & \ -\frac{\mu_B}{6} \sum_{n=0}^{2L-1} \sum_{f=0}^{1} \sum_{c=0}^{2}  \sigma_{6n + 3f + c}^z \ ,
        \label{eq:HmuB2flav}
    \end{align}
    \begin{align}
    H_{\mu_I} = & \ -\frac{\mu_I}{4} \sum_{n=0}^{2L-1} \sum_{f=0}^{1} \sum_{c=0}^{2} (-1)^{f} \sigma_{6n + 3f + c}^z  \ ,
        \label{eq:HmuI2flav}
    \end{align}
\end{subequations}

where now repeated adjoint color indices, $(a)$, are summed over,
the flavor indices, $f=0,1$, correspond to $u$- and $d$-quark flavors and $\sigma^\pm = (\sigma^x \pm i \sigma^y)/2$.

Products of charges are given in terms of spin operators as
\begin{align}
    Q_{n,f}^{(a)} \, Q_{n,f}^{(a)} =& \frac{1}{3}(3 - \sigma^z_{6n+3f} \sigma^z_{6n+3f+1} - \sigma^z_{6n+3f} \sigma^z_{6n+3f+2} - \sigma^z_{6n+3f+1} \sigma^z_{6n+3f+2}) \ ,  \nonumber \\[4pt]
    Q_{n,f}^{(a)} \, Q_{m,f'}^{(a)} =& \frac{1}{4}\bigg [2\big (\sigma^+_{6n+3f}\sigma^-_{6n+3f+1}\sigma^-_{6m+3f'}\sigma^+_{6m+3f'+1} \nonumber \\[4pt]
    &+ \sigma^+_{6n+3f}\sigma^z_{6n+3f+1}\sigma^-_{6n+3f+2}\sigma^-_{6m+3f'}\sigma^z_{6m+3f'+1}\sigma^+_{6m+3f'+2} \nonumber \\[4pt]
    &+\sigma^+_{6n+3f+1}\sigma^-_{6n+3f+2}\sigma^-_{6m+3f'+1}\sigma^+_{6m+3f'+2} + { \rm h.c.}\big ) \nonumber \\[4pt]
    &+ \frac{1}{6}\sum_{c=0}^{2} \sum_{c'=0}^2( 3 \delta_{c c'} - 1 ) \sigma^z_{6n+3f+c}\sigma^z_{6m+3f'+c'} \bigg ] \ .
    \label{eq:QnfQmfp}
\end{align}
A constant has been added to $H_m$ to ensure that all basis states contribute positive mass. The Hamiltonian for $SU(N_c)$ gauge theory with $N_f$ flavors in the fundamental representation is presented in Sec.~\ref{sec:NcNf}. 
Note that choosing $A^{(a)}_x = 0$ gauge and enforcing Gauss's law has resulted in all-to-all interactions, the double lattice sum in $H_{el}$.

For any finite lattice system, there are color non-singlet states in the spectrum, which are unphysical and have infinite energy in the continuum and infinite-volume limits.
For a large but finite system, OBCs can also support finite-energy color non-singlet states which are localized to the end of the lattice (color edge-states).\footnote{Low-energy edge-states that have global charge in a confining theory can also be found in the simpler setting of the Schwinger model. 
Through exact and approximate tensor methods, we have verified that these states exist on lattices up to length $L=13$, and they are expected to persist for larger $L$.} 
The existence of such states in the spectrum is independent of the choice of gauge or fermion mapping.
The naive ways to systematically examine basis states and preclude such configurations is found to be impractical due to the non-Abelian nature of the gauge charges
and the resulting entanglement between states required for color neutrality.
A practical way to deal with this problem is to add a term to the Hamiltonian that 
raises the energy of color non-singlet states.
This can be accomplished by including the energy density in the chromo-electric field beyond the end of the lattice with a large coefficient $h$.  
This effectively adds the energy density in a finite chromo-electric field over a large spatial extent beyond the end of the lattice.
In the limit $h\rightarrow\infty$, only states with a vanishing chromo-electric field beyond the end of the lattice remain at finite energy, rendering the system within the lattice to be a color singlet.
This new term in the Hamiltonian is
\begin{equation}
    H_{\bf 1} = \frac{h^2}{2} \sum_{n=0}^{2L-1}
    \left( \sum_{f=0}^{1} Q_{n,f}^{(a)} \, Q_{n,f}^{(a)} \ + \  
    2 Q_{n,0}^{(a)} \, Q_{n,1}^{(a)}
     \right) \ + \  h^2 \sum_{n=0}^{2L-2} \sum_{m=n+1}^{2L-1}\sum_{f=0}^1 \sum_{f'=0}^1 Q_{n,f}^{(a)} \, Q_{m,f'}^{(a)} \ ,
     \label{eq:Hpen}
\end{equation}
which makes a vanishing contribution when the sum of charges over the whole lattice is zero; otherwise, it makes a contribution $\sim h^2$.

\subsection{Spectra for \texorpdfstring{\boldmath$L=1, 2$}{L=1,2} Spatial Sites}
\label{sec:Exact}
\noindent
The spectra and wavefunctions of systems with a small number of lattice sites can be determined by diagonalization of the Hamiltonian.
In terms of spin operators, the $N_f=2$ Hamiltonian in Eq.~(\ref{eq:H2flav}) decomposes into sums of tensor products of Pauli matrices. The tensor product factorization can be exploited to perform an exact diagonalization relatively efficiently. 
This is accomplished by first constructing a basis
by projecting onto states with specific quantum numbers, and then building the Hamiltonian in that subspace. 
There are four mutually commuting symmetry generators that allow states to be labelled by $(r,g,b,I_3)$: redness, greenness, blueness and the third component of isospin.
In the computational (occupation) basis, states are represented by bit strings of $0$s and $1$s. For example, the $L=1$ state with no occupation is $\ket{000000111111}$.\footnote{Qubits are read from right to left, e.g., $\ket{q_{11} \, q_{10}\, \ldots \, q_{1}\, q_{0}}$. Spin up is $\ket{0}$ and spin down is $\ket{1}$.} Projecting onto eigenstates of $(r,g,b,I_3)$ amounts to fixing the total number of $1$s in a substring of a state.
The Hamiltonian is formed by evaluating matrix elements of Pauli strings between states in the basis, and only involves $2\times 2$ matrix multiplication.
The Hamiltonian matrix is found to be sparse, as expected, and the low energy eigenvalues and eigenstates can be found straightforwardly.
As the dimension of the Hamiltonian grows exponentially with the spatial extent of the lattice, this method becomes intractable for large system sizes, as is well known.

\subsubsection{Exact Diagonalizations, Color Edge-States and Mass Decompositions of the Hadrons}
\noindent
For small enough systems, an exact diagonalization of the Hamiltonian matrix in the previously described basis can be performed.
Without chiral symmetry and its spontaneous breaking, the energy spectrum in $1+1$D does not contain a massless isovector state (corresponding to the QCD pion) in the limit of vanishing quark masses.
In the absence of 
chemical potentials for baryon number, $\mu_B=0$, or isospin, $\mu_I=0$,
the vacuum, $\ket{\Omega}$, has $B=0$
(baryon number zero) and $I=0$ (zero total isospin). 
The $I=0$
$\sigma$-meson is the lightest meson,
while the $I=1$ $\pi$-meson is the next lightest.
The lowest-lying eigenstates in the 
$B=0$ spectra for $L=1,2$
(obtained from
exact diagonalization of the Hamiltonian) 
are given in Table~\ref{tab:specB0}. 
The masses are defined by their energy gap to the vacuum, 
and all results in this section are for $m_u=m_d=m=1$.
\begin{table}[!ht]
\renewcommand{\arraystretch}{1.2}
\begin{tabularx}{0.4\textwidth}{||c | Y | Y | Y ||} 
\hline
\multicolumn{4}{||c||}{$L=1 $} \\
 \hline
 $g^2$ & $E_{\Omega}$ & $M_{\sigma}$ & $M_{\pi}$ \\
 \hline\hline
 8 & -0.205 & 5.73 & 5.82 \\ 
 \hline
 4 & -0.321 & 4.37 & 4.47\\
 \hline
 2 & -0.445 & 3.26 & 3.30 \\
 \hline
 1 & -0.549 & 2.73 & 2.74\\
 \hline
 1/2 & -0.619 & 2.48 & 2.48\\
 \hline
 1/4 & -0.661 & 2.35 & 2.36 \\
 \hline
 1/8 & -0.684 &  2.29 & 2.30 \\
 \hline
\end{tabularx}
\renewcommand{\arraystretch}{1}
\qquad\qquad\qquad
\renewcommand{\arraystretch}{1.2}
\begin{tabularx}{0.4\textwidth}{||c | Y | Y | Y ||} 
\hline
\multicolumn{4}{||c||}{$L=2 $} \\
 \hline
 $g^2$ & $E_{\Omega}$ & $M_{\sigma}$ & $M_{\pi}$\\
 \hline\hline
 8 & -0.611 & 5.82 & 5.92 \\ 
 \hline
 4 & -0.949 & 4.41 & 4.49 \\
 \hline
 2 & -1.30 & 3.27 & 3.31 \\
 \hline
 1 & -1.58 & 2.72 & 2.74 \\
 \hline
 1/2 & -1.77 & 2.45 & 2.46 \\
 \hline
 1/4 & -1.88 & 2.30 & 2.31 \\
 \hline
 1/8 & -1.94 & 2.22 & 2.22 \\
 \hline
\end{tabularx}
\renewcommand{\arraystretch}{1}
\caption{
The vacuum energy and the masses of the $\sigma$- and $\pi$-mesons for $1+1$D QCD with $N_f=2$ for systems with $L=1,2$
spatial sites. These results are insensitive to $h$ as they are color singlets.}
\label{tab:specB0}
\end{table}
By examining the vacuum energy density 
$E_{\Omega}/L$, it is clear that, as expected, this number of lattice sites is insufficient to fully contain hadronic correlation lengths.  
While Table~\ref{tab:specB0} shows the energies of color-singlet states, there are also non-singlet states in the spectra with similar masses,
which become increasingly localized near the end of the lattice, as discussed in the previous section.

\begin{figure}[!ht]
    \centering
    \includegraphics[width=14cm]{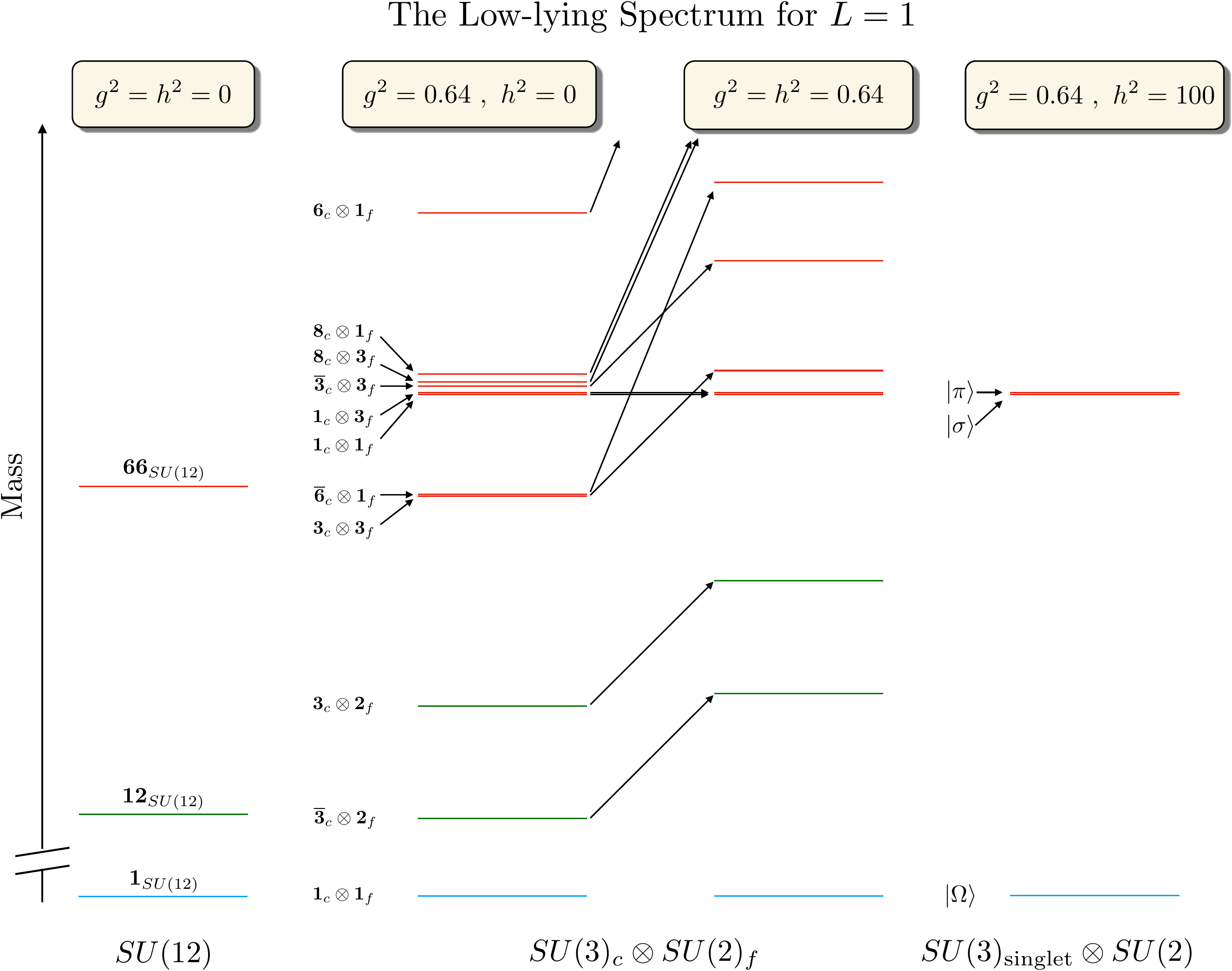}
    \caption{
    The spectrum of the Hamiltonian as the couplings $g$ and $h$ increase. 
    For $g=h=0$ there is an exact $SU(12)$ symmetry and the color-singlet 
    $\sigma$- and $\pi$-mesons are a part of the antisymmetric ${\bf 66}$ irrep. 
    When $g>0$ and $h=0$, the spectrum splits into irreps of global $SU(3)_c \otimes SU(2)_f$ with 
    color non-singlet states among the low-lying states.
    Increasing $h>0$ pushes non-singlet color states out of the low-lying spectrum. 
    Notice that the $\sigma$ and $\pi$ masses are insensitive to $h$, as expected.}
    \label{fig:specDegenh}
\end{figure}

It is informative to examine the spectrum of the $L=1$ system as both $g$ and $h$ are slowly increased and, in particular, take note of the relevant symmetries. For $g=h=0$,
with contributions from only the hopping and mass terms,
the system exhibits a global $SU(12)$ symmetry 
where the spectrum is that of free quasi-particles; see App.~\ref{app:freeSym}.
The enhanced global symmetry at this special point restricts the structure of the spectrum to the ${\bf 1}$ and ${\bf 12}$ of $SU(12)$ as well as the antisymmetric combinations of fundamental irreps, ${\bf 66}, {\bf 220}, \ldots$.  
For $g>0$, these $SU(12)$ irreps split into irreps of color $SU(3)_c$ and flavor $SU(2)_f$. 
The ${\bf 12}$ corresponds to single quark ($q$) or antiquark ($\overline{q}$) excitations
(with fractional baryon number), and splits into ${\bf 3}_c\otimes {\bf 2}_f$ for quarks and $\overline{\bf 3}_c\otimes
{\bf 2}_f$ for antiquarks.  In the absence of OBCs, these states would remain degenerate, but the boundary condition of vanishing background
chromo-electric field is not invariant under 
$q\leftrightarrow \overline{q}$ and the quarks get pushed to higher mass. As there is no chromo-electric energy associated with exciting an
antiquark at the end of the lattice in this mapping, the $\overline{\bf 3}_c\otimes {\bf 2}_f$ states remains low in the spectrum until $h\gg0$.
The ${\bf 66}$ corresponds to two-particle excitations, and contains all combinations of $qq$, $\overline{q}q$ and 
$\overline{q} \overline{q}$ excitations. 
The mixed color symmetry (i.e., neither symmetric or antisymmetric) of $\overline{q}q$ excitations allows for states with
${\bf 1}_c\otimes {\bf 1}_f
\oplus
{\bf 1}_c\otimes {\bf 3}_f
\oplus
{\bf 8}_c\otimes {\bf 1}_f
\oplus
{\bf 8}_c\otimes {\bf 3}_f
$,
while the $qq$ excitations with definite color symmetry allow for 
${\bf 6}_c\otimes {\bf 1}_f
\oplus \overline{\bf 3}_c\otimes {\bf 3}_f
$
and
$\overline{q}\overline{q}$ excitations allow for
$\overline{\bf 6}_c\otimes {\bf 1}_f
\oplus {\bf 3}_c\otimes {\bf 3}_f$,
saturating the $66$ states in the multiplet.
When $g>0$, these different configurations split in energy, and when
$h\gg0$, only color-singlet states are left in the low-lying spectrum. Figure~\ref{fig:specDegenh} shows the evolution of the spectrum as 
$g$ and $h$ increase.
The increase in mass of non-singlet color states  with $h$ is proportional to the Casimir of the $SU(3)_c$ representation which is evident in Fig.~\ref{fig:specDegenh} where, for example, the increase in the mass of the ${\bf 3}_c$s and $\overline{{\bf 3}}_c$s between $h^2 = 0$ and $h^2=0.64$ are the same.

\begin{figure}[!ht]
    \centering
    \includegraphics[width=12cm]{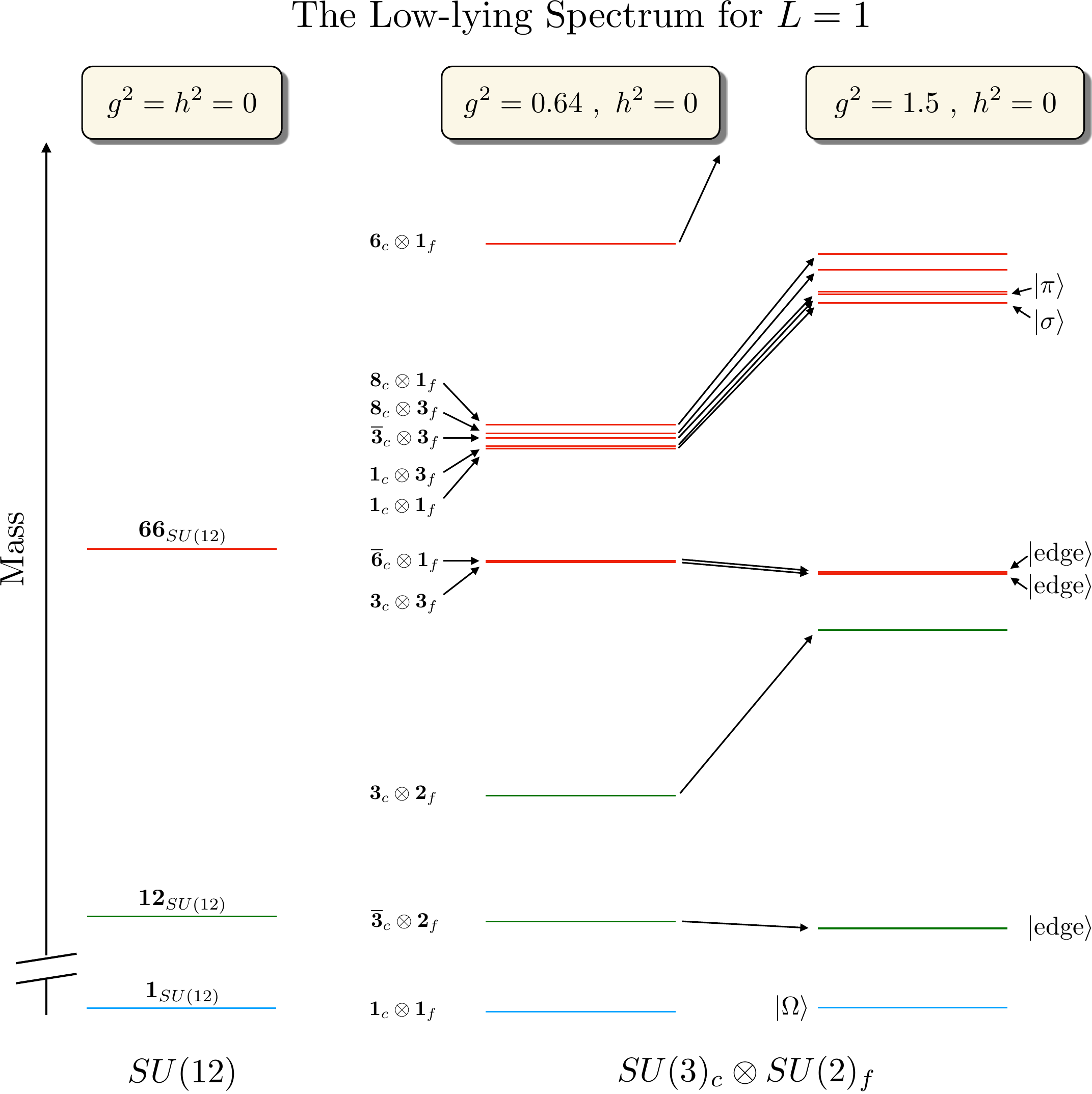}
    \caption{
    The spectrum of the Hamiltonian as $g$ increases for $h=0$. 
    When $g=h=0$ there is an exact $SU(12)$ symmetry and the $\sigma$- and $\pi$-mesons are a part of the antisymmetric ${\bf 66}$ irrep. 
    When $g>0$  but $h=0$ the spectrum splits into irreps of global $SU(3)_c \otimes SU(2)_f$, and non-singlet color states remain in the low-lying spectrum. 
    Increasing $g$ shifts all but the antiquark $\ket{\text{edge}}$ (states) to higher mass.
    }
    \label{fig:specDegeng}
\end{figure}

The antiquark states are particularly interesting as they correspond to edge states that are not ``penalized" in energy by the chromo-electric field when $h=0$. 
These states have an approximate
$SU(6)$ symmetry where the $6$ antiquarks transform in the fundamental.  
This is evident in the spectrum shown in 
Fig.~\ref{fig:specDegeng}
by the presence of a $\overline{{\bf 3}}_c \otimes {\bf 2}_f$ 
and nearly degenerate 
$\overline{\bf 6}_c\otimes {\bf 1}_f$ 
and 
${\bf 3}_c\otimes {\bf 3}_f$
which are identified as states of a 
${\bf 15}$ 
(an antisymmetric irrep of $SU(6)$)
that do not increase in mass as $g$ increases.
This edge-state $SU(6)$ symmetry is not exact 
due to interactions from the hopping term that couple the edge $\overline{q}$s to the rest of the lattice. 
These colored edge states are artifacts of OBCs and will persist in the low-lying spectrum for larger lattices.

Figures~\ref{fig:specDegenh} and \ref{fig:specDegeng}
reveal the near-degeneracy of the  $\sigma$- and $\pi$-mesons throughout the range of couplings $g$ and $h$, suggesting another approximate symmetry, which
can be understood in the small and large $g$ limits.
For small $g^2$, the effect of $H_{el} = \frac{g^2}{2}(Q_{0,u}^{(a)} + Q_{0,d}^{(a)})^2$ on the the $SU(12)$-symmetric spectrum can be obtained through perturbation theory.
To first order in $g^2$, the shift in the energy of any state is equal to the expectation value of $H_{el}$.
The $\sigma$- and $\pi$-meson states are both quark-antiquark states in the {\bf 66} irrep of $SU(12)$, and therefore, both have a ${\bf 3}_c$ color charge on the quark site and receive the same mass shift.\footnote{This also explains why
there are three other states nearly degenerate with the mesons, as seen in Fig.~\ref{fig:specDegenh}. 
Each of these states carry a ${\bf 3}_c$ or $\overline{{\bf 3}}_c$ color charge on the quark site and consequently have the same energy at first order in perturbation theory.
}
For large $g^2$, the only finite-energy excitations of the trivial vacuum (all sites unoccupied) are bare baryons and antibaryons,
and the spectrum is one of non-interacting color-singlet baryons.
Each quark (antiquark) site hosts $4$ distinct baryons (antibaryons) in correspondence with the multiplicity of the $I=3/2$ irrep.
As a result, the $\sigma$, $\pi$, $I=2,3$ mesons, deuteron and antideuteron are all degenerate.

\begin{figure}[!ht]
    \centering
    \includegraphics[width=14cm]{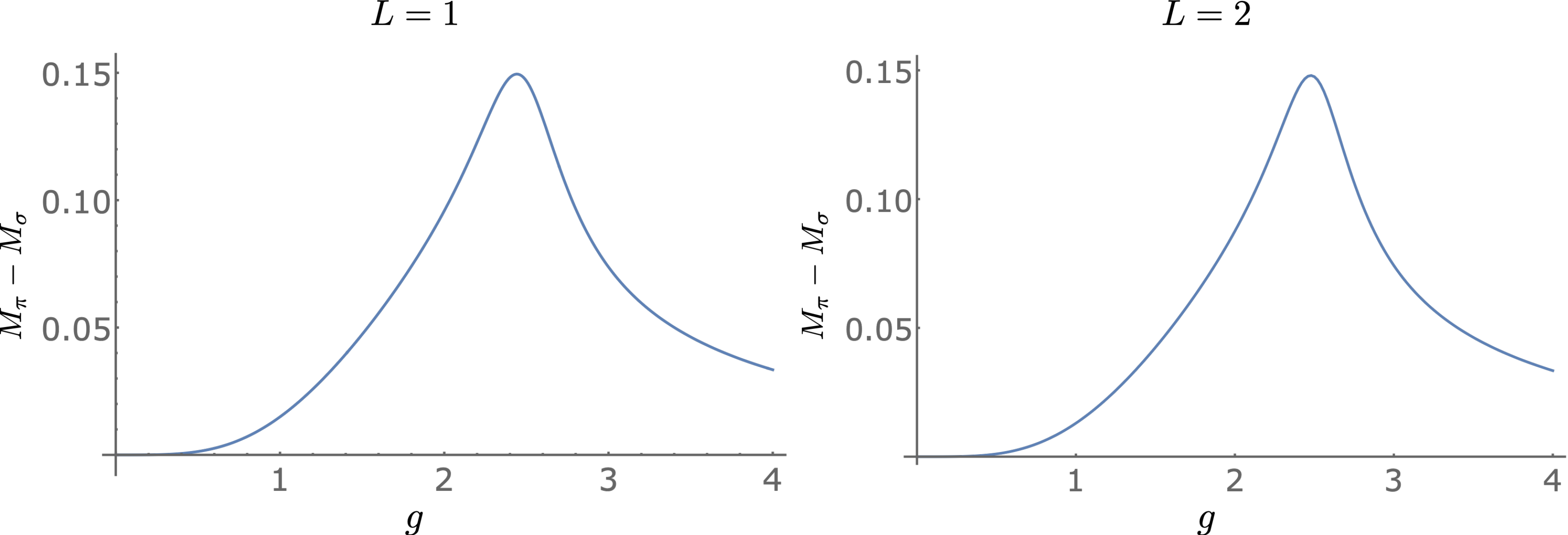}
    \caption{
    The mass splitting between the $\sigma$- and $\pi$-mesons for $L=1$ (left panel) and $L=2$ (right panel).}
    \label{fig:PiSigSplit}
\end{figure}

The $\sigma$- and $\pi$-meson mass splitting is shown in Fig.~\ref{fig:PiSigSplit} and has a clear maxima for $g \sim 2.4$. 
Intriguingly, this corresponds to the maximum of the linear entropy between quark and antiquarks (as discussed in Sec.~\ref{sec:eigenent}),
and suggests a connection between symmetry, via degeneracies in the spectrum, and entanglement.
This shares similarities with the correspondence between Wigner's $SU(4)$ spin-flavor
symmetry~\cite{physrev.51.106,physrev.51.947,PhysRev.56.519},
which becomes manifest in low-energy nuclear forces in the large-$N_c$ limit of QCD~\cite{kaplan:1995yg,Kaplan:1996rk},
and entanglement suppression in
nucleon-nucleon scattering found in Ref.~\cite{beane:2018oxh} (see also Refs.~\cite{beane:2021zvo,Low:2021ufv,Beane:2020wjl}).

Color singlet baryons are also present in this system, formed by contracting the color indices of three quarks with a Levi-Civita tensor (and antibaryons are formed from three antiquarks).
A baryon is composed of three $I=1/2$ quarks in the (symmetric) $I=3/2$ configuration and in a (antisymmetric) color singlet.
It will be referred to as the $\Delta$, highlighting its similarity to the $\Delta$-resonance in $3+1$D QCD. 
Interestingly, there is an isoscalar $\Delta\Delta$ bound state, which will be referred to as the deuteron. 
The existence of a deuteron makes this system valuable from the standpoint of quantum simulations of the formation of nuclei in a model of reduced complexity.
The mass of the $\Delta$, $M_{\Delta}$, and the binding energy of the deuteron, $B_{\Delta \Delta} = 2 M_{\Delta} - M_{\Delta \Delta}$, are shown in Table~\ref{tab:specBneq0} for a range of strong couplings.
\begin{table}[!ht]
\renewcommand{\arraystretch}{1.2}
\begin{tabularx}{0.4\textwidth}{||c | Y | Y ||} 
\hline
\multicolumn{3}{||c||}{$L=1 $} \\
 \hline
 $g^2$ & $M_{\Delta}$ & $B_{\Delta \Delta}$\\
 \hline\hline
 8 & 3.10 & $2.61\times 10^{-4}$\\ 
 \hline
 4 & 3.16 & $5.48\times 10^{-4}$\\
 \hline
 2 & 3.22 & $6.12\times 10^{-4}$\\
 \hline
 1 & 3.27 & $3.84\times 10^{-4}$\\
 \hline
 1/2 & 3.31 & $1.61\times 10^{-4}$\\
 \hline
 1/4 & 3.33 & $5.27\times 10^{-5}$\\
 \hline
 1/8 & 3.34 & $1.52\times 10^{-5}$\\
 \hline
\end{tabularx}
\renewcommand{\arraystretch}{1}
\qquad
\renewcommand{\arraystretch}{1.2}
\begin{tabularx}{0.5\textwidth}{|| c | Y | Y ||} 
\hline
\multicolumn{3}{||c||}{$L=2 $} \\
 \hline
 $g^2$ & $M_{\Delta}$ & $B_{\Delta \Delta}$\\
 \hline\hline
 8  & 3.10 & $2.50\times 10^{-4}$\\ 
 \hline
 4  & 3.16 & $4.95\times 10^{-4}$\\
 \hline
 2  & 3.21 & $5.07\times 10^{-4}$\\
 \hline
 1  & 3.24 & $4.60\times 10^{-4}$\\
 \hline
 1/2  & 3.25 & $1.53\times 10^{-3}$\\
 \hline
 1/4 & 3.23 & $3.91\times 10^{-3}$\\
 \hline
 1/8 & 3.20 & $3.35\times 10^{-3}$\\
 \hline
\end{tabularx}
\renewcommand{\arraystretch}{1}
\caption{
The mass of the $\Delta$ and the binding energy of the deuteron 
in $1+1$D QCD with $N_f=2$ for systems with $L=1,2$ spatial sites.}
\label{tab:specBneq0}
\end{table}

\begin{figure}[!ht]
    \centering
    \includegraphics[width=\columnwidth]{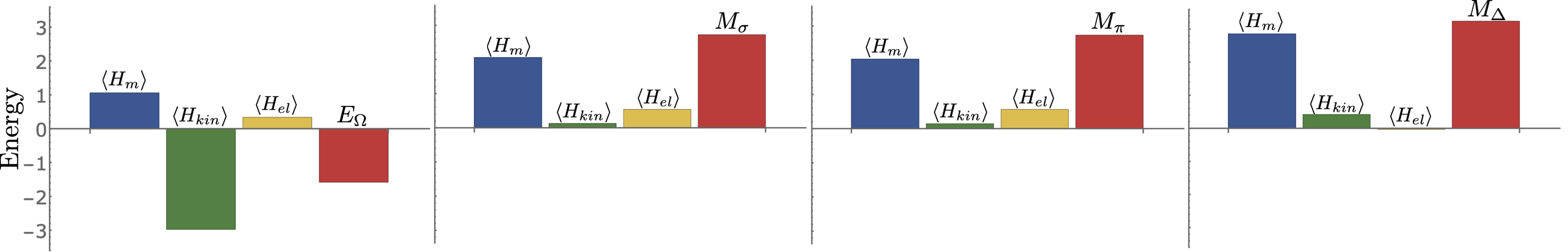}
    \caption{
    The decomposition of vacuum energy ($E_{\Omega}$) and the masses of the lightest hadrons ($M_{\sigma}$, $M_{\pi}$ and $M_{\Delta}$) 
    into contributions from the mass, the kinetic and the chromo-electric field terms in the Hamiltonian, defined in axial gauge, for $1+1$D QCD with $N_f=L=2$ and $m=g=1$.}
    \label{fig:massdeco}
\end{figure}

Understanding and quantifying the structure of the lowest-lying hadrons is a priority for nuclear physics research~\cite{LongRangePlan}.
Great progress has been made, experimentally, analytically and computationally,
in dissecting the mass and angular momentum of the proton (see, for example, Refs.~\cite{deFlorian:2009vb,Nocera:2014gqa,COMPASS:2015mhb,Yang:2018nqn,Alexandrou:2020sml,Ji:2021mtz,Wang:2021vqy,Lorce:2021xku}). 
This provides, in part, the foundation for anticipated precision studies at the future electron-ion collider (EIC)~\cite{Boer:2011fh,Accardi:2012qut} at Brookhaven National Laboratory.
Decompositions of the vacuum energy and the masses of the $\sigma$, $\pi$ and $\Delta$ are shown in Fig.~\ref{fig:massdeco} where, for example, the chromo-electric contribution to the
$\sigma$ is $\langle H_{el} \rangle = \bra{\sigma}  H_{el} \ket{\sigma} - \bra{\Omega}  H_{el} \ket{\Omega}$.
These calculations demonstrate the potential of future quantum simulations in being able to quantify decompositions of properties of the nucleon,
including in dense matter.
For the baryon states, it is $H_{el}$ that is responsible for the system coalescing into localized color singlets in order to minimize the energy in the chromo-electric field (between spatial sites).

\begin{figure}[!ht]
    \centering
    \includegraphics[width=14cm]{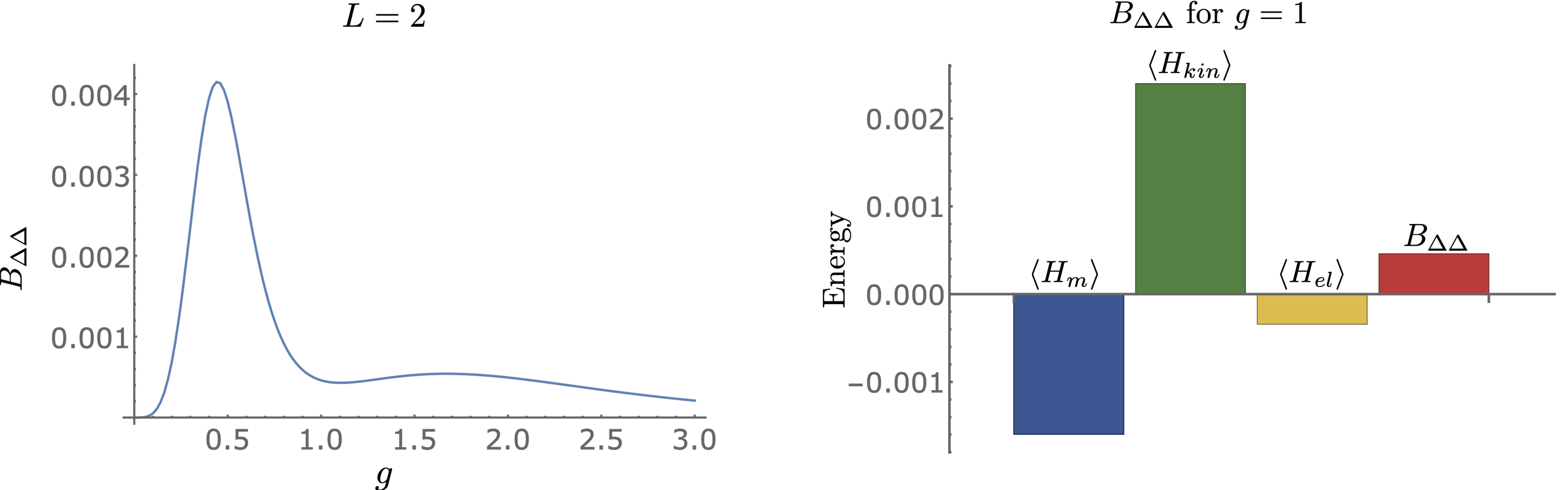}
    \caption{The left panel shows the deuteron binding energy, $B_{\Delta \Delta}$, for $m=1$ and $L=2$. The right panel shows the decomposition of $B_{\Delta \Delta}$ into contributions from the Hamiltonian for $g=1$.}
    \label{fig:deutBE}
\end{figure}

The deuteron binding energy is shown in the left panel of Fig.~\ref{fig:deutBE} as a function of $g$.
While the deuteron is unbound at $g=0$ for obvious reasons, it is also unbound at large $g$ because the spectrum is that of non-interacting color-singlet (anti)baryons.
Therefore, the non-trivial aspects of deuteron binding for these systems is for intermediate 
values of $g$. The decomposition of $B_{\Delta \Delta}$ is shown in the right panel of Fig.~\ref{fig:deutBE}, where, for example, the chromo-electric contribution is
\begin{equation}
    \langle H_{el} \rangle = 2 \big ( \bra{\Delta}  H_{el} \ket{\Delta} - \bra{\Omega}  H_{el} \ket{\Omega} \big ) - \big (\bra{\Delta \Delta}  H_{el} \ket{\Delta \Delta} - \bra{\Omega}  H_{el} \ket{\Omega} \big ) \ .
    \label{eq:deutBEdec}
\end{equation}
The largest contribution to the binding energy is $\langle H_{kin} \rangle$, which is the term responsible for creating $q \overline{q}$ pairs.
This suggests that meson-exchange may play a significant role in the attraction between baryons,
as is the case in $3+1$D QCD, but larger systems will need to be studied before 
definitive conclusions can be drawn.
One consequence of the lightest baryon 
being $I=3/2$ is that, for $L=1$, 
the $I_3=+3/2$ state completely occupies the up-quark sites.
Thus the system factorizes into an inert up-quark sector and a dynamic down-quark sector, and the absolute energy of the lowest-lying baryon state can be written as 
$E_{\Delta} =
M_{\Delta} + E_{\Omega}^{2 f} = 
3m + E_{\Omega}^{1 f}$,
where
$E_{\Omega}^{1,2 f}$ is the
vacuum energy of the 
$N_f=1,2$ flavor systems.
Analogously, the deuteron absolute energy is 
$E_{\Delta \Delta}=6m$, and therefore the 
deuteron binding energy can be written as
$B_{\Delta \Delta}= 2(3m+E_{\Omega}^{1 f}-E_{\Omega}^{2 f}) - (6m-E_{\Omega}^{2 f})
= 2E_{\Omega}^{1 f}-E_{\Omega}^{2 f}$.
This is quite a remarkable result because, in this system, the deuteron binding energy depends only on the difference between
the $N_f=1$ and $N_f=2$ vacuum energies, being bound when $2 E_{\Omega}^{1 f} - E_{\Omega}^{2 f} > 0$.
As has been discussed previously, it is the 
$q\overline{q}$ contribution from this difference that dominates the binding.
%

\subsubsection{The Low-Lying Spectrum Using D-Wave's Quantum Annealers}
\label{sec:dwave_spectrum}
\noindent
The low-lying spectrum of this system can also be determined through annealing by using
D-Wave's quantum annealer (QA) {\tt Advantage}~\cite{DwaveLeap}, 
a device with 5627 superconducting flux qubits, with a 15-way qubit connectivity via Josephson junctions rf-SQUID couplers~\cite{PhysRevB.80.052506}. 
Not only did this enable the determination of the energies of low-lying states, but it also assessed the ability of this quantum device to isolate nearly degenerate states.
The time-dependent Hamiltonian
of the device, which our systems are to be mapped,
are of the form of an Ising model, with the freedom to specify the single- and two-qubit coefficients. Alternatively, the Ising model can be rewritten in a quadratic unconstrained binary
optimization (QUBO) form, $f_Q(x)=\sum_{ij} Q_{ij}x_i x_j$, 
where $x_i$ are binary variables 
and $Q_{ij}$ is a QUBO matrix, which contains the coefficients of single-qubit ($i=j$) and two-qubit ($i\neq j$) terms. 
The QUBO matrix is the input that is submitted to 
{\tt Advantage}, with the output being a bit-string that minimizes $f_Q$.
Due to the qubit connectivity of {\tt Advantage},
multiple physical qubits are chained together to recover the required connectivity, limiting the system size that can be annealed.

The QA {\tt Advantage} was used to
determine the lowest three states in the $B=0$ sector of the $L=1$ system, with $m=g=1$ and $h=2$, following techniques presented in Ref.~\cite{Illa:2022jqb}. 
In that work, the objective function to be minimized is defined as $F=\langle \Psi \rvert \tilde{H} \lvert \Psi \rangle -\eta \langle \Psi| \Psi \rangle$~\cite{doi:10.1021/acs.jctc.9b00402}, where $\eta$ is a parameter that is included to avoid the null solution, and its optimal value 
can be iteratively tuned to be as close to the ground-state energy as possible. 
The wavefunction is expanded in a finite dimensional orthonormal basis $\psi_{\alpha}$, $\lvert \Psi \rangle =\sum^{n_s}_{\alpha} a_\alpha |\psi_{\alpha}\rangle$, which in this case reduces the dimensionality of $H$ to $88$, defining $\tilde{H}$, thus making it feasible to study with {\tt Advantage}. 
The procedure to write the objective function in a QUBO form can be found in Ref.~\cite{Illa:2022jqb} (and briefly described in App.~\ref{app:dwave}), where the coefficients $a_{\alpha}$ are
digitized using $K$ binary variables~\cite{doi:10.1021/acs.jctc.9b00402}, and the adaptive QA eigenvalue solver is implemented by using the zooming method~\cite{Chang:2019,arahman:2021ktn}. To reduce the uncertainty in the resulting energy and wavefunction, due to the noisy nature of this QA, the iterative procedure
described in Ref.~\cite{Illa:2022jqb} was used, where the (low-precision) solution obtained from the machine after several zooming steps constituted the starting point of a new anneal. This led to a reduction of the uncertainty by an order of magnitude (while effectively only doubling the resources used).

\begin{figure}[!ht]
    \centering
    \includegraphics[width=\columnwidth]{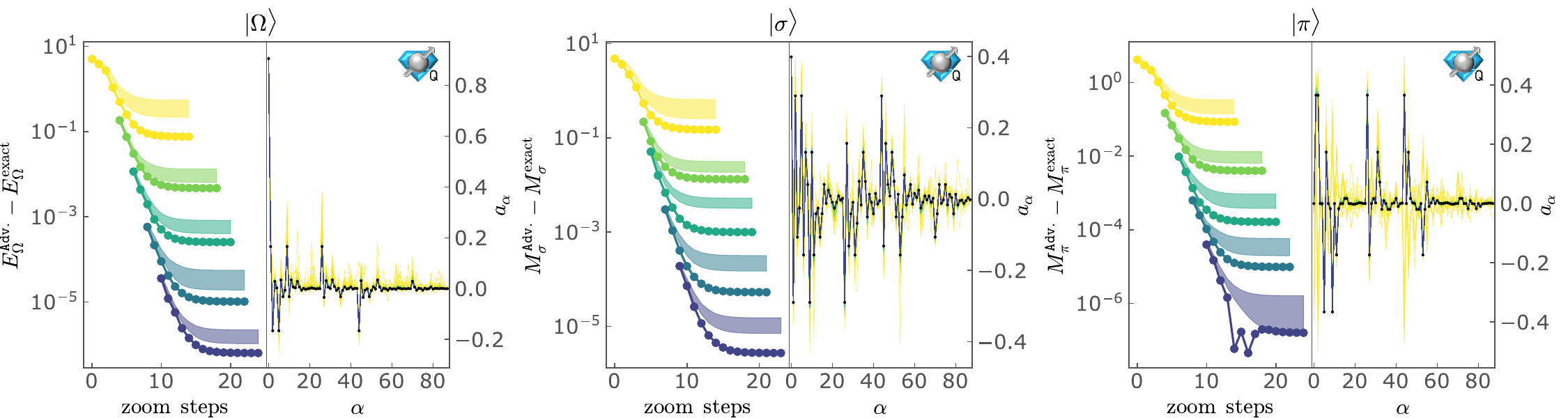}
    \caption{
    Iterative convergence of the energy, masses and wavefunctions for the three lowest-lying states in the $B=0$ sector of $1+1$D QCD with $N_f=2$ and $m=g=L=1$: vacuum (left), $\sigma$-meson (center) and $\pi$-meson (right). 
    The different colors correspond to different steps of the iterative procedure that is described in the main text. 
    The oscillatory behavior seen in the right panel around the 15th zoom step is discussed in App.~\ref{app:dwave}.
    The blue icons in the upper right indicate that this calculation was done on a quantum device~\cite{klco:2019xro}.}
    \label{fig:QAresults}
\end{figure}

Results obtained using {\tt Advantage} are shown in Fig.~\ref{fig:QAresults}, where the three panels show the convergence of the energy of the 
vacuum state (left), the mass of the $\sigma$-meson (center) and the mass of the $\pi$-meson (right) as a function of zoom steps, as well as comparisons to the exact wavefunctions. The bands in the plot correspond to 68\% confidence intervals determined from 20 independent runs of the annealing workflow, where each corresponds to $10^3$ anneals with an annealing time of $t_A=20$ $\mu$s, and the points correspond to the lowest energy found by the QA. The parameter $K$ in the digitization of $a_{\alpha}$ is set to $K=2$. The parameter $\eta$ is first set close enough to the corresponding energy (e.g., $\eta=0$ for the ground-state), and for the subsequent iterative steps it is set to the lowest energy found in the previous step. The first two excited states are nearly degenerate, and after projecting out the ground state, {\tt Advantage} finds both states in the first step of the iterative procedure (as shown by the yellow lines in the $\pi$ wavefunction of Fig.~\ref{fig:QAresults}). 
However, after one iterative step, the QA converges to one of the two excited states. 
It first finds the second excited state (the $\pi$-meson), and once this state is known with sufficient precision, it can be projected out to study the other excited state. 
The converged values for the energies and masses of these states are shown in Table~\ref{tab:QAresults},
along with the exact results. The uncertainties in these values should be understood as uncertainties on an upper bound of the energy (as they result from a variational calculation). For more details see App.~\ref{app:dwave}.
\begin{table}[!ht]
\renewcommand{\arraystretch}{1.2}
\begin{tabularx}{\textwidth}{|| c | c | Y | Y | Y ||} 
\hline
\multicolumn{5}{||c||}{$L=1$} \\
 \hline
 \multicolumn{2}{||c|}{} & $\ket{\Omega}$ & $\ket{\sigma}$ & $\ket{\pi}$ \\
 \hline\hline
 \multirow{2}{*}{Exact} & Energy & $-0.5491067$ & $2.177749$ & $2.1926786$ \\ 
 \cline{2-5}
 & Mass & - & $2.726855$ & $2.7417853$\\
 \hline
 \multirow{2}{*}{\tt Advantage} &Energy & $-0.5491051(6)$ & $2.177760(4)$ & $2.1926809(7)$ \\ 
 \cline{2-5}
 & Mass & - & $2.726865(4)$ & $2.7417860(9)$\\
 \hline
\end{tabularx}
\renewcommand{\arraystretch}{1}
\caption{Energies and masses of the three lowest-lying states in the $B=0$ sector of $1+1$D QCD with $N_f=2$ and $m=g=L=1$. Shown are the exact results from diagonalization of the Hamiltonian matrix and those obtained from D-Wave's {\tt Advantage}.}
\label{tab:QAresults}
\end{table}
%

\subsubsection{Quark-Antiquark Entanglement in the Spectra via Exact Diagonalization}
\label{sec:eigenent}
\begin{figure}[!ht]
    \centering
    \includegraphics[width=14cm]{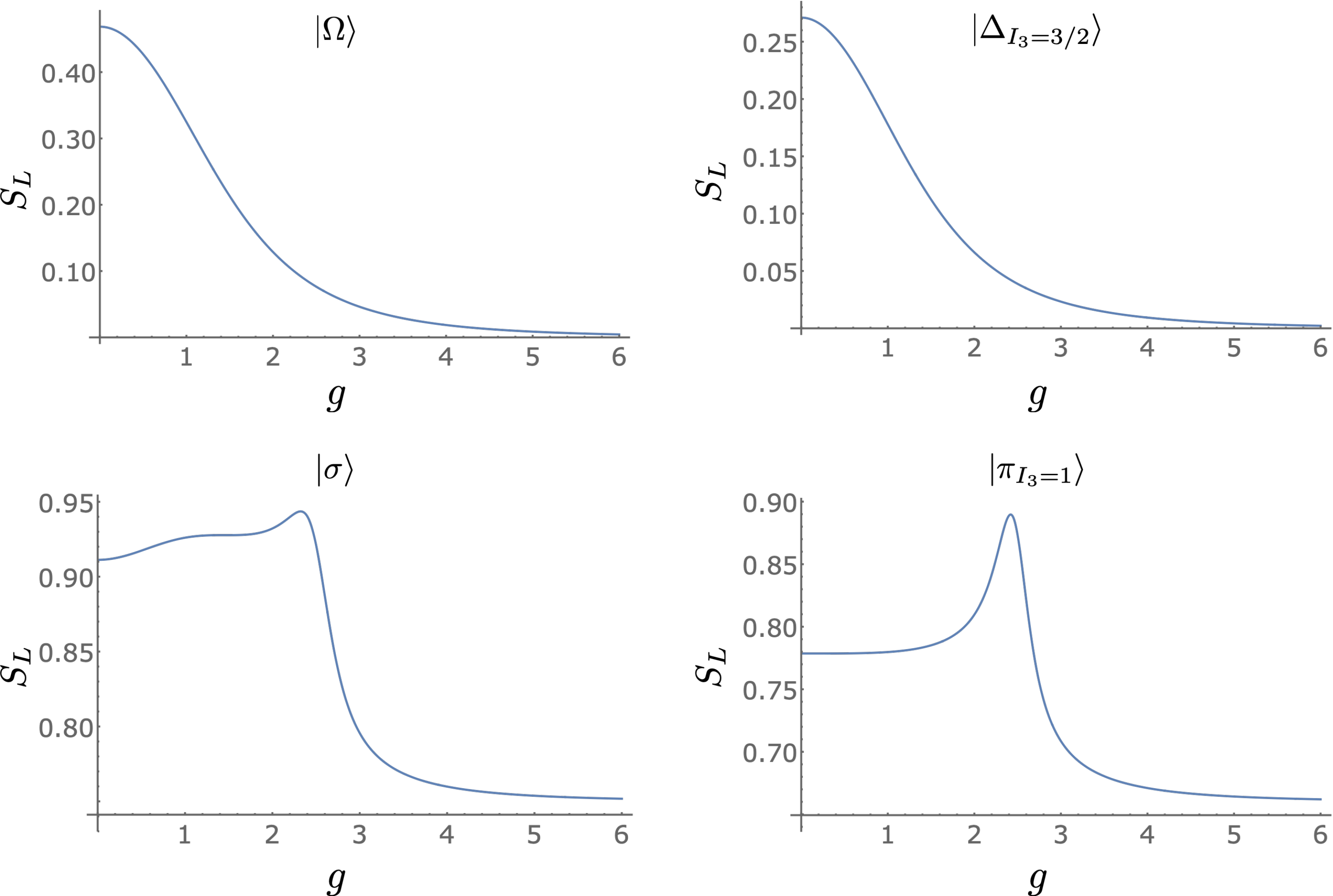}
    \caption{The linear entropy between quarks and antiquarks in $\ket{\Omega}$, $\ket{\Delta_{I_3=3/2}}$, $\ket{\sigma}$ and $\ket{\pi_{I_3=1}}$ for $m=L=1$.}
    \label{fig:m1N2Linent}
\end{figure}

\noindent
With $h \gg g$, the eigenstates of the Hamiltonian are color singlets and irreps of isospin. 
As these are global quantum numbers (summed over the lattice) the eigenstates are generically entangled among the color and isospin components at each lattice site.  With the hope of gaining insight into $3+1$D QCD, aspects of the entanglement structure of the $L=1$ wavefunctions are explored via exact methods.
An interesting measure of entanglement for these systems is the linear entropy between quarks and antiquarks, defined as
\begin{equation}
    S_L = 1 - \Tr [\rho_q^2]
    \ ,
\end{equation}
where $\rho_q =  \Tr_{\overline{q}} [\rho]$ and $\rho$ is a density matrix of the system. Shown in Fig.~\ref{fig:m1N2Linent} is the linear entropy between quarks and antiquarks in
$\ket{\Omega}$, $\ket{\sigma}$, $\ket{\pi_{I_3=1}}$ and $\ket{\Delta_{I_3=3/2}}$ as a function of $g$.
The deuteron is not shown as there is only one basis state contributing for $L=1$.

\begin{figure}
    \centering
    \includegraphics[width=14cm]{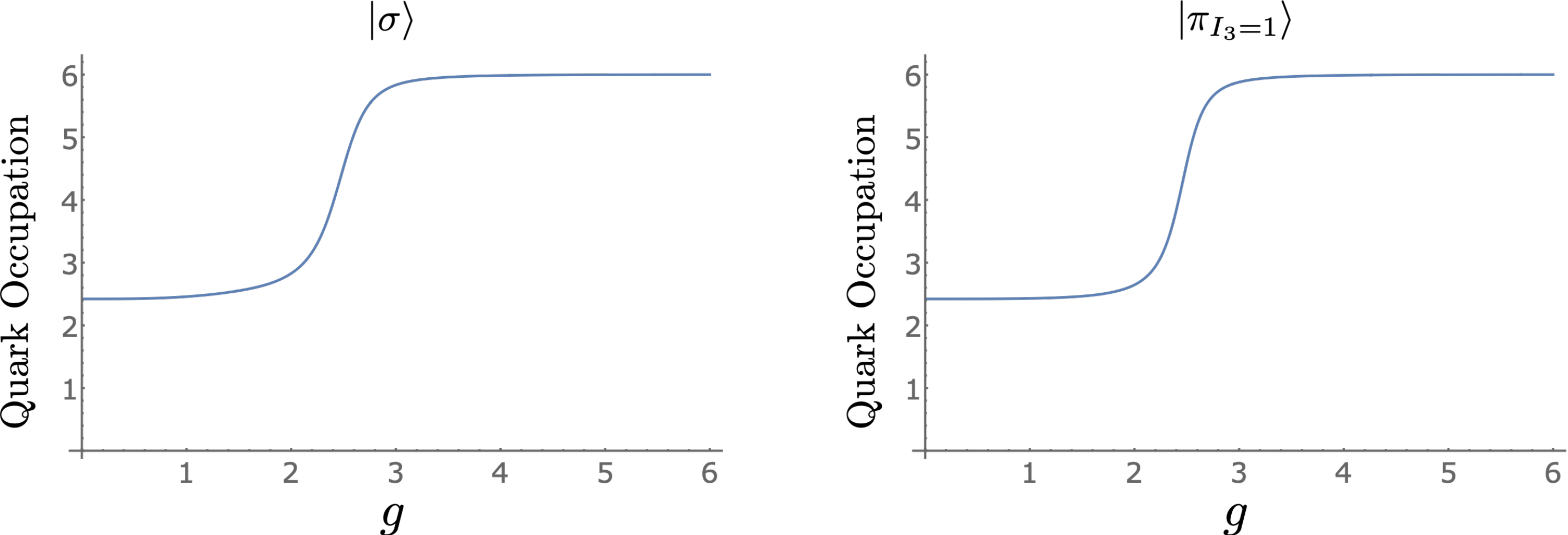}
    \caption{The expectation value of quark occupation in the $\ket{\sigma}$ and $\ket{\pi_{I_3 = 1}}$ for $m=L=1$.}
    \label{fig:m1N2Occ}
\end{figure}

The scaling of the linear entropy in the vacuum and baryon with $g$ can be understood as follows. 
As $g$ increases, color singlets on each site have the least energy density.
The vacuum becomes dominated by the unoccupied state and the $\Delta$ becomes dominated by the ``bare" $\Delta$ with all three quarks located on one site in a color singlet. 
As the entropy generically scales with the number of available states, 
the vacuum and baryon have decreasing entropy for increasing $g$.
The situation for the $\pi$ and $\sigma$ is somewhat more interesting. 
For small $g$, their wavefunctions are dominated by $q \overline{q}$ excitations on top of the trivial vacuum,  which minimizes the contributions from the mass term. 
However, color singlets are preferred as $g$ increases, 
and the mesons become primarily composed of baryon-antibaryon ($B \overline{B}$) excitations.
There are more $q \overline{q}$ states than 
$B \overline{B}$ states with a given $I_3$, 
and therefore there is more entropy at small $g$ than large $g$. 
The peak at intermediate $g$ occurs at the crossover between these two regimes where the meson has a sizable contribution from both $q \overline{q}$ and $B \overline{B}$ excitations. 
To illustrate this, 
the expectation value of total 
quark occupation (number of quarks plus the number of antiquarks) is shown in Fig.~\ref{fig:m1N2Occ}. 
For small $g$, the occupation is near $2$ since the state is mostly composed of $q \overline{q}$,
while for large $g$ it approaches $6$ as the state mostly consists of $B \overline{B}$. 
This is a transition from the excitations being 
``color-flux tubes" between quark and antiquark of the same color to bound states of color-singlet baryons and antibaryons.
%

\subsection{Digital Quantum Circuits}
\label{sec:Circuits}
\noindent
The Hamiltonian for $1+1$D QCD with arbitrary $N_c$ and $N_f$, when written in terms of spin operators, can be naturally mapped onto a quantum device with qubit registers. In this section the time evolution for systems with $N_c = 3$ and $N_f=2$ are developed.

\subsubsection{Time Evolution}
\begin{figure}[!ht]
    \centering
    \includegraphics[height=0.3\textheight]{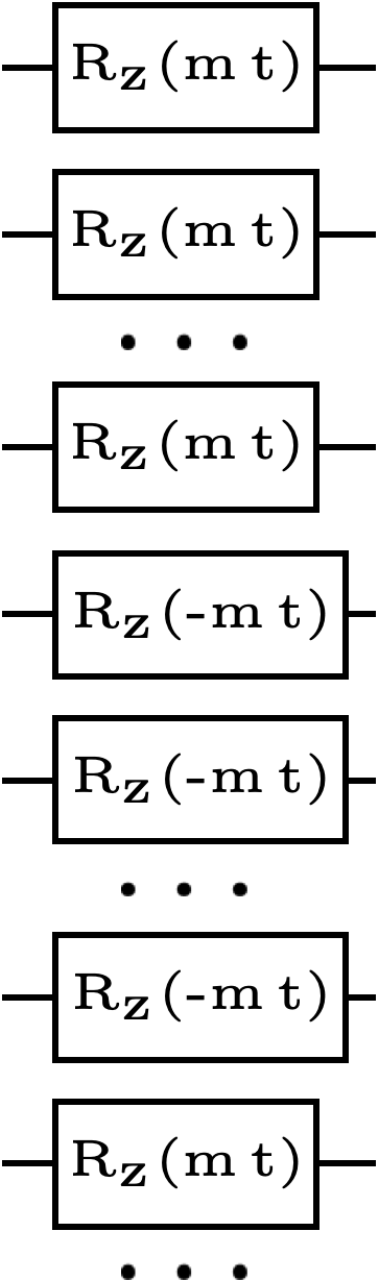}
    \caption{The quantum circuit that implements
    time evolution by the mass term,
    $U_m(t) = \exp(- i H_m t)$.}
    \label{circ:Um}
\end{figure}

\noindent
To perform time evolution on a quantum computer, the operator $U(t) = \exp(-i H t)$ is reproduced by a sequence of gates applied to the qubit register.
Generally, a Hamiltonian cannot be directly mapped to such a sequence efficiently, but each of the elements in a Trotter decomposition can, with systematically reducible errors.
Typically, the Hamiltonian is divided into Pauli strings whose unitary evolution can be implemented with quantum circuits that are readily constructed.
For a Trotter step of size $t$, the circuit that implements the time evolution from the mass term, $U_m(t) = \exp(- i H_m t)$, is shown in Fig.~\ref{circ:Um}.
The staggered mass leads to quarks being rotated by a positive angle and antiquarks being rotated by a negative angle. 
Only single qubit rotations about the z-axis are required for its implementation, with 
$R_Z(\theta) = \exp(-i \theta Z/2)$.
The circuit that implements
the evolution from the baryon chemical potential, $\mu_B$,
$U_{\mu_B}(t) = \exp(- i H_{\mu_B} t)$, 
is similar to $U_m(t)$  with 
$m \to \mu_B/3$, and with both quarks and antiquarks rotated by the same angle. 
Similarly, the circuit that implements the evolution from
the isospin chemical potential, $\mu_I$,
$U_{\mu_I}(t) = \exp(- i H_{\mu_I} t)$, 
is similar to $U_m(t)$ with $m \to
\mu_I/2$ and up (down) quarks rotated by a negative (positive) angle.

\begin{figure}[!ht]
    \centering
    \includegraphics[width=17cm]{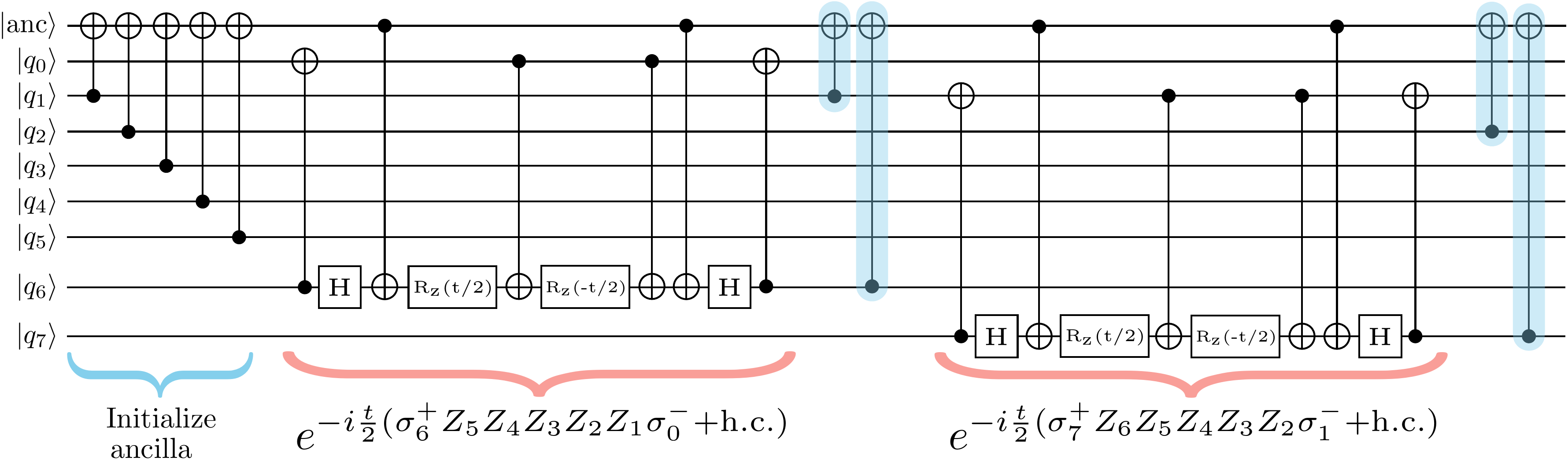}
    \caption{
    A circuit that implements the time evolution from two sequential hopping terms.
    Implementing $\exp(-i H_{kin} t)$ in Eq.~(\ref{eq:Hkin2flav}) is a straightforward extension of this circuit.}
    \label{circ:UkinAnc}
\end{figure}

The kinetic piece of the Hamiltonian, Eq.~(\ref{eq:Hkin2flav}), is composed of hopping terms of the form
\begin{equation}
H_{kin} \ \sim \ 
    \sigma^+ ZZZZZ \sigma^- + \rm{h.c.} \ .
    \label{eq:hop}
\end{equation}
The $\sigma^+$ and $\sigma^-$ operators enable quarks and antiquarks to move between sites with the same color and flavor 
(create $\overline{q}^\alpha_i q_\alpha^i$ pairs)
and the string of $Z$ operators 
incorporates the signs from Pauli statistics.
The circuits for Trotterizing these terms are 
based on circuits in Ref.~\cite{stetina:2020abi}. We introduce an ancilla to 
accumulate the parity of the JW string of $Z$s.
This provides a mechanism for the 
different hopping terms to re-use 
previously computed
(partial-)parity.\footnote{An ancilla was used similarly in Ref.~\cite{qchem2014}.}
The circuit for the first two hopping terms is shown in Fig.~\ref{circ:UkinAnc}.
The first circuit operations initialize 
the ancilla to store the parity of the string of $Z$s between the first and last qubit of the string. Next, the system is evolved
by the exponential of the hopping term. After the exponential of each hopping term, the ancilla is modified for the parity of the subsequent hopping term 
(the CNOTs highlighted in blue).
Note that the hopping of quarks, or antiquarks, of different flavors and colors commute, and the Trotter decomposition is exact (without Trotterization errors) over a single spatial site.

\begin{figure}[!ht]
    \centering
    \includegraphics[width=10cm]{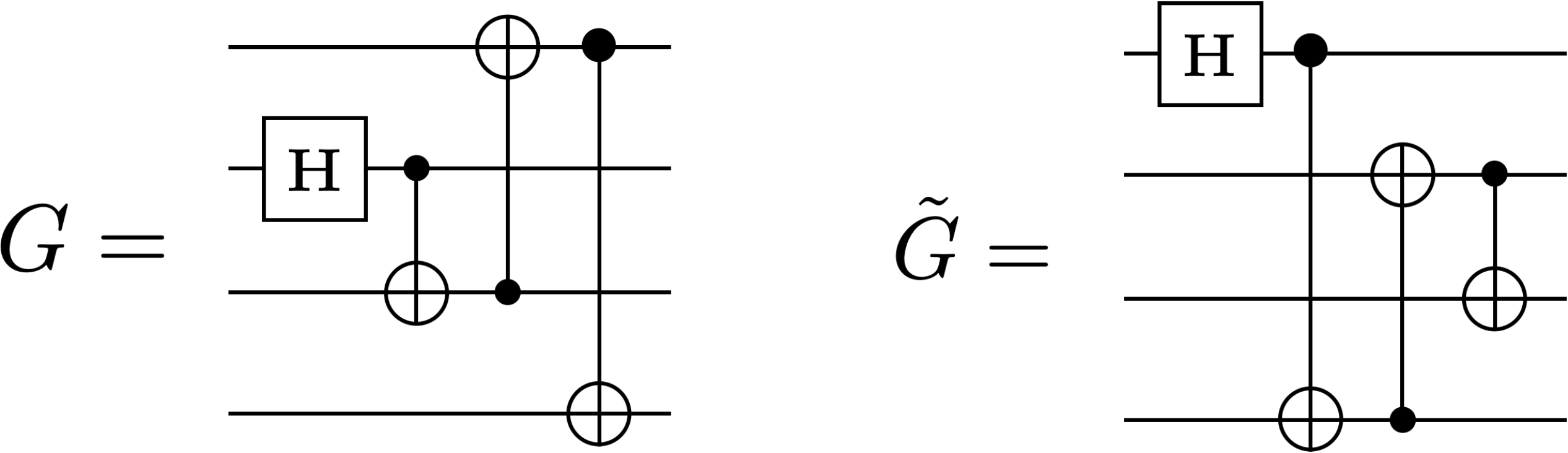}
    \caption{Two GHZ state-preparation circuits.}
    \label{circ:GHZ}
\end{figure}

\begin{figure}[!ht]
    \centering
    \includegraphics[width=15cm]{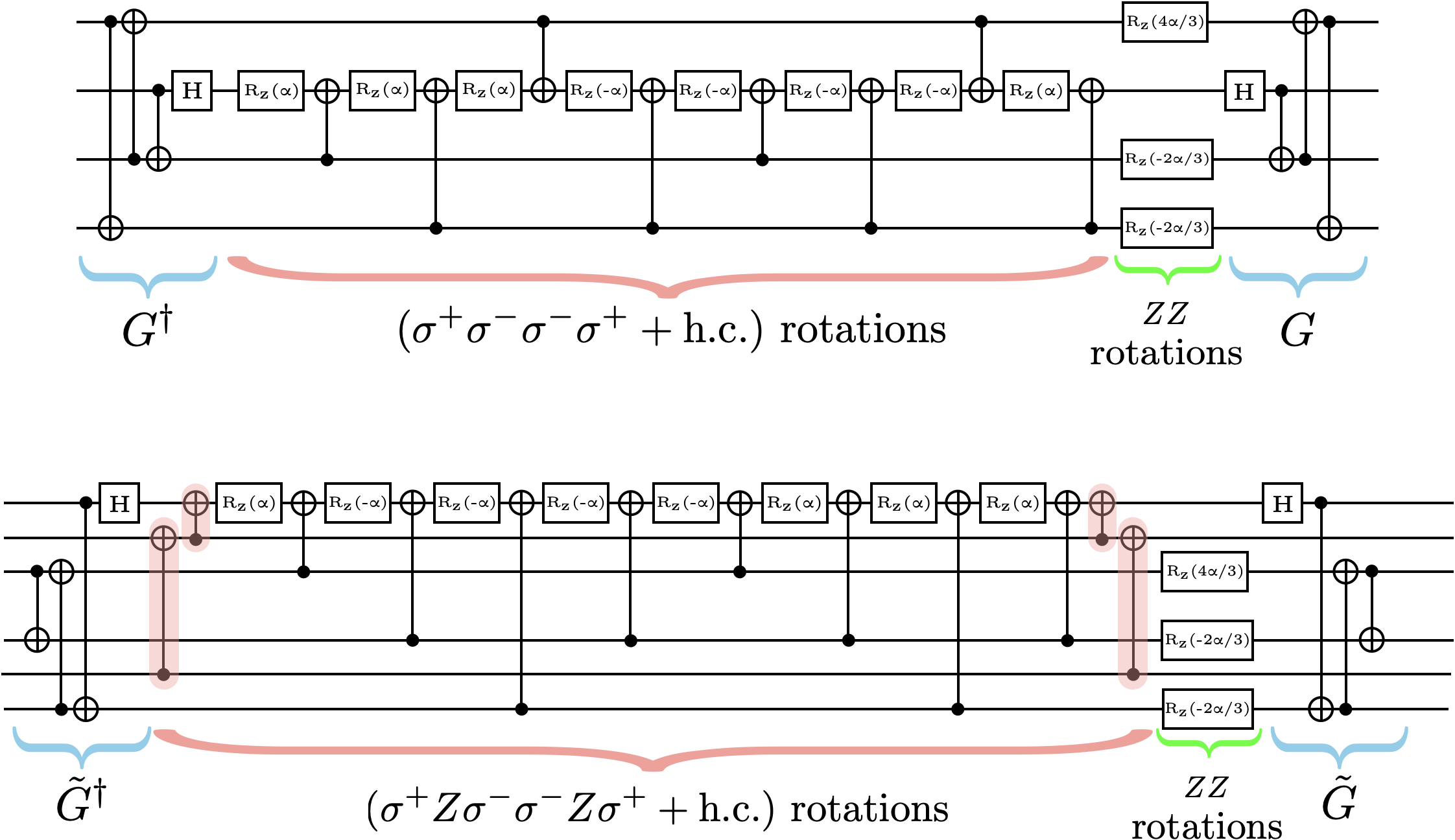}
    \caption{The circuits that implement the time evolution of ${\exp}(-8 i \alpha Q_{n,f}^{(a)} \, Q_{m,f'}^{(a)})$. 
    Specifically, the upper circuit implements
    exp$\{-i 4 \alpha [ (\sigma^+\sigma^-\sigma^-\sigma^+ + {\rm h.c.}) +  \frac{1}{12}(2 IZIZ -IZZI - ZIIZ) ]\}$, 
    while the lower circuit implements
    exp$\{-i 4 \alpha[ (\sigma^+ Z \sigma^-\sigma^- Z \sigma^- + {\rm h.c.}) +  \frac{1}{12}(2 ZIIZII -IIZZII - ZIIIIZ) ]\}$.
    The CNOTs highlighted in red account for the $Z$s in $\sigma^+ Z \sigma^- \sigma^- Z \sigma^+$.
    For 
    $SU(3)$ with $N_f=2$ and 
    $L=1$, the required evolution operators have $\alpha = t g^2 /8$.}
    \label{circ:UpmmpZZ}
\end{figure}

Implementation of the time-evolution 
induced by the energy density in the 
chromo-electric field, $H_{el}$, 
given in Eq.~(\ref{eq:QnfQmfp}),
is the most challenging due to its 
inherent non-locality in axial gauge.
There are two distinct types of contributions: One is from same-site interactions and the other from interactions between different sites.
For the same-site interactions, the operator is the product of charges 
$Q_{n,f}^{(a)} \, Q_{n,f}^{(a)}$, which contains only $ZZ$ operators, and is digitized with the standard two CNOT circuit.\footnote{Using the native $ZX$ gate on IBM's devices allows this to be done with a single two-qubit entangling gate~\cite{Kim2021ScalableEM}.}
The $Q_{n,f}^{(a)} \, Q_{m,f'}^{(a)}$ operators contain 4-qubit interactions of the form 
$(\sigma^+ \sigma^- \sigma^- \sigma^+ + \ {\rm h.c.})$
and 
6-qubit interactions of the form 
$(\sigma^+ Z \sigma^- \sigma^- Z \sigma^+ + \ {\rm h.c.})$,
in addition to $ZZ$ contributions.
The manipulations required to implement the 6-qubit operators parallel those required for the 4-qubit operators, and here only the latter is discussed in detail.
These operators can be decomposed into eight mutually commuting terms,
\begin{align}
\sigma^+ \sigma^- \sigma^- \sigma^+ + {\rm h.c.} =& \frac{1}{8}(XXXX + YYXX + YXYX - YXXY -\nonumber \\
& XYYX + XYXY + XXYY + YYYY) \ .
\label{eq:pmmp}
\end{align}
The strategy for identifying the corresponding time evolution circuit is to first apply a unitary that diagonalizes every term, apply the diagonal rotations, and finally, act with the inverse unitary to return to the computational basis. 
By only applying diagonal rotations, 
many of the CNOTs can be arranged to cancel.
Each of the eight Pauli strings
in Eq.~(\ref{eq:pmmp})
takes a state in the computational basis to the corresponding bit-flipped state (up to a phase). 
This suggests that the desired eigenbasis
pairs together states with their bit-flipped counterpart, which is an inherent property of the GHZ basis~\cite{stetina:2020abi}. 
In fact, any permutation of the GHZ state-preparation circuit diagonalizes the interaction. 
The two that will be used,
denoted by $G$ and $\tilde G$,
are shown in Fig.~\ref{circ:GHZ}.
In the diagonal bases, the Pauli strings
in Eq.~(\ref{eq:pmmp}) become
\begin{align}
    G^{\dagger}\ (\sigma^+ \sigma^- \sigma^- \sigma^+ + {\rm h.c.})\ G = \frac{1}{8} (IIZI - ZIZZ - ZZZZ + ZIZI & \nonumber \\
     + IZZI - IIZZ - IZZZ + ZZZI  ) \ ,& \nonumber \\
    \tilde{G}^{\dagger}\ (\sigma^+ \sigma^- \sigma^- \sigma^+ + {\rm h.c.})\ \tilde{G} = \frac{1}{8} (IIIZ - IZZZ - IIZZ + ZIIZ & \nonumber \\
    + IZIZ - ZZZZ - ZIZZ + ZZIZ ) \ .&
    \label{eq:pmmpdiag}
\end{align}
Another simplification comes from the fact that 
$ZZ$ in the computational basis becomes  
a single $Z$ in a GHZ basis if the GHZ state-preparation circuit has a CNOT connecting the two $Z$s. 
For the case at hand, this implies
\begin{align}
    G^{\dagger}\ (IZZI + IZIZ + ZIIZ)\ G =& \ IZII + IIIZ + ZIII \ , \nonumber \\[4pt]
    \tilde{G}^{\dagger}\ (ZIZI + IZZI + ZIIZ)\ \tilde{G} =& \ IIZI + IZII + ZIII  \ .
    \label{eq:ZZGHZ}
\end{align}
As a consequence, all nine $ZZ$ terms in $Q_{n,f}^{(a)} \, Q_{m,f'}^{(a)}$ 
become single $Z$s in a GHZ basis, thus requiring no additional CNOT gates to implement. 
Central elements of the circuits 
required to implement time evolution of the chromo-electric energy density
are shown in Fig.~\ref{circ:UpmmpZZ},
which extends the circuit presented in Fig.~4 of Ref.~\cite{stetina:2020abi} to non-Abelian gauge theories.
More details on these circuits can be found in App.~\ref{app:circ}.

\subsubsection{Trotterization, Color Symmetry and Color Twirling}
\label{sec:colorBreak}
\noindent
After fixing the gauge, the Hamiltonian is no longer manifestly invariant under local  $SU(3)$ gauge transformations. 
However, as is well known, observables of the theory are correctly computed from such a gauge-fixed Hamiltonian, which possesses a remnant global $SU(3)$ symmetry.
This section addresses the extent to which this symmetry is preserved by Trotterization of the time-evolution operator. 
The focus will be on 
the $N_f=1$ theory as including additional flavors
does not introduce new complications.

Trotterization of the mass and kinetic parts of the Hamiltonian,
while having  non-zero commutators between some terms, preserves the global $SU(3)$ symmetry.
The time evolution of $Q_{n}^{(a)} \, Q_{n}^{(a)}$
can be implemented in a unitary operator without Trotter errors, and, therefore, does not break $SU(3)$. 
On the other hand,
the time evolution induced by 
$Q_{n}^{(a)} \, Q_{m}^{(a)}$
is implemented by the operator being divided into 
four terms:
$(Q^{(1)}_n \, Q^{(1)}_{m} + Q^{(2)}_n \, Q^{(2)}_m)$,  $(Q^{(4)}_n \, Q^{(4)}_m + Q^{(5)}_n \, Q^{(5)}_m)$, $(Q^{(6)}_n \, Q^{(6)}_m + Q^{(7)}_n \, Q^{(7)}_m)$ and $(Q^{(3)}_n \, Q^{(3)}_m + Q^{(8)}_n \, Q^{(8)}_m)$. In order for global $SU(3)$ to be unbroken, 
the sum over the entire lattice
of each of the 8 gauge charges must be unchanged under time evolution. 
Therefore, 
the object of interest is the commutator
\begin{equation}
\mathcal{C} = \left [ \sum_{n=0}^{2L-1}Q^{(a)}_{n} \ , \ Q^{(\tilde{b})}_{m}\cdot Q^{(\tilde{b})}_{l} \right ] \ ,
\label{eq:Qcomm}
\end{equation}
where $\tilde{b}$ is summed over the elements of one of the pairs in $\{(1,2),\,(4,5),\,(6,7),\,(3,8)\}$. 
It is found that this commutator only vanishes if $a=3$ or $a=8$, or if $\tilde{b}$ is summed over all $8$ values (as is the case for the exact time evolution operator). 
Therefore, Trotter time evolution does not preserve the global off-diagonal $SU(3)$ charges and, for example, color singlets can evolve into non-color singlets.
Equivalently, the Trotterized time evolution operator is not in the trivial representation of $SU(3)$.
To understand this point in more detail,
consider the transformation of 
$\left(T^a\right)^i_j  \ \left(T^a\right)^k_l$ for any given $a$. 
Because of the symmetry of this product of operators, each transforming as an ${\bf 8}$, the product must decompose into ${\bf 1}\oplus {\bf 8}\oplus {\bf 27} $,
where the elements of each of the irreps can be found from
\begin{multline}
\left(T^a\right)^i_j  \ \left(T^a\right)^k_l = 
\left(\hat {\cal O}_{27}^a\right)^{ik}_{jl} 
-\frac{2}{5}\left[ \delta^i_j \left(\hat {\cal O}^a_8\right)^k_l + \delta^k_l \left(\hat {\cal O}^a_8\right)^i_j \right] \\
+\frac{3}{5}
\left[\delta^i_l \left(\hat {\cal O}^a_8\right)^k_j + \delta^k_j \left(\hat {\cal O}^a_8\right)^i_l \right]
+ \frac{1}{8} \left( \delta^i_l \delta^k_j\ -\ \frac{1}{3} \delta^i_j \delta^k_l \right) \hat {\cal O}^a_1
\  ,
\end{multline}
where
\begin{align}
\left(\hat {\cal O}^a_{27}\right)^{ik}_{jl} 
 = & \
\frac{1}{2}
\left[ \left(T^a\right)^i_j \left(T^a\right)^k_l + \left(T^a\right)^i_l \left(T^a\right)^k_j \right]
\nonumber\\
& -
\frac{1}{10}\left[ 
\delta^i_j \left(\hat {\cal O}^a_8\right)^k_l 
+ \delta^i_l \left(\hat {\cal O}^a_8\right)^k_j 
+ \delta^k_j \left(\hat {\cal O}^a_8\right)^i_l 
+ \delta^k_l \left(\hat {\cal O}^a_8\right)^i_j \right]  -\frac{1}{24} \left( \delta^i_j \delta^k_l + \delta^i_l \delta^k_j \right) \hat {\cal O}^a_1 \ ,
\nonumber\\
\left(\hat {\cal O}^a_8\right)^i_j 
 = & \
\left(T^a\right)^i_\beta \left(T^a\right)_j^\beta\ -\ \frac{1}{3} \delta^i_j \hat {\cal O}^a_1
\ ,\ \ 
\hat {\cal O}^a_1 \ =\  \left(T^a\right)^\alpha_\beta \left(T^a\right)_\alpha^\beta \ =\ \frac{1}{2}
\  .
\end{align}
When summed over $a=1,\ldots,8$, the contributions from the ${\bf 8}$ and ${\bf 27}$ vanish, leaving the familiar contribution from the ${\bf 1}$.
When only partials sums are available, as is the situation with individual contributions to the Trotterized evolution, 
each of the contributions is the exponential of 
${\bf 1}\oplus {\bf 8}\oplus 
{\bf 27} $, with only the singlet contributions leaving the lattice a color singlet.
The leading term in the expansion of the product of the four pairs of Trotterized evolution operators sum to leave only the singlet contribution. 
In contrast, higher-order terms do not cancel and 
non-singlet contributions are present.

This is a generic problem 
that will be encountered when satisfying Gauss's law
leads to non-local charge-charge interactions. 
This is not a problem for $U(1)$, and 
surprisingly, is not a problem for 
$SU(2)$ because $(Q^{(1)}_n \, Q^{(1)}_m, Q^{(2)}_n \, Q^{(2)}_m, Q^{(3)}_n \, Q^{(3)}_m )$ are in the Cartan sub-algebra of $SU(4)$ and therefore mutually commuting.  However, it is a problem for $N_c>2$.
One way around the breaking of global $SU(N_c)$ 
is through the co-design 
of unitaries that directly (natively) implement
${\exp}( i \alpha Q^{(a)}_n \, Q^{(a)}_m)$; see Sec.~\ref{sec:codesign}.
Without such a native unitary,
the breaking of $SU(N_c)$ appears as any other Trotter error, and can be systematically reduced in the same way. A potential caveat to this is if the time evolution operator took the system into a different phase, but our studies of $L=1$ show no evidence of this.

It is interesting to note that the terms generated by the Trotter commutators form a closed algebra.
In principle, a finite number of terms could be
included to define an effective Hamiltonian whose Trotterization exactly maps onto the desired evolution operator (without the extra terms).
It is straightforward to work out the terms generated order-by-order in the Baker-Campbell-Hausdorff formula.
Aside from re-normalizing the existing charges, there are $9$ new operator structures produced. 
For example, the leading-order commutators generate the three operators, ${\cal O}_i$, in Eq.~(\ref{eq:BCHOp}),
\begin{align}
{\cal O}_i = 
\begin{cases} 
      (\sigma^+ I \sigma^- \sigma^- Z \sigma^+ - \sigma^+ Z \sigma^- \sigma^- I \sigma^+) - {\rm h.c.}  \ ,\\
      (I \sigma^- \sigma^+ Z \sigma^+ \sigma^- - Z \sigma^- \sigma^+ I \sigma^+ \sigma^-) - {\rm h.c.}  \ , \\
       (\sigma^+ \sigma^- Z \sigma^- \sigma^+ I - \sigma^+ \sigma^- I \sigma^- \sigma^+ Z) - {\rm h.c.} \ .
\end{cases}
\label{eq:BCHOp}
\end{align}
In general, additional operators are constrained only by (anti)hermiticity, 
symmetry with respect to $n \leftrightarrow m$ and preservation of $(r,g,b)$, and should generically be included in the same spirit as terms in the  Symanzik-action~\cite{Symanzik:1983dc,Symanzik:1983gh} for lattice QCD.

With Trotterization of the gauge field introducing violations of gauge symmetry, and the presence of bit- and phase-flip errors within the device register, it is worth briefly considering a potential mitigation strategy. A single
bit-flip error will change isospin by $|\Delta I_3|=1/2$ and color charge by one unit of red or green or blue.
After each Trotter step on a real quantum device, such errors will be encountered and a mitigation or correction scheme is required.
Without the explicit gauge-field degrees of freedom and local charge conservation checks enabled by Gauss's law, such errors can only be detected globally, and hence, cannot be actively corrected during the evolution.\footnote{When local gauge fields are present,
previous works have found that including a quadratic ``penalty-term" in the Hamiltonian is effective in mitigating violation of Gauss's law~\cite{Hauke:2013jga,zohar:2015hwa,Dalmonte:2016alw,Halimeh:2019svu}. See also Refs.~\cite{physrevlett.112.120406,Kasper:2020owz}. \label{foot:penaltyterm}}
Motivated by this, consider introducing a twirling phase factor into the evolution, $\exp(-i \theta^a {\cal Q}^{(a)})$, where ${\cal Q}^{(a)}$ is the total charge on the lattice.
If applied after each Trotter step, with a randomly selected set of eight angles, $\theta^a$, 
the phases of color-nonsinglet states become random for each member of an ensemble, mitigating errors in some observables. 
Similar twirling phase factors could be included for the other charges that are conserved or approximately conserved.

\subsubsection{Quantum Resource Requirements for Time Evolution}
\noindent
It is straightforward to extend the circuits presented in the previous section to arbitrary $N_c$ and $N_f$. The quantum
resources required for time evolution can be quantified
for small, modest and asymptotically large systems. As discussed previously, a quantum register with $N_q=2 L N_c N_f$ qubits\footnote{The inclusion of an ancilla for the kinetic term increases the qubit requirement to $N_q = 2L N_c N_f + 1$.} is required to encode one-dimensional $SU(N_c)$ gauge theory
with $N_f$ flavors on $L$ spatial lattice sites using the JW transformation. For $SU(3)$ gauge theory, this leads to, for example, $N_q = 6L$ with only $u$-quarks and $N_q = 18L$ with $u,d,s$-quarks.
The five distinct contributions to the resource requirements, 
corresponding to application of the unitary operators providing 
a single Trotter step associated with the quark mass, $U_m$, the baryon chemical potential, $U_{\mu_B}$, the isospin chemical potential, $U_{\mu_I}$, the kinetic term, $U_{kin}$, and the chromo-electric field, $U_{el}$, are 
given in terms of the number of 
single-qubit rotations, denoted by ``$R_Z$'', the number of Hadamard gates, denoted by ``Hadamard'', and the number of CNOT gates, denoted by ``CNOT''.
It is found that\footnote{
For $N_c = 2$ only three of the $ZZ$ terms can be combined into $Q_{n,f}^{(a)} \, Q_{m,f'}^{(a)}$ and the number of CNOTs for one Trotter step of $U_{el}$ is 
\begin{equation}
U_{el}  \ : \ (2 L-1) N_f [9 (2 L-1) N_f-7] \ \ \ | \ \text{CNOT} \ .
\label{eq:Nc2CNOT}
\end{equation}
Additionally, for $N_c N_f < 4$, the Trotterization of $U_{{\rm kin}}$ is more efficient without an ancilla and the number of CNOTs required is
\begin{equation}
U_{ kin}  \ : \ 2 (2 L-1) N_c (N_c + 1) \ \ \ | \ \text{CNOT} \ .
\label{eq:UkinNoanc}
\end{equation}
The construction of the circuit that implements the time evolution of the hopping term for $N_c=3$ 
and $N_f=1$
is shown in Fig.~\ref{fig:Ukin1flavTrot}.
}
\begin{align}
   U_m  \ :& \ \ 2 N_c N_f L \ \ \ | \ R_Z \ ,\nonumber \\[5pt]
   U_{\mu_B}  \ :& \ \ 2 N_c N_f L \ \ \ | \ R_Z \ ,\nonumber \\[5pt]
   U_{\mu_I}  \ :& \ \ 2 N_c N_f L \ \ \ | \ R_Z \ ,\nonumber \\[5pt]
   U_{kin}  \ :& \ \ 2 N_c N_f(2L-1) \ \ \ | \ R_Z \ ,\nonumber \\
                & \ \ 2 N_c N_f (2L-1) \ \ \ | \ \text{Hadamard} \ ,\nonumber \\
                & \ \ 2 N_c N_f (8L-3) -4 \ \ \ | \ \text{CNOT} \ ,\nonumber \\[5pt]
    U_{el}  \ :& \ \ \frac{1}{2}(2L-1)N_c N_f\left [3-4N_c+N_f(2L-1)(5N_c-4)\right ] \ \ \ | \ R_Z \ ,\nonumber \\
                & \ \ \frac{1}{2}(2L-1)(N_c-1) N_c N_f\left [N_f(2L-1)-1\right ] \ \ \ | \ \text{Hadamard} \ ,\nonumber \\
                & \ \ \frac{1}{6} (2 L -1) (N_c-1) N_c N_f [(2 L-1) (2 N_c+17) N_f-2 N_c-11] \ \ \ | \ \text{CNOT} \ .
\label{eq:RHCN}
\end{align}

It is interesting to note the scaling of each of the contributions.  The mass, chemical potential and kinetic terms scale as ${\cal O}(L^1)$, while the non-local gauge-field contribution is ${\cal O}(L^2)$.
As anticipated from the outset, using Gauss's law to constrain the energy in the gauge field via the quark occupation has given rise to circuit depths that scale quadratically with the lattice extent, naively violating one of the criteria for quantum simulations at scale~\cite{feynman:1981tf,DiVincenzo2000ThePI}.
This volume-scaling is absent for formulations that explicitly include the 
gauge-field locally, 
but with the trade-off of requiring a volume-scaling increase in the number of 
qubits or qudits or bosonic modes.\footnote{
The local basis on each link is spanned by the possible color irreps 
and the states of the left and right Hilbert spaces (see footnote~\ref{foot:irrep}). 
The possible irreps are built from the charges of the preceding fermion sites, 
and therefore the dimension of the link basis grows polynomially in $L$. 
This can be encoded in $\mathcal{O}(\log L)$ qubits per link and 
$\mathcal{O}(L \log L)$ qubits in total. 
The hopping and chromo-electric terms in the Hamiltonian are local, 
and therefore one Trotter step will require $\mathcal{O}(L)$ gate operations up to logarithmic corrections.} 
We expect that the architecture of quantum devices used for simulation 
and the resource requirements for the local construction will determine 
the selection of local versus non-local implementations.

For QCD with $N_f=2$, the total requirements are
\begin{align}
   R_Z  \ :& \ \ (2L-1)\left( 132 L -63 \right)+18 \ ,\nonumber \\
   \text{Hadamard} \ :& \ \ (2L-1)\left( 24L - 6 \right)  \ ,\nonumber \\
    \text{CNOT} \ :& \ \ (2L-1)\left( 184L - 78 \right) + 8 \ ,
\end{align}
and further, the CNOT requirements for a single Trotter step 
of $SU(2)$ and $SU(3)$ for $N_f = 1,2,3$ are shown in Table~\ref{tab:cnotA}.
\begin{table}
\renewcommand{\arraystretch}{1.2}
\begin{tabularx}{0.5\textwidth}{||c | Y | Y | Y ||}
\hline
\multicolumn{4}{||c||}{\small \# of CNOT gates for 1 Trotter step of $SU(2)$} \\
 \hline
 $L$ & $N_f=1$ & $N_f=2$ & $N_f=3$ \\
 \hline\hline
 1 & 14 & 58 & 116 \\ 
 \hline
 2 & 96 & 382 & 818\\
 \hline
 5 & 774 & 3,082 & 6,812 \\
 \hline
 10 & 3,344 & 13,342 & 29,762 \\
 \hline
 100 & 357,404 & 1,429,222 & 3,213,062 \\
 \hline
\end{tabularx}
\renewcommand{\arraystretch}{1}
\ \
\renewcommand{\arraystretch}{1.2}
\begin{tabularx}{0.5\textwidth}{||c | Y | Y | Y ||}
\hline
\multicolumn{4}{||c||}{\small \# of CNOT gates for 1 Trotter step of $SU(3)$ } \\
 \hline
 $L$ & $N_f=1$ & $N_f=2$ & $N_f=3$ \\
 \hline\hline
 1 & 30 & 114 & 242 \\ 
 \hline
 2 & 228 & 878 & 1,940\\
 \hline
 5 & 1,926 & 7,586 & 16,970 \\
 \hline
 10 & 8,436 & 33,486 & 75,140 \\
 \hline
 100 & 912,216 & 3,646,086 & 8,201,600 \\
 \hline
\end{tabularx}
\renewcommand{\arraystretch}{1}
\caption{The CNOT requirements to perform one Trotter step of time evolution for a selection of simulation parameters.}
 \label{tab:cnotA}
 \end{table}
These resource requirements suggest that systems with up to $L=5$ could be simulated, with appropriate error mitigation protocols, using this non-local framework in the near future. Simulations beyond $L=5$ appear to present a challenge in the near term.

The resource requirements in Table~\ref{tab:cnotA} do not include those for a gauge-link beyond the end of the lattice.  As discussed previously, such additions to the time evolution could be used to move color-nonsinglet contributions to high frequency, allowing the possibility that they are filtered from observables.
Such terms contribute further to the quadratic volume scaling of resources.
Including chemical potentials in the time evolution does not increase the number of required entangling gates per Trotter step.  Their impact upon resource requirements may arise in preparing the initial state of the system.

\subsubsection{Elements for Future Co-Design Efforts}
\label{sec:codesign}
\noindent
Recent work has shown the capability of creating many-body entangling gates natively~\cite{andrade:2021pil,Katz:2022czu} which have similar fidelity to two qubit gates.
This has multiple benefits. First, it allows for (effectively) deeper circuits to be run within coherence times. 
Second, it can eliminate some of the Trotter errors due to non-commuting terms. 
The possibility of using native gates for these calculations is particularly interesting from the standpoint of eliminating or mitigating the Trotter errors that violate the global $SU(3)$ symmetry, as discussed in Sec.~\ref{sec:colorBreak}.
Specifically, 
it would be advantageous to have a ``black box" unitary operation of the form,
\begin{align}
   e^{-i \alpha Q_n^{(a)} \, Q_m^{(a)}} =& \ \exp \bigg \{-i \frac{\alpha}{2} \bigg [\sigma^+_{n} \sigma^-_{n+1}\sigma^-_{m}\sigma^+_{m+1} + \sigma^-_{n}\sigma^+_{n+1}\sigma^+_{m}\sigma^-_{m+1} +  \sigma^+_{n+1}\sigma^-_{n+2}\sigma^-_{m+1}\sigma^+_{m+2} \nonumber \\
   &+ \sigma^-_{n+1}\sigma^+_{n+2}\sigma^+_{m+1}\sigma^-_{m+2} + \sigma^+_{n}\sigma^z_{n+1}\sigma^-_{n+2}\sigma^-_{m}\sigma^z_{m+1}\sigma^+_{m+2} + \sigma^-_{n}\sigma^z_{n+1}\sigma^+_{n+2}\sigma^+_{m}\sigma^z_{m+1}\sigma^-_{m+2} \nonumber \\
   &+ \frac{1}{6}(\sigma^z_n \sigma^z_m + \sigma^z_{n+1} \sigma^z_{m+1} + \sigma^z_{n+2} \sigma^z_{m+2}) - \frac{1}{12}(\sigma^z_n \sigma^z_{m+1} + \sigma^z_n \sigma^z_{m+2} + \sigma^z_{n+1} \sigma^z_m \nonumber \\
   &+ \sigma^z_{n+1} \sigma^z_{m+2} + \sigma^z_{n+2} \sigma^z_m + \sigma^z_{n+2 }\sigma^z_{m+1})
   \bigg ] \bigg \} \ ,
\end{align}
for arbitrary $\alpha$ and pairs of sites, $n$ and $m$ (sum on $a$ is implied).
A more detailed discussion of co-designing 
interactions for quantum simulations of these theories is clearly warranted.

\subsection{Results from Quantum Simulators}
\begin{figure}[!ht]
    \centering
    \includegraphics[width=\columnwidth]{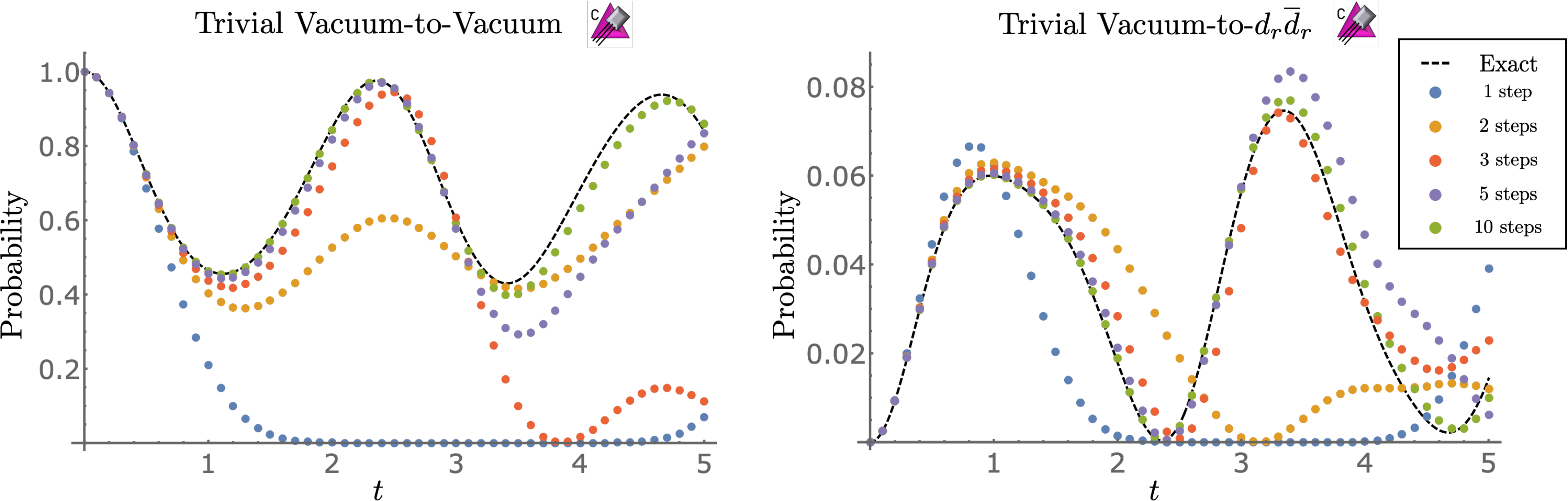}
    \caption{
    The trivial vacuum-to-vacuum 
    and trivial vacuum-to-$d_r \overline{d}_r$ probabilities in QCD with $N_f=2$
    for $m=g=L=1$. 
    Shown are the results obtained from exact exponentiation of the Hamiltonian (dashed black curve) and from the Trotterized implementation with $1$, $2$, $3$, $5$ and $10$ Trotter steps using the (classical) quantum simulators in {\tt cirq} and {\tt qiskit} (denoted by the purple icons~\cite{klco:2019xro}).
    }
    \label{fig:VacTo}
\end{figure}

\noindent
The circuits laid out in Sec.~\ref{sec:Circuits} are too deep to be executed on currently available quantum devices,  
but can be readily implemented with quantum simulators such as {\tt cirq} and {\tt qiskit}.
This allows for an estimate of the number of Trotter steps required to achieve a desired precision in the determination of any given observable as a function of time.
Figure~\ref{fig:VacTo} shows results for the 
trivial vacuum-to-vacuum and trivial vacuum-to-$d_r \overline{d}_r$ probabilities as a function of time for $L=1$. See App.~\ref{app:Nf1SU3circs} for the full circuit which
implements a single Trotter step, and App.~\ref{app:MDTD} for the decomposition of the energy starting in the trivial vacuum.

The number of Trotter steps, 
$N_{\rm Trott}$, required to evolve out to a given $t$ within a specified (systematic) error, 
$\epsilon_{\rm Trott}$, was also investigated. 
$\epsilon_{\rm Trott}$ is defined as the 
maximum fractional error between the 
Trotterized and exact time evolution in two quantities, the vacuum-to-vacuum persistence probability 
and the vacuum-to-$d_r\overline{d}_r$ transition probability. For demonstrative purposes, an analysis at leading order in the Trotter expansion is sufficient.
Naive expectations based upon global properties of the Hamiltonian defining the evolution operators indicate that an upper bound for $\epsilon_{\rm Trott}$ scales as 
\begin{equation}
\Big\lvert \Big\lvert e^{-i H t} - \left[ U_1\left (\frac{t}{N_{\rm Trott}}\right ) \right]^{N_{\rm Trott}} \Big\rvert \Big\rvert \ \le \ \frac{1}{2} \sum_i \sum_{j>i} \Big\lvert \Big\lvert \left[ H_i , H_j \right] \Big\rvert \Big\rvert \frac{t^2}{N_{\rm Trott}} \ ,
\label{eq:LOTrottbound}
\end{equation}
where the Hamiltonian has been divided into sets of mutually commuting terms, $H = \sum_i H_i$. This upper bound indicates that the required number of Trotter steps to maintain a fixed error scales as  $N_{\rm Trott}\sim t^2$~\cite{Childs_2021}.

\begin{figure}[!ht]
    \centering
    \includegraphics[width=\columnwidth]{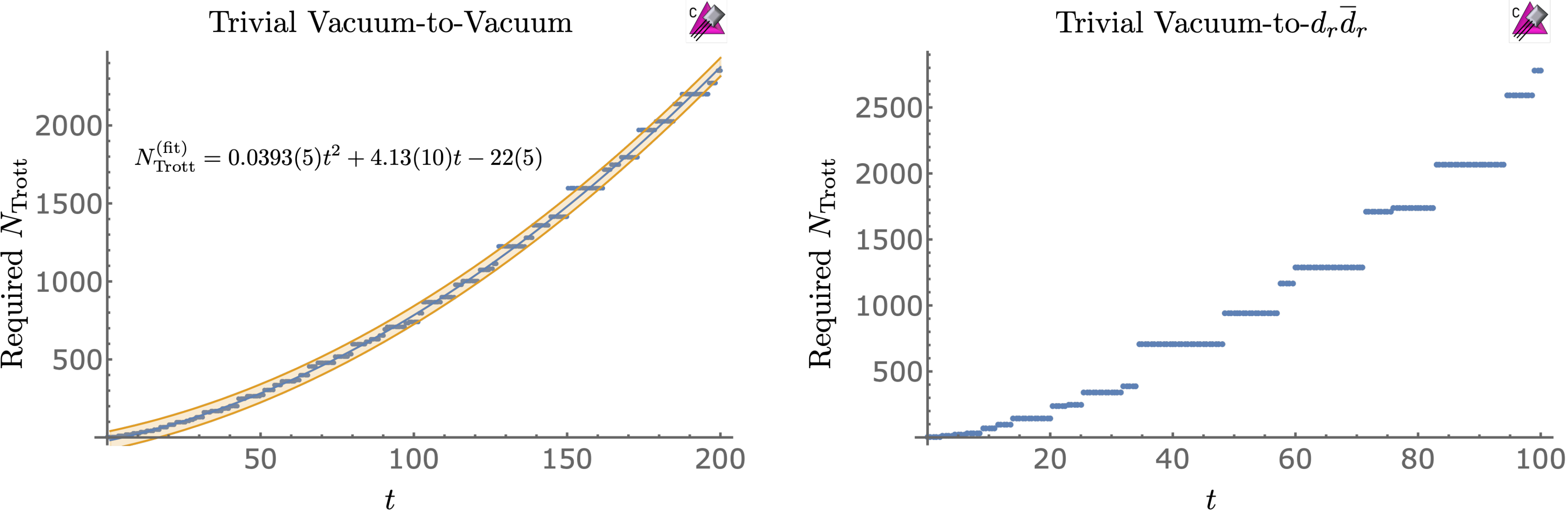}
    \caption{
    The number of Trotter steps, 
    $N_{\rm Trott}$,
   required to 
   achieve a systematic fractional error of 
   $\epsilon_{{\rm Trott}}\le 0.1$
    at time $t$ in the trivial vacuum-to-vacuum  probability (left panel) and the trivial vacuum-to-$d_r\overline{d}_r$ probability
    (right panel) for QCD with $N_f=2$ and
$m=g=L=1$. The blue points are results obtained by direct calculation.
   }
    \label{fig:TrotErrorB}
\end{figure}

To explore the resource requirements for simulation based upon explicit calculations between exclusive states, as opposed to upper bounds for inclusive processes, given in Eq.~(\ref{eq:LOTrottbound}), 
a series of calculations was performed requiring $\epsilon_{\rm Trott}\le0.1$ for a range of times, $t$.
Figure~\ref{fig:TrotErrorB} shows the required 
$N_{\rm Trott}$ as a function of $t$ for $m=g=L=1$.
The plateaus 
observed in Fig.~\ref{fig:TrotErrorB} arise from
resolving upper bounds from oscillating functions, 
and introduce a limitation in fitting to extract scaling behavior.  This is less of a limitation 
for the larger vacuum-to-vacuum probabilities which are fit well by a quadratic polynomial, starting from $t=1$, with coefficients,
\begin{equation}
    N_{{\rm Trott}} = 0.0393(5) t^2 + 4.13(10) t - 22(5) \ .
    \label{eq:TrotErrorFit}
\end{equation}
The uncertainty represents a 95\% confidence interval in the fit parameters and corresponds to the shaded orange region in
Fig.~\ref{fig:TrotErrorB}. The weak quadratic scaling with $t$ implies that, even out to $t \sim 100$, the number of Trotter
steps scales approximately linearly, and a constant error in the observables can be achieved with a fixed Trotter step size. 
We have been unable to distinguish between fits with and without logarithmic terms.

These results can be contrasted with those obtained for the Schwinger model in Weyl gauge. The authors of Ref.~\cite{shaw:2020udc} estimate a resource
requirement, as quantified by the number of $T$-gates, that scales as $\sim (L t)^{3/2}\log L t$, increasing to
$\sim L^{5/2} t^{3/2}\log L t \log L$ if the maximal value of the gauge fields is accommodated within the Hilbert space.
The results obtained in this section suggest that resource requirements in axial gauge, as quantified by the number of CNOTs,
effectively scale as $\sim L^2 t$ up to intermediate times and as $\sim L^2 t^2$ asymptotically. In a scattering process with localized
wave-packets, it is appropriate to take $L\sim t$ 
(for the speed of light taken to be $c=1$),
as the relevant non-trivial time evolution is bounded by the light cone. 
This suggests that the required resources scale asymptotically as $\sim t^4$, independent of the chosen gauge to define the simulation. 
This could have been
anticipated at the outset by assuming that the minimum change in complexity for a process has physical meaning~\cite{https://doi.org/10.48550/arxiv.quant-ph/0701004,doi:10.1126/science.1121541,https://doi.org/10.48550/arxiv.quant-ph/0502070,Jefferson:2017sdb}.

\section{Simulating \texorpdfstring{\boldmath$1+1$}{1+1}D QCD with \texorpdfstring{\boldmath$N_f=1$}{Nf=1} and \texorpdfstring{\boldmath$L=1$}{L=1}}
\label{sec:Nc3Nf1}
\noindent
With the advances in quantum devices, algorithms and mitigation strategies, quantum simulations of $1+1$D QCD can now begin, and this section presents results for $N_f=1$ and $L=1$. Both state preparation and time evolution will be discussed.

\subsection{State Preparation with VQE}
\begin{figure}[!ht]
    \centering
    \includegraphics[width=6cm]{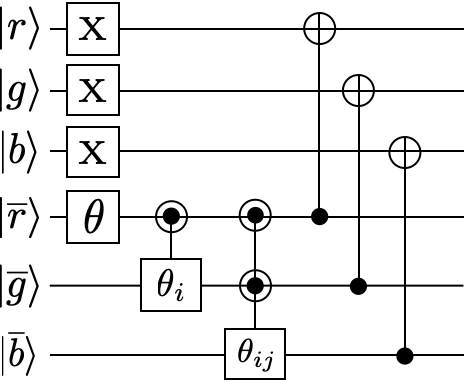}
    \caption{
    Building upon the trivial vacuum, 
    this circuit initializes the 
    most general real wavefunction 
    (with 7 independent rotation angles for 3 qubits)
    in the $\overline{q}$-sector, which is subsequently
    mirrored into the $q$-sector by 3 CNOTs. Gates labelled by ``$\theta$" are shorthand for $R_Y(\theta)$ and half-filled circles denote a control on $0$ 
    and a different control on $1$.
    }
    \label{circ:GeneralVQE}
\end{figure}

\noindent
Restricting the states of the lattice to be color singlets reduces the complexity of state preparation significantly.  
Transformations in the quark sector are mirrored in the antiquark sector.
A circuit that
prepares the most general state with $r=g=b=0$ is shown in Fig.~\ref{circ:GeneralVQE}. 
The (multiply-)controlled $\theta$ gates are short-hand for (multiply-)controlled $R_Y(\theta)$ gates with half-filled circles denoting a control on $0$ 
and a different control on $1$.
The subscripts on $\theta_{ij}$ signify that there are different angles for each controlled rotation. For example,
$\theta_{i}$ has two components, $\theta_{0}$ and $\theta_{1}$, corresponding to a rotation controlled on $0$ and $1$, respectively. 
The CNOTs at the end of the circuit
enforce that there are equal numbers of quarks and antiquarks with the same color,
i.e., that $r=g=b=0$. 
This circuit can be further simplified by constraining the angles to only parameterize color singlet states. The color singlet subspace is spanned by\footnote{
The apparent asymmetry between $q_r,q_g,q_b$ is due to the charge operators generating hops over different numbers of quarks or antiquarks. 
For example, $Q^{(1)}$ hops $q_r$ to $q_g$ without passing over any intermediate quarks, but $Q^{(4)}$ hops $q_r$ to $q_b$ passing over $q_g$. 
Also note that when $m=0$ the $\mathbb{Z}_2$ spin-flip symmetry reduces the space of states to be two-dimensional.}
\begin{align}
    \ket{{\Omega_0}} \ &, \ \ \frac{1}{\sqrt{3}}\left (\ket{q_r \overline{q}_r} - \ket{q_g \overline{q}_g} + \ket{q_b \overline{q}_b}\right ) \ , \nonumber \\
    \ket{q_r \overline{q}_r \, q_g \overline{q}_g \, q_b \overline{q}_b} \ &, \ \  \frac{1}{\sqrt{3}} \left (\ket{q_r \overline{q}_r \, q_g\overline{q}_g} - \ket{q_r \overline{q}_r \, q_b\overline{q}_b} + \ket{q_g \overline{q}_g \, q_b \overline{q}_b}\right ) \ , 
    \label{eq:vacBasis}
\end{align}
where $\ket{{\Omega_0}} = \ket{{000111}}$ is the trivial vacuum.
This leads to the following relations between angles,
\begin{align}
    \theta_{10} &= \theta_{01}\ ,  & \theta_{00} &= -2 \sin^{-1}\left[ \tan(\theta_{0}/2) \, \cos(\theta_{01}/2) \right] \ ,\nonumber \\
    \theta_{01} &= -2 \sin^{-1}\left[ \cos(\theta_{11}/2)\, \tan(\theta_{1}/2) \right]\ , & \theta_0 &= -2 \sin^{-1} \left[ \tan(\theta/2) \, \cos(\theta_{1}/2) \right] \ .
    \label{eq:angleconst}
\end{align}

\begin{figure}[!ht]
    \centering
    \includegraphics[width=\columnwidth]{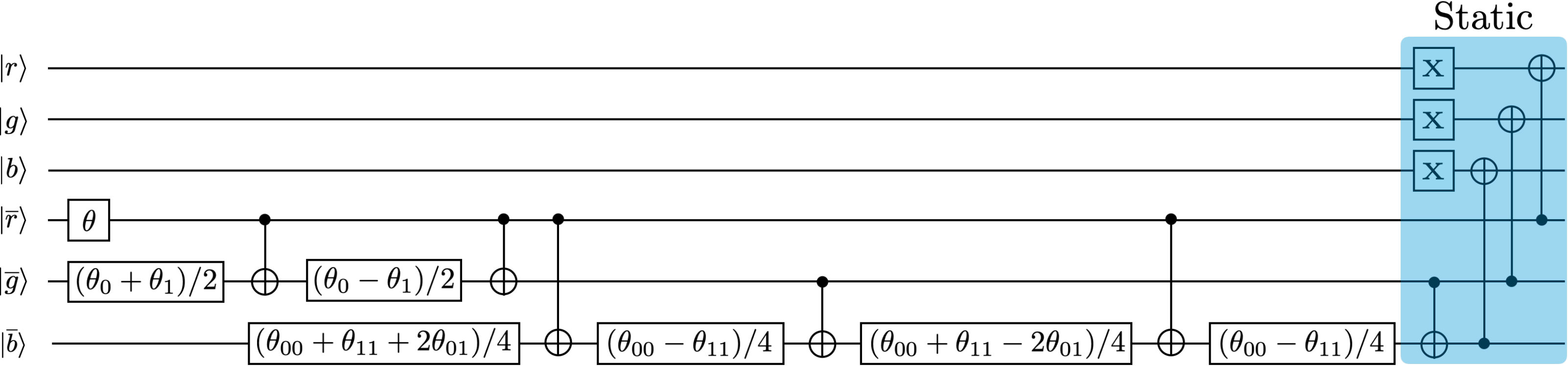}
    \caption{
    A circuit that initializes the most general 
    $B=0$ color singlet state for $N_f=1$ and $L=1$. Gates labelled by ``$\theta$" are shorthand for $R_Y(\theta)$
    and the $X$s at the end are to initialize the trivial vacuum. The color singlet constraint, $\theta_{10} = \theta_{01}$, has been used and the other angles are related by Eq.~(\ref{eq:angleconst}).}
    \label{circ:VQEImp}
\end{figure}

The circuit in Fig.~\ref{circ:GeneralVQE} can utilize the strategy outlined in Ref.~\cite{atas:2021ext}  to
separate into a ``variational" part and a ``static" part. 
If the VQE circuit can be written as 
$U_{var}(\theta) U_s$, 
where $U_s$ is independent
of the variational parameters, 
then $U_s$ can be absorbed by a redefinition of
the Hamiltonian. 
Specifically, matrix elements of the Hamiltonian can be written as 
\begin{equation}
    \bra{{\Omega_0}} U_{var}^{\dagger}(\theta) \tilde{H} U_{var}(\theta) \ket{{\Omega_0}}
    \ ,
\end{equation}
where $\tilde{H}= U_s^{\dagger} H U_s$.
Table~\ref{tab:PUHU} shows the 
transformations of various Pauli strings under conjugation by a CNOT controlled on the smaller index qubit. 
Note that the $\mathbb{Z}_2$ nature of this transformation is manifest.
\begin{table}[!ht]
\renewcommand{\arraystretch}{1.2}
\centering
\begin{tabular}{||c | c | c | c ||} 
 \hline
 $XX \to IX$ & $XY \to IY$ & $YX \to ZY$ & $XZ \to XZ$ \\
 \hline
 $ZZ \to ZI$ & $YZ \to YI$ & $ZY \to YX$ & $YY \to (-)ZX$ \\
 \hline
 $IX \to XX$ & $IY \to XY$ & $XI \to XI$  & $IZ \to IZ$ \\
 \hline
 $ZI \to ZZ$  & $YI \to YZ$ &  $ZX \to (-)YY$  & $II \to II$\\
 \hline
\end{tabular}
\renewcommand{\arraystretch}{1}
\caption{The transformation of Pauli strings under conjugation by a CNOT controlled on the smaller index qubit.}
\label{tab:PUHU}
\end{table}
In essence, entanglement is
traded for a larger number of  correlated measurements.  
Applying the techniques in Ref.~\cite{klco:2019xro}, the VQE circuit of Fig.~\ref{circ:GeneralVQE} can be put into the form of Fig.~\ref{circ:VQEImp},
which requires $5$ CNOTs along with all-to-all connectivity between the three $\overline{q}$s.
%

\subsection{Time Evolution Using IBM's 7-Qubit Quantum Computers}
\begin{figure}[!ht]
    \centering
    \includegraphics[height=0.25\textheight]{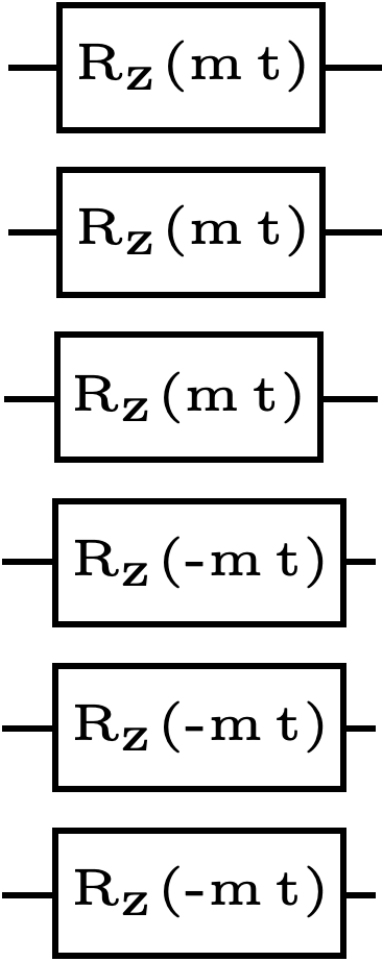}
    \caption{A circuit that implements 
    $U_m(t) = \exp(- i H_m t)$ for $N_f=1$ and $L=1$.}
    \label{circ:Um1flav}
\end{figure}

\begin{figure}[!ht]
    \centering
    \includegraphics[width=12cm]{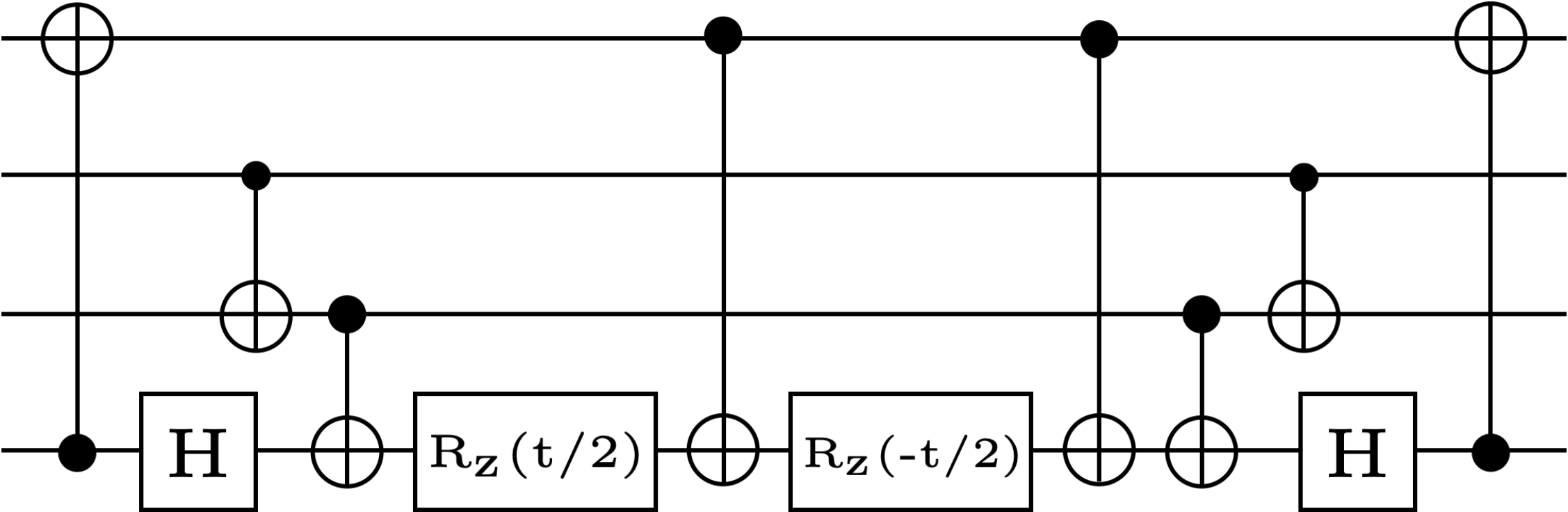}
    \caption{A circuit that implements $\exp[-i \frac{t}{2} (\sigma^+ Z Z \sigma^- + {\rm h.c.})]$.}
    \label{fig:Ukin1flavTrot}
\end{figure}

\noindent
A single leading-order Trotter step of $N_f=1$ QCD with $L=1$ requires 28 CNOTs.\footnote{By evolving with $U_{el}$ before $U_{kin}$ in the Trotterized time evolution, two of the CNOTs
become adjacent in the circuit and can be canceled.}
A circuit that implements one Trotter step of the mass term is shown in Fig.~\ref{circ:Um1flav}.
As discussed around Eq.~(\ref{eq:UkinNoanc}), it is more efficient to not use an ancilla qubit in the Trotterization of the kinetic part of the Hamiltonian. 
A circuit that implements one Trotter step of a single hopping term is shown in Fig.~\ref{fig:Ukin1flavTrot}~\cite{stetina:2020abi}.
Similarly, for this system,
the only contribution to $H_{el}$ is $Q^{(a)}_n \, Q^{(a)}_n$, which contains three $ZZ$ terms that are Trotterized using the standard two CNOT implementation.
The complete set of circuits required for Trotterized time evolution are given in App.~\ref{app:Nf1SU3circs}.

\begin{figure}[!ht]
    \centering
    \includegraphics[width=12cm]{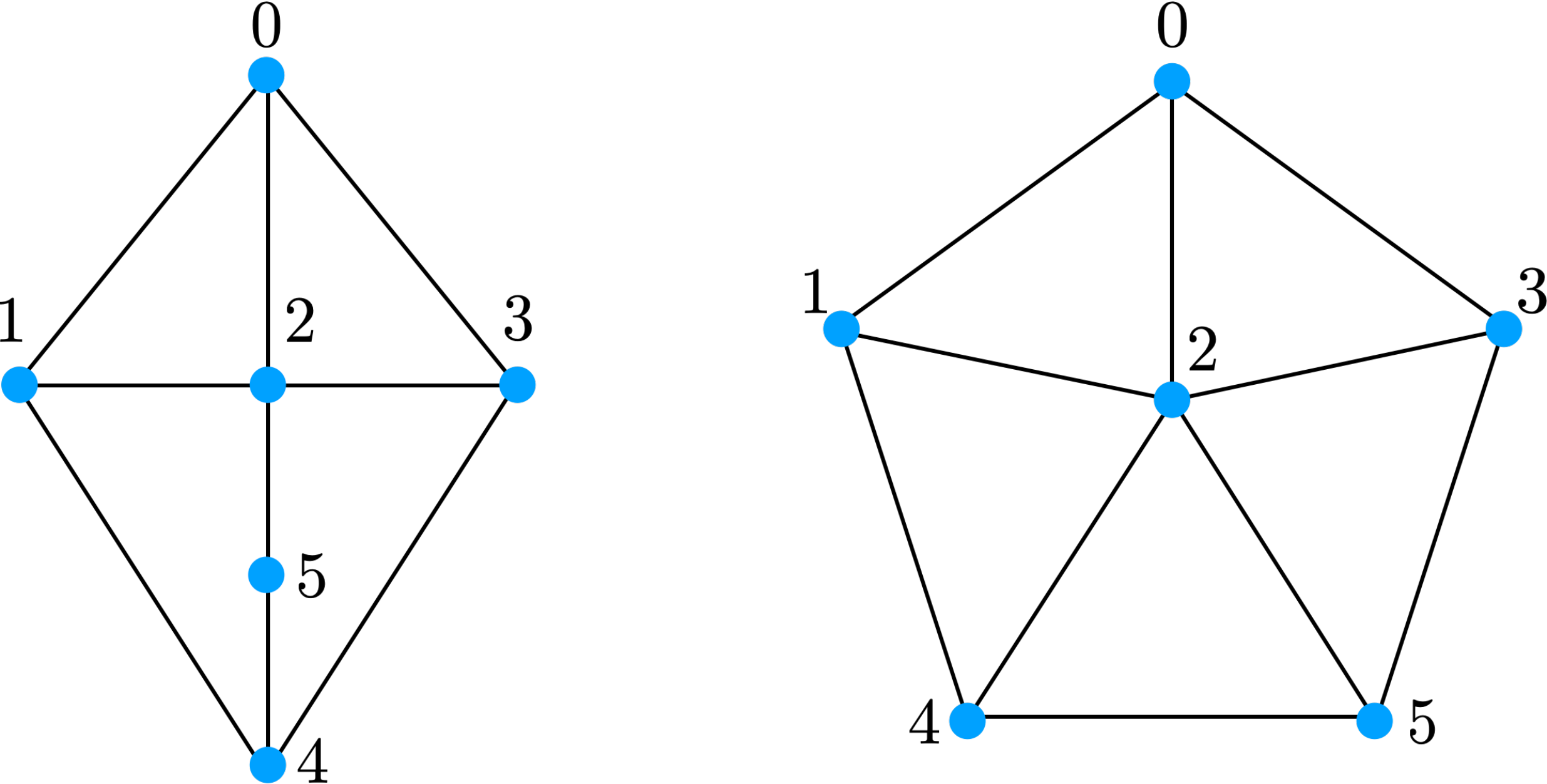}
    \caption{
    Two potential quantum device topologies for the  implementation of Trotterized time evolution.
    }
    \label{fig:TrotTopo}
\end{figure}

To map the system onto a quantum device, it is necessary to understand the required connectivity for efficient simulation. 
Together, the hopping and chromo-electric terms require connectivity between nearest neighbors as well as between $q_r$ and $q_b$ and
$q$s and $\overline{q}$s of the same color. 
The required device topology is planar and two embedding options are
shown in Fig.~\ref{fig:TrotTopo}.
The ``kite'' topology follows from the above circuits, 
while the ``wagon wheel'' topology makes use of the identities $CX(q_a,q_b) \cdot CX(q_b,q_c) = CX(q_a,q_c) \cdot CX(q_b,q_c) = CX(q_b,q_a) \cdot CX(q_a,q_c)$ where $CX(q_a,q_b)$ denotes a CNOT controlled on qubit $q_a$.
Both topologies can be employed on devices with all-to-all connectivity, such as trapped-ion systems, but
neither topology exists natively on available superconducting-qubit devices.

We performed leading-order Trotter evolution to study the trivial vacuum persistence and transition probability using IBM's quantum computers {\tt ibmq\_jakarta} and {\tt ibm\_perth}, each a {\tt r5.11H} quantum processor with 7 qubits and ``H"-connectivity.
The circuits developed for this system require a higher degree of connectivity than available with these devices, and so SWAP-gates were necessary for implementation.   
The IBM {\tt transpiler} was used to first compile the circuit for the H-connectivity and then again to compile the Pauli twirling (discussed next).
An efficient use of SWAP-gates allows for a single Trotter step to be executed with 34 CNOTs.

A number of error-mitigation techniques were employed to minimize associated systematic uncertainties in our calculations: randomized compiling of the CNOTs (Pauli twirling)~\cite{physreva.94.052325} combined with decoherence renormalization~\cite{urbanek:2021oej,rahman:2022rlg}, measurement error mitigation, post-selecting on physical states and dynamical decoupling~\cite{physreva.58.2733,duan1999139,zanardi199977,physrevlett.82.2417}.\footnote{A recent detailed study of the stability of some of IBM's quantum devices using a system of physical interest can be found in Ref.~\cite{Yeter-Aydeniz:2022vuy}.}
The circuits were randomly complied with each CNOT Pauli-twirled as a mechanism to transform coherent errors in the CNOT gates into statistical noise in the ensemble.
This has been shown to be effective in improving the quality of results in other simulations, for example, Refs.~\cite{Kim2021ScalableEM,rahman:2022rlg}.
Pauli twirling involves multiplying the right side of each CNOT by a randomly chosen
element of the two-qubit Pauli group, $G_2$,
and the left side by $G'_2$ 
such that $G'_2 \, CX \, G_2 = CX$ (up to a phase). 
For an ideal CNOT gate, this would have no effect on the circuit. 
A table of required CNOT identities is given,
for example,
in an appendix in Ref.~\cite{rahman:2022rlg}.
Randomized Pauli-twirling is combined with performing measurements of a ``non-physics", mitigation circuit, which is the time evolution circuit evaluated at $t=0$, and is the identity in the absence of noise.
Assuming that the randomized-compiling of the Pauli-twirled CNOTs transforms coherent noise into depolarizing noise, 
the fractional deviation of the noiseless and computed results 
from the asymptotic limit of complete decoherence
are expected to be approximately equal for both the physics and mitigation ensembles. Assuming linearity, it follows that
\begin{equation}
\left ( P_{\text{pred}}^{(\text{phys})}-\frac{1}{8} \right ) = \left ( P_{\text{meas}}^{(\text{phys})}-\frac{1}{8} \right ) \times \left ( \frac{1-\frac{1}{8}}{ P_{\text{meas}}^{(\text{mit})}-\frac{1}{8} } \right )\ ,
\label{eq:mit}
\end{equation}
where $P_{\text{meas}}^{(\text{phys})}$ and $P_{\text{meas}}^{(\text{mit})}$ are post-processed probabilities and
$P_{\text{pred}}^{(\text{phys})}$ is an estimate of the probability once the effects of depolarizing noise have been removed. 
The ``$\frac{1}{8}$" represents the fully decohered probability after post-selecting on physical states (described next) and the ``$1$" is the probability of measuring the initial state from the mitigation circuit in the absence of noise. 

The computational basis of 6 qubits contains $2^6$ states but time evolution only connects those with the same $r$, $g$ and $b$. Starting from the trivial vacuum, this
implies that only the $8$ states with $r=g=b=0$ are accessible through time evolution.
The results off the quantum computer were post-processed to only select events that populated 1 of the 8 physically
allowed states, discarding outcomes that were unphysical. Typically, this resulted in a retention rate of $\sim 30\%$. The
workflow interspersed physics and mitigation circuits to provide a correlated calibration of the quantum devices. This enabled the detection (and removal) of
out-of-specs device performance during post-processing. We explored using the same twirling sequences for both physics and
mitigation circuits and found that it had no significant impact. 
The impact of dynamical decoupling of idle qubits using {\tt qiskit}'s built in functionality was also investigated and found to have little effect. 
The results of each run were corrected for measurement error using IBM's available function, {\tt TensoredMeasFitter}, and associated downstream operations.

\begin{figure}[!ht]
    \centering
    \includegraphics[width=\columnwidth]{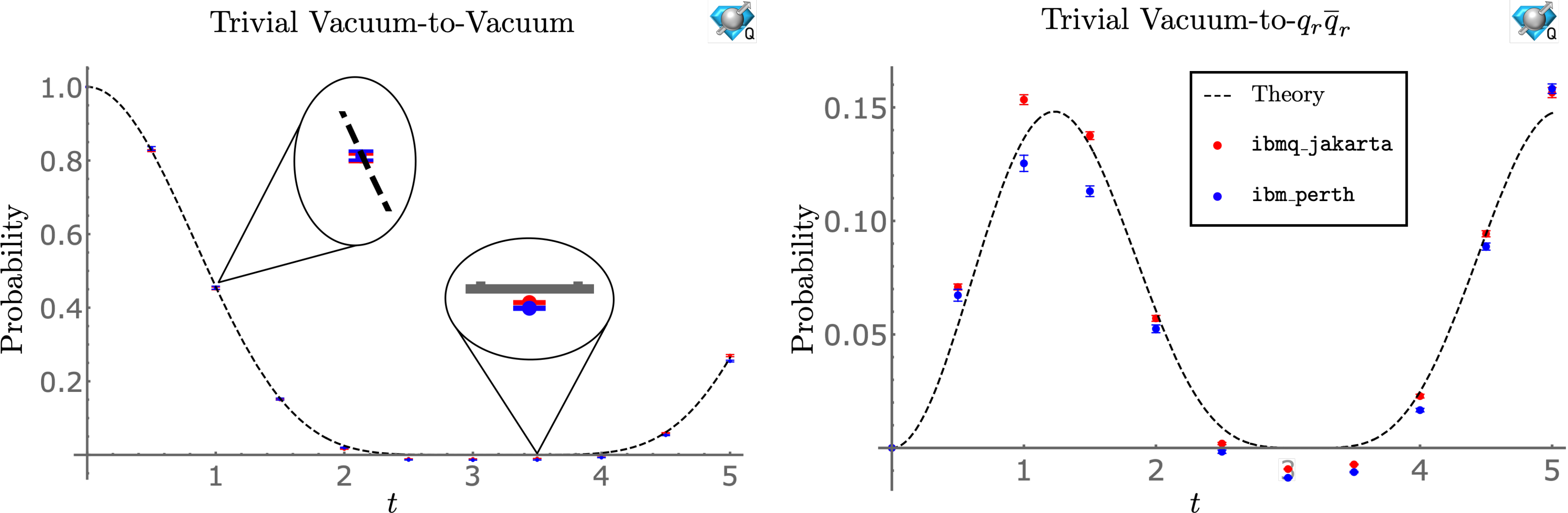}
    \caption{
    The trivial vacuum-to-vacuum (left panel) and trivial vacuum-to-$q_r \overline{q}_r$ (right panel) probabilities for $N_f=1$ QCD and $m=g=L=1$. 
    The dashed-black curve shows the expected result from one step of leading-order Trotter evolution.
    The results, given in Tables~\ref{tab:IBMvacresultsLO} and~\ref{tab:IBMrrbresultsLO}, were obtained by using $10^3$ shots for 447 Pauli-twirled circuits using IBM's quantum computers {\tt ibmq\_jakarta} (red) and {\tt ibm\_perth} (blue).
    }
    \label{fig:IBMresults}
\end{figure}

The results obtained for the trivial vacuum-to-vacuum and trivial vacuum-to-$q_r \overline{q}_r$ probabilities from one step of leading-order Trotter time evolution are shown in Fig.~\ref{fig:IBMresults}.
For each time, 447 Pauli-twirled physics circuits 
and 447 differently twirled circuits with zeroed angles (mitigation) were analyzed using $10^3$ shots on both {\tt ibmq\_jakarta} and {\tt ibm\_perth} (to estimate device systematics).
After post-selecting on physical states, correlated Bootstrap Resampling was used to form the final result.\footnote{As the mitigation and physics circuits were executed as adjacent jobs on the devices, the same Bootstrap sample was used to select results from both ensembles to account for temporal correlations.}
Tables~\ref{tab:IBMvacresultsLO} and~\ref{tab:IBMrrbresultsLO} display the results of the calculations performed using {\tt ibmq\_jakarta} and {\tt ibm\_perth} quantum computers.
The same mitigation data was used for both the trivial vacuum-to-vacuum and trivial vacuum-to-$q_r\overline{q}_r$ calculations, and is provided in columns 2 and 4 of Table~\ref{tab:IBMvacresultsLO}. 
See App.~\ref{app:LOTrott} for an extended discussion of leading-order Trotter. 
Note that the negative probabilities seen in Fig.~\ref{fig:IBMresults} indicate that additional non-linear terms are needed in Eq.~(\ref{eq:mit}).
\begin{table}[!ht]
\renewcommand{\arraystretch}{1.2}
\hspace*{-0.0cm}\begin{tabularx} {1.0\textwidth}{||c | Y | Y | Y | Y | Y | Y | Y ||}
\hline
\multicolumn{8}{||c||}{Vacuum-to-Vacuum Probabilities for $N_f=1$ QCD from {\tt ibmq\_jakarta} and {\tt ibm\_perth}} \\
 \hline
 $t$ & 
 \makecell{Mitigation \\ {\tt jakarta}} & 
 \makecell{Physics \\ {\tt jakarta}} & 
 \makecell{Mitigation \\ {\tt perth}} & 
 \makecell{Physics \\ {\tt perth}} & 
 \makecell{Results \\ {\tt jakarta}} & 
 \makecell{Results \\ {\tt perth}} & 
 Theory
 \\
 \hline\hline
 0 & - & - & - & - & - & - & 1  \\
 \hline
 0.5 & 
 0.9176(10) & 0.7607(24)  &
 0.8744(23) & 0.7310(42)  & 
 0.8268(27) & 0.8326(52) & 0.8274 \\
 \hline
 1.0 & 
 0.9059(12) & 0.4171(32) & 
 0.9118(16) & 0.4211(39) & 
 0.4523(36) & 0.4543(43) & 0.4568 \\
 \hline
 1.5 & 
 0.9180(12) & 0.1483(16) & 
 0.9077(17) & 0.1489(23) &  
 0.1507(17)  & 0.1518(25) & 0.1534 \\
 \hline
 2.0 & 
 0.8953(15) & 0.0292(08) & 
 0.8953(21) & 0.0324(10) &  
 0.0162(09)  &  0.0198(11) & 0.0249 \\
 \hline
 2.5 & 
 0.9169(12) & 0.0020(01) &
 0.8938(21) & 0.0032(02) & 
 $-1.09(3)\times 10^{-2}$   &  $-1.36(4)\times 10^{-2}$ & 0.0010 \\
 \hline
 3.0 &
 0.9282(13) & 0.00010(2) &
 0.9100(13) & 0.00017(3) &  
 $-1.11(2)\times 10^{-2}$ &  $-1.40(2)\times 10^{-2}$ & $1.3\times 10^{-7}$ \\
 \hline
 3.5 &
 0.9357(10)& 0.00017(3) & 
 0.9109(14) & 0.00037(4)& 
 $-9.7(2)\times 10^{-3}$ &  $-1.38(2)\times 10^{-2}$ & $3.2\times 10^{-5}$ \\
 \hline
 4.0 & 
 0.9267(13) & 0.0081(03) & 
 0.9023(14) & 0.0076(03) & 
 $-2.6(4)\times 10^{-3}$ &  $-7.2(4)\times 10^{-3}$ & 0.0052 \\
 \hline
 4.5 & 
 0.9213(12) & 0.0653(10) &
 0.8995(16) & 0.0619(11) & 
 0.0594(11)  & 0.0537(13) & 0.0614 \\
 \hline
 5.0 & 
 0.9105(12) & 0.2550(26) & 
 0.9031(14) & 0.2405(21) &  
 0.2698(29)  & 0.2550(23) & 0.2644 \\
 \hline
\end{tabularx}
\renewcommand{\arraystretch}{1}
\caption{
Trivial vacuum-to-vacuum probabilities for $m=g=L=1$ using IBM's {\tt ibmq\_jakarta} and {\tt ibm\_perth}, the underlying distributions of which are displayed in Fig.~\ref{fig:IBMhistos}.
Columns $2$ through $5$ are results after selecting only physical states and columns $6$ and $7$ are results after using the mitigation circuit to account for depolarizing noise.
}
\label{tab:IBMvacresultsLO}
\end{table}
\begin{table}[!ht]
\renewcommand{\arraystretch}{1.2}
\begin{tabularx}{1.0\textwidth}{||c | Y | Y | Y | Y | Y ||}
\hline
\multicolumn{6}{||c||}{ Vacuum-to-$q_r\overline{q}_r$ Probabilities for $N_f=1$ QCD from {\tt ibmq\_jakarta} and {\tt ibm\_perth}} \\
 \hline
 $t$ & 
 \makecell{Physics \\ {\tt jakarta}} & 
 \makecell{Physics \\ {\tt perth}} & 
 \makecell{Results \\ {\tt jakarta}} & 
 \makecell{Results \\ {\tt perth}} & 
 Theory
 \\
 \hline\hline
 0 & - & - & - & - & 0  \\
 \hline
 0.5 & 
 0.0760(12) & 0.0756(22)  & 
 0.0709(13) & 0.0673(26) & 0.0539 \\
 \hline
 1.0 & 
 0.1504(19) & 0.1253(32) & 
 0.1534(22) & 0.1254(36) & 0.1363 \\
 \hline
 1.5 & 
 0.1364(15) & 0.1144(21) &  
 0.1376(17)  & 0.1131(23) & 0.1332 \\
 \hline
 2.0 & 
 0.0652(11) & 0.0611(15) &  
 0.0571(13)  &  0.0525(17) & 0.0603 \\
 \hline
 2.5 & 
 0.0136(04) & 0.0137(06) & 
 0.0019(05)   &  -0.0017(07) & 0.0089 \\
 \hline
 3.0 &
 0.0017(01) & 0.0011(01) &  
 -0.0093(02)  &  -0.0132(02) & $2.5\times 10^{-5}$ \\
 \hline
 3.5 &
 0.0024(01) & 0.0032(02)& 
 -0.0073(02) &  -0.0107(03) & 0.0010 \\
 \hline
 4.0 & 
 0.0314(07) & 0.0288(07) & 
 0.0228(08) &  0.0167(08) & 0.0248 \\
 \hline
 4.5 & 
 0.0971(12) & 0.0929(14) & 
 0.0943(13)  & 0.0887(16) & 0.0943 \\
 \hline
 5.0 & 
 0.1534(20) & 0.1546(19) &  
 0.1566(22)  & 0.1583(21) & 0.1475 \\
 \hline
\end{tabularx}
\renewcommand{\arraystretch}{1}
\caption{
The trivial vacuum-to-$q_r \overline{q}_r$ probabilities for $m=g=L=1$ using IBM's {\tt ibmq\_jakarta} and {\tt ibm\_perth}.
The 2nd and 3rd columns are the results after selecting only physical states and columns 4 and 5 are the results after using the mitigation circuit to account for depolarizing noise.
}
\label{tab:IBMrrbresultsLO}
\end{table}

\begin{figure}[!ht]
    \centering
    \includegraphics[width=0.9 \columnwidth]{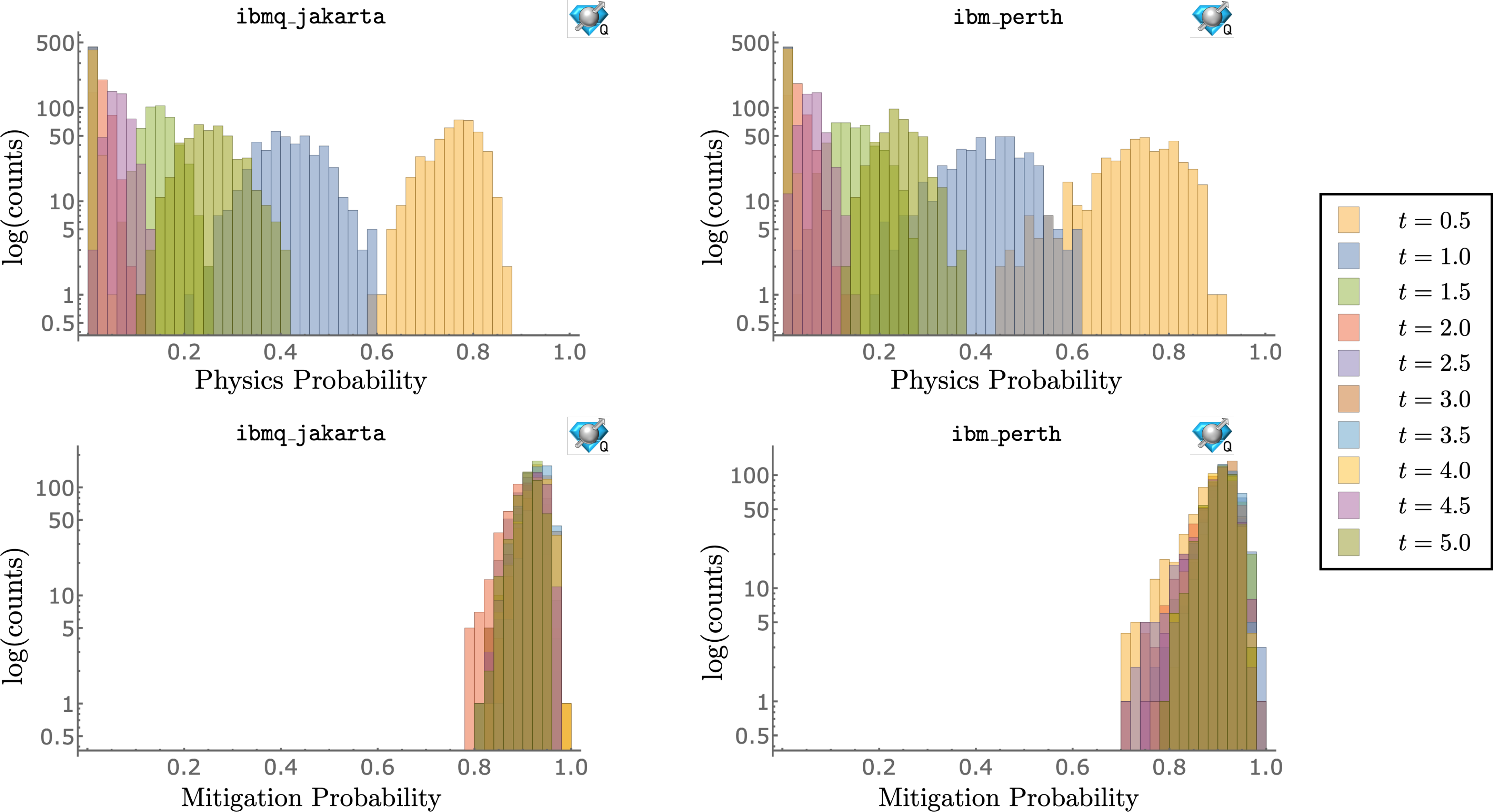}
    \caption{
    Histograms of the post-processed vacuum-to-vacuum results obtained using {\tt ibmq\_jakarta} and {\tt ibm\_perth}.
    The horizontal axes show the value of the vacuum-to-vacuum probability, and the vertical axes show bin counts on a log-scale.
    The top panels display the results obtained from the physics circuits for the range of evolution times and the bottom panels display the results obtained for the corresponding mitigation circuits.
    }
    \label{fig:IBMhistos}
\end{figure}

It is interesting to consider the distributions of events obtained from the Pauli-twirled circuits, as shown in Fig.~\ref{fig:IBMhistos}.
The distributions are not Gaussian and, in a number of instances, exhibit heavy tails particularly near the boundaries.\footnote{For a study of heavy-tailed distributions in Euclidean-space lattice QCD calculations, see Refs.~\cite{Wagman:2016bam,Wagman:2017gqi}.}
The spread of the distributions, associated with non-ideal CNOT gates, is seen to reach a maximum of $\sim 0.4$, but with a full-width at half-max that is $\sim 0.2$. These distributions are already broad with a 34 CNOT circuit, and we probed the limit
of these devices by time-evolving with two first-order Trotter steps,\footnote{Under a particular ordering of terms, two steps of first- and second-order Trotter time evolution are equivalent.}
which requires 91 CNOTs after accounting for SWAPs. 
Using the aforementioned techniques, this was found to be beyond the capabilities of {\tt ibmq\_jakarta}, {\tt ibmq\_lagos} and {\tt ibm\_perth}.

\section{Arbitrary \texorpdfstring{\boldmath$N_c$}{Nc} and \texorpdfstring{\boldmath$N_f$}{Nf}}
\label{sec:NcNf}
\noindent
In this section, the structure of the Hamiltonian for $N_f$ flavors of quarks in the fundamental representation of $SU(N_c)$ is developed. 
The mapping to spins has the same structure as for 
$N_f=2$ QCD, but now, there are $N_c\times N_f$ $q$s and $N_c\times N_f$ $\overline{q}$s per spatial lattice site.
While the mass and kinetic terms generalize straightforwardly, the energy in the chromo-electric field is more tricky.
After enforcing Gauss's law, it is 
\begin{equation}
    H_{el} = \frac{g^2}{2} \sum_{n=0}^{2L-2} \left ( \sum_{m \leq n} Q^{(a)}_m \right ) ^2
    \ ,\ \ 
    Q^{(a)}_m = \phi^{\dagger}_m T^a \phi_m
    \ ,
\end{equation}
where $T^a$ are now the generators of $SU(N_c)$.
The Hamiltonian,
including chemical potentials for baryon number (chemical potentials for other flavor combinations can be included as needed), is found to be
\begin{subequations}
    \label{eq:HNfNc}
    \begin{align}
        H = & \ H_{kin}\ +\ H_m\ +\ H_{el} \ +\ H_{\mu_B}  \ , \\[4pt]
        H_{kin} =& \ \frac{1}{2}\sum_{n=0}^{2L-2}\sum_{f=0}^{N_f-1}\sum_{c=0}^{N_c-1} \left[ \sigma_{i(n,f,c)}^+ \left ( \bigotimes_{j=1}^{N_cN_f-1}(-\sigma_{i(n,f,c)+j}^z ) \right ) \sigma_{i(n,f,c) + N_c N_f}^- +\rm{h.c.} \right] \ ,
        \label{eq:HkinN}\\[4pt]
        H_m =& \ \frac{1}{2} \sum_{n=0}^{2L-1} \sum_{f=0}^{N_f-1}\sum_{c=0}^{N_c-1} m_f \left[ (-1)^n \sigma^z_{i(n,f,c)} + 1 \right] \ ,
        \label{eq:HmN}\\[4pt]
        H_{el} =& \ \frac{g^2}{2} \sum_{n=0}^{2L-2}(2L-1-n)\left( \sum_{f=0}^{N_f-1} Q_{n,f}^{(a)} \, Q_{n,f}^{(a)} \ \ + \ \  
            2 \sum_{f=0}^{N_f-2} \sum_{f'=f+1}^{N_f-1}Q_{n,f}^{(a)} \, Q_{n,f'}^{(a)}
             \right)   \nonumber \\[4pt]
            & + g^2 \sum_{n=0}^{2L-3} \sum_{m=n+1}^{2L-2}(2L-1-m) \sum_{f=0}^{N_f-1} \sum_{f'=0}^{N_f-1} Q_{n,f}^{(a)} \, Q_{m,f'}^{(a)}  \ ,
        \label{eq:HelN}
        \end{align}
        \begin{align}
        H_{\mu_B} =& \ -\frac{\mu_B}{2 N_c} \sum_{n=0}^{2L-1} \sum_{f=0}^{N_f-1} \sum_{c=0}^{N_c-1} \sigma^z_{i(n,f,c)}   \ ,
        \label{eq:HmuBN}
    \end{align}
\end{subequations}
where, $i(n,f,c) = (N_c N_f n + N_f f + c)$, 
and the products of the charges are
\begin{align}
    4 Q_{n,f}^{(a)} \, Q_{n,f}^{(a)} =& \ \frac{N_c^2-1}{2} - \left (1+\frac{1}{N_c} \right )\sum_{c=0}^{N_c-2}\sum_{c' = c+1}^{N_c-1} \sigma^z_{i(n,f,c)} \sigma^z_{i(n,f,c')} \ ,  \nonumber \\[4pt]
    8 Q_{n,f}^{(a)} \, Q_{m,f'}^{(a)} =& \ 4 \sum_{c=0}^{N_c-2}\sum_{c'=c+1}^{N_c-1} \Bigl[ \sigma^+_{i(n,f,c)} \ \left(\otimes Z \right)_{(n,f,c,c')} \ \sigma^-_{i(n,f,c')} \sigma^-_{i(m,f',c)} \ \left(\otimes Z \right)_{(m,f',c,c')} \ \sigma^+_{i(m,f',c')}  \nonumber \\
    &+ {\rm h.c.}
    \Bigr] \ + \sum_{c=0}^{N_c-1} \sum_{c'=0}^{N_c-1}\left (\delta_{cc'} - \frac{1}{N_c} \right)\sigma^z_{i(n,f,c)}\sigma^z_{i(m,f',c')}\ , \nonumber \\
    \left(\otimes Z \right)_{(n,f,c,c')}  \equiv & \ \bigotimes_{k=1}^{c'-c-1} \sigma^z_{i(n,f,c)+k}
    \ .
    \label{eq:QnfQmfpN}
\end{align}
The resource requirements for implementing Trotterized time evolution 
using generalizations of the circuits in Sec.~\ref{sec:Circuits} are given in Eq.~(\ref{eq:RHCN}). 

It is interesting to consider the large-$N_c$ limit of the Hamiltonian, 
where quark loops are parametrically suppressed and 
the system can be described semi-classically~\cite{tHooft:1973alw,tHooft:1974pnl,Witten:1979kh,RevModPhys.54.407}.
Unitarity requires rescaling the strong coupling, 
$g^2 \to g^2/N_c$ and leading terms in the Hamiltonian scale as $\mathcal{O}(N_c)$.
The leading order contribution to the product of charges is
\begin{align}
    4 Q_{n,f}^{(a)} \, Q_{n,f}^{(a)} =&\ \sum_{c=0}^{N_c-2}\sum_{c' = c+1}^{N_c-1} \left (1 - \sigma^z_{i(n,f,c)} \sigma^z_{i(n,f,c')}\right ) \ ,  \nonumber \\[4pt]
    8 Q_{n,f}^{(a)} \, Q_{m,f'}^{(a)} =&\ 4 \sum_{c=0}^{N_c-2}\sum_{c'=c+1}^{N_c-1} 
    \Bigl[ \sigma^+_{i(n,f,c)} \ 
    \left(\otimes Z \right)_{(n,f,c,c')} \ 
    \sigma^-_{i(n,f,c')} \sigma^-_{i(m,f',c)} \ 
    \left(\otimes Z \right)_{(m,f',c,c')} \ 
    \sigma^+_{i(m,f',c')} \nonumber \\ & + {\rm h.c.}
    \Bigr] \ .
\end{align}
Assuming that the number of $q\overline{q}$ pairs that contribute to the meson wavefunctions do not scale with $N_c$, 
as expected in the large-$N_c$ limit,
$H_{el} \propto N_c$ 
and mesons are non-interacting, a well known consequence of the large-$N_c$ limit~\cite{tHooft:1973alw,tHooft:1974pnl}.
Baryons on the other hand are expected to have strong interactions at leading order in $N_c$~\cite{Witten:1979kh}. This is a semi-classical limit and we expect that there exists a basis
where states factorize into localized tensor products, and the time evolution operator is non-entangling.
The latter result has been observed in the large-$N_c$ limit of hadronic scattering~\cite{beane:2018oxh,beane:2021zvo,Low:2021ufv,Aoude:2020mlg,cervera-lierta:2017tdt}.

\section{Summary and Discussion}
\label{sec:SandC}
\noindent
Important for future quantum simulations of processes that can be meaningfully compared to experiment, the real-time dynamics of strongly-interacting systems are predicted to be efficiently computable with quantum computers of sufficient capability.
Building upon foundational work in quantum chemistry and in low-dimensional $U(1)$ and $SU(2)$ gauge theories, this work has developed the tools necessary for the quantum simulation of
$1+1$D QCD (in axial gauge) using open boundary conditions, with arbitrary numbers of quark flavors and colors and including chemical potentials for baryon number and isospin.
Focusing largely on QCD with $N_f=2$, which shares many of the complexities of QCD in $3+1$D, we have performed a detailed analysis of the required quantum resources for simulation of real-time dynamics, including efficient quantum circuits and associated gate counts, and the scaling of the number of Trotter steps for a fixed-precision time evolution.
The structure and dynamics of small systems, with $L=1,2$ for $N_c=3$ and $N_f=1,2$ have been detailed using classical computation, quantum simulators, D-Wave's {\tt Advantage} and IBM's 7-qubit devices {\tt ibmq\_jakarta} and {\tt ibm\_perth}. Using recently developed error mitigation strategies, relatively small uncertainties were obtained for a single Trotter step with $34$ CNOT gates after transpilation onto the QPU connectivity.

Through a detailed study of the low-lying spectrum, both the relevant symmetries and the color-singlets in the mesonic and baryonic sectors, including a bound two-baryon nucleus, have been identified.  
Open boundary conditions also permit low-lying color edge-states that penetrate into the lattice volume by a distance set by the confinement scale.
By examining quark entanglement in the hadrons, a transition from the mesons being primarily composed of quark-antiquarks to baryon-antibaryons was found.
We have presented the relative contributions of each of the terms in the Hamiltonian to the energy of the vacuum, mesons and baryons.  

This work has provided an estimate for the number of CNOT-gates required to implement one Trotter step in $N_f=2$, $1+1$D axial-gauge QCD. For $L = 10$ spatial sites, $\sim 3 \times 10^4$ CNOTs
are required, while $\sim 4 \times 10^6$ CNOTs are required for $L = 100$.
Realistically, quantum simulations with $L=10$ are a beginning toward providing results with a complete quantification of uncertainties, including lattice-spacing and finite-volume
artifacts, and $L=100$ will likely yield high-precision results. It was found that, in the axial-gauge formulation, resources for time evolution effectively scale as $L^2 t$ for intermediate times and $L^2 t^2$ for
asymptotic times. With $L\sim t$, this asymptotic scaling is the same as in the Schwinger model, suggesting no differences in scaling between Weyl and axial gauges.


\clearpage
\begin{subappendices}

\section{Mapping to Qubits}
\label{app:hamConst}
\noindent 
This appendix outlines how the qubit Hamiltonian in Eq.~(\ref{eq:H2flav}) is obtained from the lattice Hamiltonian in Eq.~(\ref{eq:GFHam}). 
For this system, 
the constraint of Gauss's law is sufficient to uniquely determine the chromo-electric field carried by the links between lattice sites in terms of a background chromo-electric field and the distribution of color charges.  The difference between adjacent chromo-electric fields at a site with charge 
$Q^{(a)}$ 
is
\begin{equation}
    {\bf E}^{(a)}_{n+1} - {\bf E}^{(a)}_n = Q^{(a)}_n \ ,
    \label{eq:GaussLaw}
\end{equation}
for $a=1$ to $8$, resulting in a
chromo-electric field 
\begin{equation}
    {\bf E}^{(a)}_{n} = {\bf F}^{(a)}
    \: + \: \sum_{i\leq n} Q^{(a)}_i \  .
    \label{eq:GaussLawSol}
\end{equation}
In general, there can be a non-zero background chromo-electric field, ${\bf F}^{(a)}$,
which in this paper has been set to zero.
Inserting the chromo-electric field in terms of the charges into Eq.~(\ref{eq:KSHam}) yields Eq.~(\ref{eq:GFHam}). 

The color and flavor degrees of freedom of each $q$ and $\overline{q}$ are then distributed over 
$6$ ($=N_c N_f$) sites as illustrated in Fig.~(\ref{fig:2flavLayout}). 
There are now creation and annihilation operators for each quark, and the Hamiltonian is
\begin{align}
   H =&\ \sum_{n=0}^{2L-1} \sum_{f=0}^1 \sum_{c=0}^2  \big[ m_f (-1)^n \psi^{\dagger}_{6n+3f+c} \psi_{6n+3f+c} \nonumber \\[4pt]
   & \: - \: \frac{\mu_B}{3} \psi^{\dagger}_{6n+3f+c} \psi_{6n+3f+c} \: - \: \frac{\mu_I}{2}(-1)^f \psi^{\dagger}_{6n+3f+c} \psi_{6n+3f+c} \big] \nonumber \\[4pt]
   &+  \frac{1}{2} \sum_{n=0}^{2L-2}\sum_{f=0}^1 \sum_{c=0}^2  \left (\psi^{\dagger}_{6n+3f+c} \psi_{6(n+1)+3f+c} + \: {\rm h.c.} \right ) \: + \: \frac{g^2}{2} \sum_{n=0}^{2L-2} \left ( \sum_{m\leq n} \sum_{f=0}^1 Q^{(a)}_{m,f} \right ) ^2 \ ,
    \label{eq:FockHam}
\end{align}
where the color charge is evaluated over three $(r,g,b)$ occupation sites with the same flavor,
\begin{equation}
   Q_{m,f}^{(a)} = \sum_{c=0}^{2}\sum_{c'=0}^2 \psi^{\dagger}_{6m+3f+c} \ T^a_{cc'}\  \psi_{6m+3f+c'} \ ,
\end{equation}
and the $T^a$ are the eight generators of $SU(3)$. 
The fermionic  operators in Fock space are mapped onto spin operators via the JW transformation,
\begin{equation}
    \psi_n =  \bigotimes_{l<n}( -\sigma^z_l ) \sigma^-_n \ , \ \ \psi_n^{\dagger} =  \bigotimes_{l<n}( -\sigma^z_l ) \sigma^+_n \ .
    \label{eq:JW}
\end{equation}
In terms of spins, the eight $SU(3)$ charge operators become\footnote{Calculations of quadratics of the gauge charges are simplified by the Fierz identity,
\begin{equation}
    \left ( T^{(a)} \right )^{\alpha}_{\beta} \, \left (T^{(a)}\right )^{\gamma}_{\delta} = \frac{1}{2} (\delta^{\alpha}_{\delta} \delta^{\gamma}_{\beta} - \frac{1}{N_c} \delta^{\alpha}_{\beta}\delta^{\gamma}_{\delta}) \ .
    \label{eq:Fierz}
\end{equation}}
\begin{align}
    Q_{m,f}^{(1)} = & \ \frac{1}{2}\sigma^+_{6m+3f} \sigma^-_{6m+3f+1} + \rm{h.c.} \ , \nonumber \\
    Q_{m,f}^{(2)} = & \ -\frac{i}{2}\sigma^+_{6m+3f} \sigma^-_{6m+3f+1} + \rm{h.c.} \ , \nonumber \\
    Q_{m,f}^{(3)} = & \ \frac{1}{4}(\sigma^z_{6m+3f} - \sigma^z_{6m+3f+1}) \ , \nonumber \\
    Q_{m,f}^{(4)} = & \ -\frac{1}{2}\sigma^+_{6m+3f} \sigma^z_{6m+3f+1} \sigma^-_{6m+3f+2} + \rm{h.c.} \ , \nonumber \\
    Q_{m,f}^{(5)} = & \ \frac{i}{2}\sigma^+_{6m+3f} \sigma^z_{6m+3f+1} \sigma^-_{6m+3f+2} + \rm{h.c.} \ , \nonumber \\
    Q_{m,f}^{(6)} = & \ \frac{1}{2}\sigma^+_{6m+3f+1} \sigma^-_{6m+3f+2} + \rm{h.c.} \ , \nonumber \\
    Q_{m,f}^{(7)} = & \ -\frac{i}{2}\sigma^+_{6m+3f+1} \sigma^-_{6m+3f+2} + \rm{h.c.} \ , \nonumber \\
    Q_{m,f}^{(8)} = & \ \frac{1}{4 \sqrt{3}}(\sigma^z_{6m+3f} + \sigma^z_{6m+3f+1} - 2\sigma^z_{6m+3f+2})  \ .
    \label{eq:SU3chargesFull}
\end{align}
Substituting Eqs.~(\ref{eq:JW}) and~(\ref{eq:SU3chargesFull}) into Eq.~(\ref{eq:FockHam}) gives the Hamiltonian in Eq.~(\ref{eq:H2flav}). For reference, the expanded Hamiltonian for $L=1$ is
\begin{subequations}
    \label{eq:H2flavL1}
    \begin{align}
    H = & \ H_{kin}\ +\ H_m\ +\ H_{el} \ +\ 
    H_{\mu_B}\ +\ H_{\mu_I}\ ,\\[4pt]
    H_{kin} = & \ -\frac{1}{2} (\sigma^+_6 \sigma^z_5 \sigma^z_4 \sigma^z_3 \sigma^z_2 \sigma^z_1 \sigma^-_0 + \sigma^-_6 \sigma^z_5 \sigma^z_4 \sigma^z_3 \sigma^z_2 \sigma^z_1 \sigma^+_0 + \sigma^+_7 \sigma^z_6 \sigma^z_5 \sigma^z_4 \sigma^z_3 \sigma^z_2 \sigma^-_1 \nonumber \\
    &+ \sigma^-_7 \sigma^z_6 \sigma^z_5 \sigma^z_4 \sigma^z_3 \sigma^z_2 \sigma^+_1 +\, \sigma^+_8 \sigma^z_7 \sigma^z_6 \sigma^z_5 \sigma^z_4 \sigma^z_3 \sigma^-_2 + \sigma^-_8 \sigma^z_7 \sigma^z_6 \sigma^z_5 \sigma^z_4 \sigma^z_3 \sigma^+_2 \nonumber \\
    &+ \sigma^+_9 \sigma^z_8 \sigma^z_7 \sigma^z_6 \sigma^z_5 \sigma^z_4 \sigma^-_3 + \sigma^-_9 \sigma^z_8 \sigma^z_7 \sigma^z_6 \sigma^z_5 \sigma^z_4 \sigma^+_3 +\, \sigma^+_{10} \sigma^z_9 \sigma^z_8 \sigma^z_7 \sigma^z_6 \sigma^z_5 \sigma^-_4 \nonumber \\
    &+ \sigma^-_{10} \sigma^z_9 \sigma^z_8 \sigma^z_7 \sigma^z_6 \sigma^z_5 \sigma^+_4 + \sigma^+_{11} \sigma^z_{10} \sigma^z_9 \sigma^z_8 \sigma^z_7 \sigma^z_6 \sigma^-_5 + \sigma^-_{11} \sigma^z_{10} \sigma^z_9 \sigma^z_8 \sigma^z_7 \sigma^z_6 \sigma^+_5 ) \ ,
        \label{eq:Hkin2flavL1}\\[4pt]
    H_m = & \ \frac{1}{2} \bigl [ m_u\left (\sigma^z_0 + \sigma^z_1 + \sigma^z_2 -\sigma^z_6 - \sigma^z_7 - \sigma^z_8 + 6\right )\nonumber \\
    &+ m_d\left (\sigma^z_3 + \sigma^z_4 + \sigma^z_5 -\sigma^z_9 - \sigma^z_{10} - \sigma^z_{11} + 6\right ) \bigr]\ ,
        \label{eq:Hm2flavL1}\\[4pt]
    H_{el} = & \ \frac{g^2}{2} \bigg [ \frac{1}{3}(3 - \sigma^z_1 \sigma^z_0 - \sigma^z_2 \sigma^z_0 - \sigma^z_2 \sigma^z_1) + \sigma^+_4\sigma^-_3\sigma^-_1\sigma^+_0  + \sigma^-_4\sigma^+_3\sigma^+_1\sigma^-_0 \nonumber \\
    & + \sigma^+_5\sigma^z_4\sigma^-_3\sigma^-_2\sigma^z_1\sigma^+_0  + \sigma^-_5\sigma^z_4\sigma^+_3\sigma^+_2\sigma^z_1\sigma^-_0 +\, \sigma^+_5\sigma^-_4\sigma^-_2\sigma^+_1 + \sigma^-_5\sigma^+_4\sigma^+_2\sigma^-_1 \nonumber \\ +\,\frac{1}{12} (2 \sigma^z_3 \sigma^z_0 + & 2\sigma^z_4 \sigma^z_1 +  2\sigma^z_5 \sigma^z_2 - \sigma^z_5 \sigma^z_0 - \sigma^z_5 \sigma^z_1 - \sigma^z_4 \sigma^z_2 - \sigma^z_4 \sigma^z_0 - \sigma^z_3 \sigma^z_1  - \sigma^z_3 \sigma^z_2  ) \bigg ] \ ,
    \label{eq:Hel2flavL1}\\[4pt] 
    H_{\mu_B} = \ -\frac{\mu_B}{6} & \left ( \sigma^z_0 + \sigma^z_1 + \sigma^z_2 + \sigma^z_3 + \sigma^z_4 + \sigma^z_5
    - \sigma^z_6 + \sigma^z_7 + \sigma^z_8 + \sigma^z_9 + \sigma^z_{10} + \sigma^z_{11} \right )  \ ,
        \label{eq:HmuB2flavL1}\\[4pt]
     H_{\mu_I} =  \ -\frac{\mu_I}{4} & \left( \sigma^z_0 + \sigma^z_1 + \sigma^z_2 - \sigma^z_3 - \sigma^z_4 - \sigma^z_5
    + \sigma^z_6 + \sigma^z_7 + \sigma^z_8 - \sigma^z_9 - \sigma^z_{10} - \sigma^z_{11} \right) \ .
        \label{eq:HmuI2flavL1}
    \end{align}
\end{subequations}
%

\section{Symmetries of the Free-Quark Hamiltonian}
\label{app:freeSym}
\noindent 
Here the symmetries of the free-quark Hamiltonian are identified to better understand the degeneracies observed in the spectrum of $1+1$D QCD with $N_f=2$ and $L=1$ as displayed in Figs.~\ref{fig:specDegenh} and~\ref{fig:specDegeng}.
Specifically, the Hamiltonian with $g=h=\mu_B=\mu_I=0$, leaving only the hopping and mass terms ($m = m_u = m_d$), is
\begin{equation}
  \hspace{-2ex} H = \sum_{f=0}^1 \sum_{c=0}^2  \left [ m \sum_{n=0}^{2L-1} (-1)^n \psi^{\dagger}_{6n+3f+c} \psi_{6n+3f+c} + \frac{1}{2} \sum_{n=0}^{2L-2} \left (\psi^{\dagger}_{6n+3f+c} \psi_{6(n+1)+3f+c} + \: {\rm h.c.} \right ) \right ] .
\end{equation}
The mapping of degrees of freedom is taken to be as shown in Fig.~\ref{fig:2flavLayout}, but it will be convenient to work with Fock-space quark operators instead of spin operators. 
In what follows the focus will be on $L=1$, and larger systems follow similarly.

The creation operators can be assembled into a 12-component vector, 

\noindent $\Psi^{\dagger}_i = (\psi_0^\dagger, \, \psi_1^\dagger, \ldots ,\psi_{10}^{\dagger}, \, \psi_{11}^{\dagger})$, 
in terms of which the Hamiltonian  becomes
\begin{equation}
    H = \Psi^{\dagger}_i M_{ij} \Psi_j \ ,
\end{equation}
where $M$ is a $12 \times 12$ block matrix of the form,
\begin{equation}
   M  = 
   \left[
   \begin{array}{c|c}
       m  & 1/2 \\
       \hline
    1/2  & -m 
\end{array}
\right]
\ ,
\end{equation}
with each block a $6 \times 6$  diagonal matrix. 
Diagonalizing $M$, gives rise to 
\begin{equation}
   \tilde M = 
   \left[
   \begin{array}{c|c}
       \lambda  & 0 \\
       \hline
    0  & -\lambda 
\end{array}
\right]
\ ,\ \ 
\lambda = \frac{1}{2}\sqrt{1+4m^2} \ ,
\end{equation}
with associated eigenvectors,
\begin{align}
    \tilde{\psi}_i =& \frac{1}{\sqrt{2}}\left (\sqrt{1+\frac{\lambda}{m}}\, \psi_i \ + \ \sqrt{1-\frac{\lambda}{m}}\, \psi_{6+i} \right ) \ , \nonumber \\ \ \tilde{\psi}_{6+i} =& \frac{1}{\sqrt{2}}\left (-\sqrt{1-\frac{\lambda}{m}}\,\psi_i \ + \ \sqrt{1+\frac{\lambda}{m}}\, \psi_{6+i} \right )
    \ ,
\end{align}
where $\tilde{\psi}_i$ ($\tilde{\psi}_{6+i}$) corresponds to the positive (negative) eigenvalue
and the index $i$ takes values $0$ to $5$.
These eigenvectors create superpositions of quarks and antiquarks with the same color and flavor, which are the OBC analogs of momentum plane-waves. 
In this basis, the Hamiltonian becomes
\begin{equation}
    H = \sum_{i=0}^{5} \lambda\left ( \tilde{\psi}^{\dagger}_i \tilde{\psi}_i - \tilde{\psi}^{\dagger}_{6+i} \tilde{\psi}_{6+i} \right )
    \ ,
    \label{eq:Hamifree0}
\end{equation}
which has a vacuum state,
\begin{equation}
    \lvert \Omega_0 \rangle = \prod_{i=0}^{i=5}\tilde{\psi}^{\dagger}_{6+i} \ket{\omega_0} \ ,
\end{equation}
where $\ket{\omega_0}$ is the unoccupied state,
and
$\lvert \Omega_0 \rangle$ corresponds to 
$\lvert 000000111111 \rangle$ (in binary)
in this transformed basis.
Excited states are formed by acting with either $\tilde{\psi}^{\dagger}_i$ or $\tilde{\psi}_{6+i}$ on 
$\lvert \Omega_0 \rangle$ which raises the energy of the system by $\lambda$. 
A further transformation is required for the $SU(12)$ symmetry to be manifest.
In terms of the 12-component vector, $\tilde{\Psi}^{\dagger} = (\tilde{\psi}^{\dagger}_0, \, \ldots, \, \tilde{\psi}^{\dagger}_5, \, \tilde{\psi}_6, \, \ldots, \, \tilde{\psi}_{11})$, the Hamiltonian in Eq.~(\ref{eq:Hamifree0}) becomes,
\begin{equation}
    H = 
    \sum_{i=0}^{5} \lambda\left ( \tilde{\psi}^{\dagger}_i \tilde{\psi}_i - \tilde{\psi}^{\dagger}_{6+i} \tilde{\psi}_{6+i} \right )
    \ =\ 
    \lambda\left( 
    \tilde{\Psi}^{\dagger} \tilde{\Psi} - 6 
    \right)
    \ ,
\end{equation}
where the canonical anticommutation relations have been used to obtain the final equality.
This is invariant under a $SU(12)$ symmetry, where $\tilde{\Psi}$ transforms in the fundamental representation. 
The free-quark spectrum ($g=h=0$) is therefore described by states with degeneracies corresponding to the ${\bf 1}$ and ${\bf 12}$ of $SU(12)$ as well as
the antisymmetric combinations of fundamental irreps, ${\bf 66}, {\bf 220}, \ldots$ as illustrated in Figs.~\ref{fig:specDegenh} and~\ref{fig:specDegeng}.
The vacuum state corresponds to the singlet of $SU(12)$. The lowest-lying {\bf 12} corresponds to single quark or antiquark excitations, which are color ${\bf 3}_c$s for quarks and $\overline{\bf 3}_c$s for antiquarks and will each appear as isodoublets, i.e., ${\bf 12}\rightarrow {\bf 3}_c\otimes {\bf 2}_f \oplus \overline{\bf 3}_c\otimes {\bf 2}_f$.
The {\bf 66} arises from double excitations of quarks and antiquarks.  The possible color-isospin configurations are, based upon totally-antisymmetric wavefunctions for $qq$, $\overline{q}\overline{q}$ and $\overline{q}q$,  
${\bf 66} =
{\bf 1}_c\otimes {\bf 1}_f
\oplus
{\bf 1}_c\otimes {\bf 3}_f
\oplus
{\bf 8}_c\otimes {\bf 1}_f
\oplus
{\bf 8}_c\otimes {\bf 3}_f
\oplus
{\bf 6}_c\otimes {\bf 1}_f
\oplus
\overline{\bf 6}_c\otimes {\bf 1}_f
\oplus
{\bf 3}_c\otimes {\bf 3}_f
\oplus
\overline{\bf 3}_c\otimes {\bf 3}_f
$.
The OBCs split the naive symmetry between quarks and antiquarks and, for $g\ne 0$, the lowest-lying color edge-states are from the antiquark sector with degeneracies $6$ from a single excitation and $6,9$ from double excitations. 
Larger lattices possess an analogous global 
$SU(12)$ symmetry, coupled between spatial sites by the hopping term, and the spectrum is again one of non-interacting quasi-particles.

\section{Details of the D-Wave Implementations}
\label{app:dwave}
\noindent
In this appendix, additional details are provided on the procedure used in Sec.~\ref{sec:dwave_spectrum} to extract the lowest three eigenstates and corresponding energies using D-Wave's {\tt Advantage}, (a more complete description can be found in Ref.~\cite{Illa:2022jqb}). The objective function $F$ to be minimized can be written in terms of binary variables and put into QUBO form. Defining $F=\langle \Psi \rvert \tilde{H} \lvert \Psi \rangle -\eta \langle \Psi| \Psi \rangle$~\cite{doi:10.1021/acs.jctc.9b00402}, and expanding the wavefunction with a finite dimensional orthonormal basis $\psi_{\alpha}$, $\lvert \Psi \rangle =\sum^{n_s}_{\alpha} a_\alpha |\psi_{\alpha}\rangle$, it is found
\begin{align}
    F=&\langle \Psi \rvert \tilde{H} \lvert \Psi \rangle -\eta \langle \Psi| \Psi \rangle = \sum_{\alpha\beta}^{n_s} a_\alpha a_\beta[\langle \psi_\alpha \rvert \tilde{H} \lvert \psi_\beta \rangle -\eta \langle \psi_\alpha| \psi_\beta \rangle] \nonumber \\ 
    =&\sum_{\alpha\beta}^{n_s} a_\alpha a_\beta (\tilde{H}_{\alpha\beta} -\eta \delta_{\alpha\beta})=\sum_{\alpha\beta}^{n_s} a_\alpha a_\beta h_{\alpha\beta}\ ,
\end{align}
where $h_{\alpha\beta}$ are the matrix elements of the Hamiltonian that can be computed classically. The coefficients $a_\alpha$ are then expanded in a fixed-point representation using $K$ bits~\cite{doi:10.1021/acs.jctc.9b00402,Chang:2019,arahman:2021ktn}, 
\begin{equation}
    a^{(z+1)}_\alpha=a^{(z)}_\alpha+\sum_{i=1}^{K}2^{i-K-z}(-1)^{\delta_{iK}}q^{\alpha}_i \ ,
\end{equation}
where $z$ is the zoom parameter. The starting point is $a_\alpha^{(z=0)}=0$, and for each consecutive value of $z$, the range of values that $a_\alpha^{(z+1)}$ is allowed to explore is reduced by a factor of $2$, centered around the previous solution $a_\alpha^{(z)}$. Now $F$ takes the following form,
\begin{align}
    F=&\sum_{\alpha,\beta}^{n_s}\sum_{i,j}^K Q_{\alpha,i;\beta,j} q^{\alpha}_i q^{\beta}_j\ , \ \nonumber \\
    Q_{\alpha,i;\beta,j}=&2^{i+j-2K-2z} (-1)^{\delta_{iK}+\delta_{jK}} h_{\alpha\beta} + 2 \delta_{\alpha\beta} \delta_{ij} 2^{i-K-z} (-1)^{\delta_{iK}} \sum_\gamma^{n_s} a^{(z)}_\gamma h_{\gamma\beta} \ .
\end{align}
The iterative procedure used to improve the precision of the results is based on the value $a^{(z)}_\alpha$ obtained after $14$ zoom steps (starting from $a_\alpha^{(z_0=0)}=0$), and then launching a new annealing workflow with $z_1 \neq 0$ (e.g., $z_1=4$), with $a^{(z=z_0+14)}_\alpha$ as the starting point. After another 14 zoom steps, the final value $a^{(z=z_1+14)}_\alpha$ can be used as the new starting point for $a^{(z=z_2)}_\alpha$, with $z_2 > z_1$. This process can be repeated until no further improvement is seen in the convergence of the energy and wavefunction.

In Table~\ref{tab:QAresults2}, the difference between the exact energy of the vacuum and masses of the $\sigma$- and $\pi$-mesons and the ones computed with the QA, for each iteration of this procedure after 14 zoom steps, are given, together with the overlap of the wavefunctions $1-|\langle \Psi^{\rm exact}| \Psi^{\tt Adv.}\rangle|^2$. See also Fig.~\ref{fig:QAresults}.
\begin{table}[!ht]
\centering
\renewcommand{\arraystretch}{1.2}
\begin{tabular}{|| c | r | r@{\ \ \ \ \ } | r | r@{\ \ \ \ \ } ||} 
\hline
  & \multicolumn{2}{c|}{$\ket{\Omega}$} & \multicolumn{2}{c|}{$\ket{\sigma}$} \\ \hline
  Step & \multicolumn{1}{c|}{$\delta E_\Omega$} & \multicolumn{1}{c|}{$1-|\langle \Psi_\Omega^{\rm exact}| \Psi_\Omega^{\tt Adv.}\rangle|^2$} & \multicolumn{1}{c|}{$\delta M_\sigma$} & \multicolumn{1}{c|}{$1-|\langle \Psi_\sigma^{\rm exact}| \Psi_\sigma^{\tt Adv.}\rangle|^2$} \\
 \hline \hline 
 0 & $4^{\,+2}_{\,-2}\times 10^{-1}$ & $10^{\,+3}_{\,-5}\times 10^{-2}$ & $4^{\,+2}_{\,-2}\times 10^{-1}$ & $11^{\,+7}_{\,-5}\times 10^{-2}$\\[0.2ex]
 1 & $9^{\,+4}_{\,-3}\times 10^{-3}$ & $2^{\,+6}_{\,-5}\times 10^{-3}$  & $3^{\,+1}_{\,-1}\times 10^{-2}$ & $7^{\,+2}_{\,-2}\times 10^{-3}$ \\[0.2ex]
 2 & $6^{\,+2}_{\,-2}\times 10^{-4}$ & $12^{\,+3}_{\,-5}\times 10^{-5}$ & $4^{\,+1}_{\,-1}\times 10^{-3}$ & $12^{\,+3}_{\,-4}\times 10^{-4}$\\[0.2ex]
 3 & $4^{\,+1}_{\,-2}\times 10^{-5}$ & $9^{\,+3}_{\,-4}\times 10^{-6}$  & $2^{\,+1}_{\,-1}\times 10^{-4}$ & $6^{\,+1}_{\,-2}\times 10^{-5}$\\[0.2ex]
 4 & $16^{\,+6}_{\,-6}\times 10^{-7}$& $3^{\,+2}_{\,-1}\times 10^{-7}$  & $10^{\,+6}_{\,-3}\times 10^{-6}$& $9^{\,+1}_{\,-1}\times 10^{-6}$\\
 \hline
\end{tabular}

\begin{tabular}{|| c | r | r@{\ \ \ \ \ } ||}
\hline
  & \multicolumn{2}{c|}{$\ket{\pi}$} \\ \hline
  Step & \multicolumn{1}{c|}{$\delta M_\pi$} & \multicolumn{1}{c|}{$1-|\langle \Psi_\pi^{\rm exact}| \Psi_\pi^{\tt Adv.}\rangle|^2$} \\
 \hline \hline 
 0 & $3^{\,+1}_{\,-1}\times 10^{-1}$ & $11^{\,+71}_{\,-4}\times 10^{-2}$\\[0.2ex]
 1 & $9^{\,+4}_{\,-3}\times 10^{-3}$ & $3^{\,+3}_{\,-1}\times 10^{-3}$\\[0.2ex]
 2 & $7^{\,+2}_{\,-3}\times 10^{-4}$ & $3^{\,+2}_{\,-2}\times 10^{-4}$\\[0.2ex]
 3 & $4^{\,+2}_{\,-2}\times 10^{-5}$ & $12^{\,+6}_{\,-3}\times 10^{-6}$\\[0.2ex]
 4 & $7^{\,+9}_{\,-5}\times 10^{-7}$ & $8^{\,+2}_{\,-2}\times 10^{-6}$\\
 \hline
\end{tabular}
\renewcommand{\arraystretch}{1}
\caption{Convergence of the energy, masses and wavefunctions of the three lowest-lying states in the $B=0$ sector of $1+1$D QCD with $N_f=2$ and $m=g=L=1$, between exact results from diagonalization of the Hamiltonian and those 
obtained from D-Wave's {\tt Advantage}.}
\label{tab:QAresults2}
\end{table}

\begin{figure}[!ht]
    \centering
    \includegraphics[width=\columnwidth]{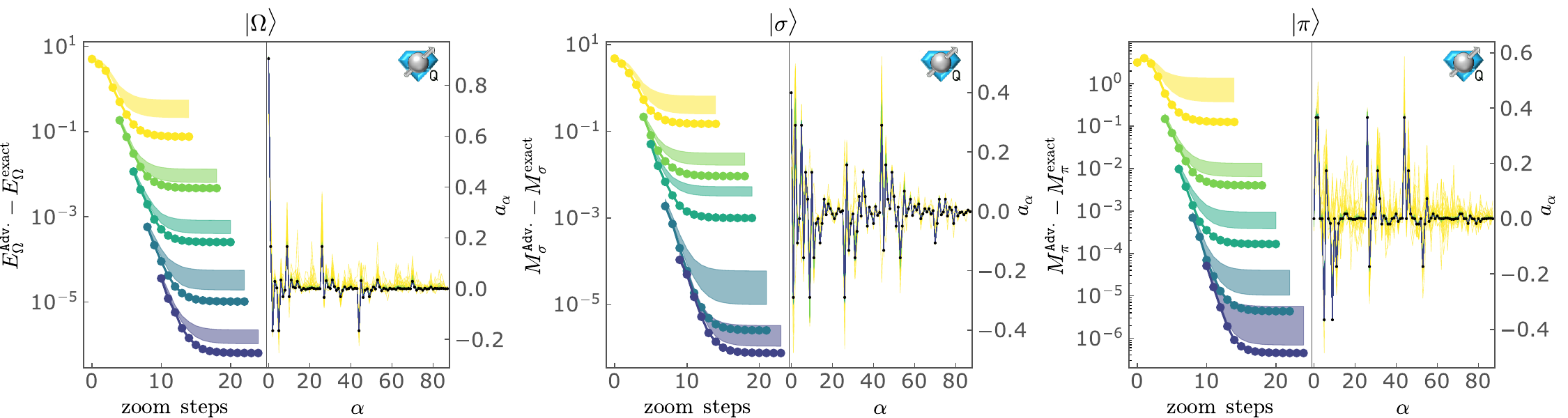}
    \caption{
    Iterative convergence of the energy, masses and wavefunctions for the three lowest-lying states in the $B=0$ sector of $1+1$D QCD with $N_f=2$ and $m=g=L=1$: vacuum (left), $\sigma$-meson (center) and $\pi$-meson (right). Compared to Fig.~\ref{fig:QAresults}, the exact wavefunctions are used when projecting the Hamiltonian to study the excited states.}
    \label{fig:QAresults_exact}
\end{figure}

Focusing on the lowest line of the last panel of Fig.~\ref{fig:QAresults}, which shows the convergence as a function of zoom steps for the pion mass, it can be seen that it displays some oscillatory behavior compared to the rest, which are smooth. This is expected, since the wavefunctions used to project out the lower eigenstates from the Hamiltonian are known with a finite precision (obtained from previous runs). For example, the vacuum state is extracted at the $10^{-6}$ precision level. Then, when looking at the excited states with increased precision (like for the pion, around $10^{-7}$), the variational principle might not hold, and the computed energy level might be below the ``true'' one (and not above). To support this argument, the same calculation has been pursued, but using the exact wavefunctions when projecting the Hamiltonian to study the excited states (instead of the ones computed using {\tt Advantage}), and no oscillatory behavior is observed, as displayed in Fig.~\ref{fig:QAresults_exact}.

\section{Quantum Circuits Required for Time Evolution by the Gauge-Field Interaction}
\label{app:circ}
\begin{figure}[!ht]
    \centering
    \includegraphics[width=8cm]{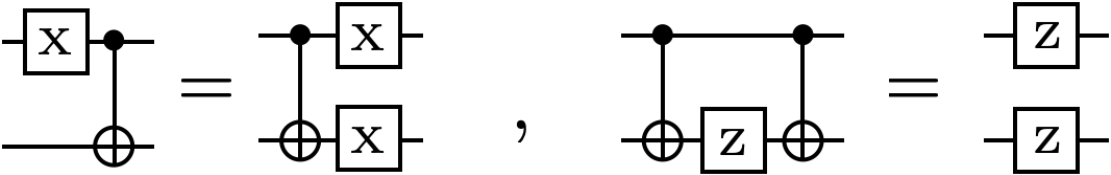}
    \caption{The $X$ and $Z$ circuit identities.}
    \label{circ:XZError}
\end{figure}

\begin{figure}
    \centering
    \includegraphics[width=16cm]{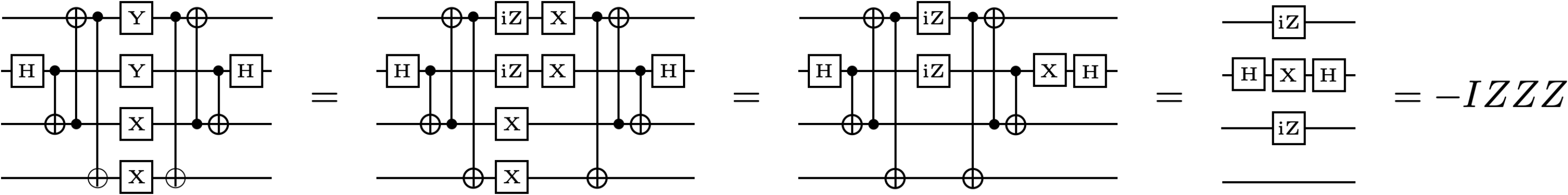}
    \caption{The diagonalization of $XXYY$ via a GHZ state-preparation circuit.}
    \label{fig:XXYYDiag}
\end{figure}

\noindent
This appendix provides more detail about the construction of the quantum circuits which implement the Trotterized time evolution of the chromo-electric terms of the Hamiltonian. 
It closely follows the presentation in the appendix of Ref.~\cite{stetina:2020abi}.
The four-qubit interaction in $H_{el}$ has the form
\begin{align}
\sigma^+ \sigma^- \sigma^- \sigma^+ + {\rm h.c.} =& \frac{1}{8}(XXXX + XXYY + XYXY - XYYX + \nonumber \\ & YXYX - YXXY +YYXX + YYYY) \ .
\label{eq:pmmpApp}
\end{align}
Since the 8 Pauli strings are mutually commuting, they can be simultaneously diagonalized by a unitary transformation. The strategy for
identifying the quantum circuit(s) to implement this term will be to first change to a basis where every term is diagonal, then apply the diagonal unitaries and finally
return back to the computational basis.
The GHZ state-preparation circuits,
shown in Fig.~\ref{circ:GHZ}, 
diagonalize all 8 of the Pauli strings, for example,
\begin{align}
G^{\dagger} \ &( XXXX + YYXX + YXYX - YXXY - XYYX + XYXY + XXYY + YYYY) \ G \ \nonumber \\[4pt]
 & = IIZI - ZIZZ - ZZZZ + ZIZI + IZZI - IIZZ - IZZZ + ZZZI  \ .
\label{eq:GHZDiag}
\end{align}
This can be verified by using the identities that are shown in Fig.~\ref{circ:XZError} to simplify the circuits formed by conjugating each Pauli string by $G$. 
As an example, the diagonalization of $XXYY$ is displayed in Fig.~\ref{fig:XXYYDiag}. 
The first equality uses $Y = i Z X$ and the second equality uses the $X$
circuit identity to move all $X$s past the CNOTs. The third equality moves the $Z$s past
the controls of the CNOTs and uses the $Z$ circuit identity. The other Pauli strings are diagonalized in a similar manner.

It is also straightforward to show that, for example, 
\begin{equation}
    G^{\dagger}(IZZI + IZIZ + ZIIZ)G = IZII + IIIZ + ZIII \ .
    \label{eq:ZZGHZG}
\end{equation}
In general, a $ZZ$ in the computational basis becomes a single $Z$ in the GHZ basis if the state-preparation circuit has a CNOT that connects the
original two $Z$s. The two GHZ state-preparation circuits, $G$ and $\tilde{G}$, were chosen so that all $9$ of the $ZZ$ terms in Eq.~(\ref{eq:QnfQmfp}) are mapped to single qubit rotations.
Once in the GHZ basis, the diagonal unitaries are performed, e.g., $\exp(-i IZZZ)$. 
They are arranged to minimize the number of CNOTs required, and the optimal circuit layouts are shown in Fig.~\ref{circ:UpmmpZZ}.

\section{Complete Circuits for \texorpdfstring{\boldmath$N_f=1,2$}{Nf=1,2} QCD with \texorpdfstring{\boldmath$L=1$}{L=1}}
\label{app:Nf1SU3circs}
\begin{figure}[!ht]
    \centering
    \includegraphics[width=\columnwidth]{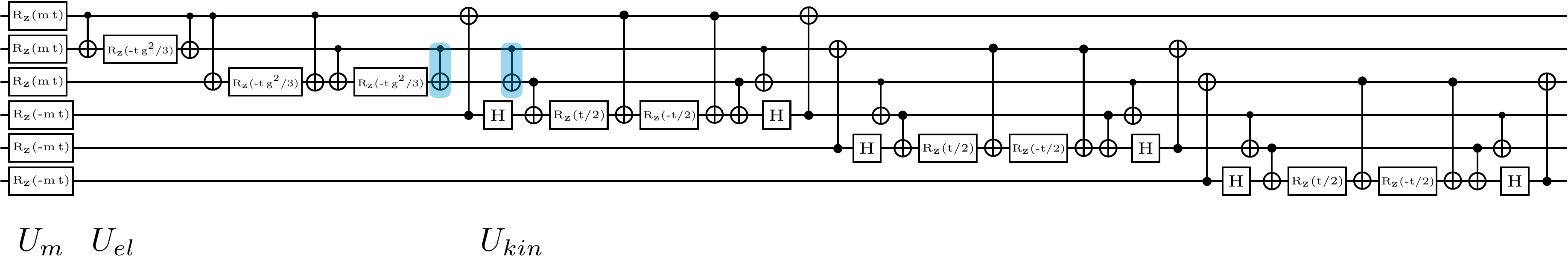}
    \caption{The complete circuit that implements a single Trotter step for $N_f=1$ QCD with $L=1$.
    }
    \label{fig:Nf1Trot}
\end{figure}

\begin{figure}[!ht]
    \centering

    \includegraphics[height=0.14\textheight]{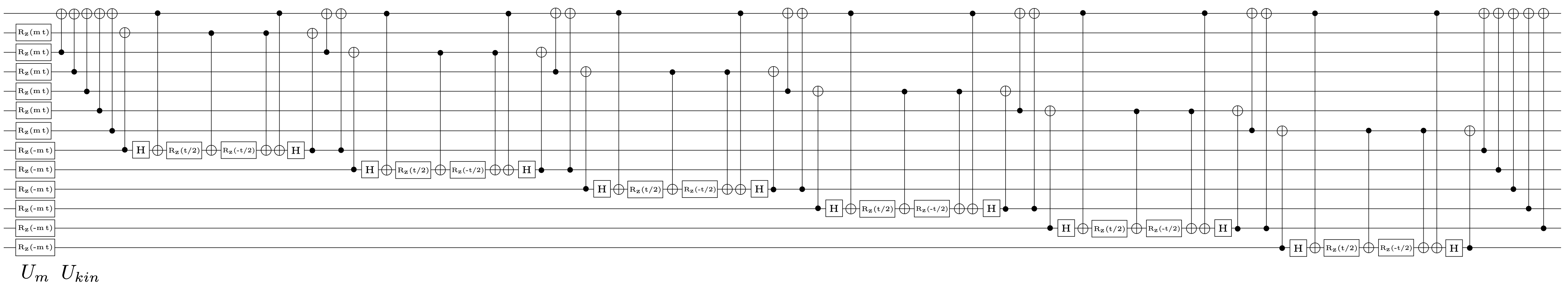} \\
    \includegraphics[height=0.14\textheight]{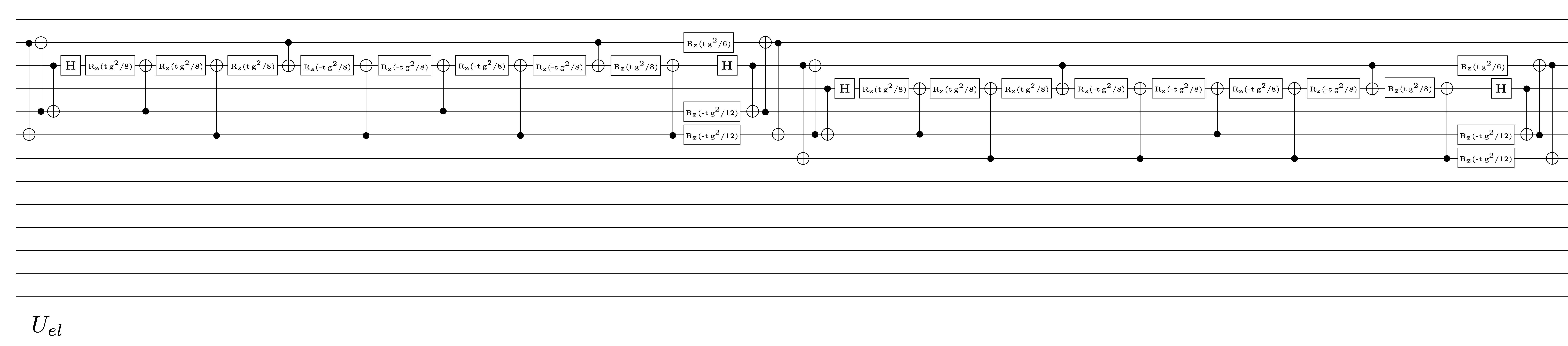} \\
    \includegraphics[height=0.14\textheight]{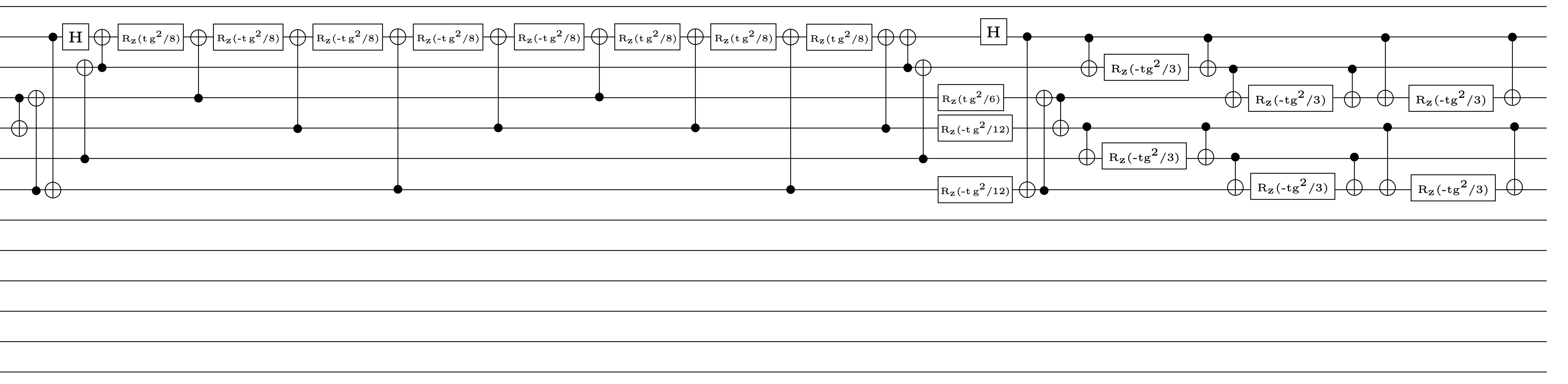}
    \caption{The complete circuit that implements a single Trotter step for $N_f=2$ QCD with $L=1$.
    }
    \label{fig:Nf2Trot}
\end{figure}

\noindent 
This appendix provides the complete set of circuits required to
implement one Trotter step for 
$N_f=1$ and $N_f=2$ QCD with $L=1$. 
The composite circuit for $N_f=1$ is shown in 
Fig.~\ref{fig:Nf1Trot} where, by ordering $U_{el}$ before $U_{kin}$, the CNOTs highlighted in blue cancel. The composite circuit for $N_f=2$ is shown in 
Fig.~\ref{fig:Nf2Trot}, 
where the ordering in the Trotterization
is $U_m$
followed by $U_{kin}$ 
and then by $U_{el}$.

%

\section{Energy Decomposition Associated with Time Evolution from the Trivial Vacuum}
\label{app:MDTD}
\begin{figure}[!ht]
    \centering
    \includegraphics[width=0.9\columnwidth]{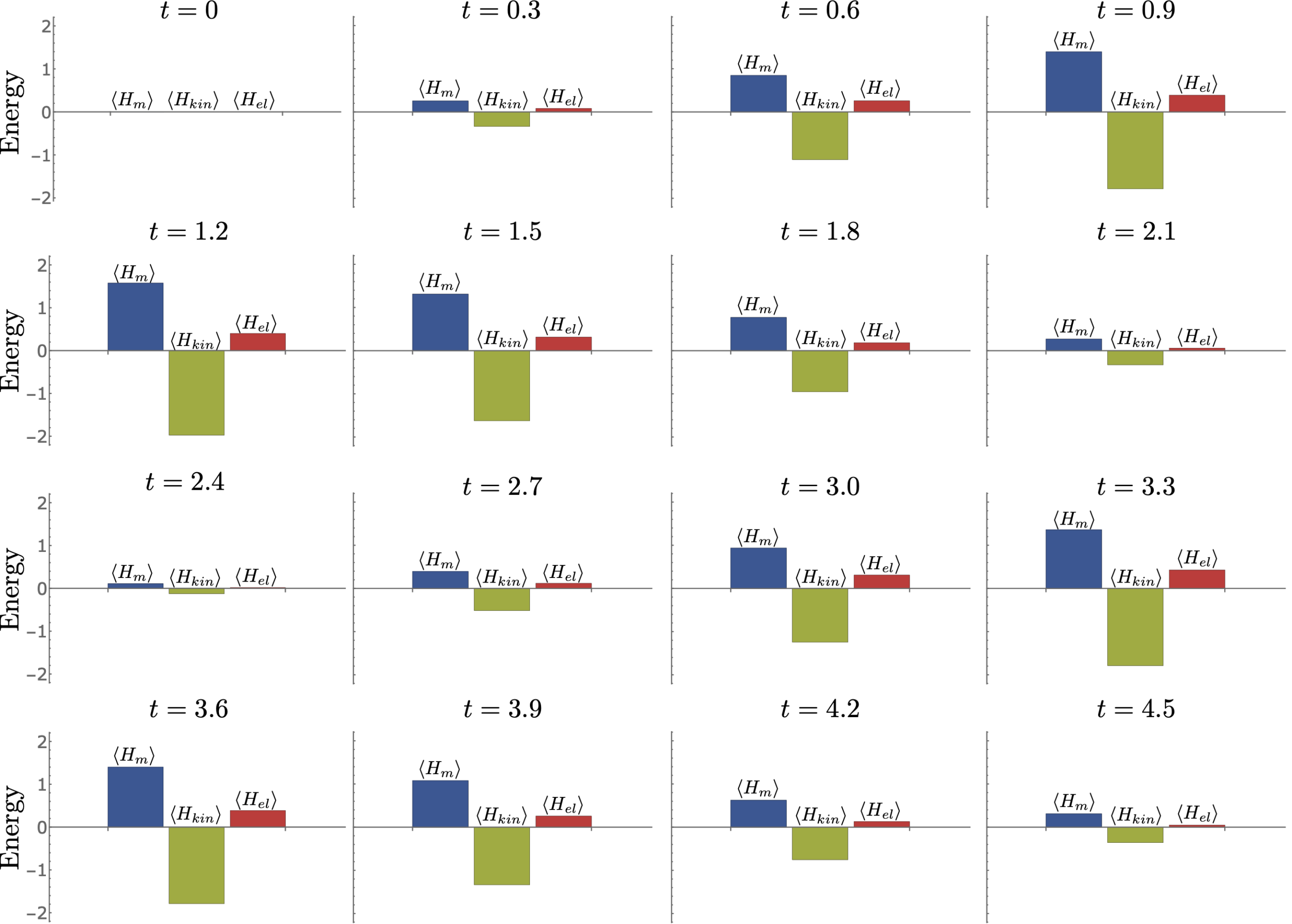}
    \caption{The time evolution of the decomposition of the energy starting from the trivial vacuum starting at $t=0$ for $N_f=2$ QCD with $m=g=L=1$.
    }
    \label{fig:Hanim}
\end{figure}

\noindent 
This appendix shows,
in Fig.~\ref{fig:Hanim}, the time evolution of the decomposition of the expectation value of the Hamiltonian starting with the trivial vacuum at $t=0$ for $N_f=2$ QCD with $m=g=L=1$. 
Notice that the sum of all three terms equals zero for all times as required by energy conservation and that the period of oscillations is the same as the period of the persistence amplitude shown in Fig.~\ref{fig:VacTo}. 
%

\section{Details on One First-Order Trotter Step of \texorpdfstring{\boldmath$N_f=1$}{Nf=1} QCD with \texorpdfstring{\boldmath$L=1$}{L=1}}
\label{app:LOTrott}
\begin{figure}[!ht]
    \centering
    \includegraphics[width=0.6\columnwidth]{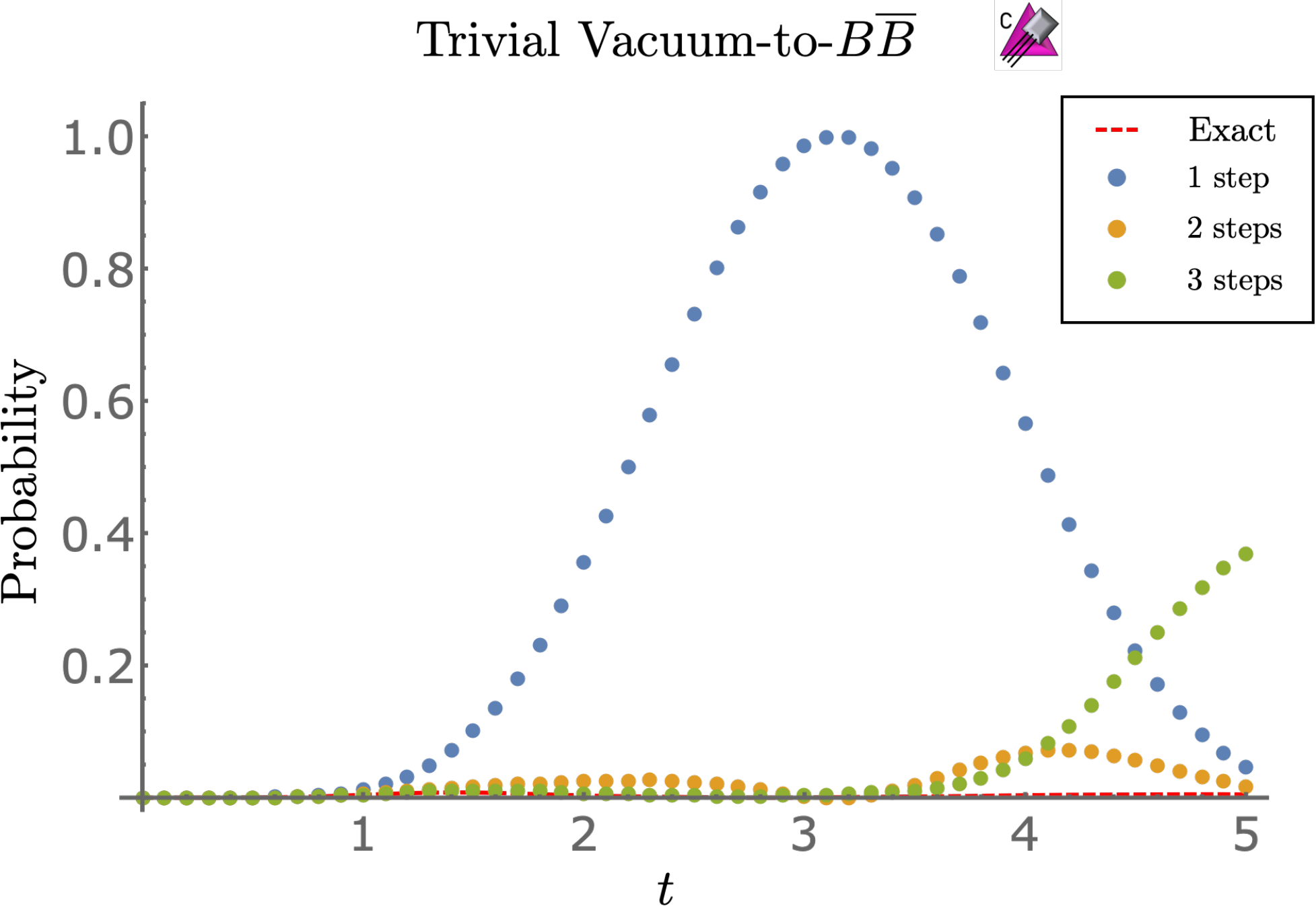}
    \caption{The trivial vacuum-to-$B\overline{B}$ probability for $1+1$D QCD with $m=g=L=1$. Shown are the results obtained from exact exponentiation of the Hamiltonian (dashed red curve) and from the Trotterized implementation with $1$, $2$ and $3$ Trotter steps.}
    \label{fig:VacToBBbar}
\end{figure}

\noindent 
This appendix discusses the theoretical expectations for one step of first-order Trotter time evolution for $N_f=1$ QCD with $L=1$.
The time evolution operator 
is decomposed
into $U_1(t) = U_{kin}(t) U_{el}(t) U_m(t)$ where the subscript ``$1$'' is to denote first-order Trotter. Both the trivial vacuum-to-vacuum and trivial vacuum-to-$q_r\overline{q}_r$ probabilities involve measurements in the computational basis where $U_m(t)$ and $U_{el}(t)$ are diagonal and have no effect. 
Thus, the time-evolution operator is effectively $U_1(t) = U_{kin}(t)$, which is exact (no Trotter errors) over a single spatial site. The trivial vacuum-to-vacuum, trivial vacuum-to-$q_r \overline{q}_r$ and trivial vacuum-to-$B \overline{B}$ probabilities are found to be,
\begin{align}
\lvert \langle \Omega_0 \rvert e^{-i H_{kin} t} \lvert \Omega_0 \rangle\rvert ^2 =& \cos^6(t/2) \ , \nonumber \\ \ \lvert\langle q_r \overline{q}_r \rvert e^{-i H_{kin} t} \lvert \Omega_0 \rangle\rvert ^2 =& \cos^4(t/2)\sin^2(t/2) \ , \nonumber \\ \ \lvert\langle B \overline{B} \rvert e^{-i H_{kin} t} \lvert \Omega_0 \rangle\rvert ^2 =& \sin^6(t/2) \ .
\end{align}
For large periods of the evolution, the wavefunction is dominated by $B\overline{B}$ as shown in Fig.~\ref{fig:VacToBBbar}. Exact time evolution, on the other hand, has a  small probability of $B\overline{B}$
which suggests that detecting  
$B \overline{B}$ could lead to an additional way to mitigate Trotter errors.
It is interesting that the kinetic term alone favors transitioning the trivial vacuum into color singlets on each site. This same behavior holds
for $N_f=2$ where the dominant transition is to $\Delta \Delta \overline{\Delta} \overline{\Delta}$.

\section{Supplementary Data}
\label{app:SuppData}
\noindent 
This appendix contains the tabulated data used to produce the figures in the text. The splitting between the $\pi$- and $\sigma$-meson is given in Table~\ref{tab:MpiMsigSplitData}.
\begin{table}[!ht]
\centering
\renewcommand{\arraystretch}{1.2}
\begin{tabularx}{0.8\textwidth}{||c | Y | Y ||c | Y | Y ||}
\hline
\multicolumn{6}{||c||}{$M_{\pi} - M_{\sigma}$ } \\
 \hline
 $g$ & 
 \makecell{$L=1$} & 
 \makecell{$L=2$} &
 $g$ & 
 \makecell{$L=1$} & 
 \makecell{$L=2$} 
 \\
 \hline\hline 
0 & 0.0 & 0.0 & 
2.1 & 0.1085 & 0.09954\\ \hline
 0.2 & 0.0000374 & 0.00002335 & 
 2.2 & 0.1227 & 0.1133\\ \hline
 0.3 & 0.0001847 & 0.0001285 & 
 2.3 & 0.1374 & 0.1287\\ \hline
 0.4 & 0.0005637 & 0.0004306 & 
 2.4 & 0.1482 & 0.1431\\ \hline
 0.5 & 0.001317 & 0.001066 & 
 2.5 & 0.1468 & 0.1475\\ \hline
 0.6 & 0.002589 & 0.002163 & 
 2.6 & 0.1321 & 0.1352\\ \hline
 0.7 & 0.004508 & 0.00383 & 
 2.7 & 0.1131 & 0.1156\\ \hline
 0.8 & 0.007173 & 0.006158 & 
 2.8 & 0.09662 & 0.09813\\ \hline
 0.9 & 0.01064 & 0.009207 & 
 2.9 & 0.08372 & 0.08456\\ \hline
 1.0 & 0.01493 & 0.01301 & 
 3.0 & 0.07375 & 0.07422\\ \hline
 1.1 & 0.02002 & 0.01755 & 
 3.1 & 0.06592 & 0.06617\\ \hline
 1.2 & 0.02586 & 0.02281 & 
 3.2 & 0.05961 & 0.05975\\ \hline
 1.3 & 0.0324 & 0.02874 & 
 3.3 & 0.05442 & 0.05448\\ \hline
 1.4 & 0.03956 & 0.03529 & 
 3.4 & 0.05005 & 0.05008\\ \hline
 1.5 & 0.04731 & 0.04243 & 
 3.5 & 0.04631 & 0.04632\\ \hline
 1.6 & 0.05562 & 0.05011 & 
 3.6 & 0.04307 & 0.04306\\ \hline
 1.7 & 0.0645 & 0.05837 & 
 3.7 & 0.04022 & 0.04021\\ \hline
 1.8 & 0.07405 & 0.06725 & 
 3.8 & 0.0377 & 0.03769\\ \hline
 1.9 & 0.0844 & 0.0769 & 
 3.9 & 0.03545 & 0.03543\\ \hline
 2.0 & 0.0958 & 0.08755 & 
 4.0 & 0.03342 & 0.0334\\ \hline
\end{tabularx}
\renewcommand{\arraystretch}{1}
\caption{The mass splitting between the $\sigma$- and $\pi$-mesons for $m=1$ and $L=1,2$.
}
\label{tab:MpiMsigSplitData}
\end{table}
The decomposition of the vacuum energy, hadronic masses and deuteron binding energy is given in Table~\ref{tab:SpecBreakdownData}.
\begin{table}[!ht]
\renewcommand{\arraystretch}{1.2}
\begin{tabularx}{1.0\textwidth}{||c | Y | Y | Y | Y | Y ||}
\hline
\multicolumn{6}{||c||}{Decomposition of the vacuum energy, hadronic masses and deuteron binding energy} \\
 \hline
 & 
 \makecell{$E_{\Omega}$} & 
 \makecell{$M_{\sigma}$} & 
 \makecell{$M_{\pi}$} &
 \makecell{$M_{\Delta}$} & 
 \makecell{$B_{\Delta \Delta}$}
 \\
 \hline\hline
 $\langle H_m \rangle$ & 1.0566 & 2.056 & 2.032 & 2.855 & -0.001596 \\
 \hline
 $\langle H_{kin} \rangle$ & -2.975 & 0.1271 & 0.1425 & 0.4182 & 0.002399\\
 \hline
 $\langle H_{el} \rangle$ & 0.3374 & 0.5401 & 0.5609 & -0.03099 & -0.0003429\\
 \hline
\end{tabularx}
\renewcommand{\arraystretch}{1}
\caption{The decomposition of vacuum energy ($E_{\Omega}$), the masses of the lightest hadrons ($M_{\sigma}$, $M_{\pi}$ and $M_{\Delta}$) and the deuteron binding energy ($B_{\Delta \Delta}$) into contributions from the mass, kinetic and chromo-electric field terms in the Hamiltonian for $L=2$ and $m=g=1$.
}
\label{tab:SpecBreakdownData}
\end{table}
The binding energy of the deuteron is given in Table~\ref{tab:deutBE}.
\begin{table}[!ht]
\renewcommand{\arraystretch}{1.2}
\centering
\begin{tabularx}{0.4\textwidth}{||c | Y || c | Y ||}
 \hline
 $g$ & 
 \makecell{$B_{\Delta \Delta}$} &
 $g$ & 
 \makecell{$B_{\Delta \Delta}$}
 \\
 \hline\hline
0 & 0.0  & 1.6 & 0.0005388 \\ \hline
 0.1 & 0.00005099 & 1.7 & 0.000541 \\ \hline
 0.2 & 0.0006768 & 1.8 & 0.0005332\\ \hline
 0.3 & 0.002351 & 1.9 & 0.0005172\\ \hline
 0.4 & 0.003947 & 2.0 & 0.0004948 \\ \hline
 0.5 & 0.003905 & 2.1 & 0.0004677\\ \hline
 0.6 & 0.002716 & 2.2 & 0.0004378\\ \hline
 0.7 & 0.001592 & 2.3 & 0.0004063\\ \hline
 0.8 & 0.0009178 & 2.4 & 0.0003745\\ \hline
 0.9 & 0.0005902 & 2.5 & 0.0003432\\ \hline
 1.0 & 0.0004599 & 2.6 & 0.0003129\\ \hline
 1.1 & 0.000429 & 2.7 & 0.0002842\\ \hline
 1.2 & 0.000443 & 2.8 & 0.0002574\\ \hline
 1.3 & 0.0004727 & 2.9 & 0.0002324\\ \hline
 1.4 & 0.0005029 & 3.0 & 0.0002095\\ \hline
 1.5 & 0.000526 & & \\
 \hline
\end{tabularx}
\renewcommand{\arraystretch}{1}
\caption{The binding energy of the deuteron, $B_{\Delta \Delta}$, for $m=1$ and $L=2$.
}
\label{tab:deutBE}
\end{table}

The linear entropy between quarks and antiquarks in the vacuum, the $\sigma$- and $\pi$-meson and the $\Delta$ are given in Table~\ref{tab:linentData}.
\begin{table}[!ht]
\centering
\renewcommand{\arraystretch}{0.93}

\begin{tabular}{||c | c  | c  | c  | c  | c | c | c | c | c ||}
\hline
\multicolumn{10}{||c||}{The linear entropy between quarks and antiquarks} \\
 \hline
 $g$ & 
 \makecell{$\ket{\Omega}$}
 & 
 \makecell{$\ket{\sigma}$}
 & 
 \makecell{$\ket{\pi_{I_3=1}}$}
 & 
 \makecell{$\ket{\Delta_{I_3=3/2}}$} & $g$ & 
 \makecell{$\ket{\Omega}$}
 & 
 \makecell{$\ket{\sigma}$}
 & 
 \makecell{$\ket{\pi_{I_3=1}}$}
 & 
 \makecell{$\ket{\Delta_{I_3=3/2}}$}
 \\
 \hline\hline
 0.2 & 0.4617 & 0.9124 & 0.7786 & 0.2663 & 3.2 & 0.03807 & 0.7803 & 0.6928 & 0.01914 \\ \hline
 0.4 & 0.4416 & 0.9154 & 0.7786 & 0.2527 & 3.4 & 0.03157 & 0.7716 & 0.6837 & 0.01586 \\ \hline
 0.6 & 0.41 & 0.9194 & 0.7787 & 0.2318 & 3.6 & 0.02633 & 0.7662 & 0.6779 & 0.01321 \\ \hline
 0.8 & 0.3699 & 0.9232 & 0.7791 & 0.206 & 3.8 & 0.02208 & 0.7625 & 0.6739 & 0.01107 \\ \hline
 1.0 & 0.3248 & 0.926 & 0.7798 & 0.1779 & 
 4.0 & 0.01863 & 0.7599 & 0.671 & 0.009334\\ \hline
 1.2 & 0.2784 & 0.9274 & 0.7811 & 0.15 & 4.2 & 0.0158 & 0.7579 & 0.6689 & 0.007913 \\ \hline
 1.4 & 0.2339 & 0.9277 & 0.7834 & 0.1241 & 4.4 & 0.01347 & 0.7565 & 0.6673 & 0.006745 \\ \hline
 1.6 & 0.1935 & 0.9278 & 0.7875 & 0.1014 & 4.6 & 0.01154 & 0.7553 & 0.666 & 0.005779\\ \hline
 1.8 & 0.1584 & 0.9287 & 0.7949 & 0.0821 & 4.8 & 0.00994 & 0.7545 & 0.665 & 0.004975 \\ \hline
 2.0 & 0.1288 & 0.9322 & 0.8097 & 0.06622 & 5.0 & 0.0086 & 0.7538 & 0.6642 & 0.004304 \\ \hline
 2.2 & 0.1045 & 0.9398 & 0.8416 & 0.05337 & 5.2 & 0.007473 & 0.7532 & 0.6636 & 0.003739 \\ \hline
 2.4 & 0.08479 & 0.9402 & 0.8889 & 0.04309 & 5.4 & 0.006522 & 0.7527 & 0.6631 & 0.003263\\ \hline
 2.6 & 0.06896 & 0.8872 & 0.8209 & 0.03491 & 5.6 & 0.005714 & 0.7524 & 0.6627 & 0.002858\\ \hline
 2.8 & 0.05629 & 0.8263 & 0.7414 & 0.02842 & 5.8 & 0.005026 & 0.752 & 0.6623 & 0.002514\\ \hline
 3.0 & 0.04617 & 0.7956 & 0.7086 & 0.02326 &  6.0 & 0.004436 & 0.7518 & 0.662 & 0.002219\\ \hline
 \hline
\end{tabular}
\renewcommand{\arraystretch}{1}
\caption{The linear entropy between quarks and antiquarks in the vacuum, $\ket{\Delta_{I_3=3/2}}$, $\ket{\sigma}$ and $\ket{\pi_{I_3=1}}$ for $m=L=1$.}
\label{tab:linentData}
\end{table}
The quark occupation (total number of quarks plus antiquarks) in the $\sigma$- and $\pi$-mesons is given in Table~\ref{tab:qoccData}.
\begin{table}[!ht]
\centering
\renewcommand{\arraystretch}{1.2}
\begin{tabular}{||c | c  | c ||c | c  | c ||c | c  | c || c | c | c ||}
\hline
\multicolumn{12}{||c||}{The quark occupation} \\
 \hline
 $g$ 
 & 
 \makecell{$\ket{\sigma}$}
 & 
 \makecell{$\ket{\pi_{I_3 = 1}}$}
 &
 $g$ 
 & 
 \makecell{$\ket{\sigma}$}
 & 
 \makecell{$\ket{\pi_{I_3 = 1}}$}
 &
  $g$ 
 & 
 \makecell{$\ket{\sigma}$}
 & 
 \makecell{$\ket{\pi_{I_3 = 1}}$}
 &
 $g$ 
 & 
 \makecell{$\ket{\sigma}$}
 & 
 \makecell{$\ket{\pi_{I_3 = 1}}$}
 \\
 \hline\hline
\ \ 0 \ \ & 2.422 & 2.422 &  1.6 & 2.586 & 2.483 & 
\ 3.1\  & 5.884 & 5.915 & 
 4.6 & 5.994 & 5.995\\ \hline
 0.1 & 2.422 & 2.422 & 1.7 & 2.626 & 2.505 &  
 3.2 & 5.913 & 5.936 & 
 4.7 & 5.995 & 5.996\\ \hline
 0.2 & 2.422 & 2.422 &  1.8 & 2.676 & 2.537 & 
 3.3 & 5.934 & 5.95 & 
 4.8 & 5.995 & 5.996\\ \hline
 0.3 & 2.423 & 2.422 & 1.9 & 2.744 & 2.584 & 
 3.4 & 5.948 & 5.961 & 
 4.9 & 5.996 & 5.997\\ \hline
 0.4 & 2.424 & 2.423 &  2.0 & 2.839 & 2.655 &
 3.5 & 5.959 & 5.968 & 
 5.0 & 5.996 & 5.997\\ \hline
 0.5 & 2.426 & 2.423 & 2.1 & 2.979 & 2.769 & 
 3.6 & 5.967 & 5.974 & 
 5.1 & 5.997 & 5.997\\ \hline
 0.6 & 2.429 & 2.423 & 2.2 & 3.193 & 2.966 & 
 3.7 & 5.973 & 5.979 & 
 5.2 & 5.997 & 5.998\\ \hline
 0.7 & 2.434 & 2.424 &  2.3 & 3.524 & 3.318 & 
 3.8 & 5.978 & 5.982 & 
 5.3 & 5.997 & 5.998\\ \hline
 0.8 & 2.44 & 2.425 &  2.4 & 4.004 & 3.911 & 
 3.9 & 5.981 & 5.985 & 
 5.4 & 5.998 & 5.998\\ \hline
 0.9 & 2.449 & 2.427 &  2.5 & 4.579 & 4.66 &
 4.0 & 5.984 & 5.988 & 
 5.5 & 5.998 & 5.998\\ \hline
 1.0 & 2.46 & 2.43 &  2.6 & 5.091 & 5.249 & 
 4.1 & 5.987 & 5.989 & 
 5.6 & 5.998 & 5.999\\ \hline
 1.1 & 2.473 & 2.434 &  2.7 & 5.439 & 5.577 & 
 4.2 & 5.989 & 5.991 & 
 5.7 & 5.998 & 5.999\\ \hline
 1.2 & 2.489 & 2.439 & 2.8 & 5.646 & 5.743 & 
 4.3 & 5.99 & 5.992 & 
 5.8 & 5.999 & 5.999\\ \hline
 1.3 & 2.507 & 2.445 &  2.9 & 5.766 & 5.832 & 
 4.4 & 5.992 & 5.993 & 
 5.9 & 5.999 & 5.999\\ \hline
 1.4 & 2.529 & 2.454 &  3.0 & 5.838 & 5.883 &
 4.5 & 5.993 & 5.994 & 
 6.0 & 5.999 & 5.999\\ \hline
 1.5 & 2.555 & 2.466 & & & & & & & & & \\ 
 \hline
\end{tabular}
\renewcommand{\arraystretch}{1}
\caption{The expectation value of quark occupation in the $\ket{\sigma}$ and $\ket{\pi_{I_3 = 1}}$ for $m=L=1$.}
\label{tab:qoccData}
\end{table}

The trivial vacuum-to-vacuum probabilities, as obtained by {\tt circ} and {\tt qiskit}, are given in Table~\ref{tab:vactovacData}. 
\begin{table}[!ht]
\centering
\renewcommand{\arraystretch}{1.06}
\begin{tabularx}{1.0\textwidth}{||c | Y  | Y  | Y  | Y | Y ||}
\hline
\multicolumn{6}{||c||}{The trivial vacuum-to-vacuum probabilities} \\
 \hline
 $t$ & 
 \makecell{\text{1 Step}}
 & 
 \makecell{\text{2 Steps}}
 & 
 \makecell{\text{3 Steps}}
 & 
 \makecell{\text{5 Steps}}
 & 
 \makecell{\text{10 Steps}}
 \\
 \hline\hline
\ \ 0 \ \  & 1 & 1 & 1 & 1 & 1 \\ \hline
 0.3 & 0.8733 & 0.878 & 0.8789 & 0.8793 & 0.8795 \\ \hline
 0.6 & 0.5779 & 0.6309 & 0.6401 & 0.6447 & 0.6466
   \\ \hline
 0.9 & 0.2841 & 0.4405 & 0.4678 & 0.4815 & 0.4872
   \\ \hline
 1.2 & 0.0999 & 0.3658 & 0.4188 & 0.4454 & 0.4565
   \\ \hline
 1.5 & 0.0235 & 0.3843 & 0.4775 & 0.5225 & 0.5408
   \\ \hline
 1.8 & 0.0033 & 0.4616 & 0.6218 & 0.6873 & 0.712 \\ \hline
 2.1 & 0.0002 & 0.5546 & 0.8075 & 0.8775 & 0.8992
   \\ \hline
 2.4 & 0.0000 & 0.6055 & 0.9374 & 0.969 & 0.9727 \\ \hline
 2.7 & 0.0000 & 0.5798 & 0.8839 & 0.8513 & 0.8418 \\ \hline
 3.0 & 0.0000 & 0.5026 & 0.6077 & 0.5778 & 0.5925 \\ \hline
 3.3 & 0.0000 & 0.4329 & 0.2629 & 0.3477 & 0.4179 \\ \hline
 3.6 & 0.0000 & 0.4179 & 0.0488 & 0.2967 & 0.424 \\ \hline
 3.9 & 0.0000 & 0.4689 & 0.0039 & 0.4009 & 0.5714 \\ \hline
 4.2 & 0.0003 & 0.5578 & 0.062 & 0.5373 & 0.7496 \\ \hline
 4.5 & 0.0038 & 0.6498 & 0.1339 & 0.6489 & 0.8833
   \\ \hline
 4.8 & 0.0258 & 0.7389 & 0.1422 & 0.7671 & 0.9177
   \\ \hline
 5.0 & 0.0699 & 0.7986 & 0.1125 & 0.8339 & 0.8586 \\
 \hline
\end{tabularx}
\renewcommand{\arraystretch}{1}
\caption{The trivial vacuum-to-vacuum probabilities for $N_f=2$ and $m=g=L=1$. Results are shown for 1, 2, 3, 5 and 10 Trotter steps.}
\label{tab:vactovacData}
\end{table}
The trivial vacuum-to-$d_r \overline{d}_r$ probabilities, as obtained by {\tt circ} and {\tt qiskit}, are given in Table~\ref{tab:vactodrdrbData}. 
\begin{table}[!ht]
\centering
\renewcommand{\arraystretch}{1.06}
\begin{tabular}{||c | c | c  | c | c | c ||}
\hline
\multicolumn{6}{||c||}{The trivial vacuum-to-$d_r \overline{d}_r$ probabilities} \\
 \hline
 $t$ & 
 \makecell{\text{1 Step}}
 & 
 \makecell{\text{2 Steps}}
 & 
 \makecell{\text{3 Steps}}
 & 
 \makecell{\text{5 Steps}}
 & 
 \makecell{\text{10 Steps}}
 \\
 \hline\hline
\ \ 0\ \  & 0 & 0 & 0 & 0 & 0 \\ \hline
 0.3 & 0.0199 & 0.0192 & 0.0191 & 0.019 & 0.019 \\ \hline
 0.6 & 0.0553 & 0.05 & 0.049 & 0.0486 & 0.0483 \\ \hline
 0.9 & 0.0663 & 0.0625 & 0.061 & 0.0601 & 0.0597 \\ \hline
 1.2 & 0.0468 & 0.0612 & 0.0597 & 0.0586 & 0.058 \\ \hline
 1.5 & 0.0204 & 0.0569 & 0.0537 & 0.0512 & 0.0501
   \\ \hline
 1.8 & 0.0053 & 0.0505 & 0.0414 & 0.0361 & 0.0339
   \\ \hline
 2.1 & 0.0007 & 0.039 & 0.0208 & 0.0137 & 0.0113 \\ \hline
 2.4 & 0.0000 & 0.0239 & 0.0024 & 0.0005 & 0.0006 \\ \hline
 2.7 & 0.0000 & 0.0099 & 0.0098 & 0.0183 & 0.0206 \\ \hline
 3.0 & 0.0000 & 0.0011 & 0.0482 & 0.0574 & 0.0568 \\ \hline
 3.3 & 0.0000 & 0.0012 & 0.0741 & 0.0818 & 0.0766 \\ \hline
 3.6 & 0.0000 & 0.0074 & 0.0594 & 0.0777 & 0.0687 \\ \hline
 3.9 & 0.0000 & 0.0117 & 0.0366 & 0.0546 & 0.0439 \\ \hline
 4.2 & 0.0008 & 0.0121 & 0.024 & 0.0351 & 0.0221 \\ \hline
 4.5 & 0.0058 & 0.0126 & 0.0166 & 0.0262 & 0.008 \\ \hline
 4.8 & 0.0217 & 0.0131 & 0.0185 & 0.0142 & 0.003 \\ \hline
 5.0 & 0.039 & 0.0119 & 0.0228 & 0.0062 & 0.0099 \\
 \hline
\end{tabular}
\renewcommand{\arraystretch}{1}
\caption{The trivial vacuum-to-$d_r \overline{d}_r$ probabilities for $N_f=2$ and $m=g=L=1$. Results are shown for 1, 2, 3, 5 and 10 Trotter steps.}
\label{tab:vactodrdrbData}
\end{table}
The required $N$\textsubscript{Trott} for a $\epsilon_{{\rm Trott}} < 0.1$ in the trivial vacuum-to-vacuum probability is given in Table~\ref{tab:vacNTrottData}.
\begin{table}[!ht]
\renewcommand{\arraystretch}{1.06}
\centering
\begin{tabular}{||c | c | c | c | c | c | c | c | c  | c ||}
\hline
\multicolumn{10}{||c||}{Required number of Trotter steps, $N$\textsubscript{Trott}} \\
 \hline
 $t$ & $N$\textsubscript{Trott} &
 $t$ & $N$\textsubscript{Trott} &
 $t$ & $N$\textsubscript{Trott} &
 $t$ & $N$\textsubscript{Trott} &
 $t$ & $N$\textsubscript{Trott}
 \\
 \hline\hline
0 & 1 & 40 & 197 & 80 & 597 & 120 & 1000 & 160 & 1598 \\ \hline 
2 & 4 & 42 & 197 & 82 & 597 & 122 & 1075 & 162 & 1719 \\ \hline 
4 & 10 & 44 & 248 & 84 & 597 & 124 & 1075 & 164 & 1719 \\ \hline 
6 & 19 & 46 & 264 & 86 & 631 & 126 & 1113 & 166 & 1747 \\ \hline 
8 & 19 & 48 & 264 & 88 & 631 & 128 & 1224 & 168 & 1798 \\ \hline 
10 & 27 & 50 & 271 & 90 & 650 & 130 & 1224 & 170 & 1798 \\ \hline 
12 & 34 & 52 & 302 & 92 & 705 & 132 & 1224 & 172 & 1798 \\ \hline 
14 & 42 & 54 & 302 & 94 & 707 & 134 & 1224 & 174 & 1967 \\ \hline 
16 & 51 & 56 & 335 & 96 & 707 & 136 & 1224 & 176 & 1967 \\ \hline 
18 & 67 & 58 & 363 & 98 & 735 & 138 & 1279 & 178 & 1967 \\ \hline 
20 & 85 & 60 & 363 & 100 & 740 &140 & 1356 & 180 & 2023 \\ \hline 
22 & 85 & 62 & 370 & 102 & 794 & 142 & 1356 & 182 & 2023 \\ \hline 
24 & 95 & 64 & 397 & 104 & 868 & 144 & 1356 & 184 & 2023 \\ \hline 
26 & 103 & 66 & 453 & 106 & 868 & 146 & 1417 & 186 & 2137 \\ \hline 
28 & 114 & 68 & 453 & 108 & 868 &148 & 1417 & 188 & 2203 \\ \hline 
30 & 130 & 70 & 475 & 110 & 896 & 150 & 1417 & 190 & 2203 \\ \hline 
32 & 159 & 72 & 475 & 112 & 896 & 152 & 1598 & 192 & 2203 \\ \hline 
34 & 167 & 74 & 475 & 114 & 975 & 154 & 1598 & 194 & 2203 \\ \hline 
36 & 167 & 76 & 521 & 116 & 1000 & 156 & 1598 & 196 & 2203 \\ \hline 
38 & 182 & 78 & 521 & 118 & 1000 & 158 & 1598 & 198 & 2273 \\ \hline

\end{tabular}
\renewcommand{\arraystretch}{1}
\caption{The required $N$\textsubscript{Trott} for a $\epsilon_{{\rm Trott}} < 0.1$ in the trivial vacuum-to-vacuum probability.}
\label{tab:vacNTrottData}
\end{table}
The required $N$\textsubscript{Trott} for a $\epsilon_{{\rm Trott}} < 0.1$ in the trivial vacuum-to-$d_r \overline{d}_r$ probability is given in Table~\ref{tab:drdrbNTrottData}.
\begin{table}[!ht]
\centering
\renewcommand{\arraystretch}{1.06}
\begin{tabular}{||c | c | c | c | c | c | c | c | c | c ||}
\hline
\multicolumn{10}{||c||}{Required number of Trotter steps, $N$\textsubscript{Trott}} \\
 \hline
 $t$ & $N$\textsubscript{Trott} &
 $t$ & $N$\textsubscript{Trott} &
 $t$ & $N$\textsubscript{Trott} &
 $t$ & $N$\textsubscript{Trott} &
 $t$ & $N$\textsubscript{Trott} 
 \\
 \hline\hline
 0 & 1 & 20 & 148 & 40 & 709 & 60 & 1285 & 80 & 1739 \\ \hline 
 1 & 2 & 21 & 233 & 41 & 709 & 61 & 1285 & 81 & 1739 \\ \hline 
 2 & 7 & 22 & 233 & 42 & 709 & 62 & 1285 & 82 & 1739 \\ \hline 
 3 & 10 & 23 & 243 & 43 & 709 & 63 & 1285 & 83 & 2064 \\ \hline 
 4 & 12 & 24 & 243 & 44 & 709 & 64 & 1285 & 84 & 2064 \\ \hline 
 5 & 24 & 25 & 243 & 45 & 709 & 65 & 1285 & 85 & 2064\\ \hline 
 6 & 24 & 26 & 337 & 46 & 709 & 66 & 1285 & 86 & 2064 \\ \hline 
 7 & 32 & 27 & 337 & 47 & 709 & 67 & 1285 & 87 & 2064\\ \hline 
 8 & 32 & 28 & 337 & 48 & 709 & 68 & 1285 & 88 & 2064 \\ \hline 
 9 & 72 & 29 & 337 & 49 & 945 & 69 & 1285 & 89 & 2064\\ \hline 
 10 & 72 & 30 & 337 & 50 & 945 & 70 & 1285 & 90 & 2064 \\ \hline 
 11 & 72 & 31 & 337 & 51 & 945 & 71 & 1285 & 91 & 2064\\ \hline 
 12 & 98 & 32 & 384 & 52 & 945 & 72 & 1706 & 92 & 2064\\ \hline 
 13 & 98 & 33 & 384 & 53 & 945 & 73 & 1706 & 93 & 2064 \\ \hline 
 14 & 148 & 34 & 384 & 54 & 945 & 74 & 1706 & 94 & 2064\\ \hline 
 15 & 148 & 35 & 709 & 55 & 945 & 75 & 1706 & 95 & 2590\\ \hline 
 16 & 148 & 36 & 709 & 56 & 945 & 76 & 1739 & 96 & 2590 \\ \hline 
 17 & 148 & 37 & 709 & 57 & 945 & 77 & 1739 & 97 & 2590 \\ \hline 
 18 & 148 & 38 & 709 & 58 & 1168 & 78 & 1739 & 98 & 2590 \\ \hline 
 19 & 148 & 39 & 709 & 59 & 1168 & 79 & 1739 & 99 & 2780 \\ \hline
 \hline
\end{tabular}
\renewcommand{\arraystretch}{1}
\caption{The required $N$\textsubscript{Trott} for a $\epsilon_{{\rm Trott}} < 0.1$ in the trivial vacuum-to-$d_r \overline{d}_r$ probability.}
\label{tab:drdrbNTrottData}
\end{table}
The decomposition of the energy, starting from trivial vacuum at $t=0$, is given in Table~\ref{tab:vacAnimData}.
\begin{table}[!ht]
\centering
\renewcommand{\arraystretch}{1.2}
\begin{tabularx}{.8\textwidth}{||c | Y  | Y  | Y ||}
\hline
\multicolumn{4}{||c||}{Decomposition of the energy starting from the trivial vacuum} \\
 \hline
 $t$ & $\langle H_m \rangle$ 
 & $\langle H_{kin} \rangle$ 
 & $\langle H_{el} \rangle$ 
 \\
 \hline\hline
  \ \ 0 \ \  & 0 & 0 & 0 \\ \hline
 0.3 & 0.254 & -0.3369 & 0.08281 \\ \hline
 0.6 & 0.8455 & -1.104 & 0.2584 \\ \hline
 0.9 & 1.393 & -1.781 & 0.3879 \\ \hline
 1.2 & 1.575 & -1.974 & 0.3997 \\ \hline
 1.5 & 1.31 & -1.625 & 0.3151 \\ \hline
 1.8 & 0.769 & -0.9532 & 0.1842 \\ \hline
 2.1 & 0.2684 & -0.3282 & 0.05983 \\ \hline
 2.4 & 0.1081 & -0.1256 & 0.01747 \\ \hline
 2.7 & 0.3939 & -0.5129 & 0.119 \\ \hline
 3.0 & 0.9349 & -1.245 & 0.31 \\ \hline
 3.3 & 1.359 & -1.788 & 0.4285 \\ \hline
 3.6 & 1.401 & -1.788 & 0.3873 \\ \hline
 3.9 & 1.079 & -1.335 & 0.2565 \\ \hline
 4.2 & 0.6237 & -0.7542 & 0.1305 \\ \hline
 4.5 & 0.3094 & -0.3586 & 0.0492 \\
 \hline
\end{tabularx}
\renewcommand{\arraystretch}{1}
\caption{Decomposition of the energy starting from the trivial vacuum at $t=0$.}
\label{tab:vacAnimData}
\end{table}
The trivial vacuum to $B \overline{B}$ probabilities are given in Table~\ref{tab:BBarprobData}.
\begin{table}[!ht]
\centering
\renewcommand{\arraystretch}{1.0}
\begin{tabular}{||c | c  | c  | c | c ||}
\hline
\multicolumn{5}{||c||}{Trivial vacuum to $B \overline{B}$ probability} \\
 \hline
 $t$ & 1 Step
 & 2 Steps
 & 3 Steps
 & Exact
 \\
 \hline\hline
\ \ 0 \ \  & 0 & 0 & 0 & 0 \\ \hline
 0.3 & 0.0000 & 0.0000 & 0.0000 & 0.0000 \\ \hline
 0.6 & 0.0007 & 0.0005 & 0.0005 & 0.0005 \\ \hline
 0.9 & 0.0068 & 0.0036 & 0.0032 & 0.0029 \\ \hline
 1.2 & 0.0324 & 0.0104 & 0.0081 & 0.0067 \\ \hline
 1.5 & 0.1003 & 0.0175 & 0.0109 & 0.0078 \\ \hline
 1.8 & 0.231 & 0.0218 & 0.0092 & 0.0052 \\ \hline
 2.1 & 0.426 & 0.0252 & 0.0059 & 0.0025 \\ \hline
 2.4 & 0.6555 & 0.0257 & 0.0036 & 0.0011 \\ \hline
 2.7 & 0.8629 & 0.0166 & 0.0024 & 0.0001 \\ \hline
 3.0 & 0.9851 & 0.0025 & 0.0036 & 0.0002 \\ \hline
 3.3 & 0.9813 & 0.0037 & 0.007 & 0.0011 \\ \hline
 3.6 & 0.853 & 0.0295 & 0.0148 & 0.0019 \\ \hline
 3.9 & 0.6427 & 0.061 & 0.042 & 0.0032 \\ \hline
 4.2 & 0.4137 & 0.0719 & 0.1085 & 0.0043 \\ \hline
 4.5 & 0.2219 & 0.057 & 0.2117 & 0.0047 \\ \hline
 4.8 & 0.095 & 0.0317 & 0.3178 & 0.0051 \\ \hline
 5.0 & 0.0459 & 0.0174 & 0.3695 & 0.0048 \\
 \hline
\end{tabular}
\renewcommand{\arraystretch}{1}
\caption{Trivial vacuum to $B \overline{B}$ probability for $m=g=L=N_f=1$.}
\label{tab:BBarprobData}
\end{table}
\end{subappendices}
%






\chapter{Preparations for Quantum Simulations of Quantum Chromodynamics in \texorpdfstring{\boldmath$1+1$}{1+1} Dimensions:
(II) Single-Baryon 
\texorpdfstring{$\beta$}{Beta}-Decay in Real Time}
\label{chap:1p1dSM}

{\it This chapter is associated with Ref. \cite{physrevd.107.054513}:}

{\it ``Preparations for quantum simulations of quantum chromodynamics in $1+1$ dimensions. II. Single-baryon $\ensuremath{\beta}$-decay in real time" by Roland C. Farrell, Ivan A. Chernyshev, Sarah J. M. Powell, Nikita A. Zemlevskiy, Marc Illa, and Martin J. Savage}



\section{Introduction}
\noindent
As mentioned in Sec. \ref{sec:intro_tnm}, simulation of out-of-equilibrium real-time dynamics is one of the main areas within Standard Model physics where quantum computation holds promise.
Perhaps the simplest non-trivial class of processes to begin attempting to simulate the real-time dynamics of is the
$\beta$-decay of low-lying hadrons and nuclei.
Single $\beta$-decay rates of nuclei have played a central role in defining the 
Standard Model (SM) of strong and electroweak processes~\cite{Glashow:1961tr,Higgs:1964pj,Weinberg:1967tq,Salam:1968rm}. They 
initially provided evidence that the weak (charged-current) 
quark eigenstates differ from the strong eigenstates, and, more recently, are
providing stringent tests of the unitarity of the Cabibbo-Kobayashi-Maskawa (CKM) matrix~\cite{Cabibbo:1963yz,Kobayashi:1973fv}.
For recent reviews of $\beta$-decay, see, e.g.,  Refs.~\cite{Gonz_lez_Alonso_2019,Hassan:2020hrj,Algora:2021,PhysRevC.102.045501}.
The four-Fermi operators responsible for $\beta$-decay~\cite{Feynman:1958ty} in the SM
emerge from operator production expansions (OPEs) 
of the non-local operators coming from the exchange of a charged-gauge boson ($W^-$) between quarks and leptons.
Of relevance to this work is the four-Fermi operator, which gives rise to the flavor changing quark process $d\rightarrow u e^-\overline{\nu}$.
In the absence of higher-order electroweak processes, including electromagnetism,
matrix elements of these operators factorize between the hadronic and leptonic sectors. This leaves, for example, a non-perturbative evaluation of $n\rightarrow p e^-\overline{\nu}$ for neutron decay, which is constrained significantly by the approximate global flavor symmetries of QCD.
Only recently have the observed systematics of $\beta$-decay rates of nuclei been understood without the need for phenomenological re-scalings of the axial coupling constant, 
$g_A$. 
As has long been anticipated, the correct decay rates are recovered when two-nucleon and higher-body interactions are included within the
effective field theories (EFTs) (or meson-exchange currents)~\cite{Baroni_2016,Krebs:2016rqz,Gysbers:2019uyb,Baroni_2021}. 
This was preceded by successes of EFTs in describing electroweak processes of few-nucleon systems through the inclusion of higher-body electroweak operators (not constrained by strong interactions alone), 
e.g., Refs.~\cite{Chen:1999tn,Butler:1999sv,Butler:2002cw,Baroni:2016xll,Li:2017udr,Baroni:2018fdn}.
The EFT framework describing nuclear $\beta$-decays involves contributions from ``potential-pion" and ``radiation-pion" exchanges~\cite{Kaplan:1998tg,Kaplan:1998we} 
(an artifact of a system of relativistic and non-relativistic particles~\cite{Grinstein:1997gv,Luke:1997ys})
and real-time simulations of these processes are expected to be able to isolate these distinct contributions.
Recently, 
the first Euclidean-space lattice QCD calculations of Gamow-Teller matrix elements in light nuclei 
(at unphysical light quark masses and without fully-quantified uncertainties) 
have been performed~\cite{Parreno:2021ovq}, 
finding results that are consistent with nature.

While $\beta$-decay is a well-studied and foundational area of sub-atomic physics, 
the double-$\beta$-decay of nuclei continues to present a theoretical challenge in the 
the search for physics beyond the SM.
For a recent review of the ``status and prospects" of $\beta\beta$-decay, see Ref.~\cite{Dolinski_2019}.
Although $2\nu\beta\beta$-decay is allowed in the SM, and is 
a second order $\beta$-decay process, 
$0\nu\beta\beta$-decay requires the violation of lepton number.
Strong interactions clearly play an essential role 
in the experimental detection of the $\beta\beta$-decay of nuclei, but
such contributions are non-perturbative and complex, and, for example, 
the EFT descriptions involve contributions from two- and higher-body correlated operators~\cite{Savage:1998yh,Shanahan:2017bgi,physrevd.96.054505,Cirigliano:2018hja,Cirigliano:2019vdj}.
The ability to study the real-time dynamics of such decay process in nuclei would likely 
provide valuable insight into the underlying strong-interaction mechanisms, and potentially offer first principles constraints beyond those from Euclidean-space lattice QCD.\footnote{
For discussions of the potential of lattice QCD to impact $\beta\beta$-decay, see, e.g., Refs.~\cite{Shanahan:2017bgi,physrevd.96.054505,Monge-Camacho:2019nby,Davoudi:2021noh,Cirigliano:2022oqy,USQCD:2022mmc,Detmold:2022jwu,Cirigliano:2022rmf}.}

This chapter adapts the work in Chapter \ref{chap:1p1dQCD} 
to include flavor-changing 
weak interactions via a four-Fermi operator that generates the $\beta$-decay 
of hadrons and nuclei. The terms in the lattice Hamiltonian that generate a Majorana mass for the neutrinos are also given, although not included in the simulations.
Applying the Jordan-Wigner (JW) mapping, it is found that a single generation of the SM (quarks and leptons) maps onto $16$ qubits per spatial lattice site. 
Using Quantinuum's {\tt H1-1} 20-qubit trapped ion quantum computer, the initial state of a baryon is both prepared and evolved with one and two Trotter steps on a single lattice site. 
Despite only employing a minimal amount of error mitigation, results at the 
$\sim 5\%$-level are obtained, consistent with the expectations.
Finally, we briefly comment on the potential of such hierarchical dynamics for error-correction purposes in quantum simulations.

\section{The \texorpdfstring{$\beta$}{Beta}-Decay Hamiltonian for Quantum Simulations in 1+1 Dimensions}
\noindent
In nature, the $\beta$-decays of neutrons and nuclei involve energy and momentum transfers related to the energy scales of nuclear forces and of isospin breaking. 
As these are much below the electroweak scale,
$\beta$-decay rates are well reproduced by matrix elements of 
four-Fermi effective interactions with $V-A$ structure~\cite{Feynman:1958ty,Sudarshan:1958vf}, of the form
\begin{equation}
    {\cal H}_\beta  = 
    \frac{G_F}{\sqrt{2}} \ V_{ud} \ 
    \overline{\psi}_u\gamma^\mu (1-\gamma_5)\psi_d\ 
    \overline{\psi}_e\gamma_\mu (1-\gamma_5)\psi_{\nu_e} 
    \ +\ {\rm h.c.}
    \ ,
    \label{eq:HbetaC}
\end{equation}
where $V_{ud}$ is the element of the CKM matrix for $d\rightarrow u$ transitions,
and $G_F$ is Fermi's coupling constant that is 
measured to be $G_F=1.1663787 (6) \times 10^{-5}~{\rm GeV}^{-2}$~\cite{Tiesinga:2021myr}.
This is the leading order (LO) SM result, obtained by matching amplitudes at 
tree-level, 
where $G_F/\sqrt{2} = g_2^2/(8 M_W^2)$ 
with $M_W$ the mass of the $W^\pm$ gauge boson
and $g_2$ the SU(2)$_L$ coupling constant.
Toward simulating the SM in $3+1$D, we consider
$1+1$D QCD containing $u$-quarks, $d$-quarks, electrons and electron neutrinos. 
For simplicity,
we model $\beta$-decay through a vector-like four-Fermi operator,
\begin{equation}
    {\cal H}_\beta^{1+1} =
    \frac{G}{\sqrt{2}} \
    \overline{\psi}_u\gamma^\mu \psi_d\ 
    \overline{\psi}_e\gamma_\mu \mathcal{C} \psi_{\nu}  
        \ +\ {\rm h.c.}
    \ ,
    \label{eq:HbetaC1}
\end{equation}
where $\mathcal{C} = \gamma_1$ is the charge-conjugation operator 
whose purpose will become clear. 
Appendices~\ref{app:betaSM} and~\ref{app:beta1p1} provide details on
calculating the single-baryon $\beta$-decay rates in the
infinite volume and continuum limits in the SM and in the $1+1$D model considered here.

The strong and weak interactions can be mapped
onto the finite-dimensional Hilbert space provided by a quantum computer
by using the same Kogut-Susskind (KS) Hamiltonian formulation of
lattice gauge theory used in Chapter \ref{chap:1p1dQCD}. 
For the $\beta$-decay of baryons, the strong and the weak KS Hamiltonian (in axial gauge) 
has the form~\cite{banuls:2017ena,atas:2021ext,atas:2022dqm}
\begin{equation}
    H = H_{{\rm quarks}} + H_{{\rm leptons}} +  H_{el} + H_{\beta}
    \ ,
\end{equation}
where
\begin{align}
    H_{\rm{quarks}} 
    =&\ 
    H_m + H_{kin}
    \nonumber\\
    H_{\rm{leptons}} 
    =&\ 
        \sum_{f=e,\nu}\left[
    \frac{1}{2 a} \sum_{n=0}^{2L-2} \left ( \chi_n^{(f)\dagger} \chi_{n+1}^{(f)}
        \ +\ {\rm h.c.} \right ) 
    \: + \: 
    m_f \sum_{n=0}^{2L-1} (-1)^{n} \chi_n^{(f)\dagger} \chi_n^{(f)} 
\right]  \ ,
    \nonumber\\
    H_{{\rm \beta}}
    =&\
    \frac{G}{a \sqrt{2}} 
    \sum_{l=0}^{L-1} \bigg [
    \left (\phi_{2l}^{(u)\dagger} \phi_{2l}^{(d)} + \phi_{2l+1}^{(u)\dagger} \phi_{2l+1}^{(d)} \right ) \left (\chi_{2l}^{(e)\dagger} \chi_{2l+1}^{(\nu)} - \chi_{2l+1}^{(e)\dagger} \chi_{2l}^{(\nu)}\right ) 
    \nonumber\\
    &+
    \left ( \phi_{2l}^{(u)\dagger} \phi_{2l+1}^{(d)} + \phi_{2l+1}^{(u)\dagger} \phi_{2l}^{(d)} \right )
    \left (\chi_{2l}^{(e)\dagger} \chi_{2l}^{(\nu)} - \chi_{2l+1}^{(e)\dagger} \chi_{2l+1}^{(\nu)}\right )+
    {\rm h.c.} \bigg ] \ 
    \label{eq:KSHam_SM}
\end{align}
and $H_{m}$, $H_{kin}$, and $H_{el}$ are defined in Sec. \ref{sec:1p1dQCD_mapping1p1DQCDtoqubits}. The masses of the $u$-, $d$-quarks, electron and neutrino (Dirac) are $m_{u,d,e,\nu}$,
and the strong and weak coupling constants are $g$ and $G$. 
$\phi^{(u,d)}_n$ are the $u$- and $d$-quark field operators (which both transform in the fundamental representation of $SU(3)$, and hence the sum over color indices has been suppressed). The electron and  neutrino field operators are
$\chi^{(e,\nu)}_n$.
We emphasize that the absence of gluon fields is due to the choice of axial gauge, whereas the lack of weak gauge fields is due to the
consideration of a low energy effective theory in which the heavy weak gauge bosons have been integrated out. 
This results in, for example, the absence of parallel transporters in the fermion kinetic terms.
\begin{figure}[ht]
    \centering
    \includegraphics[width=15cm]{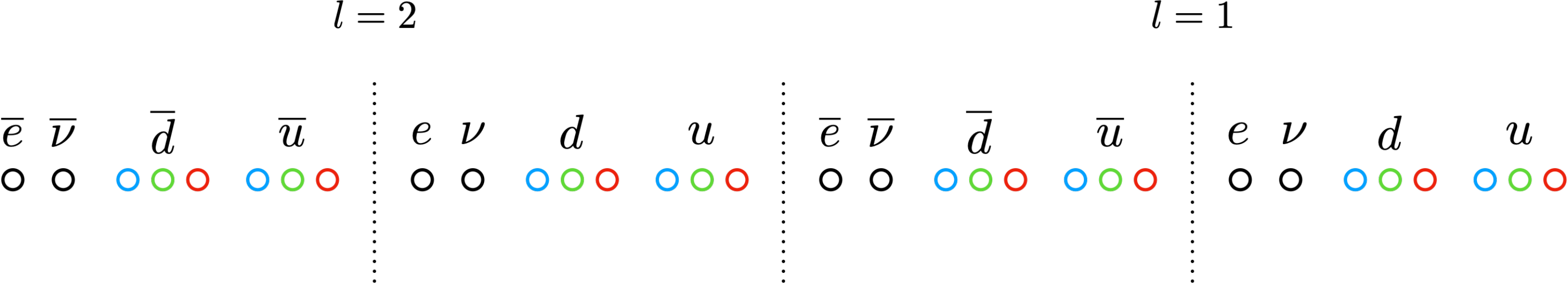}
    \caption{
    The qubit layout of a $L=2$ lattice,
    where fermions and anti-fermions are grouped together (which will be preferred if electromagnetism is included).  This layout extends straightforwardly to $L>2$.
}
    \label{fig:L1layout}
\end{figure}

The JW mapping of the Hamiltonian in Eq.~(\ref{eq:KSHam_SM}) to qubits, 
arranged as shown in Fig.~\ref{fig:L1layout},
is given by 
\begin{align}
    H_{\rm{quarks}} 
     \rightarrow &\
    \frac{1}{2} \sum_{l=0}^{L-1} \sum_{f=u,d}\sum_{c=0}^{2}  m_f\left ( Z_{l,f,c} - Z_{l,\overline{f},c} + 2\right )  \nonumber \\
    & -\frac{1}{2} \sum_{l=0}^{L-1} \sum_{f=u,d} \sum_{c=0}^{2}\left [ \sigma^+_{l,f,c} Z^7 \sigma^-_{l,\overline{f},c}  + (1-\delta_{l,L-1})  \sigma^+_{l,\overline{f},c} Z^7 \sigma^-_{l+1,f,c} + {\rm h.c.} \right ]\ , \nonumber \\[4pt]
    H_{\rm{leptons}} 
    \rightarrow  &\ \frac{1}{2} \sum_{l=0}^{L-1} \sum_{f=e,\nu}  m_f\left ( Z_{l,f} - Z_{l,\overline{f}} + 2\right )  \nonumber \\[4pt] & \: - \: \frac{1}{2} \sum_{l=0}^{L-1} \sum_{f=e,\nu} \left [ \sigma^+_{l,f} Z^7 \sigma^-_{l,\overline{f}}  + (1-\delta_{l,L-1})  \sigma^+_{l,\overline{f}} Z^7 \sigma^-_{l+1,f} + {\rm h.c.} \right ] \ ,
    \nonumber\\[4pt]
    H_{{\rm \beta}}
    \rightarrow  &\
    \frac{G}{\sqrt{2}}\sum_{l = 0}^{L-1}\sum_{c=0}^2\bigg ( \sigma^-_{l,\nub} Z^6 \sigma^+_{l,e} \sigma^-_{l,d,c} Z^2 \sigma^+_{l,u,c} \: - \: \sigma^+_{l,\eb} Z^8 \sigma_{l,\nu}^- \sigma_{l,d,c}^- Z^2 \sigma^+_{l,u,c} \nonumber \\
    \: - \: \sigma^-_{l,\nub} & Z^{2-c} \sigma^-_{l,\db,c} \sigma^+_{l,\ub,c} Z^c \sigma^+_{l,e} 
     + \: \sigma^+_{l,\eb}Z^{3-c}\sigma^-_{l,\db,c} \sigma^+_{l,\ub,c} Z^{1+c} \sigma^-_{l,\nu} \: - \: \sigma^-_{l,\db,c} Z^{3+c} \sigma^+_{l,e} \sigma^-_{l,\nu} Z^{5-c} \sigma^+_{l,u,c} \nonumber \\ \: - \: \sigma^+_{l,\eb} & \sigma^-_{l,\nub} \sigma^-_{l,\db,c}Z^{10}\sigma^+_{l,u,c}  
     - \: \sigma^+_{l,\ub,c} Z^c \sigma^+_{l,e} \sigma^-_{l,\nu} Z^{2-c} \sigma^-_{l,d,c} \: - \: \sigma^+_{l,\eb} \sigma^-_{l,\nub} \sigma^+_{l,\ub,c} Z^4 \sigma^-_{l,d,c} \: + \: {\rm h.c.}\bigg )
     \ ,
\label{eq:KSHamJWmap}
\end{align}
%
%
%
and the JW transformation used for $H_{el}$ is the same one used for $H_{el}$ in Sec. \ref{sec:1p1dQCD_mapping1p1DQCDtoqubits}. The index $l$ labels the spatial lattice site, $f \ (\overline{f})$ labels the (anti)fermion flavor and $c=0,1,2$ corresponds to red, green and blue colors.  
In the staggered mapping, there are gauge-field links every half of a spatial site and, as a result, the color charges are labelled by a half site index, $n$.
The spin raising and lowering operators are $\sigma^{\pm} = \frac{1}{2}(\sigma^x \pm i \sigma^y)$, $Z=\sigma^z$ and
unlabelled $Z$s act on the sites between the $\sigma^{\pm}$, e.g., $\sigma^-_{l,d,r} Z^2 \sigma^+_{l,u,r} = \sigma^-_{l,d,r} Z_{l,u,b} Z_{l,u,g} \sigma^+_{l,u,r}$. 
Constants have been added to the mass terms to ensure that all basis
states contribute positive mass.

\subsection{Efficiently Mapping the \texorpdfstring{$L=1$}{L=1} Hamiltonian to Qubits}
\begin{figure}[!t]
    \centering
    \includegraphics[width=15cm]{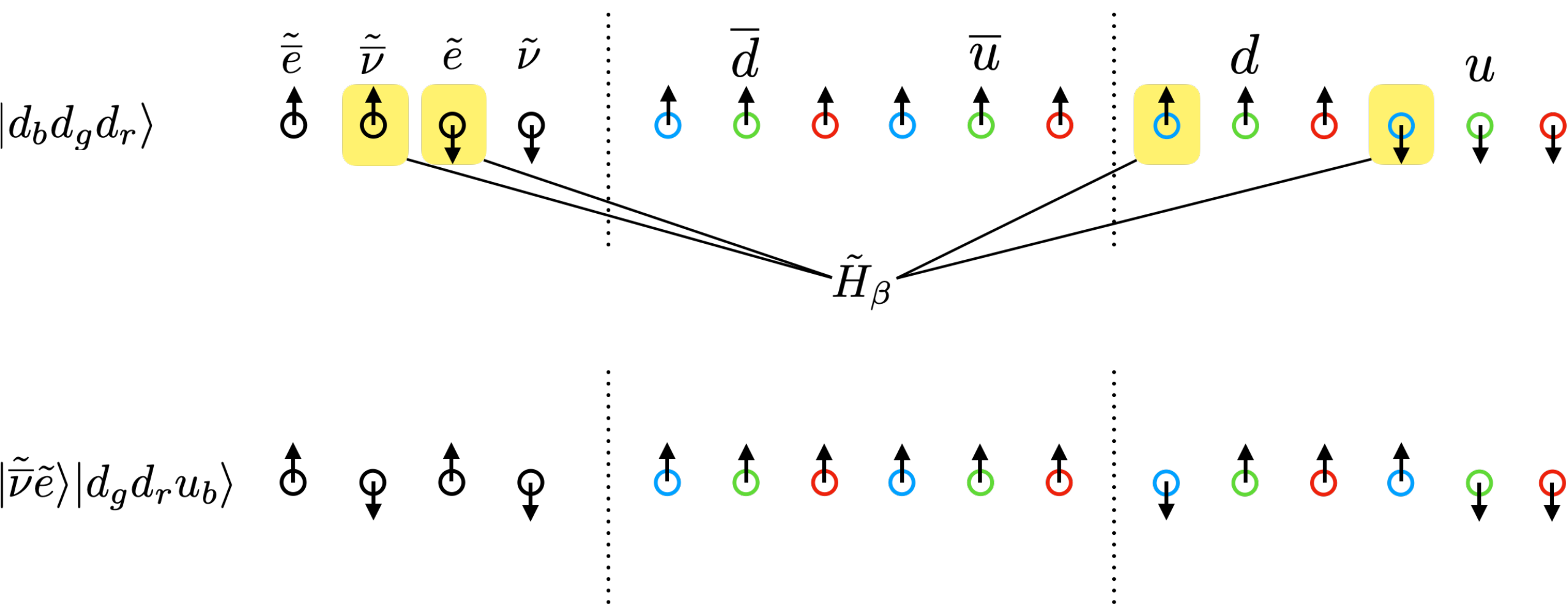}
    \caption{
    The $L=1$ lattice qubit layout of one generation of the SM that is used in this paper for quantum simulation. 
    Fermion (anti-fermion) sites are occupied when the spin is up (down), and the spins at the lepton sites represent occupation in the tilde basis.
    Specifically, 
    the example of $\ket{d_b d_g d_r}$ (upper lattice) decaying to $\ket{ \tilde{\overline{\nu}} \tilde{e}}\ket{d_g d_r u_b}$ (lower lattice) through one application of $\tilde{H}_{\beta}$ 
    in Eq.~(\ref{eq:tildeBeta}) is shown.
    }
    \label{fig:EWlayoutTilde}
\end{figure}

\noindent
To accommodate the capabilities of current devices, 
the quantum simulations performed in this work involve only a single spatial site, $L=1$, 
where the structure of the Hamiltonian can be simplified.
In particular, without interactions between leptons, it is convenient to work with field operators that create and annihilate eigenstates of the free lepton Hamiltonian, $H_{{\rm leptons}}$.
These are denoted by  ``tilde operators",
which create the open-boundary-condition (OBC) analogs of plane waves. 
In the tilde basis with the JW mapping, the lepton Hamiltonian is diagonal and becomes
\begin{align}
    \tilde{H}_{{\rm leptons}} =& \lambda_{\nu}(\tilde{\chi}^{(\nu) \dagger}_0 \tilde{\chi}^{(\nu)}_0-\tilde{\chi}^{(\nu) \dagger}_1 \tilde{\chi}^{(\nu)}_1)  + \lambda_{e}(\tilde{\chi}^{(e) \dagger}_0 \tilde{\chi}^{(e)}_0-\tilde{\chi}^{(e) \dagger}_1 \tilde{\chi}^{(e)}_1) \nonumber \\ \ \rightarrow \ & \frac{\lambda_{\nu}}{2}(Z_{\nu} - Z_{\overline{\nu}})  + \frac{\lambda_{e}}{2}(Z_{e} - Z_{\overline{e}}) 
    \ ,
    \label{eq:tildeLep}
\end{align}
where $\lambda_{\nu,e} = \frac{1}{2}\sqrt{1+4m_{\nu,e}^2}$. 
The $\beta$-decay operator in Eq.~(\ref{eq:KSHam_SM}) becomes
\begin{alignat}{2}
    \tilde{H}_{\beta} = \frac{G}{\sqrt{2}}\Bigg \{ \big( \phi_0^{(u)\dagger} \phi_{0}^{(d)}  + & \phi_{1}^{(u)\dagger} \phi_{1}^{(d)} \big) \bigg [\frac{1}{2}(s_+^e s_-^{\nu} \ - \ s_-^e s_+^{\nu})\left (\tilde{\chi}_0^{(e)\dagger} \tilde{\chi}_{0}^{(\nu)}  + \tilde{\chi}_1^{(e)\dagger} \tilde{\chi}_1^{(\nu)}\right ) \nonumber \\
     + \ &\frac{1}{2}( s_+^e s_+^{\nu} \ + \ s_-^e s_-^{\nu})\left (\tilde{\chi}_0^{(e)\dagger} \tilde{\chi}_{1}^{(\nu)} - \tilde{\chi}_1^{(e)\dagger} \tilde{\chi}_0^{(\nu)}\right )\bigg] \nonumber \\
    + \ \big( \phi_0^{(u)\dagger} \phi_{1}^{(d)}  + & \phi_{1}^{(u)\dagger} \phi_{0}^{(d)} \big)\bigg [\frac{1}{2}(s_+^e s_+^{\nu} \ -\ s_-^e s_-^{\nu})\left (\tilde{\chi}_0^{(e)\dagger} \tilde{\chi}_{0}^{(\nu)}  - \tilde{\chi}_1^{(e)\dagger} \tilde{\chi}_1^{(\nu)}\right ) \nonumber \\ 
    - \ &\frac{1}{2}(s_+^e s_-^{\nu} \ +\  s_-^e s_+^{\nu})\left (\tilde{\chi}_0^{(e)\dagger} \tilde{\chi}_{1}^{(\nu)} +  \tilde{\chi}_1^{(e)\dagger} \tilde{\chi}_0^{(\nu)}\right ) \bigg ] \ + \ {\rm h.c.} \Bigg \}  \ ,
\label{eq:HbetaTilNoJW}
\end{alignat}
where $s^{\nu,e}_\pm = \sqrt{1\pm m_{\nu,e}/\lambda_{\nu,e}}$. 
In our simulations, the initial state of the quark-lepton 
system is prepared in a strong eigenstate with baryon number $B=+1$ in the quark sector 
and the vacuum, $\lvert \Omega \rangle_{{\rm lepton}}$, in the lepton sector.
One of the benefits of working in the tilde basis is that the vacuum satisfies $\tilde{\chi}^{(e,v)}_0\lvert \Omega \rangle_{{\rm lepton}} = \tilde{\chi}^{(e,v) \dagger}_1 \lvert \Omega \rangle_{{\rm lepton}} = 0$, and the terms in the first and third lines of Eq.~(\ref{eq:HbetaTilNoJW}) do not contribute to $\beta$-decay. For the processes we are interested in, this results in an effective $\beta$-decay operator of the form
\begin{alignat}{2}
    \tilde{H}_{\beta} = \frac{G}{\sqrt{2}}\Bigg \{ \left ( \phi_0^{(u)\dagger} \phi_{0}^{(d)}  +  \phi_{1}^{(u)\dagger} \phi_{1}^{(d)} \right ) \bigg [&\frac{1}{2}( s_+^e s_+^{\nu} \ + \ s_-^e s_-^{\nu})\left (\tilde{\chi}_0^{(e)\dagger} \tilde{\chi}_{1}^{(\nu)}  - \tilde{\chi}_1^{(e)\dagger} \tilde{\chi}_0^{(\nu)}\right )\bigg] \nonumber \\
    - \ \left ( \phi_0^{(u)\dagger} \phi_{1}^{(d)} + \phi_{1}^{(u)\dagger} \phi_{0}^{(d)} \right )\bigg [ \frac{1}{2}(s_+^e s_-^{\nu} \ & +\  s_-^e s_+^{\nu})\left (\tilde{\chi}_0^{(e)\dagger} \tilde{\chi}_{1}^{(\nu)} +  \tilde{\chi}_1^{(e)\dagger} \tilde{\chi}_0^{(\nu)}\right ) \bigg ] \ + \ {\rm h.c.} \Bigg \}  \ .
\label{eq:HbetaTilNoVac}
\end{alignat}
The insertion of 
the charge-conjugation matrix,
$\mathcal{C}$, in the continuum operator, Eq.~(\ref{eq:HbetaC1}),
is
necessary to obtain a $\beta$-decay operator that does not annihilate the lepton vacuum.
To minimize the length of the string of $Z$s in the JW mapping, the lattice layout in Fig.~\ref{fig:EWlayoutTilde} is used.
In this layout, the hopping piece of $H_{{\rm quarks}}$ has only 5 $Z$s between the quark and antiquark raising and lowering operators and the $\beta$-decay operator is
\begin{align}
\!\!\!\!\!\! \tilde{H}_{\beta }
    \rightarrow &
    \frac{G}{\sqrt{2}}
      \sum_{c=r,g,b} \bigg [\frac{1}{2}( s_+^e s_+^{\nu} \ + \ s_-^e s_-^{\nu})\left ( \sigma^-_{\overline{\nu}} \sigma^+_e  \ - \  \sigma^+_{\overline{e}} Z^2 \sigma^-_{\nu} \right )\left (\sigma^-_{d,c}Z^2 \sigma^+_{u,c} \: + \: \sigma^-_{\overline{d},c} Z^2 \sigma^+_{\overline{u},c}\right ) \nonumber \\
     - \ &\frac{1}{2}(s_+^e s_-^{\nu} \ +\  s_-^e s_+^{\nu})\left (\sigma^-_{\overline{\nu}} \sigma^+_e \ + \  \sigma^+_{\overline{e}} Z^2 \sigma^-_{\nu} \right )\left ( \sigma^-_{\overline{d},c} Z^8 \sigma^+_{u,c} \: + \: \sigma^+_{\overline{u},c} Z^2 \sigma^-_{d,c} \right ) \ + \  {\rm h.c.} \bigg ]  \ .
     \label{eq:tildeBeta}
\end{align}
In total, the $L=1$ system requires $16$ ($12$ quark and $4$ lepton) 
qubits. See App.~\ref{app:fullHam} for the complete $L=1$ Hamiltonian in terms of qubits.

\subsection{A Majorana Mass for the Neutrino}
\label{sec:majmassforneutrino_SM1p1D}
\noindent
Although not relevant to the simulations performed in Sec.~\ref{sec:BetaSim}, it is of current interest to consider the inclusion of a Majorana mass term for the neutrinos.
A Majorana mass requires and induces the violation of lepton number by
$|\Delta L| = 2$, and is not present in the minimal SM, defined by dim-4 operators.
However, the Weinberg operator~\cite{Weinberg:1979sa} enters at dim-5 and generates an effective Majorana mass for the neutrinos,
\begin{align}
{\cal L}^{{\rm Weinberg}} 
&= \frac{1}{ 2\Lambda} 
\left( \overline{L}^c \epsilon \phi \right)
\left(\phi^T \epsilon L \right)
    \ +\ {\rm h.c.}
\ , \nonumber\\ \ \ &
L = \left( \nu , e \right)^T_L
\ ,\ \ 
\phi = \left( \phi^+ , \phi^0 \right)^T
\ , \ \  \langle \phi \rangle = \left( 0 , v/\sqrt{2} \right)^T
\ ,\ \ \epsilon = i\sigma_2
\ \ ,
\nonumber\\
&\rightarrow
-\frac{v^2}{4\Lambda}  \overline{\nu}^c_L \nu_L
    \ +\ {\rm h.c.}
\ +\ .... 
\label{eq:maja}
\end{align}
where $\phi$ is the Higgs doublet, 
$L^c$ denotes the charge-conjugated left-handed lepton doublet,
$v$ is the Higgs vacuum expectation value and $\Lambda$ is a high energy scale characterizing physics beyond the SM. 
The ellipsis denote interaction terms involving components of the Higgs doublet fields and the leptons.
This is the leading contribution beyond the minimal SM, 
but does not preclude contributions from other sources.
On a $1+1$D lattice there is only a single 
$\lvert \Delta L \rvert = 2$ local operator 
with the structure of a mass term
and, using the JW mapping along with the qubit layout in Fig.~\ref{fig:L1layout}, is of the form
\begin{equation}
H_{\rm Majorana} =
\frac{1}{2} m_M 
     \sum_{n={\rm even}}^{2L-2} 
\left( 
\chi_n^{(\nu)}
\chi_{n+1}^{(\nu)} \:
+ \: {\rm h.c.}
\right)
 \ \rightarrow
 \frac{1}{2} m_M  \sum_{l = 0}^{L-1} 
\left( 
\sigma^+_{l,\nu}\ 
Z^7
\sigma^+_{l,\nub} \: + \: {\rm h.c.}
\right)
\ .
\label{eq:majaHami}
\end{equation}
While the operator has support on a single spatial lattice site, it does not contribute to 
$0\nu\beta \beta$-decay on a lattice with only a single spatial site.   
This is because the processes that it could potentially induce, such as
$\Delta^-\Delta^-\rightarrow \Delta^0\Delta^0 e^- e^-$,
are Pauli-blocked by the single electron site.  
At least two spatial sites are required for any such process producing two electrons in the final state.

\section{Quantum Simulations of the \texorpdfstring{$\beta$}{Beta}-Decay of One Baryon on One Lattice Site}
\label{sec:BetaSim}
\noindent
In this section, quantum simulations of the $\beta$-decay of a single baryon are performed
in $N_f=2$ flavor QCD with $L=1$ spatial lattice site.
The required quantum circuits to perform one and two Trotter steps of time evolution were developed and run on the Quantinuum {\tt H1-1} $20$ qubit trapped ion quantum computer and its simulator {\tt H1-1E}~\cite{quantinuum,h1-1e}.

\subsection{Preparing to Simulate \texorpdfstring{$\beta$}{Beta}-Decay}
\label{sec:BetaSimA}
\noindent
It is well known that, because of confinement, the  energy eigenstates (asymptotic states) of QCD 
are color-singlet hadrons, which are composite objects of quarks and gluons.
On the other hand, 
the operators responsible for $\beta$-decay, given in Eq.~(\ref{eq:tildeBeta}), 
generate transitions between $d$- and $u$-quarks.
As a result,  observable effects of $\tilde{H}_{\beta}$, in part, 
are found in transitions between 
hadronic states whose matrix elements depend on the distribution of the quarks within. 
Toward quantum simulations of the $\beta$-decay of neutrons and nuclei more generally, 
the present work focuses on the decay of a single baryon.

Generically, three elements are required for real-time quantum simulations of 
the $\beta$-decay of baryons:
\begin{enumerate}
    \item 
    Prepare the initial hadronic state that will subsequently undergo $\beta$-decay. 
    In this work, this is one of the single-baryon states (appropriately selected in the spectrum) 
    that is an eigenstate of the strong Hamiltonian alone, 
    i.e., the weak coupling constant is set equal to $G=0$.
    \item 
    Perform (Trotterized) time-evolution using the full ($G\neq0$) Hamiltonian.
    \item 
    Measure one or more of the lepton qubits. 
    If leptons are detected, then $\beta$-decay has occurred.
\end{enumerate}
In $1+1$D, Fermi statistics 
preclude the existence of a light isospin $I=1/2$ nucleon,
and the lightest baryons are in an $I=3/2$ multiplet 
$(\Delta^{++}, \Delta^+, \Delta^0, \Delta^-)$
(using the standard electric charge assignments of the up and down quarks). 
We have  chosen to simulate the decay 
$\Delta^- \to \Delta^0 + e + \overline{\nu}$, which, at the quark level, involves 
baryon-interpolating operators with the quantum numbers of
$ddd\rightarrow udd$.

In order for $\beta$-decay to be kinematically allowed, 
the input-parameters of the theory must be such that 
$M_{\Delta^-} > M_{\Delta^0} + M_{\overline{\nu}} + M_{e}$. 
This is accomplished through tuning the parameters of the Hamiltonian.
The degeneracy in the iso-multiplet is lifted 
by using different values for the  up and down quark masses. 
It is found that the choice of parameters, $m_u=0.9$, $m_d=2.1$, $g=2$ and $m_{e,\nu} = 0$ 
results in the desired hierarchy of baryon and lepton masses. 
The relevant part of the spectrum, obtained from an exact diagonalization of the Hamiltonian, 
is shown in Table~\ref{tab:betaMass}. Although kinematically allowed, multiple instances of $\beta$-decay cannot occur for $L=1$ as there can be at most one of each (anti)lepton.
\begin{table}[!ht]
\centering
\renewcommand{\arraystretch}{1.2}
\begin{tabularx}{0.7\textwidth}{||c | Y ||} 
\hline
\multicolumn{2}{||c||}{Energy of states relevant for $\beta$-decay (above the vacuum)} \\
 \hline
 State & Energy Gap\\
 \hline\hline
 $\Delta^{++}$ & 2.868 \\ 
 \hline
 $\Delta^{++}$ + $2 l$ & 3.868\\
 \hline
 $\Delta^+$ & 4.048 \\
 \hline
 $\Delta^{++}$ + $4l$  & 4.868\\
 \hline
 $\Delta^{+}$ + $2l$  & 5.048\\
 \hline
 $\Delta^0$ & 5.229 \\
 \hline
 $\Delta^{+}$ + $4l$& 6.048\\
 \hline
 $\Delta^0$ + $2l$ & 6.229 \\
 \hline
 $\Delta^-$ & 6.409 \\
 \hline
\end{tabularx}
\caption{
The energy gap above the vacuum of states relevant for $\beta$-decays of single baryons 
with $m_u = 0.9$, $m_d = 2.1$, $g=2$ and $m_{e,\nu} = 0$. The leptons are degenerate in energy and collectively denoted by $l$.}
\renewcommand{\arraystretch}{1}
\label{tab:betaMass}
\end{table}
Note that even though $m_{e,\nu} = 0$, 
the electron and neutrino are gapped due to the finite spatial volume.

\begin{figure}[!ht]
    \centering
    \includegraphics[width=\textwidth]{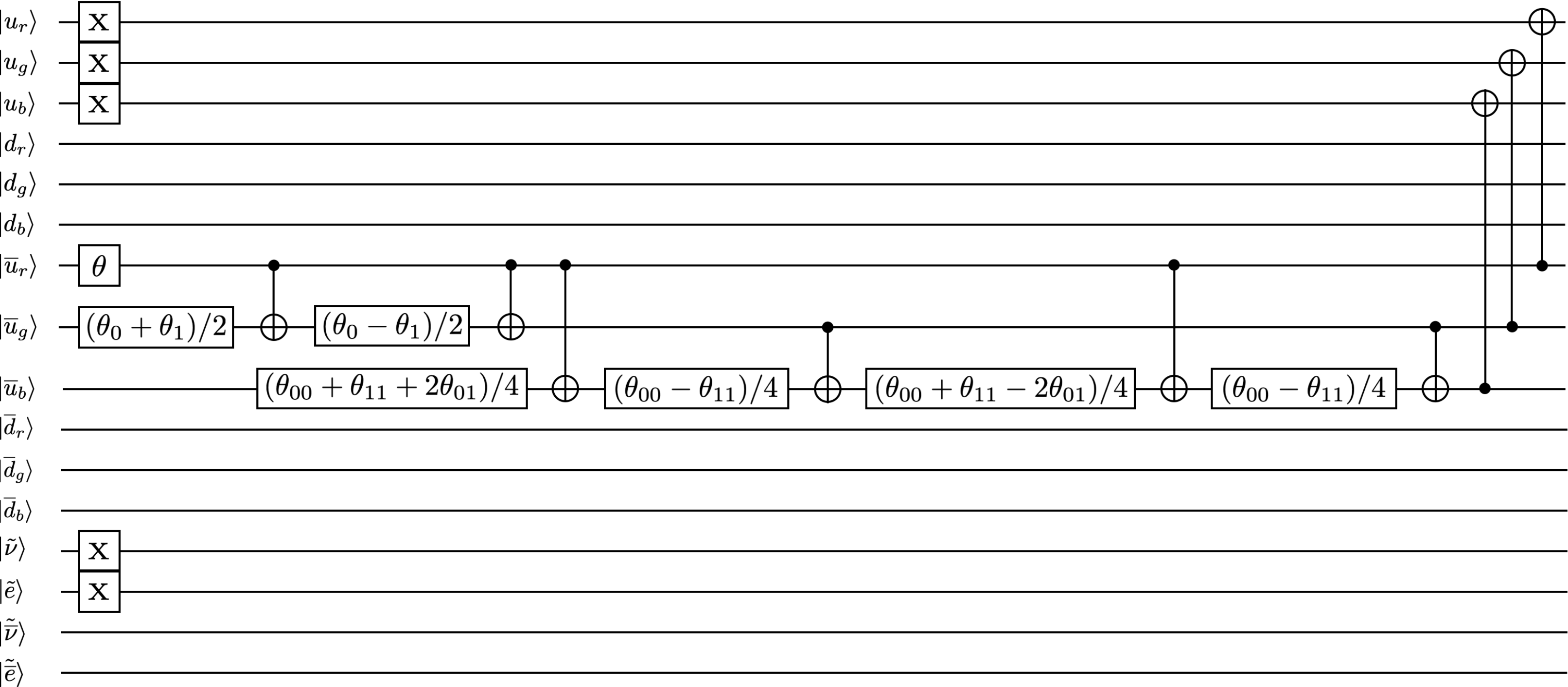}
    \caption{
    A quantum circuit for preparing the $\Delta^-$-baryon on $L=1$ spatial site.}
    \label{fig:DMVQE}
\end{figure}

To prepare the $\Delta^-$ initial state, 
we exploit the observation made in Chapter \ref{chap:1p1dQCD} 
that the stretched-isospin eigenstates of the $\Delta$-baryons, 
with third component of isospin $I_3 = \pm 3/2$,
factorize between the $u$ and $d$ flavor sectors for $L=1$. 
Therefore, the previously developed 
Variational Quantum Eigensolver (VQE)~\cite{peruzzo_2014} 
circuit used to prepare the one-flavor vacuum in Chapter \ref{chap:1p1dQCD} can be used to initialize the two-flavor $\Delta^-$ wave function.
This is done by initializing the vacuum in the lepton sector, 
preparing the state $\ket{d_r d_g d_b}$ in the $d$-sector, 
and applying the VQE circuit to produce the $u$-sector vacuum. 
In the tilde basis, the lepton vacuum is the unoccupied state (trivial vacuum), 
and the complete state-preparation circuit is shown in Fig.~\ref{fig:DMVQE}, 
where $\theta$ is shorthand for RY($\theta$). 
The rotation angles are related by
\begin{align}
    \theta_0 = -2 \sin^{-1} [ \tan(\theta/2) & \, \cos(\theta_{1}/2)  ] \ \: ,  \ \: \theta_{00} = -2 \sin^{-1}\left[ \tan(\theta_{0}/2) \, \cos(\theta_{01}/2) \right] \ \: , \ \: \nonumber \\
    \theta_{01} &= -2 \sin^{-1}\left[ \cos(\theta_{11}/2)\, \tan(\theta_{1}/2) \right]
    \label{eq:angleconst_SM}
\end{align}
and, for $m_u = 0.9$ and $g=2$,\footnote{
The $\overline{u}$ and $u$ parts of the lattice are separated by a fully packed $d$ sector which implies that the part of the wavefunctions with odd numbers of anti-up quarks have relative minus signs compared to the one-flavor vacuum wavefunction.
}
\begin{equation}
    \theta = 0.2256 \ \ , \ \ \theta_1 = 0.4794 \ \ , \ \ \theta_{11} = 0.3265 \ .
\end{equation}
In total, state preparation requires the application of $9$ CNOT gates.

\begin{figure}[!ht]
    \centering
    \includegraphics[width=12cm]{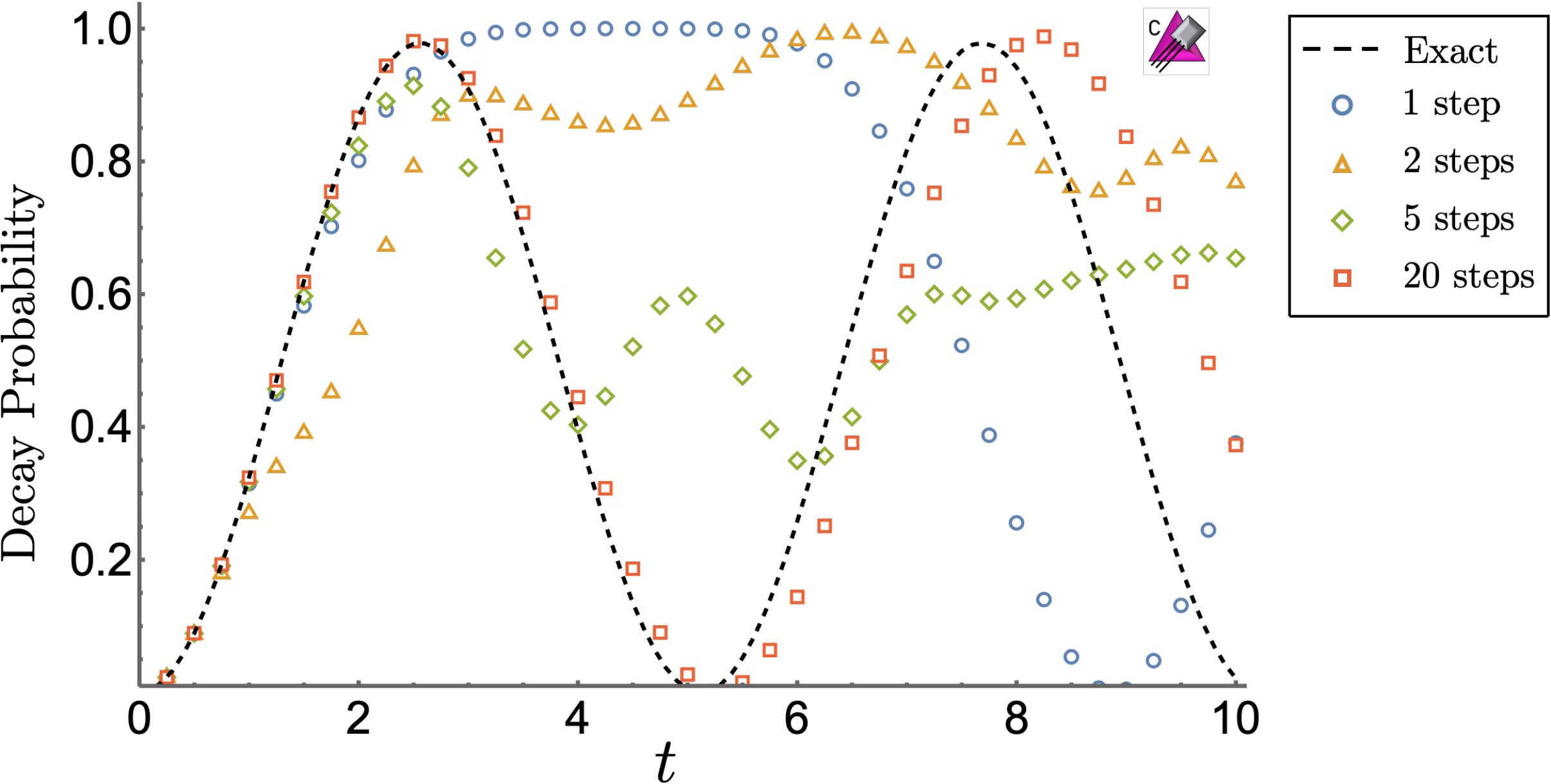}
    \caption{
The probability of $\beta$-decay, $\Delta^- \to \Delta^{0} + e + \overline{\nu}$, with $m_u = 0.9$, $m_d=2.1$, $m_{e,{\nu}} = 0$, $g=2$ and $G=0.5$ computed via exact diagonalization (dotted black line) and on the {\tt qiskit} quantum simulator~\cite{gadi_aleksandrowicz_2019_2562111} using $1,2,5,20$ Trotter steps.}
    \label{fig:BetaDecay}
\end{figure}

Once the $\Delta^-$ baryon state has been initialized 
on the register of qubits, it is then evolved in time with the full Hamiltonian.
The quantum circuits that implement the Trotterized time-evolution 
induced by $H_{{\rm quarks}}$ and $H_{el}$ were previously developed in
Chapter \ref{chap:1p1dQCD}, 
where it was found that, by using an ancilla, each Trotter step 
can be implemented using 114 CNOTs. 
The lepton Hamiltonian, $\tilde{H}_{{\rm leptons}}$, has just single $Z$s which are Trotterized with single qubit rotations.
The circuits required to implement a Trotter step from $\tilde{H}_{\beta}$ are similar to those developed in Chapter \ref{chap:1p1dQCD}, 
and their construction is outlined in App.~\ref{app:BetaCircuits}.
For the present choice of parameters, 
the main contribution to the initial ($\Delta^-$) wave function 
is $\lvert d_b d_g d_r \rangle$, 
i.e., the quark configuration associated with the ``bare" baryon in the d-sector and the trivial vacuum in the u-sector.
This implies that the dominant contribution to the $\beta$-decay 
is from the $\phi_0^{(u)\dagger} \phi_0^{(d)} \tilde{\chi}_0^{(e)\dagger} \tilde{\chi}_{1}^{(\nu)}$ 
term\footnote{Note that the $\phi_0^{(u)\dagger} \phi_0^{(d)} \tilde{\chi}_1^{(e)\dagger} \tilde{\chi}_{0}^{(\nu)}$ term is suppressed since the lepton vacuum in the tilde basis satisfies $\tilde{\chi}_1^{(e,\nu)\dagger} \lvert \Omega\rangle_{{\rm lep}} =  \tilde{\chi}_{0}^{(e,\nu)} \lvert \Omega\rangle_{{\rm lep}} = 0$.}  in Eq.~(\ref{eq:HbetaTilNoJW}), 
which acts only on valence quarks, and the $\beta$-decay operator can be approximated by 
\begin{equation}
    \tilde{H}_{\beta }^{{\rm val}}
    =
    \frac{G}{\sqrt{2}}
     \left (\sigma^-_{\overline{\nu}} \sigma^+_e  \sum_{c=r,g,b}\sigma^-_{d,c}Z^2 \sigma^+_{u,c}  + {\rm h.c.} \right  ) \ ,
     \label{eq:tildeBetaRed}
\end{equation}
for these parameter values. See App.~\ref{app:betaFull} for details on the validity of this approximation.
All of the results presented in this section implement this interaction, 
the Trotterization of which requires 50 CNOTs. 
Notice that, if the Trotterization of $\tilde{H}_{\beta}^{{\rm val}}$ is placed at the end of the first Trotter step, then\\
${U(t) = \exp(-i \tilde{H}_{\beta}^{{\rm val}} t) \times \exp \left [ -i (\tilde{H}_{{\rm leptons}} + H_{{\rm quarks}} + H_{el})t\right ]}$ and the initial exponential (corresponding to strong-interaction evolution) 
can be omitted as it acts on an eigenstate (the $\Delta^-$). 
This reduces the CNOTs required for one and two Trotter steps to $50$ and  $214$, respectively.
For an estimate of the number of CNOTs required to time evolve with the $\beta$-decay Hamiltonian on larger lattices see App.~\ref{app:LongJW}.
The probability of $\beta$-decay, 
as computed both through exact diagonalization of the Hamiltonian 
and through Trotterized time-evolution using the {\tt qiskit} classical simulator~\cite{gadi_aleksandrowicz_2019_2562111}, 
is shown in Fig.~\ref{fig:BetaDecay}.
The periodic structure is a finite volume effect, and the probability of 
$\beta$-decay is expected to tend to an exponential in time as $L$ increases, 
see App.~\ref{app:beta1p1aL}.

\begin{figure}[!ht]
    \centering
    \includegraphics[width=10cm]{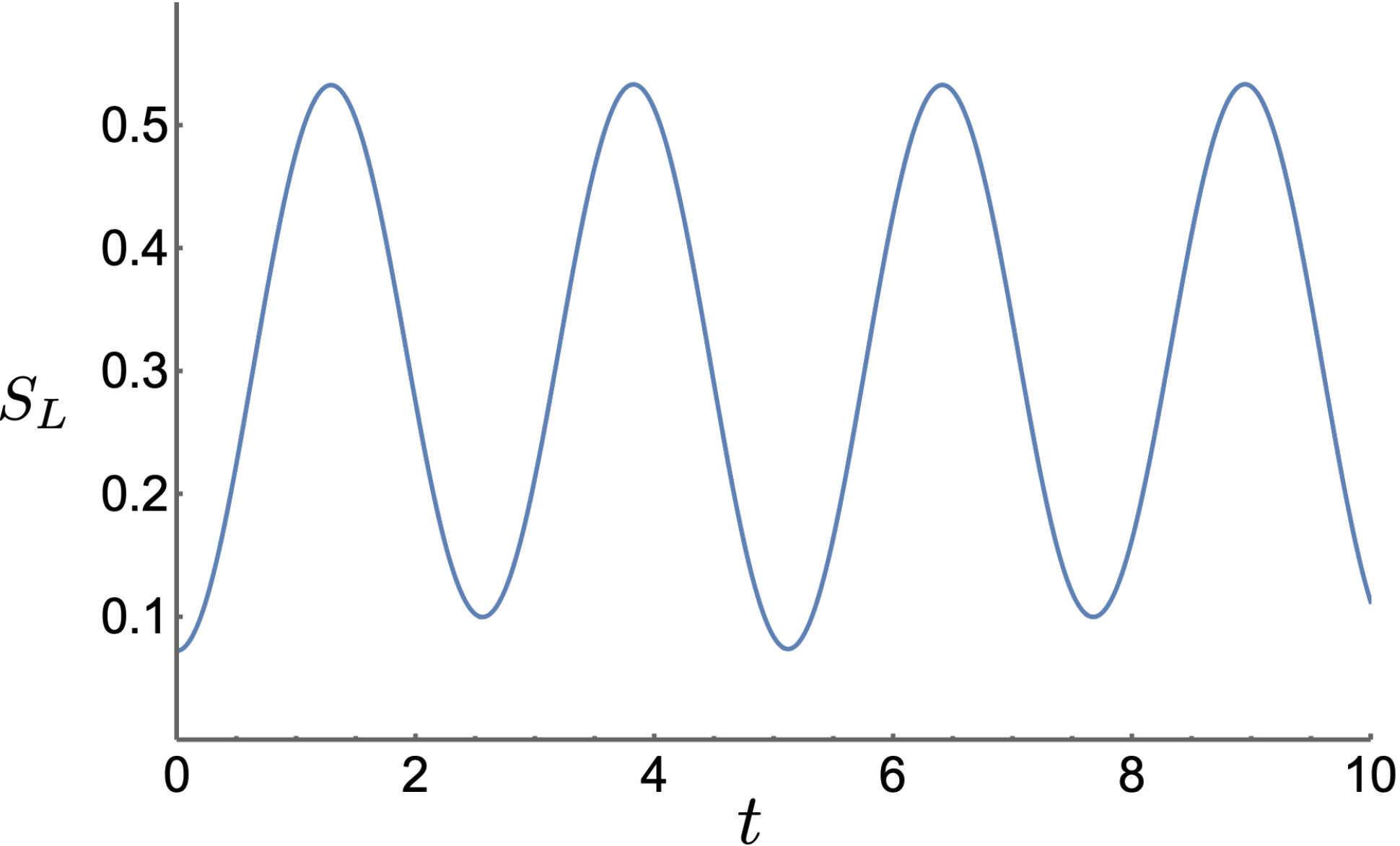}
    \caption{
    The linear entanglement entropy, $S_L$, between quarks and antiquarks plus leptons during the $\beta$-decay of an initial $\Delta^-$-baryon.}
    \label{fig:linEnt}
\end{figure}

Entanglement in quantum simulations of lattice gauge theories 
is a growing area of focus, 
see, e.g., Refs.~\cite{Ghosh:2015iwa,Soni:2015yga,Panizza:2022gvd,Rigobello:2021fxw}, and 
it is interesting to examine the evolution of entanglement during the $\beta$-decay process.
Before the decay, the quarks and antiquarks are together in a pure state as the leptons are in the vacuum, and subsequent time evolution of the state introduces components into the wavefunction that have non-zero population of the lepton states.
One measure of entanglement is the linear entropy,  
\begin{equation}
    S_L = 1 - \Tr[\rho_q^2]
    \ ,
\end{equation}
between the quarks and antiquarks plus leptons.
It is constructed by  tracing the full density matrix, $\rho$, 
over the antiquark and lepton sector to form the reduced density matrix 
$\rho_q = \Tr_{\overline{q}, {\rm leptons}} [\rho]$.
Figure~\ref{fig:linEnt} shows the linear entropy computed through exact diagonalization 
of the Hamiltonian with the parameters discussed previously in the text. 
By comparing with the persistence probability in Fig.~\ref{fig:BetaDecay}, 
it is seen that the entanglement entropy evolves at twice the frequency 
of the $\beta$-decay probability. 
This is because $\beta$-decay primarily transitions the baryon between the ground state of the $\Delta^-$ and $\Delta^0$. 
It is expected that these states will have a comparable amount of entanglement, 
and so the entanglement is approximately the same when the decay probabilities are $0$ and $1$. 
While this makes this particular example somewhat uninteresting, it does demonstrate that when multiple final states are accessible, the time-dependence of the entanglement structure might be revealing.

\subsection{Simulations Using Quantinuum's {\tt H1-1} 20 Qubit Trapped Ion Quantum Computer}
\label{sec:BetaSimB}
\noindent
Both the initial state preparation and one and two steps of Trotterized time evolution were executed
using Quantinuum's {\tt H1-1} 20 qubit trapped ion quantum computer~\cite{quantinuum} and its simulator {\tt H1-1E}\footnote{The classical simulator {\tt H1-1E} includes depolarizing gate noise, leakage errors, crosstalk noise and dephasing noise due to transport and qubit idling~\cite{h1-1e}.} (for details on the specifications of {\tt H1-1}, see App.~\ref{app:H1specs}).
After transpilation onto the native gate set of {\tt H1-1}, a single Trotter step requires 59 $ZZ$ gates, while two Trotter steps requires 212 $ZZ$ gates.\footnote{The number of $ZZ$ gates could be further reduced by 5 by not resetting the ancilla.}
By post-selecting results on ``physical" states with baryon number $B=1$ and lepton number $L=0$ 
to mitigate single-qubit errors (e.g., Ref.~\cite{physrevd.101.074512}),
approximately 90\% (50\%) of the total events from the one (two) Trotter step circuit remained. Additionally, for the two Trotter step circuit, results were selected where the ancilla qubit was in the $|0\rangle$ state (around 95\%).\footnote{For this type of error, the mid-circuit measurement and re-initialization option available for {\tt H1-1} could have been used to identify the case where the bit-flip occurred after the ancilla was used and the error had no effect on the final results.}

\begin{figure}[!t]
    \centering
    \includegraphics[width=\textwidth]{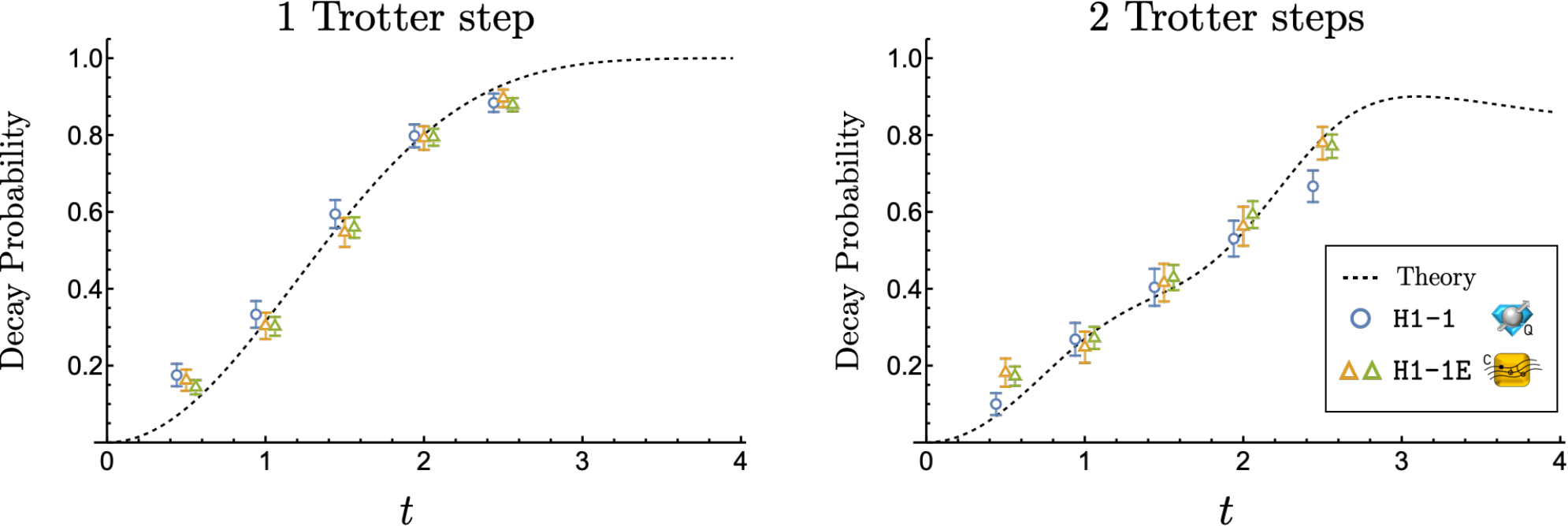}
    \caption{
    The probability of $\beta$-decay, $\Delta^- \to \Delta^{0} + e + \overline{\nu}$, with $m_u = 0.9$, $m_d=2.1$, $m_{e,{\nu}} = 0$, $g=2$ and $G=0.5$, using one (left panel) and two (right panel) Trotter steps (requiring 59 and 212 $ZZ$ gates, respectively), as given in Table~\ref{tab:H1results}.
    The dashed-black curves show the expected result from Trotterized time evolution, corresponding to the blue circles (orange triangles) in Fig.~\ref{fig:BetaDecay} for one (two) Trotter steps.
    The blue circles correspond to the data obtained on the {\tt H1-1} machine, 
    and the orange (green) triangles to the {\tt H1-1E} emulator, 
    each obtained from 200 shots (400 shots). The points have been shifted slightly along the $t$-axis for clarity.
    Error mitigation beyond physical-state post-selection has not been performed.
    The weak Hamiltonian in the time-evolution responsible for the decay is given in Eq.~(\ref{eq:tildeBetaRed}).
    }
    \label{fig:BetaDecayH1}
\end{figure}

The results of the simulations 
are shown in Fig.~\ref{fig:BetaDecayH1} and given in Table~\ref{tab:H1results}. 
By comparing the results from {\tt H1-1} and {\tt H1-1E} (using 200 shots) it is seen that the simulator is able to faithfully reproduce the behavior of the quantum computer.
The emulator was also run with 400 shots and clearly shows convergence to the expected value, verifying that the agreement between data and theory was not an artifact due to low statistics (and large error bars). 
Compared with the results presented in Chapter \ref{chap:1p1dQCD} that were performed using IBM's {\tt ibmq$\_$jakarta} and {\tt ibm$\_$perth}, 
error mitigation techniques were not 
applied to the present simulations due to the overhead in resource requirements.
Specifically, Pauli twirling, dynamical decoupling, decoherence renormalization and measurement error mitigation 
were not performed. This is practical because the two-qubit gate, state preparation and measurement (SPAM) errors are an order of magnitude smaller on Quantinuum's trapped ion system 
compared to those of IBM's superconducting qubit systems (and a similar error rate on the single-qubit gates)~\cite{Pelofske:2022vyy}. 
\begin{table}[!ht]
\renewcommand{\arraystretch}{1.2}
\begin{tabularx}{1.0\textwidth}{||c | Y | Y | Y | Y | Y | Y | Y | Y ||}
\hline
\multicolumn{9}{||c||}{
Single-Baryon Decay Probabilities using Quantinuum's {\tt H1-1} and {\tt H1-1E}} \\
\hline
 & \multicolumn{4}{c|}{1 Trotter step} & \multicolumn{4}{c||}{2 Trotter steps} \\
 \hline
 $t$ & 
 {\tt H1-1} & {\tt H1-1E} & \makecell{{\tt H1-1E}, \\$\times 2$ stats} & Theory & 
 {\tt H1-1} & {\tt H1-1E} & \makecell{{\tt H1-1E}, \\$\times 2$ stats} & Theory
 \\
 \hline
 0.5 & 
 0.175(29) & 0.162(28) & 0.144(19) & 0.089 & 
 0.100(29) & 0.182(37) & 0.173(25) & 0.088\\
 \hline
 1.0 & 
 0.333(35) & 0.303(34) & 0.302(25) & 0.315 & 
 0.269(43) & 0.248(41) & 0.272(29) & 0.270 \\
 \hline
 1.5 & 
 0.594(37) & 0.547(38) & 0.559(27) & 0.582 & 
 0.404(48) & 0.416(49) & 0.429(33)  & 0.391 \\
 \hline
 2.0 & 
 0.798(30) & 0.792(30) & 0.794(22) & 0.801 & 
 0.530(47) & 0.563(51) & 0.593(35) & 0.547 \\
 \hline
 2.5 & 
 0.884(24) & 0.896(23) & 0.879(17) & 0.931 & 
 0.667(41) & 0.779(43) & 0.771(30) & 0.792 \\
 \hline
\end{tabularx}
\renewcommand{\arraystretch}{1}
\caption{
The probability of $\beta$-decay, 
$\Delta^- \to \Delta^{0} + e + \overline{\nu}$, on $L=1$ spatial lattice with $m_u = 0.9$, $m_d=2.1$, $m_{e,{\nu}} = 0$, $g=2$ and $G=0.5$.
These simulations
were performed using Quantinuum's {\tt H1-1} and {\tt H1-1E} and included the initial state preparation and subsequent time evolution under 1 and 2 Trotter steps. The results are displayed in Fig.~\ref{fig:BetaDecayH1}. 
The columns labeled ($\times 2$ stats) were obtained using 400 shots, compared to the rest, that used 200 shots, and
uncertainties were computed assuming the results follow a binomial distribution.
}
\label{tab:H1results}
\end{table}
%

\section{Speculation about Quantum Simulations with a Hierarchy of Length Scales}
\noindent
It is interesting to consider how a hierarchy of length scales, 
as present in the SM, may be helpful in error correction.
In the system we have examined, the low energy strong sector is composed of mesons, baryons and nuclei, with both color singlet and non-singlet excitations (existing at higher energies).
As observed in Chapter \ref{chap:1p1dQCD}, OBCs allow for 
relatively low-energy colored ``edge" states to exist near the boundary of the lattice.
The energy of a color non-singlet grows linearly with its distance 
from the boundary, leading to a force on colored objects.
This will cause colored errors in the bulk to migrate to the edge of the lattice where they could be detected and possibly removed.
This is one benefit of using axial gauge, where Gauss's law is automatically enforced, 
and a colored ``error" in the bulk generates a color flux tube that extends to the boundary.

Localized two-bit-flip errors  can create color-singlet 
excitations that do not experience a force toward the boundary, but which are
vulnerable to weak decay. 
For sufficiently large lattices,  color singlet excitations will decay weakly down to stable states
enabled by the near continuum of lepton states. 
In many ways, this resembles the
quantum imaginary-time evolution (QITE)~\cite{Kamakari:2021nmf,Hubisz:2020vhx,turro:2021vbk}
algorithm, which is a special case of coupling to open systems,
where quantum systems are driven into their ground state by embedding them in a larger system that acts as a heat reservoir.
One can speculate that, in the future, quantum simulations of QCD 
will benefit from also including electroweak interactions as a mechanism to cool the strongly-interacting sector from particular classes of errors.

This particular line of investigation is currently at a ``schematic'' level, and significantly more work is required to quantify its utility.
Given the quantum resource requirements, it is likely that the Schwinger model will 
provide a suitable system to explore such scenarios.

\section{Summary and Conclusions}
\noindent
Quantum simulations of SM physics is in its infancy and, for practical reasons, has been previously
limited to either QCD or QED in one or two spatial dimensions. 
In this work, we have started the integration of the electroweak sector into quantum simulations of QCD by examining the time-evolution of the $\beta$-decay of one baryon.
In addition to the general framework that allows for 
simulations of arbitrary numbers of lattice sites in one dimension, 
we present results for $L=1$ spatial lattice site, which requires 16 qubits.
Explicitly, this work considered quantum simulations of 
$\Delta^-\rightarrow\Delta^0 e \overline{\nu}$
in two flavor $1+1$D QCD for $L=1$ spatial lattice site.
Simulations were performed using Quantinuum's {\tt H1-1} 20-qubit trapped ion quantum computer
and classical simulator {\tt H1-1E}, 
requiring 17 (16 system and 1 ancilla) qubits. 
Results were presented for both one and two Trotter steps, including the state preparation of the initial baryon, requiring 59 and 212 two-qubit gates respectively.
Even with 212 two-qubit gates, {\tt H1-1} provided results that 
are consistent with the expected results, within uncertainties, without error-mitigation beyond physical-state post selection.  
While not representative of $\beta$-decay in the continuum, 
these results demonstrate the potential of quantum simulations to determine 
the real-time evolution of decay and reaction processes in nuclear and 
high-energy processes.
High temporal-resolution studies of the evolution of the quarks and gluons
during hadronic decays and nuclear reactions
are expected to provide new insights into the mechanisms responsible for these processes, 
and lead to new strategies for further reducing systematic errors in their prediction.


\clearpage
\begin{subappendices}


\section{The Complete Spin Hamiltonian for \texorpdfstring{$L=1$}{L=1}}
\label{app:fullHam}
\noindent
After the JW mapping of the Hamiltonian to qubits, and using the tilde-basis for the leptons, 
the four contributing terms are
\begin{subequations}
    \label{eq:H2flavL1_SM}
    \begin{align}
    H &= \ H_{{\rm quarks}}\ +\ \tilde{H}_{{\rm leptons}}\ +\ H_{{\rm glue}} \ +\ \tilde{H}_{\beta} ,\\[4pt]
    H_{{\rm quarks}} &= \ \frac{1}{2} [ m_u\left (Z_0 + Z_1 + Z_2 -Z_6 - Z_7 - Z_8 + 6\right )+ m_d (Z_3 + Z_4 + Z_5 -Z_9 -
    Z_{10} \nonumber \\ - & Z_{11} + 6 ) ] -\, \frac{1}{2} (\sigma^+_6 Z_5 Z_4 Z_3 Z_2 Z_1 \sigma^-_0 + \sigma^-_6 Z_5 Z_4 Z_3 Z_2 Z_1 \sigma^+_0 + \sigma^+_7 Z_6 Z_5 Z_4 Z_3 Z_2 \sigma^-_1 \nonumber \\
    &+ \sigma^-_7 Z_6 Z_5 Z_4 Z_3 Z_2 \sigma^+_1  +\, \sigma^+_8 Z_7 Z_6 Z_5 Z_4 Z_3 \sigma^-_2 + \sigma^-_8 Z_7 Z_6 Z_5 Z_4 Z_3 \sigma^+_2 \nonumber \\
    &+ \sigma^+_9 Z_8 Z_7 Z_6 Z_5 Z_4 \sigma^-_3 + \sigma^-_9 Z_8 Z_7 Z_6 Z_5 Z_4 \sigma^+_3 +\, \sigma^+_{10} Z_9 Z_8 Z_7 Z_6 Z_5 \sigma^-_4 \nonumber \\
    &+ \sigma^-_{10} Z_9 Z_8 Z_7 Z_6 Z_5 \sigma^+_4 + \sigma^+_{11} Z_{10} Z_9 Z_8 Z_7 Z_6 \sigma^-_5 + \sigma^-_{11} Z_{10} Z_9 Z_8 Z_7 Z_6 \sigma^+_5 ) \ ,
        \label{eq:Hkin2flavL1_SM}\\[4pt]
    \tilde{H}_{{\rm leptons}} &= \ \frac{1}{4}\sqrt{1 +4 m_e^2}(Z_{13} - Z_{15}) + \frac{1}{4}\sqrt{1 +4 m_{\nu}^2}(Z_{12} - Z_{14}) \nonumber \\[4pt]
    H_{{\rm glue}} &= \ \frac{g^2}{2} \bigg [ \frac{1}{3}(3 - Z_1 Z_0 - Z_2 Z_0 - Z_2 Z_1) + \sigma^+_4\sigma^-_3\sigma^-_1\sigma^+_0  + \sigma^-_4\sigma^+_3\sigma^+_1\sigma^-_0  + \sigma^+_5Z_4\sigma^-_3\sigma^-_2Z_1\sigma^+_0 \nonumber \\
    & +\, \sigma^-_5Z_4\sigma^+_3\sigma^+_2Z_1\sigma^-_0 + \sigma^+_5\sigma^-_4\sigma^-_2\sigma^+_1 + \sigma^-_5\sigma^+_4\sigma^+_2\sigma^-_1 +\,\frac{1}{12} (2 Z_3 Z_0 + 2Z_4 Z_1 \nonumber \\
    &+ 2Z_5 Z_2 - Z_5 Z_0 - Z_5 Z_1 - Z_4 Z_2 - Z_4 Z_0 - Z_3 Z_1  - Z_3 Z_2  ) \bigg ] \ ,
    \label{eq:Hel2flavL1_SM} \\
     \tilde{H}_{\beta} = & \frac{G}{\sqrt{2}} \Bigg( \frac{1}{2} ( s_+^e s_+^{\nu} + s_-^e s_-^{\nu}) \big[ (\textcolor{blue}{\sigma^-_{14} \sigma^+_{13}}  -  \sigma^+_{15} Z_{14} Z_{13}\sigma^-_{12}) \big( \textcolor{blue}{\sigma^-_{3} Z_2 Z_1 \sigma^+_0 + \sigma^-_{4} Z_3 Z_2 \sigma^+_1} \nonumber \\
     \textcolor{blue}{+} & \textcolor{blue}{\sigma^-_5 Z_4 Z_3 \sigma^+_2} + \sigma^-_{9} Z_8 Z_7 \sigma^+_6 + \ \sigma^-_{10} Z_9 Z_8 \sigma^+_7 + \sigma^-_{11} Z_{10} Z_9 \sigma^+_8 \big) + (\textcolor{blue}{\sigma^+_{14} \sigma^-_{13}} - \sigma^-_{15} Z_{14} Z_{13}\sigma^+_{12}) \nonumber \\
    & \big (\textcolor{blue}{\sigma^+_{3} Z_2 Z_1 \sigma^-_0 + \sigma^+_{4} Z_3 Z_2 \sigma^-_1 + \sigma^+_5 Z_4 Z_3 \sigma^-_2} + \sigma^+_{9} Z_8 Z_7 \sigma^-_6 + \sigma^+_{10} Z_9 Z_8 \sigma^-_7  + \sigma^+_{11} Z_{10} Z_9 \sigma^-_8 ) \big] \nonumber \\
    &- \ \frac{1}{2}(s_+^e s_-^{\nu}  +  s_-^e s_+^{\nu}) \big [ (\sigma^-_{14} \sigma^+_{13} +  \sigma^+_{15} Z_{14} Z_{13}\sigma^-_{12}) \big( \sigma^-_{9} Z_8 Z_7 Z_6 Z_5 Z_4 Z_3 Z_2 Z_1 \sigma^+_0 \nonumber \\
    &+ \sigma^-_{10} Z_9 Z_8 Z_7 Z_6 Z_5 Z_4 Z_3 Z_2 \sigma^+_1 + \ \sigma^-_{11} Z_{10} Z_9 Z_8 Z_7 Z_6 Z_5 Z_4 Z_3 \sigma^+_2 + \sigma^+_{6} Z_5 Z_4 \sigma^-_3 \nonumber \\
    &+ \sigma^+_{7} Z_6 Z_5 \sigma^-_4 \big ) + \sigma^+_{8} Z_7 Z_6 \sigma^-_5 + \ (\sigma^+_{14} \sigma^-_{13} + \sigma^-_{15} Z_{14} Z_{13}\sigma^+_{12}) \big( \sigma^+_{9} Z_8 Z_7 Z_6 Z_5 Z_4 Z_3 Z_2 Z_1 \sigma^-_0 \nonumber \\
    &+ \sigma^+_{10} Z_9 Z_8 Z_7 Z_6 Z_5 Z_4 Z_3 Z_2 \sigma^-_1 + \  \sigma^+_{11} Z_{10} Z_9 Z_8 Z_7 Z_6 Z_5 Z_4 Z_3 \sigma^-_2 \nonumber \\
    &+ \sigma^-_{6} Z_5 Z_4 \sigma^+_3 + \sigma^-_{7} Z_6 Z_5 \sigma^+_4 + \sigma^-_{8} Z_7 Z_6 \sigma^+_5 \big ) \big] \Bigg)
    \end{align}
\end{subequations}
In the mapping, the qubits are indexed right-to-left and, 
for example, qubit zero (one) corresponds to a red (green) up-quark.
The terms highlighted in blue provide the leading contribution to the $\beta$-decay process 
for the parameters used in the text and make up the operator used for the simulations performed in Sec.~\ref{sec:BetaSim}.

\section{\texorpdfstring{$\beta$}{Beta}-Decay in the Standard Model}
\label{app:betaSM}
\noindent 
To put our simulations in $1+1$D into context, 
it is helpful to  outline relevant aspects of single-hadron $\beta$-decays 
in the SM in $3+1$D.
Far below the electroweak symmetry-breaking scale,  
charged-current interactions can be included as an infinite 
set of effective operators in a systematic EFT description, ordered by their contributions in powers of low-energy scales divided by appropriate powers of $M_W$.
For instance, $\beta$-decay rates between hadrons scale as 
$\sim \Lambda (G_F \Lambda^2 )^2 (\Lambda / M_W )^n$, 
where $\Lambda$ denotes the low-energy scales,
$\frac{G_F}{\sqrt{2}} = \frac{g_2^2}{8 M_W^2}$ is Fermi's constant and 
LO (in $\Lambda / M_W$)
corresponds to $n=0$.
By matching operators at LO in SM interactions, the $\beta$-decay of the neutron is induced by an effective Hamiltonian density of the form~\cite{Feynman:1958ty,Sudarshan:1958vf}
\begin{equation}
    {\cal H}_\beta = 
    \frac{G_F}{\sqrt{2}} \ V_{ud} \ 
    \overline{\psi}_u\gamma^\mu (1-\gamma_5)\psi_d\ 
    \overline{\psi}_e\gamma_\mu (1-\gamma_5)\psi_{\nu_e}  
    \ +\ {\rm h.c.}
    \ ,
    \label{eq:betaHamiSM}
\end{equation}
where $V_{ud}$ is the element of the CKM matrix for $d\rightarrow u$ transitions.
As ${\cal H}_\beta $ factors into contributions from lepton and quark operators, the matrix element factorizes into a plane-wave lepton contribution and a non-perturbative hadronic component requiring matrix elements of the quark operator between hadronic states. 
With the mass hierarchies and symmetries in nature, 
there are two dominant form factors, so that,
\begin{equation}
    \langle p(p_p) | \overline{\psi}_u\gamma^\mu (1-\gamma_5)\psi_d | n(p_n)\rangle = 
    \overline{U}_p \left[\ g_V(q^2) \gamma^\mu - g_A(q^2) \gamma^\mu \gamma_5\ \right] U_n \ ,
\end{equation}
where $q$ is the four-momentum transfer of the process, $g_V(0) = 1$ in the isospin limit and $g_A(0)=1.2754(13)$~\cite{Workman:2022ynf} as measured in experiment.
The matrix element for $n\rightarrow p e^- \overline{\nu}_e$
calculated from the Hamiltonian in Eq.~(\ref{eq:betaHamiSM})
is
\begin{equation}
    \lvert \mathcal{M} \rvert^2 = 16 G_F^2 \lvert V_{ud}\rvert^2 M_n M_p (g_V^2 + 3 g_A^2)(E_{\nu}E_e + \frac{g_V^2-g_A^2}{g_V^2 + 3 g_A^2} {\bf p}_e \cdot {\bf p}_{\nub}) \ ,
\end{equation}
which leads to a neutron width of 
(at LO in $(M_n-M_p)/M_n$ and $m_e/M_n$)
\begin{equation}
    \Gamma_n =
    \frac{G_F^2 |V_{ud}|^2 (M_n-M_p)^5}{60\pi^3}
    \ \left( g_V^2 + 3 g_A^2 \right)\ f^\prime(y) \ ,
\end{equation}
where $f^\prime(y)$ is a phase-space factor, 
\begin{equation}
    f^\prime (y) = \sqrt{1-y^2}\left(1 - \frac{9}{2}y^2 - 4 y^4\right)
    \ -\ \frac{15}{2}y^4 \log\left[ \frac{y}{\sqrt{1-y^2}+1}\right] \ ,
\end{equation}
and $y=m_e/(M_n-M_p)$.
Radiative effects, recoil effects and other higher-order contributions have been neglected.

\section{\texorpdfstring{$\beta$}{Beta}-Decay in \texorpdfstring{$1+1$}{1+1} Dimensions: The \texorpdfstring{$L=\infty$}{L to infinity} and Continuum Limits}
\label{app:beta1p1}
\noindent 
In $1+1$D, the fermion field has dimensions 
$\left[\psi\right] = \frac{1}{2}$, 
and a four-Fermi operator has dimension 
$[\hat \theta ] = 2$.  
Therefore, while in $3+1$D 
$\left[G_F \right] = -2$, 
in $1+1$D, the coupling has dimension $\left[G \right] = 0$.
For our purposes, to describe the $\beta-$decay of a $\Delta^-$-baryon in $1+1$D,
we have chosen to work with an effective Hamiltonian of the form
\begin{equation}
    {\cal H}_\beta^{1+1} \ = \ 
    \frac{G}{\sqrt{2}} \
    \overline{\psi}_u\gamma^\mu \psi_d\ 
    \overline{\psi}_e\gamma_\mu\psi_{\overline{\nu}} 
    \ +\ {\rm h.c.}
    \ =\ 
    \frac{G}{\sqrt{2}} \
    \overline{\psi}_u\gamma^\mu \psi_d\ 
    \overline{\psi}_e\gamma_\mu \mathcal{C} \psi_{\nu} 
        \ +\ {\rm h.c.}
\ ,
\label{eq:betaHami1p1}
\end{equation}
where we have chosen the basis
\begin{align}
    \gamma_0 \ = \ &
    \left(
    \begin{array}{cc}
    1&0 \\ 0&-1
    \end{array}
    \right)
    \ \ ,\ \ 
    \gamma_1 \ = \ 
    \left(
    \begin{array}{cc}
    0&1 \\ -1&0
    \end{array}
    \right)
    \ =\ \mathcal{C}
    \ ,\ \nonumber \\
    \gamma_0\gamma_\mu^\dagger \gamma_0\ =\ & \gamma_\mu
    \ ,\ 
    \gamma_0 \mathcal{C} ^\dagger \gamma_0\ =\ \mathcal{C}
    \ \ ,\ \ 
    \{\gamma_\mu , \gamma_\nu \} \ =\ 2 g_{\mu\nu}
    \ .
\label{eq:gammaMats1p1}
\end{align}
For simplicity, the CKM matrix element is set equal to unity 
as only one generation of particles is considered.

In the limit of exact isospin symmetry, which we assume to be approximately valid in this appendix, 
the four $\Delta$ baryons form an isospin quartet 
and can be embedded in a tensor $T^{abc}$ (as is the case for the $\Delta$ resonances in nature)
as 
$T^{111}=\Delta^{++}$,
$T^{112}=T^{121}=T^{211}=\Delta^{+}/\sqrt{3}$,
$T^{122}=T^{221}=T^{212}=\Delta^{0}/\sqrt{3}$,
$T^{222}=\Delta^{-}$.
Matrix elements of the isospin generators
are reproduced by an effective operator of the form
\begin{equation}
    \overline\psi_q \gamma^\mu \tau^\alpha \psi_q \ \rightarrow \
    3 \overline{T}_{abc} \gamma^\mu \left(\tau^\alpha \right)^c_d T^{abd} \ ,
\label{eq:DeltaIgens}
\end{equation}
which provides a Clebsch-Gordan coefficient for isospin raising operators,
\begin{equation}
    \overline\psi_q \gamma^\mu \tau^+ \psi_q \ \rightarrow \
    \sqrt{3}\  \overline{\Delta^{++}}\gamma^\mu\Delta^+ 
    \ +\ 2 \ \overline{\Delta^{+}}\gamma^\mu\Delta^0
    \ +\ 
    \sqrt{3}\  \overline{\Delta^{0}}\gamma^\mu\Delta^- \ .
\label{eq:DeltaIgens2}
\end{equation}

The matrix element for $\beta$-decay factorizes at LO in the electroweak interactions.
The hadronic component of the matrix element is given by 
\begin{align}
\langle \Delta^0(p_0) | \overline{\psi}_u & \gamma^\alpha \psi_d | \Delta^-(p_-)\rangle = \sqrt{3} g_V(q^2) \ \overline{U}_{\Delta^0}  \gamma^\alpha  U_{\Delta^-} \ =\ H^\alpha \ , \nonumber\\
H^\alpha H^{\beta\ \dagger} & = 3 |g_V(q^2) |^2 {\rm Tr}\left[ \gamma^\alpha \left( \pslash_- + M_{\Delta^-}\right) \gamma^\beta 
\left( \pslash_0 + M_{\Delta^0}\right) \right] \nonumber\\
& = 6 |g_V(q^2) |^2  
\left[p_-^\alpha p_0^\beta  + p_0^\alpha p_-^\beta  - g^{\alpha\beta} (p_-\cdot p_0)
\ +\ M_{\Delta^-} M_{\Delta^0} g^{\alpha\beta}\right] \ =\ H^{\alpha\beta} \ ,
\end{align}
and the leptonic component of the matrix element is given by, assuming that the electron and neutrino are massless, 
\begin{align}
\langle e^- \overline{\nu}_e | \overline{\psi}_e \gamma^\alpha C \psi_{\nu}  | 0\rangle 
& = \overline{U}_e\gamma^\alpha C V_\nu\ =\ L^\alpha \ ,
\nonumber\\
L^\alpha L^{\beta\ \dagger} & = {\rm Tr}\left[\ 
\gamma^\alpha C \pslash_\nu C \gamma^\beta \pslash_e \right] \ =\  {\rm Tr}\left[\ \gamma^\alpha \overline{\pslash}_\nu \gamma^\beta \pslash_e \right] \nonumber\\
& = 
2 \left[ \overline{p}_\nu^\alpha p_e^\beta + \overline{p}_\nu^\beta p_e^\alpha  -  g^{\alpha\beta} (\overline{p}_\nu\cdot p_e) \right] \ =\ L^{\alpha\beta}\ ,
\end{align}
where $p = (p^0, +p^1)$ and 
$\overline{p}=(p^0, -p^1)$.
Therefore, the squared matrix element of the process is
\begin{equation}
    |{\cal M}|^2 \ = \
    \frac{G^2}{2}
    H^{\alpha\beta} L_{\alpha\beta} \ = \ 
    12 G^2 g_V^2  M_{\Delta^-} 
    \left( M_{\Delta^-} - 2 E_{\overline{\nu}}\right)
    \left( E_e E_{\overline{\nu}} - {\bf p}_e\cdot {\bf p}_{\overline{\nu}} \right) \ ,
\end{equation}
from which
the delta decay width can be determined by standard methods,
\begin{align}
\Gamma_{\Delta^-} & =
\frac{1}{2M_{\Delta^-}}
\int 
\frac{d{\bf p}_e}{4\pi E_e}
\frac{d{\bf p}_{\overline{\nu}}}{4\pi E_{\overline{\nu}}}
\frac{d{\bf p}_0}{4\pi E_0} (2\pi)^2 \delta^2(p_- - p_0 - p_e - p_{\overline{\nu}})
|{\cal M}|^2 
\nonumber\\
& =
3 \frac{G^2 g_V^2}{2\pi}
\int\  dE_e\  dE_{\overline{\nu}}\ \delta(Q - E_e - E_{\overline{\nu}})
\ +\ {\cal O}\left(Q^n/M_\Delta^n\right)
\nonumber\\
& =
3 \frac{G^2 g_V^2 Q}{2\pi}
\ +\ {\cal O}\left(Q^n/M_\Delta^n\right)
\ ,
\label{eq:1p1decaywidth}
\end{align}
where $Q= M_{\Delta^-} - M_{\Delta^0}$ and we have retained only the leading terms in
an expansion in $Q/M_\Delta$ and evaluated the vector form factor at $g_V(q^2=0) \equiv g_V$. 
The electron and neutrino masses have been set to zero, and the inclusion of non-zero masses will lead to a phase-space factor, $f_1$, 
reducing the width shown in Eq.~(\ref{eq:1p1decaywidth}), 
and which becomes $f_1=1$ in the massless limit.

\section{\texorpdfstring{$\beta$}{Beta}-Decay in \texorpdfstring{$1+1$}{1+1} Dimensions: Finite \texorpdfstring{$L$}{L} and Non-zero Spatial Lattice Spacing}
\label{app:beta1p1aL}
\noindent 
The previous appendix computed the $\beta$-decay rate 
in $1+1$D in infinite volume and in the continuum.
However, lattice  calculations of such processes will necessarily be performed with a non-zero lattice spacing and a finite number of lattice points.
For calculations done on a Euclidean-space lattice, significant work has been done to develop the machinery used to extract physically meaningful results. 
This formalism was initially pioneered by L\"{u}scher~\cite{Luscher:1985dn,Luscher:1986pf,Luscher:1990ux} for hadron masses and two-particle scattering, and has been extended to more complex systems relevant to electroweak processes (Lellouch-L\"{u}scher)~\cite{Lellouch:2000pv,Detmold:2004qn,Christ:2005gi,Kim:2005gf,Hansen:2012tf,Meyer:2011um,Briceno:2012yi,Feng:2014gba,Briceno:2012yi,Meyer:2012wk,Bernard:2012bi,Agadjanov:2014kha,Briceno:2014uqa,Briceno:2015csa,Briceno:2015tza,Briceno:2019opb,Briceno:2020vgp} and 
to nuclear physics~\cite{Beane:2003da,Detmold:2004qn,Beane:2007qr,Beane:2007es,Luu:2010hw,Luu:2011ep,Davoudi:2011md,Meyer:2012wk,Briceno:2012rv,Briceno:2013lba,Briceno:2013bda,Briceno:2013hya,Briceno:2014oea,Grabowska:2021qqz}.   
L\"{u}scher's method was originally derived from an analysis of Hamiltonian dynamics in Euclidean space and later from a field theoretic point of view directly from correlation functions. 
The challenge is working around the Maiani-Testa theorem~\cite{Maiani:1990ca} and reliably determining Minkowski-space matrix elements from Euclidean-space observables.
This formalism has been used successfully for a number of important quantities, and continues to be the workhorse for Euclidean-space computations.

As quantum simulations provide observables directly in Minkowski space,
understanding the finite-volume and non-zero lattice spacing artifacts requires a similar but different analysis than in Euclidean space.\footnote{Estimates of such effects in model 1+1 dimensional simulations can be found in Ref.~\cite{PhysRevD.103.014506}.}  
While the method used in Euclidean space of determining S-matrix elements for scattering processes from energy eigenvalues can still be applied, Minkowski space simulations will also allow for a direct evaluation of scattering processes, removing some of the modeling that remains in Euclidean-space calculations.\footnote{
For example, the energies of states in different volumes are different, 
and so the elements of the scattering matrix are constrained over 
a range of energies and not at one single energy,
and {\it a priori} unknown interpolations are modeled.
}
Neglecting electroweak interactions beyond $\beta$-decay means that the final state leptons are non-interacting (plane-waves when using periodic boundary conditions),
and therefore the modifications to the density of states due to interactions, as encapsulated within the L\"{u}scher formalism, are absent.

With Hamiltonian evolution of a system described within a finite-dimensional Hilbert space, the persistence amplitude of the initial state coupled to final states via the weak Hamiltonian will be determined by the sum over oscillatory amplitudes.  
For a small number of final states, the amplitude will return to unity after some finite period of time.  
As the density of final states near the energy of the initial state becomes large, there will be cancellations among the oscillatory amplitudes, and the persistence probability will begin to approximate the ``classic" exponential decay over some time interval.  
This time interval will extend  to infinity as the density of states tends to a continuous spectrum. 
It is important to understand how to reliably extract an estimate of the decay rate, with a quantification of systematic errors, from the amplitudes measured in a quantum simulation.  
This is the subject of future work, but here a simple model will be used to demonstrate some of the relevant issues.

\begin{figure}[!ht]
    \centering
    \includegraphics[width=12cm]{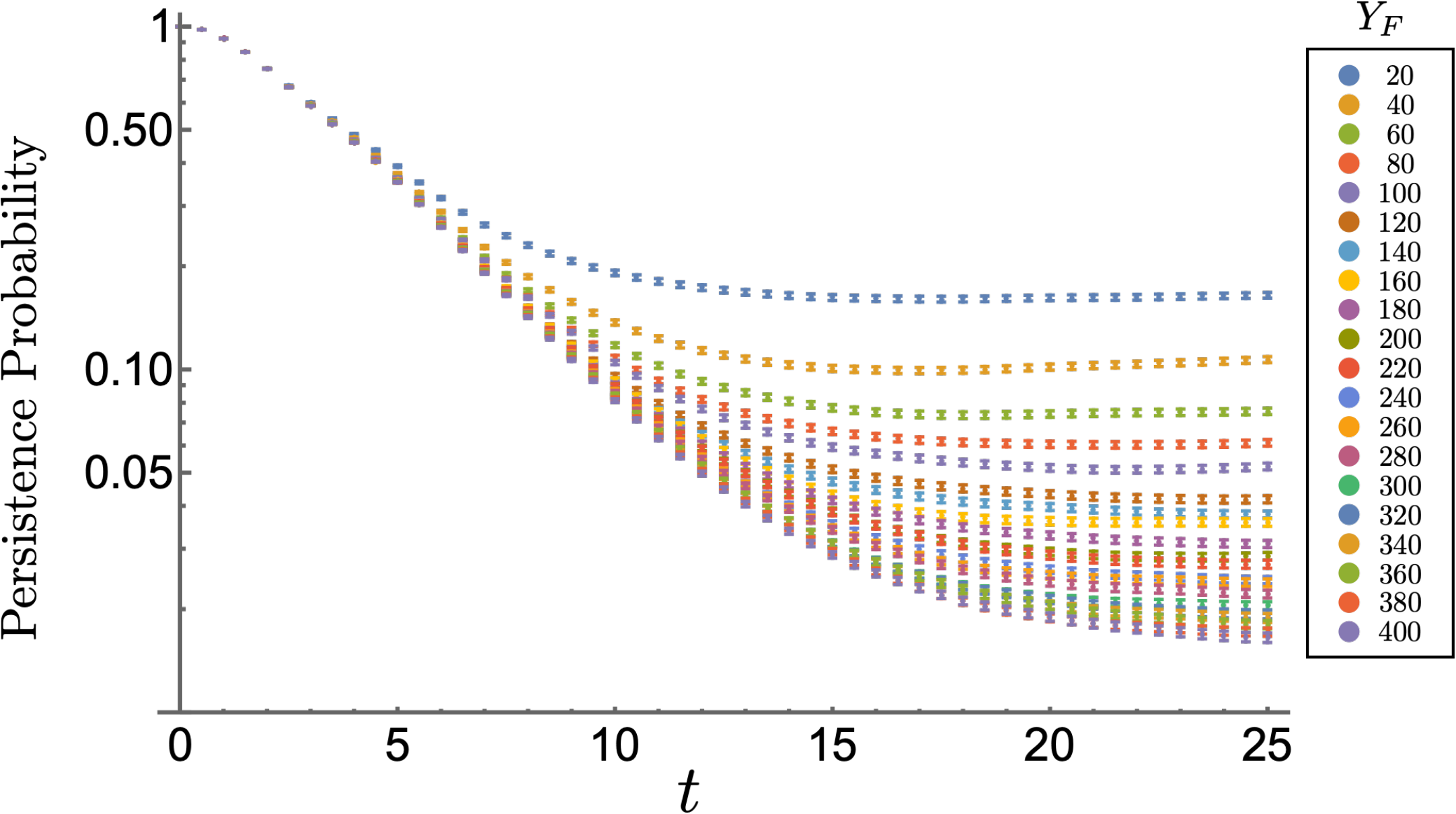}
    \caption{
    Ensemble averages (over 2000 random samples) of the persistence probability of an initial state in one sector of a strong Hamiltonian undergoing weak decay to states in a different sector, as described in this appendix. 
    The different colored points are results from calculations with an increasing number of final states, $Y_F$. 
    The weak coupling scales so that the decay probability converges to a well-defined value as the density of final states tends to a continuum.
    }
    \label{fig:ExpDecay}
\end{figure}

Consider the weak decay of a strong eigenstate in one sector to a strong eigenstate in a different sector (a sector is defined by its strong quantum numbers).
For this demonstration, 
we calculate the persistence probability of the initial state, averaged over random weak and strong Hamiltonians and initial states, as the number of states  below a given energy increases (i.e. increasing density of states).
Concretely, the energy eigenvalues of the initial strong sector range from 0 to 1.1, and 10 are selected randomly within this interval.
The initial state is chosen to be the one with the fifth lowest energy.
The eigenvalues in the final strong sector range between 0 and 2.03, and $Y_f = $ 20 to 400 are selected.
The weak Hamiltonian that induces transitions between the 10 initial states to the $Y_f$ final states is a dense matrix with each element selected randomly.
The weak coupling constant is scaled so that
$G^2\rho_f$ is independent of the number of states, where $\rho_f$ is the density of states.
This allows for a well-defined persistence probability as $Y_f \to \infty$.
For this example, the elements of the weak Hamiltonian were chosen between $\pm w_f$, 
where $w_f = 1/(2 \sqrt{Y_F})$.
Figure~\ref{fig:ExpDecay} shows the emergence of the expected exponential decay as the number of available final states tends toward a continuous spectrum.
In a quantum simulation of a lattice theory, the density of states increases with $L$, and the late-time deviation from exponential decay will exhibit oscillatory behavior, as opposed to the plateaus found in this statistically averaged model.
The very early time behavior of the probability is interesting to note, and exhibits a well-known behavior, e.g., Refs.~\cite{Urbanowski_2017,Giacosa_2017}.  
It is, as expected, not falling exponentially, which sets in over time scales set by the energy spectrum of final states.

Only small lattices are practical for near-term simulation and lattice artifacts will be important to quantify. Relative to the continuum, a finite lattice spacing modifies the energy-momentum relation and introduce a momentum cut-off on the spectra.  
However, if the initial particle has a mass that is much less than the cut-off, these effects should be minimal as the energy of each final state particle is bounded above by the mass of the initial particle.
As has been shown in this appendix, working on a small lattice with its associated sparse number of final states, will lead to significant systematic errors when extracting the decay rates directly from the persistence probabilities.
Further work will be necessary to determine how to reliably estimate these errors.

\section{\texorpdfstring{$\beta$}{Beta}-Decay Circuits}
\label{app:BetaCircuits}
\begin{figure}[!ht]
    \centering
    \includegraphics[width=10cm]{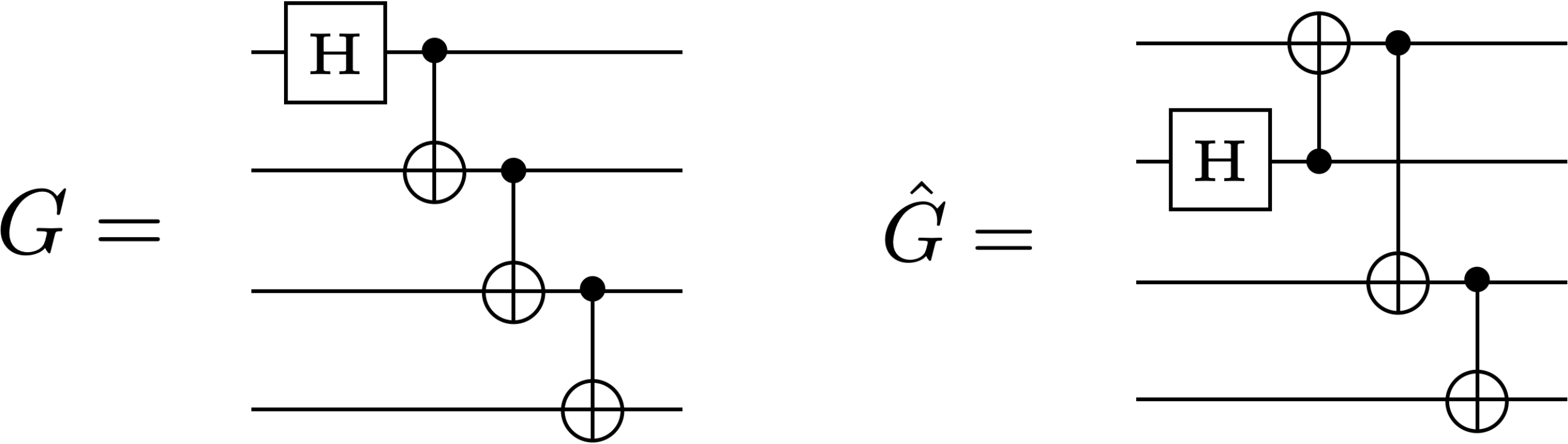}
    \caption{
    Two GHZ state preparation circuits.}
    \label{fig:GHZCirc}
\end{figure}

\begin{figure}[!ht]
    \centering
    \includegraphics[width=\columnwidth]{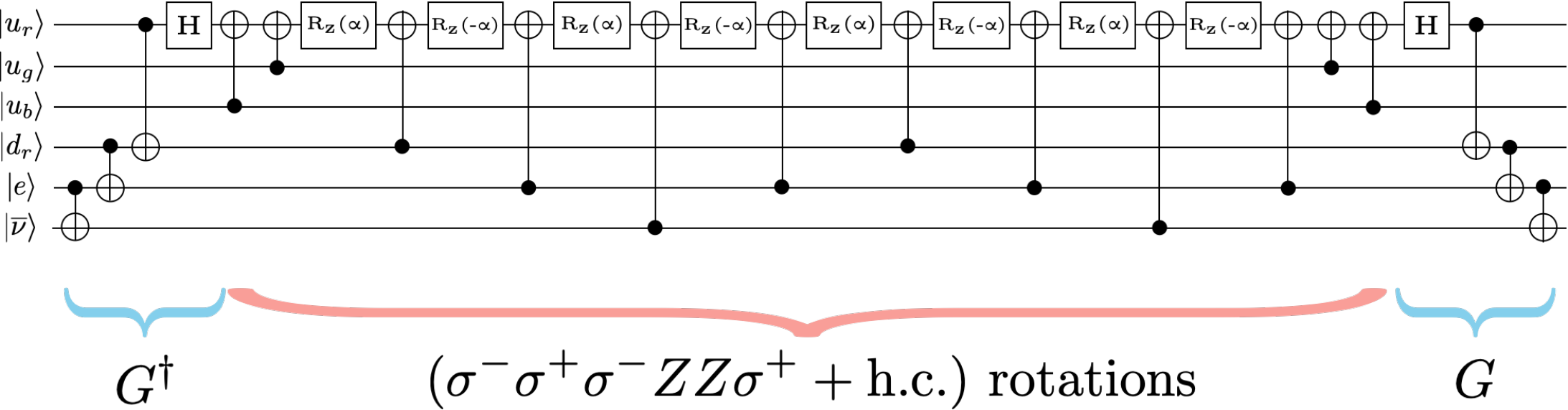}
    \caption{
    A quantum circuit that provides the time evolution associated with the 
    $\sigma^-_{\overline{\nu}} \sigma^+_e \sigma^-_{d,r}$ $ Z_{u,b} Z_{u,g} \sigma^+_{u,r}$ operator in the $\beta$-decay Hamiltonian, with 
    $\alpha = \sqrt{2} G t/8$.
    }
    \label{fig:BetaCirc}
\end{figure}

\noindent
The quantum circuits that implement the Trotterized time-evolution of the 
$\beta$-decay Hamiltonian are  similar to those
presented in Chapter \ref{chap:1p1dQCD} 
to implement the strong-interaction dynamics,
and here the differences between the two will be highlighted.
The $\beta$-decay Hamiltonian in both the standard and tilde layouts, Eqs.~(\ref{eq:KSHamJWmap}) and~(\ref{eq:tildeBeta}), contains terms of the form
\begin{align}
    H_{\beta} \sim&\ (\sigma^- \sigma^+ \sigma^- \sigma^+ + {\rm h.c.}) + (\sigma^- \sigma^+ \sigma^+ \sigma^- + {\rm h.c.}) 
    =\ \frac{1}{8}(XXXX+YYXX-YXYX\nonumber \\
    &+YXXY+XYYX-XYXY+XXYY+YYYY) + \frac{1}{8}(XXXX + YYXX \nonumber \\
    &+ YXYX - YXXY - XYYX + XYXY + XXYY + YYYY)
    \ ,
\end{align}
which can be diagonalized by the GHZ state-preparation circuits, $G$ and $\hat{G}$, shown in Fig.~\ref{fig:GHZCirc}. 
In the GHZ basis, it is found that
\begin{align}
    &G^{\dagger}(XXXX+YYXX-YXYX+YXXY+XYYX-XYXY+XXYY+YYYY)G \nonumber \\ 
    &= IIIZ - ZIIZ + ZZIZ - ZZZZ -IZIZ + IZZZ - IIZZ + ZIZZ 
    \ ,
\end{align}
and 
\begin{align}
    &\hat{G}^{\dagger}(XXXX + YYXX + YXYX - YXXY - XYYX + XYXY + XXYY + YYYY)\hat{G} \nonumber \\ 
    &= IIZI - ZIZI -ZZZZ+ZZZI+IZZZ-IZZI-IIZZ+ZIZZ
    \ .
\end{align}
Once diagonalized the circuit is a product of diagonal rotations, see Fig.~\ref{fig:BetaCirc} for an example of the quantum circuit that provides the time evolution associated with
$\sigma^-_{\overline{\nu}} \sigma^+_e \sigma^-_{d,r} Z_{u,b} Z_{u,g} \sigma^+_{u,r}$.
By diagonalizing with both $G$ and $\hat{G}$ and arranging terms in the Trotterization so that operators that act on the same quarks are next to each other, 
many of the CNOTs can be made to cancel.
Also, an ancilla can be used to efficiently store the parity of the string of $Z$s between the $\sigma^{\pm}$.
%

\section{Resource Estimates for Simulating \texorpdfstring{$\beta$}{Beta}-Decay Dynamics}
\label{app:LongJW}
\begin{figure}[!ht]
    \centering
    \includegraphics[width=15cm]{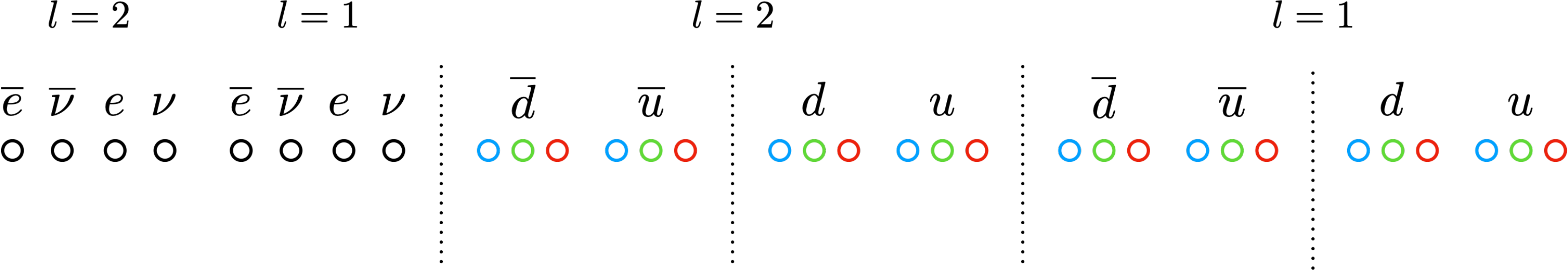}
    \caption{
    A qubit layout that is efficient for the simulation of $\beta$-decay. Shown is an example for $L=2$.}
    \label{fig:L2BetaLayout}
\end{figure}

\noindent 
For multiple lattice sites, it is inefficient to work with leptons in the tilde basis. 
This is due to the mismatch between the local four-Fermi interaction 
and the non-local tilde basis eigenstates. 
As a result, the number of terms in the $\beta$-decay component of the Hamiltonian will scale as $\mathcal{O}(L^2)$ in the tilde basis, 
as opposed to $\mathcal{O}(L)$ in the local occupation basis.
This appendix explores a layout different from the one in Fig.~\ref{fig:L1layout}, which is optimized for the simulation of $\beta$-decay on larger lattices.
To minimize the length of JW $Z$ strings, all leptons are placed at the end of the lattice, see Fig.~\ref{fig:L2BetaLayout}.
After applying the JW mapping, the $\beta$-decay operator becomes
\begin{align}
   H_{\beta} \rightarrow \frac{G}{\sqrt{2}}\sum_{l = 0}^{L-1}\sum_{c=0}^2\bigg (&\sigma^-_{l,\nub} \sigma^+_{l,e} \sigma^-_{l,d,c} Z^2 \sigma^+_{l,u,c} \: - \: \sigma^+_{l,\eb}Z^2\sigma^-_{l,\nu} \sigma^-_{l,d,c} Z^2 \sigma^+_{l,u,c} \: + \: \sigma^-_{l,\nub} \sigma^+_{l,e} \sigma^-_{l,\db,c} Z^2 \sigma^+_{l,\ub,c}  \nonumber \\ 
   &- \: \sigma^+_{l,\eb} Z^2 \sigma^-_{l,\nu} \sigma^-_{l,\db,c} Z^2 \sigma^+_{l,\ub,c} \: + \: \sigma^+_{l,e} \sigma^-_{l,\nu} \sigma^-_{l,\db,c} Z^8 \sigma^+_{l,u,c} \: - \: \sigma^+_{l,\eb} \sigma^-_{l,\nub} \sigma^-_{l,\db,c}Z^8 \sigma^+_{l,u,c} \nonumber \\
   &+ \: \sigma^+_{l,e} \sigma^-_{l,\nu} \sigma^+_{l,\ub,c} Z^2 \sigma^-_{l,d,c} \: - \: \sigma^+_{l,\eb} \sigma^-_{l,\nub} \sigma^+_{l,\ub,c} Z^2 \sigma^-_{l,d,c} \: + \: {\rm h.c.} \bigg ) \ .
   \label{eq:HBlow}
\end{align}

Using the techniques outlined in App.~\ref{app:BetaCircuits} 
to construct the relevant quantum circuits, 
the resources required per Trotter 
step of 
$H_{\beta}$ are estimated to be
\begin{align}
    R_Z  \ :& \ \ 192L \ ,\nonumber \\
   \text{Hadamard} \ :& \ \ 48L  \ ,\nonumber \\
    \text{CNOT} \ :& \ \ 436 L \ .
\end{align}
For small lattices, $L\lesssim 5$, it is expected that use of the tilde basis will be more efficient and these estimates should be taken as an upper bound.
Combining this with the resources required to time evolve with the rest of the Hamiltonian, see Chapter \ref{chap:1p1dQCD}, the total resource requirements per Trotter step are estimated to be
\begin{align}
    R_Z  \ :& \ \ 264L^2 -54L +77 \ ,\nonumber \\
   \text{Hadamard} \ :& \ \ 48L^2 + 20L +2 \ ,\nonumber \\
    \text{CNOT} \ :& \ \ 368L^2 + 120L+74 \ .
\end{align}
It is important to note that the addition of $H_{\beta}$ does not contribute to the quadratic scaling of resources as it is a local operator. 
Recently, the capability to produce multi-qubit gates natively with similar fidelities to two-qubit gates has also been demonstrated~\cite{Katz:2022ajk,Katz:2022czu,andrade:2021pil}.
This could lead to dramatic reductions in the resources required and, for example, the number of multi-qubit terms in the Hamiltonian scales as
\begin{equation}
    \text{Multi-qubit terms} \ : \ \ 96 L^2 -68L+22 \ .
\end{equation}
The required number of CNOTs and, for comparison, the number of multi-qubit terms in the Hamiltonian, for a selection of different lattice sizes are given in Table~\ref{tab:cnot}. 
\begin{table}[!ht]
\centering
\renewcommand{\arraystretch}{1.2}
\begin{tabularx}{0.48\textwidth}{||c | Y | Y ||}
 \hline
 $L$ & CNOTS & Multi-Qubit Terms \\
 \hline\hline
 5 & 9874 & 2082 \\
 \hline
 10 & 38,074 & 8942\\
 \hline
 50 & 926,074 & 236,622\\
 \hline
 100 & 3,692,074 & 953,222\\
 \hline
\end{tabularx}
\renewcommand{\arraystretch}{1}
\caption{The CNOT-gate requirements to perform one Trotter step of time evolution 
of $\beta$-decay for a selection of lattice sizes. For comparison, the number of multi-qubit terms in the Hamiltonian is also given.}
\label{tab:cnot}
\end{table}
Note that these estimates do not include the resources required to prepare the initial state.

\section{Technical Details on the Quantinuum {\tt H1-1} Quantum Computer}
\label{app:H1specs}
\noindent 
For completeness, this appendix contains a brief description of Quantinuum's {\tt H1-1} 20 
trapped ion quantum computer (more details can be found in~\cite{h1-1}). 
The {\tt H1-1} system uses the System Model {\tt H1} design, 
where unitary operations act on a single line of ${}^{172}$Y${}^+$ ions induced by lasers. 
The qubits are defined as the two hyperfine clock states in the ${}^2S_{1/2}$ ground state of ${}^{172}$Y${}^+$. 
Since the physical position of the ions can be modified, 
it is possible to apply two-qubit gates to any pair of qubits, 
endowing the device with all-to-all connectivity. 
Moreover, there are five different physical regions where these gates can be applied in parallel. Although we did not use this feature, it is also possible to perform a mid-circuit measurement of a qubit, i.e., initialize it and reuse it (if necessary).

The native gate set for {\tt H1-1} is the following,
\begin{equation}
    U_{1q}(\theta,\phi)=e^{-i\frac{\theta}{2}[\cos(\phi) X+\sin(\phi) Y]}\ , \quad R_Z(\lambda)=e^{-i\frac{\lambda}{2}Z}\ , \quad ZZ=e^{-i\frac{\pi}{4}ZZ} \ ,
\end{equation}
where $\theta$ in $U_{1q}(\theta,\phi)$
can only take the values $\{\frac{\pi}{2},\pi\}$, 
and arbitrary values of $\theta$ can be obtained by combining several single-qubit gates, $\tilde{U}_{1q}(\theta,\phi)=U_{1q}(\frac{\pi}{2},\phi+\frac{\pi}{2}) . R_Z(\theta) . U_{1q}(\frac{\pi}{2},\phi-\frac{\pi}{2})$. 
Translations between the gates used in the circuits shown in the main text and appendices to the native ones are performed automatically by {\tt pytket}~\cite{sivarajah2020t}.
The infidelity of the single- and two-qubit gates, as well as the error of the SPAM operations, are shown in Table~\ref{tab:H1err}.
\begin{table}[!ht]
\centering
\renewcommand{\arraystretch}{1.2}
\begin{tabularx}{0.7\textwidth}{||r | Y | Y | Y ||}
 \hline
  & Min & Average & Max \\
 \hline\hline
 Single-qubit infidelity & $2\times 10^{-5}$ & $5\times 10^{-5}$ & $3\times 10^{-4}$ \\
 \hline
 Two-qubit infidelity & $2\times 10^{-3}$ & $3\times 10^{-3}$ & $5\times 10^{-3}$ \\
 \hline
 SPAM error & $2\times 10^{-3}$ & $3\times 10^{-3}$ & $5\times 10^{-3}$ \\
 \hline
\end{tabularx}
\renewcommand{\arraystretch}{1}
\caption{
Errors on the single-qubit, two-qubit and SPAM operations, with their minimum, average and maximum values.
}
\label{tab:H1err}
\end{table}
%

\section{Time Evolution Under the Full \texorpdfstring{$\beta$}{Beta}-Decay Operator}
\label{app:betaFull}
\begin{figure}[!ht]
    \centering
    \includegraphics[width=12cm]{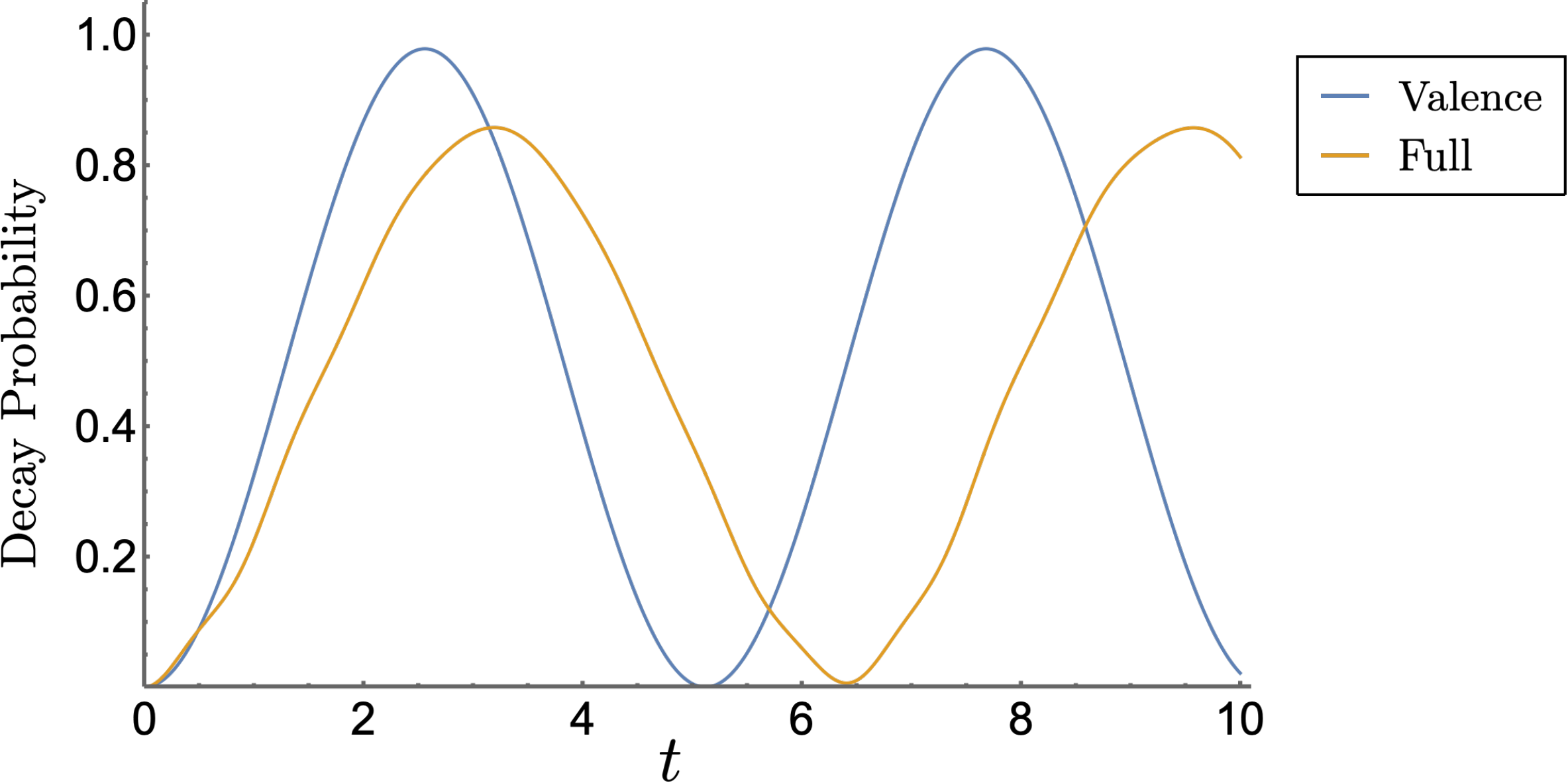}
    \caption{The probability of $\beta$-decay using both the approximate $\beta$-decay operator which only acts on valence quarks (blue) and the full operator (orange).}
    \label{fig:valfull}
\end{figure}

\noindent
The simulations performed in Sec.~\ref{sec:BetaSim} kept only the terms in the $\beta$-decay Hamiltonian which act on valence quarks, see Eq.~(\ref{eq:tildeBetaRed}). 
This appendix examines how well this valence quark $\beta$-decay operator approximates the full operator, Eq.~(\ref{eq:tildeBeta}), for the parameters used in the main text.
Shown in Fig.~\ref{fig:valfull} is the decay probability when evolved with both the approximate and full operator as calculated through exact diagonalization of the Hamiltonian.
The full $\beta$-decay operator has multiple terms that can interfere leading to a more jagged decay probability.
The simulations ran on {\tt H1-1} only went out to $t=2.5$ where the error of the approximate operator is $\sim 20\%$.

\end{subappendices}
\chapter{State-preparation and Time-evolution circuits for simulation of neutrinoless double-beta decay on a 1+1D lattice and QCD on larger lattices}
\label{chap:0vBBandlargeQCD}

\section{Introduction}


As discussed in Sec. \ref{sec:majmassforneutrino_SM1p1D}, the Jordan-Wigner mapped 1+1D SU(3) Kogut-Susskind Hamiltonian found in Eq. \ref{eq:KSHamJWmap} of Chapter \ref{chap:1p1dSM} can be combined with the Majorana mass term for the neutrinos laid out in Eq. \ref{eq:majaHami} to simulate neutrinoless double beta decay ($0 \nu \beta \beta$). Ref. \cite{davoudi2024long} has used the reactions $n n \rightarrow p p e^- e^-$ and $\Sigma^- \rightarrow \Sigma^{+} e^- e^-$ to compute the matrix elements of $0 \nu \beta \beta$ in a multi-nucleon system . In this chapter, the protocols for state-preparation and real-time evolution on quantum devices for the 1+1D analogs of these reactions, $\Delta^- \Delta^- \rightarrow \Delta^0 \Delta^0 e^- e^-$ and $\Delta^- \rightarrow \Delta^+ e^- e^-$, are discussed. 

Additionally, IBM has recently produced several devices publicly accessible through the cloud that have 100 - 150 qubits\cite{ibmq}. These devices have been used to run state-preparation and real-time dynamics, which has included the formulation and testing of SC-ADAPT-VQE on the lattice Schwinger model \cite{Farrell:2023fgd}, real-time hadron dynamics in the lattice Schwinger model \cite{Farrell:2024fit}, dynamics of an SU(3) gauge field on a 2+1D lattice without fermions \cite{Ciavarella:2024fzw}, dynamics of quark-pair string-breaking in an SU(2) Kogut-Susskind Hamiltonian theory on a 1+1D lattice \cite{ciavarella2024string}, and scalar wavepacket scattering in 1+1D \cite{zemlevskiy2024scalable}. 100-150 qubits can theoretically produce a lattice large enough to simulate scattering simulations under a 1+1D SU(3) Kogut-Susskind Hamiltonian, so in Sec. \ref{sec:0vBB_timeevcircforsupercond} I propose designs for circuits that can Trotterize time-evolution under the 1+1D Kogut-Susskind Hamiltonian on superconducting devices with a minimal SWAP gate overhead. 

\section{Hamiltonians}
\label{sec:0vBB_stateprep_Hamiltonians}
\subsection{Open Boundary Conditions}

The Hamiltonians used in Chapters \ref{chap:1p1dQCD} and \ref{chap:1p1dSM} represent QCD in open boundary conditions (OBC). 
That is, the invariance with respect to a translation by an even number of lattice inherent to the staggered lattice is broken only at the edges. As a result, as discussed in Ref. \cite{rahman:2022rlg} and Chapters \ref{chap:1p1dQCD} and \ref{chap:1p1dSM}, the gauge fields can be moved to one of the edges of the lattice through fixing to the axial gauge, where it could be set to zero by convention. 

The actual Hamiltonian used for neutrinoless double beta decay on an OBC lattice in this chapter is the same as the one used in Eq. \ref{eq:KSHamJWmap}, with two extra terms. To recap, the lattice is the same as the one in Fig. \ref{fig:EWlayoutTilde}, with one register of qubits standing in for the quarks and being governed by an SU(3) Kogut-Susskind Hamiltonian (labeled $H_{quarks} + H_{el}$) and another register of qubits standing in for the leptons and being governed by a mass term and a hopping/kinetic term (together labeled $H_{leptons}$). A $\beta$-decay term is the only interaction between these two registers. The two extra terms are the Majorana mass term, $H_{majorana}$, laid out in Eq. \ref{eq:majaHami} which acts on the neutrino qubits and makes it possible for a double-$\beta$ decay to be neutrinoless and a chemical potential term, $H_{\mu_{II}}$, which acts on the quarks and creates a mass-hierarchy intended to make single-$\beta$ decay energetically unfavorable but double-$\beta$ decay energetically favorable. This is done by making the ground state energy of the former higher than the initial state's energy and the ground state energy of the latter lower than the initial state's energy. The full Hamiltonian can be written as so:

\begin{align}
    H = &\
    H_{\rm{quarks}} + H_{el} + H_{\rm {\mu_{II}}} + H_{\rm{leptons}} + H_{{\rm Majorana}} + H_{\beta}, 
    \label{eq:0vBB_stateprep_fullhamiltonian}
\end{align}
\noindent where

\begin{align}
    H_{\rm{quarks}} 
     \rightarrow &\
    \frac{1}{2} \sum_{l=0}^{L-1} \sum_{f=u,d}\sum_{c=0}^{2}  m_f\left ( Z_{l,f,c} - Z_{l,\overline{f},c} + 2\right )  \nonumber \\
    & -\frac{1}{2} \sum_{l=0}^{L-1} \sum_{f=u,d} \sum_{c=0}^{2}\left [ \sigma^+_{l,f,c} Z^5 \sigma^-_{l,\overline{f},c}  + (1-\delta_{l,L-1})  \sigma^+_{l,\overline{f},c} Z^5 \sigma^-_{l+1,f,c} + {\rm h.c.} \right ]\ , \nonumber \\[4pt]
    H_{el}
    \rightarrow  &\ \frac{g^2}{2} \sum_{n=0}^{2L-2}(2L-1-n)\left( \sum_{f=u,d} Q_{n,f}^{(a)} \, Q_{n,f}^{(a)} \ + \
        2 Q_{n,u}^{(a)} \, Q_{n,d}^{(a)}
         \right)    \nonumber \\
         &+ \: g^2 \sum_{n=0}^{2L-3} \sum_{m=n+1}^{2L-2}(2L-1-m) \sum_{f=u,d} \sum_{f'=u,d} Q_{n,f}^{(a)} \, Q_{m,f'}^{(a)} \ , \nonumber \\[4pt]
H_{\mu_{II}} \rightarrow &\ -\frac{\mu_{II}}{16}\sum_{n,m=0}^{2L-1} \sum_{f,f'=u,d} \sum_{c,c'=0}^2 (-1)^{f+f'}(\sigma^z_{n,f,c} \sigma^z_{m,f',c'})\ ,
    \nonumber\\[4pt]
    H_{\rm{leptons}} 
    \rightarrow  &\ \frac{1}{2} \sum_{l=0}^{L-1} \sum_{f=e,\nu}  m_f\left ( Z_{l,f} - Z_{l,\overline{f}} + 2\right ) \nonumber \\ \: - \: & \frac{1}{2} \sum_{l=0}^{L-1} \sum_{f=e,\nu} \left [ \sigma^+_{l,f} Z \sigma^-_{l,\overline{f}}  + (1-\delta_{l,L-1})  \sigma^+_{l,\overline{f}} Z \sigma^-_{l+1,f} + {\rm h.c.} \right ] \ ,
    \nonumber\\[4pt]
    H_{\rm{Majorana}} 
    \rightarrow  &\ \frac{m_M}{2} \sum_{l=0}^{L-1} \left [ \sigma^+_{l,\nu} Z_{l,\bar{e}} \sigma^+_{l,\bar{\nu}} + {\rm h.c.} \right ] \ ,
    \nonumber\\[4pt]
    H_{{\rm \beta}}
    \rightarrow  \
    \frac{G_F}{\sqrt{2}}\sum_{l = 0}^{L-1}\sum_{c=0}^2 & \bigg ( \sigma^-_{l,\nub} Z^6 \sigma^+_{l,e} \sigma^-_{l,d,c} Z^2 \sigma^+_{l,u,c} \: - \: \sigma^+_{l,\eb} Z^8 \sigma_{l,\nu}^- \sigma_{l,d,c}^- Z^2 \sigma^+_{l,u,c} 
    \: - \: \sigma^-_{l,\nub} Z^{2-c} \sigma^-_{l,\db,c} \sigma^+_{l,\ub,c} Z^c \sigma^+_{l,e} \nonumber \\ + \: \sigma^+_{l,\eb}Z^{3-c}\sigma^-_{l,\db,c} & \sigma^+_{l,\ub,c} Z^{1+c} \sigma^-_{l,\nu} \: - \: \sigma^-_{l,\db,c} Z^{3+c} \sigma^+_{l,e} \sigma^-_{l,\nu} Z^{5-c} \sigma^+_{l,u,c} \: - \: \sigma^+_{l,\eb} \sigma^-_{l,\nub} \sigma^-_{l,\db,c}Z^{10}\sigma^+_{l,u,c} \nonumber \\ 
     - \: \sigma^+_{l,\ub,c} Z^c \sigma^+_{l,e} & \sigma^-_{l,\nu} Z^{2-c} \sigma^-_{l,d,c} \: - \: \sigma^+_{l,\eb} \sigma^-_{l,\nub} \sigma^+_{l,\ub,c} Z^4 \sigma^-_{l,d,c} \: + \: {\rm h.c.}\bigg )
     \ ,
     \label{eq:OBC_0nuBB_fullham}
\end{align}

\noindent where $\mu_{II}$ is a constant tuned to whatever value is necessary to fix the desired mass hierarchy and $m_{M}$ the neutrino Majorana mass. The color charge operator $Q_{n,f}^{(a)}$ is defined in Eq. \ref{eq:QnfQmfp}. 


\subsection{Periodic Boundary Conditions}
\label{sec:0vBB_stateprep_Hamiltonians_periodic_boundary_conditions}

Open boundary conditions come with boundary effects from the broken translational symmetries, and to avoid these boundary effects one can utilize periodic boundary conditions (PBC). Practically, this means that there is a link between the first and last qubit on the register, across which there is a kinetic term and through which color charges on one end can affect gauge fields on the other end via Gauss's Law. In this chapter, the periodicity is applied separately to the lepton and quark registers. 

The mass terms, $H_{majorana}$, and $H_{\beta}$ are not affected by this change as they do not act outside their physical lattice site. Neither is $H_{\mu_{II}}$, as it is simply an all-to-all interaction. For $H_{el}$, one cannot simply remove the need to explicitly represent the gauge field by moving it to the boundary through gauge-fixing the way it is done for OBC. This is because due to the link between the first and last sites, such a process applied to PBC will always leave at least on explicit gauge degree of freedom, which corresponds to to an overall contribution to the gauge field on every link on the lattice. For the Schwinger model, it has been found that this degree of freedom manifests itself in the form of the phenomenon of flux unwinding, where the overall energy of the lattice can be incremented by a charge-pair circumnavigating the lattice\cite{nagele2019flux}. To address the ambiguity caused by this, the overall contribution degree of freedom is held to be constant and $H_{el}$ is rewritten accordingly, to reflect only the variable parts of the chromoelctric Hamiltonian\cite{farrell2024betabeta}:

\begin{equation}
    H_{el} \rightarrow \frac{g^2}{2} \sum_{n = 0}^{2L - 1} \sum_{d = 1}^{\lambda} (-d + \frac{d^2}{2L}) \Bigg( \Big( \sum_f Q^{(a)}_{n, f} \Big) \Big( \sum_f Q^{(a)}_{n + d, f} \Big) \Bigg) . 
    \label{eq:HchromoPBC_0vBB}
\end{equation}

\noindent The motivation behind the form of the re-write is that the coefficient of $Q^{(a)}_{n, f} Q^{(a)}_{m, f'}$ should be directly proportional to the distance $d$ on the staggered lattice between $n$ and $m$. This is because under Gauss's Law, a gauge field excitation created by a non-color-singlet site propagates to all lattice sites until it reaches the correct non-color-singlet site to cancel the excitation. Under the Weyl gauge, such behavior would manifestly mean that the magnitude of this excitation's effect on the chromoelectric term would need to be directly proportional to the distance between the two excitations, so the same should be true for the axial gauge. The minus sign was chosen because it was found that it was necessary to make the $H_{el}$ from Eq. \ref{eq:HchromoPBC_0vBB} equivalent to its counterpart in \ref{eq:OBC_0nuBB_fullham} for a hypothetical excitation arbitrarily far away from the boundaries, where PBC and OBC should be equivalent. The factor of $\frac{d^2}{2L}$ was added because my colleague Roland Farrell found that it was necessary to preserve the expected correlation between color-charges separated by distance $d$ \cite{farrell2024betabeta},

\begin{align}
\langle \hat{Q}_n^{(a)} \hat{Q}_{n+d}^{(a)}\rangle \sim \ \exp^{-c_1 d \, m_{\text{hadron}}} .
\label{eq:0vBB_expectedchargecorrelation}
\end{align}

\noindent The expected direct proportionality between the $H_{el}$ contribution of an excitation and the distance between color charges is preserved in the infinite volume limit. 

For the kinetic Hamiltonian, naively one would find the PBC Hamiltonian simply by adding a fermionic hopping term between the last staggered lattice site and the first. However, PBC means there are two ways of getting from one qubit on the lattice to the other, and if the number of qubits in the ``down" state on the lattice is even, the Jordan-Wigner mapping would result in opposite signs between the two choices. To resolve this, the choice of path that does not go between the last site and the first is held to have the ``correct" sign, and a factor of $(-1)^{L + B + 1}$ is appended to the terms between the last staggered lattice site and the first to enforce this\cite{farrell2024betabeta}. Just like with the $\frac{d^2}{2L}$ term in $H_{el}$'s case, the naively expected behavior at the boundary is reduced to what happens in the infinite volume limit. This change results in $H_{quarks}$ and $H_{leptons}$ becoming

\begin{align}
    H_{\rm{quarks}} 
     \rightarrow &\
    \frac{1}{2} \sum_{l=0}^{L-1} \sum_{f=u,d}\sum_{c=0}^{2}  m_f\left ( Z_{l,f,c} - Z_{l,\overline{f},c} + 2\right ) -  \nonumber \\
    \frac{1}{2} & \sum_{l=0}^{L-1} \sum_{f=u,d} \sum_{c=0}^{2}\left [ \sigma^+_{l,f,c} Z^5 \sigma^-_{l,\overline{f},c}  + (1-(1 + (-1)^{L + B})\delta_{l,L-1})  \sigma^+_{l,\overline{f},c} Z^5 \sigma^-_{l+1,f,c} + {\rm h.c.} \right ]\ , \nonumber \\[4pt]
    H_{\rm{leptons}} 
    \rightarrow  &\ \frac{1}{2} \sum_{l=0}^{L-1} \sum_{f=e,\nu}  m_f\left ( Z_{l,f} - Z_{l,\overline{f}} + 2\right )  \: - \: \nonumber \\[4pt]
    \frac{1}{2} & \sum_{l=0}^{L-1} \sum_{f=e,\nu} \left [ \sigma^+_{l,f} Z \sigma^-_{l,\overline{f}}  + (1-(1 + (-1)^{L + B})\delta_{l,L-1})  \sigma^+_{l,\overline{f}} Z \sigma^-_{l+1,f} + {\rm h.c.} \right ] .\
    \label{eq:0vBB_HquarkleptonPBC}
\end{align}

\section{Lepton state-preparation}
\label{sec:0vBB_leptonstateprep}

First, the lepton state initializer can be simplified by applying the fact that the Jordan-Wigner transformation can be applied by first running the appropriate circuits without the Jordan-Wigner transformations and then applying a network of fermionic SWAP (FSWAP) gates. This has been found before in Ref. \cite{cervera2018exact}, and can be understood to be a consequence of the fact that an FSWAP gate effectively acts as a combination of the CZ and SWAP gates, and that the action of a CZ gate is as follows:

\begin{align}
 &CZ (\hat{Y} \otimes \hat{I} ) CZ = \hat{Y} \otimes \hat{Z} \nonumber &CZ (\hat{Y} \otimes \hat{Z} ) CZ = \hat{Y} \otimes \hat{I} \nonumber & \hspace{4ex} CZ(\hat{X} \otimes \hat{X}) CZ = \hat{Y} \otimes \hat{Y}& \\
 &CZ (\hat{X} \otimes \hat{I} ) CZ = \hat{X} \otimes \hat{Z} \nonumber &CZ (\hat{X} \otimes \hat{Z} ) CZ = \hat{X} \otimes \hat{I} \nonumber & \hspace{4ex} CZ(\hat{X} \otimes \hat{Y}) CZ = -\hat{Y} \otimes \hat{X}& \\
 &CZ (\hat{I} \otimes \hat{X} ) CZ = \hat{Z} \otimes \hat{X} &CZ (\hat{Z} \otimes \hat{X} ) CZ = \hat{I} \otimes \hat{X} & \hspace{4ex} CZ(\hat{Y} \otimes \hat{X}) CZ = -\hat{X} \otimes \hat{Y}& \label{eq:CZidentities} \\
 &CZ (\hat{I} \otimes \hat{Y} ) CZ = \hat{Z} \otimes \hat{Y} \nonumber &CZ (\hat{Z} \otimes \hat{Y} ) CZ = \hat{I} \otimes \hat{Y} \nonumber & \hspace{4ex} CZ(\hat{Y} \otimes \hat{Y}) CZ = \hat{X} \otimes \hat{X}&  \\
  &CZ (\hat{Z} \otimes \hat{I} ) CZ = \hat{Z} \otimes \hat{I} \nonumber & CZ (\hat{I}\otimes \hat{Z} ) CZ = \hat{I} \otimes \hat{Z} \nonumber & \hspace{4ex} CZ (\hat{Z}\otimes \hat{Z} ) CZ = \hat{Z} \otimes \hat{Z}& \ 
\end{align}

\begin{figure}
    \centering
    \includegraphics[width=0.49\textwidth]{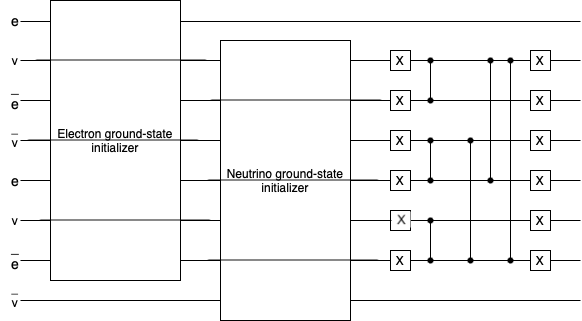}
    \includegraphics[width=0.49\textwidth]{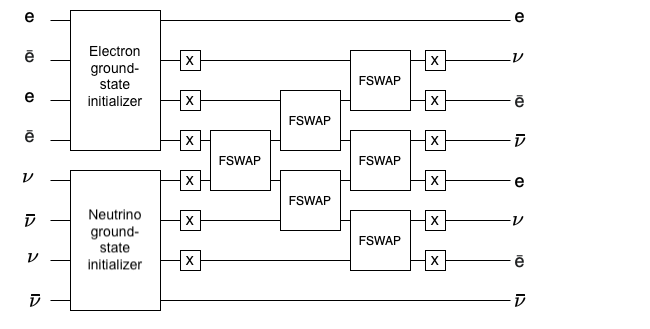}
    \caption{The decomposition of the lepton-qubit register state preparation circuits into state-preparation circuits for the electron and neutrino subcomponents of the register and a CZ or FSWAP network. To the left: the implementation for devices with all-to-all connectivity; to the right: the implementation for devices with nearest-neighbor connectivity}
    \label{fig:general_lepinit_plan_0vBB}
\end{figure}

\noindent Thus, the neutrino and electron qubits can be initialized separately, followed by a network of either FSWAP or CZ gates, as shown in Fig. \ref{fig:general_lepinit_plan_0vBB}. The initializers for the electron and neutrino subcomponents differ between the OBC and PBC Hamiltonians. Second, because the Hamiltonian is real, its eigenstates and by extension its ground state must also be real. Thus, we can restrict ourselves to circuits that only initialize real states.

\subsection{OBC Hamiltonian}

\begin{figure}
    \centering
    \includegraphics[width=0.8\linewidth]{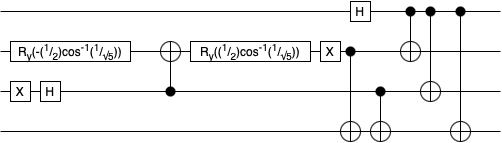}
    \caption{The circuit for insertion into the electron ground-state initializer subcomponent of the circuits in Fig. \ref{fig:general_lepinit_plan_0vBB} and the neutrino ground-state initializer subcomponent of the circuits in Fig. \ref{fig:general_lepinit_plan_0vBB} for the case of $m_{M} = 0$.This is for the open boundary conditions (OBC) Hamiltonian.}
    \label{fig:OBC_lepinit_mmaj0}
\end{figure}

\begin{figure}
    \centering
    \includegraphics[width=0.5\linewidth]{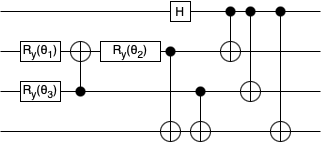}
    \caption{The circuit for insertion into the neutrino ground-state initializer subcomponent of the circuits in Fig. \ref{fig:general_lepinit_plan_0vBB} for the case of $m_{M} \neq 0$. $\theta_1$, $\theta_2$, and $\theta_3$ are variational parameters set to a different value for each $m_{M}$. This is for the open boundary conditions (OBC) Hamiltonian.}
    \label{fig:OBC_lepinit_mmajnot0}
\end{figure}

Fig. \ref{fig:OBC_lepinit_mmaj0} shows the initialization circuit for the electron subcomponent and for the neutrino subcomponent in the case where $m_{M} = 0$. Fig. \ref{fig:OBC_lepinit_mmajnot0}, which is controlled by three parameters, shows the initialization circuit for the neutrino subcomponent in the case where $m_{M} \neq 0$. The former is a generalization of the latter, which is a generalized 2-qubit real-state initializer on the middle two qubits, with couplings to the outer two qubits that are static components set by the symmetries in the lepton ground states. The Hadamard gate on the first qubit followed by the three CNOT gates controlled on it is based on the symmetrized exponential wavefunction initializer from Ref. \cite{klco:2019xro}. The CNOTs from the middle qubit to the last qubit were obtained by identifying patterns within lepton ground states obtained using exact classical diagonalization and mapped to bitstrings.

\subsection{PBC Hamiltonian}
\label{sec:pbclephamiltonianinitializatiocircuits}
\begin{figure}
    \centering
    \includegraphics[width=0.5\linewidth]{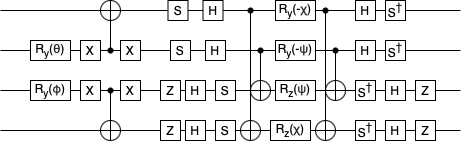}
    \caption{The circuit for insertion into the electron ground-state initializer subcomponent of the circuits in Fig. \ref{fig:general_lepinit_plan_0vBB} and the neutrino ground-state initializer subcomponent of the circuits in Fig. \ref{fig:general_lepinit_plan_0vBB} for the case of $|m_{M}| \leq 1$. $\theta = -2 \tan^{-1}{(\sqrt{2} - 1)}$, $\phi = 2 \tan^{-1}{(-\sqrt{2} - 1)}$, $\chi = -\tan^{-1}{(\sqrt{2} - 1)}$, $\psi = -\tan^{-1}{(\sqrt{2} - 1)}$. The subcomponents with the parameters $\chi$ and $\psi$ are based on the circuit for the operation $e^{(i \frac{\theta}{2} (XY \pm YX))}$ from Ref. \cite{Farrell:2023fgd}. This is for the periodic boundary conditions (PBC) Hamiltonian on a device with all-to-all connectivity.}
    \label{fig:PBC_lepinit_mmaj_atoa_absvalleq1}
\end{figure}

\begin{figure}
    \centering
    \includegraphics[width=0.7\linewidth]{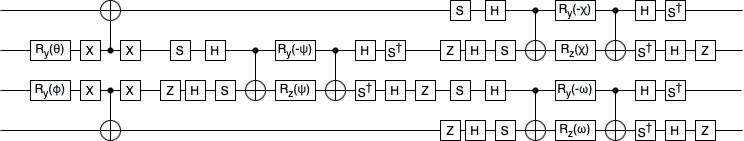}
    \caption{The circuit for insertion into the electron ground-state initializer subcomponent of the circuits in Fig. \ref{fig:general_lepinit_plan_0vBB} and the neutrino ground-state initializer subcomponent of the circuits in Fig. \ref{fig:general_lepinit_plan_0vBB} for the case of $|m_{M}| \leq 1$. $\theta$ = -0.8301450945867347, $\phi$ = -0.8301450945867347, $\chi$ = 0.24090007687244258, $\omega$ = -1.026298240269891, and $\psi$ = -1.2802979760585618. The subcomponents with the parameters $\chi$, $\psi$, and $\omega$ are based on the circuit for the operation $e^{(i \frac{\theta}{2} (XY \pm YX))}$ from Ref. \cite{Farrell:2023fgd}. This is for the periodic boundary conditions (PBC) Hamiltonian on a device with nearest-neighbor connectivity.}
    \label{fig:PBC_lepinit_mmaj_nn_absvalleq1}
\end{figure}

\begin{figure}
    \centering
    \includegraphics[width=0.2\linewidth]{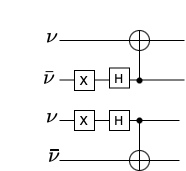}
    \includegraphics[width=0.2\linewidth]{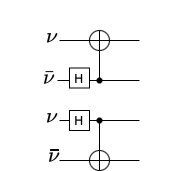}
    \caption{Left: The circuit for insertion into the neutrino ground-state initializer subcomponent of the circuits in Fig. \ref{fig:general_lepinit_plan_0vBB} for the case of $m_{M} \leq -1$. Right: The circuit for insertion into the neutrino ground-state initializer subcomponent of the circuits in Fig. \ref{fig:general_lepinit_plan_0vBB} for the case of $m_{M} \geq 1$. This is for the periodic boundary conditions (PBC) Hamiltonian.}
    \label{fig:PBC_lepinit_mmaj_absvalgreq1}
\end{figure}

There are 4 initialization circuits in this case. One, in Fig. \ref{fig:PBC_lepinit_mmaj_atoa_absvalleq1}, is for the electron subcomponent and for the neutrino subcomponent in the case where $-1 < m_{M} < 1$ on devices with all-to-all connectivity. One, in Fig. \ref{fig:PBC_lepinit_mmaj_nn_absvalleq1}, is for the electron subcomponent and for the neutrino subcomponent in the case where $-1 < m_{M} < 1$ on devices with nearest-neighbor connectivity. The other two, in Fig. \ref{fig:PBC_lepinit_mmaj_absvalgreq1}, are for the neutrino subcomponent in the case where $|m_{M}| \geq 1$. All but the ones in Fig.  \ref{fig:PBC_lepinit_mmaj_absvalgreq1} are built out of components based on the $e^{i \theta (XY \pm YX)}$ building-block component from Ref. \cite{Farrell:2023fgd}.

\section{Baryon state-preparation}
\label{sec:0vBB_stateprep_baryons}

The $B = 1$ and $B = 2$ ground states for the subspace with a Z-component of isospin of $-\frac{3}{2}$ for the baryon-register qubits. These states are the initial state for the reactions $\Delta^- \rightarrow \Delta^+ e^- e^-$ and $\Delta^- \Delta^- \rightarrow \Delta^0 \Delta^0 e^- e^-$, respectively, and so must be initialized in order for the real-time evolution of these reactions to be simulated. My colleague Roland Farrell has found that it is possible to use SC-ADAPT-VQE using the same operator-pool used in Ref. \cite{Farrell:2023fgd}, after removing terms that simultaneously apply creation and annihilation operators to qubits of different colors, to obtain the U(1) Kogut-Susskind lattice gauge theory ground state to obtain the aforementioned ground states. In this section I present a simplified PBC SU(3) Kogut-Susskind Hamiltonian that makes the VQE optimization required for this application of SC-ADAPT-VQE less resource-intensive.

\subsection{Construction of the color-singlet space}
\label{sec:colorsingletconstruction}

The possible occupation-states of the quarks of each flavor on each staggered site the quark lattice characterized by the $H_{{\rm quarks}} + H_{el}$ subcomponent of one of the Hamiltonians discussed in Sec. \ref{sec:0vBB_stateprep_Hamiltonians} is shown in Table \ref{tab:Su3sitestates}. A similar theory can be constructed for an SU(2) gauge field, with two colors and hence two quarks (and therefore qubits) per flavor per staggered lattice site. The possible occupation-states of the quarks of each flavor on each staggered lattice site in this simpler SU(2) theory are shown in Table \ref{tab:Su2sitestates}. The irreducible representations (``irreps") for SU(2) and SU(3) are named using conventions from Ref. \cite{ physrevd.103.094501}. 


The aim of Sec. \ref{sec:colorsingletconstruction} is to prove that all SU(3) color singlet fermion states can be expressed as a sum of tensor products of the following states:

\begin{multline}
    \ket{0}, \ket{7}, \frac{1}{\sqrt{3}}(\ket{6} \otimes \ket{1} - \ket{5} \otimes \ket{2} + \ket{3} \otimes \ket{4}), \frac{1}{\sqrt{6}}(\ket{1} \otimes \ket{2} \otimes \ket{4} - \ket{2} \otimes \ket{1} \otimes \ket{4} - \ket{1} \otimes \ket{4} \otimes \ket{2} + \\
    \ket{4} \otimes \ket{1} \otimes \ket{2} + \ket{2} \otimes \ket{4} \otimes \ket{1} - \ket{4} \otimes \ket{2} \otimes \ket{1}), \frac{1}{\sqrt{6}} (\ket{3} \otimes \ket{5} \otimes \ket{6} - \ket{5} \otimes \ket{3} \otimes \ket{6} - \\
    \ket{3} \otimes \ket{6} \otimes \ket{5} + \ket{6} \otimes \ket{3} \otimes \ket{5} + \ket{5} \otimes \ket{6} \otimes \ket{3} - \ket{6} \otimes \ket{5} \otimes \ket{3}) \hspace{8ex}
    \label{eq:SU3colorsingletcomponents}
\end{multline}

\begin{table}[h]
    \centering
    \begin{tabular}{c|c|c}
        Qubit-state & Qu8it notation  & SU(3) irrep\\
    \hline
        $\ket{\downarrow \downarrow \downarrow}$ & $\ket{0}$ & $\bf{1}$ \\
        $\ket{\downarrow \downarrow \uparrow}$ & $\ket{1}$ & $\bf{3}$ \\
        $\ket{\downarrow \uparrow \downarrow}$ & $\ket{2}$ & $\bf{3}$ \\
        $\ket{\downarrow \uparrow \uparrow}$ & $\ket{3}$ & $\bf{\bar{3}}$ \\
        $\ket{\uparrow \downarrow \downarrow}$ & $\ket{4}$ & $\bf{3}$ \\
        $\ket{\uparrow \downarrow \uparrow}$ & $\ket{5}$ & $\bf{\bar{3}}$ \\
        $\ket{\uparrow \uparrow \downarrow}$ & $\ket{6}$ & $\bf{\bar{3}}$ \\
        $\ket{\uparrow \uparrow \uparrow}$ & $\ket{7}$ & $\bf{1}$ 
    \end{tabular}
    \caption{Left: the possible states, expressed in the form of spins, of the qubits representing the three quarks of a given flavor on a given staggered lattice site. Center: the notation that will be used in the rest of Sec. \ref{sec:colorsingletconstruction} to discuss SU(3) lattice gauge theory fermion states. It is similar to the one used in Ref. \cite{ Illa:2024kmf} to encode SU(3) lattice gauge theory on qu8its. Right: the irreducible representation of SU(3) that the state falls into, denoted by dimensionality. }
    \label{tab:Su3sitestates}
\end{table}

\begin{table}[h]
    \centering
    \begin{tabular}{c|c|c}
        Qubit-state & Ququart notation  & SU(2) irrep\\
    \hline
        $\ket{\downarrow \downarrow}$ & $\ket{0}$ & $\bf{0}$ \\
        $\ket{\downarrow \uparrow}$ & $\ket{1}$ & $\bf{\frac{1}{2}}$ \\
        $\ket{\uparrow \downarrow}$ & $\ket{2}$ & $\bf{\frac{1}{2}}$ \\
        $\ket{\uparrow \uparrow}$ & $\ket{3}$ & $\bf{0}$ 
    \end{tabular}
    \caption{Left: the possible states, expressed in the form of spins, of the qubits representing the two quarks of a given flavor on a given staggered lattice site; Center: the notation that will be used in the rest of Sec. \ref{sec:colorsingletconstruction} to discuss SU(2) lattice gauge theory fermion states. Right: the irreducible representation of SU(2) that the state falls into, denoted by the total spin quantum number. }
    \label{tab:Su2sitestates}
\end{table}

\subsubsection{SU(2) singlet space}
\label{sec:SU2colorsingletspaceproof}

The first step of the proof that Eq. \ref{eq:SU3colorsingletcomponents} contains all of the basic components of the fermion SU(3) color singlet space is a similar proof for the SU(2) color singlet space. Thus, this section details a proof that all SU(2) color singlet fermion states can be expressed as a sum of tensor products of the following states:

\begin{equation}
    \ket{0}, \ket{3}, \frac{1}{\sqrt{2}}(\ket{1} \otimes \ket{2} - \ket{2} \otimes \ket{1})
    \label{eq:SU2colorsingletcomponents}
\end{equation}

\noindent $\ket{0}$ and $\ket{3}$ are color singlet states, so any states purely composed of sums of tensor products of them are color singlet states. Thus, the only question that remains is which, if any, combinations of the states $\ket{1}$ and $\ket{2}$ are color singlets.

The following facts can be used as a starting point:

\begin{enumerate}
    \item The overall z-component of the spin of a color singlet must be 0, hence the number of $\ket{1}$'s and $\ket{2}$'s in a color singlet state must be equal.
    \item Applying a raising or lowering operator to any SU(2) color singlet state returns a value of $\bf 0$.
\end{enumerate}

\noindent A raising ($J^{+}$) or lowering ($J^{-}$) operator on a state can be represented in terms of a sum of raising and lowering operators on its irrep-$\bf \frac{1}{2}$ components ($J^{+}_{\bf 1/2}$ and $J^{-}_{\bf 1/2}$, respectively) like so:

\begin{equation}
    J^{+} = J^{+}_{\bf 1/2} \otimes \mathbb{1} \otimes \cdots \otimes \mathbb{1} + \mathbb{1} \otimes J^{+}_{\bf 1/2} \otimes \cdots \otimes \mathbb{1} + \cdots +  \mathbb{1} \otimes \mathbb{1} \otimes \cdots \otimes J^{+}_{\bf 1/2}
    \label{eq:Su2raisingoperatordef}
\end{equation}

\begin{equation}
    J^{-} = J^{-}_{\bf 1/2} \otimes \mathbb{1} \otimes \cdots \otimes \mathbb{1} + \mathbb{1} \otimes J^{-}_{\bf 1/2} \otimes \cdots \otimes \mathbb{1} + \cdots +  \mathbb{1} \otimes \mathbb{1} \otimes \cdots \otimes J^{-}_{\bf 1/2}
    \label{eq:Su2loweringoperatordef}
\end{equation}

\noindent Given that 

\begin{equation}
    J^{+}_{\bf 1/2} \ket{\uparrow} = {\bf 0} ; J^{+}_{\bf 1/2} \ket{\downarrow} = \ket{\uparrow} ;
    J^{-}_{\bf 1/2} \ket{\uparrow} = \ket{\downarrow} ;
    J^{-}_{\bf 1/2} \ket{\downarrow} = {\bf 0} ,
    \label{eq:fundamentalsu2raisingandlowering}
\end{equation}

\noindent applying $J^{+}$ results in a sum of states that have one more $\ket{\uparrow}$ state than the original state and applying $J^{-}$ results in a sum of states that have one more $\ket{\uparrow}$ state than the original $\ket{\downarrow}$. In both cases the sign is the same as it was in the original state. 

Thus, to satisfy Fact \# 2's stipulation that applying $J^{+}$ ($J^{-}$) result in a state of $\bf 0$, each term in a color-singlet state composed of irrep-$\bf \frac{1}{2}$ components must have a counterpart with the opposite sign that will produce the same result when $J^{+}$ ($J^{-}$) is acted on it. Since different initial states must be acted on by different operators in order to achieve the same final state, this must be accomplished by means of $J^{+}_{\bf 1/2}$ ($J^{-}_{\bf 1/2}$) acted on site $a$ of term $A$ producing the additive inverse of the result of $J^{+}_{\bf 1/2}$ ($J^{-}_{\bf 1/2}$) acted on site $b$ of term $B$. (Here $A$, $B$, $a$, and $b$ are arbitrary indicators for a given term or site, respectively). The only way to meet this requirement is to select $a$ and $b$ so that site $a$ of term $A$ is in state $\ket{\downarrow}$ and site $b$ of term $A$ is in state $\ket{\uparrow}$, and mandate that term $B$ is the same as term $A$ except that site $a$ is in state $\ket{\uparrow}$ and site $b$ is in state $\ket{\downarrow}$ and its amplitude is the additive inverse of term $A$'s. Then Terms $A$ and $B$ must satisfy the following requirement:

\begin{equation}
    \ket{A} + \ket{B} = C \ket{\downarrow}_a \otimes \ket{\uparrow}_b \otimes \ket{other} - C \ket{\uparrow}_a \otimes \ket{\downarrow}_b \otimes \ket{other} \equiv C \sqrt{2} \frac{1}{\sqrt{2}}(\ket{1} \otimes \ket{2} - \ket{2} \otimes \ket{1}) \otimes \ket{other}
    \label{eq:Su2singletdecomposition}
\end{equation}

\noindent where $C$ is an arbitrary amplitude and $\ket{other}$ are the parts of Terms $A$ and $B$ that do not lie on sites $a$ and $b$. Since $\frac{1}{\sqrt{2}}(\ket{1} \otimes \ket{2} - \ket{2} \otimes \ket{1})$ is an SU(2) singlet state, $\ket{other}$ must also be an SU(2) singlet state, so the process in Eq. \ref{eq:Su2singletdecomposition} can be repeated for all pairs of sites occupied by a $\ket{1}$ or a $\ket{2}$. Thus, the only states composed of $\ket{1}$'s and $\ket{2}$'s that are SU(2) singlets are sums of tensor products of $\frac{1}{\sqrt{2}}(\ket{1} \otimes \ket{2} - \ket{2} \otimes \ket{1})$. Since $\ket{1}$, $\ket{2}$, $\ket{0}$, and $\ket{3}$ are the only states in a register of fermions obeying an SU(2) lattice gauge theory, sums of tensor products of the states listed in Eq. \ref{eq:SU2colorsingletcomponents} encompass the entire space of singlet states possible among fermions obeying SU(2) lattice gauge theory.

\centerline{QED}

\subsubsection{Extension to SU(3) singlet space}
\label{sec:proofextensiontoSU3}

\begin{figure}
    \centering
    \includegraphics[width=0.5\linewidth]{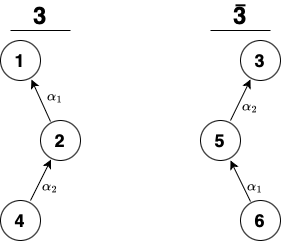}
    \caption{The actions of the SU(3) annihilation operators on the $\bf 3$ and $\bf \bar{3}$ states. (creation operators go in the direction opposite the arrows)}
    \label{fig:SU3creationoperatorchart}
\end{figure}

The next step is to extrapolate the proof for SU(2) singlet states in Sec. \ref{sec:SU2colorsingletspaceproof} to a proof that all SU(3) singlet states composed of quarks can be expressed as a sum of tensor products of the states in Eq. \ref{eq:SU3colorsingletcomponents}. The first difference between SU(2) and SU(3) that needs to be accounted for is that while quarks in SU(2) lattice gauge theory only come in one irreducible representation besides the singlet ($\bf \frac{1}{2}$), while quarks in SU(3) lattice gauge theory come in two ($\bf 3$ and $\bf \bar{3}$). The second is that while SU(2) has one set of creation and annihilation operators, SU(3) has two (I will denote $\alpha_1$ and $\alpha_2$ as the annihilation operators and $\alpha_1^{\dagger}$ and $\alpha_2^{\dagger}$ as their corresponding creation operators). For $\bf 3$ and $\bf \bar{3}$, the actions of $\alpha_1$ and $\alpha_2$ can be shown in Fig. \ref{fig:SU3creationoperatorchart} \cite{georgi2000lie}.

Thus, the ($\ket{1}$, $\ket{4}$) and ($\ket{4}$, $\ket{2}$) state-pairs within the $\bf 3$ irrep and the ($\ket{5}$, $\ket{6}$) and ($\ket{3}$, $\ket{5}$) pairs within the $\bf \bar{3}$ irrep each form SU(2) subgroups. Combining this fact with the fact that all SU(2) singlets composed entirely of irrep-$\frac{1}{2}$ states are sums of tensor products of the state $\frac{1}{\sqrt{2}}(\ket{1} \otimes \ket{2} - \ket{2} \otimes \ket{1})$, one can come to the conclusion that all SU(3) singlets composed entirely of irrep-$\bf 3$ states are sums of tensor products of the state

\begin{align}
    \frac{1}{\sqrt{6}}&(\ket{1} \otimes \ket{2} \otimes \ket{4} - \ket{2} \otimes \ket{1} \otimes \ket{4} - \ket{1} \otimes \ket{4} \otimes \ket{2} \nonumber \\+& \ket{4} \otimes \ket{1} \otimes \ket{2} + \ket{2} \otimes \ket{4} \otimes \ket{1} - \ket{4} \otimes \ket{2} \otimes \ket{1})
    \label{eq:su3irrep3colorsinglet}
\end{align}

\noindent and all SU(3) singlets composed entirely of irrep-$\bf \bar{3}$ states are sums of tensor products of the state

\begin{align}
    \frac{1}{\sqrt{6}}&(\ket{3} \otimes \ket{5} \otimes \ket{6} - \ket{5} \otimes \ket{3} \otimes \ket{6} - \ket{3} \otimes \ket{6} \otimes \ket{5} \nonumber \\+& \ket{6} \otimes \ket{3} \otimes \ket{5} + \ket{5} \otimes \ket{6} \otimes \ket{3} - \ket{6} \otimes \ket{5} \otimes \ket{3})
    \label{eq:su3barirrep3colorsinglet}
\end{align}

Besides the color-singlets composed of states in the $\bf 1$ irrep, those composed of states in the $\bf 3$ irrep, and those composed of states in the $\bf \bar{3}$ irrep, one should also consider combinations of states in the $\bf 3$ and $\bf \bar{3}$ irreps. Considering these combinations is fortunately simple: states in the $\bf 3$ irrep can be written as combinations of two states in the $\bf \bar{3}$ irrep, and vice versa. They can be derived from SU(3)'s weight-charts, found in Ref. \cite{georgi2000lie}, and are as follows:

\begin{align}
\ket{1} \equiv \frac{1}{\sqrt{2}} (\ket{3} \otimes \ket{5} - \ket{5} \otimes \ket{3}) & \hspace{6ex} \ket{6} \equiv \frac{1}{\sqrt{2}} (\ket{2} \otimes \ket{4} - \ket{4} \otimes \ket{2}) \nonumber \\
\ket{2} \equiv \frac{1}{\sqrt{2}} (\ket{3} \otimes \ket{6} - \ket{6} \otimes \ket{3}) & \hspace{6ex} \ket{5} \equiv \frac{1}{\sqrt{2}} (\ket{1} \otimes \ket{4} - \ket{4} \otimes \ket{1}) \label{eq:3and3barintermsofeachother} \\
\ket{4} \equiv \frac{1}{\sqrt{2}} (\ket{5} \otimes \ket{6} - \ket{6} \otimes \ket{5}) & \hspace{6ex} \ket{3} \equiv \frac{1}{\sqrt{2}} (\ket{1} \otimes \ket{2} - \ket{2} \otimes \ket{1}) \nonumber
\end{align}

\noindent Applying the relations in Eq. \ref{eq:3and3barintermsofeachother} to Eqs. \ref{eq:su3irrep3colorsinglet} and \ref{eq:su3barirrep3colorsinglet} gives the additional SU(3) singlet basis state enabled by combining states with irreps $\bf 3$ and $\bf \bar{3}$: 

\begin{equation}
    \frac{1}{\sqrt{3}}(\ket{6} \otimes \ket{1} - \ket{5} \otimes \ket{2} + \ket{3} \otimes \ket{4})
    \label{eq:33barsu3singlet}
\end{equation}

\noindent Adding in the states in Eqs. \ref{eq:su3irrep3colorsinglet}, \ref{eq:su3barirrep3colorsinglet}, and \ref{eq:33barsu3singlet} with the trivial color singlet states $\ket{0}$ and $\ket{7}$ gives Eq. \ref{eq:SU3colorsingletcomponents}, and one has proven that all SU(3) color singlet quark combinations are sums of tensor products of the states in Eq. \ref{eq:SU3colorsingletcomponents}.

\centerline{QED}

As a side-note, the basis-states of the SU(3) color singlet space are the natural consequence of the fact pointed out in Chapters 13 - 16 of Ref. \cite{georgi2000lie} that color-singlets can be created by taking a group of quarks and antiquarks and applying and summing over Levi-Civita symbols and/or Kronecker deltas, combined with the Jordan-Wigner mapping. 

\subsection{Simplifying the Chromoelectric Hamiltonian within the color-singlet space}

Since the only combinations of quarks that will occur physically are color-singlets, the next question is: how can one take advantage of the fact that all color-singlets are sums of tensor-products of states in Eq. \ref{eq:SU3colorsingletcomponents} (which in this section I will call ``fundamental singlet states") to simplify the action of the  $Q^{(a)}_{n,f} Q^{(a)}_{m,f'}$ terms present in the $H_{el}$ Hamiltonian terms from Eqs. \ref{eq:OBC_0nuBB_fullham} and \ref{eq:HchromoPBC_0vBB} on states within the color-singlet space? This will have two components: $Q^{(a)}_{n,f}$ and $Q^{(a)}_{m,f'}$ where $(n,f)$ and $(m,f')$ are from different fundamental singlet states and where $(n,f)$ and $(m,f')$ are from the same fundamental singlet state.

First, any $Q^{(a)}_{n,f} Q^{(a)}_{m,f'}$ where one of the $Q$'s is applied to a $\ket{0}$ or a $\ket{7}$ state is 0. This is true separately for the diagonal and off-diagonal components of $Q^{(a)}_{n,f} Q^{(a)}_{m,f'}$. For the off-diagonal component, applying a $\sigma^{+}\sigma^{-}$ to a register where all qubits have the same spin will produce a result of $\bf 0$. As for the diagonal component, it can be re-written as $\frac{1}{6}\sum_{c = 0}^2 \sigma^{(z)}_{3 N_f n + 3 f + c} \Big( 2 \sigma^{(z)}_{3 N_f m + 3 f' + c} - \sigma^{(z)}_{3 N_f m + 3 f' + ((c - 1) \mod 3)} - \sigma^{(z)}_{3 N_f m + 3 f' + ((c + 1) \mod 3)} \Big)$. Let's say $(m, f')$ is the site with the $\ket{0}$ or $\ket{7}$ state. Since the spin is the same for all 3 colors, the sum $\Big( 2 \sigma^{(z)}_{3 N_f m + 3 f' + c} - \sigma^{(z)}_{3 N_f m + 3 f' + ((c - 1) \mod 3)} - \sigma^{(z)}_{3 N_f m + 3 f' + ((c + 1) \mod 3)} \Big)$, to which the diagonal part of $Q^{(a)}_{n,f} Q^{(a)}_{m,f'}$ can now be reduced, is 0. This leaves pairs of sites from the states from Eqs. \ref{eq:su3irrep3colorsinglet}, \ref{eq:su3barirrep3colorsinglet}, and \ref{eq:33barsu3singlet}.

\subsubsection{Chromoelectric term within a fundamental singlet state}

Applying the off-diagonal portion of $Q^{(a)}_{n,f} Q^{(a)}_{m,{f'}}$ 
to the two sites occupied by ``meson excitation", i.e. $\frac{1}{\sqrt{3}}(\ket{6} \otimes \ket{1} - \ket{5} \otimes \ket{2} + \ket{3} \otimes \ket{4})$ results in the state $-\frac{1}{\sqrt{3}}(\ket{6} \otimes \ket{1} - \ket{5} \otimes \ket{2} + \ket{3} \otimes \ket{4})$. Applying the diagonal portion of $Q^{(a)}_{n,f} Q^{(a)}_{m,{f'}}$ to the same site-pair produces $-\frac{1}{3}*\frac{1}{\sqrt{3}}(\ket{6} \otimes \ket{1} - \ket{5} \otimes \ket{2} + \ket{3} \otimes \ket{4})$. Thus, the action of the chromoelectric term on the sites of the meson excitation can be represented simply by the diagonal portion scaled up by a factor of 4. 

Similarly, applying the off-diagonal portion of $Q^{(a)}_{n,f} Q^{(a)}_{m,{f'}}$ to a pair of sites occupied by either the state in Eqs. \ref{eq:su3irrep3colorsinglet} or the state in \ref{eq:su3barirrep3colorsinglet} results in an overall $-\frac{1}{2}$ factor being multiplied into the state, and applying the on-diagonal term results in a $-\frac{1}{6}$ factor being multiplied into the state. Thus, the diagonal portion of $Q^{(a)}_{n,f} Q^{(a)}_{m,{f'}}$ scaled up by a factor of 4 is equivalent to the chromoelectric term for all pairs of sites occupied by the same fundamental singlet state. 

Combining this with the fact that SU(3) color singlet states are sums of tensor products of the states in Eq. \ref{eq:SU3colorsingletcomponents}, scaling the diagonal portion of $Q^{(a)}_{n,f} Q^{(a)}_{m,{f'}}$ up by a factor of 4 is equivalent to the entirety of $Q^{(a)}_{n,f} Q^{(a)}_{m,{f'}}$ if one is working with two sites from the same color-singlet space.  Therefore, 


\begin{multline}
    Q^{(a)}_{n,f} Q^{(a)}_{m,{f'}} \equiv (Q^{(a)}_{n,f} Q^{(a)}_{m,{f'}})_{trun1} = \frac{1}{3} \Big( \sum_{c = 0}^2 \sigma^{(z)}_{3 N_f n + 3 f + c} \sigma^{(z)}_{3 N_f m + 3 f' + c} - \frac{1}{2}\sigma^{(z)}_{3 N_f n + 3 f + 1} \sigma^{(z)}_{3 N_f m + 3 f'} \\
    - \frac{1}{2} \sigma^{(z)}_{3 N_f n + 3 f + 2} \sigma^{(z)}_{3 N_f m + 3 f'} - \frac{1}{2} \sigma^{(z)}_{3 N_f n + 3 f + 2} \sigma^{(z)}_{3 N_f m + 3 f' + 1} - \frac{1}{2}\sigma^{(z)}_{3 N_f n + 3 f} \sigma^{(z)}_{3 N_f m + 3 f' + 1} \\
    - \frac{1}{2} \sigma^{(z)}_{3 N_f n + 3 f} \sigma^{(z)}_{3 N_f m + 3 f' + 2} - \frac{1}{2} \sigma^{(z)}_{3 N_f n + 3 f + 1} \sigma^{(z)}_{3 N_f m + 3 f' + 2} \Big)
    \label{eq:simplifiedsitepairchromoterm}
\end{multline}

\noindent for $Q^{(a)}_{n,f} Q^{(a)}_{m,{f'}}$ within a color-singlet subspace. An attempt to simplify Eq. \ref{eq:simplifiedsitepairchromoterm} even further was made using the conjecture that $\sigma^{(z)}_{3 N_f n + 3 f + 1} \sigma^{(z)}_{3 N_f m + 3 f'} + \sigma^{(z)}_{3 N_f n + 3 f + 2} \sigma^{(z)}_{3 N_f m + 3 f'} + \sigma^{(z)}_{3 N_f n + 3 f + 2} \sigma^{(z)}_{3 N_f m + 3 f' + 1}$ is equivalent to that of $\sigma^{(z)}_{3 N_f n + 3 f} \sigma^{(z)}_{3 N_f m + 3 f' + 1} + \sigma^{(z)}_{3 N_f n + 3 f} \sigma^{(z)}_{3 N_f m + 3 f' + 2} + \sigma^{(z)}_{3 N_f n + 3 f + 1} \sigma^{(z)}_{3 N_f m + 3 f' + 2}$, so one can trim 3 Pauli-Z pairs from $Q^{(a)}_{n,f} Q^{(a)}_{m,{f'}}$. Therefore, this should be true:

\begin{multline}
    (Q^{(a)}_{n,f} Q^{(a)}_{m,{f'}})_{trun1} \equiv (Q^{(a)}_{n,f} Q^{(a)}_{m,{f'}})_{trun2} = \frac{1}{3} \Big( \sum_{c = 0}^2 \sigma^{(z)}_{3 N_f n + 3 f + c} \sigma^{(z)}_{3 N_f m + 3 f' + c} - \sigma^{(z)}_{3 N_f n + 3 f + 1} \sigma^{(z)}_{3 N_f m + 3 f'} \\
    - \sigma^{(z)}_{3 N_f n + 3 f + 2} \sigma^{(z)}_{3 N_f m + 3 f'} - \sigma^{(z)}_{3 N_f n + 3 f + 2} \sigma^{(z)}_{3 N_f m + 3 f' + 1} \Big)
    \label{eq:simplifiedsitepairchromoterm_minimalZZs}
\end{multline}

The next step is to find out whether Eqs. \ref{eq:simplifiedsitepairchromoterm} and \ref{eq:simplifiedsitepairchromoterm_minimalZZs} hold for $Q^{(a)}_{n,f} Q^{(a)}_{m,{f'}}$ terms applied to site-pairs that are from different fundamental singlet states in the general case.

\subsubsection{Chromoelectric term between fundamental singlet states}

\begin{figure}
    \centering
    \includegraphics[width=1.0\linewidth]{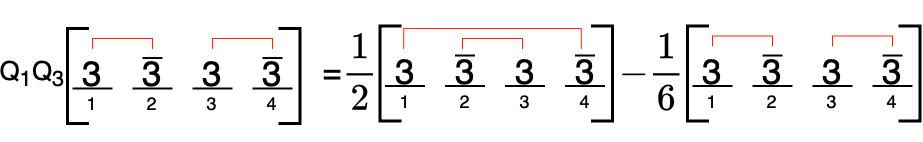}
    \caption{The result of acting the $Q^{(a)}_{n,f} Q^{(a)}_{m,f'}$ on two sites from two different meson excitations but of the same irrep. The $\bf{3}$ and $\bf{\bar{3}}$ denote the irreps on the sites; the red connections denote which sites are part of the same meson excitation.}
    \label{fig:QnQmdifferentmesonexcsameirrep}
\end{figure}
\begin{figure}
    \centering
    \includegraphics[width=1.0\linewidth]{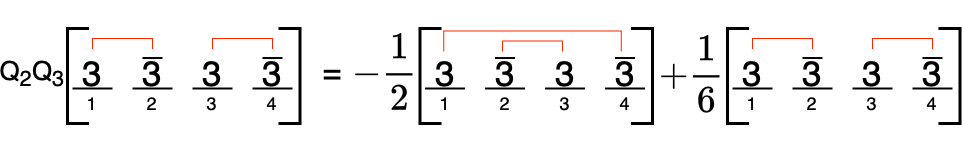}
    \caption{The result of acting the $Q^{(a)}_{n,f} Q^{(a)}_{m,f'}$ on two sites from two different meson excitations and of different irreps. The $\bf{3}$ and $\bf{\bar{3}}$ denote the irreps on the sites; the red connections denote which sites are part of the same meson excitation.}
    \label{fig:QnQmdifferentmesonexcdifferentirreps}
\end{figure}

The action of a $Q^{(a)}_{n,f} Q^{(a)}_{m,f'}$ on a pair of sites from different fundamental singlet states is shown in Fig. \ref{fig:QnQmdifferentmesonexcsameirrep} for a site-pair where both sites are either in irrep $\bf{3}$ or $\bf{\bar{3}}$ or in Fig. \ref{fig:QnQmdifferentmesonexcdifferentirreps} for a site pair where one site is in irrep $\bf{3}$ and the other is in irrep $\bf{\bar{3}}$. Figs. \ref{fig:QnQmdifferentmesonexcsameirrep} and \ref{fig:QnQmdifferentmesonexcdifferentirreps} show the results for the case of a pair of  $\frac{1}{\sqrt{3}}(\ket{6} \otimes \ket{1} - \ket{5} \otimes \ket{2} + \ket{3} \otimes \ket{4})$ (which we will call ``meson excitations"), but all other cases can be derived using the relations in Eq. \ref{eq:3and3barintermsofeachother}.

\paragraph{Intuitive ``proof" from locality of zero $Q^{(a)}_{n,f} Q^{(a)}_{m,f'}$ action}

Imagine that one has a lattice which contains two meson excitations that are separated from each other and otherwise is composed of entirely of $\ket{0}$ and $\ket{7}$ states. Since the meson excitations are self-contained color singlet states and could be produced locally by applying the $e^{i (XY - YX)}$ operator from Ref. \cite{Farrell:2023fgd} to $\ket{0}$ and $\ket{7}$ states, the only places where the gauge field would be a non-color-singlet and would thus contribute to the chromoelectric term should be on the links separating the sites that are members of the same meson excitation. Thus, the different-excitation $Q^{(a)}_{n,f} Q^{(a)}_{m,f'}$ terms, which act between the two meson excitations, must be suppressed. This logic can be applied to all possible pairs of states from Eqs. \ref{eq:su3irrep3colorsinglet}, \ref{eq:su3barirrep3colorsinglet}, and \ref{eq:33barsu3singlet}, not just meson excitation pairs.

However, the question remains of what happens when the links excited by the two meson excitations overlap. One such system is visible on the right side of the equations in Figs. \ref{fig:QnQmdifferentmesonexcsameirrep} and \ref{fig:QnQmdifferentmesonexcdifferentirreps}, where it has a coefficient of $\pm \frac{1}{2}$. From the equations in Figs. \ref{fig:QnQmdifferentmesonexcsameirrep} and \ref{fig:QnQmdifferentmesonexcdifferentirreps}, it is straightforward to deduce that the different-excitation $Q^{(a)}_{n,f} Q^{(a)}_{m,f'}$ terms acting on such a state would produce a superposition of the original state and a state with the meson excitations spatially separated. However, by the logic in the previous paragraph, the mapping via different-excitation $Q^{(a)}_{n,f} Q^{(a)}_{m,f'}$ terms from the spatially separated meson excitations to the overlapping ones is suppressed. In order for the operator to be Hermitian, the the converse mapping must be suppressed as well. Thus, the only permitted action of the different-excitation $Q^{(a)}_{n,f} Q^{(a)}_{m,f'}$ terms on such a system is a diagonal interaction. However, if one does the algebra on the interactions of the forms of the equations in Figs. \ref{fig:QnQmdifferentmesonexcsameirrep} and \ref{fig:QnQmdifferentmesonexcdifferentirreps}, the suppression of the non-diagonal interaction will also require the suppression of the diagonal term.

\begin{figure}[h]
    \centering
    \includegraphics[width=0.25\linewidth]{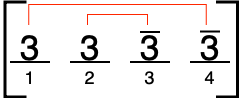}
    \includegraphics[width=0.25\linewidth]{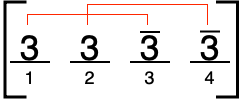}
    \caption{The two remaining overlapping two-meson-excitation states.}
    \label{fig:remainingoverlappingmesonexcs}
\end{figure}

The remaining overlapping-link states with two meson excitations can be seen in Fig. \ref{fig:remainingoverlappingmesonexcs}. They can both be formed from an overlapping state that can be mapped by a $Q^{(a)}_{n,f} Q^{(a)}_{m,f'}$ term to a spatially separated meson excitation pair by swapping the $\bf{3}$ and $\bf{\bar{3}}$ sites of a given meson excitation. From Figs. \ref{fig:QnQmdifferentmesonexcsameirrep} and \ref{fig:QnQmdifferentmesonexcdifferentirreps}, the action of a $Q^{(a)}_{n,f} Q^{(a)}_{m,f'}$ term between a $\bf{3}-\bf{3}$ and a $\bf{\bar{3}}-\bf{\bar{3}}$ site pair on different meson excitations is the additive inverse of the action of a $Q^{(a)}_{n,f} Q^{(a)}_{m,f'}$ term between a $\bf{3}-\bf{\bar{3}}$ site pair on different meson excitations. Thus, the aforementioned swap's effects are (1) to change the non-initial state in all of the different-excitation $Q^{(a)}_{n,f} Q^{(a)}_{m,f'}$ actions in the same way for all terms and (2) to flip the sign of all different-excitation $Q^{(a)}_{n,f} Q^{(a)}_{m,f'}$ actions. Neither of these effects changes the fact that the different-excitation $Q^{(a)}_{n,f} Q^{(a)}_{m,f'}$ would be suppressed.

\noindent Thus, for all states, $Q^{(a)}_{n,f} Q^{(a)}_{m,f'}$ terms between different fundamental color singlet states should be suppressed. This proof, however, depends on the locality of the theory. This conjecture is accurate for open boundary conditions. However, in the treatment of periodic boundary conditions present in Sec. \ref{sec:0vBB_stateprep_Hamiltonians_periodic_boundary_conditions}, it is only accurate in the infinite-volume limit. The next two proofs cover in a more mathematically-rigorous manner the case of expectation values of the chromoelectric portion of the Hamiltonian for all chromoelectric terms and the case chromoelectric terms independent of or linearly dependent on distance between sites. 

\paragraph{General proof of zero action for expectation values of $H_{el}$}
\label{sec:generalexpvalueprooffordifferentfundamentalsingletsites}


The chromoelectric term is equivalent to the SU(3) Casimir operator \cite{ physrevd.103.094501}. The Casimir operator ($T_a^2$) of a state that is composed of two excitations on a singlet can be expressed in terms of the Casimir operators of the individual excitations (which we can call ``A" and ``B") like so \cite{georgi2000lie}:

\begin{equation}
    T_a^2 = {T_a^A}^2 + {T_a^B}^2 + 2 T_a^A T_a^B
    \label{eq:Casimirdecomposition}
\end{equation}

\noindent with $a$ denoting which of the eight Gell-Mann matrices the term that it subscripts stands in for. Since in this case excitations A and B are entangled with the states of quarks on the lattice-sites (which are themselves superpositions of states with an equal probability of measuring any of the states in either the $\bf 3$ or $\bf \bar{3}$ irrep), any local operations on the link-state will have an expectation value equivalent to that obtained from applying said operations to the maximally mixed state of all basis states of the irreps that the link is in. $T_a^A T_a^B$ has an expectation value of 0 for such a state, so one can obtain the expectation value of the Casimir of the link from the expectation values of the Casimirs of the two excitations.

Given this fact, the expectation value of $H_{el}$ of a system is always equivalent to the sum of expectation values of $H_{el}$ of the system's component fundamental color singlets. Thus, if one's goal is to compute the expectation value of $H_{el}$, and by extension the expectation value of the full SU(3) Kogut-Susskind Hamiltonian, it is an acceptable approximation to assume that

\paragraph{Assessment for Eqs. \ref{eq:simplifiedsitepairchromoterm} and \ref{eq:simplifiedsitepairchromoterm_minimalZZs}}

\begin{figure}
    \centering
    \includegraphics[width=1.0\linewidth]{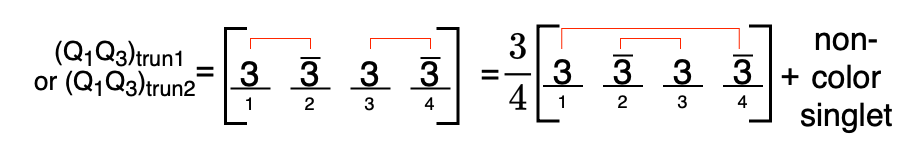}
    \caption{The result of acting $(Q^{(a)}_{n,f} Q^{(a)}_{m,f'})_{trun1}$ or $(Q^{(a)}_{n,f} Q^{(a)}_{m,f'})_{trun2}$ on two sites from two different meson excitations but of the same irrep. The $\bf{3}$ and $\bf{\bar{3}}$ denote the irreps on the sites; the red connections denote which sites are part of the same meson excitation.}
    \label{fig:QnQmtrundifferentmesonexcsameirrep}
\end{figure}

\begin{figure}
    \centering
    \includegraphics[width=1.0\linewidth]{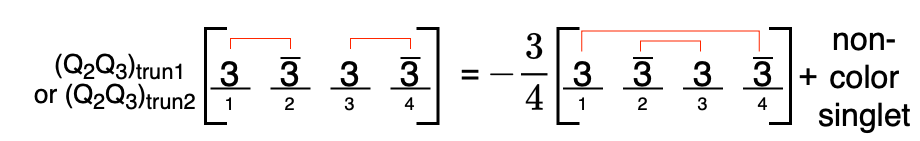}
    \caption{The result of acting $(Q^{(a)}_{n,f} Q^{(a)}_{m,f'})_{trun1}$ or $(Q^{(a)}_{n,f} Q^{(a)}_{m,f'})_{trun2}$ on two sites from two different meson excitations and of different irreps. The $\bf{3}$ and $\bf{\bar{3}}$ denote the irreps on the sites; the red connections denote which sites are part of the same meson excitation.}
    \label{fig:QnQmtrundifferentmesonexcdifferentirreps}
\end{figure}

Fig. \ref{fig:QnQmtrundifferentmesonexcsameirrep} shows the action of $(Q^{(a)}_{n,f} Q^{(a)}_{m,{f'}})_{trun1}$ and $(Q^{(a)}_{n,f} Q^{(a)}_{m,{f'}})_{trun2}$ on two lattice sites from two different fundamental singlet states that are either both in irrep $\bf{3}$ or both in irrep $\bf{\bar{3}}$., and Fig. \ref{fig:QnQmtrundifferentmesonexcdifferentirreps} shows the action of $(Q^{(a)}_{n,f} Q^{(a)}_{m,{f'}})_{trun1}$ and $(Q^{(a)}_{n,f} Q^{(a)}_{m,{f'}})_{trun2}$ on two lattice sites from two different fundamental singlet states, one of which is in irrep $\bf{3}$ and the other one of which is in irrep $\bf{\bar{3}}$.

The existence of the non-color singlet terms in the actions of $(Q^{(a)}_{n,f} Q^{(a)}_{m,f'})_{trun1}$ and $(Q^{(a)}_{n,f} Q^{(a)}_{m,f'})_{trun2}$ on sites belonging to different color singlet states mean that in general, these terms are not viable substitutes for $Q^{(a)}_{n,f} Q^{(a)}_{m,f'}$ unless the Hilbert space is artificially restricted to the color singlet subspace. 

Both the actions in Figs. \ref{fig:QnQmdifferentmesonexcsameirrep} and \ref{fig:QnQmdifferentmesonexcdifferentirreps} and the actions in Figs. \ref{fig:QnQmtrundifferentmesonexcsameirrep} and \ref{fig:QnQmtrundifferentmesonexcdifferentirreps} contain a component orthogonal to the input that is only produced by applying a chromoelectric term to site-pairs belonging to different fundamental singlet states. In the latter, the components not orthogonal to the input are directly proportional to the orthogonal components. Thus, a proof that a sum of $Q^{(a)}_{n,f} Q^{(a)}_{m,f'}$ terms applied to site-pairs belonging to different fundamental singlet states is zero is also sufficient to prove that the sum of the orthogonal components of the actions in Figs. \ref{fig:QnQmtrundifferentmesonexcsameirrep} and \ref{fig:QnQmtrundifferentmesonexcdifferentirreps}, and by extension those in Figs.  \ref{fig:QnQmdifferentmesonexcsameirrep} and \ref{fig:QnQmdifferentmesonexcdifferentirreps}, sum up to 0. Since this orthogonal component is the only color singlet component of the actions in Figs. \ref{fig:QnQmtrundifferentmesonexcsameirrep} and \ref{fig:QnQmtrundifferentmesonexcdifferentirreps}, in a situation where the sum of the full $Q^{(a)}_{n,f} Q^{(a)}_{m,{f'}}$ terms applied to site pairs belonging to different fundamental singlet states is 0 and the Hilbert space is restricted to the color singlet subspace, the sum of the $(Q^{(a)}_{n,f} Q^{(a)}_{m,{f'}})_{trun1}$ or the $(Q^{(a)}_{n,f} Q^{(a)}_{m,{f'}})_{trun2}$ terms applied to site-pairs belonging to different fundamental singlet states equal 0. This, combined with . Because SC-ADAPT-VQE is dependent on the expectation value of the Hamiltonian, for which it has already been proven that the sum of $Q^{(a)}_{n,f} Q^{(a)}_{m,{f'}}$ terms applied to site pairs belonging to different fundamental singlet states is 0, the former condition is met. Since SC-ADAPT-VQE, as specified at the end of the first paragraph of Sec. \ref{sec:0vBB_stateprep_baryons}, is automatically restricted to producing states within the color singlet subspace due to the extent of its operator pool.

Thus, for all intents and purposes pertaining to the application of SC-ADAPT-VQE as specified at the end of the first paragraph of Sec. \ref{sec:0vBB_stateprep_baryons} (and by extension all instances of VQE and its variants whose ansatze and operator pools map elements of the color singlet space only to other elements of the color singlet space), $(Q^{(a)}_{n,f} Q^{(a)}_{m,{f'}})_{trun1}$ and $(Q^{(a)}_{n,f} Q^{(a)}_{m,{f'}})_{trun2}$ are equivalent to $Q^{(a)}_{n,f} Q^{(a)}_{m,{f'}}$.

\subsection{Numerical testing and discussion}
\label{sec:0vBB_numtestanddisc}

The $H_{{\rm quarks}} + H_{el}$ subcomponent of the full Kogut-Susskind Hamiltonian from Eq. \ref{eq:0vBB_stateprep_fullhamiltonian} both in its original open boundary conditions case and in the periodic boundary conditions case laid out in Eqs. \ref{eq:HchromoPBC_0vBB} and \ref{eq:0vBB_HquarkleptonPBC} in both the open boundary conditions and periodic boundary conditions scenarios were both initialized both in the case where $Q^{(a)}_{n,f} Q^{(a)}_{m,{f'}}$ takes on its original value from Eq. \ref{eq:QnfQmfp} and in the cases where $Q^{(a)}_{n,f} Q^{(a)}_{m,{f'}}$ is swapped out for $(Q^{(a)}_{n,f} Q^{(a)}_{m,{f'}})_{trun1}$ and $(Q^{(a)}_{n,f} Q^{(a)}_{m,{f'}})_{trun2}$. All initializations were then mapped to the SU(3) color singlet subspace defined in Sec. \ref{sec:colorsingletconstruction}, and the cases where $Q^{(a)}_{n,f} Q^{(a)}_{m,{f'}}$ takes on its original value were found to be equivalent to their counterparts where $Q^{(a)}_{n,f} Q^{(a)}_{m,{f'}}$ is swapped out for $(Q^{(a)}_{n,f} Q^{(a)}_{m,{f'}})_{trun1}$ and $(Q^{(a)}_{n,f} Q^{(a)}_{m,{f'}})_{trun2}$. However, a similar test where after the initialization the penalty term discussed in Footnote \ref{foot:penaltyterm} of  Sec. \ref{sec:colorBreak} is added to each Hamiltonian and the eigenvectors of each Hamiltonian are taken and tested for their overlaps with the color singlet subspace, the original $Q^{(a)}_{n,f} Q^{(a)}_{m,{f'}}$ value Hamiltonian was found to not be equivalent to the $(Q^{(a)}_{n,f} Q^{(a)}_{m,{f'}})_{trun1}$ and $(Q^{(a)}_{n,f} Q^{(a)}_{m,{f'}})_{trun2}$ Hamiltonians. This confirms the result in previous sections that while in general, $(Q^{(a)}_{n,f} Q^{(a)}_{m,{f'}})_{trun1}$ and $(Q^{(a)}_{n,f} Q^{(a)}_{m,{f'}})_{trun2}$ is not a viable substitute for $Q^{(a)}_{n,f} Q^{(a)}_{m,{f'}}$, they are a viable substitute for cases where the Hilbert space is restricted to the color singlet subspace, such as as SC-ADAPT-VQE as specified at the end of the first paragraph of Sec. \ref{sec:0vBB_stateprep_baryons}.

As discussed in Sec. \ref{sec:intro_stateprep}, SC-ADAPT-VQE has a step in which ADAPT-VQE is implemented, which in turn has a step where VQE is implemented. As seen in Ref. \cite{peruzzo_2014}, when VQE is applied to the problem of optimizing the expectation value of the Hamiltonian (and finding the baryon ground state is one of these cases), the Hamiltonian is split into parts, each consisting entirely of terms which commute with each other. The quantum circuit is then run separately for each of these parts. Substituting in  $(Q^{(a)}_{n,f} Q^{(a)}_{m,{f'}})_{trun2}$ for $Q^{(a)}_{n,f} Q^{(a)}_{m,{f'}}$ (the two of which, as proven above, are equivalent for purposes of VQE with a color singlet space conserving operator pool) removes all of the off-diagonal terms from the chromoelectric term of the Hamiltonian. Thus, if the chromoelectric term of the 1+1D SU(3) Kogut-Susskind Hamiltonian term is written using this substitution, for purposes of VQE it could be split into one part for both the mass and chromoelectric term and a few other parts for the kinetc terms, while without the substitution multiple parts would be needed for the chromoelectric term alone. The former is therefore much less expensive to take the expectation value of, and thus is much less expensive to run VQE on.

\section{Time-evolution circuits for superconducting devices}
\label{sec:0vBB_timeevcircforsupercond}

The circuits used to effect the time-evolution in Chapters \ref{chap:1p1dQCD} and \ref{chap:1p1dSM} require a large SWAP gate overhead. For instance, the 
circuit developed in Ref. \cite{klco:2019xro} and used in Chapters \ref{chap:1p1dQCD} and \ref{chap:1p1dSM} for the Trotterization of the chromoelectric term require at least 2-4 SWAP gates per CNOT gate between non-adjacent qubits. In this section, I devise circuits which, at least for the lepton lattice from Chapter \ref{chap:1p1dSM} and quark lattices with one quark flavor, execute the Trotterization of the kinetic and chromoelectric terms on nearest-neighbor devices with a CNOT overhead that requires only 1.5 times the CNOT gate depth for implementation on nearest-neighbor connectivity devices as it does for implementation on all-to-all connectivity devices. These circuits are presented for the periodic boundary conditions (PBC) Hamiltonian, but they are trivial to adapt to the OBC Hamiltonian. Additionally, in Refs. \cite{Farrell:2023fgd} and \cite{Farrell:2024fit} it was found that color charge correlations decay exponentially over distance between the color charges, so truncating terms of $H_{el}$ that couple sites farther away from each other than a cutoff distance $\lambda$ would lower circuit depth and gate count and bring a semblance of locality to the Hamiltonian without causing a significant deviation of the final results from those that would be obtained using the full physical picture. For $\lambda = 1$, which is used in this section, the chromoelectric term becomes

\begin{equation}
    H_{el_{\lambda = 1}} \rightarrow K \frac{g^2}{2} (-1 + \frac{1}{2L}) \sum_{n = 0}^{L - 1} Q^{a}_{n} Q^{a}_{n+1}; 
    \left\{
    \begin{array}{lr}
        K = \frac{1}{2} & \text{if } L = 1 \\
        K = 1 & \text{if } L > 1
    \end{array}
    \right .
    \label{eq:0vBB_latticeHchromolambda1}
\end{equation}

\noindent for the case where there is only one flavor of quarks. 

One of the IBM superconducting devices' main advantages is that unlike trapped ion devices they can trivially run gates on different qubits in parallel, and one of their main disadvantages is that their devices are, for most practical intents and purposes, restricted to nearest-neighbor connectivity\cite{ibmq}. Thus, in order for circuits to be efficiently implementable on IBM's superconducting devices, they should require a minimal number of SWAP gates in order to be transpiled onto a device with nearest-neighbor connectivity and besides that should have the lowest possible circuit depth, but aside from that the number of gates on the device is not important. The circuits in Sec. \ref{sec:0vBB_timeevcircforsupercond} are designed with these considerations in mind, and thus are not always optimal for trapped ion devices, whose ability to run gates in parallel is limited \cite{quantinuum, moses2023race, IonQ}. 

\begin{figure}
    \centering

    \includegraphics[width=0.4\linewidth]{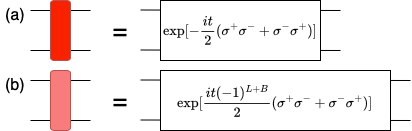}
    \includegraphics[width=0.4\linewidth]{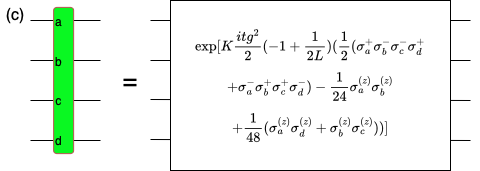} \\

    \includegraphics[width=0.49\linewidth]{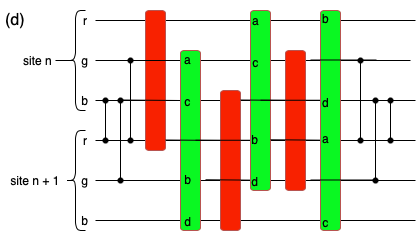} 
    \includegraphics[width=0.49\linewidth]{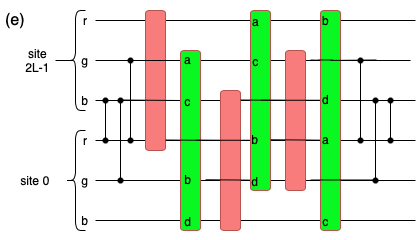}
    
    \caption{A Trotter decomposition of the kinetic and chromoelectric terms from the PBC SU(3) Kogut-Susskind Hamiltonian on one pair of 3-color staggered lattice sites using the CZ-operations in Eq. \ref{eq:CZidentities}. (a), (b): Trotterized terms from the kinetic and chromoelectric portions of the Hamiltonian, respectively, (c): a Trotterized term from the chromoelectric part of the Hamiltonian, (d) the Trotterization of all kinetic and chromoelectric Hamiltonian terms acting between two adjacent lattice sites other than the site-pair (n = 2L - 1, n = 0), (e) the Trotterization of all kinetic and chromoelectric Hamiltonian terms acting between the site-pair (n = 2L - 1, n = 0). } 
    \label{fig:CZimplemented_kineticChromo}
\end{figure}

\subsection{Implementation of the Fermionic Z-strings}

The technique of building out Jordan-Wigner transformed operations by implementing them as if they were not Jordan-Wigner mapped and then applying a network of fermionic SWAP gates (FSWAPs) discussed in Sec. \ref{sec:0vBB_leptonstateprep} can be applied to the kinetic and chromoelectric parts of the lepton and quark parts of the Hamiltonians discussed in Sec. \ref{sec:0vBB_stateprep_Hamiltonians}.
The one difference is that the CZ or FSWAP network needs to be applied on both sides of the non-Jordan-Wigner mapped rotations. Taking advantage this, the Trotterized time-evolution of any given pair of adjacent staggered lattice sites with the kinetic and chromoelectric terms in the Hamiltonian can be implemented using the procedure in Fig. \ref{fig:CZimplemented_kineticChromo}. It is important to note that Fig.  \ref{fig:CZimplemented_kineticChromo} uses a CZ network designed for the kinetic term for both the kinetic and chromoelectric terms. This works, but has the effect of flipping the sign of the off-diagonal portion of the chromoelectric term, which has to be adjusted for in the Trotterization.

\subsection{Adaptation to nearest-neighbor circuits}

In its raw form, the Trotter decomposition in Fig. \ref{fig:CZimplemented_kineticChromo} requires a significant number of SWAP gates to implement on a nearest-neighbor device. One can mitigate this requirement by replacing the CZ gates with FSWAP gates, which simultaneously implement a CZ gate and a SWAP gate while having a depth of 2 two-qubit gates. An FSWAP gate can be implemented using single-qubit gates and CNOT gates as shown in Fig. \ref{fig:FSWAPimplementation}:

\begin{figure}[h!]
    \centering
    \includegraphics[width=0.5\linewidth]{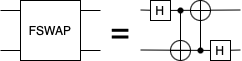}
    \caption{The implementation of the FSWAP gate, which simulataneously executes a SWAP and a CZ gate}
    \label{fig:FSWAPimplementation}
\end{figure}

\begin{figure}[h!]
    \centering

    \includegraphics[width=0.49\linewidth]{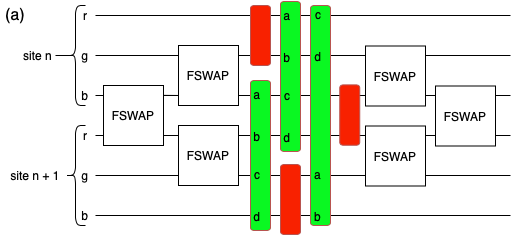} 
    \includegraphics[width=0.49\linewidth]{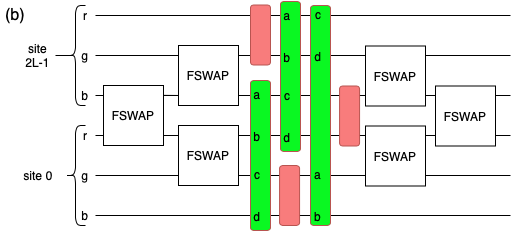}
    
    \caption{A Trotter decomposition of the kinetic and chromoelectric terms from the PBC SU(3) Kogut-Susskind Hamiltonian on one pair of 3-color staggered lattice sites using the CZ-operations in Eq. \ref{eq:CZidentities} implemented using the FSWAP operation shown in Fig. \ref{fig:FSWAPimplementation}. (a) the Trotterization of all kinetic and chromoelectric Hamiltonian terms acting between two adjacent lattice sites other than the site-pair (n = 2L - 1, n = 0), (b) the Trotterization of all kinetic and chromoelectric Hamiltonian terms acting between the site-pair (n = 2L - 1, n = 0). Definitions of the red and green operators can be found in Fig. \ref{fig:CZimplemented_kineticChromo}.} 
    \label{fig:FSWAPimplemented_kineticChromo}
\end{figure}

Replacing the CZ gates in Fig. \ref{fig:CZimplemented_kineticChromo} with FSWAPs causes the qubits to be rearranged in order of color. That is, the two color-red qubits are on top, the two color-green qubits are in the middle, and the two color-blue qubits are on the bottom. The benefit of this is that the qubits that the kinetic and chromoelectric Hamiltonian's terms act on are, with the exception of the chromoelectric term between the red and blue qubits, adjacent to each other. This greatly reduces the number of SWAP operations needed when implementing the Trotterization on a device with nearest neighbor connectivity. The Trotterization in Fig. \ref{fig:CZimplemented_kineticChromo} can then be replaced with the one in Fig. \ref{fig:FSWAPimplemented_kineticChromo}.

\begin{figure}
    \centering
    \includegraphics[width=0.5\linewidth]{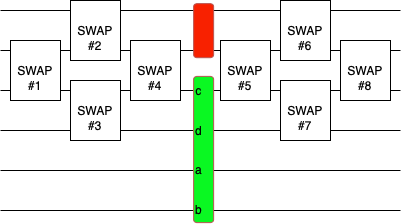}
    \caption{The SWAP-network needed to execute the portion of the chromoelectric term in the circuit in Fig. \ref{fig:FSWAPimplemented_kineticChromo} that couples red quarks with blue quarks. SWAP gates are numbered for future reference.}
    \label{fig:redblueSWAPnet}
\end{figure}

The Trotterization in Fig. \ref{fig:FSWAPimplemented_kineticChromo} still requires a network of SWAP gates (na\"{i}vely 8 SWAP gates with a depth of 18 CNOTs) in order to implment the part of the chromoelectric term that acts between qubits representing red quarks and those representing blue quarks. This SWAP network is illustrated in Fig. \ref{fig:redblueSWAPnet}. Additionally, the chromoelectric term implementation derived in Sec. \ref{sec:chromotermimplementation} does not include diagonal parts that couple qubits representing quarks of different colors, so these have to be implemented separately. Thus, the next step is to find ways of incorporating the SWAPs necessary to meet these two requirements in as compact a manner as possible. 

\begin{figure}
    \centering
    \includegraphics[width=1.0\linewidth]{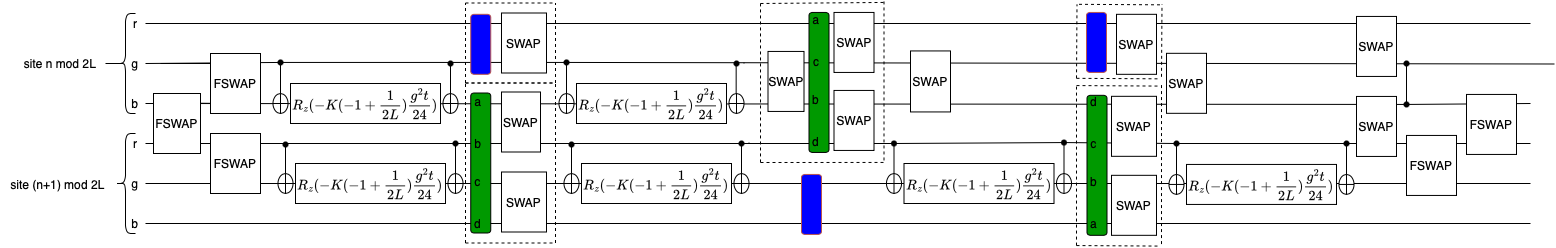}\\
    \includegraphics[width=0.33\linewidth]{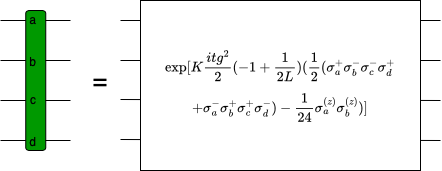}
    \hspace{16ex}
     \includegraphics[width=0.16\linewidth]{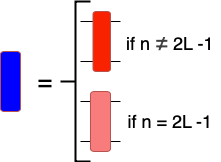}
    \caption{Top: the Trotterization of the kinetic term and the $\lambda$=1 chromoelectric term truncation of the Hamiltonian between two adjacent sites that requires no extra SWAP gates to run on a nearest neighbor device. Dashed line enclosures indicate circuit components that are executed together. Bottom left: a definition of the chromoelectric term between four quarks of two different colors without the diagonal part which couples different colors. Bottom right: the definition of the piecewise kinetic term, meant to handle both the loop-around of the PBC lattice and the rest of it.}
    \label{fig:completeNNTrotterization}
\end{figure}

In order to implement the diagonal terms that pair the red quark from site n with the green quark from site n + 1, the green quark from site n with the blue quark from site n + 1, and the red quark from site n with the blue quark from sign n + 1, SWAP gates between qubits representing quarks of the same color will be needed. Luckily, as is demonstrated in Secs. \ref{sec:kintermimplementation} and \ref{sec:chromotermimplementation}, it is possible to incorporate these SWAPs into the kinetic and chromoelectric term at the cost of an additional depth of 1 CNOT. The specific places where these SWAPs are incorporated are at the first and third layers (not counting the FSWAPs) of the Trotterization in Fig. \ref{fig:FSWAPimplemented_kineticChromo}. After these SWAPs are applied, it is straightforward to add ZZ rotations directly into the circuit. For the SWAP-network in Fig. \ref{fig:redblueSWAPnet}, two optimizations can be applied. First, SWAPs {\#} 2 and {\#} 3 can be incorporated into the second layer's chromoelectric term using the aforementioned technique. Second, SWAP {\#} 8 can be merged with one of the FSWAP gates at the end of the Trotterization in Fig. \ref{fig:FSWAPimplemented_kineticChromo} by replacing both with a CZ gate. The circuit after these steps is illustrated in Fig. \ref{fig:completeNNTrotterization}. The rest of the section will concern implementation of the kinetic and chromoelectric terms.

\subsection{Kinetic term implementation}
\label{sec:kintermimplementation}
\begin{figure}
    \centering
    \includegraphics[width=0.4\linewidth]{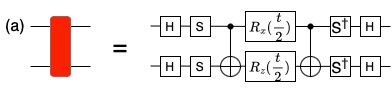}
    \includegraphics[width=0.59\linewidth]{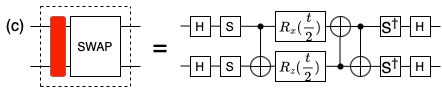} \\
    \includegraphics[width=0.4\linewidth]{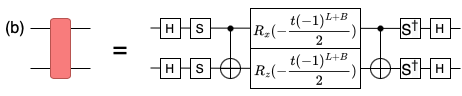}
    \includegraphics[width=0.59\linewidth]{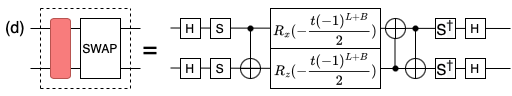}
    \caption{The explicit implmenentations of the kinetic terms used in Figs. \ref{fig:CZimplemented_kineticChromo} and \ref{fig:FSWAPimplementation} in addition to variations with a SWAP operation appended. Dashed line enclosures indicate circuit components that are executed together.}
    \label{fig:kinetic_detailedimplementations}
\end{figure}

The optimal Trotterization of a term in the kinetic portion of the Hamiltonian is presented in Ref. \cite{Farrell:2024fit}. Additionally (this will be useful in future steps), a SWAP can be implemented simultaneously with the kinetic Trotterization at the cost of one additional CNOT gate. The optimal Trotterizations and the appending of the SWAP term are detailed in Fig. \ref{fig:kinetic_detailedimplementations}.

\subsection{Chromoelectric term implementation}
\label{sec:chromotermimplementation}
In Chapters \ref{chap:1p1dQCD} and \ref{chap:1p1dSM}, a chromoelectric term based on the $\beta$-angle implementation in Ref. \cite{klco:2019xro} is used. However, if implemented on a nearest-neighbor connectivity device, it would require 12 SWAP gates per term, even after applying the FSWAP network in Fig. \ref{fig:FSWAPimplemented_kineticChromo}. Thus, in order to efficiently implement the chromoelectric term on a nearest-neighbor connectivity devices such as IBM's superconducting devices, one should find ways of cutting down on this SWAP overhead. 

As seen in Eq. \ref{eq:QnfQmfp}, the off-diagonal component of the chromoelectric term is composed of terms $\sigma^{+}\sigma^{-}\sigma^{-}\sigma^{+} + \sigma^{-}\sigma^{+}\sigma^{+}\sigma^{-}$, which, as found in Chapter \ref{chap:1p1dQCD}, can be broken down like so:

\begin{align}
    \sigma^{+}\sigma^{-}\sigma^{-}\sigma^{+} + h.c. =& \frac{1}{8}(XXXX + YYXX \nonumber \\ +& YXYX - YXXY - XYYX + XYXY + XXYY + YYYY)
    \label{eq:offdiagonalchromotermXYbreakdown}
\end{align}

\begin{figure}
    \centering
    \includegraphics[width=1.0\linewidth]{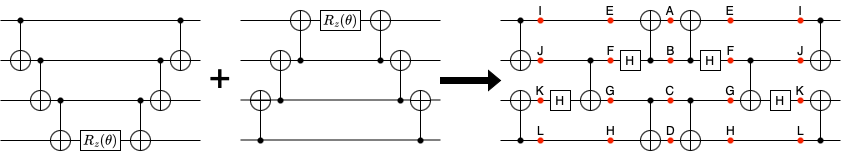}
    \caption{A naive combination of two implementations of an $e^{-i \theta ZZZZ}$ rotation in an attempt to pack as many rotations with respect to four-element Pauli strings into a quantum circuit with a depth of 6 CNOT gates. The Hadamard gates are an attempt to make the CNOTs between the middle two qubits bidirectional. The red points are possible insertion-locations for single-qubit rotations, and the circuit rotations produced by inserting a single-qubit rotation at each point are found in Tab. \ref{tab:singlequbitrotmaptable}. Red points with the same label produce the same circuit rotations out of the same single-qubit rotations.}
    \label{fig:ZZZZcombination}
\end{figure}

One can change the bases of the individual qubits using single-qubit transformations, so in principle it is possible to encode the off-diagonal component of the chromoelectric term by having a sequence of ZZZZ rotations, each basis-shifted in order to match a term in Eq. \ref{eq:offdiagonalchromotermXYbreakdown}. Given that each ZZZZ rotation has 6 CNOTs, this would require 48 CNOTs per term, which, given that the diagonal components of the chromoelectric term must be implemented separately, is not an improvement in circuit depth over the approach in Chapters \ref{chap:1p1dQCD} and \ref{chap:1p1dSM}. Thus, in order for this approach to succeed, one must squeeze as many rotations of separate length-4 Pauli strings into a depth-6 CNOT circuit. Here, it is attempted via an overlap of two implementations of the ZZZZ term, as shown in Fig. \ref{fig:ZZZZcombination}. The $R_x$, $R_y$, and $R_z$ rotations are applied to each one of the red points in Fig. \ref{fig:ZZZZcombination}, and the rotation exhibited by the circuit for an insertion of $R_x$, $R_y$, and $R_z$ at each point is shown in Tab. \ref{tab:singlequbitrotmaptable}.

\begin{table}[h!]
    \centering
    \scalebox{0.9}{\begin{tabular}{l|c c c c c c c c c c c c}
          & A & B & C & D & E & F & G & H & I & J & K & L \\
         \hline
         $R_x$ & XXII & YYII & IIYY & IIXX & XXII & ZZII & IIZZ & IIXX & XXII & IXII & IIXI & IIXX \\ 
         $R_y$ & YIZZ & YZZZ & ZZZY & ZZIY & YXII & ZYZZ & ZZYZ & IIXY & YXII & ZYII & IIYZ & IIXY \\ 
         $R_z$ & ZXZZ & IXZZ & ZZXI & ZZXZ & ZIII & IXZZ & ZZXI & IIIZ & ZIII & ZZII & IIZZ & IIIZ \\ 
\end{tabular}}
    \caption{A table of the Pauli strings of the rotations that the circuit on the right of Figure \ref{fig:ZZZZcombination} maps each one of the single-qubit rotations $R_x$, $R_y$, and $R_{z}$ to if they are applied to a red point labeled by each letter in the leftmost column.}
    \label{tab:singlequbitrotmaptable}
\end{table}

As seen in Tab. \ref{tab:singlequbitrotmaptable}, four different rotations with respect to a length-4 Pauli string are possible on the circuit on the right of Fig. \ref{fig:ZZZZcombination}. Eq. \ref{eq:offdiagonalchromotermXYbreakdown} has 8 terms, so in principle it should be possible to implement a rotation with respect to Eq. \ref{eq:offdiagonalchromotermXYbreakdown} by applying the correct single-qubit basis transformations to a sequence of two repetitions of the circuit on the right of Fig. \ref{fig:ZZZZcombination}. The set of single-qubit basis transformations used for this purpose is shown in Table \ref{tab:basistransformationset}. 

\begin{table}[h!]
    \centering
    \begin{tabular}{l|c c c}
         & X & Y & Z \\
         \hline
         $H S^{\dagger}$ & Y & Z & X \\
         $S H$ & Z & X & Y \\
         $S H S^{\dagger}$ & -X & Z & Y \\
         $HX$ & Z & Y & -X \\
         $XH$ & -Z & Y & X 
    \end{tabular}
    \caption{A table of the basis that each transformation in the leftmost column takes maps the basis in the topmost row onto. Negative signs denote that a change in sign took place.}
    \label{tab:basistransformationset}
\end{table}

\begin{figure}
    \centering
    \includegraphics[width=1.0\linewidth]{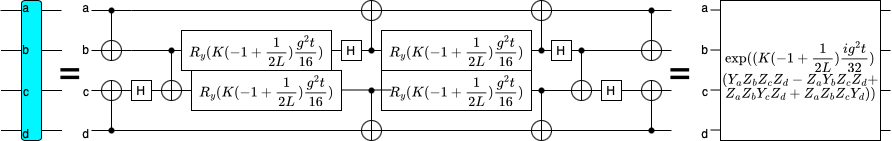} \\
    \includegraphics[width=1.0\linewidth]{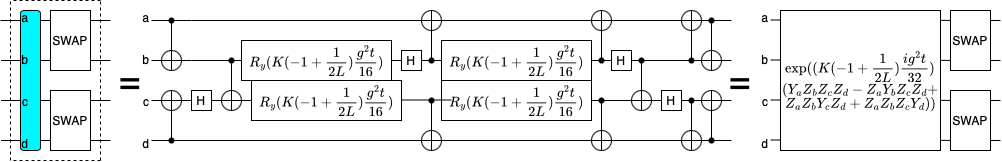} \\
    \includegraphics[width=1.0\linewidth]{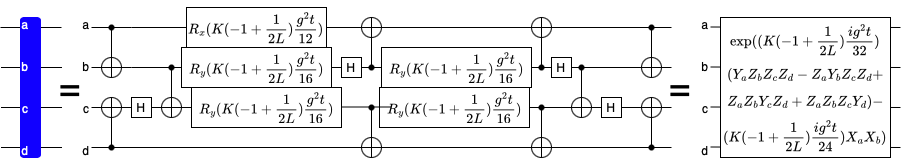} \\
    \includegraphics[width=1.0\linewidth]{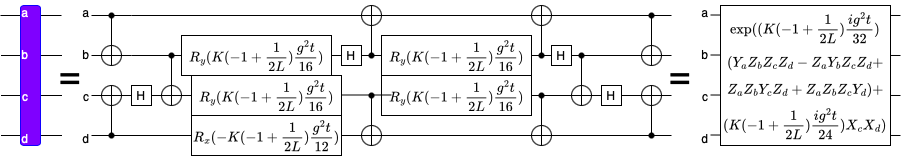}
    \caption{The definitions of four circuits based on the mappings in Tab. \ref{tab:singlequbitrotmaptable} which the construction of the Trotterization of the chromoelectric term is based on. Dashed line enclosures denote components that are executed together}
    \label{fig:ZZZZcombinationchromoelectricbasiscircuits}
\end{figure}

\begin{figure}
    \centering
    \includegraphics[width=1.0\linewidth]{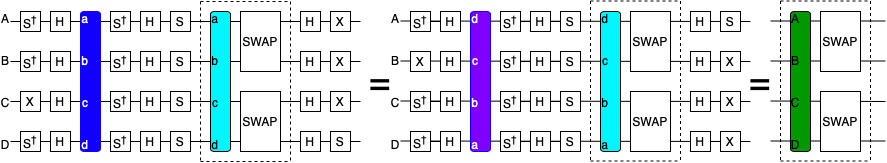} \\
    \includegraphics[width=0.7\linewidth]{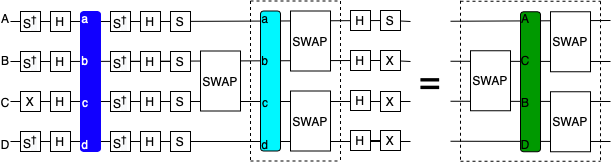}
    \caption{Constructions of the 4-qubit chromoelectric term defined in Fig. \ref{fig:completeNNTrotterization} using the circuits defined in Fig. \ref{fig:ZZZZcombinationchromoelectricbasiscircuits} and the transformations detailed in Tab. \ref{tab:basistransformationset}. Top: the terms with just the same-color SWAP; Bottom: the term that partially implements the SWAP network in Fig. \ref{fig:redblueSWAPnet}.}
    \label{fig:roughgreenchromotermconstruction}
\end{figure}

Additionally, based on the mappings laid out in Tab. \ref{tab:singlequbitrotmaptable} the circuits in Fig. \ref{fig:ZZZZcombinationchromoelectricbasiscircuits} are defined. These can then be combined with the mappings in Tab. \ref{tab:basistransformationset} to produce a construction of the chromoelectric 4-qubit term defined in Fig. \ref{fig:CZimplemented_kineticChromo}. This construction is shown in Fig. \ref{fig:roughgreenchromotermconstruction}.

\subsection{Quantum resource count for the new chromoelectric implementation}
The circuit in Fig. \ref{fig:roughgreenchromotermconstruction} has a CNOT gate count of 112 and a CNOT gate depth of 63, and would have a CNOT gate count of 84 and a CNOT gate depth of 48 if adapted to an all-to-all connectivity device by having all of the SWAP operations in the circuit removed and all the FSWAP gates substituted for CZ's. The counterpart to this circuit from Chapter \ref{chap:1p1dQCD}, the chromoelectric and kinetic term Trotterization for a pair of adjacent lattice sites, would have both a CNOT gate count and circuit depth of 64 for all-to-all connectivity. This is lower in terms of CNOT gate count but higher in terms of depth, and it cannot as easily be adapted to trapped ion devices. Thus, the old chromoelectric term Trotterization circuit from Chapters \ref{chap:1p1dQCD} and \ref{chap:1p1dSM} is optimal for trapped ion devices, which have all-to-all connectivity but limited parallelism, so gate count matters more for them. However, the new circuits laid out in this section are more optimal for superconducting devices, as they require mapping to nearest-neighbor connectivity and can trivially run gates in parallel.

The circuits in this section were successfully tested via (1) using linear algebra to create exponentiations of the Hamiltonian's terms and comparing them to the actions of the circuit elements and (2) time-evolving using exact classical methods on an SU(3) Kogut-Susskind lattice with 2 physical sites and 1 quark flavor and replicating the time-evolution through a Trotterization using the circuit in Fig. \ref{fig:completeNNTrotterization} with the expectation that the state-overlaps at the end of the time evolution would converge to 1 in the limit of many Trotter step.

\section{Lepton term Trotterization}

The Trotterization on the lepton register is mostly the same for both types of devices. This is done on four qubits at a time, as follows. First, the Jordan-Wigner mapping is implemented by having a pair of CZ gates for an all-to-all connectivity device or a pair of FSWAP gates for a nearest-neighbor device. Then, one $e^{(-i \frac{1}{2} (\theta_1 XY + \theta_2 YX))}$ operation using from Ref. \cite{Farrell:2023fgd} between the electron qubits and one $e^{(-i \frac{1}{2} (\theta_1 XY + \theta_2 YX))}$  operation between the neutrino qubits would be wedged between the CZs or the FSWAPs to implement the Majorana mass and kinetic Hamiltonian terms without the Jordan-Wigner mapping.

\section{Time-evolution circuits for trapped ion devices}
\label{sec:0vBBtimeevfortrappedion}

On trapped ion devices, the optimal Trotterization is mostly the same as in Chapters \ref{chap:1p1dQCD} and \ref{chap:1p1dSM}, with a couple of modifications. 

\begin{figure}
    \centering
    \includegraphics[width=\linewidth]{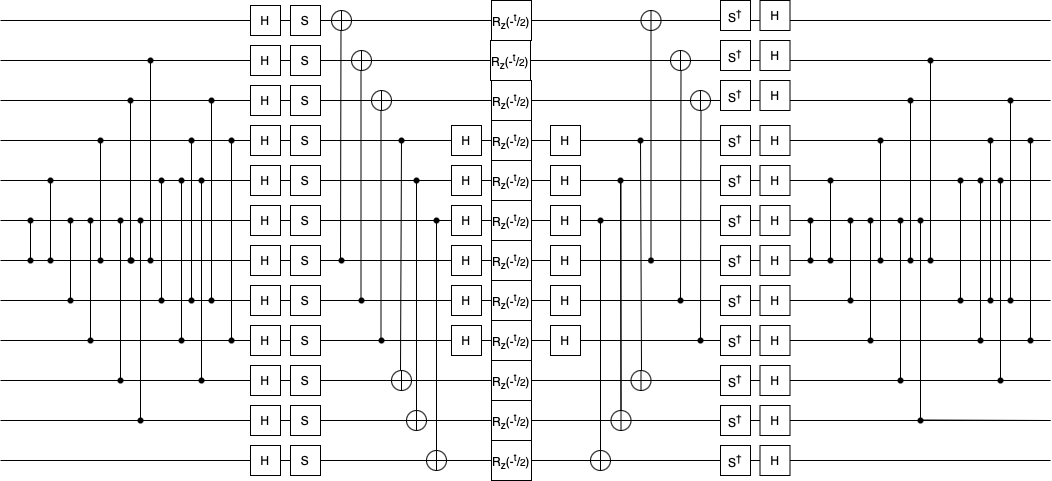}
    \caption{The implementation of the Trotterization of the kinetic portion of the quark term of the SU(3) Kogut-Susskind Hamiltonian in the case of 2 quark flavors, using the procedure in Sec. \ref{sec:0vBB_timeevcircforsupercond}.}
    \label{fig:kinetic2flavorhamiltoniantrotter}
\end{figure}

First,the kinetic term is implemented the same way it would be on a nearest-neighbor device, as described in Sec. \ref{sec:0vBB_timeevcircforsupercond} This is preferable for two reasons. First, this way requires less resources than the method used in Chapters \ref{chap:1p1dQCD} and \ref{chap:1p1dSM}. The Trotterization of a kinetic term between two adjacent staggered sites implemented using the procedure outlined in Sec. \ref{sec:0vBB_timeevcircforsupercond} would have a CNOT gate count of 12 and a CNOT gate depth of 6 for a one-flavor quark lattice, and a CNOT gate count of 42 and a CNOT gate depth of 12 for a two-flavor quark lattice. The corresponding resource counts from Trotterizing the kinetic term using the approach in Chapter \ref{chap:1p1dQCD} are a CNOT gate count of 24 and a CNOT gate depth of 18 for the one-flavor lattice case and a CNOT gate count of 56 and a gate depth of 56 for the two-flavor lattice case. Second, this avoids the need for ancilla qubits in a situation where the number of qubits may be scarce. Third, Quantinuum's devices can run up to 4 gates in parallel \cite{moses2023race}, and the circuit-building procedure described in Sec. \ref{sec:0vBB_timeevcircforsupercond} can help utilize this parallelism better than the kinetic term in Chapters \ref{chap:1p1dQCD} and \ref{chap:1p1dSM}, which runs the entire quark kinetic term sequentially. Such a kinetic term for the case of two quark flavors is illustrated in Fig. \ref{fig:kinetic2flavorhamiltoniantrotter}.
 
Additionally, in the $\beta$ decay component, for each permutation of the neutrino, electron, up quark, and down quark staggered sites, the terms for each of the three quark colors are executed immediately after each other. This avoids unnecessary repetition of the diagonalization of the electron portion of the $\beta$-decay Hamiltonian. 

\section{Classical simulation results}
\label{sec:0vBB_classicalsim}
\begin{figure}
    \centering
    \includegraphics[width=0.5\linewidth]{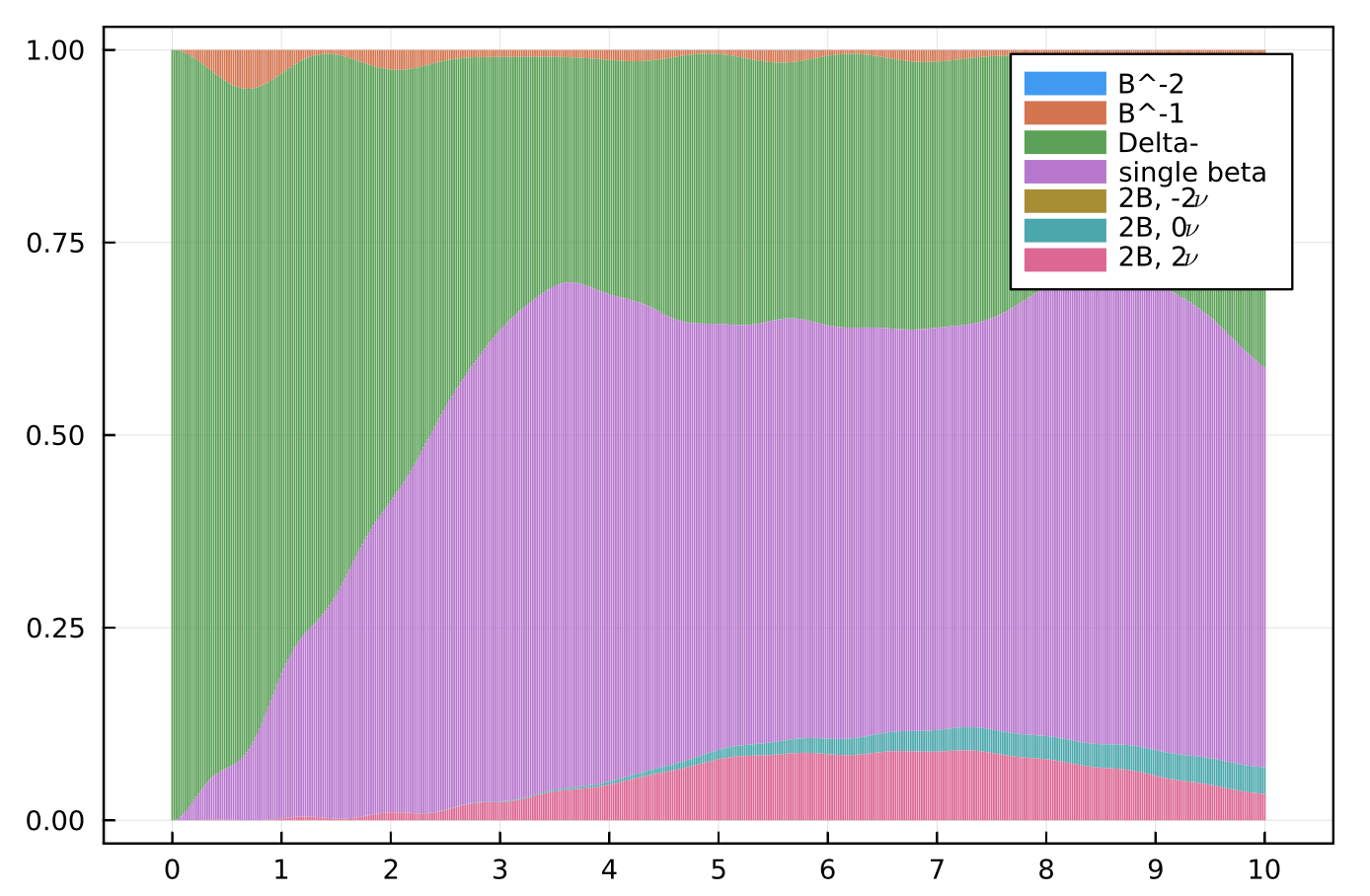}
    \caption{The results of time evolution under the Hamiltonian in Eq. \ref{eq:0vBB_stateprep_fullhamiltonian} on an initial state of $\Delta^-$, obtained using exact classical simulation}
    \label{fig:exact0vBBcolorsingletsim}
\end{figure}

In Fig. \ref{fig:exact0vBBcolorsingletsim}, I present exact classical results of time evolution under the Hamiltonian in Eq. \ref{eq:0vBB_stateprep_fullhamiltonian} of a state initially in the ground state of $\Delta^-$. To do so, I constructed the exact matrix expression of the full Hamiltonian in Eq. \ref{eq:0vBB_stateprep_fullhamiltonian}. I then use the method outlined in Sec. \ref{sec:colorsingletconstruction} to create a list of all possible color singlet states with baryon number 1 and sorted them by Z-component of isospin. Out of that list, I created a mapping matrix from the full Hilbert space to the baryon-number-1 color-singlet subspace and used it to map the exact matrix expression of the Hamiltonian. I then created a similar mapping matrix to the baryon-number-1, Z-component of isospin $-\frac{3}{2}$, used it to map just the $H_{quarks}$ component in Eq. \ref{eq:OBC_0nuBB_fullham}. I then used exact eigendecomposition to find the $H_{quarks} + H_{el}$ ground state within the baryon-number-1, Z-component of isospin $-\frac{3}{2}$ color singlet subspace, which I append to the lepton vacuum (found using exact eigendecomposition of $H_{leptons}$ from Eq. \ref{eq:OBC_0nuBB_fullham})  and choose as the $\Delta^{-}$ initial state in the $0 \nu \beta \beta$ surrogate reaction $\Delta^{-} \rightarrow \Delta^{+} e^- e^-$. Using the Expokit library \cite{sidje1998expokit} implemented in the Julia language \cite{julia-2017, expokit} I evolve my choice for $\Delta^{-}$ under the color-singlet-subspace-mapped full Hamiltonian, and sort the possible states in the result by isospin and lepton number to obtain results, an example of which are shown in Fig. \ref{fig:exact0vBBcolorsingletsim}.

The results in Fig. \ref{fig:exact0vBBcolorsingletsim} had Hamiltonian parameters of $m_u = 1.7$, $m_d = 3.6$, $g = 1$, $G_F = 0.5$, $\mu_{II} = -0.72$, $m_{M} = 0.2$. Besides these values I tested values of $G_F$ between 0.1 and 1.0, values of $g$ between 1 and 2, and values of $m_{M}$ between 0.0 and 0.2. I tested various other combinations of $m_u$ and $m_d$, notably $m_u = 2.9$ and $m_d = 6.2$.
$\mu_{II}$ was fixed to whatever value would ensure that the ($H_{quarks} + H_{el} + H_{leptons} + H_{Majorana}$) ground state energy of the baryon-number-1, Z-component of isospin $-\frac{3}{2}$ subspace would be less than the ground state energy of the baryon-number-1, Z-component of isospin $-\frac{1}{2}$ subspace but greater than the ground state energy of the baryon-number-1, Z-component of isospin $\frac{1}{2}$ subspace (-0.54 for $m_u = 2.9$ and $m_d = 6.2$, for instance). In all of these samples, I find the following patterns:
\begin{enumerate}
    \item Increasing the $m_{M}$ from 0.0 to 0.2 makes the interaction $\Delta^- \rightarrow \Delta^{+} e^- e^-$ possible, as expected.
    \item The parameters chosen for Fig. \ref{fig:exact0vBBcolorsingletsim} produced the highest probability of the interaction $\Delta^- \rightarrow \Delta^{+} e^- e^-$ happening.
    \item No matter the choice of parameters $m_u$, $m_d$, and $\mu_{II}$, single-beta decay is much more prevalent than double-beta decay, neutrinoless or neutrinoful.
\end{enumerate}

The third of these observations is likely because the lattice used in the aforementioned simulations has only 2 physical, or 4 staggered, lattice sites, which is the smallest possible lattice on which neutrinoless double beta decay can happen. This is significant, because the mass-hierarchy enforced by $m_u$, $m_d$, and $\mu_{II}$ is of ground states of the sub-Hamiltonian $(H_{quarks} + H_{el} + H_{leptons} + H_{Majorana})$. $H_{\beta}$, being outside this sub-Hamiltonian can cause excitations from a ground state to an excited state of the sub-Hamiltonian that would violate conservation of energy if $(H_{quarks} + H_{el} + H_{leptons} + H_{Majorana})$ was the only part of the Hamiltonian. As a result, only the mean of the expectation value of $(H_{quarks} + H_{el} + H_{leptons} + H_{Majorana})$ would be conserved. This is not as much of a problem in the continuum limit, as since as the lattice increases in size, as (1) the spacing of the states on the lattice decreases as the lattice gets larger and (2) excitations from one part of the lattice are more likely to be cancelled out on another part of the lattice, under the central limit theorem. 

Thus, while on a large lattice the mass hierarchy should still permit double-beta decay while forbidding the single-beta decay, it's natural that for the 2-physical-site lattice, perturbations from $H_{\beta}$ would cause single-beta decay to happen despite the mass hierarchy. Thus, for small lattices the chemical potential term $H_{\mu_{II}}$ is omitted and no circuit is devised for Trotterizing it.

\section{Discussion}

In summary, my colleague Roland Farrell and I have adapted the Hamiltonians from Chapter \ref{chap:1p1dSM} to a formulation that, given a device with sufficient qubits and circuit depth, could be used to simulate the interactions $\Delta^- \rightarrow \Delta^+ e^- e^-$ and $\Delta^- \Delta^- \rightarrow \Delta^0 \Delta^0 e^- e^-$, which are surrogates for neutrinoless double beta decay on a 1+1D lattice. Part of this effort was a failed attempt to use a chemical potential term to impose a mass hierarchy that would permit double beta decay but not single beta decay. We expect this mass hierarchy to work for much larger lattices. More important was the design of the state-preparation and Trotterization circuits tailored to the simulation of the neutrinoless double beta decay surrogates. The state-preparation circuits are based on the SC-ADAPT VQE circuits from Ref. \cite{Farrell:2023fgd}. As for time-evolution, because trapped ion devices have all-to-all connectivity but limited parallelizability, the old circuits from Chapters \ref{chap:1p1dQCD} and \ref{chap:1p1dSM} are, for the most part, the best for trapped ion devices due to their lower qubit count than the circuits in Sec. \ref{sec:0vBB_timeevcircforsupercond}, though there are a few small modifications that can make them more efficient. There is a situational exception to this for the Trotterization of the kinetic component of the Hamiltonian.

Nonetheless, the circuits in Sec. \ref{sec:0vBB_timeevcircforsupercond} will likely be essential to the use of superconducting devices, which have considerably more qubits than the trapped ion devices that have been the default for lattices greater in size than 1 quark flavor and 1 physical lattice site, in simulation of the SU(3) Kogut-Susskind Hamiltonian. I have created a circuit for implementing the Trotterization of the kinetic and chromoelectric terms of the SU(3) Kogut-Susskind Hamiltonian with one quark flavor with a depth of 63 CNOTs for an adjacent pair of staggered lattice sites or 126 CNOTs for a single Trotter step. Considering that the circuit-depth achieved in Ref. \cite{Farrell:2024fit} is 370, a time evolution circuit applied to the trivial vacuum can be done with up to 3 steps of either the first or second-order Trotterization. Furthermore, given that, as proven in Sec. \ref{sec:0vBB_stateprep_baryons}, the same operators used for SC-ADAPT-VQE on the Schwinger model in Ref. \cite{Farrell:2023fgd} can also be used for the SU(3) lattice Schwinger model, it may be possible to both prepare and evolve more complex systems, perhaps an extension of the wavepacket scattering studies in Refs. \cite{Farrell:2024fit, zemlevskiy2024scalable} to propagation and scattering of baryon bound states in 1+1D SU(3) lattice gauge theory, especially as the quantum volume of devices improve. 

The alternative chromoelectric term in \ref{sec:0vBB_stateprep_baryons} has more implications worth further study. One straightforward application is to replicate it for an SU(2) Kogut-Susskind Hamiltonian, as its chromoelectric term has the same $(\sigma^+ \sigma^- \sigma^- \sigma^+ + h.c.)$ term as its SU(3) counterpart \cite{atas:2021ext}.  Also, this proof created a much simpler verion of the SU(3) Kogut-Susskind Hamiltonian for the case where the Hilbert space can be restricted to just the color singlet subspace, so one potential topic of future research could be a search for scenarios besides state preparation using variagional quantum algorithms. Additionally, the color singlet subspace formulation found in Sec. \ref{sec:colorsingletconstruction} has proven invaluable for extending the reach of classical simulations in this study. For instance, it was instrumental in obtaining the results found in Sec. \ref{sec:0vBB_classicalsim}. The formulation was also invaluable in shedding light on the structure of the color singlet subspace. Firstly and most obviously, it was essential to the circuit-construction in Sec. \ref{sec:0vBB_stateprep_baryons}. Second, it could be used to prove that $H_{el}$ from Eq. \ref{eq:HchromoPBC_0vBB} is equivalent on sites arbitrarily far from the lattice to its counterpart from Eq. \ref{eq:OBC_0nuBB_fullham} despite the fact that unlike the latter, the former lacks chromoelectric terms applied to the same site. It is possible that insights like these could help guide future circuit design. 

\chapter{Qutrit and Qubit Circuits for Three-Flavor Collective Neutrino Oscillations}
\label{chap:qutritqubitneutrino}

{\it This chapter is associated with Ref. \cite{turro2024qutrit}:}

{\it ``Qutrit and Qubit Circuits for Three-Flavor Collective Neutrino Oscillations" by Francesco Turro and Ivan Chernyshev, Ramya Bhaskar and Marc Illa}

\section{Introduction }
\label{sec:neutrinos_on_qutrits_intro}

The use of qudits~\cite{Gottesman:1998se} for simulating nuclear and high-energy physics systems has generated significant interest~\cite{physrevd.103.094501,Gustafson:2021qbt,Calixto_2021,Gustafson:2022xlj,Gonzalez-Cuadra:2022hxt,gustafson:2022xdt,Gustafson:2023swx,Zache:2023cfj,Illa:2023scc,Popov:2023xft,Meth:2023wzd,Calajo:2024qrc,Carena:2024dzu,Illa:2024kmf,Gustafson:2024kym}, as a result of recent advancements in experimental realizations of qudit-based platforms, including trapped-ion systems~\cite{Low:2019,Ringbauer:2021lhi,Low:2023dlg,Zalivako:2024bjm,Nikolaeva:2024wxl}, superconducting circuits~\cite{Blok:2020may,Seifert:2023ous,Nguyen:2023svc,Champion:2024ufp}, superconducting radio-frequency cavities~\cite{Roy:2024uro}, and photonic systems~\cite{Chi:2022}.
Multilevel quantum devices can efficiently map to high-dimensional systems, which is advantageous for the quantum simulation of such systems, as well as quantum algorithm performance~\cite{Cerf_2002,Gedik:2015,3307650,Baker:2020,Lim:2023hkb} (see Ref.~\cite{10.3389} for a review).
Three-level quantum systems (qutrits)~\cite{Blok:2020may,Yurtalan:2020,Morvan:2021qju,Cervera-Lierta:2021nhp,Ringbauer:2021lhi,Hrmo:2022bvo,Goss:2022bqd,Subramanian:2023xzi,Nikolaeva:2024wxl} are particularly attractive for simulating three-flavor neutrino systems.


As mentioned in Sec. \ref{sec:intro_magic_entanglement_quantumadvantage}, collective neutrino oscillations happen in high-density environments such as core-collapse supenovae (CCSNe) and mergers involving compact stellar objects. At distance $\lesssim 100$ km away from the center of a CCSN, self-interacting neutrino-neutrino currents \cite{Samuel:1993uw,Kostelecky:1993yt,Kostelecky:1993dm,Kostelecky:1993ys,Kostelecky:1995xc} predominate the dynamics, while at distance $\gtrsim 100$ km, neutrino-vacuum oscillations and the Mikeheyev-Smirnov-Wolfenstein (MSW) effect~\cite{Wolfenstein:1977ue,Wolfenstein:1979ni,Mikheyev:1985zog,Mikheyev:1989dy} become the primary mechanisms driving neutrino flavor evolution. 
The Hamiltonian describing the flavor dynamics of neutrinos propagating through a CCSN
environment therefore contains three terms: the one-body vacuum oscillation term, the one-body background matter interaction term modeled by the MSW effect, and the two-body neutrino-neutrino self-interaction term describing the coherent forward scattering \cite{1987apj...322..795f,Notzold:1987ik,savage:1990by,Pantaleone:1992xh,Pantaleone:1992eq,Sigl:1993ctk,Samuel:1995ri,hoffman1997nucleosynthesis,Balantekin:2006tg,Dasgupta:2017oko,Fiorillo:2024wej,Cirigliano:2024pnm} that gives rise to the quantum phenomenon of coherent collective flavor oscillations.

While mean-field studies have evidenced collective flavor dynamics 
\cite{Qian:1994wh,Duan:2006jv,Duan:2006an,Izaguirre:2016gsx,Capozzi:2020kge,Fiorillo:2023mze,Fiorillo:2023hlk,Fiorillo:2024fnl}, there is growing interest in collective dynamics beyond the mean-field approximation, such as when nontrivial neutrino-neutrino two-body correlations are taken into account~\cite{Roggero:2021asb,Xiong:2021evk,Roggero:2022hpy,Bhaskar:2023sta,Kost:2024esc}. 
The all-to-all connectivity of the two-body operator requires exponentially-growing classical resources when attempting to simulate neutrino dynamics from the exact many-body Hamiltonian at physically relevant scales, mainly due to the rapid entanglement growth in such systems.\footnote{See Refs.~\cite{Shalgar:2023ooi,Johns:2023ewj} for discussions regarding the apparent differences between the mean-field and many-body approaches.}

Recent progress in quantum simulations of two flavor neutrino systems~\cite{Hall:2021rbv,Yeter-Aydeniz:2021olz,Illa:2022jqb,Amitrano:2022yyn,Illa:2022zgu,Siwach:2023wzy} has demonstrated quantum devices' potential~\cite{feynman:1981tf} to efficiently capture the non-trivial entanglement structure present in many-body neutrino systems. 

As mentioned in Sec. \ref{sec:intro_magic_entanglement_quantumadvantage}, there have been several simulations on digital quantum devices of collective neutrino oscillations in recent years. Classical and quantum simulations of relatively small-sized two-flavor neutrino systems have uncovered a variety of uniquely quantum phenomena~\cite{Rrapaj:2019pxz,Patwardhan:2019zta,Patwardhan:2021rej,Roggero:2021asb,Xiong:2021evk,Illa:2022zgu,Bhaskar:2023sta,Martin:2023gbo,Neill:2024klc}, further motivating quantum simulations of three-flavor self-interacting neutrino systems~\cite{Siwach:2022xhx,Chernyshev:2024kpu}.

In this work, we introduce qubit and qutrit-based quantum circuits for simulating the time evolution of the three-flavor neutrino system. 
Simulation of the system dynamics for $N=2,4$ and $8$ neutrinos is demonstrated on the IBM \texttt{ibm\_torino}~\cite{ibmq} and Quantinuum {\tt H1-1} qubit platforms~\cite{quantinuum} and time evolved observables, such as  single-neutrino flavor probabilities and entanglement entropy, are studied.

\section{Hamiltonian}
\label{sec:hamiltonian}

Within 100 km from the CCSN core, contributions from background matter (MSW effect) can be assumed negligible as similarly approximated in Refs.~\cite{Pehlivan:2011hp,Patwardhan:2019zta,Patwardhan:2021rej,Roggero:2021asb,Cervia:2022pro}), and antineutrinos are not considered in this work. 
The Hamiltonian governing the evolution of the neutrino system studied is then 
\begin{equation}
    H= H_{\nu}+H_{\nu\nu}\ ,
    \label{eq:simpleHam}
\end{equation}
where $H_{\nu}$  describes the one-body neutrino Hamiltonian given by the vacuum oscillation and $H_{\nu\nu}$ the  neutrino-neutrino interactions resulting from coherent forward scattering.

The vacuum oscillations can be described in the mass basis as~\cite{Balantekin:2006tg}
\begin{equation}
H_{\nu}=\sum_i^N H^{(i)}_{\nu}=\sum_i^N -\frac{\omega}{2}\lambda^{(i)}_3 + \frac{\omega-2\Omega}{2\sqrt{3}} \lambda^{(i)}_8\,,
\label{eq:single_body_a}
\end{equation}
where the index $i$ sums over $N$ neutrinos in the system, $\lambda^{(i)}_n$ is the $n^{\rm th}$ Gell-Mann matrix acting on the $i^{\rm th}$ neutrino (the Gell-Mann matrices are detailed in App.~\ref{app:Gell_Mann}), and $\omega$ and $\Omega$ are the oscillation frequencies, defined as
\begin{equation}
\omega= \frac{1}{2E} \Delta m_{21}^2 \ , \quad \Omega=\frac{1}{2E} \Delta m_{31}^2=\omega \frac{\Delta m_{31}^2}{\Delta m_{21}^2} \ , 
\end{equation}
with $\Delta m^2_{ij}=m^2_i-m^2_j$, and $m^2_i$ being the squared mass of the $i^{\rm th}$ mass-eigenstate neutrino, with the neutrinos taken to have the same energy.
The $i^{\rm th}$ one-body Hamiltonian can also be written as
\begin{equation}
    H_\nu^{(i)}=\begin{pmatrix}
    0 & 0 & 0 \\
    0 & \omega & 0 \\
0 & 0 & \Omega 
\end{pmatrix} \ .
\label{eq:single_body_matrix}
\end{equation}
Here, identity contributions that correspond to global phases in the real-time evolution operator have been neglected.

To operate in the flavor basis, the Pontecorvo-Maki-Nakagawa-Sakata (PMNS) matrix is used to transform between mass and flavor basis,
\begin{align}
&\qquad\qquad\qquad\qquad\begin{pmatrix}
\nu_e \\ \nu_\mu \\ \nu_\tau
\end{pmatrix} = U_{\rm PMNS} \begin{pmatrix}
\nu_1 \\ \nu_2 \\ \nu_3
\end{pmatrix}, \\
    & U_{\rm PMNS} = \scalebox{0.83}{$\begin{pmatrix}
        1 & 0 & 0 \\
        0 & c_{23} & s_{23} \\
        0 & -s_{23} & c_{23}
    \end{pmatrix}\begin{pmatrix}
        c_{13} & 0 & s_{13}e^{-i\delta_{\rm CP}} \\
        0 & 1 & 0 \\
        -s_{13} & 0 & c_{13}
    \end{pmatrix}\begin{pmatrix}
        c_{12} & s_{12} & 0 \\
        -s_{12} & c_{12} & 0 \\
        0 & 0 & 1
    \end{pmatrix},$}
\end{align}
where $c_{ij}\equiv \cos(\theta_{ij})$ and $s_{ij}\equiv\sin(\theta_{ij})$, and the mixing angles $\theta_{ij}$ and phase $\delta_{\rm CP}$ are taken from NuFIT v5.3~\cite{Esteban:2020cvm,nufit} (with other groups recovering consistent results~\cite{deSalas:2020pgw,Capozzi:2021fjo}), and are tabulated in Table~\ref{tab:parameters}.
\begin{table}[t!]
    \centering
    \renewcommand{\arraystretch}{1.4}
    \begin{tabularx}{0.7\columnwidth}{|Y|Y|}
    \hline
        \multicolumn{2}{|c|}{PMNS parameters (deg.)}  \\
        \hline
        $\theta_{12}$ & $33.67^{+0.74}_{-0.71}$\\
         $\theta_{23}$ & $42.3^{+1.1}_{-0.9}$\\
         $\theta_{13}$ & $8.58^{+0.11}_{-0.11}$\\
         $\delta_{\rm CP}$ & $232^{+39}_{-25}$\\
        \hline \hline
        \multicolumn{2}{|c|}{Mass parameters ${\rm MeV}^2$}   \\
        \hline
        $\Delta m_{21}^2\; (\times 10^{17})$ & $7.41^{+0.21}_{-0.20}$ \\
        $\Delta m_{31}^2\; (\times 10^{15})$ & $2.505^{+0.024}_{-0.026}$ \\
        \hline
    \end{tabularx}
    \renewcommand{\arraystretch}{1.0}
    \caption{PNMS mixing parameters and mass differences taken from Refs.~\cite{Esteban:2020cvm,nufit}, assuming normal ordering.}
    \label{tab:parameters}
\end{table}

The three-flavor coherent neutrino-neutrino interaction can be described by the following Hamiltonian~\cite{Balantekin:2006tg},
\begin{equation}
  H_{\nu\nu} = 
  \sum_{i<j}  J_{ij}\bm{\lambda}^{(i)} \cdot \bm{\lambda}^{(j)} \ ,
  \label{eq:twobody}
\end{equation}
with $\bm{\lambda}^{(i)}=(\lambda^{(i)}_1,\lambda^{(i)}_2,\ldots,\lambda^{(i)}_8)$, and the coupling coefficient $J_{ij}$ is defined as 
\begin{equation}
    J_{ij}=\frac{G_F\rho_\nu}{\sqrt{2}N}(1-\cos\theta_{ij}) \ , 
\end{equation} 
with $G_F$ being Fermi's constant and $\rho_\nu$ the number-density of neutrinos.
The angle $\theta_{ij}$ is the angle between the momentum of the $i^{\rm th}$ and $j^{\rm th}$ neutrino, and for demonstration purposes, we use the simple model introduced in Ref.~\cite{Hall:2021rbv}, where it is sampled from a cone-shaped distribution, $\theta_{ij}=\frac{|i-j|}{N-1} \arccos(0.9)$, as done in Refs.~\cite{Hall:2021rbv,Illa:2022jqb,Amitrano:2022yyn,Illa:2022zgu,Chernyshev:2024kpu}.
The neutrino density coupling constant $\mu$ is defined such that the one-body and two-body terms have the same magnitude,

\begin{equation}
    \mu = \frac{G_F \rho_\nu}{\sqrt{2}}=\frac{\Delta m_{31}^2 N}{2E}\, ,
\end{equation}
and it is assumed time independent. However,  since the density of neutrinos decreases as a function of time, a realistic simulation should consider a time-dependent strength of the two-body term
(see, e.g., Refs.~\cite{Cervia:2022pro,Siwach:2022xhx,Siwach:2023wzy}).

The final Hamiltonian in the mass basis as a function of $\mu$ is then given by 
\begin{equation}
\begin{split}
    H & = \frac{\mu}{N}  \sum_i \left[-\frac{ \omega}{2\Omega} \lambda^{(i)}_3 +\frac{\omega-2\Omega}{2\sqrt{3}\Omega} \lambda^{(i)}_8 \right]\\ 
    &\qquad\qquad\qquad+\frac{\mu}{N} \sum_{i<j} [1-\cos(\theta_{ij})] \bm{\lambda}^{(i)} \cdot \bm{\lambda}^{(j)}\;.
\end{split}
\end{equation}
While the two-body term is basis-independent, to transform the one-body term into flavor space the PNMS matrix is applied to the one-body term (Eq.~\eqref{eq:single_body_a}) as follows
\begin{equation}
H^{(i)}_\nu|_{\rm flavor}= U_{\rm PNMS} \cdot H^{(i)}_\nu \cdot U^\dagger_{\rm PNMS}\,.
\label{eq:singleneut_flavor}
\end{equation}

\section{Qutrit mapping}
\label{sec:qutrit}

Motivated by neutrinos' natural three-level structure, we present qutrit-based quantum circuits that can be used for simulating the time evolution of a many-body three-flavored neutrino system.  Our proposed qutrit circuit follows the native qutrit gate set and the notation of the transmon qudits in Ref.~\cite{Goss:2022bqd} (see App.~\ref{app:qutrit_gate} for more details). 

The one-body term $H_{v}$, first implemented in Ref.~\cite{Nguyen:2022snr} on an IBM quantum computer, only involves single-qutrit gates. Equation~\eqref{eq:single_body_matrix} can be implemented with a single phase gate, 
\begin{equation}
e^{-itH^{(i)}_\nu}={\rm Ph}(0,-\omega\, t,-\Omega\, t)={\rm diag}(1,e^{-i\omega\,t},e^{-\Omega\,t})\,,    
\end{equation}
 and the PNMS matrix can be decomposed as
\begin{equation}
\begin{split}
    U_{\rm PNMS} \ = \ & R^{12}_y(\theta_{23})    
    R^{02}_Z\left(\frac{-\pi+\delta_{CP}}{2}\right) R^{02}_y(\theta_{13})\\ & \times R^{02}_Z\left(-\frac{-\pi+\delta_{CP}}{2}\right) R^{01}_y(\theta_{12}) \ ,    
\end{split}
\end{equation}
where App.~\ref{app:qutrit_gate} shows  our gate definitions.

\begin{figure}[!t]
\centering
$$\Qcircuit @C=1em @R=1em { & \ctrl{1}& \gate{X} & \qw &\gate{X^{\dagger}} &\qw & \ctrl{1} &  \qw \\
& \gate{X} & \ctrl{-1}  & \gate{X^{12}_{-2  t J_{ij}}} & \ctrl{-1} & \gate{{\rm Ph}(-2 t J_{ij},0,0)} & \gate{X^{\dagger}} &\qw\\
} $$
\caption{Quantum circuit implementing the term $e^{-i t J_{ij}\bm{\lambda}^{(i)}\cdot \bm{\lambda}^{(j)}}$ from the two-neutrino part $H_{\nu\nu}$. Definitions of the gates can be found in App. ~\ref{app:qutrit_gate}.}
\label{fig:qutrit_qc}
\end{figure}
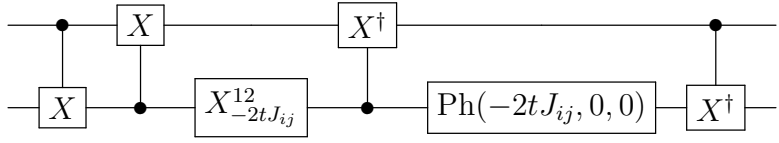

For the two-body term, while a numerical compilation requires at most 6 $CX$ (or $CX^\dagger$) gates~\cite{Goss:2022bqd}, two controlled-shift gates $CX$ and two $CX^\dagger$  were found to be sufficient to apply the specific SU(9) rotation described in Eq.~\eqref{eq:twobody}, as shown in Fig.~\ref{fig:qutrit_qc}.
The qutrit $CX$ gate is defined as $CX\ket{x,y}=\ket{x,\rm{mod}(x+y,3)}$, and the $X^{12}_{2tJ_{ij}}$ gate is given by

\begin{equation}
    X^{12}_{-2tJ_{ij}}=\begin{pmatrix}
        1 & 0 & 0\\
        0 & \cos(J_{ij}\,t) & -i\sin(J_{ij}\,t)\\
        0 & -i\sin(J_{ij}\,t) & \cos(J_{ij}\,t)
    \end{pmatrix}\ .
\end{equation}

\subsection{Swap network \label{sec:swap_network}}

\begin{figure}
    \raggedright
    (a)
    \includegraphics[width=0.95\columnwidth]{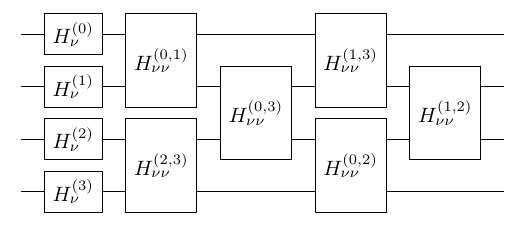}
    (b)
     \includegraphics[width=\columnwidth]{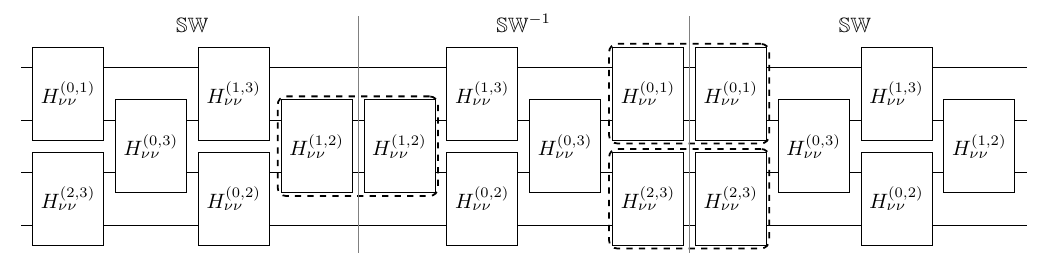}
    
    \caption{(a) Quantum circuit implementing a single LO Trotterized time evolution step via the swap network for four neutrinos. (b) Simplification when using three steps of the NLO$^*$ Trotterized time evolution operator for four neutrinos, where the highlighted operations can be simplified by a single two-qubit box with twice the time step.}
    \label{fig:circuitSWAP}
\end{figure}

Since strategic ordering of application of Eq.~\eqref{eq:twobody} is required to minmize circuit depth, we implement the swap network, $\mathbb{SW}$, which was first proposed in Ref.~\cite{Hall:2021rbv} and is similar to the fermionic swap  from Refs.~\cite{Kivlichan2018prl,OGorman:2019mll}. 
This network limits the depth to $N$ layers of Eq.~\eqref{eq:twobody} for a system with $N$ neutrinos (assuming the qubits are in a linear network and gates can be applied in parallel). 
In the original $\mathbb{SW}$, SWAP gates are needed between each layer to achieve the all-to-all connectivity. 
However, for our implementation, they can be absorbed into the two-body term,
\begin{align}
    {\rm SWAP}_{ij}\cdot e^{-it J_{ij}\bm{\lambda}^{(i)}\cdot \bm{\lambda}^{(j)}}&=e^{-i\frac{\pi}{4}\bm{\lambda}^{(i)}\cdot \bm{\lambda}^{(j)}}\cdot e^{-it J_{ij}\bm{\lambda}^{(i)}\cdot \bm{\lambda}^{(j)}} \nonumber \\
    &=e^{-i(t J_{ij}+\frac{\pi}{4})\bm{\lambda}^{(i)}\cdot \bm{\lambda}^{(j)}} \ ,
\end{align}
substantially reducing the number of entangling gates, particularly in hardware with limited connectivity such as superconducting systems.\footnote{The same simplification can be applied on the two-flavor case, where the SWAP gate can be written as $e^{-i\frac{\pi}{4}\bm{\sigma}^{(i)}\cdot \bm{\sigma}^{(j)}}$.} 
The schematic of such a network can be seen in Fig.~\ref{fig:circuitSWAP}a for four neutrinos.
While SWAP gates might not be required in devices with all-to-all connectivity (such as trapped ions), this new strategy 
can also improve the circuit fidelity in devices where qubits have to be physically moved around, such as in the Quantinuum devices.

The one- and two-body terms in Eqs.~\eqref{eq:single_body_a} and~\eqref{eq:twobody} commute, $[H_{\nu},H_{\nu\nu}]=0$, so they can be Trotterized independently. 
The leading-order (LO) Trotterized time evolution operator is therefore
\begin{equation}
    U(t)_{\rm LO}=e^{-itH_\nu}\prod_{i,j\, \in\, \mathbb{SW}}e^{-i t J_{ij} \bm{\lambda}^{(i)}\cdot \bm{\lambda}^{(j)}} \ .
\end{equation}
When applying multiple Trotter steps, it is more efficient to use the second order (NLO) Suzuki-Trotter formula~\cite{Suzuki1990,Suzuki1991}, as used in Ref.~\cite{Amitrano:2022yyn},
\begin{align}
    U(t)_{\rm NLO} \ = \ & e^{-itH_\nu} \prod_{i,j \, \in \, \mathbb{SW}} e^{-i \frac{t}{2} J_{ij} \bm{\lambda}^{(i)} \cdot \bm{\lambda}^{(j)}} \nonumber \\
    & \times \prod_{i,j \, \in \, \mathbb{SW}^{-1}} e^{-i \frac{t}{2} J_{ij} \bm{\lambda}^{(i)} \cdot  \bm{\lambda}^{(j)}} \ ,
\end{align}
with $\mathbb{SW}^{-1}$  applying the gates in reverse order. With this method, additional simplifications are possible. In particular, the last and first layer in $\mathbb{SW}$ and $\mathbb{SW}^{-1}$, respectively, can be merged into a single operation with twice the time step, as depicted in Fig.~\ref{fig:circuitSWAP}b.

We can generalize this to multiple Trotter steps by interleaving layers ordered in the $\mathbb{SW}$ and $\mathbb{SW}^{-1}$ network sequences, allowing for cancellations between these layers (referred as NLO$^*$).\footnote{This can be extended to higher-order Suzuki-Trotter formulas, although further simplifications are not expected to compensate for the increase in circuit depth.}
Then, the total number of $CX$ gates for $k$ number of Trotter steps (with $k \geq 2$) is
\begin{align}
    \left[U(\tfrac{t}{k})_{\rm LO}\right]^k\; & : \; N_{CX} k\frac{N(N-1)}{2}\ , \\
    U(t)_{{\rm NLO}^*_k}\; & : \; N_{CX}k\frac{N(N-1)}{2} -N_{CX}\left\lfloor \frac{k}{2} \right\rfloor \left(\left\lceil\frac{N}{2}\right\rceil-1\right) \nonumber\\
    & \ \ -N_{CX}\left\lfloor \frac{k-1}{2}\right\rfloor \cdot \left\lfloor\frac{N}{2}\right\rfloor     \ ,
\end{align}
where $\lceil \cdot \rceil$ ($\lfloor \cdot \rfloor$) is the ceiling (floor) function, $N$ is the number of neutrinos, and $N_{CX}$ is the number of $CX$ gates needed to compile a single neutrino-neutrino term (for the case in Fig.~\ref{fig:qutrit_qc}, $N_{CX}=4$).
A reduction in two-qubit gates is seen when using NLO$^*$ compared to LO with increasing number of Trotter steps, as well as improved convergence. 

\section{Qubit mapping}
\label{sec:qubit}

In the absence of qutrit-based platforms, an alternate approach involves mapping a three-flavor neutrino to two qubits~\cite{Arguelles:2019phs}
, with each flavor encoded as: $\ket{\nu_e} \rightleftharpoons \ket{00}$, $\ket{\nu_\mu} \rightleftharpoons \ket{01}$,  $\ket{\nu_\tau} \rightleftharpoons \ket{10}$, and the unassigned state $\ket{11}$ designated as the unphysical state (assuming no sterile neutrinos).
Therefore, no matrix elements in the unitaries that implement the one- and two-neutrino terms are allowed to mix states between the physical and the unphysical sub-spaces. However, we have the freedom to allow arbitrary mixing between unphysical states if the resulting quantum circuits are shallower.

For the one-neutrino term $H_{v}$, its time evolution operator in the mass basis is diagonal, and can be implemented with single-qubit $R_z$ gates,
\begin{equation}
    e^{-itH^{(i)}_{v}}=R_z(-\omega t)_{2i}\otimes R_z(-\Omega t)_{2i+1}\ .
\end{equation}

In the flavor basis, the $3\times 3$ matrix can be embedded into a $4\times 4$ one, and the resulting SU(4) matrix can be transpiled into the three-CNOT circuits from Refs.~\cite{Vatan:2004nmz,PhysRevA.69.010301,PhysRevA.77.066301} (or use the circuits from Refs.~\cite{Arguelles:2019phs,Molewski:2021ogs}).

\begin{figure*}[ht]
    \centering
$\resizebox{\textwidth}{!}{
\Qcircuit @C=0.3em @R=0.6em {
    & \qw & \qw & \targ & \gate{R_z(\frac{\pi}{2}+\alpha)} & \qw & \ctrl{2} & \qw & \qw & \targ & \gate{R_z^+} & \qw & \ctrl{3} & \gate{H} & \gate{S} & \ctrl{2} & \qw & \gate{R_x(\alpha)} & \ctrl{2} & \qw & \gate{R_z^-} & \gate{H} & \qw & \ctrl{3} & \gate{R_z^-} & \gate{H} & \gate{S} & \qw & \ctrl{2} & \gate{ R_x(\alpha)} & \qw & \ctrl{2} & \gate{R_z^-} & \gate{H} & \qw & \ctrl{3} & \gate{R_z^-} & \gate{H} & \gate{S} & \ctrl{2} & \gate{R_x(\alpha)} & \ctrl{2} & \gate{R_z^-} & \gate{H} & \ctrl{3} & \qw & \qw & \qw & \qw \\
    & \qw & \targ & \qw & \gate{R_z(\frac{\pi}{2}+\alpha)} & \ctrl{2} & \qw & \qw & \targ & \qw & \gate{R_z^+} & \ctrl{1} & \qw & \gate{H} & \gate{S} & \qw & \ctrl{2} & \gate{R_x( {\alpha})} & \qw & \ctrl{2} & \qw & \qw & \targ & \qw  & \qw & \gate{H} & \gate{S} & \ctrl{2} & \qw & \gate{R_x(\alpha)} & \ctrl{2} & \qw & \qw & \qw & \targ & \qw & \qw & \qw & \qw & \qw & \qw & \qw & \qw & \qw & \qw & \ctrl{1} & \gate{H} & \gate{S} & \qw \\
    & \gate{R_z^-} & \qw & \ctrl{-2} & \gate{R_y(-\frac{\pi}{2}-\alpha)} & \qw & \targ & \gate{ R_y(\frac{\pi}{2}+\alpha)} & \qw & \ctrl{-2} & \qw & \control\qw & \qw & \gate{H} & \gate{S} & \targ & \qw & \gate{R_z(\alpha)} & \targ & \qw & \gate{R_z^-} & \gate{H} & \ctrl{-1} & \qw & \gate{R_z^-} & \gate{H} & \gate{S} & \qw & \targ & \gate{R_z( {\alpha})} & \qw & \targ & \gate{R_z^-} & \gate{H} & \ctrl{-1} & \qw & \gate{R_z^-} & \gate{H} & \gate{S} & \targ & \gate{R_z(\alpha)} & \targ & \gate{R_z^-} & \gate{H} & \qw & \control\qw & \qw & \qw & \qw \\
    & \gate{R_z^-} & \ctrl{-2} & \qw & \gate{R_y(-\frac{\pi}{2}-\alpha)}  & \targ & \qw & \gate{ R_y(\frac{\pi}{2}+\alpha)} & \ctrl{-2} & \qw & \qw & \qw & \control\qw & \gate{H} & \gate{S} & \qw & \targ & \gate{R_z(\alpha)} & \qw & \targ & \qw & \qw & \qw & \targ & \qw & \gate{H} & \gate{S} & \targ & \qw & \qw & \targ & \qw & \qw & \qw & \qw & \targ & \qw & \qw & \qw & \qw & \qw & \qw & \qw & \qw & \control\qw & \qw & \gate{H} & \gate{S} & \qw \\}
}
$
\caption{Circuit A implementing $e^{-i\alpha \bm{\lambda}^{(i)} \cdot \bm{\lambda}^{(j)}}$ in the physical subspace, using 24 CNOTs. The gates $R_z^\pm$ represents the short-hand version of $R_z(\pm\frac{\pi}{2})$.}
\label{fig:2neutA}
\end{figure*}
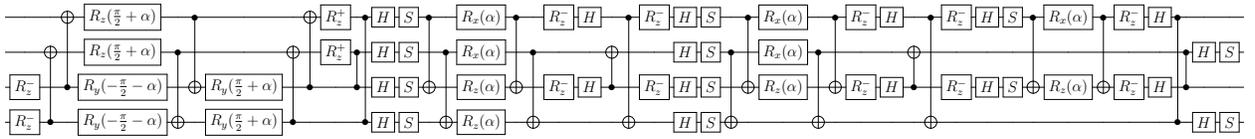
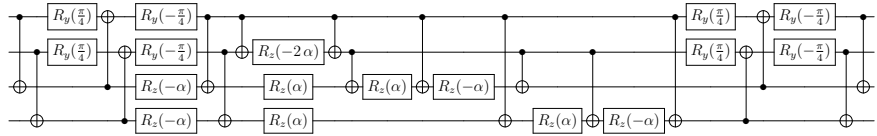
\begin{figure*}[ht]
    \centering 
$\resizebox{0.7\textwidth}{!}{
\Qcircuit @C=0.3em @R=0.6em { \\
	& \ctrl{2} & \qw & \gate{R_y(\frac{\pi}{4})} & \targ & \qw & \gate{R_y(-\frac{\pi}{4})}  & \ctrl{2} & \qw & \ctrl{1} & \qw & \ctrl{1} & \qw & \qw & \ctrl{2} & \qw & \ctrl{3} & \qw & \qw & \qw & \qw & \ctrl{3} & \gate{R_y(\frac{\pi}{4})} & \qw & \targ & \gate{R_y(-\frac{\pi}{4})} & \qw & \ctrl{2} & \qw \\
	& \qw & \ctrl{2} & \gate{R_y(\frac{\pi}{4})} & \qw & \targ & \gate{R_y(-\frac{\pi}{4})} & \qw & \ctrl{2} & \targ & \gate{R_z(-2\,\alpha)} & \targ & \ctrl{1} & \qw & \qw & \qw & \qw & \ctrl{1} & \qw & \ctrl{2} & \qw & \qw & \gate{R_y(\frac{\pi}{4})} & \targ & \qw & \gate{R_y(-\frac{\pi}{4})} & \ctrl{2} & \qw & \qw \\
	& \targ & \qw & \qw & \ctrl{-2} & \qw & \gate{R_z(-\alpha)} & \targ & \qw & \qw & \gate{R_z(\alpha)} & \qw & \targ & \gate{R_z(\alpha)} & \targ & \gate{R_z(-\alpha)} & \qw & \targ & \qw & \qw & \qw & \qw & \qw & \qw & \ctrl{-2} & \qw & \qw & \targ & \qw \\
	& \qw & \targ & \qw & \qw & \ctrl{-2} & \gate{ R_z(-\alpha)} & \qw & \targ & \qw & \gate{ R_z(\alpha)} & \qw & \qw & \qw & \qw & \qw & \targ & \qw & \gate{ R_z(\alpha)} & \targ & \gate{ R_z (-\alpha)} & \targ  & \qw & \ctrl{-2} & \qw & \qw & \targ & \qw & \qw \\
}
}$
    \caption{Circuit B implementing $e^{-i\alpha \bm{\lambda}^{(i)} \cdot \bm{\lambda}^{(j)}}$ in the physical subspace, using 18 CNOTs.}
    \label{fig:2neutB}
\end{figure*}
The two-neutrino term $H_{\nu\nu}$ is more delicate in this case, compared to the qutrit implementation. As mentioned, while the physical subspace is fixed, we have the freedom on the unphysical one to implement any rotation, as long as the two subspaces do not get mixed. Here we propose two circuits, A (shown in Fig.~\ref{fig:2neutA}) and B (shown in Fig.~\ref{fig:2neutB}), both using qiskit conventions~\cite{qiskit} for the gate definitions. The difference between these two circuits can be seen by looking at the unitary they implement,
\begin{align}
    & \left. e^{-i\alpha \bm{\lambda}^{(i)} \cdot \bm{\lambda}^{(j)}}\right|_A=\left(\vcenter{\hbox{\includegraphics[width=0.4\columnwidth]{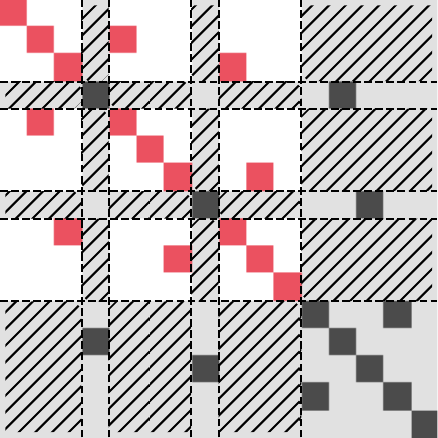}}}\right) \ , \label{eq:circA}\\[1em]
    & \left. e^{-i\alpha \bm{\lambda}^{(i)} \cdot \bm{\lambda}^{(j)}}\right|_B=\left(\vcenter{\hbox{\includegraphics[width=0.4\columnwidth]{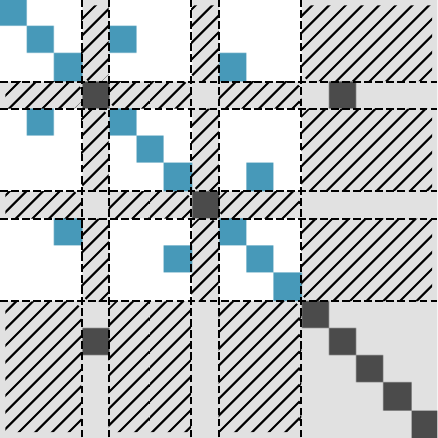}}}\right) \label{eq:circB} \ .
\end{align}
While both circuits have the same effect in the physical subspace, shown with white background in Eqs.~\eqref{eq:circA}-\eqref{eq:circB}, the rotations in the unphysical subspace, shown with a gray background, are different (with the dashed region being the transition between the two subspaces). 
For example, circuit A, in the full four-qubit space, implements the more general gate $e^{-i\alpha \tilde{\bm{\lambda}}^{(i)} \cdot \tilde{\bm{\lambda}}^{(j)}}$ (with $\tilde{\bm{\lambda}}^{(i)}$ being the set of SU(4) generators that act on the $i^{\rm th}$ neutrino), which can also be written as $e^{-i\frac{\alpha}{2}\sum_{a,b}(\sigma_a\otimes \sigma_b)^i\otimes (\sigma_a\otimes \sigma_b)^j}$ (with $\sigma_a$ being all possible elements of the Pauli group and the superscripts denoting neutrino-indices). 
While circuit A does not perform the most general SU(16) gate, it contains significantly less CNOT gates than what one would obtain for an SU(16) gate using currently available operator-to-circuit transpilers such as qiskit~\cite{qiskit,Javadi-Abhari:2024kbf}, tket~\cite{sivarajah2020t}, or other decomposition approaches~\cite{Shende:2006,Mottonen:2006,Mansky:2022bai,Krol:2024taf}, which result in circuits containing on the order of $\mathcal{O}(100)$ CNOTs.

In both circuits A and B, the method discussed in Sec.~\ref{sec:swap_network} can absorb the SWAP operation into the two-body term, modifying $\alpha\rightarrow\alpha+\frac{\pi}{4}$.
For this case, where each neutrino is composed of two qubits, the action of the SWAP gate is ${\rm SWAP}_{ij} \ket{ab}_i\otimes\ket{cd}_j=\ket{cd}_i\otimes\ket{ab}_j$.
Table~\ref{tab:cnot_counts} reports the number of required $CX$ or CNOT gates and its corresponding depth for each quantum circuit (both qubit and qutrit) after compilation in both an-all-to-all and linear-chain architecture.
\begin{table}[h!]
\centering
\renewcommand{\arraystretch}{1.2}
\begin{tabularx}{\columnwidth}{|Y|c|YY|YY|}
\hline
\multirow{2}{*}[-0.5em]{Qudit} &\multirow{2}{*}[-0.5em]{Circuit} & \multicolumn{2}{c|}{All-to-all} & \multicolumn{2}{c|}{Linear chain}  \\
& & 2-q gate count & 2-q gate depth & 2-q gate count & 2-q gate depth \\
\hline \hline
Qutrit & Fig.~\ref{fig:qutrit_qc} & 4 & 4 & 4 & 4 \\ \hline
\multirow{2}{*}{Qubit} & A (Fig.~\ref{fig:2neutA}) & 24 & 13 & 42 & 31 \\ 
& B (Fig.~\ref{fig:2neutB})  & 18 & 12 & 30 & 25 \\
\hline
\end{tabularx}
\renewcommand{\arraystretch}{1.0}
\caption{The two-qudit entangling gate count and depth for the two-neutrino quantum circuits proposed, involving two qutrits or four qubits.}
\label{tab:cnot_counts}
\end{table}

\section{Results}
\label{sec:results}

Dynamics for $N=\{2,4,8\}$ neutrinos were simulated with the \texttt{H1-1} Quantinuum trapped-ion and \texttt{ibm\_torino} IBM superconducting quantum computers (device parameters can be found in App.~\ref{app:device_parameters}). We use the circuits described in Sec.~\ref{sec:qubit}, in particular circuit B in Fig.~\ref{fig:2neutB}, to implement the two-neutrino interaction. For two and four neutrinos, the time step was fixed (while increasing the number of Trotter steps), while for eight neutrinos number of Trotter steps was fixed (while increasing the time step).

For current noisy intermediate-scale quantum (NISQ) devices~\cite{preskill2018quantumcomputingin} error mitigation techniques are critical for reliable results.
Well-known techniques, such as zero-noise extrapolation~\cite{physrevx.7.021050,physrevlett.119.180509,2020arXiv200510921G} and probabilistic error cancellation~\cite{physrevlett.119.180509,Berg:2022ugn}, require a large overhead in circuit sampling or an increase in circuit depth. 
For this study, we decided to use a more resource-friendly algorithm, decoherence renormalization (DR), first introduced in Refs.~\cite{urbanek:2021oej,rahman:2022rlg}, and later implemented in increasingly larger simulations in Refs.~\cite{Ciavarella:2023mfc,Farrell:2023fgd,Farrell:2024fit,Ciavarella:2024fzw} and Chapters \ref{chap:1p1dQCD} and \ref{chap:1p1dSM}, to mitigate decoherent errors. 
For each time step, two different quantum circuits were ran: the first one (which we call the \textit{physics} quantum circuit)
implements the correct dynamics, and the second one (which we call the \textit{identity} quantum circuit) runs a quantum circuit with the same structure as the \textit{physics} circuit, except its action on the initial state is the identity. For first-order Trotter, this can be achieved by setting the time step to zero. For second-order Trotter, the sign in the time step for the second half of the circuit is flipped. 
Through the following rescaling formula, the experimental noiseless probability $P_{phys}^{\rm ex}(t)$ can be computed as:
\begin{equation}
    P_{phys}^{\rm ex}(t)-d_n = \frac{ P_{id}^{\rm ex}-d_n}{    P_{id}^{\rm noisy}-d_n
} \left( P_{phys}^{\rm noisy}(t)-d_n\right) \,,
\label{eq:DR}
\end{equation}
where $P_{phys}^{\rm noisy}(t)$ indicates the obtained probability value from the \textit{physics} quantum circuit and $P_{id}^{\rm noisy}$ indicates the obtained probability from the \textit{identity} quantum circuit. $P_{id}^{\rm ex}$ corresponds to the noiseless result of the \textit{identity} quantum circuit. $d_n$ represents the decoherence value of the quantity computed $P$, which, for a generic $n$-neutrino measurement probability, is $d_n=1/4^n$.

Equation~\eqref{eq:DR} assumes all noise from the device is depolarizing. While this seems a reasonable assumption for the Quantinuum device, as observed in other trapped ion devices~\cite{Nguyen:2021hyk}, the IBM quantum computer requires additional steps. To ensure all noise to be depolarizing, Pauli twirling~\cite{physreva.94.052325,Hashim:2020cop} was applied to transform coherent noise into incoherent noise, as well as dynamical decoupling~\cite{physreva.58.2733,2012RSPTA.370.4748S,Ezzell:2022uat} to suppress cross-talk and idling errors. Moreover, the matrix-free measurement mitigation (M3)~\cite{Nation:2021kye} provided by the {\tt Sampler} function from {\tt qiskit}~\cite{qiskit} was used to correct readout errors.

\begin{figure}[t]
    \centering
    \includegraphics[width=\columnwidth]{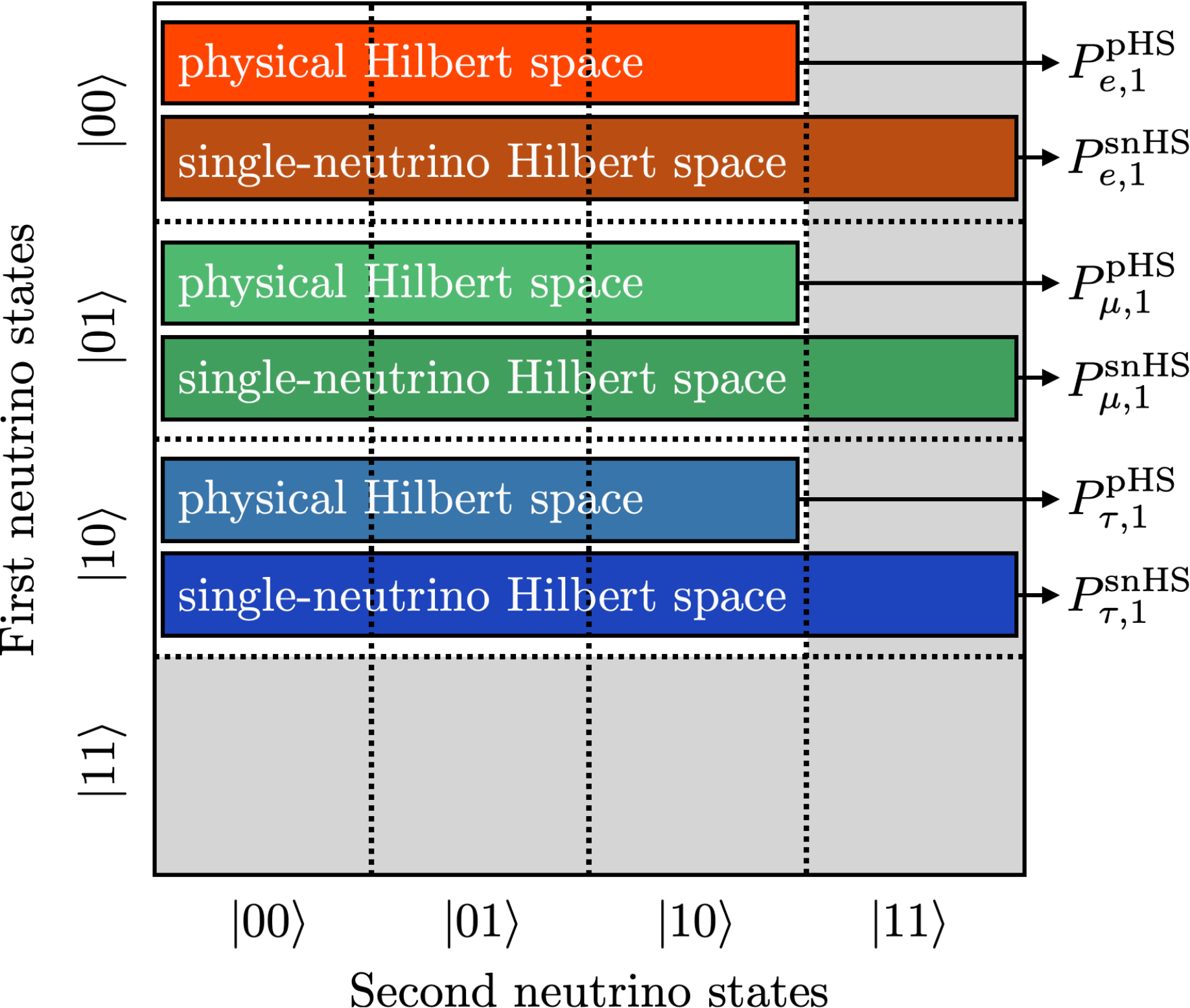}
    \caption{Different post-selecting procedures for computing the single-neutrino flavor probability, for a system of two neutrinos. The physical Hilbert space (pHS) approach only accounts for the physical flavor states; the single neutrino Hilbert space (snHS) approach keeps all contributions from the other neutrino states (both physical and unphysical).}
    \label{fig:postselecting_scheme}
\end{figure}

Two observables for each system were computed: the probability of a single neutrino in a particular flavor state $P_{\nu}$ and the persistence probability of the initial state, $|\langle \psi(0)|\psi(t)\rangle|^2$. Since device errors will populate the unphysical Hilbert space, we implement two different strategies for computing the single-neutrino probabilities. The first method uses the full physical Hilbert space (pHS) where all nonphysical states $\ket{11}$ are discarded. The second method involves summing over the remaining states in the single-neutrino Hilbert space (snHS), mimicking a single measurement of the qubits representing the $i^{\rm th}$ neutrino. While unphysical states for other neutrinos contribute to the probability, after being corrected with DR their contribution can be small.\footnote{One should notice that the two methods have two different decoherence values in Eq.~\eqref{eq:DR}. When computing $P_\nu$, $d^{\rm pHS}_{1} = 3^{N-1}/{4^N}$ (with $N$ the total number of neutrinos) and $d^{\rm snHS}_{1} = 1/4$.}
The two methods are illustrated in Fig.~\ref{fig:postselecting_scheme} for the two neutrino case, where the big squares represent all possible 16 states.

\subsection{Quantinuum}
\label{sec:quantinuum}

Due to the all-to-all architecture of the Quantinuum device~\cite{quantinuum}, the circuit in Fig.~\ref{fig:2neutB} for implementing the neutrino-neutrino term can be used without having to rewrite the CNOTs connecting distant qubits. 
After transpiling the circuit to the native gate-set (single-qubit gates and ZZ$(\theta)=e^{-i(\theta/2) Z\otimes Z}$), the entangling gate count and depth gets reduced by one unit, compared to the numbers in Table~\ref{tab:cnot_counts}.\footnote{The simplification occurs in the $R_z(-2\alpha)$ rotation and its two neighbouring CNOTs in Fig.~\ref{fig:2neutB}, which get transformed into a single ZZ$(-2\alpha)$ gate.}
We use the trick of adding a $\pi/4$ phase to the neutrino-neutrino terms to incorporate the SWAP gate into the latter. This is because although it is not necessary for this hardware, as mentioned in Sec.~\ref{sec:swap_network}, it should reduce the shuffling of trapped ions.

\begin{figure*}[htb!]
    \centering
    \includegraphics[width=\textwidth]{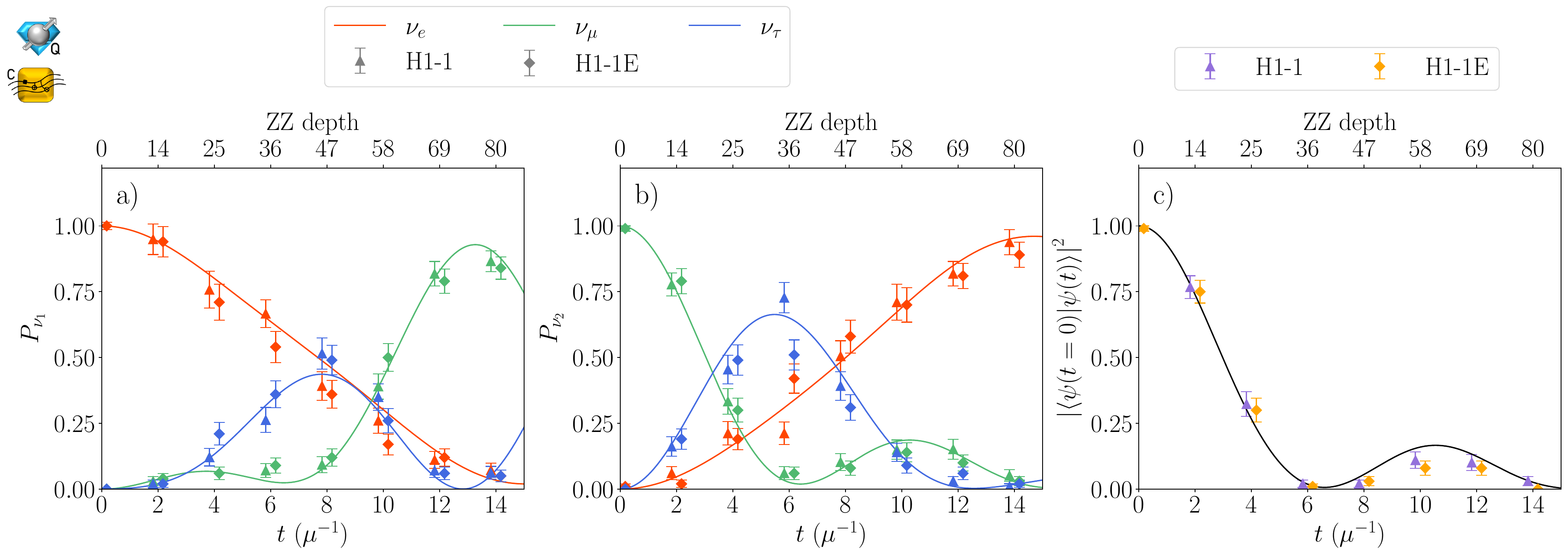}
    \caption{Flavor evolution of a two-neutrino system as a function of time. Panels (a) and (b) show the flavor evolution of the first and second neutrinos, respectively. Panel (c) shows the persistence probability of the initial state. Triangles and diamonds indicate results from the device (\texttt{H1-1}) and its emulator (\texttt{H1-1E}), respectively, slightly shifting the points for ease of readability; solid lines show the exact result. The top axis measures the ZZ depth for each point}
    \label{fig:quantinnum_2_neutrinos}
\end{figure*}
First, the evolution of two neutrinos was simulated, using multiple numbers of Trotter time steps to study the propagation at long times and the noise sources for deep quantum circuits.\footnote{Notice that for two neutrinos, there are no Trotter errors in the decomposition of the time evolution circuits, therefore the multiple Trotter steps are a way to increase the circuit depth and benchmark the quantum computer.} 
Figure~\ref{fig:quantinnum_2_neutrinos} shows the results when we start from the $\ket{\nu_e \nu_\mu}$ state from the device  \texttt{H1-1} and the emulator \texttt{H1-1E}, and use 100 shots per circuit. In this simulation, no error-mitigation techniques were implemented. Also, a 1-3\% contribution from unphysical states was found, supporting the pHS approach to be a good approximation. 
The results, even for longer times, ($t=14 \mu^{-1}$), are compatible with the exact evolution (represented with solid lines). The icons in the top-left corner that appear in this and subsequent plots, identify if the noisy simulator (yellow icon) or quantum device (blue icon) was used~\cite{klco:2019xro}.

\begin{figure*}[htb!]
\centering
\includegraphics[width=\textwidth]{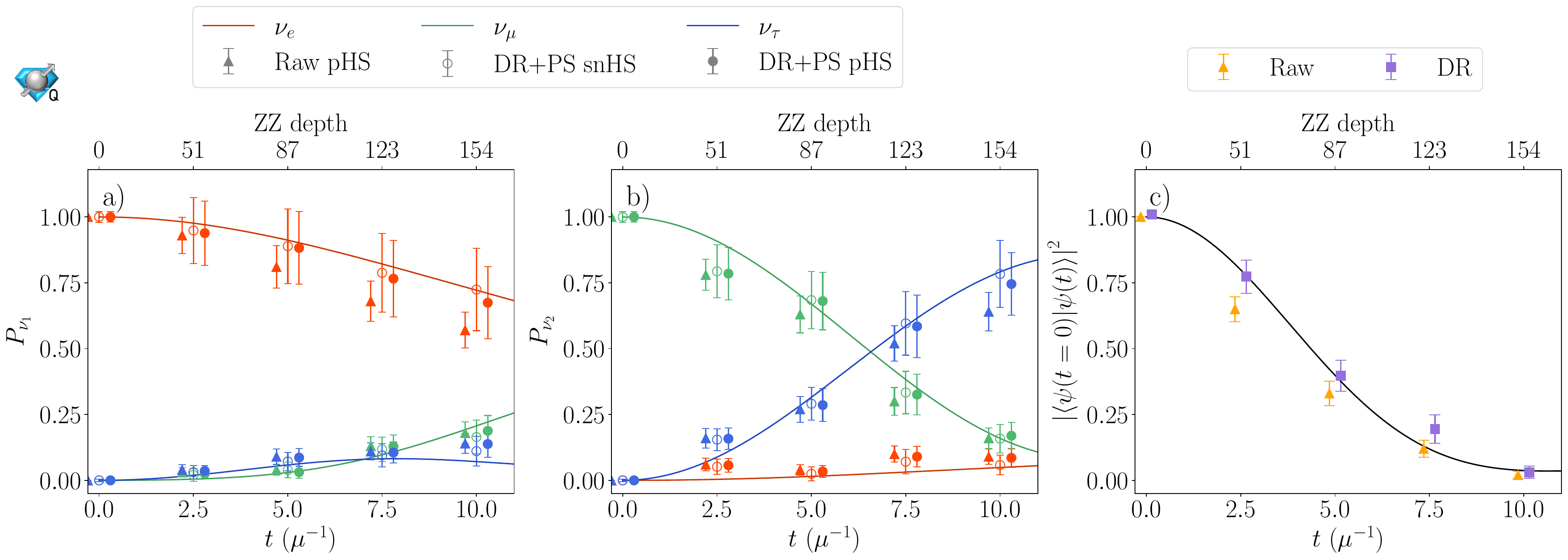}
\caption{Flavor evolution of a four-neutrino system as a function of time, using the \texttt{H1-1} device.  Panels (a) and (b) show the flavor evolution of the first and second neutrinos, respectively. Triangles indicate the raw pHS results. Empty and solid circles represent the results of applying DR,  snHS, and pHS post-selecting procedures, respectively.  Panel (c) shows the persistence probability of the initial state. The triangle and square markers represent the results without and with applying DR, respectively. 
In all panels, the solid lines show the exact evolution, and the points have been slightly shifted to ease the readability.
The top axis measures the ZZ depth for each point.}
    \label{fig:quantinuum_4_neutrinos}
\end{figure*}
The dynamics of four neutrinos starting from the $\ket{\nu_e \nu_\mu \nu_e \nu_\tau
}$ state on the \texttt{H1-1} device was then simulated.
Figure~\ref{fig:quantinuum_4_neutrinos} shows the obtained results, using 100 shots per circuit. Like in Fig.~\ref{fig:quantinnum_2_neutrinos}, panels (a) and (b) illustrate the flavor evolution of the first (the neutrino starting as $\nu_e$) and second neutrino (the neutrino starting as $\nu_\mu$), respectively. In this case DR was used for error-mitigation, and after post-selection via snHS and pHS, the probabilities were normalized to sum to $1$.
Panel (c) shows the persistence probability of the initial state. For this larger system, an improvement after performing error mitigation was observed.

In App.~\ref{app:4neutrinos} the emulator \texttt{H1-1E} and device \texttt{H1-1} results is compared. Generally the emulator and device results are observed to be compatible within reasonable uncertainties. Nevertheless, we observe that the \texttt{H1-1E} emulator gives more pessimistic results than the actual machine.  
\begin{figure*}[ht]
    \centering
    \includegraphics[width=\textwidth]{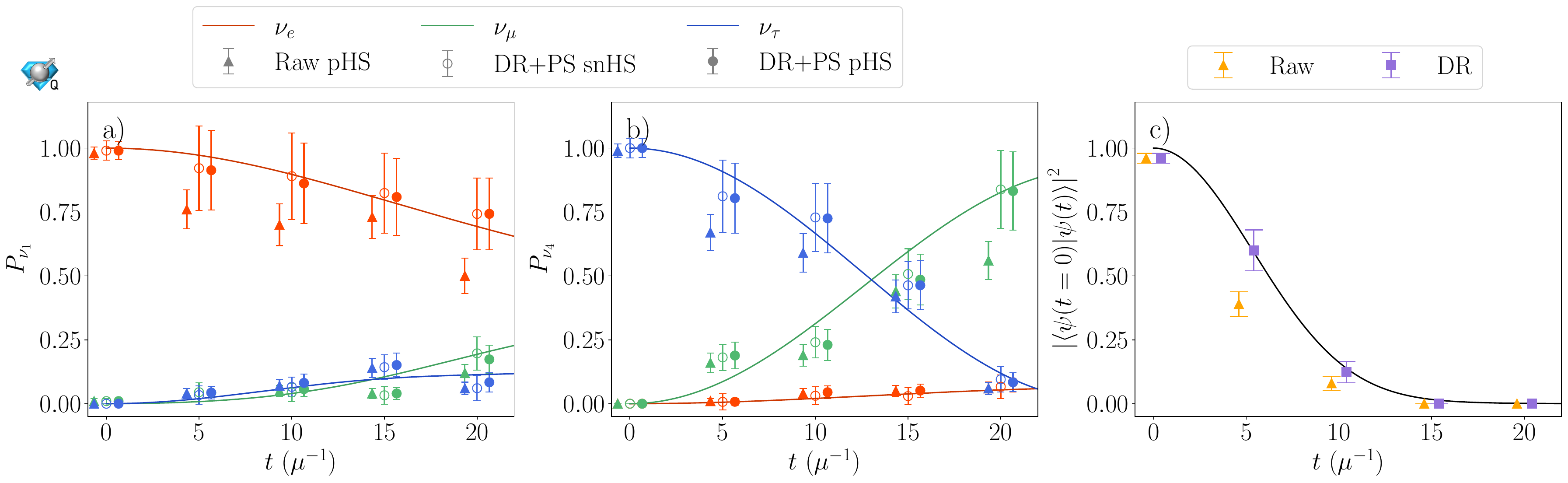}
\caption{Flavor evolution for an eight-neutrino system as a function of time, using the \texttt{H1-1} device. Panels (a) and (b) show the flavor evolution of the first and fourth neutrinos, respectively. Panel (c) shows the persistence probability of the initial state. 
Details are as in Fig.~\ref{fig:quantinuum_4_neutrinos}. 
}
\label{fig:quantinuum_8_neutrinos}
\end{figure*}
Figure~\ref{fig:quantinuum_8_neutrinos} shows the results for an eight-neutrino system, starting from $\ket{\nu_{e}\nu_{\mu}\nu_{e}\nu_{\tau}\nu_{e}\nu_{\mu}\nu_{e}\nu_{\tau}}$. In this case only a single Trotter time step was performed, increasing the time step. 
The implemented quantum circuits have a ZZ depth of 91, and 100 shots were used (except for $t=10\mu^{-1}$, which used 79 shots).
All results are compatible within $2\sigma$ with the exact evolution, albeit with larger uncertainties than the cases for two and four neutrinos.

\subsection{IBM}

Compared to Quantinuum, the IBM hardware has the constraint of linear connectivity between qubits (a one-dimensional chain was selected from the heavy-hex lattice), necessitating the circuit in Fig.~\ref{fig:2neutB} to be compiled into a linear chain architecture. All results were obtained from implementations on \texttt{ibm\_torino}, where the native entangling gate is the controlled-Z (CZ) gate, leading to the same depth and number of gates as in Table~\ref{tab:cnot_counts}.

As discussed at the beginning of 
in Sec.~\ref{sec:results}, $10\times N$ different Pauli-twirled quantum circuits were ran for each time step in order to average out the coherence noise (with $N$ being the number of neutrinos), using 8000 shots per circuit. 
After applying DR and the post-selecting procedures it was noted that the resulting single-neutrino probabilities could become unphysical (either negative or greater than 1), an issue not encountered when using the Quantinuum device in Sec.~\ref{sec:quantinuum}. To fix this, the algorithm of 
Ref.~\cite{PhysRevLett.108.070502} was applied to find the closest probability distribution. 
The uncertainties from combining the different twirled circuits were computed via bootstrap resampling.

\begin{figure*}[ht]
    \centering
    \includegraphics[width=\textwidth]{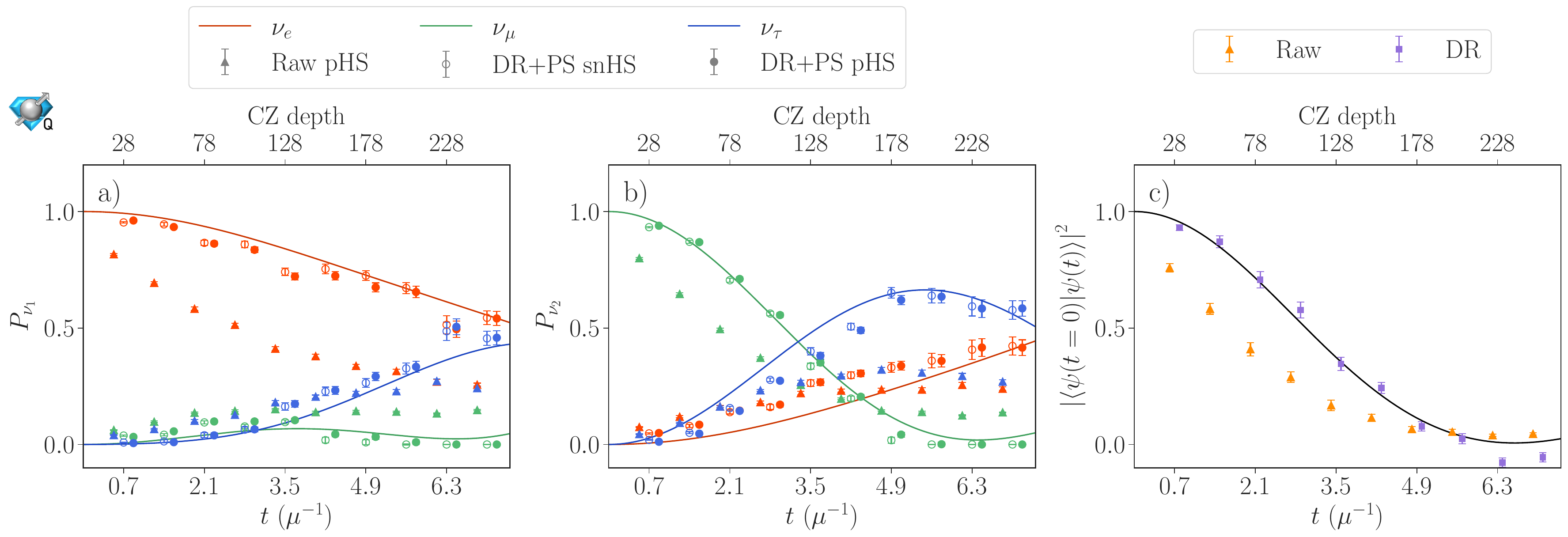}
\caption{Flavor evolution for two neutrinos as a function of time obtained from \texttt{ibm\_torino} device.  
Panels (a) and (b) show the flavor evolution of the first and second neutrinos, respectively. 
Panel (c) shows the persistence probability of the initial state.
Details are as in Fig.~\ref{fig:quantinuum_4_neutrinos}. 
The top axis measures the CZ depth at every other point.}
    \label{fig:ibm_2_neutrinos}
\end{figure*}
Figure~\ref{fig:ibm_2_neutrinos} depicts the evolution for a two-neutrino system, starting from the $\ket{\nu_{e}\nu_{\mu}}$ state. Unlike for Quantinuum (Fig.~\ref{fig:quantinnum_2_neutrinos}), in this case error mitigation is essential for the results to be compatible with the exact evolution.
While most points are within $1\sigma$ and $3\sigma$, it seems that the initial state persistence is more robust against errors than the single-neutrino probabilities.

\begin{figure*}[ht]
    \centering
    \includegraphics[width=\textwidth]{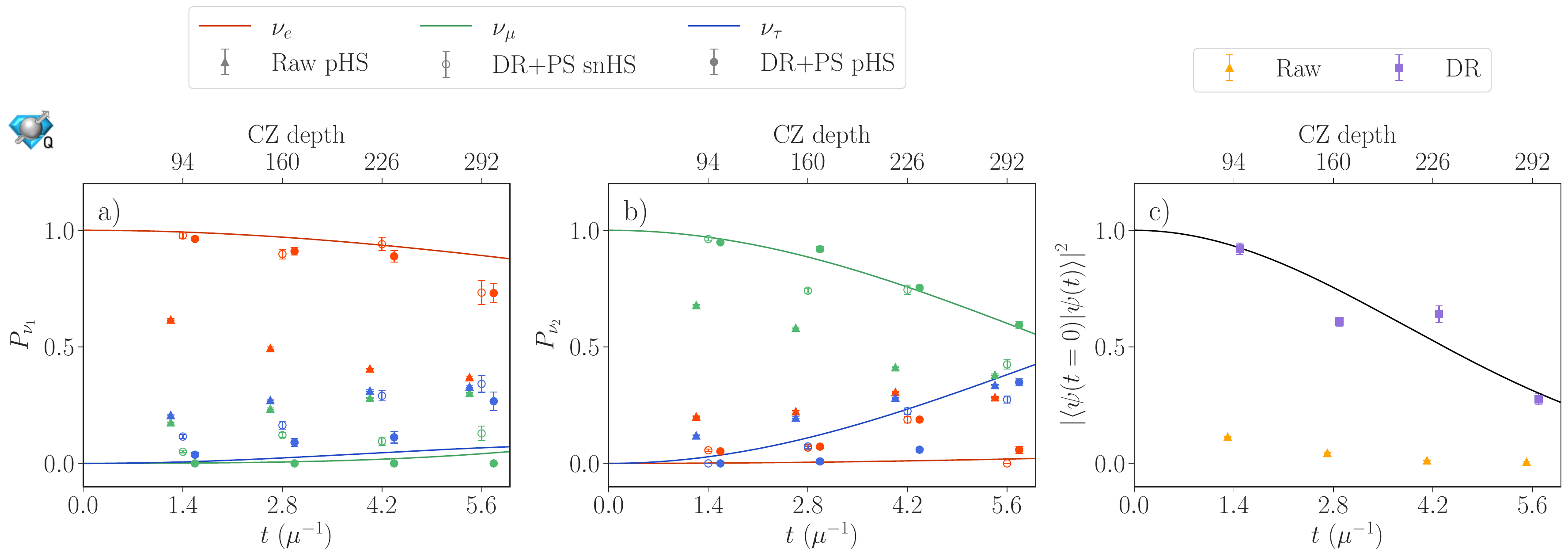}
\caption{Flavor evolution for four neutrinos as a function of time obtained from the \texttt{ibm\_torino} device.  
Panels (a) and (b) show the flavor evolution of the first and second neutrinos, respectively. 
Panel (c) shows the persistence probability of the initial state.
Details are as in Fig.~\ref{fig:quantinuum_4_neutrinos}. 
The top axis measures the CZ depth at each point.}
\label{fig:ibm_4_neutrinos}
\end{figure*}
Figure~\ref{fig:ibm_4_neutrinos} shows the evolution for the four-neutrino system, starting from the $\ket{\nu_{e}\nu_{\mu}\nu_{e}\nu_{\tau}}$ state. 
Like in the two-neutrino case, the obtained results follow the analytical evolution, although in some cases there is a difference of more than $5\sigma$.
This growing tension is investigated further in the eight-neutrino system.
\begin{figure*}[ht]
    \centering
\includegraphics[width=\textwidth]{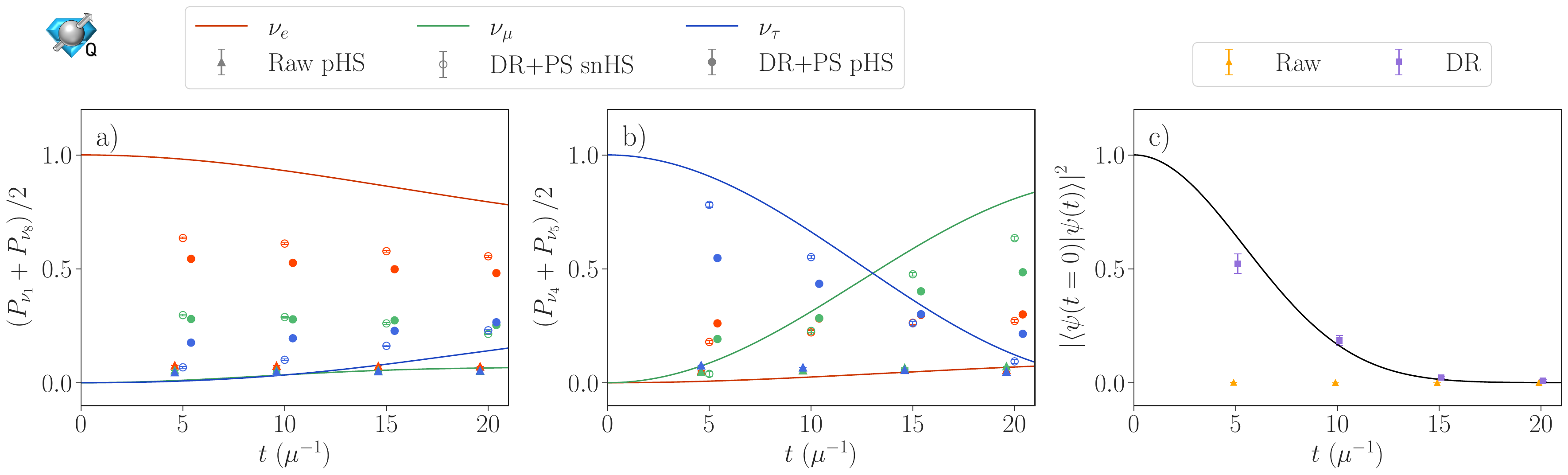}
\caption{Flavor evolution for eight neutrinos as a function of time obtained from the \texttt{ibm\_torino} device.  Panels (a) and (b) show the flavor evolution of the (symmetrized) first and fourth neutrinos, respectively. 
Panel (c) shows the persistence probability of the initial state.
Details are as in Fig.~\ref{fig:quantinuum_4_neutrinos}.
}
\label{fig:ibm_8_neutrinos}
\end{figure*}
Figure~\ref{fig:ibm_8_neutrinos} shows the evolution for the eight-neutrino system, with quantum circuits that have a CZ depth of 182.
In contrast to the previous results, here the initial state was the symmetric state $\ket{\nu_{e}\nu_{\mu}\nu_{e}\nu_{\tau}\nu_{\tau}\nu_{e}\nu_{\mu}\nu_{e}}$. 
This change enables averaging of the single-neutrino probability between the $i^{\rm th}$ and $(N+1-i)^{\rm th}$ neutrino, improving the quality of the obtained results (as seen by the degradation in the four neutrino system in Fig.~\ref{fig:ibm_4_neutrinos}).\footnote{A similar exchange symmetry has been observed for the two-flavor case~\cite{Hall:2021rbv,Amitrano:2022yyn,Illa:2022zgu}, although in the three-flavor case it is only manifested for symmetric initial states.} 
After performing the symmetrization, the noise contributions can be averaged out and reduced (for the non-symmetric initial state, see App.~\ref{app:8ibm}). 
Despite performing error mitigation,  the results in Fig.~\ref{fig:ibm_8_neutrinos} for the single-neutrino probabilities have large deviations from the expected values (more noticeable with pHS post-selection than with snHS). 
As before, the initial state persistence seems to be more effectively recovered after DR.

\begin{figure*}[ht]
    \centering
\includegraphics[width=\textwidth]{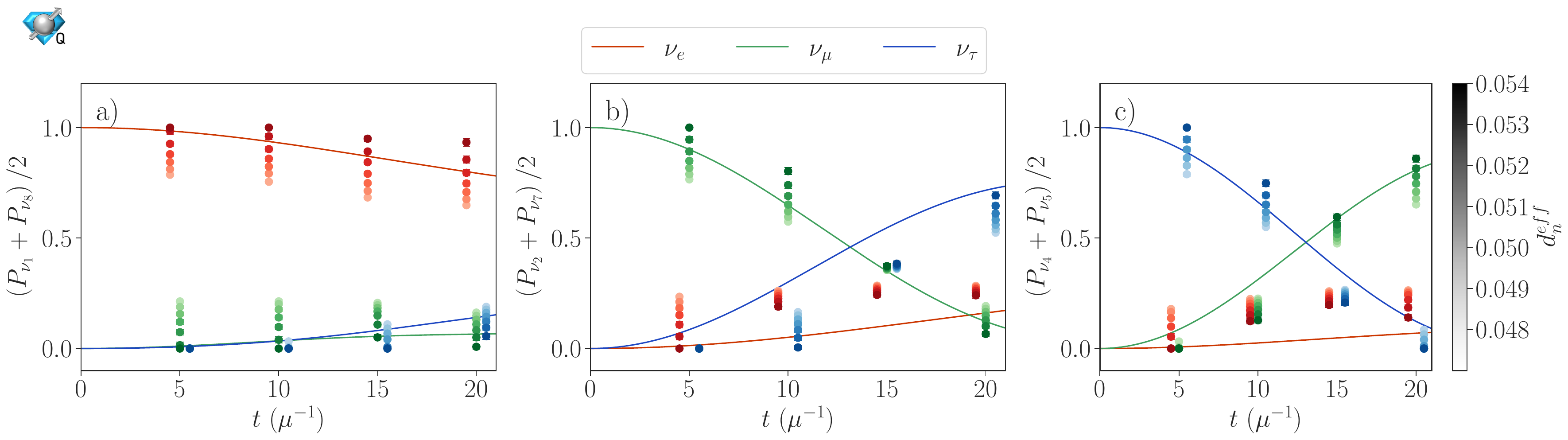}
\caption{Flavor evolution for an eight-neutrino system as a function of time obtained from \texttt{ibm\_torino} device as in Fig.~\ref{fig:ibm_8_neutrinos}, scanning over effective decoherence values $d^{(eff)}_n$ after applying DR and pHS post-selection procedure. Panels (a), (b), and (c) show the flavor evolution of the (symmetrized) first, second and fourth neutrinos, respectively. As shown in the color bar on the far-right, $d^{(eff)}_n$ increases from lighter to darker colors.
}
\label{fig:ibm_8_neutrinos_dn}
\end{figure*}
\begin{table}[ht]
\centering
\renewcommand{\arraystretch}{1.2}
\begin{tabularx}{\columnwidth}{|c|YYY|c|}
\hline
Neutrino  & $P_e$ & $P_\mu$ & $P_\tau$ & $d_n^{(id)}$ \\
\hline
$(P_1+P_8)/2$ & \textbf{0.079(4)} &0.054(3) &0.047(3) &0.033\\
$(P_2+P_7)/2$& 0.055(3) &\textbf{0.081(4)} &0.043(2) &0.033\\
$(P_3+P_6)/2$ & \textbf{0.081(4)} &0.052(3) &0.046(2) &0.033\\
$(P_4+P_5)/2$ & 0.051(3) &0.043(2) &\textbf{0.085(4)} &0.033\\
\hline
\end{tabularx}
\renewcommand{\arraystretch}{1}
\caption{pHS probabilities obtained from implementing the \textit{identity} quantum circuit on \texttt{ibm\_torino} averaging the $i^{\rm th}$ and $(N+1-i)^{\rm th}$ neutrinos. The last column shows the theoretical value for the decoherence line, $d_n^{(id)}=3^7/4^8$. In the noiseless case, the probabilities in bold should be 1.}
\label{tab:identityDR_8neutrinos}
\end{table}

The single-flavor probability results obtained from running the \textit{identity} quantum circuits in the DR method, shown in Table~\ref{tab:identityDR_8neutrinos}, suggest a shift of the decoherence value $d_n$ in Eq.~\eqref{eq:DR}. 
After applying the identity operator, the decoherence line $d_n$ is expected to be the plateau value of the probability that decays due to noise sources, i.e., the initial state probability value goes from 1 to $d_n$, while the other states' probabilities go from $0$ to $d_n$. Therefore, in the ideal case, the state probability results never cross the decoherence line. Instead, Table~\ref{tab:identityDR_8neutrinos} reports that all the obtained probabilities are greater than the theoretical decoherence value $d_n^{(id)}$ (given by $3^7/4^8$).
A possible explanation is the simple depolarizing noise model, assumed when applying DR method, is not enough to describe the noise contributions, and different qubits are subjected to different noise sources. 
Moreover, a non-negligible contribution from relaxation process is observed, increasing the probability of being in the $|0\rangle$ state.

This implies that the decoherence value $d_n$ in Eq.~\eqref{eq:DR} should be empirically changed. Looking at the obtained values, the ``effective'' experimental decoherence value can be estimated to be in the range $d^{(eff)}_n\in[0.048,0.054]$. 
If DR is applied using $d^{(eff)}_n$, the flavor probabilities for the first, second and fourth neutrino are closer to the analytical curves, as shown in Fig.~\ref{fig:ibm_8_neutrinos_dn}.
Also noted was that running a deeper quantum circuit for two Trotter steps causes the empirical decoherence line to move closer to the theoretical decoherence line.

\subsubsection{Entropy and tomography calculations}

\begin{figure*}[th]
    \centering
    \includegraphics[width=\textwidth]{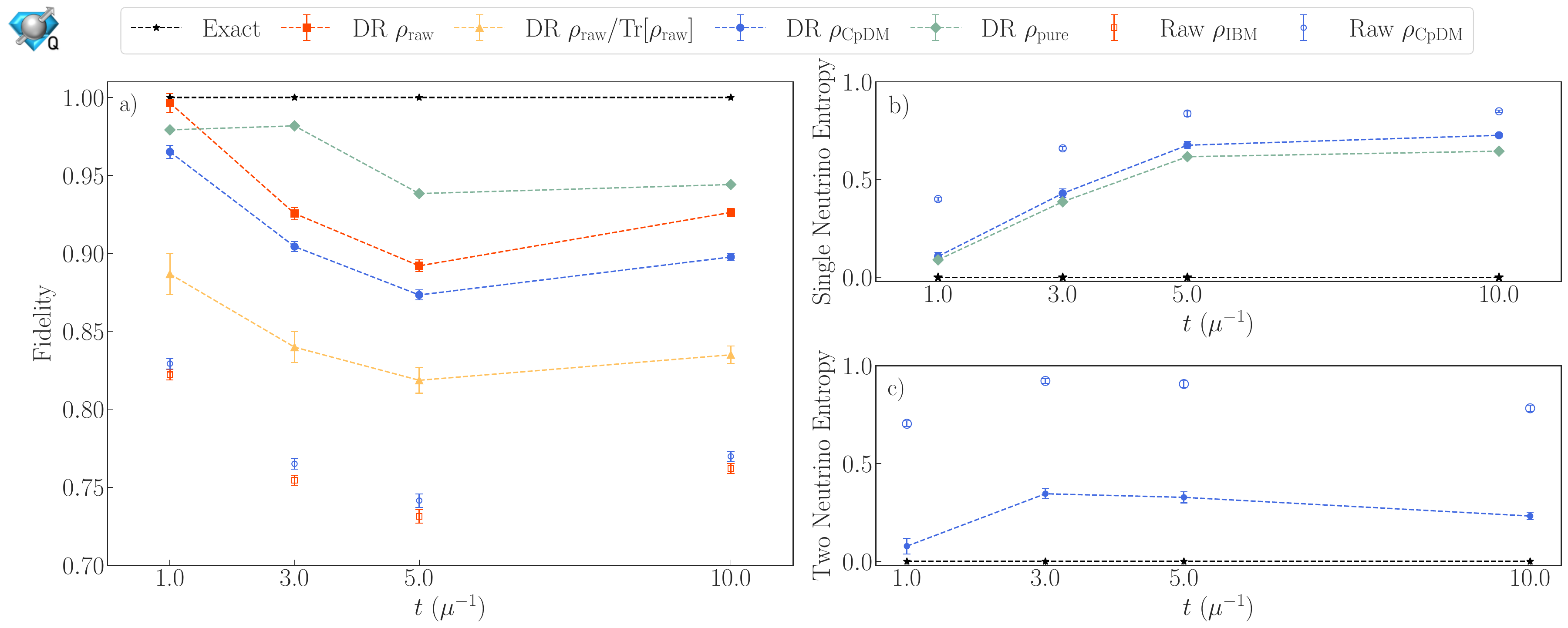}
    \caption{(a) Fidelity and (b) single-neutrino entropy for different $\Delta t$ for the two neutrino system starting from $\ket{e\mu}$ state. The different points correspond to different methods in computing the density matrix (see the main text for details). (c) Two-neutrino entropy computed from $\rho_{\rm CDM}$ density matrix. }
    \label{fig:fidelity}
\end{figure*}
To compute the entanglement entropy and other entanglement witnesses, the density matrix of $n\leq N$ neutrinos is evaluated. 
One approach uses classical shadows~\cite{Huang:2020tih}. 
Here, full state tomography is performed, as in Ref.~\cite{Hall:2021rbv}. 
Since not all $2^{2n}$ states are physical in the qubit mapping, the number of measurements needed to estimate the physical density matrix can be reduced.
For example, for a single neutrino the state-tomography operator pool can be reduced from 15 to 7 (independent) different operators that describe the Gell-Mann decomposition in the qutrit physical space.
The reduced density matrix for one neutrino is given by
\begin{equation}
\rho_\nu= \sum_{i=1}^9 c_i \lambda_i  \ ,
\label{eq:rho_1neut} 
\end{equation}
where $\lambda_9=\mathbb{I}$ and $c_{i}=\Tr(\lambda_i\: \rho)/\mathcal{A}_{i}$, with $\mathcal{A}_{i}=\Tr(\lambda^2_i)$. 
Tables~\ref{tab:tomography_pool_part1} and ~\ref{tab:tomography_pool_part2} in App.~\ref{app:tom} contains the explicit operator pool needed to extract each coefficient $c_i$, where the last column formulates the $c_i$ coefficient from the resulting measurement probabilities.
For a generic system of $N$ neutrinos, all combinations of the 7 operators should be implemented to obtain the corresponding density matrix.
Once these coefficients are fixed, the single-neutrino entanglement von Neumann entropy can be obtained using $S = -\Tr\left[ \rho_\nu \log(\rho_\nu)\right]$.

This three-flavor state tomography procedure is implemented on \texttt{ibm\_torino} for the two-neutrino system using a single Trotter time step. This requires running 49 different quantum circuits (that describe all possible independent combinations of Tables~\ref{tab:tomography_pool_part1} and ~\ref{tab:tomography_pool_part2}). Then, after applying the error mitigation methods, the coefficients $c_{ij}=\Tr(\lambda_i\otimes \lambda_j\: \rho)/\mathcal{A}_{ij}$ are evaluated, with $\mathcal{A}_{ij}=\Tr(\lambda^2_i\otimes \lambda^2_j)$, via the probabilities listed in Table~\ref{tab:tomography_pool_part1} and ~\ref{tab:tomography_pool_part2}. 
The corresponding density matrix, labeled $\rho_{\rm IBM}$, is obtained in a similar manner to Eq.~\eqref{eq:rho_1neut}, 
\begin{equation}
\rho_{\rm IBM}= \sum_{i,j={1}}^9 c_{ij} \lambda_i \otimes \lambda_j \ .
\end{equation}

Due to hardware noise, $\rho_{\rm IBM}$ is generally not positive semi-definite, therefore it does not represent a physical density matrix. Here, the algorithm from Ref.~\cite{Acharya:2021byb} is applied to find the closest physical density matrix, labeled $\rho_{\rm CpDM}$, via a rescaling of the eigenvalues (more details in App.~\ref{app:tom}). Moreover, because in this case the density matrix of the whole system is computed, the final state is expected to be a pure quantum state. This can be enforced by using the eigenstate with the largest eigenvalue of $\rho_{\rm CpDM}$. This state will correspond to the closest pure quantum state (it has the highest contribution in the Schmidt decomposition), and is labeled as $\rho_{\rm pure}$.

The fidelity between the obtained $\rho_{\rm IBM}$, $\rho_{\rm CpDM}$, $\rho_{\rm pure}$ and the exact one, $\zeta=|\Psi(t)\rangle \langle \Psi(t)|$, is:
\begin{equation}
    F(\rho,\zeta)=\left({\rm Tr}\sqrt{\sqrt{\zeta}\rho\sqrt{\zeta}}\right)^2\, .
\end{equation}
Panel (a) of Fig.~\ref{fig:fidelity} shows the results. It is interesting to note that due to the error mitigation used here (the coefficients $c_{ij}$ are all normalized using the same $P_{id}$ quantity), the raw and DR $\rho_{\rm IBM}/{\rm Tr}(\rho_{\rm IBM})$ completely overlap, since the DR renormalization factor is cancelled out when enforcing the unity of the trace. Similarly, the $\rho_{\rm pure}$ does not depend on whether error mitigation is applied or not, since in this case the largest eigenvalue is the same.

Single neutrino entanglement entropy is also copmuted using the $\rho_{\rm CpDM}$ and $\rho_{\rm pure}$ density matrices,\footnote{The matrix $\rho_{\rm IBM}$ cannot be used since it can have negative eigenvalues.} shown in panel (b) of Fig.~\ref{fig:fidelity}. 
Results from $\rho_{\rm pure}$ (dark blue points) are observed to be closer to the exact entropy behavior than the results from $\rho_{\rm CpDM}$ (green triangles). 
As a further test on the fidelity, the two-neutrino entropy for $\rho_{\rm CpDM}$ is computed to diagnose the effect of noise, since the analytical two-body entropy remains zero. These results are reported in panel (c) of Fig.~\ref{fig:fidelity}, where while the fidelity is seen to be $>90\%$, the density matrix $\rho_{\rm CpDM}$ still exhibits features of a mixed state.

\section{Conclusions}

In this work, we introduce new quantum circuits for simulating the collective dynamics of three-flavor neutrinos on gate-based quantum computers, and also provide an implementation of the two-neutrino flavor-exchange operator on qutrit-based computers using 4 $CX$ gates.
The implementation of the same dynamics on qubit-based platforms is demonstrated, where each neutrino is mapped to two qubits. The corresponding circuits require at least 18 CNOT gates.
Additionally, we introduce a simplification that allows the realization of the required all-to-all connectivity in a linear chain without any additional computational costs (same circuit depth).

We have executed the qubit-based quantum circuits with up to eight neutrinos on the Quantinuum \texttt{H1-1} and IBM \texttt{ibm\_torino} devices, and computed the probabilities of finding each neutrino in a specific flavor state, as well as the initial state persistence. These are key observables used to study the thermalization and equilibration of such systems~\cite{Martin:2023gbo}.
For the smaller systems, with two and four neutrinos, the circuit depth is small enough that using multiple Trotter steps to perform time evolution is feasible.
For eight neutrinos, while only one Trotter step was used, the resulting Trotter errors were smaller than the statistical and systematic uncertainties from the device.
Hardware noise was corrected through various error mitigation techniques and post-selection procedures.
The quality of the results (after mitigation) are higher in the trapped-ion device than the superconducting one, though the different number of shots used in both hinders a direct comparison.

Using the IBM quantum computer, it was also possible to test the proposed partial state tomography, which required implementing 49 operators, allowing us to evaluate the full density matrix of two neutrinos and the entanglement entropy.

The algorithms needed to perform realistic simulations require quantum computers with longer coherence times. That is because one might need to start from a thermal state (and not a pure state)~\cite{motta_2019,sagastizabal2021variational,turro2023quantum,Davoudi:2022uzo}, include the time-dependence in the two-neutrino term in the Hamiltonian~\cite{Rajput:2021khs,Siwach:2023wzy}, or include the effect of anti-neutrinos.




\clearpage
\begin{subappendices}

\section{Gell-Mann matrices}
\label{app:Gell_Mann}

Our notation for the Gell-Mann matrices is as follows,
\begin{align}
&\lambda_1=\begin{pmatrix}
    0 & 1 & 0\\
    1 & 0 & 0\\
    0 & 0 & 0\\
\end{pmatrix} , \
\lambda_2=\begin{pmatrix}
    0 & i & 0\\
    -i & 0 & 0\\
    0 & 0 & 0\\
\end{pmatrix} , \ 
\lambda_3=\begin{pmatrix}
    1 & 0 & 0\\
    0 & -1 & 0\\
    0 & 0 & 0\\
\end{pmatrix} , \nonumber \\
&\lambda_4=\begin{pmatrix}
    0 & 0 & 1\\
    0 & 0 & 0\\
    1 & 0 & 0\\
\end{pmatrix} ,  \
\lambda_5=\begin{pmatrix}
    0 & 0 & -i\\
    0 & 0 & 0\\
    i & 0 & 0\\
\end{pmatrix} , \nonumber    \\
&\lambda_6=\begin{pmatrix}
    0 & 0 & 0\\
    0 & 0 & 1\\
    0 & 1& 0\\
\end{pmatrix} ,  \  
\lambda_7=\begin{pmatrix}
    0 & 0 & 0\\
    0 & 0 & -i\\
    0 & i & 0\\
\end{pmatrix}  ,  \nonumber    \\
&\lambda_8=\sqrt{\frac{1}{3}}\,\begin{pmatrix}
    1 & 0 & 0\\
    0 & 1 & 0\\
    0 & 0 & -2\\
\end{pmatrix} .
\end{align}

\section{Qutrit gates}
\label{app:qutrit_gate}

This App. writes the explicit matrix representation of the qutrit gates.
The single-qutrit gates used in the circuits described in the main text are~\cite{Goss:2022bqd}
\begin{subequations}
\begin{align}
X^{12}_{\alpha} & = \begin{pmatrix}
     1 & 0 & 0\\
     0 & \cos \tfrac{\alpha}{2} & -i\sin \tfrac{\alpha}{2}\\
     0 & -i\sin \tfrac{\alpha}{2} & \cos \tfrac{\alpha}{2}\\
 \end{pmatrix} \ ,  
 R_y^{01}(\alpha) & = \begin{pmatrix}
     \cos\tfrac{\alpha}{2} & -\sin\tfrac{\alpha}{2} & 0\\
     \sin\tfrac{\alpha}{2} & \cos\tfrac{\alpha}{2} & 0\\
     0 & 0 & 1\\
 \end{pmatrix} \ ,  
 \end{align}
\begin{align}
 R_y^{12}(\alpha) & = \begin{pmatrix}
     1 & 0 & 0\\
     0 & \cos \tfrac{\alpha}{2} & -\sin \tfrac{\alpha}{2}\\
     0 & \sin \tfrac{\alpha}{2} & \cos \tfrac{\alpha}{2}\\
 \end{pmatrix} \ ,  
 \end{align}
\begin{align}
  {\rm Ph}(\theta,\phi,\lambda) & = \begin{pmatrix}
    e^{i\theta} & 0 &0\\
    0 & e^{i\phi} &0\\
    0 & 0 &e^{i\lambda}\\
\end{pmatrix} \ ,
\end{align}
\begin{align}
R_Z^{01}(\theta)&=  {\rm Ph}(-\frac{\theta}{2},\frac{\theta}{2},0)\ ,\\
R_Z^{12}(\phi)&= {\rm Ph}(0,-\frac{\phi}{2},\frac{\phi}{2})\ .
\end{align}
\end{subequations}

The two-qutrit $CX$ gate, whose action is given by $CX\ket{x,y}=\ket{x,{\rm mod}(x+y,3)}$, implements the following operation,
\begin{equation}
    CX = \begin{pmatrix}
1 & 0 & 0 & 0 & 0 & 0 & 0 & 0 & 0\\0 & 1 & 0 & 0 & 0 & 0 & 0 & 0 & 0\\0 & 0 & 1 & 0 & 0 & 0 & 0 & 0 & 0\\0 & 0 & 0 & 0 & 1 & 0 & 0 & 0 & 0\\0 & 0 & 0 & 0 & 0 & 1 & 0 & 0 & 0\\0 & 0 & 0 & 1 & 0 & 0 & 0 & 0 & 0\\0 & 0 & 0 & 0 & 0 & 0 & 0 & 0 & 1\\0 & 0 & 0 & 0 & 0 & 0 & 1 & 0 & 0\\0 & 0 & 0 & 0 & 0 & 0 & 0 & 1 & 0
    \end{pmatrix} \ .
\end{equation}
While the $CX$ gate might not be a native two-qutrit gate on current qutrit quantum devices, it is straightforward to transform to alternative entangling gates, like the $CZ$ gate~\cite{Goss:2022bqd},
\begin{equation}
    CZ = \sum_{i,j\in\{0,1,2\}}\tilde{\omega}^{ij}|ij\rangle \langle ij | \ , \quad  \tilde{\omega}= e^{i\frac{2\pi}{3}} \ .
\end{equation}
Using the qutrit Hadamard gate~\cite{Morvan:2021qju}, the $CX$ gate can transformed into a $CZ$ gate,
\begin{equation}
    CZ = (1 \otimes H^\dag) \cdot  CX \cdot  (1 \otimes H) \ , \quad H = \frac{1}{\sqrt{3}}\begin{pmatrix}
        1 & 1 & 1 \\
        1 & \tilde{\omega} & \tilde{\omega}^2 \\
        1 & \tilde{\omega}^2 & \tilde{\omega}
    \end{pmatrix} \ ,
\end{equation}
with a similar transformation for $CX^\dag$.

\section{Quantinuum emulator \label{app:4neutrinos}}

\begin{figure*}[ht]
\centering
\includegraphics[width=\textwidth]{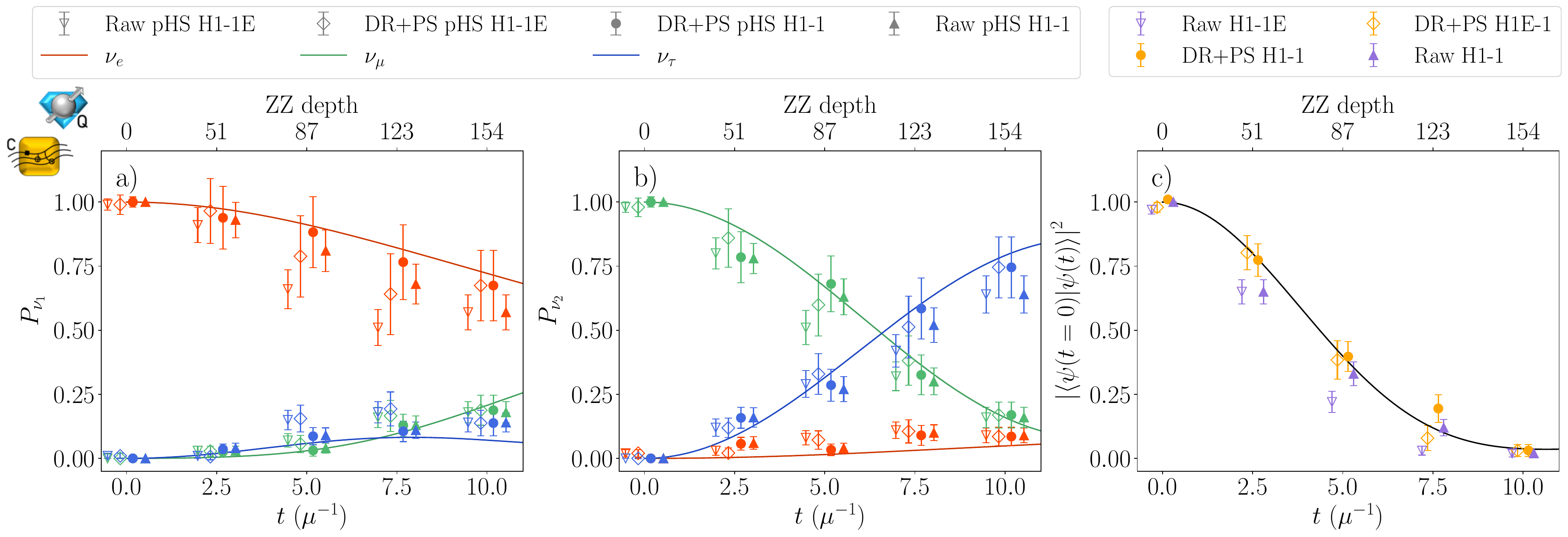}
\caption{Flavor evolution for four neutrinos as a function of time from the \texttt{H1-1E} emulator and \texttt{H1-1} device. Empty triangles and circles represent the emulator's raw and DR+PS results, respectively. Solid symbols show the corresponding results from \texttt{H1-1}. We use the same conventions as in Fig.~\ref{fig:quantinuum_4_neutrinos}.}

\label{fig:emulator_4_neutrinos}
\end{figure*}

Results from \texttt{H1-1} and its emulator for the four neutrino dynamics is reported in Fig.~\ref{fig:emulator_4_neutrinos}. DR and the pHS post-selecting procedure were implemented in both cases. 
The emulator's results (empty symbols) were observed to be compatible with the \texttt{H1-1} device's results (solid symbols).

\section{Simulations of eight neutrinos on IBM quantum computer\label{app:8ibm} }

\begin{figure*}[ht]
    \centering
    \includegraphics[width=\textwidth]{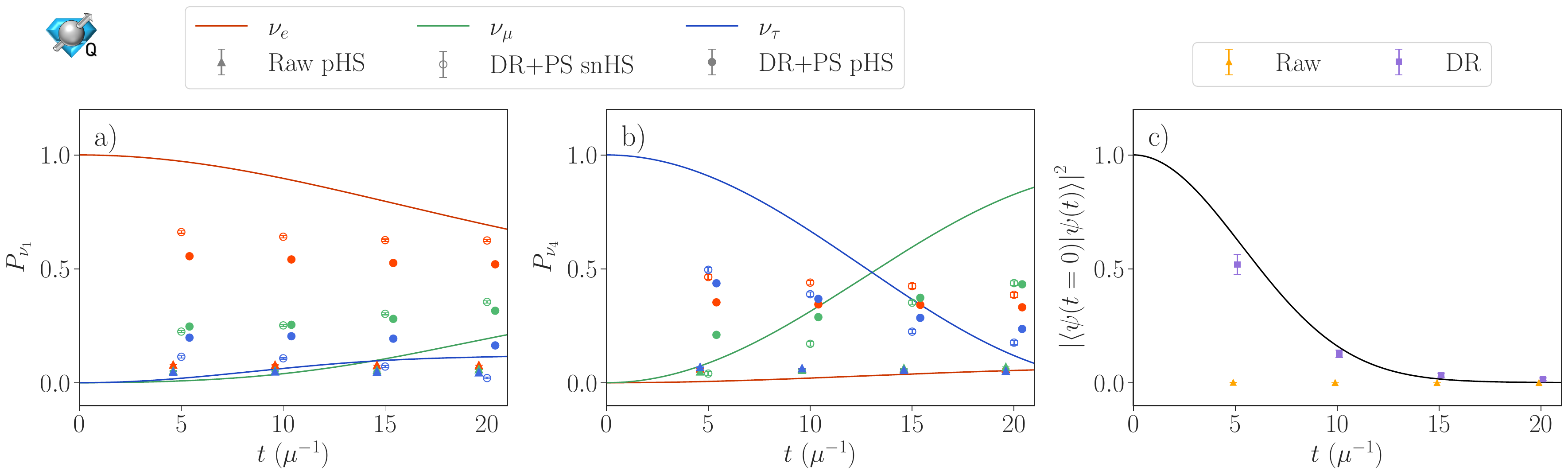}
    \caption{Flavor evolution for eight neutrinos as a function of time obtained from the \texttt{ibm\_torino} device, with $\ket{\nu_e \nu_\mu \nu_e \nu_\tau \nu_e \nu_\mu \nu_e \nu_\tau}$ as the initial state.  Panels (a) and (b) show the flavor evolution of the first and fourth neutrinos, respectively. 
    Panel (c) shows the persistence probability of the initial state.
    We use the same conventions as in Fig.~\ref{fig:quantinuum_4_neutrinos}.}
    \label{fig:enter-label}
\end{figure*}

Figure~\ref{fig:enter-label} shows the results from the evolution of eight neutrinos starting from the non-symmetric state $\ket{\nu_e \nu_\mu \nu_e \nu_\tau \nu_e \nu_\mu \nu_e \nu_\tau}$. 
Compared to the results in Fig.~\ref{fig:ibm_8_neutrinos}, the averaging procedure appears helpful for averaging out device errors.
A clear example is the electron flavor evolution for the fourth neutrino, displayed in panel (b).

\section{Details about the tomography study}
\label{app:tom}

This App. contains details on the operator pool used to reconstruct the density matrix. 
Each row in Tables~\ref{tab:tomography_pool_part1} and ~\ref{tab:tomography_pool_part2} shows how to evaluate the coefficient $c_i$ from Eq.~\eqref{eq:rho_1neut}.
The second column contains the operator needed to change the basis in which to measure the two qubits. By measuring the state probability $P_i$ with the expression in the third column, the value of $c_i$ is recovered. Note that the operators for $\lambda_3$, $\lambda_8$, and $\lambda_9$ (the latter of which is not shown in the table because $c_9$ is $\frac{1}{3}$ times the sum of all 3 state-probabilities and thus is $\frac{1}{3}$ no matter the measurement-result) are identity operators, thus the operator pool is composed of 7 independent operators (instead of 9). 

The goal of the algorithm from Ref.~\cite{Acharya:2021byb} is to find the closest positive semi-definite density matrix to the one obtained from {\tt ibm\_torino}. The general idea of the algorithm is to find the the density matrix $\rho_{\rm CpDM}$ that minimizes the trace distance with $\rho_{\rm IBM}$ while having all eigenvalues positive. This is done by shifting the eigenvalues of $\rho_{\rm IBM}$ using the algorithm of Ref.~\cite{Wang:2013}, while leaving the eigenvectors of $\rho_{\rm IBM}$ unmodified.

\begin{table}[ht]
    \footnotesize
    \centering
    \begin{tabularx}{\columnwidth}{|c|Y|Y|}
    \hline
    $\lambda_i$ & Operator & Probability \\
    \hline
    $\lambda_1$ & $\frac{1}{\sqrt{2}} \begin{pmatrix}
            1 & 1 & 0 &0\\
            1 & -1 & 0 &0\\
            0 & 0 & \sqrt{2}&0\\ 
            0 &  0 &0 &\sqrt{2}\\
        \end{pmatrix}$ & $c_1= \frac{1}{2}(P_{00}-P_{01})$\\
    $\lambda_2$ & $\frac{1}{\sqrt{2}} \begin{pmatrix}
            1 & -i & 0 &0\\
            1 & i& 0 &0\\
            0 & 0 & \sqrt{2}&0\\   
            0 &  0 &0 &\sqrt{2}\\
        \end{pmatrix}$ & $c_2= \frac{1}{2}(P_{00}-P_{01})$\\
    $\lambda_3$ & $\begin{pmatrix}
            1 & 0 & 0 &0\\
           0 & 1 & 0 &0\\
            0 & 0 &1&0\\  
            0 &  0 &0 &1\\
        \end{pmatrix}$ & $c_3= \frac{1}{2}(P_{00}-P_{01})$\\            
    $\lambda_4$ & $\frac{1}{\sqrt{2}} \begin{pmatrix}
            1 & 0 & 1 &0\\
            0 & \sqrt{2} & 0 &0\\
            1 & 0 & -1&0\\
            0 &  0 &0 &\sqrt{2}\\
        \end{pmatrix}$ & $c_4=\frac{1}{2}(P_{00}-P_{10})$\\        
 \hline
    \end{tabularx}
    \caption{First half of the tomographic pool for computing $c_i=\Tr (\rho \lambda_i)/\mathcal{A}_i$, where $\lambda_i$ represents the Gell-Mann matrix. We perform the change of basis by implementing the operator shown in the second column. The coefficient $c_i$ are then given by the linear combination of the obtained probabilities $P_{ij}=\{ P_{00},P_{01},P_{10},P_{11}\}$ given in the third column.
    }
    \label{tab:tomography_pool_part1}
\end{table}

\begin{table}[ht]
    \footnotesize
    \centering
    \begin{tabularx}{\columnwidth}{|c|Y|Y|}
    \hline
    $\lambda_i$ & Operator & Probability \\
    \hline
    $\lambda_5$ & $\frac{1}{\sqrt{2}} \begin{pmatrix}
            1 & 0 & -i &0\\
            0 & \sqrt{2} & 0 &0\\
            1 & 0 & i&0\\  
            0 &  0 &0 &\sqrt{2}\\
        \end{pmatrix}$ & $c_5=\frac{1}{2}(P_{00}-P_{10})$\\
    $\lambda_6$ & $\frac{1}{\sqrt{2}} \begin{pmatrix}
           \sqrt{2} & 0 & 0 &0\\
            0 & 1 & 1 &0\\
            0 & 1 & -1&0\\ 
            0 &  0 &0 &\sqrt{2}\\
        \end{pmatrix}$ & $c_6=\frac{1}{2}(P_{01}-P_{10})$\\         
    $\lambda_7$ & $\frac{1}{\sqrt{2}} \begin{pmatrix}
           \sqrt{2} & 0 & 0 &0\\
            0 & 1 & -i&0\\
            0 & 1 & i&0\\
            0 & 0 &0 &\sqrt{2}\\
        \end{pmatrix}$ & $c_7=\frac{1}{2}(P_{01}-P_{10})$\\ 
    $\lambda_8$ & $ \begin{pmatrix}
           1 & 0 & 0 &0\\
            0 & 1 &0 &0\\
            0 & 0 & 1&0\\
            0 &  0 &0 &1\\
        \end{pmatrix}$ & $c_8=\frac{1}{2\sqrt{3}}(P_{00}+P_{01}-2\,P_{10})$\\ 
 \hline
    \end{tabularx}
    \caption{Second half of the tomographic pool for computing $c_i=\Tr (\rho \lambda_i)/\mathcal{A}_i$, where $\lambda_i$ represents the Gell-Mann matrix. We perform the change of basis by implementing the operator shown in the second column. The coefficient $c_i$ are then given by the linear combination of the obtained probabilities $P_{ij}=\{ P_{00},P_{01},P_{10},P_{11}\}$ given in the third column.
    }
    \label{tab:tomography_pool_part2}
\end{table}

\section{Device parameters\label{app:device_parameters}}

In this appendix we report the experimental parameters of the quantum computers used in this paper.

In Table~\ref{tab:quantinuum_parameters}, we report the Quantinuum \texttt{H1-1} device parameters, and in Table~\ref{tab:ibm_parameters}, we report the IBM \texttt{ibm\_torino} device parameters.

\begin{table}[ht]
\centering
\begin{tabular}{|c|c|}
\hline
Total number of qubits & 20  \\
Typical Single-qubit gate  infidelity & $2 \cdot 10^{-5}$ \\
Typical Two-qubit gate  infidelity & $1 \cdot 10^{-3}$ \\
SPAM error &  $2 \cdot 10^{-4}$\\
\hline
\end{tabular}
\caption{Quantinuum H1-1 device parameters, as reported in Ref.~\cite{quantinuum}.}
\label{tab:quantinuum_parameters}
\end{table}

\begin{table}[ht]
\centering
\begin{tabular}{|c|c|c|c|}
\hline
Total number of qubits & \multicolumn{3}{c|}{133} \\ \hline
Neutrinos & 2 & 4 & 8\\
Date accessed & 6/12/24 & 6/21/24 & 6/25/14 \\
\hline

Number of qubits used & 4 & 8 & 16 \\
Median T1 coherence time ($\mu$s) & 150 & 133 & 142 \\
Median T2 coherence time ($\mu$s) & 147 & 127 & 151 \\
Median $X$-gate error & $3.2 \cdot 10^{-4}$ & $3.3 \cdot 10^{-4}$ & $2.8 \cdot 10^{-4}$ \\
Median $CZ$-gate error & $9.4 \cdot 10^{-3}$ & $7.8 \cdot 10^{-3}$ & $4.0\cdot 10^{-3}$ \\
Median readout error &  $2.6 \cdot 10^{-2}$ &  $2.9 \cdot 10^{-2}$ &  $2.3 \cdot 10^{-2}$\\
\hline

\end{tabular}
\caption{IBM \texttt{ibm\_torino} device parameters.}
\label{tab:ibm_parameters}
\end{table}

\end{subappendices}
\chapter{Quantum Magic and Computational Complexity in the Neutrino Sector}
\label{chap:neutrinomagic}

{\it This chapter is associated with Ref. \cite{chernyshev2024quantum}:}
{\it ``Quantum Magic and Computational Complexity in the Neutrino Sector" by Ivan A. Chernyshev, Caroline E. P. Robin, and Martin J. Savage}

\noindent
To optimize the impact of quantum computers in simulating key aspects of fundamental physics, 
it is essential to understand the required balance among quantum and classical computing resources
to address specific observables.
As advances in quantum simulations feed back to improve classical simulations, 
this balance changes with time, 
and guidance from the target physical systems must be folded in with each new advance.
Robust simulations of neutrinos produced during supernova and during neutron-star binary mergers
are important, e.g., Refs.~\cite{Burrows_2021,mueller2019,wanajo2014production,hoffman1997nucleosynthesis,winteler2012magnetorotationally,Bruenn:2009ucj,Bruenn:2006oub,Foucart_2023,Cusinato_2022,Vijayan:2023bfs,George:2020veu,PhysRevD101043009,PhysRevLett.132.211001,PhysRevD109103027,Balantekin:2023ayx,Cornelius:2024zsb,Shalgar:2024gjt,Padilla-Gay:2024wyo},
not only for their evolution and for the predictions of the chemical elements in such processes, 
but also for probing the properties and interactions of neutrinos themselves and potentially discovering new physics, e.g., Ref.~\cite{Suliga:2024oby}.
As part of the integration of the neutrino processes that take place during a supernova into simulations, 
a much better understanding of the quantum complexity of coherent flavor transformations is essential.

As mentioned in Sec. \ref{sec:intro_magic_entanglement_quantumadvantage} and \ref{sec:neutrinos_on_qutrits_intro}, during core-collapse supernova (CCSN) 
the neutrino density becomes sufficiently high that self-interactions play an essential role in the evolution of lepton flavor, and
there have been numerous studies performed  to describe the impact of the
$\sim 10^{58}$ neutrinos that are produced in such events.
The range of mass-scales involved, and the interaction processes that take place, present a significant challenge to accurately describing  this evolution.
The mean-field (introduced in Sec. \ref{sec:neutrinos_on_qutrits_intro}) and many-body approaches for the dynamics continues to provide a firm foundation, 
underpinning much of what is known about these systems, e.g., Refs.~\cite{Qian:1994wh,Duan:2006jv,Duan:2006an,Izaguirre:2016gsx,Capozzi:2020kge,Fiorillo:2023mze,Fiorillo:2023hlk,Fiorillo:2024fnl,Shalgar:2023ooi,Johns:2023ewj}.
However, advances in quantum information are providing motivation and techniques 
to consider aspects of these systems beyond the currently employed approximations~\cite{Rrapaj:2019pxz,Patwardhan:2019zta,Patwardhan:2021rej,Roggero:2021asb,Xiong:2021evk,roggero2021dynamical,Martin:2021bri,Roggero:2022hpy,Illa:2022zgu,Bhaskar:2023sta,Martin:2023gbo,Martin:2023ljq,Neill:2024klc,Kost:2024esc,Cirigliano:2024pnm}.
These nascent explorations, 
that include 
the low-energy effective Hamiltonian from the Standard Model mapped to all-to-all connected spin models, 
have examined the evolution of the neutrino flavors, their entanglement entropy, 
multi-partite entanglement using $n$-tangles, and more.
Typically these have been performed using an effective two-neutrino system, 
and extensions to include three flavors, such as the one in Chapter \ref{chap:qutritqubitneutrino}, which include designs for qutrit quantum circuits, are now beginning~\cite{Balantekin:2006tg,Siwach:2022xhx,Balantekin:2023qvm,Chernyshev:2024kpu}.
The entanglement between multiple neutrinos exceeds that of systems
of Bell pairs, and hence is fundamentally multi-partite in nature~\cite{Illa:2022zgu,Martin:2023ljq}.
Simulations of modest-sized systems of neutrinos have been performed using 
superconducting-qubit and trapped-ion qubit quantum computers~\cite{Hall:2021rbv,Yeter-Aydeniz:2021olz,Illa:2022jqb,Amitrano:2022yyn,Illa:2022zgu,Siwach:2023wzy}.
Further, 
classical simulations~\cite{Balantekin:2023qvm} 
have also been recently performed.
Simulating neutrino environments of interest requires working with 
mixed states, and 
as such, these early investigations are important for  guiding 
development toward more robust simulations.

As discussed in Sec. \ref{sec:intro_magic_entanglement_quantumadvantage}, significant quantum magic (non-stabilizerness) along with large-scale entanglement is the requisite for the need for quantum computation.
Stabilizer states
can be efficiently prepared 
using a classical gate set of the Hadamard-gate, H, the phase-gate, S, 
and the CNOT-gate~\cite{gottesman:1998hu,gottesman:1997zz}
(App.~\ref{app:Stabs}).
By construction, stabilizer states have vanishing measures of magic.
Thus, both magic and entanglement determine the computational complexity 
and quantum resource requirements for simulating physical systems.
Including the T-gate to establish a universal quantum 
gate set, 
Gottesman-Knill-Aaronson~\cite{gottesman:1998hu,aaronson_2004} showed that 
the exponential-scaling (with system size) of classical resource requirements is determined 
by the minimum number of T-gates (a similar argument exists for scaling with precision, e.g., Ref.~\cite{chen2020quantum}).  
A number of measures of magic have been developed, e.g., Refs.~\cite{Emerson:2013zse,PhysRevLett.124.090505,PhysRevLett.118.090501,Bravyi2019simulationofquantum,PhysRevA.83.032317,Beverland:2019jej},
and 
the stabilizer R\'enyi entropies (SREs)~\cite{Leone:2021rzd} and Bell magic~\cite{PRXQuantum.4.010301}
have been measured in quantum simulations of some systems~\cite{Oliviero_2022,PRXQuantum.4.010301,Bluvstein:2023zmt},
and are efficiently calculable in MPS~\cite{Haug:2022vpg,Haug:2023hcs,Tarabunga:2024ugl, Lami:2023naw,lami2024quantum, lami2024unveiling}.
The magic properties of physical many-body systems and quantum field theories are less known than their entanglement structures.  Explorations in the Ising and Heisenberg models~\cite{Oliviero_2022,Haug:2023hcs,Rattacaso:2023kzm,frau2024nonstabilizerness,Catalano:2024bdh},
lattice gauge models~\cite{Tarabunga:2023ggd}, 
quantum gravity~\cite{Cepollaro:2024qln},
in nuclear and hypernuclear forces~\cite{Robin:2024bdz}, 
and in the structure of nuclei~\cite{Brokemeier:2024lhq} 
are in their earliest stages.
Interestingly, it has recently been shown that 
the entanglement and magic in
random quantum circuits doped with T-gates and measurements undergo phase transitions (between volume-law and area-law scaling) at different dopings~\cite{Fux:2023brx,Bejan:2023zqm,chen2020quantum,Li:2024ahc,Catalano:2024bdh}.

To determine the measures of magic in a wavefunction, matrix elements of  
strings of Pauli operators, $\hat{P}$, are computed, 
$c_P (t) \equiv \langle \psi (t) |\hat{P} | \psi (t) \rangle$.
For qutrits, the Pauli strings are constructed from tensor-products of the nine operators, 
$\hat \Sigma_i$~\cite{Howard_2013,Cui_2015,Wang:QutritZX},
\begin{eqnarray}
\hat \Sigma_i & \in & \{
\hat I, 
\hat X, 
\hat Z, 
\hat X^2, 
\omega \hat X \hat Z, 
\hat Z^2, 
\omega^2 \hat X \hat Z^2, 
\hat X^2 \hat Z,
\hat X^2 \hat Z^2  
\}
,
\label{eq:qutritPaulismain}
\end{eqnarray}
where 
\begin{eqnarray}
\hat X |j\rangle & = & |j+1\rangle
\ \ ,\ \ 
\hat Z |j\rangle\ =\ \omega^j |j\rangle
\ \ ,\ \ 
\omega\ =\ e^{i 2 \pi/3}
\ \ ,
\label{eq:qutritXZmain}
\end{eqnarray}
for $j=0,1,2$.
For stabilizer states of $n_Q$ qutrits,
$d=3^{n_Q}$ of the $d^2$ Pauli strings give $c_P = 1, \omega$ or $\omega^2$, 
while the other $d^2-d$ give $c_P = 0$~\cite{zhu2016clifford}.
For an arbitrary quantum state, all $d^2$ values can be non-zero.
As is the case for qubits, 
the deviation from stabilizerness 
defines the magic in a state~\cite{Leone:2021rzd}, using
\begin{eqnarray}
\Xi_P & = & |c_P|^2/d
\ ,\ 
\sum_P \Xi_P\ =\ 1
\ \ ,
\label{eq:XiP}
\end{eqnarray}
where $\Xi_P$ forms a probability distribution.
Based on our previous studies~\cite{Robin:2024bdz,Brokemeier:2024lhq},
we consider the $\alpha=2$ stabilizer Renyi entropy (SRE),
\begin{eqnarray}
    \mathcal{M}_2 & = &    -\log_2 \ d \sum_P  \Xi_P^2 
    \; ,
    \label{eq:SREM2}
\end{eqnarray}
to explore the quantum magic in a neutrino 
wavefunction. 
This SRE has been shown to satisfy properties of a proper magic measure~\cite{Haug:2023hcs,Leone:2024lfr}.
For more details, see App.~\ref{app:CompMag}.

In the case of three flavors of neutrinos, the  charged-current eigenstates are related to the mass eigenstates by the Pontecorvo–Maki–Nakagawa–Sakata (PMNS)
matrix~\cite{Pontecorvo1957,Maki1962}, 
\begin{eqnarray}
{\bm\nu}_F & = & U_{PMNS} . {\bm\nu}_M
\ \ ,
\end{eqnarray}
where,
${\bm\nu}_F=\left(\nu_e, \nu_\mu, \nu_\tau\right)^T$ and 
${\bm\nu}_M=\left(\nu_1, \nu_2, \nu_3\right)^T$.
Neglecting Majorana phases, $U_{PMNS}$ can be paramterized in terms of three angles, 
$\theta_{12}$, $\theta_{13}$, $\theta_{23}$ and one CP-violating phase $\delta$.
The experimental determinations of these angles in a commonly used parameterization of the matrix 
are taken from the Particle Data Group (PDG)~\cite{ParticleDataGroup:2024cfk},
and reproduced in App.~\ref{app:OneNu}.

This work uses the same mass-basis one-body Hamiltonian term for a neutrino of energy $E$ that Chapter \ref{chap:qutritqubitneutrino} uses,
\begin{eqnarray}
\hat H_1 & = &  
{1\over 2 E}\left(
\begin{array}{ccc}
0&0&0\\
0&\delta m_{21}^2&0\\
0&0&\Delta m_{31}^2
\end{array}
\right)
\ +\  ...
    \ ,
    \label{eq:U1bodQT}
\end{eqnarray}
where the ellipses denote terms proportional to the identity matrix or higher order in the neutrino-mass expansion of the kinetic energy.
The difference in mass-squareds, $\Delta m_{31}^2$ can be related to the experimentally measured values,
$\Delta m_{31}^2 =\Delta m_{32}^2 + \delta m_{21}^2$~\cite{ParticleDataGroup:2024cfk}.
\footnote{In this work, only the normal hierarchy of neutrino masses is considered.}
An effective two-flavor reduction of the system is typically found by retaining $\theta_{12}$ and 
$\delta m_{21}^2$ and discarding the third eigenstate.

\begin{figure}[!ht]
    \centering
    \includegraphics[width=0.9\columnwidth]{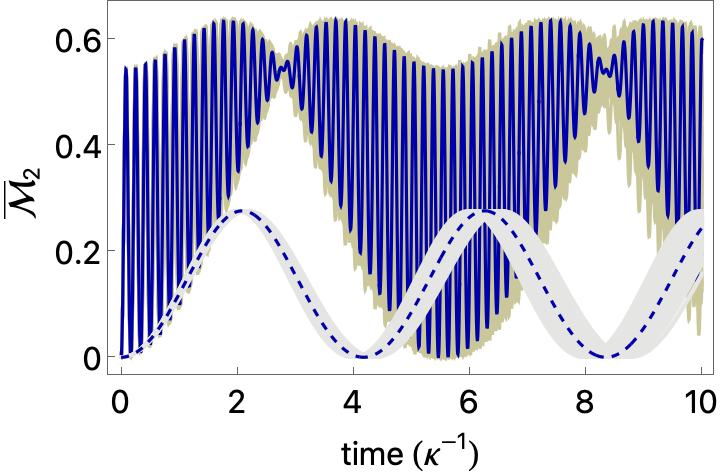}
    \caption{
    The magic power, $\overline{\mathcal{M}}_2(\hat U_1)$, of the free-space one-body evolution operator for 
    three flavors of neutrinos
    given in Eq.~(\ref{eq:UtPhi}).
 The  solid blue line shows the central value of the magic power, 
    while the khaki region corresponds to the values of magic power from a sampling 
    over the $68\%$ confidence intervals of $\Delta m_{32}^2$ and $\delta m_{21}^2$.
    The dashed-blue line line and lighter shaded region correspond to the magic power of the evolution operator in the effective two-flavor system.
    Analytic expressions for $\overline{\mathcal{M}}_2(\hat U_1)$ are provided in 
    App.~\ref{app:CompMagPow}.
    }
    \label{fig:1qT}
\end{figure}

In the mass basis, each neutrino flavor  has non-zero magic from the $U_{PMNS}$ mixing matrix.
In the case of single electron-flavored neutrino in the 
effective two-flavor system, its magic is computed to be 
${\cal M}_2=0.195(23)$,
which should be compared to a maximum value of $0.415$ for relatively real states and 
$0.585$
for complex states.
For a three-flavor neutrino,  
the magic in the single neutrino is found to be
${\cal M}_2=0.891(14)$, which should be compared with a  maximum possible value of $1$.
The presence of the third generation of neutrinos changes the magic in the single neutrino sector substantially.
To reinforce this observation, 
it is helpful to consider the magic power~\cite{Leone:2021rzd,Robin:2024bdz}
of the single-neutrino evolution operator.
The magic power of a unitary operator,
which we denote by $\overline{\mathcal{M}}_2$,
quantifies the average fluctuations in magic induced by the operator, 
based upon its action on stabilizer states $|\Phi_i\rangle$.
By considering the set of time evolved states, under the evolution of the free-space one-body Hamiltonian in Eq.~(\ref{eq:U1bodQT}),
\begin{eqnarray}
|\Phi_i \rangle (t) & = &  
\hat U_1(t) |\Phi_i\rangle\ =\ 
e^{-i \hat H_1 t} |\Phi_i\rangle
    \ ,
\label{eq:UtPhi}
\end{eqnarray}
the magic power of $e^{-i \hat H_1 t}$ is shown in Fig.~\ref{fig:1qT}.
There is a significant difference between the magic power of the free-space one-body evolution operator 
for three flavors compared with two.

There are a number of models employed to 
expose essential elements of 
the coherent evolution of neutrinos in 
supernovae.  
We select one such model, 
that has been fruitfully used to study the evolution of entanglement, 
to illustrate the corresponding behavior of magic. 
The pair-wise coherent forward interactions between neutrinos is captured by the low-energy position-dependent effective Hamiltonian~\cite{1987apj...322..795f,savage:1990by,Pantaleone:1992eq,Pantaleone:1992xh,Malaney:1993ah,Kostelecky:1993yt,DOlivo:1995qgv,Qian:1993hh,Fuller:2005ae,
Balantekin:2006tg},
combined with a 
(naively integrable)
model-dependent neutrino density profile in the single-angle limit~\cite{Balantekin:2023qvm},
\begin{eqnarray}
    \hat H_2(r) & = & \mu (r) \sum_{a=1}^8 \hat T^a \otimes \hat T^a 
    \ ,    
    \nonumber\\
    \mu(r) & = &  
    \mu_0 \left( 1-\sqrt{1- (R_\nu/r)^2}\right)^2
    \ ,
    \label{eq:2bodyevol}
\end{eqnarray}
where the $\hat T^a$ are the generators of SU(3) transformations, and at the edge of the neutrino sphere,
the model uses 
$\mu_0=3.62 \times 10^4$ MeV, 
$\kappa R_\nu=32.2$,
and $\kappa=10^{-17}$MeV.
The time evolution of multi-neutrino systems is determined by integrating the action of the evolution operator on a given initial state.
In this model,  
the radial location of the neutrinos is 
given by $r(t) = r_0+t$, 
with $r_0=210.65/\kappa$ defining $t=0$.
Using a distribution of neutrino one-body energies below $E_0=10$ MeV, scaling as $E_n=E_0/n$, the 
time-dependent Hamiltonian and wavefunction evolution
describing the coherent flavor evolution can be written as, 
assuming radial propagation, 
\begin{eqnarray}
\hat H (t) & = &  
\sum_n n \hat H_1^{(n)} 
\ +\ 
\sum_{n,n^\prime} \hat H_2^{(n,n^\prime)} (t)
\ , \nonumber\\
|\psi (t)\rangle  & = & \hat U_2 (t,0) |\psi\rangle_0
    \ =\ T \left[ e^{-i \int_{0}^{t}\ dt^\prime\ \hat H (t^\prime)} \right] |\psi\rangle_0
    \ ,
    \label{eq:NeutHt}
\end{eqnarray}
where $\hat H_1^{(n)}$ is given in Eq.~(\ref{eq:U1bodQT}) acting on the $n^{\rm th}$ neutrino, and $\hat H_2^{(n,n^\prime)} (t)$ corresponds to the two-body operator in 
Eq.~(\ref{eq:2bodyevol}) acting on the $n^{\rm th}$ and ${n^\prime}^{\rm th}$ neutrinos.

\begin{figure}[!ht]
    \centering
    \includegraphics[width=0.8\columnwidth]{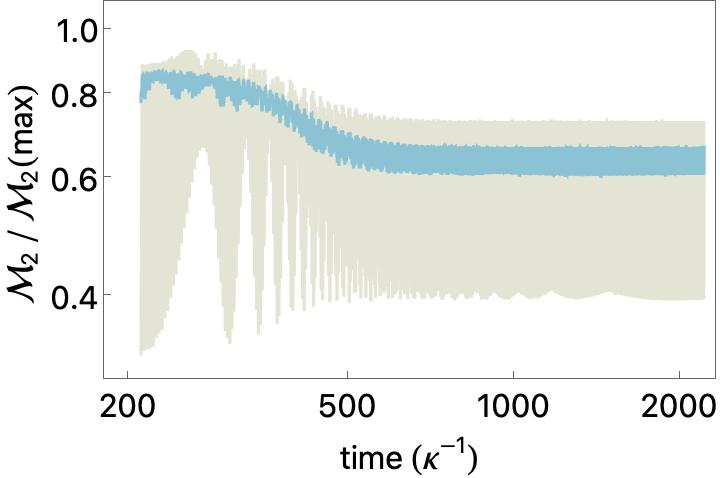}
    \caption{
    The normalized magic in the two-flavor (lighter, cream)
    and three-flavor (darker, blue)
    neutrino wavefunctions as a function of time, starting in the pure tensor-product states $|\nu_e\nu_e\rangle$.
    The $\mathcal{M}_2$ measure of magic,
    defined in Eq.~(\ref{eq:SREM2}),
    is normalized to its maximum value,
    $\mathcal{M}_2({\rm max})=1.19265$ for two flavors and 
    $\mathcal{M}_2({\rm max})=2.23379$ for three flavors. 
    }
    \label{fig:2magdivmag_ee}
\end{figure}

\begin{figure}[!ht]
    \centering
    \includegraphics[width=0.8\columnwidth]{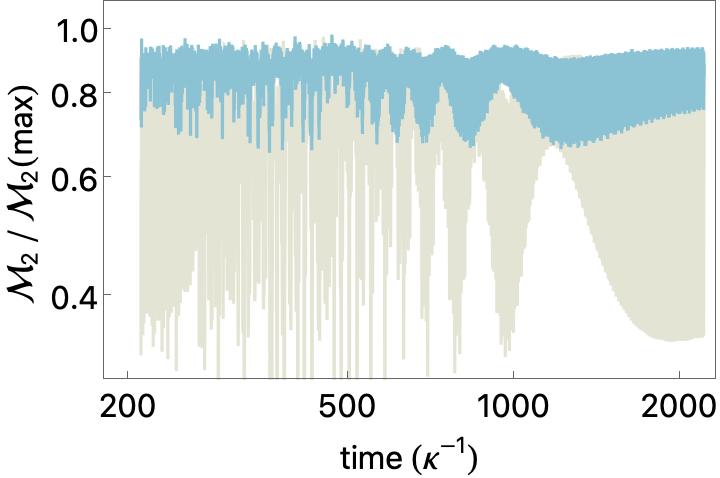}
    \caption{
    The normalized magic in the two-flavor (lighter, cream)
    and three-flavor (darker, blue)
    neutrino wavefunctions as a function of time, starting in the pure tensor-product states $|\nu_e\nu_\mu\rangle$.
    The $\mathcal{M}_2$ measure of magic,
    defined in Eq.~(\ref{eq:SREM2}),
    is normalized to its maximum value,
    $\mathcal{M}_2({\rm max})=1.19265$ for two flavors and 
    $\mathcal{M}_2({\rm max})=2.23379$ for three flavors. 
    }
    \label{fig:2magdivmag_eu}
\end{figure}

In the two-neutrino sector, we consider initial conditions of a tensor-product pure-state 
of two electron-flavor neutrinos, $|\psi\rangle_0=|\nu_e\nu_e\rangle$,
and one electron with one muon flavor neutrinos, $|\nu_e\nu_\mu\rangle$,
in the two-flavor and three-flavor frameworks.
Evolving these 
states
forward using $\hat U_2 (t,0)$ in Eq.~(\ref{eq:NeutHt}) provides (pure-state) wavefunctions at some later time, from which the flavor composition, entanglement and magic are computed.
Normalizing the magic in the wavefunction with respect to the maximum possible magic, gives the curves shown in Figs.~\ref{fig:2magdivmag_ee} and ~\ref{fig:2magdivmag_eu}.

Fluctuations in magic in the three-flavor system are significantly smaller than in the two-flavor system, but are consistent with each other.
Both systems have stabilized with regard to their overall behavior for $\kappa t \gtrsim 600$, for which 
the maximum values of magic are 0.871 (two flavors) and 1.491 (three flavors).
Interestingly, the magic in the $|\nu_e\nu_e\rangle$ systems 
decrease (on average) as the neutrinos move outward, 
while the magic in the $|\nu_e\nu_\mu\rangle$ systems do not show this trend.

Generalizing the analysis to the evolution of multi-neutrino systems is straightforward.
An initial tensor-product state of selected three-flavor structure is evolved forward in time using the evolution operator in Eq.~(\ref{eq:NeutHt}).
For a system of $N_\nu$ neutrinos, 
the magic is computed by evaluating forward matrix elements 
$c_P (t)$, defined above.
The evolution of $\mathcal{M}_2$ as a function of time,
computed using Eq.~(\ref{eq:SREM2}),
is observed to stabilize after 
$\kappa t \gtrsim 800$, 
and its asymptotic value is determined by averaging over a time interval at much later times.
\begin{figure}[!ht]
    \centering
    \includegraphics[width=0.9\columnwidth]{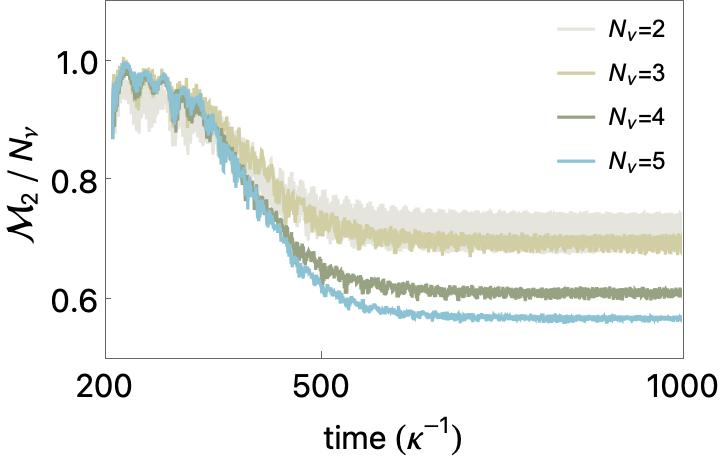}    
    \caption{$\mathcal{M}_2$ per neutrino in systems initially in tensor-product states of $|\nu_e\rangle^{\otimes N_\nu}$ only }
    \label{fig:magdensitytime_1}
\end{figure}

\begin{figure}[!ht]
    \centering
    \includegraphics[width=0.9\columnwidth]{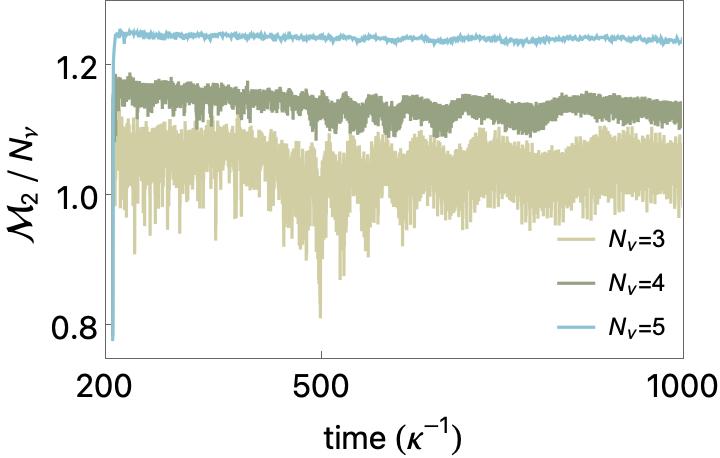}    
    \caption{$\mathcal{M}_2$ per neutrino in systems initially in tensor-products of all three 
    $|\nu_e\rangle$, $|\nu_\mu\rangle$, $|\nu_\tau\rangle$ (lower curves), as a function of time.
    Initial states with
    the maximum asymptotic values of $\mathcal{M}_2$ from the possible flavor combinations  for a given $N_\nu$ are shown, i.e.,  
    $|\nu_e\nu_\mu\nu_\tau\rangle$,
    $|\nu_e\nu_\mu\nu_\tau\nu_\tau\rangle$ and
    $|\nu_\tau\nu_\mu\nu_e\nu_\tau\nu_\mu\rangle$.
        }
    \label{fig:magdensitytime_2}
\end{figure}
\begin{figure}[!ht]
    \centering
    \includegraphics[width=0.9\columnwidth]{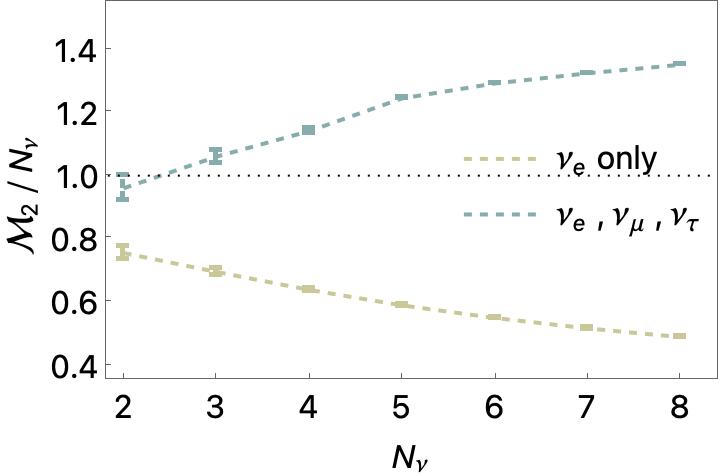}    
    \caption{The asymptotic values of  
    $\mathcal{M}_2$
    per neutrino in systems initially in a tensor-product state of 
    $|\nu_e\rangle^{\otimes N_\nu}$ 
    (brown points and dashed curve) and in systems initially in tensor-products of all three 
    $|\nu_e\rangle$, $|\nu_\mu\rangle$, $|\nu_\tau\rangle$ (blue points and dashed curve).
    The maximum value of $\mathcal{M}_2$ from the possible flavor combinations of the initial state for a given $N_\nu$ has been chosen.
    The horizontal-dotted-black line corresponds to the maximum value attainable with tensor-product states. 
    The numerical values of the displayed results 
    are given in Tables~\ref{tab:MagicPerN_part1} and~\ref{tab:MagicPerN_part2} of App.~\ref{app:results}.
    }
    \label{fig:magdensity}
\end{figure}

The time dependencies of $\mathcal{M}_2$ for 
systems with $N_\nu\le 5$ are shown in Figs.~\ref{fig:magdensitytime_1} and ~\ref{fig:magdensitytime_2}.
Interestingly, 
the wavefunctions of the $|\nu_e\rangle^{\otimes N_\nu}$ 
initial states contain less magic than the maximum possible for a  
tensor-product state, $\mathcal{M}_2\le N_\nu$, at all times.
Further, the asymptotic values are decreasing with increasing $N_\nu$.
In contrast, wavefunctions from initial states containing all three flavors support magic 
that exceeds the maximum value in tensor-product states, and hence necessarily requires entanglement between the neutrinos.
In addition, the $\mathcal{M}_2$ per neutrino is increasing with increasing numbers of neutrinos,
as displayed in Fig.~\ref{fig:magdensity} for $N_\nu\le 8$.
See App.~\ref{app:results} for the asymptotic values of $\mathcal{M}_2$ from a selection of initial states.
The evolution of the probabilities of being in the mass eigenstates,
the concurrence and generalized-concurrence, and the 
2- and 4-tangles in the wavefunctions,
are displayed 
for $|\nu_e\rangle^{\otimes 5}$ and 
$|\nu_e\nu_e\nu_\mu\nu_\mu\nu_\tau\rangle $, as examples in the $N_\nu=5$ systems, 
in App.~\ref{app:Obs}.

With the recent advances in better understanding the roles of magic and entanglement in the computational complexity of many-body systems,
this work represents a step toward quantifying the magic in dense systems of neutrinos.
The combination of large-scale entanglement and large measures of magic are both necessary 
to conclude that quantum resources are required to prepare a state.
The results that we have obtained (with the small numbers of neutrinos considered) here 
build upon previous results to further suggest that 
quantum resources will be required to prepare and evolve 
systems of dense neutrinos
due to the scaling of the magic
in the mixed-flavor channels.
Quantifying the behavior of magic and multi-partite entanglement in larger systems of neutrinos is an important next step.
However, this is only part of the challenge that lies ahead in describing these systems.  
Combining these quantum aspects of the system into realistic simulations, 
including scattering processes and full kinetics, remains to be accomplished.  
Thus, the full impact of observations made here remain to be determined.

In a broader context, there are indications that the parameters defining the Standard Model 
are such that the interactions are near extremal points in their entanglement power~\cite{cervera-lierta:2017tdt,beane:2018oxh}, 
related to emergent symmetries~\cite{beane:2018oxh,beane:2021zvo,PhysRevC.107.025204,liu2023hints,Miller:2023ujx}, and connected to flavor structures~\cite{Thaler:2024anb}.
The present work, along with what is already known about magic in strongly-interacting systems~\cite{Robin:2024bdz,Brokemeier:2024lhq}, 
is highlighting their connection to the 
computing resources required for simulating 
systems of fundamental particles.

\begin{subappendices}

\section{Stabilizer States}
\label{app:Stabs}
\noindent
Stabilizer states can be generated by repeated applications of 
the classical 
gate
set on a tensor-product state, or another stabilizer state.
The classical gate set can be defined in terms of the Hadamard gate, $H$, 
the phase gate, $S$, 
and CNOT gates.
The number of stabilizer states can be computed exactly for a given number of qudits~\cite{aaronson_2004,Gross_2006,10.5555/2638682.2638691}:
${\bf d}({\bf d}+1)$ for 1 qudit, 
${\bf d}^2({\bf d}+1)({\bf d}^2+1)$ for 2 qudits, 
${\bf d}^3({\bf d}+1)({\bf d}^2+1)({\bf d}^3+1)$ for 3 qudits, and so forth. Thus there are
6, 60, 1080, $\cdots$ stabilizer states for qubits (${\bf d}=2$), and 
12, 360, 30240, $\cdots$ stabilizer states for qutrits (${\bf d}=3$).

The single-qubit H-gate and S-gate, and the two-qubit CNOT$_{ij}$-gates 
(a two-qubit control-X entangling gate where $i$ denotes the control qubit and $j$ the target qubit), 
are given by, for example,
\begin{eqnarray}
{\rm H} & = & 
{1\over\sqrt{2}}\ \left(
\begin{array}{cc}
1&1\\ 1&-1
\end{array}
\right)
\ \ ,\ \ 
{\rm S}\ =\ 
\left(
\begin{array}{cc}
1&0\\ 0&i
\end{array}
\right)
\ \ ,\ \ 
{\rm CNOT}_{12} \ =\ 
\left(
\begin{array}{cccc}
1&0&0&0\\ 
0&1&0&0\\ 
0&0&0&1\\ 
0&0&1&0 
\end{array}
\right)
\ \ \ .
\end{eqnarray}
Generalizing to qutrits, the 
single-qutrit H-gate and S-gate, 
and the two-qutrit CNOT$_{ij}$-gates 
can be given by,
\begin{eqnarray}
\hat H & = & 
{1\over\sqrt{3}}
\left(
\begin{array}{ccc}
 1 & 1 & 1 \\
 1 & \omega & \omega^2 \\
 1 & \omega^2& \omega \\
\end{array}
\right)\ ,\ 
\hat S\ =\ 
\left(
\begin{array}{ccc}
 1 & 0 & 0 \\
 0 & 1 & 0 \\
 0 & 0 & \omega \\
\end{array}
\right)
\ \ ,\ \ 
\omega\ =\ e^{i 2\pi/3}
\ ,
\label{eq:qutrit1Cliff}
\end{eqnarray}
and
\begin{eqnarray}
{\rm CNOT}_{12} |a,b\rangle & = & |a,a+b\  {\rm mod}(3)\rangle
\ \ .
\end{eqnarray}
The latter is implemented using projectors and shift operators, as in the case of qubits.
For example, ${\rm CNOT}_{12}$ has matrix representation
\begin{eqnarray}
{\rm CNOT}_{12} & = & 
\hat\Lambda_0\otimes \hat I_3
\ +\ 
\hat\Lambda_1\otimes \hat R_1
\ +\ 
\hat\Lambda_2\otimes \hat R_2
\ \ ,
\nonumber\\
\hat\Lambda_0 & = & 
\left(
\begin{array}{ccc}
1&0&0 \\0&0&0 \\0&0&0 
\end{array}
\right)
\ ,\ 
\hat\Lambda_1\ =\  
\left(
\begin{array}{ccc}
0&0&0 \\0&1&0 \\0&0&0 
\end{array}
\right)
\ ,\  
\hat\Lambda_2\ =\  
\left(
\begin{array}{ccc}
0&0&0 \\0&0&0 \\0&0&1 
\end{array}
\right)
\ ,\
\nonumber\\
\hat R_1 & = & 
\left(
\begin{array}{ccc}
0&0&1 \\1&0&0 \\0&1&0 
\end{array}
\right)
\ ,\ 
\hat R_2 \ =\  
\left(
\begin{array}{ccc}
0&1&0 \\0&0&1 \\1&0&0 
\end{array}
\right)
\ \ .
\end{eqnarray}

A universal quantum gate set can be formed by including $T$-gates, which for qubits and qutrits are, respectively,
\begin{eqnarray}
\hat T_2 & = &   
\left(
\begin{array}{cc}
1&0 \\0&e^{i \pi/4}
\end{array}
\right)
\ \ ,\ \ 
\hat T_3 \ =\    
\left(
\begin{array}{ccc}
 1 & 0 & 0 \\
 0 & e^{\frac{2 i \pi }{9}} & 0 \\
 0 & 0 & e^{-\frac{2 i \pi }{9}} \\
\end{array}
\right)
\ \ .
\end{eqnarray}

The single-qubit stabilizer states are,
\begin{eqnarray}
\{\ 
(1,0) ,\  (0,1)
 , \ {1\over\sqrt{2}} (1,1)
 ,\  {1\over\sqrt{2}} (1,-1)
 , \ {1\over\sqrt{2}} (1,i)
 ,\  {1\over\sqrt{2}} (1,-i)
 \ \}
\  ,
\label{eq:1quStabSta}
\end{eqnarray}
and the single-qutrit stabilizer states are, 
\begin{eqnarray}
&& 
\{\ 
(1,0,0) ,\  (0,1,0),\  (0,0,1),
\nonumber\\
 &&  
 {1\over\sqrt{3}} (1,1,1)
 ,\  {1\over\sqrt{3}} (1,1,\omega)
 ,\  {1\over\sqrt{3}} (1,1,\omega^2)
 ,\  {1\over\sqrt{3}} (1,\omega, 1)
 ,\  {1\over\sqrt{3}} (1,\omega^2,1)
 \nonumber\\
 &&  {1\over\sqrt{3}} (1,\omega,\omega), \ 
{1\over\sqrt{3}} (1,\omega,\omega^2), \ 
{1\over\sqrt{3}} (1,\omega^2,\omega), \ 
{1\over\sqrt{3}} (1,\omega^2,\omega^2)
\ \}
\ .
\end{eqnarray}
%

\section{Computing Magic in a Quantum State}
\label{app:CompMag}
\noindent
The magic in a  wavefunction encoded in qudits can be straightforwardly computed in principle,
but with the 
classical computational resources
increasing exponentially with system size.
Here, we present the established ``in-principle'' 
method for qubits   and qutrits , and which can be  
extended to arbitrary ${\bf d}$.

\subsection{Qubits}
\label{app:QubitsMag}
\noindent
To quantify the magic in a qubit-supported wavefunction, we compute the stabilizer R\'enyi entropies (SREs)~\cite{Leone:2021rzd}. 
An arbitrary density matrix can be written in terms of Pauli strings,
\begin{equation}
    \hat{\rho}  
    \ =\ 
    \frac{1}{d} \sum_{\hat P \in \widetilde{\mathcal{G}}_{n_Q}} c_P \, \hat{P} 
    \; ,
\end{equation}
where $d=2^{n_Q}$ and
$c_P = {\rm Tr} \hat\rho \hat P$.
$\widetilde{\mathcal{G}}_{n_Q}$ is the subgroup of 
the generalized Pauli group
$\mathcal{G}_{n_Q}$,
\begin{equation}
    \mathcal{G}_{n_Q} = \lbrace \varphi \, \hat{\sigma}^{(1)} \otimes \hat{\sigma}^{(2)} \otimes ... \otimes \hat{\sigma}^{(n_Q)} \rbrace \; ,
    \label{eq:Gn}
\end{equation}
where $\hat{\sigma}^{(j)} \in \lbrace \mathds{1}^{(j)}, \hat{\sigma}_x^{(j)}, \hat{\sigma}_y^{(j)}, \hat{\sigma}_z^{(j)} \rbrace$ act on qubit 
$j$ and $\varphi \in \lbrace \pm 1 , \pm i \rbrace$, 
with phases chosen to be $\varphi = +1$.
It can be shown that~\cite{Leone:2021rzd} the quantity 
\begin{equation}
    \Xi_P \equiv   \frac{c_P^2 }{d} \; ,
\end{equation}
is a probability distribution for pure states,
corresponding to the probability for $\hat{\rho}$ to be in $\hat{P}$.
If $\ket{\Psi}$ is a stabilizer state, the expansion coefficients 
$c_P = \pm 1$ for $d$ commuting Pauli strings $\hat P \in \widetilde{\mathcal{G}}_{n_Q}$, 
and $c_P = 0$ for the remaining 
$d^2-d$ strings~\cite{zhu2016clifford}. 
Therefore, $\Xi_P = 1/d$ or $0$ for a 
qubit stabilizer state, and the stabilizer $\alpha$-R\'enyi entropies~\cite{Leone:2021rzd},
\begin{equation} 
\mathcal{M}_{\alpha}(\ket{\Psi})= -\log_2 d + \frac{1}{1-\alpha} \log_2 
\left( \sum_{\hat{P} \in \widetilde{\mathcal{G}}_{n_Q}} \Xi_P^{\alpha} \right) \; ,
\label{eq:Renyi_entropy_def1}
\end{equation}
which vanish for stabilizer states,
are measures of magic in the state. 
It has been shown that $\alpha \geq 2$ SREs are magic monotones for pure states, 
in contrast to those with $\alpha < 2$~\cite{Leone:2024lfr,Haug:2023hcs}.
Three commonly utilized measures of magic from the SREs are 
\begin{eqnarray}
    {\cal M}_{lin} & = & 1 - d \sum_{\hat{P} \in \widetilde{\mathcal{G}}_{n_Q}} \Xi_P^2 
\ \ ,\ \ 
    {\cal M}_1 \ = \  -\sum_{\hat{P} \in \widetilde{\mathcal{G}}_{n_Q}}      \Xi_P\log_2 d \ \Xi_P
\ \ ,\ \ \nonumber \\
    {\cal M}_2 & \ =\ &  -\log_2 \ d \sum_{\hat{P} \in \widetilde{\mathcal{G}}_{n_Q}}  \Xi_P^2 \; .
    \label{eq:SREM012}
\end{eqnarray}
%

\subsection{Qutrits}
\label{app:QutritsMag}
\noindent
The formulation of measures of magic for qutrits is similar to that for qubits.  
Instead of using the Gell-Mann matrices to define the generators of SU(3), the generalized $\hat X$ and $\hat Z$ operators are 
more widely 
used because of their properties under tracing.
Strings of Pauli operators can be written as 
\begin{equation}
    \hat P_{i_1,i_2,..., i_{n_Q}} =  \hat{\Sigma}_{i_1} \otimes \hat{\Sigma}_{i_2} \otimes ... \otimes \hat{\Sigma}_{i_{n_Q}} 
    \ \ ,
    \label{eq:GnT}
\end{equation}
where 
the nine Pauli operators for qutrits (including the identity), written in terms of $\hat X$ and $\hat Z$ operators, are  
\begin{eqnarray}
\hat \Sigma_i & \in & \{
\hat I \ , 
\hat X \ , 
\hat Z \ , 
\hat X^2 \ , 
\omega \hat X \hat Z \ , 
\hat Z^2 \ , 
\omega^2 \hat X \hat Z^2 \ , 
\hat X^2 \hat Z \ ,
\hat X^2 \hat Z^2  
\}
\ \ ,
\label{eq:qutritPaulis}
\end{eqnarray}
with
\begin{eqnarray}
\hat X |j\rangle & = & |j+1\rangle
\ \rightarrow\ 
\left(
\begin{array}{ccc}
 0 & 0 & 1 \\
 1 & 0 & 0 \\
 0 & 1 & 0 \\
\end{array}
\right)
\ \ ,\ \ 
\hat Z |j\rangle\ =\ \omega^j |j\rangle
\ \rightarrow\ 
\left(
\begin{array}{ccc}
 1 & 0 & 0 \\
 0 & \omega & 0 \\
 0 & 0 & \omega^2\\
\end{array}
\right)
\ \ .
\label{eq:qutritXZ}
\end{eqnarray}
The Pauli operators 
in Eq.~(\ref{eq:qutritPaulis})
are normalized such that
\begin{eqnarray}
{\rm Tr}
\hat \Sigma_i  \hat \Sigma_j  & = & 3 K_{ij}
\ ,
\nonumber\\
{\rm with} &  & K_{11} =  K_{24}=K_{36}=K_{42}=K_{59}=K_{63}=K_{78}=K_{87}=K_{95}=1
\ \ ,\ \ \nonumber\\
{\rm else} &  & \ \ K_{ij}\ =\ 0
\ \ ,
\label{eq:qutritPnorm}
\end{eqnarray}
where $1+\omega+\omega^2=0$ has been used.

An arbitrary density matrix 
for a wavefunction supported on $n_Q$ qutrits
can be decomposed into sums of products of Pauli strings,
\begin{eqnarray}
\hat\rho & = & 
{1\over d}\sum_{i_a,j_b}
{\rm Tr} \left[ \hat\rho.\hat P_{i_1,i_2,..., i_{n_Q}} \right]\ 
K_{i_1,j_1}
K_{i_2,j_2}
\cdots
K_{i_{n_Q},j_{n_Q}}\ 
\hat P_{j_1,j_2,..., j_{n_Q}}
\ \ ,
\end{eqnarray}
where $d=3^{n_Q}$.
\footnote{
There is a useful relation between sums of operators
\begin{eqnarray}
\sum_{a=1}^8\ \hat T^a \otimes \hat T^a
& = & 
{2\over 3}\ \sum_{a,b=2}^9\ \hat \Sigma_a \otimes \hat \Sigma_b\  K_{a,b}
\ \ .
\end{eqnarray}
}

To determine the magic in a given pure state, 
the forward matrix elements of all Pauli strings 
are formed, $c_P \equiv \langle \Psi |\hat{P} | \Psi \rangle$.
For stabilizer states,
$d$ of the strings give $c_P = 1, \omega$ or $\omega^2$, while the other $d^2-d$ give $c_P = 0$.
However, in general, for an arbitrary state, all $d^2$ values will 
{\it a priori} be nonzero.
As is the case for qubits, described above,
we can define the deviation from stabilizerness in a given state 
as the magic, using
\begin{eqnarray}
\Xi_P & = & |c_P|^2/d
\ ,\ 
\sum_P \Xi_P\ =\ 1
\ \ .
\label{eq:XiP_appendix}
\end{eqnarray}
%

\subsection{The Magic in Entangled Versus Tensor-Product States}
\label{app:MEnE}
\noindent
It is known that entangled states can support more magic than  non-entangled states
\footnote{We thank Alioscia Hamma for making this point to us.}.
As an example,
in the case of a two-qubit system, straightforward calculations demonstrate that the maximum $\mathcal{M}_2$ that a tensor-product state can contain is 
$\mathcal{M}_2=1.16993$ (consistent with twice the value for a single two-flavor neutrino), while entangled states can contain up to 
$\mathcal{M}_2=1.19265$.
For the two-qutrit system, explicit calculation gives a maximum value of magic in a tensor-product state of 
$\mathcal{M}_2=2$ 
(consistent with $2\times$ the maximum value for a single three-flavor neutrino), 
while entangled states can support a maximum value of 
$\mathcal{M}_2=2.23379$.

\section{The One Neutrino Sector}
\label{app:OneNu}
\noindent
The neutrino flavor and mass eigenstates are related by the Pontecorvo–Maki–Nakagawa– Sakata (PMNS)
matrix~\cite{Pontecorvo1957,Maki1962},
\begin{eqnarray}
{\bm\nu}_F & = & U_{PMNS} . {\bm\nu}_M
\ \ ,
\end{eqnarray}
where 
${\bm\nu}_F=\left(\nu_e, \nu_\mu, \nu_\tau\right)^T$ and 
${\bm\nu}_M=\left(\nu_1, \nu_2, \nu_3\right)^T$ 
are the three-component vectors of neutrino fields in the flavor and mass bases, respectively.
In a common paramterization,
the PMNS mixing matrix can be written as (without Majorana phases),
\begin{multline}
    U_{PMNS} = 
    \left(
    \begin{array}{ccc}
    1&0&0\\0&\cos\theta_{23}&\sin\theta_{23}\\
    0&-\sin\theta_{23}&\cos\theta_{23}
    \end{array}
    \right)\\
    *\left(
    \begin{array}{ccc}
    \cos\theta_{13}&0&e^{-i\delta}\sin\theta_{13}\\0&1&0\\-e^{+i\delta}\sin\theta_{13}&0&\cos\theta_{13}
    \end{array}
    \right)
    \left(
    \begin{array}{ccc}
    \cos\theta_{12}&\sin\theta_{12}&0\\
    -\sin\theta_{12}&\cos\theta_{12}&0\\
    0&0&1
    \end{array}
    \right)
    \ ,
    \label{eq:UPMNS}
\end{multline}
where the experimentally determined angles are~\cite{ParticleDataGroup:2024cfk},
\begin{eqnarray}
\sin^2\theta_{12} & = & 0.307\pm 0.013
,\ 
\sin^2\theta_{23} \ =\  0.553^{+0.016}_{-0.024}
,\ 
\sin^2\theta_{13} \ =\  (2.19\pm 0.07)\times 10^{-2}
,\ \nonumber \\
\delta & = &  (1.19\pm 0.22)\ \pi\  {\rm rad}
    \ .
    \label{eq:UPMNSangs}
\end{eqnarray}
The neutrino mass-squared differences are known experimentally to be~\cite{ParticleDataGroup:2024cfk},
\begin{eqnarray}
\delta m_{21}^2 & = & (7.53 \pm 0.18) \times 10^{-17}~{\rm MeV}^2
\ ,\ \nonumber \\
\Delta m_{32}^2 & \ =\ & (2.455 \pm 0.028) \times 10^{-15}~{\rm MeV}^2  \ \ [{\rm normal}]
    \ .
    \label{eq:Dm2s}
\end{eqnarray}
We are only considering the normal hierarchy of neutrino masses and not the inverted hierarchy.
While the above mixing and masses are in the case of three neutrinos, 
the (commonly considered) effective two-neutrino sector  is found by using the $\theta_{12}$ mixing angle and 
$\delta m_{21}^2$ mass-squared difference.

With these experimental values, the mixing matrices 
for the effective three-flavor and two-flavor systems
become
\begin{multline}
U_{PMNS} \\ 
=   \left(
\begin{array}{ccc}
0.8233(77) & 0.548(12) & -0.096(57)+i 0.065(71) \\
-0.311(37) + i 0.041 (44) & 0.596 (27)+i 0.027 (29) & 0.735(13)\\
0.466(33)+i 0.036(40) & -0.583(25)+i 0.024(26) & 0.661(15)
\end{array}
\right)
\ , \\
U_{2} =  
\left(
\begin{array}{cc}
0.8324(78) & 0.554(12) \\ -0.554(12) & 0.8324(78)
\end{array}
\right)
\ , 
\end{multline}
respectively.
When evaluated at the mean values of the angles and phase, the mixing matrices are,
\begin{eqnarray}
U_{PMNS}  & = & 
    \left(
    \begin{array}{ccc}
    0.823300 & 0.547975 & -0.122396+i 0.083181 \\
    -0.294674 + i 0.051493 & 0.607002+i 0.034273 & 0.735451\\
    0.480155+i 0.046295 & -0.573713+i 0.030813 & 0.661219
    \end{array}
    \right),
\nonumber\\
    U_{2} & = & 
    \left(
    \begin{array}{cc}
    0.832466 & 0.554076 \\ -0.554076 & 0.832466
    \end{array}
    \right)
    \ , 
    \end{eqnarray}
with the slight differences (within uncertainties)
resulting from $\sin^2\langle\theta\rangle \ne \langle \sin^2\theta \rangle$.

\section{Computing the Magic Power of a Unitary Operator}
\label{app:CompMagPow}
\noindent
The magic power of a unitary operator $\hat {\bf S}$, denoted by 
$\overline{\mathcal{M}}(\hat {\bf S})$, 
is defined to be the average magic induced by the operator on all $n$-qudit stabilizer states $\ket{\Phi_i}$:
\begin{align}
    \overline{\mathcal{M}}(\hat {\bf S}) \equiv \frac{1}{\mathcal{N}_{ss}} \sum_{i=1}^{\mathcal{N}_{ss}}  \mathcal{M} \left( \hat {\bf S} \ket{\Phi_i} \right) \; ,
\label{eq:Magic_Power}
\end{align}
where $\mathcal{N}_{ss}$ denotes the total number of $n$-qudit stabilizer states. 
$\mathcal{M}$ is a measure of magic, which we define
in terms of SREs
in Eq.~(\ref{eq:SREM012}).

\subsection{The Magic Power of the Single-Neutrino Evolution Operator}
\label{app:1NuMagPow}
\noindent
The magic power of the free-space  single neutrino evolution operator is computed using Eq.~(\ref{eq:Magic_Power}).
For two flavors, it is found to be
\begin{eqnarray}
\overline{\mathcal{M}}_2(\hat U)  = 
2\left[\ 
1 - {1\over 3} \log_2 \left( 7 + \cos\left( {2 \delta m^2_{21} \over E  }   t \right)\right) 
\ \right]
\ \ ,
\label{eq:2F1nuMP}
\end{eqnarray}
and for three flavors
\begin{multline}
\overline{\mathcal{M}}_2(\hat U) 
 =  
-{3\over 4}
\log_2\Bigg[ 
{1\over 81} \Bigg(
57
+ 8\cos\left( {3 \delta m^2_{21} \over E  }   t \right) \\
+ 8\cos\left( {3 \Delta m^2_{31} \over E  }   t \right) 
+ 8\cos\left( {3 (\Delta m^2_{31}-\delta m^2_{21}) \over E  }   t \right)
\Bigg)
\Bigg]
\ \ .
\label{eq:3F1nuMP}
\end{multline}
%

\section{Tables of Results}
\label{app:results}
\noindent
In this section, we provide tables of results displayed in figures in the main text.
\begin{table}[!t]
\centering
\renewcommand{\arraystretch}{1.4}

\begin{tabular}{| c | c | c | c | }
\hline
$N_\nu$
& Method
&  Asymp. $\mathcal{M}_2$  for $|\nu_e\rangle^{\otimes N_\nu}$       
&  Asymp. $\mathcal{M}_2$ for $|\nu_e\rangle, |\nu_\mu\rangle, |\nu_\tau\rangle $ \\
\hline
2 & Trotterized,   & 0.755(19)
& 0.97(5) ($e\mu$) 
\\
 &  $\Delta \kappa t=0.05$  &  &  0.96(4) ($e\tau$) 
\\
\hline 
2 & Numerical ODE  & 0.755(19) 
& 0.97(4) ($e\mu$); 0.97(3) ($e\tau$)
\\
\hline 
\hline 
3 & Trotterized,  
& 0.694(10)
& 1.06(2) ($e\mu\tau$)  \\
 & $\Delta \kappa t=0.05$ &   &   \\
\hline 
3 & Numerical ODE   & 0.695(10) & 1.06(2) ($e\mu\tau$) \\
\hline 
\hline 
4 & Trotterized,  & 0.637(5)  
& 1.125(7) ($e\mu\mu\tau$)  \\
 & $\Delta \kappa t=0.05$ &   &  1.139(8) ($e\mu\tau\tau$)  \\
\hline 
4 & Numerical ODE & 0.638(5)  
& 1.120(8) ($e\mu\mu\tau$) \\
 &  &   &  1.140(9) ($e\mu\tau\tau$)  \\
\hline
\hline 
5 &   Trotterized,  &  0.589(3)
&  1.133(6)  ($ee\mu\mu\tau$)  \\
 & $\Delta \kappa t=0.05$  &  
 &   1.154(3) ($ee\mu\tau\tau$) \\
 &   &  
 &   1.243(2)  ($\tau\mu e \tau\mu$) \\
\hline
 \hline
\end{tabular}
\caption{
The asymptotic magic per neutrino for select initial states,
as displayed in Fig.~4 of the main text, for $N_\nu$ up to 5.  
The third and fourth column headers denote the flavor composition of the initial states, i.e., either all electron-type, or a mix of all three flavors. 
The ``Numerical ODE solutions'' were performed using  $9^{th}$ order lazy and $4^{th}$ order stiffness-aware interpolation  and Tolerances: $10^{-8}$ absolute, $10^{-8}$ relative.
}
\label{tab:MagicPerN_part1}
\end{table}

\begin{table}[!t]
\centering
\renewcommand{\arraystretch}{1.4}

\begin{tabular}{| c | c | c | c | }
\hline
$N_\nu$
& Method
&  Asymp. $\mathcal{M}_2$  for $|\nu_e\rangle^{\otimes N_\nu}$      
&  Asymp. $\mathcal{M}_2$ for $|\nu_e\rangle, |\nu_\mu\rangle, |\nu_\tau\rangle $ \\
\hline 
6 &  Trotterized,  &  0.548(2) 
 &   1.236(2) ($\tau\mu\tau\mu\tau e$) \\
   &   $\Delta \kappa t=0.05$ &   
 &   1.265(1)  ($e\mu\tau e\mu\tau$) \\
  &   &  
 &   1.280(1) ($\mu \tau e  \mu \tau \mu$) \\
   &   &  
 &   1.292(1) ($\tau \mu e \tau \mu \tau $) \\
 \hline
\hline 
 7 &   Trotterized,  
 &  0.516(2)
 &  1.3072(3) ($\tau\tau e\tau\tau\tau\tau$) \\
 &   $\Delta \kappa t=0.05$ &
 &  1.3163(3) ($\tau \mu e \tau e \tau \mu$) \\
 &   &
 & 1.3190(4) ($\tau \mu e \tau \mu e \tau$) \\
 &   &
 & 1.3243(2) ($\tau \mu e \tau \mu \tau \mu$)  \\
\hline 
 \hline
 8 &  Trotterized,
 & 0.488(1)
 &  1.3292(2) ($e \mu \tau e \mu \tau e \mu$) \\
 &   $\Delta \kappa t=0.05$  &   
 &  1.3460(1) ($\tau \mu e \tau \mu \tau \mu \tau $) \\
 &  &   
 &  1.35126(7) ($\tau \mu e \tau \mu e \tau \mu$) \\
 \hline
\end{tabular}
\caption{
The asymptotic magic per neutrino for select initial states,
as displayed in Fig.~4 of the main text, for $N_\nu$ values of 6 to 8.  
The third and fourth column headers denote the flavor composition of the initial states, i.e., either all electron-type, or a mix of all three flavors. 
The ``Numerical ODE solutions'' were performed using  $9^{th}$ order lazy and $4^{th}$ order stiffness-aware interpolation  and Tolerances: $10^{-8}$ absolute, $10^{-8}$ relative.
}
\label{tab:MagicPerN_part2}
\end{table}

\section{The Evolution of Select Quantities}
\label{app:Obs}
\noindent
To illustrate the general behavior of the evolution of three-flavor neutrino systems, we present 
results for the probabilities, 
$\mathcal{M}_2$,
concurrence, generalized-concurrence, the $2$-tangle and $4$-tangle,
in systems resulting from initial states of $|\nu_e\rangle^{\otimes 5} $ 
and, as an example of mixed-flavor state,
$|\nu_e \nu_e \nu_\mu \nu_\mu \nu_\tau\rangle $.

The probabilities are found from projections of each of the neutrinos onto the mass eigenstates as a function of time.  For a system of $N_\nu$ neutrinos, this gives rise to $3 N_\nu$ curves evolving from just three values at the initial time.
The concurrence and generalized-concurrence are found by forming the 
single-neutrino reduced-density matrix for each neutrino in the state, $\hat\rho_i$, and computing its eigenvalues, $\lambda_{i1,i2,i3}$.
The concurrence for each $\hat\rho_i$ is determined by four times the sum of products of two eigenvalues, while the generalized-concurrence is the product of the three eigenvalues.  These are then summed over each of the neutrinos, i.e.,
\begin{eqnarray}
    C & = & 4 \sum_i 
    \left(\lambda_{i1}\lambda_{i2}+\lambda_{i1}\lambda_{i3}+\lambda_{i2}\lambda_{i3}\right)
    \ \ ,\ \ 
    G\ =\ \sum_i \lambda_{i1}\lambda_{i2}\lambda_{i3}
     \ \ \ .
\end{eqnarray}

The $n$-tangles are formed from matrix elements of $n$ insertions of the SO(3) generators~\cite{Chen_2012}, 
$\hat J_i^n$, where,
\begin{eqnarray}
    J_1 & = & 
    \left(
    \begin{array}{ccc}
    0&0&0 \\   0&0&-i \\ 0&i&0
    \end{array}
    \right)
    \ \ ,\ \ 
        J_2 \ =\  
    \left(
    \begin{array}{ccc}
    0&0&i \\   0&0&0 \\ -i&0&0
    \end{array}
    \right)
        J_3 \ =\  
    \left(
    \begin{array}{ccc}
    0&-i&0 \\   i&0&0 \\ 0&0&0
    \end{array}
    \right)
\ \ \ ,
\label{eq:so3gens}
\end{eqnarray}
and averaging over the squared-magnitude, i.e., 
\begin{eqnarray}
    \tau_4 & = {1\over {\cal N}_4} & \sum_i\ \sum_{a\ne b\ne c \ne d}\ 
    |\langle\psi|\ \hat J_{i,a}  \hat J_{i,b}  \hat J_{i,c}  \hat J_{i,d} |\psi\rangle|^2
    \ ,
\end{eqnarray}
where  ${\cal N}_4$ is the number of contributions to the sum.
This is the generalization of the $n$-tangles for two-flavor systems.

\subsection{Initially $|\nu_e\rangle^{\otimes 5} $}
\label{app:5e}
\begin{figure}[!ht]
    \centering
    \includegraphics[width=0.45\columnwidth]{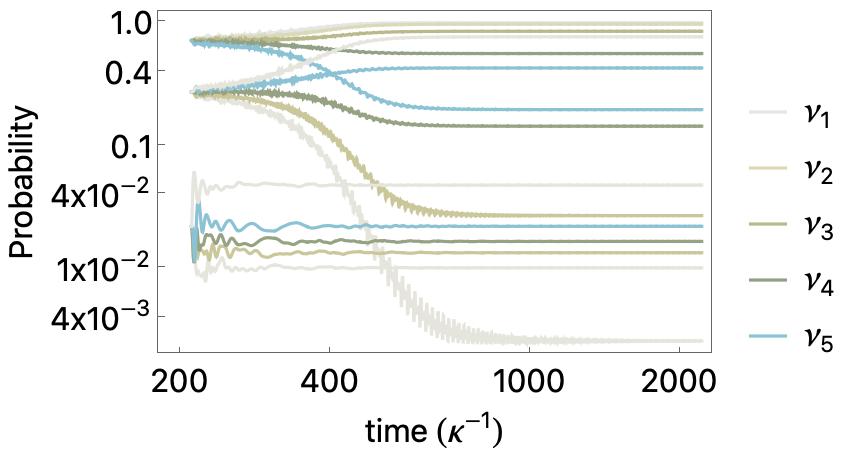}
    \includegraphics[width=0.45\columnwidth]{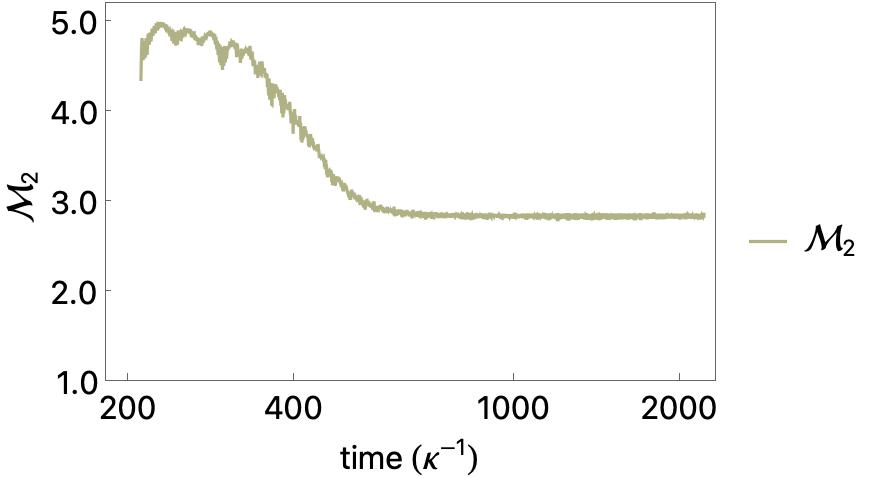}
    \caption{
    The left panel shows the probabilities of neutrinos initially in the $|\nu_e\rangle^{\otimes 5}$ state evolving into one of the three mass eigenstates, 
    while the right panel shows the evolution of the magic 
    $\mathcal{M}_2$. 
    }
    \label{fig:5nueMASS}
\end{figure}

\begin{figure}[!ht]
    \centering
    \includegraphics[width=0.49\columnwidth]{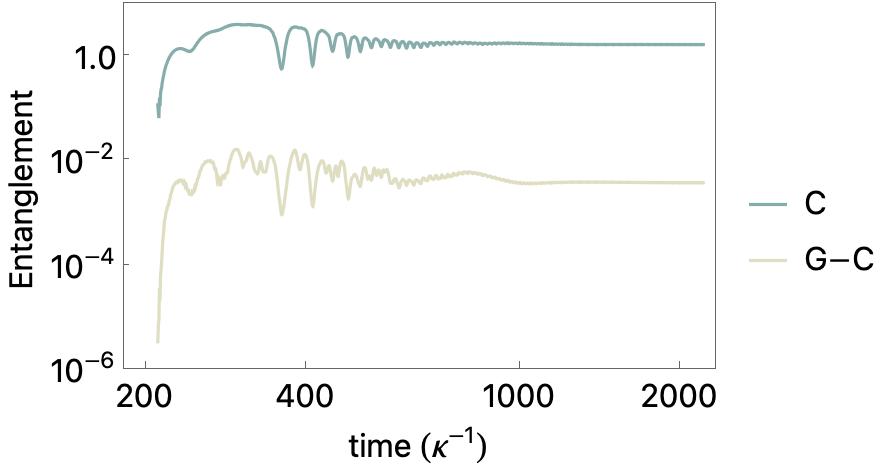}
    \includegraphics[width=0.49\columnwidth]{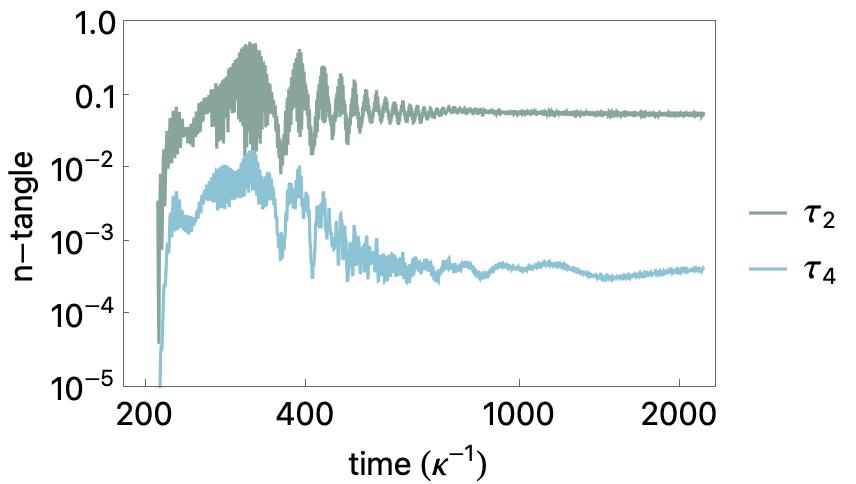}
    \caption{
    The left panel shows the sum of the 
    concurrence (C) and generalized-concurrence (G-C)
    of neutrinos initially in the $|\nu_e\rangle^{\otimes 5}$ state evolving into the three mass eigenstates, while the right panel shows the evolution of the $2$-tangle 
    $\tau_2$ and $4$-tangle $\tau_4$. 
    }
    \label{fig:5nueEntagle}
\end{figure}

\noindent
Figure~\ref{fig:5nueMASS} displays the probabilities of being in one of 
the three mass eigenstates and the magic 
as a function of time starting from an initial state of $|\nu_e\rangle^{\otimes 5}$,
while Fig.~\ref{fig:5nueEntagle} displays the concurrence, generalized-concurrence, $\tau_2$ and $\tau_4$.
It can be observed that while the eigenstate projections and magic appear 
to approach asymptotic values, the concurrences and $n$-tangles approach appears to be somewhat delayed.

\subsection{Initially $|\nu_e \nu_e \nu_\mu \nu_\mu \nu_\tau\rangle $}
\label{app:eemumutau}
\noindent
Here we display the corresponding results for an initial state of $|\nu_e \nu_e \nu_\mu \nu_\mu \nu_\tau\rangle $.
\begin{figure}[!ht]
    \centering
    \includegraphics[width=0.49\columnwidth]{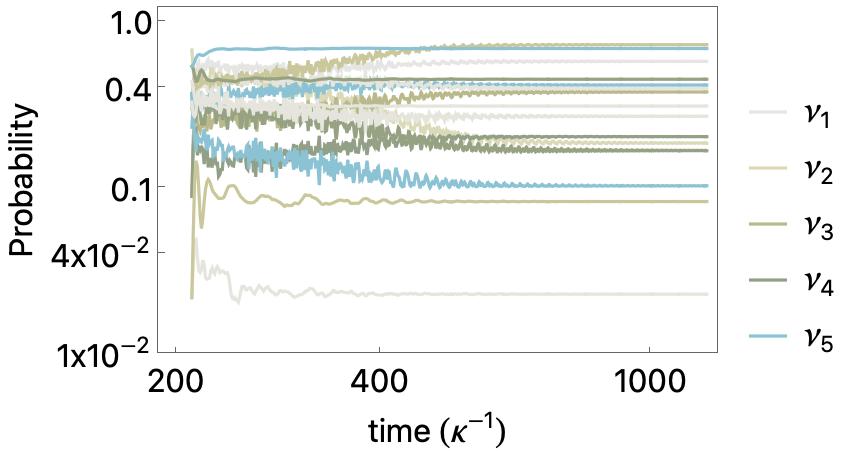}
    \includegraphics[width=0.49\columnwidth]{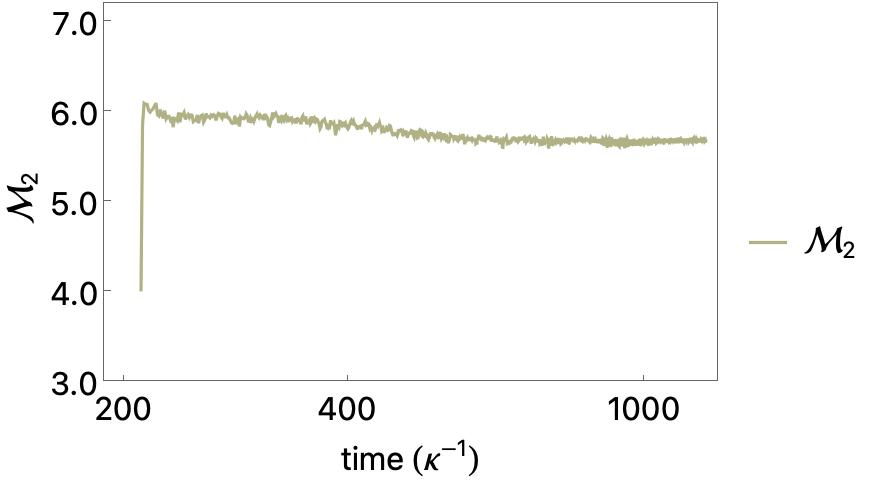}
    \caption{
    The left panel shows the probabilities of neutrinos initially in the 
    $|\nu_e \nu_e \nu_\mu \nu_\mu \nu_\tau \rangle$ state evolving into one of 
    the three mass eigenstates, while the right panel shows the evolution of the magic 
    $\mathcal{M}_2$. 
    The results were generated with a Trotter time interval of $\kappa \Delta t=0.05$, 
    and sampled every 20 time steps for display purposes.
    }
    \label{fig:5nueMASS_all3flavors}
\end{figure}
The probability of being in a mass eigenstate exhibits quite different behavior when compared with a
$|\nu_e\rangle^{\otimes 5}$ initial state.   This is also the case for $\mathcal{M}_2$, which rapidly rises to its maximum value and stays approximately near this value throughout the evolution.
The value of $\mathcal{M}_2\sim 6$ is noticeably larger than the maximum magic that a 
tensor-product state of $N_\nu=5$ can support, and thus the two-neutrino interactions are generating magic in the multi-neutrino systems.
\begin{figure}[!ht]
    \centering
    \includegraphics[width=0.49\columnwidth]{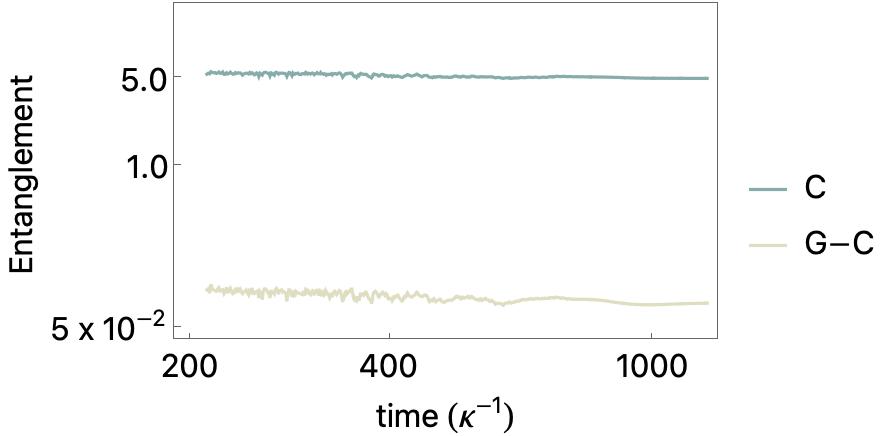}
    \includegraphics[width=0.49\columnwidth]{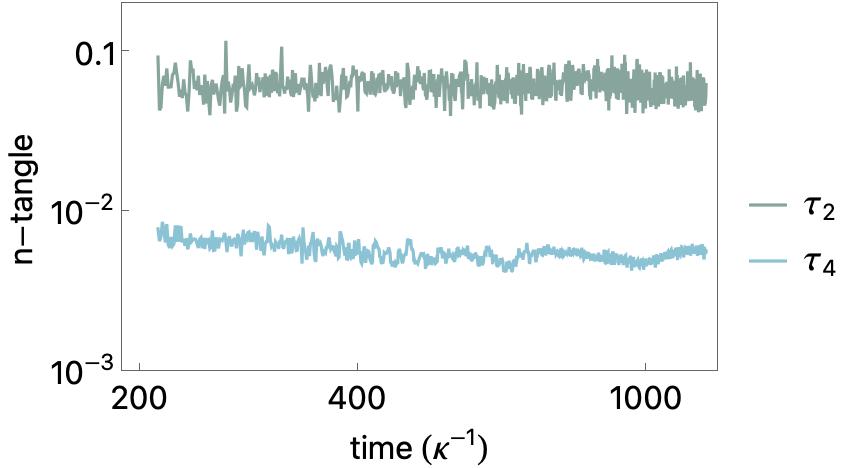}
    \caption{
    The left panel shows the sum of the 
    concurrence (C) and generalized-concurrence (G-C)
    of neutrinos initially in the $|\nu_e \nu_e \nu_\mu \nu_\mu \nu_\tau \rangle $  state evolving into the three mass eigenstates, while the right panel shows the evolution of the $2$-tangle 
    $\tau_2$ and $4$-tangle $\tau_4$. 
    The results were generated with a Trotter time interval of $\kappa \Delta t=0.05$, 
    and sampled every 20 time steps for display purposes.
    }
    \label{fig:5nueEntagle_all3flavors}
\end{figure}
The concurrence and generalized-concurrence exhibit a similar behavior and have values that are substantially larger than for the $|\nu_e\rangle^{\otimes 5}$ initial state. 
While $\tau_2$ behaves differently with time, its asymptotic value is similar.  In contrast, $\tau_4$ is substantially larger asymptotically.


\end{subappendices}

\chapter{Conclusion}


The work enclosed in this thesis has expanded the capabilities of state-preparation and time-evolution of non-Abelian lattice gauge theory and related problems from demonstrations on and extrapolations from the smallest lattices to larger-scale lattices and efficient, scalable algorithms that can begin to extract physically significant observables. It has also spotlighted opportunities within the physics where quantum simulation has particularly high potential.




Chapter \ref{chap:prepofsu3latyangmilsvacvarqumethods} favors gradient descent as the method of choice for optimizing the parameters of a VQE circuit, a result reflected in the fact that algorithms such as ADAPT-VQE \cite{grimsley_2019} and SC-ADAPT-VQE \cite{Farrell:2023fgd} select which operators they build their circuits out of using . In Chapter \ref{chap:1p1dQCD}, circuits based on those from Refs. \cite{stetina:2020abi, qchem2014, klco:2019xro} are developed and benchmarked, with the goal of laying a foundation for future simulation on quantum devices of problems that involve SU(3) gauge theory. Quantum circuits for two such problems, $\beta$-decay and $0 \nu \beta \beta$ decay, are devised in Chapters \ref{chap:1p1dSM} and \ref{chap:0vBBandlargeQCD} and the former has already been implemented on Quantinuum’s {\tt H1-1} device. Additionally, in Sec. \ref{sec:0vBB_timeevcircforsupercond}, circuits were devised that could accomplish the Trotterization of the SU(3) Hamiltonian on nearest-neighbor devices with minimal overhead. These circuits could be used in the near future to leverage the higher qubit-counts of superconducting devices. Chapter \ref{chap:qutritqubitneutrino} presents successful results for real-time flavor dynamics of 3-flavor collecive oscillations from the {\tt H1-1} and {\tt ibm\_torino} devices, as well as a qutrit Trotterization circuit with significantly less complexity than its qubit-counterpart. Additionally, the implementation of an all-to-all interaction by using the two-neutrino rotation itself to perform a SWAP can potentially be replicated on 1+1D QCD, as many of the circuits in Sec. \ref{sec:0vBB_timeevcircforsupercond} can incorporate SWAPs into their procedure at low cost. Finally, in Chapter \ref{chap:neutrinomagic}, a study of magic in collective neutrino oscillations produces results that, in the context of recent literature \cite{cervera-lierta:2017tdt, beane:2018oxh, beane:2021zvo, PhysRevC.107.025204, liu2023hints, Miller:2023ujx, Thaler:2024anb, Brokemeier:2024lhq, Robin:2024bdz} suggest that due to traits inherent to the physics itself, there are problems within the Standard Model for which there is quantum advantage \cite{chernyshev2024quantum}, and that collective oscillations of neutrinos initially in a maximal mix of neutrino flavors is a promising candidate.

Many of these developments are enabled by advances in hardware. When I began conducting research as part of the InQubator for Quantum Simulation (IQuS), quantum devices were restricted to approximately 5-30 qubits and could handle circuits of up to a depth of approximately 30 CNOT gates. Today, superconducting devices with 100-156 qubits \cite{ibmq} that can execute circuits with circuit depths of 100-430 2-qubit gates, depending on the error mitigation methods used and the resilience-to-error of the desired observables \cite{Farrell:2023fgd, Farrell:2024fit, ciavarella2024string, Ciavarella:2024fzw, zemlevskiy2024scalable, chowdhury2024enhancing, shinjo2024unveiling, kavaki2024square} and trapped-ion devices with 20-36 qubits \cite{quantinuum, IonQ} that can execute circuits with 2-qubit gate counts ranging from 400 to a little over 2000 \cite{physrevd.107.054513, moses2023race, turro2024qutrit}. These improvements to hardware have been complemented by developments in quantum algorithms. For instance, when I joined IQuS in early 2021, the state of the art of the Variational Quantum Eigensolver (VQE) for applications to nuclear and particle physics was a circuit that could access all real states within the circuit's Hilbert space, subject to the system's symmetries \cite{klco:2019xro, atas:2021ext, ciavarella:2021lel, physrevd.107.054512}. This is an expensive and not scalable implementation, as its gate count and circuit depth would generally scale exponentially with the number of qubits. The development of ADAPT-VQE \cite{grimsley_2019} and SC-ADAPT-VQE \cite{Farrell:2023fgd} has reduced circuit depth and gate-count requirements and enabled general-case scalability, proving invaluable to the construction of the circuits in Chapter \ref{chap:0vBBandlargeQCD}. I hope that the results in this thesis will provide similar benefit to other researchers in the field in the near future.

One straightforward next step is to extend these studies to SU(3) 2+1D lattice gauge theory with fermions, given that the ultimate goal is 3+1D QCD on quantum devices. In this process, new methods will be needed to encode the interactions involving gauge degrees of freedom that could no longer be removed through gauge-fixing.
Another potential direction to take is to implement the circuits from Sec. \ref{sec:0vBB_timeevcircforsupercond} on superconducting quantum devices in order to simulate hadron dynamics, in a similar vein to the work in Refs. \cite{Farrell:2024fit, zemlevskiy2024scalable}.

%
%
\bibliographystyle{plain}
\bibliography{uwthesis}
%
%
\appendix
\raggedbottom\sloppy

\end{document}